\RequirePackage{silence} % :-\
\documentclass[ twoside,openright,titlepage,numbers=noenddot,%1headlines,
                headinclude,footinclude,cleardoublepage=empty,abstract=on,
                BCOR=5mm,paper=a4,fontsize=11pt,
                ]{scrreprt}

\PassOptionsToPackage{utf8}{inputenc}
  \usepackage{inputenc}

\PassOptionsToPackage{T1}{fontenc} % T2A for cyrillics
  \usepackage{fontenc}

\PassOptionsToPackage{
  drafting=false,    % print version information on the bottom of the pages
  tocaligned=false, % the left column of the toc will be aligned (no indentation)
  dottedtoc=true,  % page numbers in ToC flushed right
  eulerchapternumbers=true, % use AMS Euler for chapter font (otherwise Palatino)
  linedheaders=true,       % chaper headers will have line above and beneath
  floatperchapter=true,     % numbering per chapter for all floats (i.e., Figure 1.1)
  eulermath=false,  % use awesome Euler fonts for mathematical formulae (only with pdfLaTeX)
  beramono=true,    % toggle a nice monospaced font (w/ bold)
  palatino=true,    % deactivate standard font for loading another one, see the last section at the end of this file for suggestions
  style=classicthesis % classicthesis, arsclassica
}{classicthesis}

\newcommand{\myTitle}{The Gravity of Classical Fields\xspace}
\newcommand{\mySubtitle}{And Its Effect on the Dynamics of Gravitational Systems\xspace}

\newcommand{\myName}{Rodrigo Luís Lourenço Vicente\xspace}

\newcommand{\myFaculty}{Put data here\xspace}

\newcommand{\myUni}{Instituto Superior Técnico\xspace}

\newcommand{\myTime}{September 2021\xspace}
\newcommand{\myVersion}{draft v1}

\providecommand{\mLyX}{L\kern-.1667em\lower.25em\hbox{Y}\kern-.125emX\@}
\newcommand{\ie}{i.\,e.}

\newcommand{\eg}{e.\,g.}

\PassOptionsToPackage{ngerman,american}{babel} % change this to your language(s), main language last
    \usepackage{babel}

\usepackage{csquotes}
\PassOptionsToPackage{%
  backend=bibtex8,bibencoding=ascii,%
  language=auto,%
  style=numeric-comp,%
  sorting=none, % name, year, title
  maxbibnames=10, % default: 3, et al.
  natbib=true % natbib compatibility mode (\citep and \citet still work)
}{biblatex}
    \usepackage{biblatex}

\PassOptionsToPackage{fleqn}{amsmath}       % math environments and more by the AMS
  \usepackage{amsmath,amssymb,lmodern,amsthm}

\usepackage{graphicx,psfrag,mathrsfs} %
\usepackage{scrhack} % fix warnings when using KOMA with listings package
\usepackage{xspace} % to get the spacing after macros right
\PassOptionsToPackage{printonlyused,smaller}{acronym}
  \usepackage{acronym} % nice macros for handling all acronyms in the thesis
\usepackage{tabularx} % better tables
\usepackage{subfig}
\usepackage{listings}
\usepackage{classicthesis}

\hypersetup{%
  colorlinks=true, linktocpage=true, pdfstartpage=3, pdfstartview=FitV,%
  breaklinks=true, pageanchor=true,%
  pdfpagemode=UseNone, %
  plainpages=false, bookmarksnumbered, bookmarksopen=true, bookmarksopenlevel=1,%
  hypertexnames=true, pdfhighlight=/O,%nesting=true,%frenchlinks,%
  urlcolor=CTurl, linkcolor=CTlink, citecolor=CTcitation, %pagecolor=RoyalBlue,%
  pdftitle={\myTitle},%
  pdfauthor={\textcopyright\ \myName, \myUni, \myFaculty},%
  pdfsubject={},%
  pdfkeywords={},%
  pdfcreator={pdfLaTeX},%
  pdfproducer={LaTeX with hyperref and classicthesis}%
}

\makeatletter
\@ifpackageloaded{babel}%
  {%
    \addto\extrasamerican{%
    }%
    \addto\extrasngerman{%
    }%
    }{\relax}
\makeatother

\begin{document}
\frenchspacing
\raggedbottom
\selectlanguage{american} % american ngerman
%\renewcommand*{\bibname}{new name}
%\setbibpreamble{}
\pagenumbering{roman}
\pagestyle{plain}
%********************************************************************
% Frontmatter
%*******************************************************
%*******************************************************
% Little Dirty Titlepage
%*******************************************************
\thispagestyle{empty}
\pdfbookmark[1]{Title}{title}
%*******************************************************
%\begin{center}
%    \spacedlowsmallcaps{\myName} \\ \medskip
%
%    \begingroup
%        \color{CTtitle}\spacedallcaps{\myTitle}
%    \endgroup
%
%
%\end{center}

\begin{addmargin}[-1cm]{-3cm}
%%% LOGO
\begin{flushleft} ~\\ \vspace{-30mm} \hspace{-12mm}  \includegraphics[width=8cm]{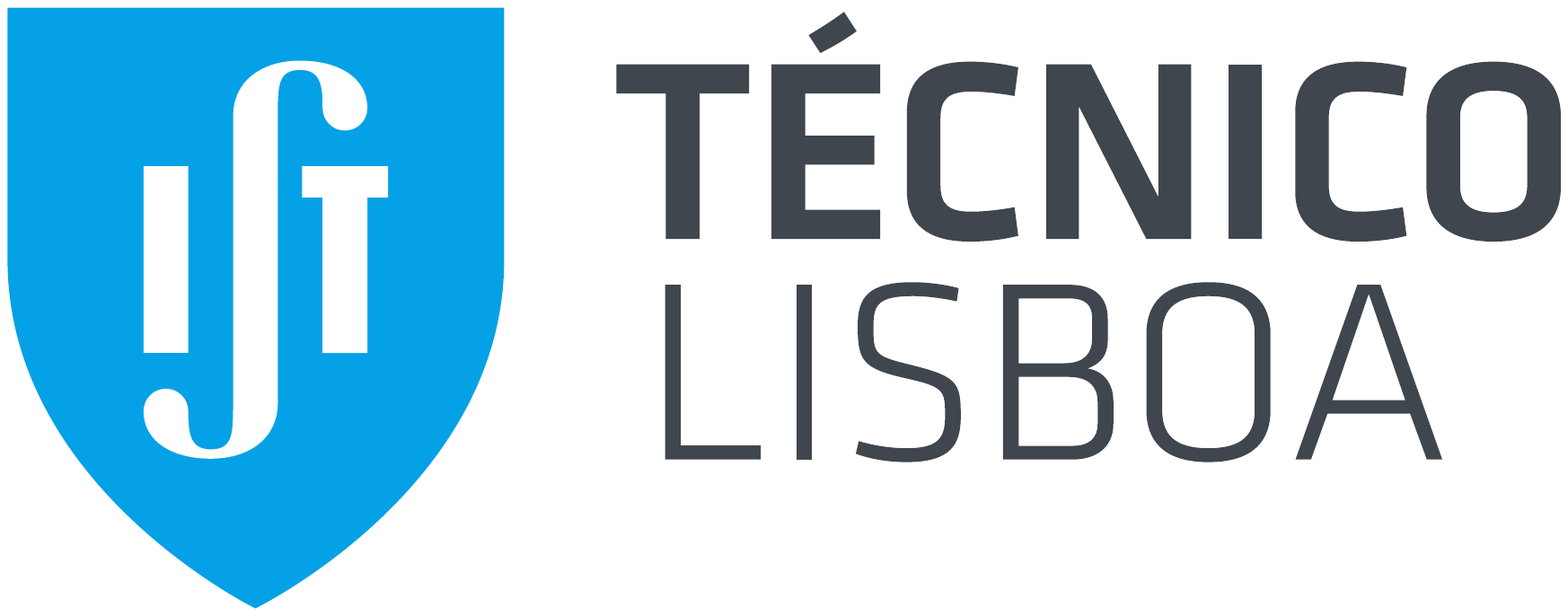} 
	
	%%% Instituição
	\centering
	\LARGE \textbf{\spacedallcaps{UNIVERSIDADE DE LISBOA \\ INSTITUTO SUPERIOR TÉCNICO}}
	%%% espaço sem gráficos
	\\ \vspace{33mm}

	%%% Optional Image
	%\vspace{10mm}
	%~\\ \vspace{50mm} % gráficos
	%\\ \begin{center} \includegraphics[height=50mm]{Cover/coverimage}  \end{center} % gráficos
	
	%%% Titulo
	\centering
	\LARGE \spacedallcaps{\myTitle}
	\\ \vspace{5mm}
	\Large \mySubtitle
	\\ \vspace{15mm}
	\LARGE \spacedlowsmallcaps{\myName} 
	\vspace{2.5cm}
	
	\begin{minipage}{\textwidth}
		\begin{tabularx}{\textwidth}{ l @{ } l }
			\large \spacedlowsmallcaps{Supervisor} : & \large Doctor Vitor Manuel dos Santos Cardoso\\
			\large \spacedlowsmallcaps{Co-Supervisor} :  & \large Doctor Carlos Alberto Ruivo Herdeiro\\
			%&    ~~~~~~~~~~~~~~~~~~~~~~~~~~ large name\\
		\end{tabularx}
		
	\end{minipage}
	\\ \vspace{26mm}
	%\vspace{12mm}
	\centering
	\large \spacedlowsmallcaps{Thesis approved in public session to obtain the PhD Degree in}\\
	\large \spacedallcaps{Physics}\\
	%\\ \vspace{2mm}
	\vspace{12mm}
	\spacedlowsmallcaps{Jury final classification}:
	\large \spacedlowsmallcaps{Pass with distinction and honour}
	
	\vspace{8mm}
	
	%\large \textbf{\todaythesis\today} \\
	\Large \textbf{2021} \\
	\let\thepage\relax
\end{flushleft}
\end{addmargin}
\pagebreak

%*******************************************************
% Little Dirty Titlepage
%*******************************************************
\thispagestyle{empty}
\pdfbookmark[1]{Title}{title}
%*******************************************************
%\begin{center}
%    \spacedlowsmallcaps{\myName} \\ \medskip
%
%    \begingroup
%        \color{CTtitle}\spacedallcaps{\myTitle}
%    \endgroup
%
%
%\end{center}

\begin{titlepage}
\begin{addmargin}[-1cm]{-3cm}
%%% LOGO
\begin{flushleft} ~\\ \vspace{-30mm} \hspace{-12mm}  \includegraphics[width=8cm]{gfx/IST_A_CMYK_POS} 
	
	%%% Instituição
	\centering
	\LARGE \textbf{\spacedallcaps{UNIVERSIDADE DE LISBOA \\ INSTITUTO SUPERIOR TÉCNICO}}
	%%% espaço sem gráficos
	\\ \vspace{10mm}

	%%% Optional Image
	%\vspace{10mm}
	%~\\ \vspace{50mm} % gráficos
	%\\ \begin{center} \includegraphics[height=50mm]{Cover/coverimage}  \end{center} % gráficos
	
	%%% Titulo
	\centering
	\Large \spacedallcaps{\myTitle}
	\\ \vspace{3mm}
	\large \mySubtitle
	\\ \vspace{6mm}
	\Large \spacedlowsmallcaps{\myName} 
	\vspace{0.5cm}
	
	\begin{minipage}{\textwidth}
		\begin{tabularx}{\textwidth}{ l @{ } l }
			\normalsize \spacedlowsmallcaps{Supervisor} : & \normalsize Doctor Vitor Manuel dos Santos Cardoso\\
			\normalsize \spacedlowsmallcaps{Co-Supervisor} :  & \normalsize Doctor Carlos Alberto Ruivo Herdeiro\\
			%&    ~~~~~~~~~~~~~~~~~~~~~~~~~~ large name\\
		\end{tabularx}
		
	\end{minipage}
	\\ \vspace{10mm}
	%\vspace{12mm}
	\centering
	\normalsize \spacedlowsmallcaps{Thesis approved in public session to obtain the PhD Degree in}\\
	\normalsize \spacedallcaps{Physics}\\
	%\\ \vspace{2mm}
	\vspace{2mm}
	\spacedlowsmallcaps{Jury final classification}:
	\large \spacedlowsmallcaps{Pass with distinction and honour}
	\vspace{10mm}
	
	\normalsize \spacedallcaps{Jury}
	
	\vspace{2mm}

			\normalsize \spacedlowsmallcaps{Chairperson} :  \\
			\vspace{1mm}  
			Doctor José Pizarro de Sande e Lemos\small, Instituto Superior Técnico, Universidade de Lisboa\\
			\vspace{2mm}
			\normalsize \spacedlowsmallcaps{Members of the Committee} :  \\
			\vspace{1mm}
			\normalsize Doctor Richard Pires Brito\small, Instituto Superior Técnico, Universidade de Lisboa\\ \vspace{1.5mm} 
			\normalsize Doctor Vitor Manuel dos Santos Cardoso\small, Instituto Superior Técnico, Universidade de Lisboa\\ \vspace{1.5mm}
			\normalsize Doctor Katy Clough\small, Astrophysics, University of Oxford, UK\\ \vspace{1.5mm}
			\normalsize Doctor Lam Hui\small, Physics Department, Columbia University, EUA\\ \vspace{1.5mm}
			\normalsize Doctor Caio Filipe Bezerra Macedo\small, Universidade Federal do Pará, Brasil\\ \vspace{1.5mm}
			\normalsize Doctor José António Maciel Natário\small, Instituto Superior Técnico, Universidade de Lisboa\\ \vspace{1.5mm}

	\vspace{10mm}
	
	{\normalsize \spacedlowsmallcaps{Funding institution:}}\\ \vspace{1mm}	
	{\large Fundação para a Ciência e Tecnologia} {\normalsize (SFRH/BD/128834/2017)} 
	
	\vspace{5mm}

	%\large \textbf{\todaythesis\today} \\
	\large \textbf{2021} \\
	\let\thepage\relax
\end{flushleft}
\end{addmargin}
\pagebreak
\end{titlepage}
%*******************************************************
% Titlepage
%*******************************************************
\begin{titlepage}
    %\pdfbookmark[1]{\myTitle}{titlepage}
    % if you want the titlepage to be centered, uncomment and fine-tune the line below (KOMA classes environment)
    \begin{addmargin}[-1cm]{-3cm}
    \begin{center}
        \large

        \hfill

        \vfill

        \begingroup
            \LARGE \color{CTtitle}\spacedallcaps{\myTitle} \\  \bigskip
            \Large \mySubtitle
        \endgroup
        
        \vfill

        \Large \spacedlowsmallcaps{\myName}

        \vfill

        \includegraphics[width=6cm]{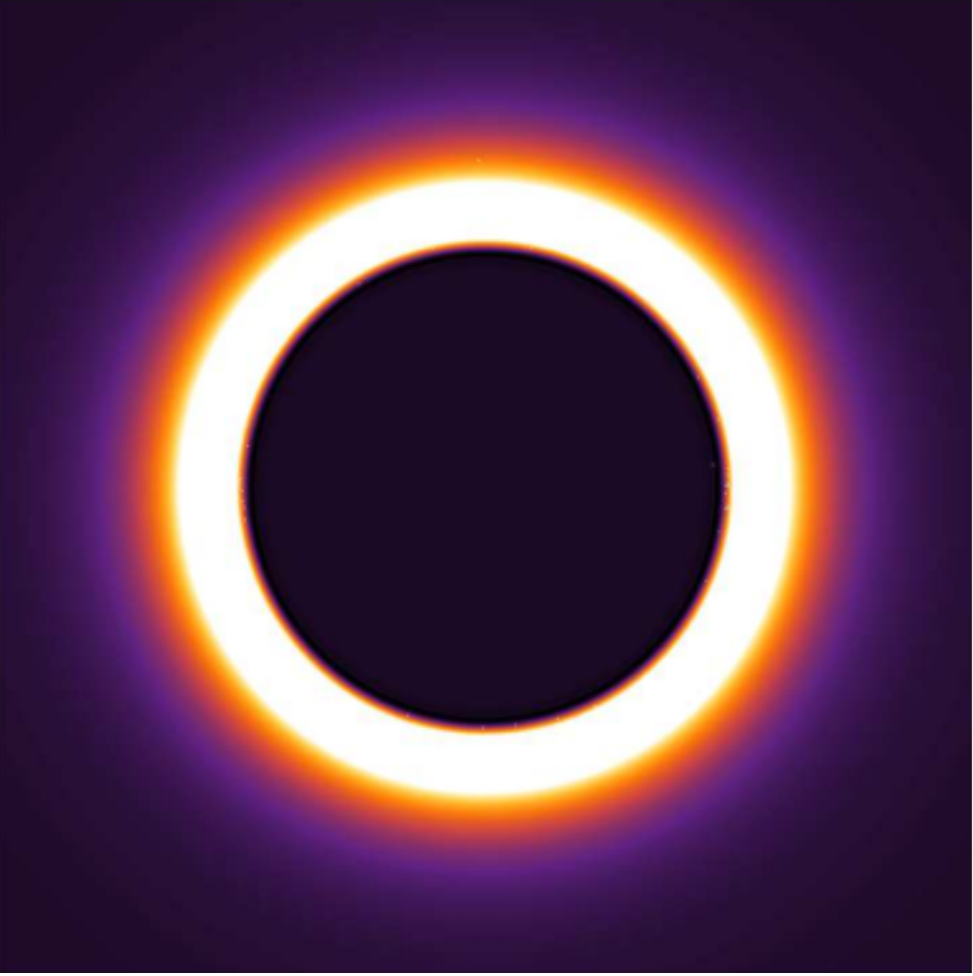} \\ \medskip

        \vfill 
        
        \spacedlowsmallcaps{\myTime}

    \end{center}
  \end{addmargin}
\end{titlepage}

\thispagestyle{empty}

\hfill

\vfill

\noindent\myName, \textit{\myTitle:} \mySubtitle, %\myDegree,
\textcopyright\ \myTime

%\bigskip
%
%\noindent\spacedlowsmallcaps{Supervisors}: \\
%\myProf \\
%\myOtherProf \\
%\mySupervisor
%
%\medskip
%
%\noindent\spacedlowsmallcaps{Location}: \\
%\myLocation
%
%\medskip
%
%\noindent\spacedlowsmallcaps{Time Frame}: \\
%\myTime

\cleardoublepage%*******************************************************
% Dedication
%*******************************************************
\thispagestyle{empty}
\phantomsection
\pdfbookmark[1]{Dedication}{Dedication}

\vspace*{1cm}
\begin{center}
	\spacedlowsmallcaps{Estética da abdicação} \\
	\textit{''Conformar-se é submeter-se e vencer é conformar-se, ser vencido. Por isso toda a vitória é uma grosseria. Os vencedores perdem sempre todas as qualidades de desalento com o presente que os levaram à luta que lhes deu a vitória. Ficam satisfeitos, e satisfeito só pode estar aquele que se conforma, que não tem a mentalidade do vencedor. Vence só quem nunca consegue. Só é forte quem desanima sempre. O melhor e o mais púrpura é abdicar. O império supremo é o do Imperador que abdica de toda a vida normal, dos outros homens, em quem o cuidado da supremacia não pesa como um fardo de jóias.''} \\ \medskip
	--- Bernardo Soares \emph{em} ''Livro do Desassossego''
\end{center}
\vspace*{1cm}

\begin{center}
    \spacedlowsmallcaps{AESTHETICS OF ABDICATION} \\
    \textit{''To conform is to submit, and to conquer is to conform, to be conquered. Thus every victory is a debasement. The conqueror inevitably loses all the virtues born of frustration with the status quo that led him to the fight that brought victory. He becomes satisfied, and only those who conform – who lack the conqueror’s mentality – are satisfied.
    Only the man who never achieves his goal conquers. Only the man who is forever discouraged is strong. The best and most regal course is to abdicate. The supreme empire belongs to the emperor who abdicates from all normal life and from other men, for the preservation of his supremacy won’t weigh on him like a load of jewels.''} \\ \medskip
    --- Bernardo Soares \emph{in} ''The Book of Disquiet''
\end{center}

\bigskip 
\vspace*{1cm}

\begin{center}
    {Dedicated to my parents.} \\ \smallskip
\end{center}

%\cleardoublepage\include{FrontBackmatter/Foreword}
\cleardoublepage%*******************************************************
% Abstract
%*******************************************************
%\renewcommand{\abstractname}{Abstract}
\pdfbookmark[1]{Abstract}{Abstract}
% \addcontentsline{toc}{chapter}{\tocEntry{Abstract}}
%\begingroup
%\let\clearpage\relax
%\let\cleardoublepage\relax
%\let\cleardoublepage\relax

\chapter*{Resumo}
Campos escalares são um conceito transversal em física teórica. Eles encontram aplicações em praticamente todas as áreas da física, desde matéria condensada e física de partículas à astrofísica e cosmologia. Campos escalares, em particular, podem dar origem a estruturas confinadas, como estrelas de bosões, “oscilatões” ou Q-balls. Estes objetos podem ser hipotéticas “estrelas de matéria escura” ainda não observadas, ou podem descrever núcleos de matéria escura no centro de halos galácticos, se os campos forem ultra-leves. Nesta tese é estudada a resposta dinâmica destas estruturas bosónicas confinadas quando excitadas por matéria externa (estrelas, planetas, ou buracos negros) na sua vizinhança. Estes objetos podem estar simplesmente a atravessar por entre a configuração bosónica, ou a mover-se periodicamente no seu centro (\eg, binárias). O nosso sistema pode também descrever  de forma eficiente a interação entre um buraco negro massivo em movimento e o seu ambiente envolvente. Também permite descrever de que forma a matéria escura é varrida por uma binária espiralando dentro de um núcleo de matéria escura.
Os resultados obtidos neste trabalho fornecem uma descrição completa da interação entre buracos negros (ou estrelas) e um núcleo de matéria escura ultra-leve envolvente, no interior do qual elas podem evoluir. Esta tese foca-se também em vários efeitos “ambientais” causados por campos clássicos que podem afetar o movimento (e, ultimamente, a própria sobrevivência) de objetos compactos, como buracos negros. Neste contexto, estudámos a interação de buracos negros em movimento num meio homogéneo de campo escalar. Obtivemos expressões analíticas através de primeiros princípios para a fricção dinâmica que atua sobre buracos negros, e mostrámos que apesar de buracos negros serem absorvedores naturais, a gravidade extrema na sua vizinhança pode torná-los, globalmente, em amplificadores (se se moverem suficientemente rápido). Estudámos também o efeito de fronteiras no fenómeno de fricção dinâmica em geometrias com uma dimensão compacta, e mostrámos que este efeito tende a suprimir a força de fricção (principalmente, no regime sub-sónico). Re-avaliámos o lema “a emissão de ondas gravitacionais tende a circularizar uma binária” quando ambientes astrofísicos (como discos de acreção) ou outras interações fundamentais estão presentes no problema. A este respeito mostrámos que (i) radiação escalar, vetorial e gravitacional contribuem ambas para circularizar o movimento orbital; (ii) pelo contrário, efeitos “ambientais” como acreção e fricção dinâmica levam ao aumento da excentricidade de binárias. Por fim, mostrámos que campos clássicos de “teste” não são capazes de destruir buracos negros extremos, desde que satisfaçam a condição de energia nula no horizonte de eventos . Este resultado é válido também para buracos negros em dimensões mais altas, e com constante cosmológica positiva ou negativa.  

\vspace{0.7cm}
\noindent \spacedlowsmallcaps{Palavras-chave:} relatividade geral; buracos negros; bosões ultra-leves; fricção dinâmica; ondas gravitacionais. 

\vfill

\newpage

\begin{otherlanguage}{ngerman}
\pdfbookmark[1]{Zusammenfassung}{Zusammenfassung}
\chapter*{Abstract}
Classical fields are ubiquitous in theoretical physics. They find applications in almost all areas of physics, from condensed matter and particle phyics to cosmology and astrophysics. Scalar fields, in particular, can give rise to confined structures, such as boson stars, oscillatons or Q-balls. These objects are interesting hypothetical new ``dark matter stars'', but also good descriptions of dark matter cores when the fields are ultralight. In this thesis, we study the dynamical response of such confined bosonic structures when excited by external matter (stars, planets or black holes) in their vicinities. Such perturbers can either be piercing through the bosonic configuration or undergoing periodic motion at its center (\eg, binaries). 
Our setup can also efficiently describe the interaction between a moving massive black hole and the surrounding environment. It also depicts dark matter depletion as a reaction to an inspiralling binary within a dark matter core. Our results provide a complete picture of the interaction between black holes or stars and the ultralight dark matter core environment where they may live in. This thesis also deals with several classical field environmental effects on the motion (or, ultimately, the survival) of compact objects, like black holes. We study the interaction of a moving black hole with a homogeneous scalar field medium. We obtain analytical expressions from first principles for the dynamical friction acting on a moving black hole in this environment, and show that, although black holes are natural absorbers, the strong pull of gravity can turn them into overall amplifiers (for large enough velocities). We study the effect of boundaries in dynamical friction for slab-like geometries, and show that they tend to lead to a suppression of the drag (more so in the subsonic case). The lemma ''gravitational-wave emission circularizes a binary'' is also re-evaluated when astrophysical environments (\eg, accretion disks) or other fundamental interactions are present. We show that (i) back-reaction from radiative mechanisms, including scalars, vectors, and gravitational waves circularizes the orbital motion; (ii) by contrast, environmental effects such as accretion and dynamical friction increase the eccentricity of binaries. Finally, we give a general proof that a test classical field cannot destroy an extremal black hole, provided it satisfies the null energy condition at the event horizon. This result also holds for black holes in higher dimensions and with positive or negative cosmological constant. 

\vspace{0.7cm}
\noindent \spacedlowsmallcaps{Key-words:} general relativity; black holes; ultralight bosons; dynamical friction; gravitational waves. 
\end{otherlanguage}

%\endgroup

\vfill

\cleardoublepage%*******************************************************
% Publications
%*******************************************************
\pdfbookmark[1]{Publications}{publications}
\chapter*{Publications}

Most of this doctoral thesis is based on the following publications:\\\\

{\small
\noindent V. Cardoso, C. F. B. Macedo, and R. Vicente, \textit{Eccentricity evolution of compact binaries and applications to gravitational-wave physics}, Phys. Rev. D103 (2021) 2, 023015;
\href{https://arxiv.org/abs/2010.15151}{\footnotesize arXiv:2010.15151 [gr-qc]}.\\

\noindent L. Annulli, V. Cardoso, and R. Vicente, \textit{The response of ultralight dark matter to supermassive black holes and binaries}, Phys. Rev. D102 (2020) 6, 063022; \href{https://arxiv.org/abs/2009.00012}{\footnotesize arXiv:2009.00012 [gr-qc]}.\\

\noindent L. Annulli, V. Cardoso, and R. Vicente, \textit{Stirred and shaken: dynamical behavior of boson stars and dark matter cores}, Phys. Lett. B811 (2020) 135944; 
\href{https://arxiv.org/abs/2007.03700}{\footnotesize arXiv:2007.03700 [astro-ph.HE]}. \\

\noindent J. Natário, and R. Vicente, \textit{Test fields cannot destroy extremal de Sitter black holes}, Gen. Rel. Grav. 52 (2020) 1, 5; 
\href{https://arxiv.org/abs/1908.09854}{\footnotesize arXiv:1908.09854 [gr-qc]}.\\

\noindent V. Cardoso, and R. Vicente, \textit{Moving black holes: energy extraction, absorption cross-section and the ring of fire}, Phys. Rev. D100 (2019) 8, 084001; \href{https://arxiv.org/abs/1906.10140}{\footnotesize arXiv:1906.10140 [gr-qc]}.\\

\noindent R. Vicente, V. Cardoso, and M. Zilhão, \textit{Dynamical friction in slab geometries and accretion discs}, Mon. Not. Roy. Astron. Soc. 489 (2019) 4, 5424-5435; \href{https://arxiv.org/abs/1905.06353}{\footnotesize arXiv:1905.06353 [astro-ph.GA]}.\\

\noindent J. Natário, L. Queimada, and R. Vicente, \textit{Test fields cannot destroy extremal black holes}, Class. Quant. Grav. 33 (2016) 17, 175002; \href{https://arxiv.org/abs/1601.06809}{\footnotesize arXiv:1601.06809 [gr-qc]}. \\\\}

\noindent During the years of my doctoral program I was also an author of the following works (not included in this thesis):\\\\

{\small
\noindent I. S. Fernández, R. Vicente, and D. Hilditch, \textit{A semi-linear wave model for critical collapse}, Phys. Rev. D103 (2021) 4, 044016; 
\href{https://arxiv.org/abs/2007.13764}{\footnotesize arXiv:2007.13764 [gr-qc]}.\\

\noindent R. Vicente, V. Cardoso, and J. C. Lopes, \textit{Penrose process, superradiance, and ergoregion instabilities}, Phys. Rev. D97 (2018) 8, 084032; \href{https://arxiv.org/abs/1803.08060}{\footnotesize arXiv:1803.08060 [gr-qc]}.\\

\noindent J. Natário, L. Queimada, and R. Vicente, \textit{Rotating elastic string loops in flat and black hole spacetimes: stability, cosmic censorship and the Penrose process}, Class. Quant. Grav. 35 (2018) 7, 075003; \href{https://arxiv.org/abs/1712.05416}{\footnotesize arXiv:1712.05416 [gr-qc]}. 
}

\cleardoublepage%*******************************************************
% Acknowledgments
%*******************************************************
\pdfbookmark[1]{Acknowledgments}{acknowledgments}

\bigskip

\begingroup
\let\clearpage\relax
\let\cleardoublepage\relax
\let\cleardoublepage\relax
\chapter*{Acknowledgments}

I am extremely grateful to my supervisor Vitor, for all the joy and excitement of the last 4 years. It was really inspirational to discuss physics and to work with such a \emph{giant} like him. I want also to thank him for the constant feedback and all the career advice that will be invaluable for my future. Thanks for always pushing us to give our best. 
\vspace{3mm}

I also want to thank my co-supervisor Carlos, for always being available to discuss physics and give advice, and for welcoming me so warmly in Aveiro. It was a great stay. 
\vspace{3mm}

I am sincerely thankful to all the members of the Jury, for all the time and care in reading this thesis, and for all the comments and suggestions that helped to improve its quality. 
\vspace{3mm}

This research has been developed at the gravity group (GRIT) of the Center for Astrophysics and Gravitation (CENTRA) in Instituto Superior Técnico (IST). There I had amazing colleagues that became my friends, with whom I learned so much and had a great time. Thank you all. A special thanks goes to my officemates and PhD fellows: Arianna, David, Francisco, Isa, Lorenzo, Rui and Thanasis, for all the laughs and companionship. This journey was much easier and joyful with all of you.
\vspace{3mm}

Lorenzo, has Vitor says, we have very different personalities (you like to cook, but I like to eat). Our friendly competitions and amazing travels made me a better and more practical person. This PhD gave me a \emph{frate} for life. Thank you.
\vspace{3mm}

I also want to thank to all my friends, specially to Gonçalo Andrade, Gonçalo Matos, Maria Inês and Marina, for being such great friends and for always being there to remind me that life is much more than physics. I'm grateful for giving me a social life.
\vspace{3mm}

I'm forever deeply grateful to my parents and sister, for all the patience, support and love. Thank you for always pushing me to follow my dreams. I dedicate this thesis to you.
\vspace{3mm}

Anita, you accompanied me in this entire journey and I simply couldn't have done this thesis without you. You are an amazing person, and the best partner I could have in this adventure called life. Thank you for your patience and love. You made me a better person. I love you.
\vspace{3mm}

I'm grateful to Fundação para a Ciência e Tecnologia which made this work possible through the PhD scholarship SFRH/BD/128834/2017. I also acknowledge partial financial support from the European Union's H2020 ERC Consolidator Grant "Matter and strong-field gravity: New frontiers in Einstein's theory" grant agreement no. MaGRaTh-646597, and networking support from GWverse COST Action CA16104, "Black holes, gravitational waves and fundamental physics". I also thank the hospitality of the Perimeter Institute and CERN's Theory Department, where part of this research was undertaken. 
%I thank to Ana Sousa Carvalho for producing some of the figures presented in this thesis.

\endgroup

\cleardoublepage%*******************************************************
% Table of Contents
%*******************************************************
\pagestyle{scrheadings}
%\phantomsection
\pdfbookmark[1]{\contentsname}{tableofcontents}
\setcounter{tocdepth}{2} % <-- 2 includes up to subsections in the ToC
\setcounter{secnumdepth}{3} % <-- 3 numbers up to subsubsections
\manualmark
\markboth{\spacedlowsmallcaps{\contentsname}}{\spacedlowsmallcaps{\contentsname}}
\tableofcontents
\automark[section]{chapter}
\renewcommand{\chaptermark}[1]{\markboth{\spacedlowsmallcaps{#1}}{\spacedlowsmallcaps{#1}}}
\renewcommand{\sectionmark}[1]{\markright{\textsc{\thesection}\enspace\spacedlowsmallcaps{#1}}}
%*******************************************************
% List of Figures and of the Tables
%*******************************************************
\clearpage
\begingroup
    \let\clearpage\relax
    \let\cleardoublepage\relax
    %*******************************************************
    % List of Figures
    %*******************************************************
    %\phantomsection
    %\addcontentsline{toc}{chapter}{\listfigurename}
    \pdfbookmark[1]{\listfigurename}{lof}
    \listoffigures

    \vspace{8ex}

    %*******************************************************
    % List of Tables
    %*******************************************************
    %\phantomsection
    %\addcontentsline{toc}{chapter}{\listtablename}
    \pdfbookmark[1]{\listtablename}{lot}
    \listoftables

    \vspace{8ex}
    % \newpage

    %*******************************************************
    % List of Listings
    %*******************************************************
    %\phantomsection
    %\addcontentsline{toc}{chapter}{\lstlistlistingname}
    \pdfbookmark[1]{Conventions and Units}{conventions}
    \markboth{\spacedlowsmallcaps{Conventions and Units}}{\spacedlowsmallcaps{Conventions and Units}}
	\chapter*{Conventions, Notation and Units}
	In this thesis I follow the conventions of Refs.~\cite{Misner:1974qy,Wald:1984rg}.
	Unless stated otherwise, I use geometrized units~$(c=G=1)$ and Lorentz-Heaviside units~$(\epsilon_0=1)$ and work with the~\emph{mostly positive} metric signature~$(-\, +\, +\; +)$. \\
	
	\begin{tabular}{lll}
		&$\alpha,\beta,\gamma, ...$          &  spacetime indices (from~$0$ to~$3$) \\
		&$i,j, k, ...$     & 3-spatial indices (from~$1$ to~$3$) \\
		%$\boldsymbol{r}, \boldsymbol{V}, \boldsymbol{\nabla}(\cdot), ...$          & spatial vectors or operators \\
		%$\boldsymbol{e}_x, \boldsymbol{e_r}, ...$ & spatial unit vector along~$x,\boldsymbol{r}$\\
		&$V_\alpha W^\alpha \equiv \sum_{\alpha=0}^{3} V_\alpha W^\alpha$ & Einstein's notation\\
		&$T_{(\alpha_1\,...\,\alpha_l)}\equiv= \frac{1}{l!}\sum_{\,\sigma} T_{\alpha_{\sigma(1)}\,...\,\alpha_{\sigma(l)}}$ & symmet. over all permutat.~$\sigma$\\
		%&$T_{[\alpha_1\,...\,\alpha_l]}\equiv \frac{1}{l!}\sum_{\,\sigma} \epsilon_\sigma T_{\alpha_{\sigma(1)}\,...\,\alpha_{\sigma(l)}}$ & anti-symmet. over all permutat.~$\sigma$\\
		&$g_{\alpha \beta} \,, \eta_{\alpha \beta}$       & curved, flat spacetime metric  \\
		&$(\,\cdot\,)_{,\alpha}=\partial_\alpha(\cdot)=\frac{\partial}{\partial x^\alpha}(\cdot)$ &  coord. derivative \\
		&$(\,\cdot\,)_{;\alpha}= \nabla_\alpha(\cdot)$  & Levi-Civita derivative \\
		&$\Box(\cdot) \equiv \nabla_\alpha \nabla^\alpha(\cdot)$ &  Levi-Civita d'Alembertian 
	\end{tabular}
	
    \vspace{8ex}

    %*******************************************************
    % Acronyms
    %*******************************************************
    %\phantomsection
    \pdfbookmark[1]{Acronyms}{acronyms}
    \markboth{\spacedlowsmallcaps{Acronyms}}{\spacedlowsmallcaps{Acronyms}}
    \chapter*{Acronyms}
    \begin{acronym}[UMLX]
    	\acro{BH}{black hole}
    	\acro{BHB}{black hole binary}
    	\acro{CDM}{cold dark matter}
    	\acro{DF}{dynamical friction}
    	\acro{DM}{dark matter}
    	\acro{EMRI}{extreme mass-ratio inspiral}
    	\acro{EOM}{equation of motion}
    	\acro{GW}{gravitational wave}
    	\acro{MACHO}{massive compact halo object}
    	\acro{NBS}{Newtonian boson star}
    	\acro{QNM}{quasi-normal mode}
    	\acro{WIMP}{weakly interacting massive particle}
    \end{acronym}

\endgroup

%********************************************************************
% Mainmatter
%*******************************************************
\cleardoublepage
\pagestyle{scrheadings}
\pagenumbering{arabic}
%\setcounter{page}{90}
% use \cleardoublepage here to avoid problems with pdfbookmark
\cleardoublepage
\part*{The Gravity of Classical Fields}\label{pt:introduction}
%************************************************
\chapter{Introduction}\label{ch:introduction}
%************************************************

\section{The dark matter problem}

After a century of investigation, the nature and properties of dark matter (\acs{DM}) remain a mystery. We have strong evidence for its existence -- all of them through \emph{gravity} --, but its nature and properties are hardly constrained. At the present moment, dark matter models range from ultralight bosons with masses~$\sim 10^{-22}\, \text{eV}$ to black holes (\acsp{BH}) with masses (of the order of) ten times the one of our sun~\cite{Barack:2018yly,Baibhav:2019rsa}. There are several proposals for the constituents of \acs{DM}:  weakly interacting massive particles (\acsp{WIMP}), massive compact halo objects (\acsp{MACHO}) and axions are some of the most famous candidates. \acsp{WIMP} could be particles predicted in supersymmetric extensions of the Standard Model of particle physics~\cite{Bertone:2004pz}, while \acsp{MACHO} would be dim (or dark) compact objects like, \eg, (primordial) black holes or neutron stars.  Interestingly, axions arise in a completely independent context in particle physics: they are light particles theorized to solve the strong CP problem in quantum chromodynamics~\cite{PecceiQuinn,PecceiQuinn1}. All of these are examples of what is broadly called cold dark matter (\acs{CDM}), which is part of the "standard model"~$\Lambda$CDM of Cosmology~\cite{Peebles,Blumenthal:1984bp}. 

The \acs{CDM} paradigm provides an excellent description of large-scale observations (on scales larger than~$\sim 10 \text{kpc}$), though it is finding several problems in describing smaller scales~\cite{Hui:2016ltb,Weinberg:2013aya,Del_Popolo_2017}. One of those is known as the "cusp-core" problem: cosmological $N$-body simulations show that \acs{CDM} leads to cuspy dark matter halos with a central density profile growing as~$\sim r^{-1}$~\cite{Flores:1994gz,Navarro:1996gj}, while galaxy rotation curves suggest the existence of a (constant density) core at the center of~\acs{DM} halos~\cite{Moore:1999gc,Weinberg:2013aya}. Other small-scale problems include the "missing satellites" problem (which may be solved by baryonic physics~\cite{Weinberg:2013aya}) and the "too big to fail" problem (similarly to the "cusp-core", \acs{CDM} simulations also predict too much mass at the center of dark matter subhalos). It is possible that some of (or all) these "problems" are solved by introducing more complicated baryonic physics (\eg, photoionization or/and stellar feedback), but it seems implausible that these effects can explain observations in \acs{DM}-dominated low-surface-brightness galaxies~\cite{Hui:2016ltb}. Alternatively, there is the interesting possibility that these (apparent) tension between \acs{CDM} and small-scale observations signals a problem of \acs{CDM} itself, and that a new \acs{DM} model is necessary.~\footnote{It can also signal a problem with the theory of gravity (\ie, general relativity) and the need for a modified theory of gravity~\cite{Del_Popolo_2017}.}

\section{The ultralight (wave) dark matter solution}

The (potential) problem with the \acs{CDM} paradigm resembles a lot the early years of quantum physics. Mechanics (with thermodynamics) and electromagnetism were believed to describe perfectly most of the macroscopic world, but they had problems in describing the microscopic world (\ie, small length-scales/high energies). The most famous problem was the \emph{ultraviolet catastrophe} of the Rayleigh-Jeans law~\footnote{Another (famous) micro-physics problem was the stability of the Hydrogen atom.}, which could not be explained by the (classical) physics of the time, and physicists were in a situation where they needed to create a new theory containing new micro-physics, without spoiling the excellent macroscopic predictions of classical physics. The solution to the ultraviolet catastrophe was found by Max Planck by introducing \emph{quanta} (\ie, discrete packets) of energy. This led, then, Albert Einstein to think of these quanta as real particles (photons, in this particular case), which inspired Louis de Broglie to go further and (reciprocally) postulate that electrons and matter have a wave nature -- this was confirmed experimentally in 1927, through the diffraction of electrons. This \emph{wave-particle} duality started a revolution in physics and led Erwin Schrödinger to formulate a theory of wave (quantum) mechanics. Interestingly, in this framework there is a natural length-scale above which the wave (quantum) effects are suppressed and waves describe (classical) particles, this is the de Broglie wavelength~\cite{landau1981quantum}
\begin{equation} \label{deBroglie}
	\lambda=\frac{h}{m v}\,,
\end{equation}
where~$m$ and~$v$ are the mass and velocity of the particle, and~$h$ is the Planck constant. This idea was transported to and play even a more profound role in quantum field theory, where the fundamental objects are fields (\ie, waves) and \emph{classical} particles~\footnote{Often in quantum field theory one calls particle to an irreducible unitary representation of the Poincaré group, which has nothing to do with what I am calling classical particle, which is just the ordinary idea of a particle with a definite position, momentum, etc (which you can usually recover in the geometrical optics limit).} are seen just as a particular type of wave (the length-scale~$\lambda$ continues to mark the transition from the quantum to the classical "worlds"). 

Inspired by the wave-particle duality, it is natural to think that a sufficiently light particle (with sufficiently large~$\lambda$) would be an ideal solution to the program of changing the small-scale physics of~\acs{CDM} (in this case, by introducing wave effects), while keeping its excellent large-scale (particle) predictions. Some important questions are then: (i) what light particle(s) could play this role? (ii) what are these wave effects? (iii) do they match better observations than~\acs{CDM}? \emph{Ultralight} (\emph{fuzzy}) \acs{DM} is a model which postulates that \acs{DM} is constituted by ultralight bosons or axion-like particles with masses~$m \sim10^{-22}-10^{-20} \,\text{eV}$~\cite{Press:1989id,Sin:1992bg,Hu:2000ke,Marsh:2015xka,Hui:2016ltb}, providing the perfect framework for pursuing the above program. These ultralight bosons are predicted generically by string theory~\cite{Arvanitaki:2009fg}, or simple Standard Model extensions~\cite{Freitas:2021cfi}. For an explanation of the particle physics motivations (independent of the \acs{DM} problem) to consider these ultralight bosons, and how they can have a relic abundance that matches today's observed \acs{DM} density, I direct the reader to the wonderful review~\cite{Hui:2021tkt}.

For typical velocities in galactic halos, these ultralight particles have a de Broglie wavelength
\begin{align}
	\lambda \simeq 0.5 \text{ kpc} \left(\frac{10^{-22} \text{ eV}}{m}\right)\left(\frac{250 \text{ km/s}}{v}\right)\,,
\end{align}
so, the uncertainty principle (through the form of a \emph{wave pressure}) can be shown to suppress the formation of structures at small-scales, solving some of the problems of~\acs{CDM}~\cite{Hu:2000ke,Hui:2021tkt}. From observations we know that the \acs{DM} density in the solar system's neighborhood is~$\sim 0.5 \text{ GeV/cm}^3$~\cite{Bovy:2012tw,McKee_2015,Sivertsson:2017rkp}, then the number of particles contained in a de Broglie volume~$\lambda^3$ is
\begin{equation}
	N \sim 10^{96} \left(\frac{10^{-22} \text{ eV}}{m}\right)^4\left(\frac{250 \text{ km/s}}{v}\right)^3\,,
\end{equation}
which means that this type of system is well described by classical fields -- in the sense that, since the number of particles is huge, the quantum fluctuations can be safely neglected (as it happens for electromagnetic waves with a large number of photons)~\cite{Hui:2021tkt}. This serves as the main motivation to the study of classical (scalar) fields in this thesis.

\section{Scalar structures and dark stars}

Observations (\eg, cusp-core problem) suggest that structures made of dark matter, like \acs{DM} cores -- regions of nearly constant \acs{DM} density -- inside galactic halos, exist in our Universe. This is one of the first tests that any alternative \acs{DM} model to \acs{CDM} must pass. Thus, the existence and stability of \acs{DM} solitonic objects in a given theory is a very relevant issue. In fact, it was famously shown by \citeauthor{Derrick:1964ww} that (self-interacting) scalar fields cannot form stable, time-independent, localized solutions~\cite{Derrick:1964ww}. However, there is a simple way to circumvent such no-go result: considering (periodic) time-dependent fields. In this way it is possible to construct, for instance, self-gravitating configurations of (possibly self-interacting) massive complex scalar fields; the resulting objects are known as \emph{boson stars} and can provide good descriptions of \acs{DM} cores~\cite{Kaup:1968zz,Membrado,Liebling:2012fv,Chavanis1,Chavanis2}. In the case of very dilute configurations these are called Newtonian boson stars (\acsp{NBS}). If the scalar is ultralight ($m\sim 10^{-22}\, \text{eV}$) these are good descriptions of most cores in \acs{DM} halos~\cite{Ruffini:1969qy,Hui:2016ltb,Liebling:2012fv}. Another alternative are the so-called \emph{oscillatons}. These are objects made of self-gravitating real scalar fields, having an oscillating gravitational potential -- which is their main difference with respect to boson stars~\cite{Seidel1994,Copeland:1995fq}. This provides them with a very rich phenomenology~\cite{Khmelnitsky:2013lxt,Boskovic:2018rub}. These structures are also interesting from the point of view of non-baryonic \acsp{MACHO}; if these \emph{dark stars} are very compact, they could be the building blocks of \acs{DM} halos~\cite{Giudice:2016zpa,Ellis:2017jgp,Barack:2018yly,Cardoso:2019rvt}. In this last case, they can also be strong gravitational-wave (\acs{GW}) sources and mimic black holes~(\acsp{BH})~\cite{Cardoso:2019rvt,Palenzuela:2017kcg}. All of the above motivates us to study~\acsp{NBS} and their response to binary black holes (\acsp{BHB}) and stars in Part~\ref{pt:ultralight} of this thesis (the same for Q-balls in Appendix~\ref{app:Qball}).

Although, not directly connected with the work in this thesis, I would like to point out here some additional remarks about recent research on ultralight bosons. Surprisingly, it can be shown that rotating \acsp{BH} may stimulate the growth of macroscopic bosonic \emph{clouds} in their vicinities~\cite{Herdeiro:2014goa,Brito:2015oca}. Although \acsp{BH} tend to absorb the matter and fields entering their event horizon -- from where they cannot ever escape (at least classically) --, they are also capable of enhancing bosonic fields through a mechanism called \emph{superradiance}~\cite{zeldovich,Misner:1972kx,Brito:2015oca}. These clouds, which are supported by the rotation of \acsp{BH}, are often thought of as the gravitational parallel of the hydrogen atom, since they also satisfy a Schrödinger equation~\cite{Arvanitaki:2009fg,Baumann:2019eav}. However, because the gravitational coupling can be much weaker than the electromagnetic one (in particular, in the case of ultralight fields), these gravitational atoms can be much larger than the usual hydrogen atom and may span over astrophysical scales.
Using the observations of masses and angular momenta of astrophysical \acsp{BH}, superradiance has been used to put bounds on the masses of ultralight particles, like the axion, the massive photon, or the massive graviton~\cite{Arvanitaki:2010sy,Brito:2014wla,Brito:2015oca,Brito:2020lup}. The gravitational atom paradigm is being intensively used to explore ultralight \acs{DM}. Very recently, it was found that \acsp{BHB} can also support global quasi-bound states of scalar field in their surroundings, resembling in many ways the hydrogen molecule~\cite{Bernard:2019nkv,Ikeda:2020xvt}.

\section{Environmental effects due to classical fields}

Drag forces of electromagnetic origin are ubiquitous in everyday life, and shape -- to some extent -- our own civilization.
On large scales, such as those of stars and galaxies, gravitational drag forces dominate the dynamics. When stars or planets move through a medium, a wake of fluctuation in the medium density is left behind. Gravitational drag -- also known as dynamical friction (\acs{DF}) -- is caused by the backreaction of the wake on the moving object. \acs{DF} determines a number of features of astrophysical systems, for example planetary migration within disks, the sinking of supermassive \acsp{BH} to the center of galaxies
or the motion of stars within galaxies on long timescales~\cite{Chandrasekhar:1943v1,Chandrasekhar:1943v2,Chandrasekhar:1943v3,Ostriker:1998fa,binney2011galactic}.
Binaries immersed in a nontrivial environment will be subjected to \textit{accretion} and \acs{DF}, which will also affect their dynamics~\cite{Macedo:2013qea,Cardoso:2019dte}. 
Merging \acsp{BHB} are now ``visible'', thanks to \acs{GW} astronomy~\cite{Abbott:2016blz,Barack:2018yly}.
A good modeling of the dynamics of such compact binaries in such environments is important to increase our ability to actually see them,
to infer the properties of the merging objects and to impose constraints on the underlying gravitational theory, or other fundamental interactions~\cite{Barack:2018yly}.
Although the environmental \acs{DM} effects are expected to be usually weak, they can result in an observable phase shift of the gravitational waveform if the binary describes many orbits in a region with large overdensities of dark matter~\cite{Macedo:2013qea,Eda:2013gg,Eda:2014kra,Barausse:2014tra,Barausse:2014pra,Yue:2017iwc}. Thus, a particularly suitable system to consider is an extreme mass ratio inspiral (\acs{EMRI}) immersed in a \acs{DM} core in a galactic halo. This is a system whose gravitational waves we will observe with Laser Interferometer Space Antenna (LISA). Surprisingly, the response of fuzzy \acs{DM} cores to the presence of compact objects and the subsequent \acs{DF} acting on these objects had not been computed until very recently; there existed only estimates based on results derived for baryonic fluid media~\cite{Macedo:2013qea,Barausse:2014tra,Boskovic:2018rub}, or neglecting the self-gravity of the scalar wake~\cite{Hui:2016ltb,Hui:2021tkt}. In this thesis we tackle this problem in a self-consistent way using (linear) perturbation theory. 

In this thesis, we will take the liberty of including a perfect fluid in what we call a classical field (in the sense that it is described by a set of continuous functions of space and time).~\footnote{Reciprocally, a non-relativistic scalar field described by the Schrödinger equation can alternatively be described by a set of "hydrodynamical" variables satisfying the Euler equations (with a certain "quantum" pressure) obtained through the so-called Madelung transformation~\cite{Feynman:1494701}.} In particular, in Chapters~\ref{ch:DF} and~\ref{ch:eccentricity} we will consider environmental effects due to an ordinary perfect fluid; these are interesting and important on its own (having applications, \eg, to \acs{GW} physics), but their conclusions can (and should) also be extended to ultralight~\acs{DM} environments.

From a formal point of view it is also interesting to study if (or under which conditions) \acsp{BH} can be destroyed (\ie, turned into a naked singularity) by accreting some classical field (or matter) with sufficiently large angular momentum (or electric charge). This would violate the weak cosmic censorship conjecture, which state that singularities resulting from gravitational collapse are, generically, hidden from observers at infinity by a \acs{BH} event horizon. To test this conjecture, Robert Wald devised a thought experiment to destroy extremal Kerr-Newman \acsp{BH} by dropping charged and/or spinning test particles into the event horizon. Both he and subsequent authors found that if the parameters of the infalling particle were suited to overspin/overcharge the \acs{BH}, then the particle would not be absorbed, in agreement with the weak cosmic censorship conjecture. Similar conclusions were obtained by analyzing scalar and electromagnetic test fields propagating in extremal Kerr-Newman \acs{BH} backgrounds. But, until recently, no general proof (or condition) was known that could guarantee the survival of extremal \acsp{BH} interacting with an arbitrary classical field. We tackle this problem in Chapter~\ref{ch:wcc}.

\section{Organization of the thesis}

Chapter~\ref{ch:theory} presents summarily the theory, equations of motion (\acsp{EOM}) and currents that will be used in great part of the thesis. It includes also the Newtonian (non-relativistic) and weak field limits of the \acsp{EOM}, which will be largely used in Part~\ref{pt:ultralight} of this work (and are explained in more detail in Appendix~\ref{app:PN}). 

Part~\ref{pt:ultralight} of the thesis deals with the response of ultralight \acs{DM} cores (modeled through \acsp{NBS}) to the presence and motion of stars, \acsp{BH} and binaries. It is based on the publications~\cite{Annulli:2020lyc} and~\cite{Annulli:2020ilw}, which were done in collaboration with my colleague Lorenzo Annulli and Prof. Vitor Cardoso. Chapter~\ref{ch:framew} describes the perturbative framework and how quantities like the energy and momenta of the scalar field radiated (\ie, depleted), or the energy loss of the moving compact objects are computed. In Chapter~\ref{ch:nbs} the background configurations modeling ultralight \acs{DM} cores are presented. We obtain the entire family of \acsp{NBS} (using an important scaling symmetry) and find the linearized system of equations describing the perturbations on this background configuration, which may be sourced by an external particle (modeling, \eg, a star or a \acs{BH}). Some of the normal modes of these configurations are also obtained. In Chapter~\ref{ch:stir} the perturbative framework is used to study several systems of astrophysical interest like massive compact objects piercing through the \acs{DM} core, or a massive \acs{BH} oscillating at the center of the core after forming with some "kick", or some binary evolving deep inside the core. We study both the response of the core (\eg, scalar depletion) to these systems as well as the evolution of these system within the core.

Part~\ref{pt:environment} deals with several classical field environmental effects on the motion (or, ultimately, on the existence) of compact objects, like \acsp{BH}. These environments are not necessarily ultralight~\acs{DM}, and some chapters deal exclusively with \acs{DF} due to perfect fluids. However, all the conclusions drawn in these chapters may, in principle, be extended (at least qualitatively) to ultralight \acs{DM} environments. 

Chapter~\ref{ch:movingBH} is in part based in the publication~\cite{Cardoso:2019dte}, done in collaboration with Prof. Vitor Cardoso, but it contains a great amount of original work to be published soon. In this chapter we study the interaction between a plane wave and a (counter-moving) \acs{BH}. We show that, overall, energy is transferred from the moving \acs{BH} to the wave, giving rise to a negative absorption cross-section. \acsp{BH} are natural absorbers, but the universal, strong pull of gravity can turn them into overall amplifiers. Due to this effect, a \acs{BH} of mass $M$ moving at relativistic speeds in a cold medium will appear to be surrounded by a bright ring of diameter~$3 \sqrt{3} G M/c^2$ and thickness~$\sim G M/c^2$. We also compute for the first time, from first principles, the \acs{DF} acting on \acsp{BH} moving at possibly relativistic speeds in light scalar field environments. We find several simple analytical expressions valid for different regimes of \acs{BH} velocity.

Chapter~\ref{ch:DF} is based on the publication~\cite{Vicente:2019ilr} with Prof. Vitor Cardoso and Dr. Miguel Zilhão, and deals with the effect of boundaries in \acs{DF}. In particular, we compute the wake and corresponding \acs{DF} in a three-dimensional gaseous medium with a slab-like geometry, finding a generic suppression (larger in the subsonic regime) of \acs{DF}. This chapter also provides a natural bridge between two different results present in the literature: time-dependent (finite), and steady-state (vanishing) subsonic \acs{DF} forces. 

Chapter~\ref{ch:eccentricity} is based on the publication~\cite{Cardoso:2020iji}, done in collaboration with Prof. Vitor Cardoso and Prof. Caio Macedo. Here we study if the lemma "\acs{GW} emission circularizes a binary" still holds when astrophysical environments (\eg, accretion disks) or other fundamental interactions are taken into account. We show that (i) back-reaction from radiative mechanisms, including scalars, vectors, and \acsp{GW} circularizes the orbital motion; (ii) by contrast, environmental effects such as accretion and \acs{DF} increase the eccentricity of binaries. We also show that this effect can be important for LISA sources.

Chapter~\ref{ch:wcc} is based partly on the publication~\cite{Natario:2016bay}, done in collaboration with Prof. José Natário and my colleague Leonel Queimada, and partly on~\cite{Natario:2019iex}, done with Prof. José Natário. Here, we give a general proof that, at linear level, a test classical field cannot destroy an extremal \acs{BH}, provided it satisfies the \emph{null energy condition} at the event horizon. Our result is very general and applies also to \acsp{BH} in higher dimensions, to the case of a negative cosmological constant, and it is valid for any type of test matter. Then, we proceed to find the correct definition of energy for test classical fields propagating in \acs{BH} spacetimes with positive cosmological constant, and use that definition to extend the previous result to those spacetimes.

%*****************************************
%*****************************************
%*****************************************
%*****************************************
%*****************************************

%************************************************
\chapter{Theory, Equations of Motion and Currents}\label{ch:theory}
%************************************************

In this chapter we present the action that generates the theories studied in this thesis. Here we also show the equations of motion~(\acsp{EOM}), energy-momentum tensors and the most important currents used in this work. Finally, this chapter ends with the Newtonian limit of the~\acsp{EOM}.

\paragraph{Action}In this thesis we consider a theory containing one (possibly complex) scalar~$\Phi$ and one (electromagnetic) covector~$A_\alpha$ -- both minimally coupled to gravity -- generated by the action
\begin{align}\label{theory_action}
	S&= \int d^4x \sqrt{-g}\left[\frac{R- 2 \Lambda}{8 \pi}- \Phi^*_ {\;\,; \alpha} \Phi^{\,;\alpha}-\mathcal{U}_S[|\Phi|^2]-\frac{F^{\alpha \beta} F_{\alpha \beta}}{2}\right] \nonumber \\
	&-\int d^4x \sqrt{-g}\left[J_S \left(\Phi+ \Phi^*\right)+2J_V^{\;\;\alpha} A_\alpha\right]+ \int d^4x \sqrt{-g}\, L_\textrm{M}\,,
\end{align}
where~$F\equiv d A$ is the Faraday 2-form,~$R$ is the Ricci scalar and~$\Lambda$ is the cosmological constant. The function~$\mathcal{U}_S$ is the self-interaction potential of the scalar fields and~$L_\textrm{M}$ is the Lagrangian density of \emph{other} matter fields (which will be assumed to be independent of~$g_{\alpha \beta}$,~$\Phi$ and~$A_\alpha$). The currents~$J_S$ and~$J_V^{\;\;\alpha}$ are arbitrary sources of scalar and vector fields, respectively, which in this work are assumed to depend \emph{only} on the matter fields.~\footnote{With the exception of Chapter~\ref{ch:wcc}, in this thesis we will not consider direct couplings between scalar and vector fields (\eg, scalar QED), but these can still interact (indirectly) through gravity.} 

\paragraph{Equations of motion} Taking the first variation of the action in~$\Phi^*$,~$A_\beta$ and~$g_{\alpha \beta}$, respectively, one finds the~\acsp{EOM}~\footnote{We used the results~$\delta (\sqrt{-g})=-\frac{1}{2}\sqrt{-g}\, g_{\alpha \beta}\,\delta g^{\alpha \beta}$ and~$g^{\alpha \beta}\delta R_{\alpha \beta}= \nabla^\alpha v_\alpha$ (the explicit form of the co-vector~$v_\alpha$ is unimportant here)~\cite{Wald:1984rg}, and performed several integrations by parts.}
\begin{align}
	&\Box \Phi= J_S+\frac{\delta \mathcal{U}_S}{\delta \Phi^*} \label{KG_EOM} \,, \\
	&\nabla_\alpha F^{\alpha \beta}= J_V^{\;\;\beta} \label{Maxw_EOM}\,, \\
	&G^{\alpha \beta}+\Lambda g^{\alpha \beta}=8 \pi \,T^{\alpha \beta} \label{Eins_EOM}\,,
\end{align}
where~$G^{\alpha \beta}\equiv R^{\alpha \beta} -\frac{1}{2} Rg^{\alpha \beta}$ is the Einstein tensor and~$T^{\alpha \beta}$ is the \emph{total} energy-momentum tensor (obtained through the first variation in~$g_{\alpha \beta}$), which is divergenceless
\begin{equation}
	\nabla_\alpha T^{\alpha \beta}=0\,
\end{equation}
due to the contracted Bianchi identities, and can be expressed as
\begin{equation}
	T^{\alpha \beta}=T_S^{\;\;\alpha \beta}+ T_V^{\;\; \alpha \beta}+T_M^{\;\; \alpha \beta}+ 2\frac{g^{\alpha \delta} g^{\beta \gamma}}{\sqrt{-g}} \frac{\delta }{\delta g^{\delta \gamma}}\left\{\sqrt{-g}\left(J_S \Re{[\Phi]}+J_V^{\;\; \alpha}A_\alpha\right)\right\}\,,
\end{equation}
with~$\Re[\Phi]$ the real part of~$\Phi$. The individual energy-momentum tensors appearing in the sum are
\begin{align}
	T_S^{\;\;\alpha \beta}&\equiv \nabla^{(\alpha}\Phi^* \nabla^{\beta)} \Phi-\frac{1}{2}g^{\alpha \beta} \left(\Phi^*_ {\;\,; \delta} \Phi^{\,;\delta}+\mathcal{U}_S[|\Phi|^2]\right)\,, \label{scalarEMT}\\
	T_V^{\;\;\alpha \beta}&\equiv F^{\alpha}_{\;\;\delta}\,F^{\beta \delta}-\frac{1}{4}g^{\alpha \beta} F_{\delta \gamma} F^{\delta \gamma} \label{vectorEMT}\,,\\
	T_\textrm{M}^{\;\;\alpha \beta}&\equiv-\frac{g^{\alpha \delta} g^{\beta \gamma}}{\sqrt{-g}} \frac{\delta \left(\sqrt{-g}\, L_\textrm{M}\right)}{\delta g^{\delta \gamma}}\,.
\end{align}

Since the focus of this thesis is on the effects of bosonic fields in astrophysics (and not so much in cosmology), hereafter we will consider~$\Lambda=0$. In most of this work we will be interested in a simple \emph{mass} potential
\begin{align}
	\mathcal{U}_S&= \mu_S^2 |\Phi|^2 \label{mass_pot_S}\,,
\end{align}
where~$m_S= \hbar \mu_S$ is the mass of the scalar particles. In some sections of this text we will restrict to massless scalars.

In some chapters we will consider that the "matter" Lagrangian density describes a system of~$N$ point particles minimally coupled with gravity, in which case it has the form~\footnote{In addition, when matter couples \emph{only} with gravity the currents~$J_S$ and~$J_V^{\;\;\,\alpha}$ vanish.}
\begin{equation} \label{matter_particles_Lagr}
\mathcal{L}_\textrm{M}=- 2\sum_{n=1}^{N} m_n \int d \tau_n \frac{\delta^{(4)}\left(x^\alpha-x_n^{\;\; \alpha}(\tau_n)\right)}{\sqrt{-g}}\,,
\end{equation}
where~$x_n^{\;\;\alpha}$ and~$\tau_n$ are, respectively, the world-line and proper time of the~$n^\textrm{th}$ particle. This Lagrangian density results in the \acsp{EOM}~\cite{poisson_will_2014}
\begin{align}
	x_n^{\;\;\alpha} \nabla_\alpha x_n^{\;\; \beta}=0\,,
\end{align}
which assert that the point particles move along spacetime \emph{geodesics}, and implies the energy-momentum tensor
\begin{align}\label{particleEMT}
	T_\textrm{M}^{\;\; \alpha \beta}=\sum_{n=1}^N m_n \int d\tau_n  \frac{\delta^{(4)}\left(x^\alpha-x_n^{\;\; \alpha}(\tau_n)\right)}{\sqrt{-g}} \,\dot{x}_n^{\;\;\alpha}\, \dot{x}_n^{\;\;\beta}\,,
\end{align}
where~$\dot{x}_n^{\;\;\alpha}\equiv d x_n^{\;\;\alpha}/d \tau_n$ is the~4-velocity of the~$n^\textrm{th}$ particle (in the coordinate basis). To obtain this energy-momentum tensor one needs to take the first variation in~$g^{\alpha \beta}$ of~$d\tau_n=\sqrt{-g_{\alpha \beta} \,dx_n^{\;\;\alpha} dx_n^{\;\;\beta}}$.

\paragraph{Currents} In most of this thesis it will be assumed that both the scalar and the vector interact only with gravity (\ie,~$J_S=J_V^{\;\;\,\alpha}=0$) through the minimal coupling in~\eqref{theory_action}. In these cases the resulting theory is not U(1) gauge invariant, but it is still invariant under \emph{global} U(1) transformations of~$\Phi$. This continuous symmetry is associated with a Noether current
\begin{align} \label{NoetherCurrent}
	J_Q^{\;\;\alpha}=\frac{i}{2 \hbar}\left(\Phi\,\partial^\alpha \Phi^*-\Phi^* \partial^\alpha \Phi\right)\,,
\end{align}
which satisfies
\begin{equation}
	\nabla_\alpha J_Q^{\;\;\alpha} =0\,,
\end{equation}
and gives rise to a \emph{conserved} Noether charge
\begin{equation} \label{NoetherCharge}
	Q= -\int_\mathcal{S} dV_3 \,J_Q^{\;\;\alpha} N_\alpha\,,
\end{equation}
where~$\mathcal{S}$ is a spacelike hypersurface (extending through all spacetime),~$N_\alpha$ is the covector associated with the future-pointing unit normal to~$\mathcal{S}$, and~$dV_3$ is the 3-volume form induced on the hypersurface.

If the spacetime admits a Killing vector field~$\xi^\alpha$, it is well-known that one can combine it with the total stress-energy tensor~$T^{\alpha \beta}$ to construct the current
\begin{equation}
	J_\xi^{\;\;\alpha}=T^{\alpha \beta} \xi_\beta\,,
\end{equation}
which is divergenceless
\begin{equation}
	\nabla_\alpha J_\xi^{\;\; \alpha} =0
\end{equation}
and results in the \emph{conserved} quantity~\footnote{In fact, this quantity only need to be conserved if~$J_\xi^{\;\;\alpha}$ decays sufficiently fast at spacelike infinity (which is not the case of monochromatic waves) and if the spacetime is geodesically complete (which is not the case for black hole spacetimes). The same remark applies to the Noether charge~\eqref{NoetherCharge}.}
\begin{equation} \label{energy_angularmom}
Q_\xi= -\int_\mathcal{S} dV_3 \,J_\xi^{\;\;\,\alpha} N_\alpha\,.
\end{equation}
In this thesis, we will always consider stationary axisymmetric spacetimes. These admit a Killing vector~$X^\alpha$ that is asymptotically timelike, and a Killing vector~$Y^\alpha$ which is asymptotically spacelike and whose integral curves are closed~\cite{Wald:1984rg}. These two Killing vector field commute~$\left[X^\alpha,Y^\beta\right]=0$, so their flows can be used as local coordinates. The vectors~$-X^\alpha=-(\partial_t)^\alpha$ and~$Y^\alpha=(\partial_\varphi)^\alpha$ will be used to define, respectively, the \emph{energy} and \emph{angular momentum} of the fields. 

\paragraph{Newtonian limit} Let us consider the simplest case of the mass potential~\eqref{mass_pot_S}, with the scalar and matter coupling only with gravity ($J_S=0$) and no electromagnetic field ($A_\alpha=J_V^\alpha=0$). Furthermore, let us restrict to a matter part consisting of~$N$ point particles minimally coupled to gravity~\eqref{matter_particles_Lagr}. In the Minkowskian (weak gravitational field~$\frac{G M}{R c^2} \ll 1$) limit and for Newtonian (small velocity~$ v/c \ll 1$) fields the spacetime metric (at leading order) in Cartesian-like coordinates is simply~(\eg, \cite{poisson_will_2014})
\begin{equation}
	ds^2=-\left(1+2 U\right) dt^2+\left(1-2 U\right) \delta_{i j}\,dx^i dx^j\,,
\end{equation}
and the~\acsp{EOM} (at leading order) reduce to
\begin{align}
	&i \partial_t \phi=-\frac{1}{2 \mu_S} \partial_i \partial_i \phi+ \mu_S U \phi \,,\\
	&\frac{d^2}{d t^2} z_n^i=-\partial_i U(z_n^j)\,, \qquad z_n^0=t\,,\\
	&\partial_k \partial_k U= 4 \pi \left[\mu_S|\phi|^2+\sum_{n=1}^N m_n \delta^{(3)}\left(x^i-z_n^{\;\;i}(t)\right)\right] \label{Poisson_Newtonian}\,,
\end{align}
with the following field redefinition
\begin{align}
	&\Phi(x^\beta)=\frac{e^{-i \mu_S t}}{\sqrt{\mu_S}}\phi(x^\beta) \,.
\end{align}

To obtain the above~\acsp{EOM} one needs to use the fact that for Newtonian fields
\begin{align}
	&\left|\frac{1}{\Phi}\partial_t \Phi\right| \sim \mu_S\left(1+\mathcal{O}(\epsilon^2)\right)\, \Leftrightarrow\, \left|\frac{1}{\phi}\partial_t \phi\right| \sim \mu_S \mathcal{O}(\epsilon^2)\,, 
\end{align}
where~$\epsilon\ll 1$ is a small expansion parameter (of the same order as the velocity of the scalar particles). The scalar and vector field \acsp{EOM} imply that for the Newtonian (small velocity) fields to be able to~\emph{feel} gravity at leading order, the gravitational potential must be of order~$U\sim \mathcal{O}(\epsilon^2)$. In that case, the Poisson equation~\eqref{Poisson_Newtonian} gives the following order of magnitude estimates
\begin{align}
	 &\frac{U}{R^2} \sim \frac{M}{R^3}\, \Rightarrow\, \frac{M}{R} \sim U \sim  \mathcal{O}(\epsilon) \,, \\
	 &\mu_S R^2 |\phi|^2 \sim \left(\frac{m_n}{R_n}\right)\left(\frac{R}{R_n}\right)^2 \sim \mathcal{O}\left(\epsilon\right)\,,
\end{align}
where~$M$ is the total mass in the system,~$R$ is the size of the system and~$R_n$ is an effective (cut-off) radius of the~$n^\textrm{th}$ particle. We need to introduce these cut-off radii, because the gravitational potential~$U$ diverges with~$|\boldsymbol{r}-\boldsymbol{r}_n|^{-1}$ close to the particles and therefore the Minkowskian approximation cannot be trusted in those regions.~\footnote{Indeed, there are no (infinitely compact) particles in the theory of general relativity; the most compact objects are black holes (which have~$M/R\sim 1$).} Often in this thesis we will simply remove those regions from our study (as we will see in Chapter~\ref{ch:movingBH}, it turns out that this approximation is very well justified in most of the cases studied here); a more rigorous analysis would require more information about the nature (stars, black holes, etc) and micro-physics of these objects (\eg, equation of state).

A more thorough treatment of the Newtonian limit can be found in Appendix~\ref{app:PN}, where we consider the first post-Newtonian order expansion of the Einstein-Klein-Gordon system.

%*****************************************
%*****************************************
%*****************************************
%*****************************************
%*****************************************

\cleardoublepage
\part{Response of ultralight fields to compact objects}\label{pt:ultralight}
%*****************************************
\chapter{Framework}\label{ch:framew}
%*****************************************

\section*{Introduction}

The existence, stability and dynamical behavior of \emph{solitons} in a given theory is relevant for a wide range of topics, from planetary science to a description of fundamental particles. 
Taking as starting point a theory of a scalar field in flat space, it can be shown that localized time-independent solutions cannot exist~\cite{Derrick:1964ww}. This powerful result limits the ability of fundamental scalars to describe possible novel objects where the scalar is confined. A promising way to circumvent such no-go result is to consider time-dependent fields.
Within this more general framework, it can be shown that black holes (\acsp{BH}) can stimulate the growth of structures in their vicinities~\cite{Herdeiro:2014goa,Brito:2015oca},
and also that new self-gravitating solutions are possible. Such objects can describe dark stars which have so far gone undetected~\cite{Barack:2018yly,Cardoso:2019rvt,Giudice:2016zpa,Ellis:2017jgp}. Surprisingly, the simplest such solutions also seem to be a good description of structures we know to exist: dark matter (\acs{DM}) cores in halos. These are often referred to as \emph{fuzzy} \acs{DM} models, and require ultralight bosonic fields (see, \eg, Refs.~\cite{Robles:2012uy,Hui:2016ltb,Bar:2019bqz,Bar:2018acw,Desjacques:2019zhf,Davoudiasl:2019nlo,Hui:2021tkt}, but the literature on the subject is very large and growing).

In Part~\ref{pt:ultralight} (and Appendix~\ref{app:Qball}), we consider two different theories of scalar fields, yielding localized objects with a static energy-density profile, but with a time-periodic scalar. The first theory describes a self-gravitating massive scalar, and the resulting objects are known as boson stars~\cite{Kaup:1968zz,Ruffini:1969qy,Liebling:2012fv}. Newtonian boson stars (\acsp{NBS}) made of very light fields (in particular, bosons with a mass~$\sim 10^{-22}\,\textrm{eV}$) are good descriptions of most cores of \acs{DM} halos; thus, this is an especially exciting simple theory to consider. 
The second theory (in Appendix~\ref{app:Qball}) describes a nonlinearly-interacting scalar in flat space, yielding solutions known as Q-balls: non-topological solitons which arise in a large family of field theories admitting a conserved charge~$Q$, associated with some continuous internal symmetry~\cite{Coleman:1985ki}. Q-balls seem to arise generically in supersymmetric field theories and may contribute significantly to the \acs{DM} content of our Universe~\cite{Kusenko:1997si,Frieman:1988}. In this thesis, they will serve just as an additional example of a scalar configuration to which the formalism presented in this chapter can be directly applied.

\paragraph{Stirring-up DM.}
The study of the dynamics of such objects is interesting for a number of reasons. As \acs{DM} candidates, it is important to understand the stability of such configurations, and the way they interact with surrounding bodies (stars, \acsp{BH}, etc)~\cite{Macedo:2013qea,Khlopov:1985}. For example, the mere \emph{presence} of a star or planet will change the local \acs{DM} density. In which way?
The motion of a compact binary can, in principle, stir the surrounding \acs{DM} to such an extent that a substantial emission of scalars takes place. How much, and how is it dependent on the binary parameters? 
When a star crosses one of these extended bosonic configurations, it may change its properties to the extent that the configuration simply collapses or disperses; in the eventuality that it settles down to a new configuration, it is important to understand the timescales involved. Such processes are specially interesting in the context of the growth of \acs{DM} halos and supermassive \acsp{BH}. Baryonic matter, in fact, tends to slowly accumulate near the center of a \acs{DM} structure, where it may eventually collapse to a massive \acs{BH}. Gravitational collapse can impart a recoil velocity $v_\text{recoil}$ to the \acs{BH} of the order of $300\,\text{km/s}$~\cite{1973ApJ...183..657B}, leaving the \acs{BH} in an damped oscillatory motion through the \acs{DM} halo, with respect to its center, with a crossing timescale
\begin{align}
\tau_\text{cross}=\sqrt{\frac{3\pi}{G\rho}}\sim 1.4\times 10^6 \,\text{yr}\sqrt{\frac{10^3M_{\odot}\,\text{pc}^{-3}}{\rho}}\,,
\end{align}
and an amplitude
\begin{align}
{\cal A}\sim 69\,\text{pc} \sqrt{\frac{10^3M_{\odot}\text{pc}^{-3}}{\rho}}\frac{v_\text{recoil}}{300 \, \text{km/s}}\,.
\end{align}
The damping is due to dynamical friction (\acs{DF}) caused by stars and \acs{DM}; our results suggest that the \acs{DM} effects may be comparable to the one of stars in galactic cores.
Finally, massive objects traveling through scalar media can deposit energy and momentum in the surrounding scalar field due to gravitational interaction~\cite{Hui:2016ltb,Bernard:2019nkv,Cardoso:2019dte,Hui:2021tkt}.
Thus, it is important to quantify the \acs{DF} that bodies are subjected to when immersed in scalar structures, and to confirm existing estimates~\cite{Hui:2016ltb,Hui:2021tkt}. 

All of this applies also in the context where scalar structures are viewed as compact, and potentially strong, gravitational-wave (\acs{GW}) sources, when they could mimic \acsp{BH}, or simply be new sources on their own right~\cite{Cardoso:2019rvt,Palenzuela:2017kcg}. Additionally, we expect some of these findings to be also valid in theories with a massive vector or tensor.

\paragraph{Gravitational-wave astronomy and DM.}
Understanding the behavior of \acs{DM} when moving perturbers drift by, or when a binary inspirals within a \acs{DM} medium
is crucial for attempts at detecting \acs{DM} via \acsp{GW}. In the presence of a nontrivial environment accretion, \acs{DF} and the self-gravity of the medium all contribute to a small, but potentially observable, change of the \acs{GW} phase~\cite{Eda:2013gg,Macedo:2013qea,Barausse:2014tra,Hannuksela:2018izj,Cardoso:2019rou,Baumann:2019ztm,Kavanagh:2020cfn}. Understanding the backreaction on the environment seems to be one crucial ingredient in this endeavor, at least for equal-mass mergers and when the Compton wavelength of \acs{DM} is very small~\cite{Kavanagh:2020cfn}.

\paragraph{Screening mechanisms.} Our results and methods can be of direct interest also for theories with screening mechanisms, where new degrees of freedom -- usually scalars --
are screened, via nonlinearities, on some scales~\cite{Babichev:2013usa}. Such mechanisms do give rise to nontrivial profiles for the new degrees of freedom,
for which many of the tools we use here should apply (see also Ref.~\cite{Brito:2014ifa}).

\begin{figure}	
	\centering
	\includegraphics[width=0.75\linewidth]{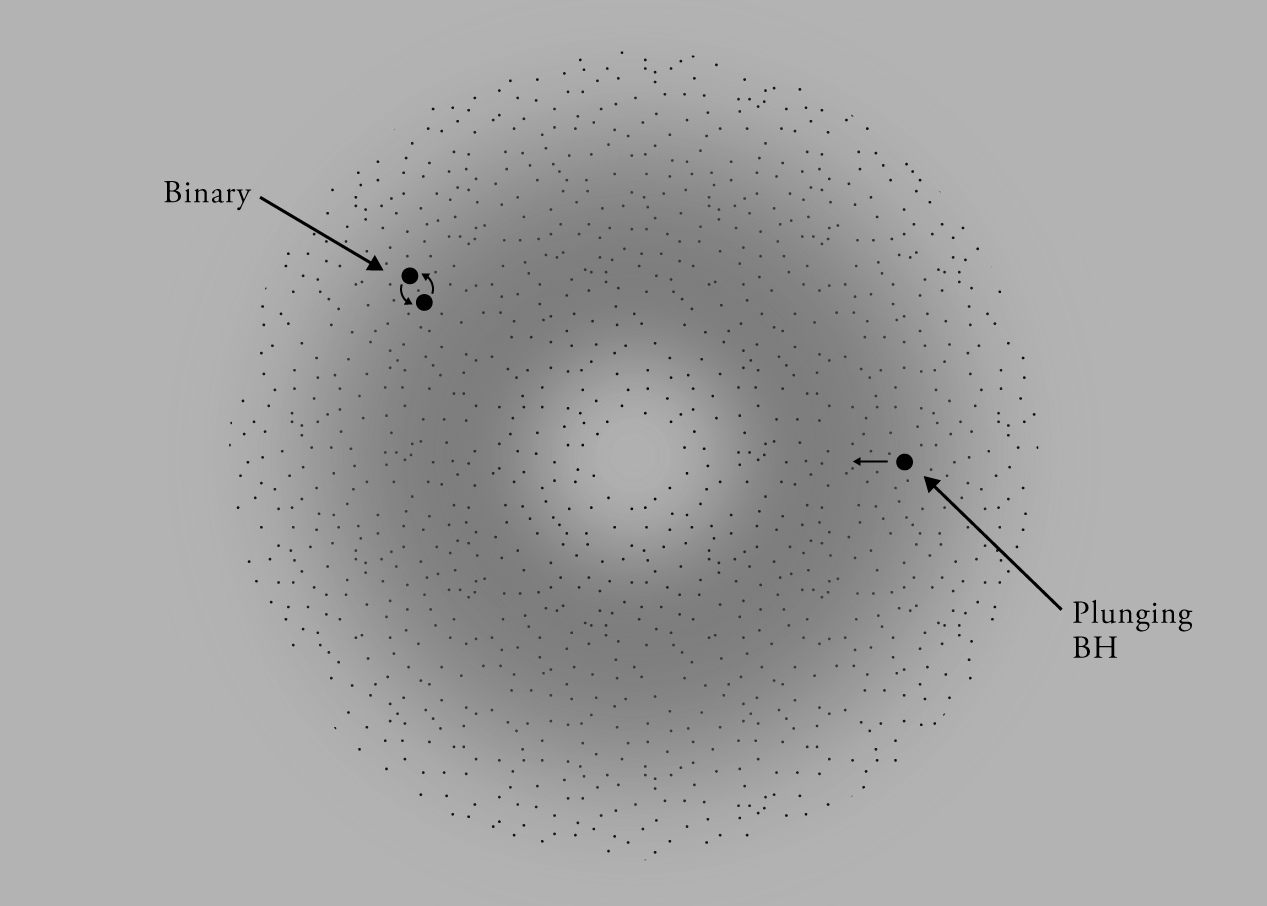} 
	\caption{An equatorial slice of our setup, where a binary of two \acsp{BH} or stars is orbiting inside a \acs{NBS}, and a single \acs{BH} is plunging through it.
	Our formalism is able to accommodate both scenarios, and others.
	The \acs{NBS} scalar field is pictured in gray dots, and forms a large spherical configuration. The motion of the binary or of the plunging \acs{BH} or star stirs the scalar profile, excites the \acs{NBS} modes and may
	eject some scalar field. All these quantities are computed in the main body of Part~\ref{pt:ultralight}.}\label{fig:anatomy}
\end{figure}

In Part~\ref{pt:ultralight}, we wish to provide the answers to the above questions. 
This work studies the response of localized scalar configurations to bodies moving in their vicinities. The setup is depicted in Fig.~\ref{fig:anatomy}. The moving external bodies are modeled as point-like.
Such approximation is a standard and successful tool in \acs{BH} perturbation theory~\cite{Zerilli:1971wd,Davis:1971gg,Barack:2018yvs}, in seismology~\cite{Ari} or in calculations of \acs{DF} by fluids~\cite{Ostriker:1998fa,Vicente:2019ilr}. In this approximation one loses
small-scale information. 
For light fields -- those we focus on -- the Compton wavelength of the field is much larger than the size of stars, planets or \acsp{BH}. In other words, we do not expect to lose important details of the physics at play (as it is confirmed by the results in Chapter~\ref{ch:movingBH}). The extrapolation of our results to moving \acsp{BH} or \acs{BH} binaries (\acsp{BHB}) should yield sensible answers.

In this chapter we explain the framework that will be used to study the issues described above.

\section*{Theory}

In Part~\ref{pt:ultralight} we consider a general \emph{global} U(1)-invariant, self-interacting, complex scalar field $\Phi(x^\alpha)$, minimally coupled to gravity, described by the action~\eqref{theory_action} (with~$J_V^{\;\;\alpha}=A^\alpha=0$ and~$J_S=0$).
Due to the invariance under global U(1) transformations this theory admits the Noether current given in Eq.~\eqref{NoetherCurrent} and the conserved Noether charge given in Eq.~\eqref{NoetherCharge}. We shall interpret this charge as the number of scalar particles in the system. 
The energy-momentum tensor of the scalar field is given in Eq.~\eqref{scalarEMT} and the energy contained in a given spacelike hypersurface~$\mathcal{S}$ is obtained using the timelike Killing vector field~$-X^\alpha=-(\partial_t)^\alpha$ through Eq.~\eqref{energy_angularmom}, resulting in
\begin{align} \label{energy_spacelike}
	E=\int_\mathcal{S} dV_ 3 T^S_{\;\;t \alpha} N^\alpha\,,
\end{align}
where~$N^\alpha$ is a future-pointing unit normal to~$\mathcal{S}$.
Analogously, the scalar field angular momentum (along~$z$) is obtained using the spacelike Killing vector field~$Y^\alpha=(\partial_\varphi)^\alpha$ and it is given by
\begin{align}
L^z=-\int_\mathcal{S} dV_ 3 T^S_{\;\;\varphi \alpha} N^\alpha\,.
\end{align}

\section*{Objects}

We are interested in spherically symmetric, time-periodic, localized solutions of the field equations. These will be describing, for example, new \acs{DM} stars or the core of \acs{DM} halos.
We take the following ansatz for the scalar in such a configuration,
\begin{equation} 
\Phi_0=\Psi_0(r)e^{-i\Omega t}\,,\label{BKG_ansatz}
\end{equation}
where~$\Psi_0$ is a real-function satisfying~$\partial_r \Psi_0(0)=0$ and~$\lim_{r\to \infty} \Psi_0=0$. 

Our primary target are self-gravitating solutions; when gravity is included, a simple minimally coupled massive field is able to self-gravitate. Thus, we consider minimal boson stars -- self-gravitating configurations of scalar field in curved spacetime with a simple mass potential~\eqref{mass_pot_S}.
In this thesis, for simplicity, we restrict to the Newtonian limit of these objects, where gravity is not very strong. So, we study \acsp{NBS}.

However, many of the technical issues in dealing with \acsp{NBS} are also present in theories where gravity is neglected.
So, we will also consider Q-balls~\cite{Coleman:1985ki} -- objects made of a nonlinearly-interacting scalar field in flat spacetime. For these objects, we use the Minkowski spacetime metric~$\eta_{\alpha \beta}$ and restrict to the class of nonlinear potentials 
\begin{equation} 
\mathcal{U}_\text{Q}=\frac{\mu_S^2}{2}|\Phi|^2\left(1-\frac{|\Phi|^2}{\Phi_c^2}\right)^2\,,\label{Potential_Qball}
\end{equation}
where $\Phi_c$ is a real free parameter of the theory.

We are ultimately interested not in the objects \emph{per se}, but rather on their dynamical response to external agents.
The response to external perturbers is taken into account by linearizing over the spherically symmetric, stationary background,
\begin{equation} 
\Phi=\left[\Psi_0(r)+\delta \Psi(t,r,\theta, \varphi)\right] e^{-i \Omega t}\,,\label{Perturbation}
\end{equation}
with the assumption~$|\delta \Psi|\ll 1$, where~$\Psi_0$ is the radial profile of the unperturbed object. Then, the perturbation~$\delta \Psi$ allows us to obtain all the physical quantities of interest, like the modes of vibration of the object, or the energy, linear and angular momenta radiated in a given process. This approach has a range of validity ($|\delta \Psi|\ll 1$) which can be controlled by selecting appropriately the perturber. As we show in the next section,~$\delta \Psi \propto m_p \mu_S$, where~$m_p$ is the rest mass of the external perturber. Since our results scale simply with~$m_p$, it is always possible to find an external source whose induced dynamics always fall in our perturbative scheme.

The energy-momentum tensor of a generic point particle perturber is given by~\eqref{particleEMT}, which in Schwarzschild (spherical) coordinates can be written as
\begin{equation}
T_p^{\;\;\alpha \beta}=m_p \frac{u^\alpha u^\beta}{u^t} \frac{\delta\left(r-r_p(t)\right)}{r^2}\frac{\delta\left(\theta-\theta_p(t)\right)}{\sin\theta} \delta\left(\varphi-\varphi_p(t)\right) \,,\label{Stress_energy_particle}
\end{equation}
where $u^\alpha \equiv dx_p^\alpha/d\tau$ is the perturber's 4-velocity and
\begin{equation}
x_p^\mu(t)=(t,r_p(t),\theta_p(t),\varphi_p(t))
\end{equation}
is a parametrization of its world-line. 

\section*{Fluxes}

The energy, linear and angular momenta carried by the radiated scalar field can be obtained by computing the flux of certain currents through a 2-sphere at spatial infinity. These currents are obtained from the scalar field energy-momentum tensor.

First, we decompose the fluctuations as
\begin{equation} 
\delta \Psi=\sum_{l,m}\int \frac{d \omega}{\sqrt{2 \pi}\, r} \left[Z_1^{\omega l m }Y_{l m} e^{-i\omega t}+\big(Z_2^{\omega l m }\big)^*Y_{l m}^* e^{i\omega t}\right]\label{Decomposition}\,,
\end{equation}
where $Y_{lm}(\theta,\varphi)$ is the spherical harmonic function of degree~$l$ and azimuthal number~$m$, and $Z_1(r)$ and $Z_2(r)$ are radial complex-functions.~\footnote{It should be noted that~$Z_1$ and~$Z_2$ are not linearly independent. In particular, for the setups considered in Part~\ref{pt:ultralight}, we can show that~$Z_1(\omega,l,m;r)=(-1)^mZ_2(-\omega,l,-m;r)^*$. For generality, we do not consider any constraint on the relation between these functions.} This decomposition can be rewritten in the equivalent form
\begin{align} 
\delta \Psi&=\sum_{l,m}\int \frac{d \omega}{\sqrt{2 \pi}\, r}Y_{l m} e^{-i \omega t} \left[Z_1(\omega,l,m;r)+(-1)^m Z_2(-\omega,l,-m;r)^*\right]\,.\label{Decompositionv2}
\end{align}
Unless strictly needed, hereafter, we omit the labels~$\omega$,~$l$ and~$m$ in the functions~$Z_1^{\omega l m}(r)$ and~$Z_2^{\omega l m}(r)$ to simplify the notation.
For a source vanishing at spatial infinity, we will see in the following chapters that one has the asymptotic form for the fields
\begin{align}
Z_1(r \to \infty) &\sim Z_1^\infty e^{i \epsilon_1 \left(\sqrt{\left(\omega+\Omega\right)^2-\mu^2}\right) r}\,, \label{Asymptotics1}\\
Z_2(r \to \infty) &\sim Z_2^\infty e^{i \epsilon_2 \left(\sqrt{\left(\omega-\Omega\right)^2-\mu^2}\right)^* r}\,,\label{Asymptotics2}
\end{align}
where~$\epsilon_1\equiv \text{sign}(\omega+\Omega+ \mu)$ and~$\epsilon_2\equiv \text{sign}(\omega-\Omega-\mu)$, and~$Z_1^\infty$ and~$Z_2^\infty$ are complex amplitudes which depend on the source. We choose the signs~$\epsilon_1$ and~$\epsilon_2$ to enforce the Sommerfeld radiation condition at large distances.~\footnote{By Sommerfeld condition we mean either: (i) outgoing group velocity for propagating frequencies; or, (ii) regularity for bound frequencies.}

A scalar field fluctuation gives rise to a perturbation in its energy-momentum tensor, which, at leading order and asymptotically, is given by
\begin{align}
\delta T^{\;\;\alpha \beta}_S(r\to \infty)\sim\partial^{(\alpha}\delta \Phi^* \partial^{\beta)}\delta \Phi-\frac{1}{2}\eta^{\alpha \beta}\left[\partial_\delta \delta \Phi^* \partial^\delta \delta \Phi +\mu_S^2|\delta \Phi|^2 \right] \,,
\end{align}
with $\delta \Phi \equiv e^{-i \Omega t} \delta \Psi$.
Then, the outgoing flux of energy (carried by the scalar field) at an instant~$t$ through a 2-sphere at infinity is
\begin{align}
\dot{E}^\text{rad}= \lim_{r\to \infty} r^2 \int d\theta d\varphi \sin \theta\, \delta T_S^{\;\;r t}  \,.
\end{align}
Plugging the asymptotic fields~\eqref{Asymptotics1} and~\eqref{Asymptotics2} in the last expression, it is straightforward to show that the total energy radiated with frequency in the range between~$\omega$ and~$\omega+d\omega$ is
\begin{align}
\frac{dE^\text{rad}}{d\omega}&=\left|\omega+\Omega\right| \Re\left[\sqrt{(\omega+\Omega)^2-\mu_S^2}\right]  \nonumber \\
&\times\sum_{l,m}\left|Z_1^{\infty}(\omega,l,m)+(-1)^m Z_2^{\infty}\left(-\omega,l,-m\right)^*\right|^2\,,\label{Energy_flux}
\end{align}
with~$\Re[z]$ the real part of a complex-number~$z$.
In deriving the last expression we considered a process in which the small perturber interacts with the background configuration during a finite amount of time. In the case of a (eternal) periodic interaction (\eg, small particle orbiting the scalar configuration) the total energy radiated is not finite. However, we can compute the averaged rate of energy emission in such processes, obtaining
\begin{align} 
\dot{E}^\text{rad}&=\int \frac{d\omega}{2\pi}\left|\omega+\Omega\right| \Re\left[\sqrt{(\omega+\Omega)^2-\mu_S^2}\right]  \nonumber \\
&\times\sum_{l,m}\left|Z_1^{\infty}(\omega,l,m)+(-1)^m Z_2^{\infty}\left(-\omega,l,-m\right)^*\right|^2\,.\label{Energy_flux_rate}
\end{align}
The last expression must be used formally; as we will see, the amplitudes~$Z_1^\infty$ and~$Z_2^\infty$ contain Dirac delta functions in frequency $\omega$. The correct way to proceed is to substitute the product of compatible delta functions by just one of them, and the incompatible by zero.~\footnote{A rigorous derivation can be done by applying the formalism directly to a specific process. For generality, we let~\eqref{Energy_flux_rate} as it is.} 
The outgoing flux of linear momentum carried by the scalar field at instant~$t$ is
\begin{equation} 
\dot{P}^\text{rad}_i= \lim_{r\to \infty} r^2 \int  d\theta d\varphi \sin \theta\, \delta T_{S\;\,\mu}^{\;\;r} \boldsymbol{e}_i^\mu\,,\label{momentum}
\end{equation}
with $i=\{x,y,z\}$ and where $\boldsymbol{e}_x$, $\boldsymbol{e}_y$, $\boldsymbol{e}_z$ are (mutually orthogonal) unit spacelike vectors in the~$x$,~$y$,~$z$ directions, respectively. These can be written as
\begin{align}
\boldsymbol{e}_x^\alpha&= \sin \theta \cos \varphi  (\partial_r)^\alpha+ \frac{\cos \theta \cos\varphi}{r} (\partial_\theta)^\alpha -\frac{\sin \varphi}{r \sin \theta} (\partial_\varphi)^\alpha\,,  \\
\boldsymbol{e}_y^\alpha&= \sin \theta \sin \varphi  (\partial_r)^\alpha+ \frac{\cos \theta \sin\varphi}{r} (\partial_\theta)^\alpha +\frac{\cos \varphi}{r \sin \theta} (\partial_\varphi)^\alpha\,,  \\
\boldsymbol{e}_z^\alpha&= \cos \theta (\partial_r)^\alpha- \frac{\sin \theta}{r} (\partial_\theta)^\alpha  \,,
\end{align}
For an axially symmetric process (with respect to the~$z$ axis) there are only modes with azimuthal number~$m=0$ composing the scalar field fluctuation~\eqref{Decomposition}. In that case, using the asymptotic fields~\eqref{Asymptotics1} and~\eqref{Asymptotics2}, one can show that the total linear momentum radiated along~$z$ with frequency in the range between $\omega$ and $\omega+d\omega$ is~\footnote{Additionally, it is straightforward to show that no linear momentum is radiated along~$x$ and~$y$ in an axially symmetric process.}
\begin{align}
\frac{d P_z^\text{rad}}{d \omega}&=\sum_l\frac{2(l+1)\Theta\left[\left(\omega+\Omega\right)^2-\mu_S^2\right]\left|(\omega+\Omega)^2-\mu_S^2\right|}{\sqrt{(2l+1)(2l+3)}} \nonumber \\
&\qquad\times\left[\Lambda_{11}(\omega,l)+2\Lambda_{12}(\omega,l)+\Lambda_{22}(\omega,l)\right]\,,\label{Momentum_flux}
\end{align}
where $\Theta(x)$ is the Heaviside step function and we defined the functions
\begin{align}
\Lambda_{11}(\omega,l)&\equiv\Re\Big[Z_1^\infty(\omega,l,0)Z_1^\infty(\omega,l+1,0)^*\Big]\,,\\
\Lambda_{12}(\omega,l)&\equiv \Re\Big[Z_1^\infty(\omega,l,0)Z_2^\infty(-\omega,l+1,0)\Big]\,,\\
\Lambda_{22}(\omega,l)&\equiv \Re\Big[Z_2^\infty(-\omega,l+1,0)Z_2^\infty(-\omega,l,0)^*\Big]\,.
\end{align}
Finally, the outgoing flux of angular momentum along~$z$ carried by the scalar field at instant~$t$ is
\begin{align} 
\dot{L}^\text{rad}_z = \lim_{r\to \infty} r^2 \int d\theta d\varphi \sin \theta\, \delta T_S^{\;\;r \varphi} \,.\label{angularm}
\end{align}
Plugging the asymptotic fields~\eqref{Asymptotics1} and~\eqref{Asymptotics2} in the last expression, it can be shown that the total angular momentum along~$z$ radiated with frequency in the range between $\omega$ and $\omega+d\omega$ is 
\begin{align}
\frac{d L_z^\text{rad}}{d\omega}&=\epsilon_ 1\Re\left[\sqrt{(\omega+\Omega)^2-\mu_S^2}\right]  \nonumber \\
&\times\sum_{l,m}m\left|Z_1^{\infty}(\omega,l,m)+(-1)^m Z_2^{\infty}\left(-\omega,l,-m\right)^*\right|^2\,.\label{AngularMomentum_flux}
\end{align}
In the case of a periodic process, the angular momentum along~$z$ is radiated at a rate given by
\begin{align}
\dot{L}_z^\text{rad}&=\int \frac{d \omega}{2\pi}\,\epsilon_ 1\Re\left[\sqrt{(\omega+\Omega)^2-\mu_S^2}\right]  \nonumber \\
&\times\sum_{l,m}m\left|Z_1^{\infty}(\omega,l,m)+(-1)^m Z_2^{\infty}\left(-\omega,l,-m\right)^*\right|^2\,.
\label{AngularMomentum_flux_rate}
\end{align}
We can also compute how many scalar particles cross a 2-sphere at spacial infinity per unit of time. This is obtained by
\begin{equation}
\dot{Q}^\text{rad}= \lim_{r\to \infty} r^2 \int d\theta d\varphi \sin \theta\, \delta J_Q^{\;\;r}  \,,\label{numberf}
\end{equation}
with
\begin{equation}
\delta J_Q^{\;\;r}(r \to \infty)\sim \frac{1}{\hbar}\,\Im\left(\delta \Phi^*\partial^r \delta \Phi\right)\,,
\end{equation}
(see Eq.~\eqref{NoetherCurrent}), with~$\Im[z]$ the imaginary part of the complex-number~$z$. Using the asymptotic fields~\eqref{Asymptotics1} and~\eqref{Asymptotics2}, we can show that the total number of particles radiated in the range between~$\omega$ and~$\omega+d\omega$ is
\begin{align}
\frac{d Q^\text{rad}}{d \omega}&=\frac{\epsilon_1}{\hbar} \Re\left[\sqrt{(\omega+\Omega)^2-\mu_S^2}\right]  \nonumber \\
&\times\sum_{l,m}\left|Z_1^{\infty}(\omega,l,m)+(-1)^m Z_2^{\infty}\left(-\omega,l,-m\right)^*\right|^2\,.\label{Particles_flux}
\end{align}
This gives us a simple interpretation for expressions~\eqref{Energy_flux} and~\eqref{AngularMomentum_flux}. The spectral flux of scalar field energy is just the product between the spectral flux of scalar particles and their individual energy $\hbar(\Omega+\omega)$; similarly, the spectral flux of scalar field angular momentum is the product between the number of scalar particles radiated with azimuthal number~$m$ and their individual angular momentum -- which is~$\hbar m$. For a periodic interaction, scalar particles are radiated at an average rate
\begin{align}
\dot{Q}^\text{rad}&=\int \frac{d \omega}{2\pi \hbar}\,\epsilon_ 1\Re\left[\sqrt{(\omega+\Omega)^2-\mu_S^2}\right]  \nonumber \\
&\times\sum_{l,m}\left|Z_1^{\infty}(\omega,l,m)+(-1)^m Z_2^{\infty}\left(-\omega,l,-m\right)^*\right|^2\,.
\label{Particles_flux_rate}
\end{align}

Now, what is the relation between the radiated fluxes and the energy and momenta lost by the massive perturber ($E^\text{lost}$,~$P_z^\text{lost}$,~$L_z^\text{lost}$). 
Noting that both the energy and momenta of the scalar configuration may change due to the interaction, by conservation of the total energy and momenta we know that
\begin{align} 
E^\text{lost}&=\Delta E+E^\text{rad}\,,  \label{LossRadE}\\
P_z^\text{lost}&=\Delta P_z+P_z^\text{rad}\,, \label{LossRadP} \\
L_z^\text{lost}&=\Delta L_z+L_z^\text{rad}\,, \label{LossRadL}	
\end{align}
where~$\Delta E$,~$\Delta P_z$ and~$\Delta L_z$ are the changes in the energy and momenta of the configuration.
So, having the radiated fluxes, the task of determining the energy and momenta lost by the perturber reduces to computing the change in the respective quantities of the scalar configuration.

In a perturbation scheme it is hard to aim at a direct calculation of these changes, because in general, at leading order, they include second order fluctuations of the scalar -- terms mixing~$\Phi_0$ with~$\delta^2 \Phi$; this issue is not present in the radiated fluxes, since~$\Phi_0$ is suppressed at infinity (so, the only terms present are of the form~$(\delta \Phi)^2$ and all we need to know is~$\delta \Phi$).
Nevertheless, for certain setups we can compute \emph{indirectly} the change in the configuration's energy~$\Delta E$. Let us see an example.
A massive perturber interacting with the scalar only through gravitation is described by a globally U(1)-invariant action; so, Noether's theorem implies that
\begin{align}
\nabla_ \mu \,J_Q^{\;\;\mu}=0\,.
\end{align}
Using the divergence theorem, we obtain that the number of scalar particles is \emph{conserved},
\begin{align} \label{DeltaQ}
\Delta Q&=-Q^\text{rad}\,, 
\end{align}
\ie, the number of particles lost by the configuration matches the number of radiated particles -- no scalar particles are created.
If, additionally, we can express the change in the configuration's energy~$\Delta E$ only in terms of the change in the number of particles~$\Delta Q$ -- as it happens to be the case of \acsp{NBS} (at leading order) -- we are able to compute~$\Delta M$ from the number of radiated particles~$Q^\text{rad}$; so, we obtain the energy lost by the perturber~$E^\text{lost}$ using only radiated fluxes. The lost momenta~$P_z^\text{lost}$ and~$L_z^\text{lost}$ can, then, be obtained through the energy-momenta relations; for example, a non-relativistic massive perturber moving along~$z$ satisfies
\begin{align} 
E^\text{lost}&= \frac{\left(m_p v_\text{i}\right)^2-\left(m_p v_\text{i}-P_z^\text{lost}\right)^2}{2 m_p} \nonumber \\
&=P_z^\text{lost} v_\text{i}-\frac{(P_z^\text{lost})^2}{2 m_p}\,, \label{EvsPloss}
\end{align}
where~$v_\text{i}$ is the initial velocity along~$z$.
Finally, we can compute the change in the scalar configuration momenta~$\Delta P_z$ and~$\Delta L_z$ using Eqs.~\eqref{LossRadP} and~\eqref{LossRadL}. 

The conservation of the number of scalar particles (\ie, Noether's theorem) plays a key role in our scheme; it allows us to compute the change in the number of particles in the scalar configuration -- a quantity that involves the second order fluctuation~$\delta^2\Phi$ -- using only the first order fluctuation~$\delta \Phi$. When the perturber couples \emph{directly} with the scalar via a scalar interaction that breaks explicitly the global U(1) symmetry -- like the coupling~$J_S=T_p$ (with $T_p= \eta_{\alpha \beta} T_p^{\alpha \beta}$) in \eqref{theory_action} -- the number of scalar particles is not conserved; the perturber can create and absorb scalar particles. In that case, our scheme fails and it is not obvious how to circumvent this issue to calculate~$\Delta E$ (without calculating the second order fluctuations). 
In the following chapters of Part~\ref{pt:ultralight} (except in Chapter~\ref{ch:movingBH}) we apply explicitly the scheme described above to compute the energy and momentum lost by massive compact objects (\eg, a \acs{BH} binary) moving through a \acs{NBS}.

%*****************************************
%*****************************************
%*****************************************
%*****************************************
%*****************************************
%************************************************
\chapter{Newtonian Boson Stars}\label{ch:nbs}
%************************************************

In this chapter we consider the simplest theory of a scalar giving rise to self-gravitating objects. The theory is that of a minimally coupled massive field,
or even with higher order interactions, but taken at Newtonian level. The objects themselves -- Newtonian boson stars (\acsp{NBS}) -- have been studied for decades, either as \acs{BH} mimickers, as toy models for more complicated exotica that could exist, or as realistic configurations that can describe \acs{DM}~\cite{Kaup:1968zz,Ruffini:1969qy,Liebling:2012fv}. Despite the intense study and the recent activity at the numerical relativity level~\cite{Cardoso:2016oxy,Helfer:2018vtq,Palenzuela:2017kcg,Sanchis-Gual:2019ljs,Bezares:2018qwa,Sanchis-Gual:2018oui,Widdicombe:2019woy}, their interaction with smaller objects (describing, for example, stars piercing through or orbiting such \acsp{NBS}) has hardly been studied. The variety and disparity of scales in the problem makes it ill-suited for full-blown numerical techniques, but ideal for perturbation theory.

%%%%%%%%%%%%%%%%%%%%%%%%%%%%%%%%%%%%%%
\section{Background configurations}
%%%%%%%%%%%%%%%%%%%%%%%%%%%%%%%%%%%%%%

The~\acs{EOM} describing the scalar field is Eq.~\eqref{KG_EOM} (with~$J_S=0$, since in Part~\ref{pt:ultralight} we are interested in a scalar interacting only through gravity),
\begin{equation}
	\Box \Phi= \mu_S^2 \Phi\,.
\end{equation}
We are using~$\mathcal{U}_S\simeq \mu_ S^2 \left| \Phi\right|^2/2$ because we want to consider a diluted (\ie, weak) scalar field. The~\acs{EOM} describing the spacetime metric is the Einstein equation~\eqref{Eins_EOM} (neglecting the cosmological constant~$\Lambda=0$ and with energy-momentum tensor~$T^{\alpha \beta}=T_S^{\;\;\alpha \beta}$ given in Eq.~\eqref{scalarEMT}),
\begin{align}
	G^{\alpha \beta}=8 \pi \,T_ S^{\;\;\alpha \beta}\,.
\end{align}
We are interested in localized solutions of this model with a scalar field of the form~\eqref{BKG_ansatz}, with frequency
\begin{equation}
\Omega = \mu_S-\gamma\,.
\end{equation}
in the limit $0<\gamma\ll\mu$. These are the so-called \acsp{NBS}. Note that the energy eigenstate~$\hbar\Omega$ (which is of the order of -- but not equal to -- the individual scalar particles energy forming the \acs{NBS}~\cite{Chavanis1,Hui:2016ltb}) is approximately given by the rest mass energy $\hbar \mu_S$.
As discussed in the end of Chapter~\ref{ch:theory}, in the Minkowskian (weak gravitational field~$U\sim \mathcal{O}(\epsilon^2)$) limit and for a Newtonian (small velocity) scalar field, the spacetime metric (at leading order) is~(\eg, \cite{poisson_will_2014})
\begin{equation}
ds^2=-\left(1+2U\right)dt^2+(1-2 U)dr^2+r^2\left(d\theta^2+\sin^2\theta d\varphi^2\right)\,,
\end{equation}
and the~\acsp{EOM} become
\begin{align}
i \partial_t\phi&=-\frac{1}{2 \mu_S} \nabla^2 \phi+ \mu_S U \phi\,,  \\
\nabla^2 U&= 4\pi \mu_S |\phi|^2\,, 
\end{align}
where the Schrödinger field~$\phi$ is related with the Klein-Gordon field~$\Phi$ through 
\begin{equation}
\Phi=\frac{e^{-i \mu_S t}}{\sqrt{\mu_S}}\phi \,.
\end{equation}
This is known as Schr\"{o}dinger-Poisson system (\eg,~\cite{Chavanis1}). 
Using ansatz~$\Phi_0$ in~\eqref{BKG_ansatz} for the scalar field $\Phi$, one finds 
\begin{align}
&\partial_r^2 \Psi_0+\frac{2}{r}\partial_r \Psi_0 - 2 \mu_S \left(\mu_S U+\gamma\right) \Psi_0=0\,, \label{EOM_BS_radial1}  \\
&\partial_r^2 U_0 +\frac{2}{r} \partial_r U_0- 4\pi \mu_S^2 \Psi_0^2=0\,,\label{EOM_BS_radial2}
\end{align}
with the constraints~$0<\gamma \ll \mu_S$,~$|U_0|\ll 1$ and~$|\Psi_0|\ll 1$. Remarkably, this system is left invariant under the transformation
\begin{align}
(\Psi_0,U_0,\gamma) \to \lambda^2 (\Psi_0, U_0, \gamma)\,,\quad  r \to r/\lambda\,.\label{eq:scaling}
\end{align}
These relations imply that the \acs{NBS} mass scales as $M_\text{NBS} \to \lambda M_\text{NBS}$ (as can be seen from~\eqref{M_NBS}).
This scale invariance is extremely useful, because it allows us to effectively ignore the constraints on~$\gamma$,~$U_ 0$ and~$\Psi_ 0$ when solving numerically Eqs.~\eqref{EOM_BS_radial1} and~\eqref{EOM_BS_radial2}; one can always rescale the obtained solution with a sufficiently small~$\lambda$, such that the constraints (\ie, the regime of validity of the Newtonian approximation) hold for the rescaled solution. Even more importantly is the fact that once a fundamental (\emph{ground state}) \acs{NBS} solution is found, all other fundamental Newtonian stars can be obtained through a rescaling of that solution; the same applies to any other particular excited state.  

\begin{figure}
\centering
\includegraphics[width=0.9\textwidth]{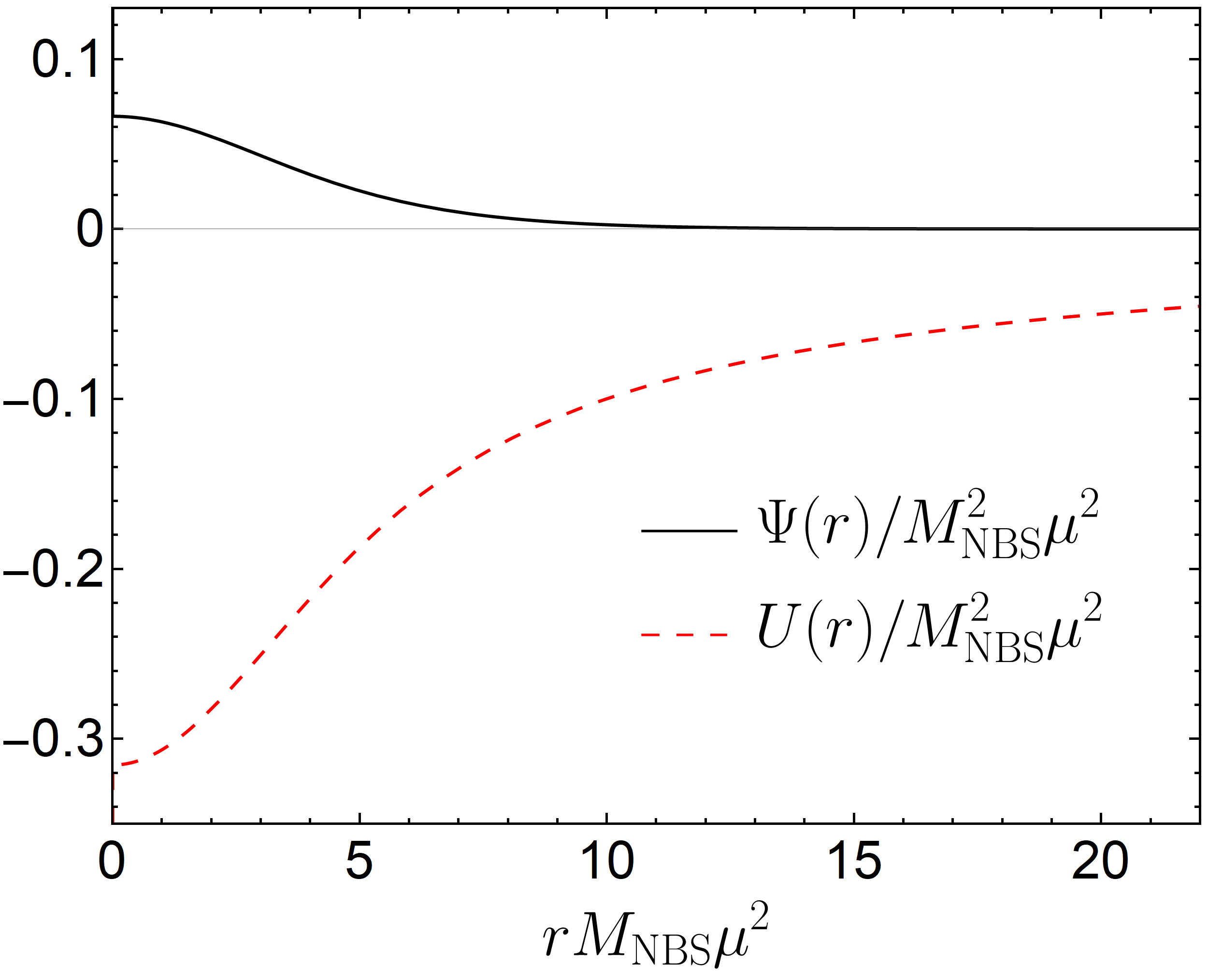} 
\caption{Universal radial profiles $\Psi_0(r)$ and $U_0(r)$ of the numerical solution of Eqs.~\eqref{EOM_BS_radial1} and~\eqref{EOM_BS_radial2} with appropriate boundary conditions. 
Due to the scaling \eqref{eq:scaling}, this profile describes \emph{all} the fundamental \acsp{NBS}. They are characterized by the re-scaling invariant quantity
$\gamma/(M_\text{NBS}^2\mu_S^3)\simeq0.162712$ and the mass-radius relation \eqref{eq:mass_radius_BS}.
}\label{fig:BS}
\end{figure}

A numerical solution of Eqs.~\eqref{EOM_BS_radial1} and~\eqref{EOM_BS_radial2}, with appropriate boundary conditions describing \emph{all} fundamental \acsp{NBS} is shown in Fig.~\ref{fig:BS}.~\footnote{In addition to the conditions on $\Phi_0$ described bellow Eq.~\eqref{BKG_ansatz},~$\partial_r \Phi_0(0)=0$ and~$\lim_{r\to \infty} \Phi_0=0$, we impose $\partial_r U_ 0(0)=0$ and $\lim_{r\to \infty} U_ 0=0$.} 
It is easy to see that, at large distances, the scalar decays exponentially as~$\Psi_0 \sim e^{-\sqrt{2\mu_S\gamma}r}/r$, whereas the Newtonian potential falls off as~$-M_\text{NBS}/r$.

Noting that (at leading order) the mass of a \acs{NBS} is given by
\begin{equation} \label{M_NBS}
M_\text{NBS}=4 \pi \mu_S^2 \int_{0}^{\infty}dr\, r^2 \left|\Psi \right|^2\,,
\end{equation}
it is possible to show numerically that, for a fundamental \acs{NBS}, 
\begin{align}
\frac{M_\text{NBS}}{M_\odot} \simeq 3 \times 10^{12}\, \lambda\left(\frac{10^{-22}\, \text{eV} }{\hbar \mu_S}\right)\,,
\end{align}
with a scaling parameter~$\lambda$, such that $\{\Psi_0,U_0,\gamma/\mu_S\} \sim \mathcal{O}(\lambda^2)$. If one is interested in describing a \acs{DM} core of mass~$M \sim 10^{10} M_\odot$, this can be achieved then via a fundamental \acs{NBS} made of self-gravitating scalar particles of mass $\hbar \mu_S \sim 10^{-22} \,\text{ eV}$, with a scaling parameter $\lambda \sim  10^{-2}$, which satisfies the Newtonian constraints. 

All the fundamental \acsp{NBS} satisfy the scaling-invariant mass-radius relation
\begin{equation}
M_\text{NBS}\mu_S=\frac{9.1}{R\mu_S}\,,\label{eq:mass_radius_BS}
\end{equation}
where the \acs{NBS} radius is defined as the radius of the sphere enclosing~$98\%$ of its mass. This result agrees well with previous
results in the literature~\cite{Liebling:2012fv,Boskovic:2018rub,Bar:2018acw,Membrado,Chavanis1,Chavanis2}. Comparing with some relevant scales, it can be written as
\begin{align}
\frac{M_\text{NBS}}{M_{\odot}}=9\times 10^9\,\frac{100\, \text{pc}}{R}\,\left(\frac{10^{-22}\,\text{eV}}{\hbar\mu_S}\right)^2\,.\label{eq:mass_radius_BS2}
\end{align}

Accurate fits for the profile of the scalar field are provided in Ref.~\cite{Kling:2017mif}. Unfortunately, these fits are defined by branches, and similar results for the gravitational potential are not discussed at length. We find that a good description of the gravitational potential of \acsp{NBS}, accurate to within~$1\%$ everywhere is the following:
\begin{align}
&U_0=\mu_S^2M_\text{NBS}^2f\,,\\
&f=\frac{a_0+11\frac{a_0}{r_1}x+\sum_{i=2}^{9}a_ix^i-x^{10}}{(x+r_1)^{11}}\,,\nonumber\\
&x=\mu_S^2M_\text{NBS}r\,, \qquad r_1=1.288\,,\nonumber\\
&a_0=-5.132\,,\quad a_2=-143.279\,,\quad a_3=-645.326\,,\nonumber\\
&a_4=277.921\,,\quad  a_5=-2024.838\,,\quad a_6=476.702\,,\nonumber\\
&a_7=-549.051,\,\quad a_8=-90.244\,,\quad a_9=-13.734\,. \nonumber
\end{align}
The (cumbersome) functional form was chosen such that it yields the correct large-$r$ behavior and the correct regular behavior at the \acs{NBS} center. For the scalar field we find the following $1\%$-accurate expression inside the star,
\begin{align}
&\Psi_0=\mu_S^2 M_\text{NBS}^2g\,,\\
&g=e^{-0.570459 x}\frac{\sum_{i=0}^{8}b_ix^i+b_fx^{9.6}}{(x+r_2)^{9}}\,, \nonumber\\
&x=\mu_S^2M_\text{NBS}r\,, \qquad r_2=1.182\,,\nonumber\\
&b_0=0.298\,,\quad b_1=2.368\,, \quad b_2=10.095\,,\nonumber\\
&b_3=12.552\,,\quad  b_4=51.469\,, \quad b_5=-8.416\,,\nonumber\\
&b_6=54.141,\,\quad b_7=-6.167\,,\quad b_8=8.089\,,\nonumber\\
&b_f=0.310 \nonumber\,.
\end{align}

Finally, for future reference, the number of particles contained in a \acs{NBS} is (at leading order)
\begin{equation}
Q_\text{NBS}=\frac{4 \pi}{\hbar} \mu_S \int_0^\infty dr\, r^2\left|\Psi_0\right|^2\,,
\end{equation}
and, then, at leading order, we can write the mass as~$M_\text{NBS}= \hbar\mu_S Q_\text{NBS}$.

%%%%%%%%%%%%%%%%%%%%%%%%%%%%%%%%%%%%%%%%%%%%%%%%%%%%
\section{Small perturbations} \label{sec:SmallPert}
%%%%%%%%%%%%%%%%%%%%%%%%%%%%%%%%%%%%%%%%%%%%%%%%%%%%

Small perturbations of the form~\eqref{Perturbation} to the scalar field, together with the \acs{NBS} perturbed gravitational potential
\begin{equation}
U=U_0(r)+\delta U (t,r,\theta, \varphi)\,,
\end{equation}
satisfy the linearized system of equations
\begin{align}
&i \partial_{t} \delta \Psi=-\frac{1}{2 \mu_S}\nabla^2 \delta \Psi+ \left(\mu_S U_0 +\gamma\right) \delta \Psi+ \mu_S \Psi_0 \delta U\,,\label{Sourced_SP_System1}\\
&\nabla^2 \delta U=4 \pi\left[2 \mu_S^2 \Psi_0 \Re\left(\delta \Psi\right)+P\right]\,,\label{Sourced_SP_System2}
\end{align}
where $U_0$ is the gravitational potential of the unperturbed star, and we have included an external point-like perturber~\footnote{We are assuming a non-relativistic external perturber. Note that~$P$ is just the non-relativistic limit of $T_p^{\,\,t t}$ given in Eq.~\eqref{Stress_energy_particle}.}
\begin{align}
P\equiv m_p \frac{\delta \left(r-r_p(t)\right)}{r^2} \frac{\delta\left(\theta-\theta_p(t)\right)}{\sin \theta} \delta\left(\varphi-\varphi_p(t)\right)\,.\label{source_BS}
\end{align}
This system of equations is valid for non-relativistic fluctuations (which satisfy $|\partial_t \delta \Psi|\ll \mu_S |\delta \Psi|$) that are sourced by a non-relativistic Newtonian perturber. If, additionally, $|\partial_t \delta \Psi|\gg \gamma |\delta \Psi|$ the above system of equation simplifies to
\begin{align}
&i \partial_{t} \delta \Psi=-\frac{1}{2 \mu_S}\nabla^2 \delta \Psi+ \mu_S \Psi_0 \delta U\,,\label{Sourced_SP_System11}\\
&\nabla^2 \delta U=4 \pi P\,,\label{Sourced_SP_System22}
\end{align}
\ie, the self-gravity of the perturbations can be neglected. 

This perturbation scheme assumes that~$\delta U \ll U_0$ and either: \textit{i)}$|\delta \Psi| \ll \Psi_0$ in the homogeneous case (without external perturber); or \textit{ii)} $m_p \mu_S\ll 1$ in the case of a perturbation induced by a point particle. In the latter case, the linear fluctuation~$\delta \Psi$ is the leading order contribution to a power expansion in~$m_p \mu_S$ of the \emph{total} perturbation to~$\Psi_0$. 
To study the homogeneous case, one can simply set $m_p=0$.
As shown in the end of Chapter~\ref{ch:theory}, in the Newtonian limit, the perturber couples to the scalar through the Poisson equation~\eqref{Poisson_Newtonian}. We neglect the backreaction on the perturber's motion and treat its world-line as given.

Now, let us decompose the fluctuations of the scalar field as in~\eqref{Decomposition}, and the gravitational potential and the source, respectively, as~\footnote{Note that the perturbation $\delta U$ must be real-valued. Again, we will omit the labels $\omega$, $l$ and $m$ in the functions $u^{\omega l m }(r)$ and~$p^{\omega l m }(r)$ to simplify the notation.}
\begin{align}
\delta U&=\sum_{l,m}\int \frac{d \omega}{\sqrt{2 \pi}\, r} \left[u^{\omega l m }Y_{lm} e^{-i\omega t}+\left(u^{\omega l m }\right)^*Y_{l m}^* e^{i\omega t}\right]\,,\\
P&= \sum_{l,m} \int \frac{d \omega}{\sqrt{2 \pi}\, r} \left[p^{\omega l m} Y_{l m} e^{-i \omega t}+\left(p^{\omega l m}\right)^* Y_{l m}^* e^{i \omega t}\right]\,,
\end{align}
where~$p^{\omega l m}$ are radial complex-functions defined by
\begin{equation} 
p^{\omega l m} \equiv \frac{r}{2 \sqrt{2 \pi}}\int dt d \theta d \varphi \sin \theta P\,  Y_{l m}^* e^{i \omega t}\,.\label{p_def}
\end{equation}
From equations~\eqref{Sourced_SP_System1} and~\eqref{Sourced_SP_System2} one can build the matrix equation
\begin{equation} 
\partial_r \boldsymbol{X} -V_\text{B}(r) \boldsymbol{X}= \boldsymbol{P}\,,\label{BS_Perturbation_Matrix_Sourced}
\end{equation}
with the column vector $\boldsymbol{X}\equiv (Z_1, Z_2, u, \partial_r Z_1, \partial_r Z_2, \partial_r u)^T$, the matrix $V_B$ given by 
\begin{equation}
\begin{pmatrix} 
	0 & 0& 0 & 1 & 0 & 0 \\
	0 & 0 & 0 & 0 & 1 & 0 \\
	0 & 0 & 0 & 0 & 0 & 1 \\
	V-2 \mu_S (\omega- \gamma) & 0 & 2 \mu_S^2\Psi_0 & 0 &0 & 0  \\
	0 & V+2 \mu_S (\omega+ \gamma) & 2 \mu_S^2\Psi_0 & 0 & 0 & 0 \\
	4\pi \mu_S^2 \Psi_0 & 4\pi \mu_S^2 \Psi_0 & V-2 \mu_S^2 U_0 & 0 & 0 & 0  
\end{pmatrix}\,.
\end{equation}
The radial potential~$V(r)$ is
\begin{equation}
V\equiv \frac{l(l+1)}{r^2}+2 \mu_S^2 U_0\,,
\end{equation}
and the (column vector) source term
\begin{equation}
\boldsymbol{P}(r)\equiv \left(0,0,0,0,0,4\pi p\right)^T\,.
\end{equation}
Note that the condition of non-relativistic fluctuations translates, here, into the simple inequality~$|\omega| \ll \mu_S$.

As suitable boundary conditions to solve for the perturbation we require both regularity at the origin,
\begin{align}
\hspace{-0.4cm}\boldsymbol{X}(r \to 0)\sim\left(a r^{l+1},b r^{l+1},c r^{l+1},a (l+1)r^l,b (l+1)r^l,c (l+1)r^l\right)^T\,,
\end{align}
with complex constants~$a$,~$b$ and~$c$, and the Sommerfeld radiation condition at infinity,
\begin{align}
\boldsymbol{X}(r \to \infty)\sim\left(Z_1^\infty e^{i k_1 r} ,Z_2^\infty e^{i k_2 r},u^\infty, i k_1 Z_1^\infty e^{i k_1r},i k_2 Z_2^\infty e^{i k_2r},0\right)^T\,, \label{BC_sommerfeld_infinity}
\end{align}
with 
\begin{align} \label{Wave_number_1}
k_1&\equiv \sqrt{2 \mu_S \left(\omega-\gamma\right)}\,, \\ \label{Wave_number_2}
k_2&\equiv -\left(\sqrt{-2 \mu_S \left(\omega+\gamma\right)}\right)^*\,,
\end{align}
where we are using the principal square root.

To calculate the perturbation we will make use of the set of independent homogeneous solutions $\{\boldsymbol{Z_{(1)}},\boldsymbol{Z_{(2)}},\boldsymbol{Z_{(3)}},\boldsymbol{Z_{(4)}},\boldsymbol{Z_{(5)}},\boldsymbol{Z_{(6)}}\}$, uniquely determined by
\begin{align}
&\boldsymbol{Z_{(1)}}(r \to 0)\sim \Big(r^{l+1},0,0,(l+1)r^l,0,0\Big)^T\,, \\
&\boldsymbol{Z_{(2)}}(r \to 0)\sim \Big(0,r^{l+1},0,0,(l+1)r^l,0\Big)^T\,, \\
&\boldsymbol{Z_{(3)}}(r \to 0)\sim \Big(0,0,r^{l+1},0,0,(l+1)r^l\Big)^T\,, \\
&\boldsymbol{Z_{(4)}}(r \to \infty)\sim \Big(e^{i k_1 r},0,0,i k_1 e^{i k_1 r},0,0\Big)^T\,, \\
&\boldsymbol{Z_{(5)}}(r \to \infty)\sim \Big(0,e^{i k_2 r},0,0,i k_2 e^{i k_2 r},0\Big)^T\,,  \\
&\boldsymbol{Z_{(6)}}(r \to \infty)\sim \Big(0,0,u^\infty,0,0,0\Big)^T\,.
\label{BC_BS}
\end{align}
Then, the square matrix
\begin{align}
F(r)\equiv\big(\boldsymbol{Z_{(1)}},\boldsymbol{Z_{(2)}},\boldsymbol{Z_{(3)}},\boldsymbol{Z_{(4)}},\boldsymbol{Z_{(5)}},\boldsymbol{Z_{(6)}}\big)
\label{eq:fundamental_matrix}
\end{align}
is known as the fundamental matrix of system~\eqref{BS_Perturbation_Matrix_Sourced}. As shown in Appendix~\ref{app:detF}, the determinant of~$F$ is independent of~$r$.

Finally, note that the homogeneous part of system~\eqref{BS_Perturbation_Matrix_Sourced} is invariant under the re-scaling
\begin{equation}
(U_0, \Psi_0,\gamma, \omega) \to \lambda^2 (U_0, \Psi_0, \gamma, \omega)\,,\quad r \to r/\lambda\,, \label{eq:scaling2}
\end{equation}
and, so, it can always be pushed into obeying the non-relativistic constraint. Additionally, for convenience, we can impose that~$\delta \Psi$ and~$\delta U$ are left invariant by the re-scaling, by performing the extra transformation
\begin{align}
(Z_{1,2}, u)\to \lambda^{-3}(Z_{1,2}, u)\,, \quad m_p\to\lambda^{-1} m_p \,.
\end{align}

It is easy to show that, for a process happening during a finite amount of time, the change in the \acs{NBS} energy, at leading order, is given by
\begin{align} \label{DeltaE}
\Delta E_\text{NBS}=\hbar \mu_S \Delta Q_\text{NBS}\,.
\end{align}
%

%%%%%%%%%%%%%%%%%%%%%%%%%%%%%%%%%%%%%%%%%%%%%%%%%%%%
\subsection{Validity of perturbation scheme}
%%%%%%%%%%%%%%%%%%%%%%%%%%%%%%%%%%%%%%%%%%%%%%%%%%%%
The perturbative scheme requires that $|\delta \Psi|\ll 1$, which can always be enforced by making $m_p \mu_S$ as small as necessary.
On the other hand, the background construction neglects higher-order post-Newtonian contributions. A self-consistent perturbative expansion requires that such neglected terms (of order $\sim U_0^2$) do not affect the dynamics of small fluctuations (of order~$\sim \delta U$). This imposes 
\begin{align}
m_p\gtrsim 10^4 M_{\odot}\left(\frac{M_\text{NBS}}{10^{10}M_\odot}\right)^3\left(\frac{\hbar \mu_S}{10^{-22}\,\text{eV}}\right)^2\,,
\end{align}
which holds true for many systems of astrophysical interest. As shown in Appendix~\ref{app:PN}, the scalar evolution equation \eqref{KG_all} is sourced by higher order Post-Newtonian terms. However,
these are nearly static, or very low frequency terms, hence will make a negligible contribution for high-energy binaries or plunges. In other words, the previous constraint can be substantially relaxed in dynamical situations, such as the ones we focus on.
Finally, the non-relativistic approximation requires the source to have a small frequency~$\lesssim 2\times 10^{-8}\text{Hz}\left(\hbar\mu_S/10^{-22}\text{eV}\right)$, in the case of a periodic motion. In Appendix~\ref{app:PN}, we show how to extend the formalism to include Newtonian but high frequency sources; in Section~\ref{BS_binaries}, we use that to calculate scalar emission by a high frequency binary. For plunges of nearly constant velocity~$v$ piercing through a \acs{NBS}, the non-relativistic approximation requires~$v\ll R\mu_S$. Fortunately, any \acs{NBS} has $R \mu_S \gg 1$ and the latter condition is easy to satisfy.

%%%%%%%%%%%%%%%%%%%%%%%%%%%%%%%%%%%%%%%%%%%%%%%%%%%%
\subsection{Sourceless perturbations}
%%%%%%%%%%%%%%%%%%%%%%%%%%%%%%%%%%%%%%%%%%%%%%%%%%%% 

\emph{Free oscillations} of \acsp{NBS} are fluctuations of the form
\begin{align}
\delta \Psi&= \frac{1}{\sqrt{2 \pi}\, r} \left[Z_1 Y_{l m} e^{-i \omega t}+Z_2^*Y_{l m}^* e^{i \omega^* t}\right]\,,  \\
\delta U&= \frac{1}{\sqrt{2 \pi}\, r} \left[u Y_{l m} e^{-i \omega t}+u^*Y_{l m}^* e^{i \omega^* t}\right]\,,
\end{align}
where~$Z_1$,~$Z_2$ and~$u$ are regular solutions of system~\eqref{BS_Perturbation_Matrix_Sourced} with~$P=0$, satisfying the Sommerfeld condition at infinity. These are also known as quasi-normal mode (\acs{QNM}) solutions, and the corresponding frequency~$\omega$ is the \acs{QNM} frequency.
Noting that the condition
\begin{equation}
\det (F)=0\,\label{eq:fundamental_matrix_condition}
\end{equation}
holds if and only if~$\omega$ is a \acs{QNM} frequency, we are able to find the \acs{NBS} proper oscillation modes by solving the sourceless system~\eqref{BS_Perturbation_Matrix_Sourced}, requiring at the same time that~\eqref{eq:fundamental_matrix_condition} is verified. These frequencies are shown in Table~\ref{table:QNM_BS_invariant}.

Additionally, notice that the sourceless system~\eqref{BS_Perturbation_Matrix_Sourced} admits also the trivial solution
\begin{align} \label{HS_trivial}
\delta \Psi_\epsilon&= \epsilon\, \Psi_0 (1+i \gamma t)\,, \nonumber \\
\delta U_\epsilon&=\epsilon\, U_0\,,	
\end{align}
with a constant $\epsilon\ll1$. This solution is valid only for a certain amount of time (while the perturbation scheme holds) and it corresponds just to an infinitesimal change of the background \acs{NBS} (\ie, an infinitesimal re-scaling of the original star) by a~$\lambda=1+\epsilon/2$. This perturbation causes a static change in the number of particles in the star
\begin{equation}
\delta Q_\epsilon= \frac{\epsilon}{2} Q_\text{NBS}\,,
\end{equation}
and in its mass
\begin{equation}
\delta M_\epsilon= \hbar \mu_S\, \delta Q_\epsilon=\frac{\epsilon}{2}  M_\text{NBS}\,.
\end{equation}

%%%%%%%%%%%%%%%%%%%%%%%%%%%%%%%%%%%%%%%%%%%%%%%%%%%%
\subsection{External perturbers}
\label{sec:External perturbers}
%%%%%%%%%%%%%%%%%%%%%%%%%%%%%%%%%%%%%%%%%%%%%%%%%%%%
In the presence of an external perturber, one can prescribe its motion through the source term~\eqref{source_BS}. The solution of system~\eqref{BS_Perturbation_Matrix_Sourced} which is regular at the origin and satisfies the Sommerfeld condition at infinity can be obtained through the method of variation of parameters, and it reads
\begin{align} \label{Z1_of_r}
Z_1(r)&= 4\pi\Bigg[\sum_{n=1}^3 F_{1,n}(r) \int_\infty^r dr' F^{-1}_{n,6} p +\sum_{n=4}^6 F_{1,n}(r) \int_0^r dr' F^{-1}_{n,6} p\Bigg]\,,\\\label{Z2_of_r}
Z_2(r)&= 4\pi\Bigg[\sum_{n=1}^3 F_{2,n}(r) \int_\infty^r dr' F^{-1}_{n,6} p +\sum_{n=4}^6 F_{2,n}(r) \int_0^r dr' F^{-1}_{n,6} p \Bigg]\,,\\\label{u_of_r}
u(r)&= 4\pi\Bigg[\sum_{n=1}^3 F_{3,n}(r) \int_\infty^r dr' F^{-1}_{n,6} p +\sum_{n=4}^6 F_{3,n}(r) \int_0^r dr' F^{-1}_{n,6} p \Bigg]\,,
\end{align}
where $F_{i,j}$ is the $(i,j)$-component of the fundamental matrix defined in Eq.~\eqref{eq:fundamental_matrix}. To obtain the total scalar field energy, linear and angular momenta radiated during a given process, all we need are the amplitudes~$Z_1^\infty$ and~$Z_2^\infty$. These are given simply by
\begin{align}
Z_1^\infty&=4\pi \int_{0}^{\infty} dr' F^{-1}_{4,6}(r')p(r') \,, \label{Z1inf_BS}\\
Z_2^\infty&=4 \pi \int_{0}^{\infty} dr' F^{-1}_{5,6}(r')p(r') \,. \label{Z2inf_BS}
\end{align}
Let us now apply our framework to some physically interesting external perturbers.

\paragraph{Plunging particle.} Consider a point particle plunging into a \acs{NBS}. Without loss of generality, one can assume its motion to take place in the~$z$-axis, being described by the world-line~$x_p^{\;\alpha}(t)=(t,0,0,z_p(t))$ in Cartesian-like coordinates.
Neglecting the backreaction of the fluctuations on the perturber's motion,
\begin{align}
\ddot{z}_p(t)=-\partial_z U_0(z_p)\,.
\end{align}
We consider that the perturber crosses the \acs{NBS} center at~$t=0$, \ie~$z_p(0)=0$, with velocity
\begin{align}
\dot{z}_p(0)=-\sqrt{2\left(U_0(R)-U_0(0)\right)+v_R^2}\,,
\end{align}
where~$v_R$ is the velocity with which the massive object enters the \acs{NBS}; in other words, it is the velocity at~$r=R$.
In spherical coordinates, the source reads
\begin{align}
&P=m_p \frac{\delta(\varphi)}{r^2 \sin \theta}\left[\delta\left(r-z_p(t)\right)\delta\left(\theta\right)+ \delta\left(r+z_p (t)\right)\delta\left(\theta-\pi\right)\right].%\label{P_plunging}
\end{align}
We do not want to be restricted to massive objects describing unbounded motions and, so, we consider also perturbers with small~$v_R$. These may not have sufficient energy to escape the \acs{NBS} gravity, being doomed to remain in a bounded oscillatory motion (\emph{see} Section~\ref{oscillating_particle_BS}). In these cases, we want to find the energy and linear momentum that is lost in one full crossing of the \acs{NBS} and, so, we shall take the above source as \emph{active} just during that time interval, vanishing whenever else.
Using Eq.~\eqref{p_def} the function~$p$ is
\begin{align}
p=-\frac{m_p}{\sqrt{2 \pi}}Y_{l,0}(0) \delta_m^0 \frac{|t'_p(r)|}{r} \left(e^{-i \omega t_p(r)}+(-1)^l e^{i \omega t_p(r)}\right) \,,
\end{align}
with~$t_p(r)\geq0$ defined by~$z_p\left[t_p(r)\right]=-r$.
This can be rewritten in the form
\begin{align}
\hspace{-0.6cm}p=\frac{m_p}{\sqrt{2\pi}}Y_{l,0}(0)\, \delta_m^0 \frac{|t_p'(r)|}{r}\left(\cos\left[\omega t_p(r)\right] \delta_l^\text{even}-i \sin\left[\omega t_p(r)\right] \delta_l^\text{odd}\right).
\end{align}
The property 
\begin{equation}
p(\omega,l,0;r)=p(-\omega,l,0;r)^*\,,
\end{equation}
together with the form of system~\eqref{BS_Perturbation_Matrix_Sourced}, implies that
\begin{align}
&Z_2(\omega,l,0;r)=Z_1(-\omega,l,0;r)^*\,,\\
&Z_2^\infty(\omega,l,0)=Z_1^\infty(-\omega,l,0)^*\,.\label{property_Z1Z2_plunge_BS}
\end{align}
So, the spectral fluxes~\eqref{Particles_flux},~\eqref{Energy_flux},~\eqref{Momentum_flux} and~\eqref{AngularMomentum_flux} become, respectively,
\begin{align}
\frac{d Q^\text{rad}}{d \omega}=\frac{4}{\hbar}\Re\left[\sqrt{2 \mu_S  (\omega-\gamma)}\right] \sum_l \left|Z_1^\infty(\omega,l,0)\right|^2\,,
\label{Particles_flux_BS}
\end{align}
\begin{align}
\frac{d E^\text{rad}}{d \omega}= \hbar\left(\mu_S-\gamma+\omega\right) \frac{d Q^\text{rad}}{d \omega}\simeq \hbar\mu_S \frac{d Q^\text{rad}}{d \omega} \,,
\label{Energy_flux_BS}
\end{align}
\begin{align}
\hspace{-0.5cm}\frac{d P_z^\text{rad}}{d \omega}=\sum_{l}\tfrac{16\mu_S(l+1)\Theta( \omega-\gamma)}{\sqrt{(2l+1)(2l+3)}}\, (\omega-\gamma)\Re\left[Z_1^\infty(\omega,l,0) Z_1^\infty(\omega,l+1,0)^*\right]\,,\label{Momentum_flux_BS}
\end{align}
and
\begin{equation}
\frac{d L_z^\text{rad}}{d \omega}=0\,.
\end{equation}
These expressions were derived assuming a perturber in an unbounded motion. However, these are also good estimates to the energy and momenta radiated during one full crossing of the \acs{NBS} by a bounded perturber, as long as its half-period is much larger than the \acs{NBS} crossing time.

To compute how much energy is lost by the perturber, we need to know the change in the \acs{NBS} energy~$\Delta E_\text{NBS}$. At leading order, this is given by
\begin{equation}
\Delta E_\text{NBS}=\hbar \mu_S\, \Delta Q_\text{NBS}=- \hbar\mu_S\, Q^\text{rad}\,, 
\end{equation}
using~\eqref{DeltaQ} in the second equality.
Conservation of total energy-momenta, expressed through Eq.~\eqref{LossRadE}, implies that the perturber loses the energy
\begin{align}
&E^\text{lost}= \Delta E_\text{NBS}+E^\text{rad}=\hbar\int d\omega(\omega- \gamma) \frac{d Q^\text{rad}}{d \omega}  \nonumber \\
&=4\sqrt{2 \mu_S}\int d\omega\Re\left[(\omega-\gamma)^{\frac{3}{2}}\right] \sum_l \left|Z_1^\infty(\omega,l,0)\right|^2\,.
\label{Energy_loss_BS}
\end{align}
The last expression should be understood only as an order of magnitude estimate. If we had considered only the leading order contribution to~$E^\text{rad}$ as we did for~$\Delta E_\text{NBS}$, we would have obtained~$E^\text{lost}=0$. In the second equality we used higher order corrections to $E^\text{rad}$ -- the factor~$(\omega-\gamma) \ll \mu_S$; but not to~$\Delta E_\text{NBS}$. There are corrections to~$\Delta E_\text{NBS}$ of the same order of those to~$E^\text{rad}$ that should be included in a rigorous calculation of~$E^\text{lost}$. We do not attempt that in this work. Interestingly, in our approximation the energy lost by the perturber matches the kinetic energy of the radiated scalar particles at infinity, as can be readily verified. The terms neglected should contain information about, for instance, the gravitational and kinetic energy of the radiated particles when they were in the unperturbed \acs{NBS}. Still, we believe that Eq.~\eqref{Energy_loss_BS} is good estimate of the order of magnitude of~$E^\text{lost}$ and that it scales correctly with the boson star and perturber's mass,~$M_\text{NBS}$ and~$m_p$, respectively.

For a small perturber~$m_p \mu_S \ll v_R$, the momentum and energy that are lost in this type of process are related through (\emph{see} Eq.~\eqref{EvsPloss})~\footnote{Using the full expression~\eqref{EvsPloss}, it is easy to see that if~$E^\text{lost}\propto m_p^2$, then~$P^\text{lost} \propto m_p^2$ in the limit $m_p \mu_S\ll v_R$. The~$E^\text{lost}\propto m_p^2$ follows from~$Z_1^\infty \propto m_p$ (\eg, Eq.~\eqref{Z1inf_BS}).}
\begin{align}
P_z^\text{lost}\simeq-\frac{E^\text{lost}}{v_R}\,.
\end{align}
Conservation of total momentum, as expressed in~\eqref{LossRadP}, implies that the \acs{NBS} acquires a momentum~\footnote{The watchful reader may wonder why the kinetic energy associated with the momentum acquired by the boson star~$\Delta P_z$ is not included in~$\Delta E_\text{NBS}$. Actually, this is one of the higher order corrections neglected in~\eqref{Energy_loss_BS}, but it is easy to check that it is subleading comparing with the correction of~$E^\text{rad}$ considered.} 
\begin{align}
P_\text{NBS}=P_z^\text{lost}-P_z^\text{rad}=-\frac{E^\text{lost}}{v_R}-P_z^\text{rad}\,.\label{eq:NBS_momentum}
\end{align}

\paragraph{Orbiting particles.}

Consider an equal-mass binary, with each component having mass $m_p$, and describing a circular orbit of radius $r_{orb}$ and angular frequency $\omega_\text{orb}$ in the equatorial plane of a \acs{NBS}. The source is modeled as
\begin{align}
\hspace{-0.4cm}P=\frac{m_p}{r_\text{orb}^2} \delta(r-r_\text{orb})\delta\left(\theta-\frac{\pi}{2}\right)\left[\delta(\varphi-\omega_\text{orb} t)+\delta(\varphi+\pi-\omega_\text{orb} t)\right]\,. \label{P_orbiting}
\end{align}
We are assuming that the center of mass of the binary is at the center of the \acs{NBS}, but in principle our results extend to all binaries sufficiently deep inside the \acs{NBS}.
Also, our methods can be applied to any binary as long as a suitable source~$P$ is given.

Using Eq.~\eqref{p_def} the above source yields
\begin{align}
p=m_p\sqrt{\frac{\pi}{2}} \frac{Y_{lm}\left(\pi/2,0\right)}{r_\text{orb}} (1+(-1)^m)  \delta\left(r-r_\text{orb}\right)\delta\left(\omega -m \omega_\text{orb}\right)\,.\label{p_orbiting}
\end{align}
The perturber's motion is fully specified by a prescription relating $r_\text{orb}$ and $\omega_\text{orb}$; we consider Keplerian orbits
$r_\text{orb}^3=M/\omega_\text{orb}^2$, where $M=2m_p$ is the total binary mass. This setup describes either stellar-mass or supermassive \acs{BH} binaries orbiting inside a \acs{NBS}.
Alternatively, applying the transformation $m_p(1+(-1)^m)\to m_p$, we obtain a source that describes an extreme mass-ratio inspiral (\acs{EMRI}). This could be, for instance, a star of mass~$m_p$ on a circular orbit around a central massive \acs{BH} of mass~$M_\text{BH}$. In such case we consider the Keplerian prescription $r_\text{orb}^3=M_\text{BH}/\omega_\text{orb}^2$. 

The symmetry
\begin{equation}
p(\omega,l,m;r)=(-1)^m p(-\omega,l,m;r)^*\,,
\end{equation}
together with the form of system~\eqref{BS_Perturbation_Matrix_Sourced}, implies
\begin{align}
&Z_2(\omega,l,m;r)=(-1)^m Z_1(-\omega,l,-m;r)^*\,,\\
&Z_2^\infty(\omega,l,m)=(-1)^m Z_1^\infty(-\omega,l,-m)^*\,.
\end{align}
These simplify the emission rate expressions~\eqref{Particles_flux_rate},~\eqref{Energy_flux_rate} and~\eqref{AngularMomentum_flux_rate}, yielding
\begin{align}
&\hspace{-0.4cm}\dot{Q}^\text{rad}=\frac{2}{\pi \hbar} \int d\omega\Re\bigg[\sqrt{2 \mu_S \left(\omega-\gamma\right)}\bigg]\sum_{l,m} \left|Z_1^\infty(\omega,l,m)\right|^2  \,, \\
&\hspace{-0.4cm}\dot{E}^\text{rad}=\frac{2}{\pi} \int d\omega(\mu_S-\gamma+\omega) \Re\bigg[\sqrt{2 \mu_S \left(\omega-\gamma\right)}\bigg]\sum_{l,m} \left|Z_1^\infty(\omega,l,m)\right|^2  \,, \\
&\hspace{-0.4cm}\dot{L}_z^\text{rad}=\frac{2}{\pi} \int d\omega \Re\bigg[\sqrt{2 \mu_S \left(\omega-\gamma\right)}\bigg]\sum_{l,m} m \left|Z_1^\infty(\omega,l,m)\right|^2\,.
\end{align}
These can be written explicitly as
\begin{align}
\dot{Q}^\text{rad}&=\frac{32 \pi}{\hbar} \widetilde{p}^2 \sum_{l,m}\Re\left[\sqrt{2 \mu_S \left(m \omega_\text{orb}-\gamma\right)}\right]\left|F_{4,6}^{-1}\left(m\omega_\text{orb};\,r_\text{orb}\right)\right|^2\,,\label{Particles_flux_orbitingBS2}\\
\dot{E}^\text{rad}&=32 \pi \, \widetilde{p}^2 \sum_{l,m}\Re\left[\sqrt{2 \mu_S \left(m \omega_\text{orb}-\gamma\right)}\right]\nonumber\\
&\qquad\times(\mu-\gamma+m \omega_\text{orb}) \left|F_{4,6}^{-1}\left(m\omega_\text{orb};\,r_\text{orb}\right)\right|^2,\label{Energy_flux_orbitingBS2}\\
\dot{L}_z^\text{rad}&=32 \pi \, \widetilde{p}^2  \sum_{l,m}m \Re\left[\sqrt{2 \mu_S \left(m \omega_\text{orb}-\gamma\right)}\right]\left|F_{4,6}^{-1}\left(m\omega_\text{orb};\,r_\text{orb}\right)\right|^2\,,\label{AngularMomentum_flux_orbitingBS}
\end{align}
where we defined
\begin{equation*}
\widetilde{p}\equiv m_p \sqrt{\frac{\pi}{2}}\frac{Y_{lm}(\pi/2,0)}{r_\text{orb}}\left(1+(-1)^m\right)\,.
\end{equation*}
Equation~\eqref{Energy_flux_orbitingBS2} can be further simplified using
\begin{equation*}
\mu_S-\gamma+m\omega_{orb}\simeq \mu_S\,,
\end{equation*}
since we are treating the scalar fluctuations as non-relativistic; that is only valid if~$\gamma \ll \mu_S$ and~$\omega_\text{orb}\ll \mu_S$.~\footnote{Large azimuthal numbers~$m$ do not spoil the approximation, because the emission is strongly suppressed by~$F_{4,6}^{-1}$ in that limit.} 

Now we follow the same procedure that we applied above to a \emph{plunging particle}, to estimate the rate at which the binary loses energy. Again, we start by computing, at leading order, the change in the \acs{NBS} energy per unit of time,
\begin{equation}\label{eq:circular_flux_Erad}
\dot{E}_\text{NBS}=\hbar\mu_S \dot{Q}_\text{NBS}=-\hbar\mu_S  \dot{Q}^\text{rad}\,,
\end{equation}
where we used Eq.~\eqref{DeltaE} in the first equality and~\eqref{DeltaQ} in the second.~\footnote{Equations~\eqref{DeltaQ} and~\eqref{DeltaE} are easy to adapt to changes happening during a finite amount of time~$\Delta t$. To get the rates of change one just needs to divide these expressions by~$\Delta t$ and take the limit~$\Delta t\to 0$.}
Conservation of the total energy implies that the energy that is lost per unit of time by the binary is 
\begin{align} \label{eq:E_loss_circular}
&\hspace{-0.9cm}\dot{E}^\text{lost}=\dot{E}^\text{rad}+ \dot{E}_ \text{NBS}\nonumber \\
&\hspace{-0.9cm}= 32 \pi \widetilde{p}^2 \sum_{l,m}\left(m \omega_\text{orb}-\gamma\right) \Re \bigg[\sqrt{2 \mu_S \left(m \omega_\text{orb}-\gamma\right)}\bigg] \left|F_{4,6}^{-1}\left(m\omega_\text{orb};\,r_\text{orb}\right)\right|^2\,.
\end{align}
Again, the last expression should be understood only as an order of magnitude estimate~(as it was discussed above when considering a \textit{plunging particle}).

For a small perturber~$m_p\ll \omega_\text{orb} r_\text{orb}^2$, the angular momentum and energy that are lost in this type of process are related through
\begin{align}
\dot{L}_z^\text{lost}\simeq \frac{\dot{E}^\text{lost}}{\omega_ \text{orb}}\,.
\end{align}
Conservation of total angular momentum, expressed through Eq.~\eqref{LossRadL}, implies that per unit of time the \acs{NBS} acquires the angular momentum
\begin{align}
\dot{L}_\text{NBS}=\dot{L}_z^\text{lost}-\dot{L}_z^\text{rad}=\frac{\dot{E}^\text{lost}}{\omega_ \text{orb}}-\dot{L}_z^\text{rad}\,.
\end{align}
%

%%%%%%%%%%%%%%%%%%%%%%%%%%%%%%
\section{Free oscillations}
%%%%%%%%%%%%%%%%%%%%%%%%%%%%%%

The characteristic, non-relativistic oscillations of \acs{NBS} are regular solutions of system~\eqref{Sourced_SP_System1}-\eqref{Sourced_SP_System2}
satisfying Sommerfeld conditions~\eqref{BC_sommerfeld_infinity} at large distances. For each multipole~$l$, there seems to be an infinite, discrete set of solutions which we label with an overtone index~$n$,~$\omega^n_\text{QNM}$. The first few characteristic frequencies, normalized to the \acs{NBS} mass, are shown in Table~\ref{table:QNM_BS_invariant}. 
They turn out to be all \emph{normal mode} solutions, confined within the \acs{NBS}. The characteristic frequencies are all purely real and cluster around~$\gamma$.
We highlight the fact that the numbers in Table~\ref{table:QNM_BS_invariant} are universal, they hold for any \acs{NBS}. The fundamental~$l=0$ mode (the first entry in the table) had been computed previously~\cite{Guzman:2004wj}, and agrees with our calculation to excellent precision (after proper normalization). Our results are also in very good agreement with the frequencies of the first two modes obtained in a recent time-domain analysis~\cite{Guzman:2018bmo}.
\begin{table}[th]
	\centering
	\begin{tabular}{c||c}
		\hline
		\hline
		$l$ &  \multicolumn{1}{c}{$\omega^{(n)}_\text{QNM}/(M_\text{NBS}^2\mu_S^3)$} \\ 
		\hline
		\hline
		0 & $0.0682\;\,\,\,    0.121\;\,\,     0.138\;\,\,    0.146\;\,\,    0.151\;\,\,  0.154\;\,\, 0.159$\\
		1 & $\,\,0.111\,\,\;\,\,     0.134\;\,\,     0.144\;\,\,    0.149\;\,\,    0.153\;\,\,  0.157\;\,\, 0.162$\\
		2 & $\,\,0.106\,\,\;\,\,     0.131\;\,\,     0.143\;\,\,    0.149\;\,\,    0.153\;\,\,  0.156\;\,\, 0.161$\\
		\hline
		\hline
	\end{tabular} 
	\caption{Normal frequencies of a \acs{NBS} of mass $M_\text{NBS}$ for the three lowest multipoles. For each multipole~$l$ we show the fundamental mode ($n=0$) and the first five overtones.
		At large overtone number the modes cluster around~$\gamma\simeq0.162712 M_\text{NBS}^2\mu_S^3$. The first mode for~$l=0$ agrees with that of Ref.~\cite{Guzman:2004wj} when properly normalized and with an ongoing fully relativistic analysis~\cite{Caio:2020comment}. The two lowest~$l=\{0,\,1,\,2\}$ modes are in good agreement with a recent time-domain analysis~\cite{Guzman:2018bmo}.}
	\label{table:QNM_BS_invariant}
\end{table}
Modes of relativistic stars have been considered in the literature~\cite{Yoshida:1994xi,Kojima:1991np,Macedo:2013jja,Macedo:2016wgh}
and should smoothly go over to the numbers in Table~\ref{table:QNM_BS_invariant}. Note that modes of relativistic boson stars are damped, due to couplings between the scalar and the metric and the possibility to lose energy via gravitational waves. Such damping -- which is small for the relevant polar fluctuations~\cite{Macedo:2013jja,Macedo:2016wgh,GRITJHU} -- should get smaller as one approaches the Newtonian regime, but a full characterization of the modes of boson stars is missing.
Our results show that \acsp{NBS} are linearly mode stable; it would be interesting to have a formal proof, perhaps following the methods of Ref.~\cite{Kimura:2018eiv,Kimura:2017uor}.
We point out that the stabilization of a perturbed boson star through the emission of scalar field -- known as \textit{gravitational cooling} -- has been studied previously~\cite{Seidel1994,Balakrishna:2006ru,Guzman:2006yc}.

%*****************************************
%*****************************************
%*****************************************
%*****************************************
%*****************************************

%************************************************
\chapter{Stirring up an ultralight dark matter core}\label{ch:stir} 
%************************************************

%In this chapter we apply the tools developed in the previous chapters (in particular in Section~\ref{sec:SmallPert}) to study several astrophysical interesting systems involving compact objects (\eg, \acsp{BH} or \acsp{BHB}) moving or sitting inside ultralight \acs{DM} cores (here modeled by \acsp{NBS}).

%%%%%%%%%%%%%%%%%%%%%%%%%%%%%%%%%%%%%%%%%%%%%%%%%%%%%%%%%%%%%%%%%%%%%%%%%%%%%
\section{A perturber sitting at the center\label{sec_sitting_bs}}
%%%%%%%%%%%%%%%%%%%%%%%%%%%%%%%%%%%%%%%%%%%%%%%%%%%%%%%%%%%%%%%%%%%%%%%%%%%

Static perturbations of \acsp{NBS}, or, more generally, of solitonic \acs{DM} cores of light fields are interesting in their own right. For perturbers localized far away, the induced tidal effects can dissipate energy and lead to distinct signatures, both in \acs{GW} signals and in the dynamics of objects close to such configurations~\cite{Mendes:2016vdr,Cardoso:2017cfl,Sennett:2017etc}.
Here we will not perform a general analysis of static tidal effects and will instead focus on perturbations due to a massive object at the center of an \acs{NBS}.
Such object can be taken to be a supermassive \acs{BH} or a neutron star, and the induced changes are important to understand how \acs{DM} distribution is affected by baryonic \emph{impurities}. 

Consider then a \acs{BH} or star, described by the source~\eqref{source_BS}, and inducing static, spherically symmetric, real perturbations on the scalar field and gravitational potential, respectively, $\delta \Psi_p(r)$ and $\delta U_p(r)$. Then, Eqs.~\eqref{Sourced_SP_System1} and~\eqref{Sourced_SP_System2} become
\begin{align}
\nabla^2\delta \Psi_p&= 2\mu_S \left(\mu_S U_0+\gamma\right)\delta \Psi_p+2\mu_S^2\Psi_0 \delta U_p\,,  \\
\nabla^2\delta U_p&= 4\pi \left(2\mu_S^2 \Psi_0\, \delta \Psi_p+P\right) \,. 
\end{align}
For a static source at the origin, it is easy to show that the matter moments are given by
\begin{align}
p=\lim_{r_p \to 0}\,\frac{1}{2 \sqrt{2}} \frac{m_p}{r_p} \,\delta_l^0 \delta_m^0 \delta(\omega) \delta(r-r_p)  \,,
\end{align}
which, through the variation of parameters, implies that
\begin{align}
\delta \Psi_p &=m_p \sum_{n=4}^{6} \frac{F_{1,n}(r)}{r}  \lim_{r_p \to 0} \left(\frac{F^{-1}_{n,6}(r_p)}{r_p}\right)\,, \nonumber \\
\delta U_p &= m_p \sum_{n=4}^{6} \frac{F_{3,n}(r)}{r}  \lim_{r_p \to 0} \left(\frac{F^{-1}_{n,6}(r_p)}{r_p}\right)\,,
\end{align}
where the components of the fundamental matrix and its inverse are evaluated at~$l=m=\omega=0$. Note that the change in the number of particles and mass of the \acs{NBS}, respectively,~$\delta Q_p$ and~$\delta M_p$, is constant, but non-zero in general. This is a consequence of the source being treated as if it was eternal. However, we expect that, if the perturber is brought in an adiabatic way to the center of the \acs{NBS}, there is no scalar radiation emitted, and, so, no change in the number of particles and mass of the scalar configuration,~$\delta Q_\text{NBS}=\delta M_\text{NBS}=0$. Fortunately, we are free to sum a trivial homogeneous solution~\eqref{HS_trivial} to enforce~$\delta Q_\text{NBS}=\delta M_\text{NBS}=0$, while keeping~$\delta \Psi=\delta \Psi_p+ \delta \Psi_\epsilon$ and~$\delta U=\delta U_p+\delta U_\epsilon$ a solution of the inhomogeneous system. 
Then, the total perturbation induced in the density of particles is given by
\begin{align}
\delta \rho_Q=\delta J_Q^{\;\;t}=\frac{2}{\hbar}\mu_S \Psi_0 \Re\left( \delta \Psi\right)=\frac{2}{\hbar}\mu_S \Psi_0\left(\delta \Psi_p+\frac{\epsilon}{2} \Psi_0\right)\,, \nonumber
\end{align}
and the one induced in the mass density by
\begin{align} \label{deltarho}
\delta \rho_M=\delta T_{tt}^S= \hbar \mu_S \delta \rho_Q= 2\mu_S^2 \Psi_0\left(\delta \Psi_p+\frac{\epsilon}{2} \Psi_0\right)\,, 
\end{align}
where $J_Q^{\;\;t}$ is the $t$-component of the Noether's current. The parameter $\epsilon$ associated with the trivial homogeneous solution must be chosen appropriately, so that
\begin{align}
4\pi \int_{0}^{\infty} dr \,r^2 \delta \rho_Q= 4\pi \int_{0}^{\infty} dr\, r^2 \delta \rho_M=0\,.
\end{align}

\begin{figure}
	\centering
	\includegraphics[width=0.9\textwidth]{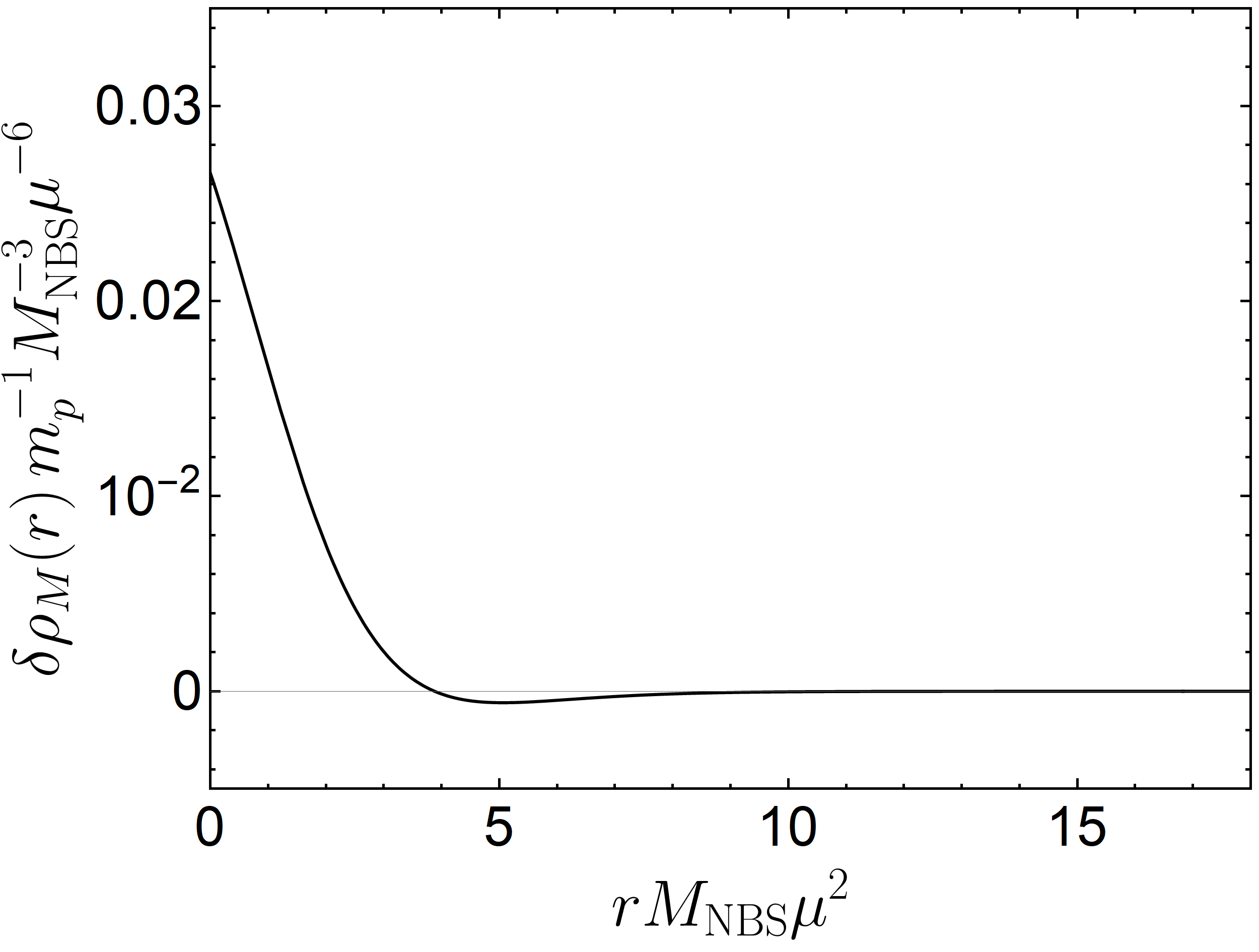} 
	\includegraphics[width=0.9\textwidth]{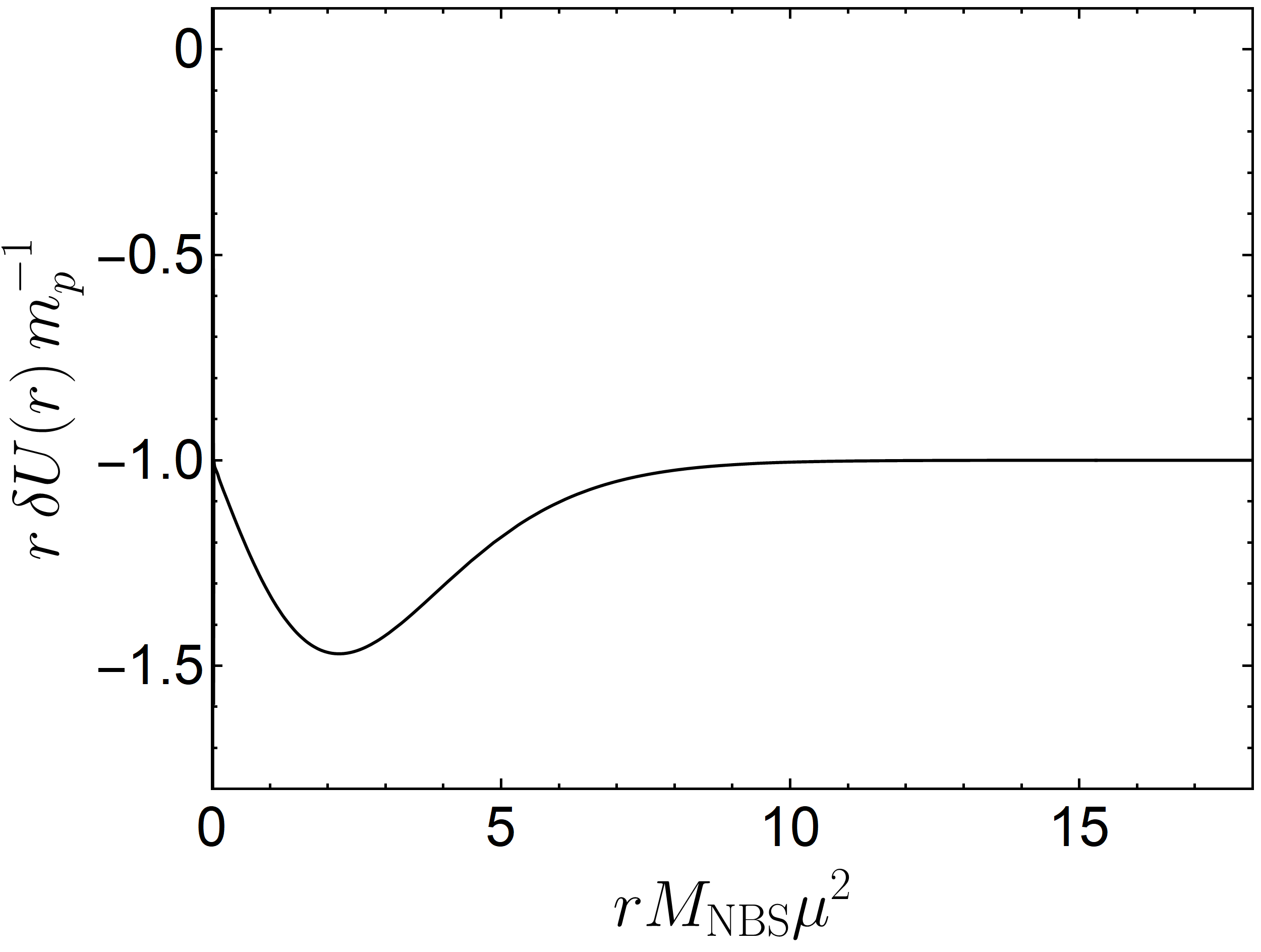}
	\caption{Universal perturbations induced by a massive object of mass~$m_p$, sitting at the center of the scalar configuration. We assume that the perturber was brought adiabatically so that~$\delta Q_\text{NBS}=\delta M_\text{NBS}=0$. Upper panel: perturbation in the mass density of the \acs{NBS} obtained using Eq.~\eqref{deltarho}. 
	Lower panel: perturbation in the gravitational potential~$r~\delta U= r \left(\delta U_p+\delta U_\epsilon\right)$. As expected, for large~$r$, one recovers the Coulombian potential~$U=-m_p/r$.
	}\label{fig:Particlesittingcenter}
\end{figure}

The perturbations in the mass density and gravitational potential of a \acs{NBS} induced by a massive object sitting at its center are shown in Fig.~\ref{fig:Particlesittingcenter}.
Our results indicate that the massive perturber attracts scalar field towards the center, where the gravitational potential corresponds solely to that of the point-like mass.
These results are consistent with those in Ref.~\cite{Bar:2018acw}. 
We find an insignificant change in the local \acs{DM} mass density, when placing a point-like perturber at the center of a \acs{NBS}; notice that~$\delta \rho_M(0)/\rho_M(0)\sim 10\, m_p/M_\text{NBS}$. Thus, a massive perturber will not enhance greatly the local \acs{DM} density, which is smooth and flat for light scalars.  
On the other hand, studies with particle-like \acs{DM} models find that its density close to supermassive \acsp{BH} increases significantly~\cite{Gondolo:1999ef,Sadeghian:2013laa}.
This is in clear contrast to our results for light fields, a perturber does not significantly alter the local ambient density, since its size is much smaller than the scalar's Compton wavelength.
Parenthetically, large overdensities seem to be in some tension with observations~\cite{Robles:2012uy}. Possible ways to ease the tension rely on scattering of \acs{DM} by stars or \acsp{BH}, accretion by the central \acs{BH}, or induced heating by its vicinities~\cite{Merritt:2002vj,Bertone:2005hw,Merritt:2003qk}.
These outcomes cannot possibly generalize to light scalars, at least not when the configuration is spherically symmetric, since there are no stationary 
\acs{BH} configurations with scalar ``hair''~\cite{Herdeiro:2015waa,Cardoso:2016ryw,Clough:2019jpm,Bamber:2021knr}. But these results do prompt the questions: what happens to an \acs{NBS} when a \acs{BH} is placed at its center? What happens to the local scalar amplitude of an \acs{NBS} when a binary is orbiting? We now turn to these issues.

%%%%%%%%%%%%%%%%%%%%%%%%%%%%%%%%%%%%%%%%%%%%%%%%%%%%%%%%%%%%%%%%%%%%%%%%%%%%%
\section{A black hole eating its host \label{sec_sitting_bh}}
%%%%%%%%%%%%%%%%%%%%%%%%%%%%%%%%%%%%%%%%%%%%%%%%%%%%%%%%%%%%%%%%%%%%%%%%%%%%%
As we noted, there are no stationary, spherically symmetric, scalar configurations when a non-spinning \acs{BH} is placed at its center.
On long timescales, the entire \acs{NBS} will be accreted by the \acs{BH}, a fraction dissipating to infinity.
This means, in particular, that our results cannot be extrapolated to the case in which the point particle is a \acs{BH}, and describe the system only at intermediate times. 
What is the \emph{lifetime} of such a system, composed of a small \acs{BH} sitting at the center of a \acs{NBS}?
Unfortunately, most of the studies on \acs{BH} growth and accretion assume a fluid-like environment~\cite{Giddings:2008gr}, an assumption that breaks down completely
here, since the Compton wavelength of the scalar is much larger than the size of the \acs{BH}. Exceptions to this rule exist~\cite{Clough:2019jpm,Hui:2019aqm}, but focus
on different aspects, and do not consider setups with the necessary difference in lengthscales.

The precise answer to this question requires full nonlinear simulations in a challenging regime, with proper initial conditions.
However, in the limit we are interested in, where the \acs{BH}, of mass~$M_\text{BH}\ll M_\text{NBS}$, is orders of magnitude smaller and lighter than the~\acs{NBS}, 
a perturbative calculation is appropriate. Consider a sphere of radius~$r_+$ centered at the origin of the \acs{NBS}. The \acs{NBS} is stationary and so there is a flux of energy crossing such a sphere inwards (details in Appendix~\ref{app:incoming_flux})
\begin{align}
\dot{E}_\text{in}\approx 10^{-3} \mu_S^7 r_+^2 M_\text{NBS}^5\,,
\end{align}
and the same amount crossing it outwards. If such a sphere defines the \acs{BH} boundary~$r_+=2M_\text{BH}$~\footnote{Actually, such a sphere should be placed outside the effective potential for wave propagation around \acsp{BH}, but the difference is not relevant here.}, a fraction will be absorbed by the \acs{BH}. Low-frequency waves (the scalar field frequency is~$\sim\mu_S$ and we are in the low frequency regime with $\mu_S M_\text{BH}\ll 1$) are poorly absorbed, and one finds that the flux into the \acs{BH} is~\cite{Unruh:1976fm}~\footnote{We are taking the limit $\omega \to \mu_S$ in the expression for the transmission amplitude. Strictly speaking, we are in the $\omega<\mu_S$ regime, but continuity of results should be valid.}
\begin{align}
\dot{E}_\text{abs}=32\pi\left(M_\text{BH}\mu_S \right)^3\dot{E}_\text{in}=\frac{16\pi}{125}\frac{M_\text{BH}^5}{M_\text{NBS}^5}\left(M_\text{NBS}\mu_S \right)^{10}\,.
\end{align}
We have tested the above physics with a series of toy models, including the study of accretion of a massive, non self-gravitating scalar confined in a spherical cavity
with a small \acs{BH} at the center (Appendix~\ref{app:bh_bomb}). This toy model conforms to the physics just outlined. One simple toy model, summarized in Appendix~\ref{app:string_toy}, suggests that all modes
of the \acs{NBS} are excited during such an accretion process, but made quasinormal (\ie, damped) by the presence of absorption. These are all low-frequency modes, and our argument should be valid even in such circumstance.

With $\dot{E}_\text{abs}=\dot{M}_\text{BH}$ and fixed \acs{NBS} mass, one finds the timescale
\begin{align}
\tau \sim \frac{1}{M_\text{BH}^4M_\text{NBS}^5\mu_S^{10}}=10^{24}\,\text{yr}\,\frac{M_\text{NBS}}{10^{10}M_{\odot}}\left(\frac{\chi}{10^4}\right)^4\left(\frac{0.1}{M_\text{NBS}\mu_S}\right)^{10}\,,
\end{align}
where $\chi\equiv M_\text{NBS}/M_\text{BH}$. 
In other words, the timescale for the \acs{BH} to increase substantially its mass -- which we take as a conservative indicative of the lifetime of the entire \acs{NBS} -- is
larger than a Hubble timescale for realistic parameters. This timescale is the result of forcing the \acs{BH} with a nearly monochromatic field from the \acs{NBS}. When the material of the star is nearly exhausted, a new timescale is relevant, that of the quasinormal modes of the \acs{BH} surrounded by a massive scalar.
This timescale is $\tau_\text{QNM}\sim M_\text{BH}(M_\text{BH}\mu_S)^{-6}<\tau$~\cite{Detweiler:1980uk,Brito:2015oca}, but still typically larger than a Hubble time.  

When rotation is included, the entire setup may become even more stable: rotation is able to provide energy, via superradiance, to the surrounding field, and sustain
nearly stationary, but non spherically-symmetric, configurations~\cite{Herdeiro:2014goa,Brito:2015oca}. We will not discuss these effects here.

%%%%%%%%%%%%%%%%%%%%%%%%%%%%%%%%%%%%%%%%%%%%%%%%%%%%%%%%%%%%%%%%%%%%%%%%%%%%%%%%%%
\section{Massive objects plunging into the core \label{Plunging_particle_BS}}
%%%%%%%%%%%%%%%%%%%%%%%%%%%%%%%%%%%%%%%%%%%%%%%%%%%%%%%%%%%%%%%%%%%%%%%%%%%%%%%%%%
%
\begin{figure}
	\centering
	\includegraphics[width= 0.9\textwidth]{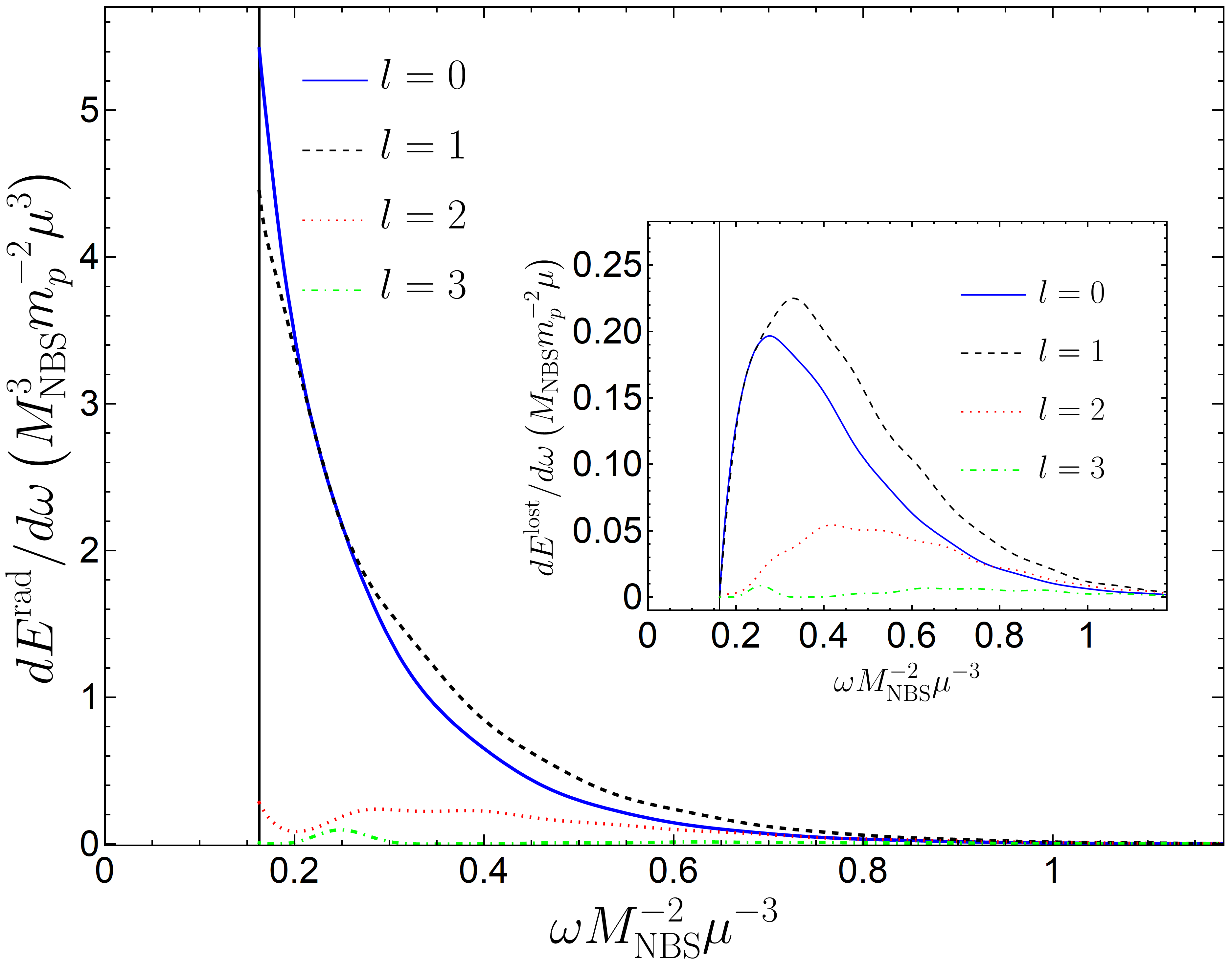}
	\includegraphics[width= 0.9\textwidth]{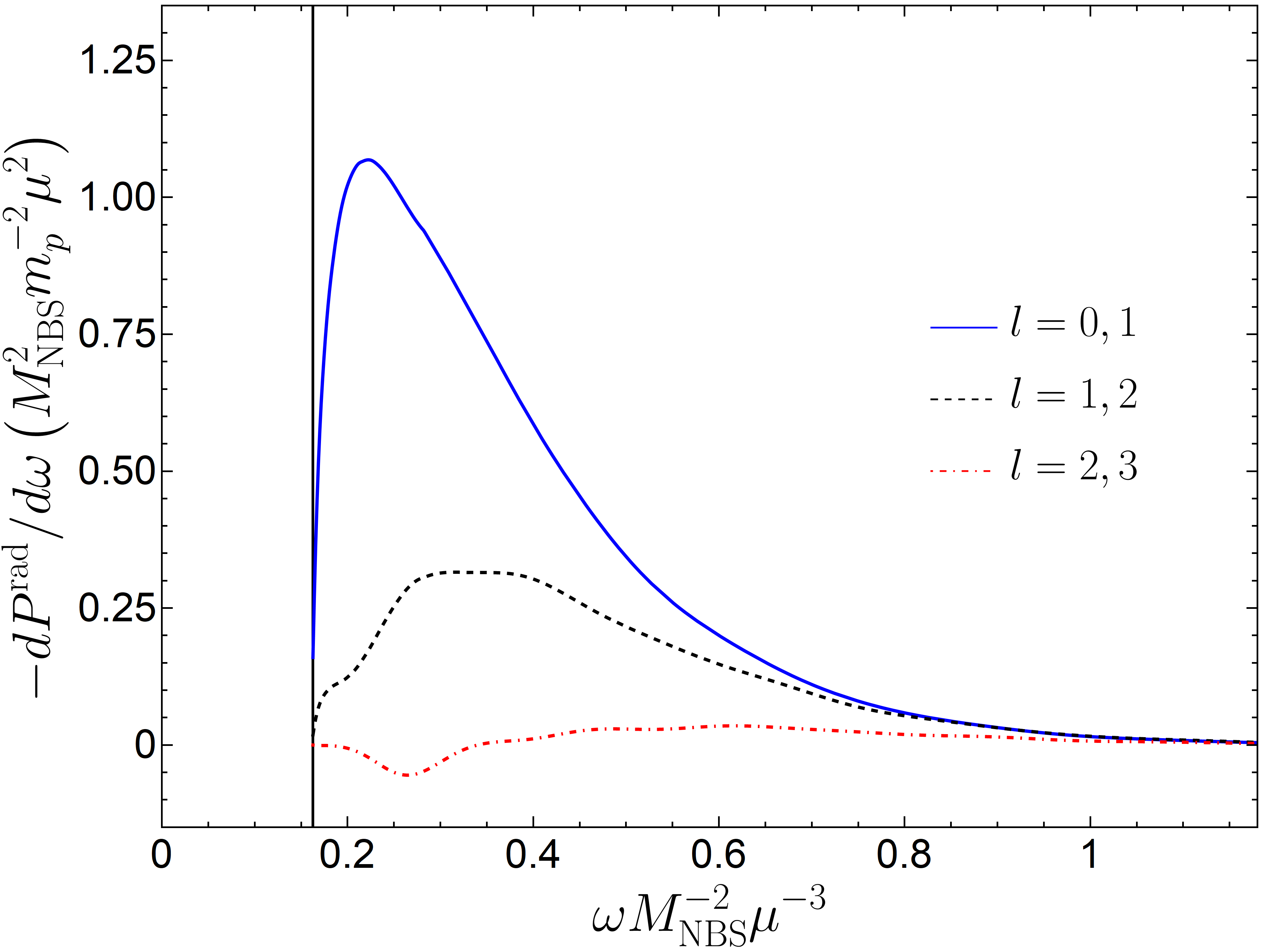}
	\caption{Spectrum of radiation released when an object of mass $m_p$ plunges through an \acs{NBS} with initial velocity $v_R\approx 0 $. Emission takes place for frequencies $\omega>\gamma$ (Eqs.~\eqref{Energy_flux_BS}-\eqref{Momentum_flux_BS}). Upper panel: lowest multipole contribution $l=\{0,1,2,3\}$ to the spectrum of total radiated energy. Inset: multipole contributions to the kinetic energy of the radiated scalar field. Lower panel: spectral fluxes of linear momentum along $z$ associated with the lowest multipoles. The results obtained for other plunging velocities are summarized in Eqs.~\eqref{eq:fit_BS_Erad_plunge_gravityon}--\eqref{eq:fit_BS_Prad_plunge_gravityon}.}
	\label{fig:PlungingSpectrabosonstar_gravityon}
\end{figure}
Consider now a massive perturber plunging, head-on, into a \acs{NBS}. The perturber is assumed to have traveled from far away, but for our purposes the only relevant quantity is the perturber velocity when it reaches the \acs{NBS} surface, $\boldsymbol{v}=-v_{R}\boldsymbol{e}_z$, with~$v_R\geq 0$. This setup is described in detail in Section~\ref{sec:External perturbers}. As we argued before, this situation could describe a massive \acs{BH} \emph{kicked} at formation, via \acs{GW} emission, in a \acs{DM} core of light fields, or, simply, stars crossing a \acs{NBS}. Our framework allow us to do the first self-consistent computation (including self-gravity of the scalar perturbation and finite-size effects of the core) of the \acs{DF} acting on perturbers in such systems. The effect of the \acs{NBS} gravitational potential on the perturber motion sets a natural critical velocity in the problem, the escape velocity $v_\text{esc}$. For the fundamental \acs{NBS} described in Fig.~\ref{fig:BS}, the velocity needed to escape from its surface is~$v_\text{esc}\sim 0.47M_\text{NBS}\mu_S$.
When the velocity is smaller than that, the crossing object will be bound to the \acs{NBS} describing an oscillatory motion. 
For now, we study a simple one-way motion and assume that after crossing the \acs{NBS} once, the particle simply ``disappears''. This will allow us to estimate the \acs{DF} acting on the perturber.
This assumption is formally correct and accurate for unbound motion. For bound oscillatory motion it is not, and we work out the full case below, in Section~\ref{oscillating_particle_BS}.

Some quantities of interest are the energy and linear momentum carried by the scalar field radiated in these processes, as well as the energy that is lost by the perturber. These are given, respectively, by Eqs.~\eqref{Energy_flux_BS}-\eqref{Momentum_flux_BS} and~\eqref{Energy_loss_BS}. The upper panel of Fig.~\ref{fig:PlungingSpectrabosonstar_gravityon} shows the contribution of the lowest multipoles to the total radiated energy spectrum~$d E^\text{rad}/d \omega$ ($d E^\text{lost}/d \omega$ in inset). This result was obtained through the numerical evaluation of expressions~\eqref{Energy_flux_BS}-\eqref{Energy_loss_BS} for a perturber plunging into a \acs{NBS}, starting the fall from rest at~$R$. The fluxes converge exponentially with multipole number~$l$, after a sufficiently large~$l$. Our results are compatible with~$E_l^\text{rad}\propto e^{-l}$, where~$E_l^\text{rad}$ is the~$l$-mode contribution to the energy radiated. Once the behavior of~$E_l^\text{rad}$ for large~$l$ is obtained, one can find the total energy radiated.
For a particle plunging with zero initial velocity into an NBS we obtain $E^\text{rad}\sim 1.28 \, m_p^2/M_\text{NBS}$ and $E^\text{lost}\sim 0.18 \, m_p^2 M_\text{NBS}\mu_S^2$. 
Applying this procedure to other velocities, we find that the following expressions provide a good description of our results,
\begin{align}
E^\text{rad}&=29\frac{m_p^2}{M_\text{NBS}}\frac{e^{-3.25/X}}{X^{17/4}}\,,\label{eq:fit_BS_Erad_plunge_gravityon}\\
E^\text{lost}&=7 m_p^2 M_\text{NBS}\mu_S^2\frac{e^{-3.54\,\left(X-0.05\right)^{-1}}}{\left(X-0.05\right)^{17/4}}\,,\label{eq:fit_BS_Ekin_plunge_gravityon}
\end{align}
accurate to within $5\%$ of error for~$0\lesssim v_R\lesssim 2.5M_\text{NBS}\mu_S$. This interval spans over astrophysical relevant velocities (\eg, $0\lesssim v_R[\text{km/s}]\lesssim 6000$, for a \acs{DM} core with mass~$\sim10^{10} M_\odot$ and a scalar mass~$\hbar \mu_S\sim 10^{-22} \text{eV}$). 
In the above expression we defined 
\begin{align}
X\equiv \frac{v_R}{M_\text{NBS}\mu_S}+0.68\,.
\end{align}

The lower panel of Fig.~\ref{fig:PlungingSpectrabosonstar_gravityon} shows the multipolar contribution to the spectrum of radiated linear momentum along $z$. These also converge exponentially in~$l$, after a sufficiently large~$l$. For a perturber starting at rest, the total linear momentum radiated along~$z$ in the whole process is $P^\text{rad}\sim -0.43 m_p^2\mu_S$.
The fitting expression
\begin{align}
	P^\text{rad}=-2.4 m_p^2\mu_S\frac{e^{-2.26\,\left(X-0.27\right)^{-1}}}{\left(X-0.27\right)^{17/4}}\,,\label{eq:fit_BS_Prad_plunge_gravityon}
\end{align}
is a good approximation to our results (within $5\%$ of error for~$0 \lesssim v_R\lesssim 2.5M_\text{NBS}\mu_S$). 
\begin{figure}
	\centering
	\includegraphics[width=0.9 \textwidth]{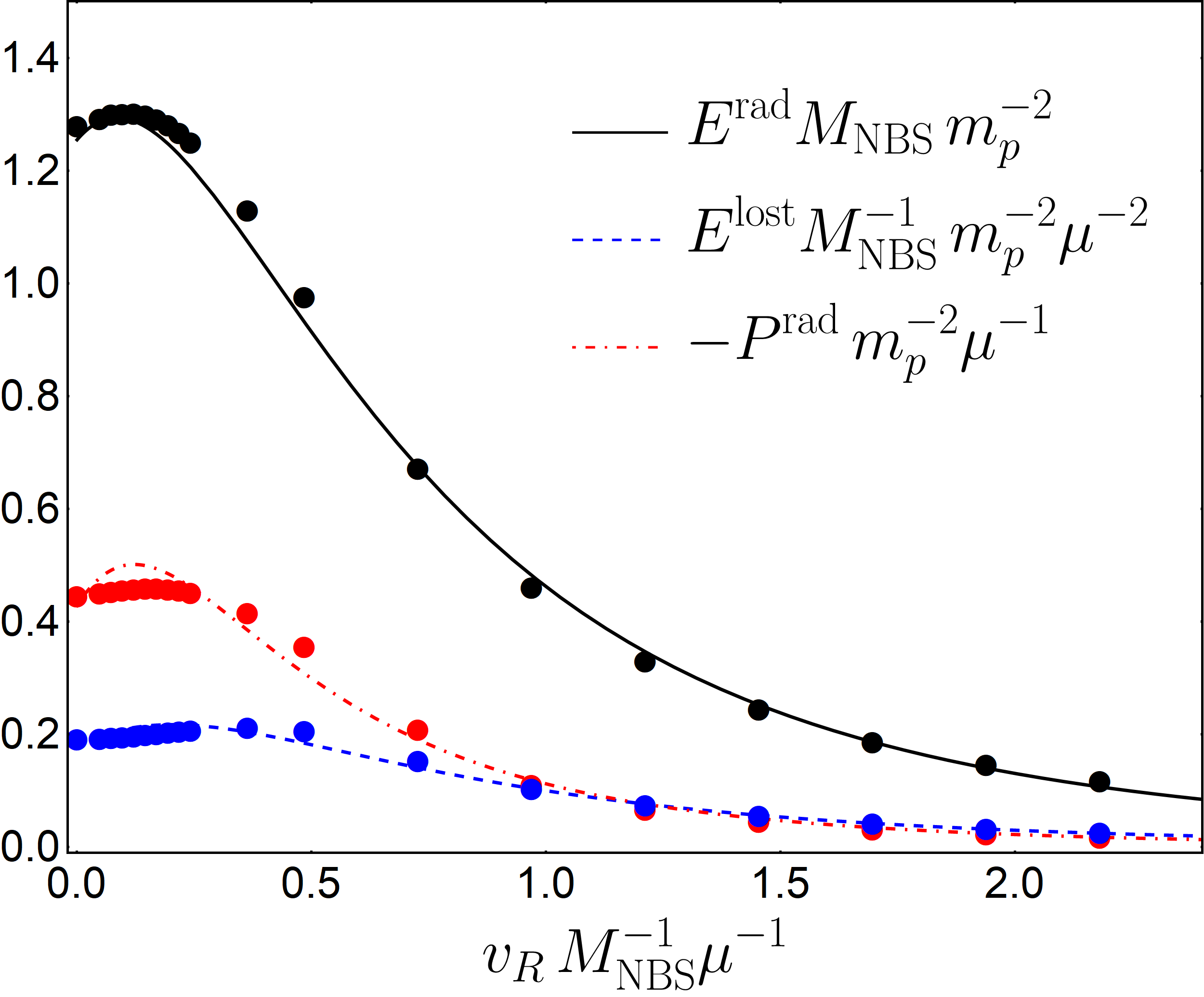} 
	\caption{Total energy, kinetic energy, and linear momentum emitted when an object of mass~$m_p$ plunges through a \acs{NBS}, as a function of the entering velocity. The dots correspond to the numerical data used to obtain the fits~\eqref{eq:fit_BS_Erad_plunge_gravityon}--\eqref{eq:fit_BS_Prad_plunge_gravityon}.}
	\label{fig:fitsBS}
\end{figure}
Figure~\ref{fig:fitsBS} shows how the total radiated energy $E^\text{rad}$, the total energy lost by the moving perturber $E^\text{lost}$, and the linear momentum radiated $P^\text{rad}$ vary with the change of entering velocity.

The momentum that is lost by a small plunging object ($m_p\mu_S \ll v_R$) is given by $P^\text{lost}=-E^\text{lost}/v_R$, as shown in Eq.~\eqref{EvsPloss}.
We have thus computed the \acs{DF} acting upon a body moving within a NBS. The quantity $E^\text{lost}$ is the actual energy that is lost by the perturber in crossing the \acs{NBS}. Note that our result for~$E^\text{lost}$ -- in particular, its sign -- indicate that there exists indeed a friction; the body will slow down. Additionally,  note that the same result~$E^\text{lost}$ together with the radiated momentum~$P^\text{rad}$ show that the \acs{NBS} will acquire a small momentum in the direction of the moving object, as described by Eq.~\eqref{eq:NBS_momentum}; the two lines cross each other close to~$v_R= M_\text{NBS} \mu_S$ as shown in Fig.~\ref{fig:fitsBS}.

Our results should be compared and contrasted with those of Refs.~\cite{Hui:2016ltb,Lancaster:2019mde}, where \acs{DF} in these structures was estimated, neglecting the self-gravity of scalar fluctuations and considering a homogeneous density medium and a constant velocity perturber (the finite size of the scalar configuration was forced through a cut-off radius). These results were recently extended to the relativistic regime -- \ie, \acsp{BH} moving at high velocities (the same setup will be considered in the next chapter, were we derive analytical expressions for the~\acs{DF} in several regimes).
In the limit~$\beta\equiv m_p \mu_s/ v_0\ll1$, with~$v_0$ the constant velocity of the perturber -- which is the limit consistent with our perturbation scheme -- the \acs{DF} force obtained in Ref.~\cite{Hui:2016ltb} is
\begin{align}
	F=-\frac{4\pi m_p^2 \rho_M}{v^2}\left(\text{Cin}(2 v R \mu_S)+ \frac{\sin(2 v R \mu_S)}{2 v R \mu_S}-1\right)\,,
\end{align}
where~$\text{Cin}(x)=\int_0^x(1-\cos x')dx'/x'$ is the cosine integral. For small velocities~$v \ll 1/(R \mu_S)$, this becomes 
\begin{align}
	F \simeq -\frac{4 \pi}{3}m_p^2 \rho_M R^2\mu_S^2\,.
\end{align}
This amounts to a loss of momentum of the order
\begin{align}
	P^\text{lost} \sim F \left(\frac{2R}{v_0}\right) \sim - \frac{2}{v_0}m_p^2\mu_S^2 M\,,
\end{align}
where~$M=(4\pi/3)\rho_M R^3$.
Our perturbative framework, at small velocities, gives (looking at Figure~\ref{fig:fitsBS})
\begin{align}
	P^\text{lost} \simeq -\frac{0.2}{v_R}m_p^2\mu_S^2 M_\text{NBS}\,.
\end{align}
If we consider~$v_0\sim v_R$ and~$M\sim M_\text{NBS}$, our result has the same form than the one in Ref.~\cite{Hui:2016ltb}, but it is one order of magnitude smaller (which is not very surprising, given the different assumptions in the two treatments). 

%%%%%%%%%%%%%%%%%%%%%%%%%%%%%%%%%%%%%%%%%%%%%%%%%%%%%%%%%%%%%%%%%%%%%%%%%%%%
\section{A perturber oscillating at the center \label{oscillating_particle_BS}}
%%%%%%%%%%%%%%%%%%%%%%%%%%%%%%%%%%%%%%%%%%%%%%%%%%%%%%%%%%%%%%%%%%%%%%%%%%%%
%
As a \acs{BH} forms through gravitational collapse in a \acs{DM} core it can be \emph{kicked}, via \acs{GW} emission, and left in an oscillatory motion around the center of the core. 
The reason for the kick is that collapse is, in general, an asymmetric process, and leads to emission of \acsp{GW} which carry some momentum. This process is known
to lead to velocities of at most a few hundred kilometers per second~\cite{1973ApJ...183..657B}, generally, smaller than the galactic escape velocity. Thus, the remnant \acs{BH} is bound to the galaxy and, in absence of dissipation, performs an oscillatory motion.

It is crucial to understand how the \acs{DM} core reacts to this motion and to quantify the energy and momentum radiated and deposited in the scalar field. Similar issues were addressed in Ref.~\cite{Gualandris:2007nm}, in the context of the interaction between a kicked supermassive \acs{BH} and stars in galaxy cores.

At the center of a \acs{NBS}, the mass density is approximately constant~$\rho_M\simeq 4\times 10^{-3} M_\text{NBS}^4\mu_S^6$. So, the motion of the perturber is
\begin{align}
&z_p(t)=	-\mathcal{A} \sin\left(\omega_\text{osc} t\right)\,,  \\
&\mathcal{A} \equiv \sqrt{\frac{3}{4 \pi} \frac{v_0^2}{\rho_ M}}\,, \qquad \omega_\text{osc}\equiv \sqrt{\frac{4 \pi \rho_M}{3}}\,,
\end{align}
where~$v_0$ is the velocity of the perturber at the center of the core.
The source is then described by
\begin{align}
P=m_p \frac{\delta(\varphi)}{r^2 \sin \theta}\left[\delta\left(r-z_p(t)\right)\delta\left(\theta\right)+ \delta\left(r+z_p (t)\right)\delta\left(\theta-\pi\right)\right].
\end{align}
Using Eq.~\eqref{p_def} the function~$p$ reads
\begin{align}
p&=\frac{m_p}{2\sqrt{2 \pi}}\frac{|\tau'_{1,n}(r)|}{r}Y_{l,0}(0) \delta_m^0 \nonumber \\
&\quad\times
\sum_{n \in \mathbb{Z}}\Big[e^{-i \omega \tau_{1,n}}+e^{-i \omega \tau_{2,n}}+(-1)^n \left(e^{i \omega \tau_{1,n}}+e^{i \omega \tau_{2,n}}\right)\Big]\,,
\end{align}
where we defined~\footnote{The functions~$\tau_{1,n}(r)$ and~$\tau_{2,n}(r)$ are the roots of~$r+z_p(\tau)=0$; the symmetric functions~$-\tau_{1,n}(r)$ and~$-\tau_{2,n}(r)$ are the roots of~$r-z_p(\tau)=0$.}
\begin{align}
\tau_{1,n}&\equiv \frac{1}{\omega_\text{osc}}\left[\arcsin\left(\frac{r}{\mathcal{A}}\right)+2n \pi\right] \,, \nonumber \\
\tau_{2,n}&\equiv \frac{1}{\omega_\text{osc}}\left[(2n+1) \pi-\arcsin\left(\frac{r}{\mathcal{A}}\right)\right] \,.
\end{align}
In the last expressions we are using the principal branch of the inverse sine function. It is easy to see that the function~$p$ can be put in the form
\begin{align} \label{p_osc_start}
p&=\frac{m_p}{\sqrt{2 \pi}}\frac{Y_{l,0}(0)}{\sqrt{\mathcal{A}^2-r^2}}\frac{\delta_m^0}{\omega_\text{osc}}\,\Theta\left(\mathcal{A}-r\right) \nonumber \\
&\quad\times \sum_{n \in \mathbb{Z}}\bigg[\delta_l^\text{even} \left(\cos\left[\omega \tau_{1,n}(r)\right]+\cos\left[\omega \tau_{2,n}(r)\right]\right) \nonumber \\
&\quad-i\, \delta_l^\text{odd} \left(\sin\left[\omega \tau_{1,n}(r)\right]+\sin\left[\omega \tau_{2,n}(r)\right]\right)\bigg]\,.
\end{align}
Using the mathematical identities
\begin{align}
&\sum_{n \in \mathbb{Z}} \sin \left(2 n \pi \frac{\omega}{\omega_\text{osc}}\right)=0 \,, \\
&\sum_{n \in \mathbb{Z}} \cos \left(2 n \pi \frac{\omega}{\omega_\text{osc}}\right)=\omega_\text{osc} \sum_{n \in \mathbb{Z}} \delta(\omega-n \omega_\text{osc})\,,
\end{align}
together with some trigonometric identities, one can rewrite~\eqref{p_osc_start} as
\begin{align} 
p&=m_p\sqrt{\frac{2}{\pi}}\frac{Y_{l,0}(0)}{\sqrt{\mathcal{A}^2-r^2}}\delta_m^0 \,\Theta\left(\mathcal{A}-r\right) \sum_{n \in \mathbb{Z}} \delta(\omega-2n\omega_\text{osc}) \nonumber \\
&\quad\times \bigg[\delta_l^\text{even} \cos\left(2 n \arcsin\frac{r}{\mathcal{A}}\right)-i\, \delta_l^\text{odd} \sin\left(2 n \arcsin\frac{r}{\mathcal{A}}\right)\bigg]\,.
\end{align}
With the help of the trigonometric identities
\begin{align}
&\cos(2n x)=\sum_{k=0}^{n}(-1)^k \binom{2n}{2k}\sin^{2k} x \cos^{2(n-k)}x\,, \\
&\sin(2n x)=\sum_{k=0}^{n-1}(-1)^k \binom{2n}{2k+1}\sin^{2k+1} x \cos^{2(n-k)-1}x\,,
\end{align}
the last expression can be written in the alternative form
\begin{align} 
p&=m_p\sqrt{\frac{2}{\pi}}Y_{l,0}(0)\delta_m^0 \,\Theta\left(\mathcal{A}-r\right) \sum_{n \in \mathbb{Z}} \frac{1}{\mathcal{A}^{2n}}\delta(\omega-2n\omega_\text{osc}) \nonumber \\
&\quad\times \bigg[-i\, \delta_l^\text{odd} \sum_{k=0}^{n-1}(-1)^k \binom{2n}{2k+1}r^{2k+1}\left(\mathcal{A}^2-r^2\right)^{n-k-1} \nonumber \\
&\quad+\delta_l^\text{even}\sum_{k=0}^{n}(-1)^k \binom{2n}{2k} r^{2k}\left(\mathcal{A}^2-r^2\right)^{n-k-\frac{1}{2}}\bigg]\,.
\end{align}

We want to calculate the energy carried by the radiated scalar field due to the oscillatory motion of the massive object. First, note that the oscillation frequency is~$\omega_\text{osc}\sim 0.135M^2_\text{NBS}\mu_S^3\lesssim \gamma$. Only the modes with~$n\geq 1$ arrive at infinity; so, only these contribute to the energy radiated.
Applying the formalism described in Section~\ref{sec:SmallPert}, we obtain
\begin{align}
&Z_1^\infty=4 \pi \int_0^\mathcal{A}dr'F_{4,6}^{-1}(r')p(r')\,,  \\
&Z_2^\infty(\omega,l,0)=Z_1^\infty(-\omega,l,0)^*\,.
\end{align}
The energy radiated per unit of time is (Eq.~\eqref{Energy_flux_rate})
\begin{align}
\dot{E}^\text{rad}&=\frac{2}{\pi} \sum_{l, n}\left(\mu_S-\gamma+2n\omega_\text{osc}\right) \Re\left[\sqrt{2 \mu_S (2 n \omega_\text{osc}-\gamma)}\right]|\widetilde{Z}_1^\infty|^2  \\
&\simeq\frac{2}{\pi}\mu_S \sum_{l, n}  \Re\left[\sqrt{2 \mu_S (2 n \omega_\text{osc}-\gamma)}\right]|\widetilde{Z}_1^\infty(2n \omega_\text{osc},l,0)|^2 \,,
\end{align}
where we used that both the~\acs{NBS} and its perturbations are non-relativistic,~$\gamma \ll \mu_S$ and~$\omega_\text{osc}\ll \mu_S$, and defined the quantity
\begin{align}
&\widetilde{Z}_1^\infty\equiv 4 \pi \int_0^\mathcal{A}dr'F_{4,6}^{-1}(r')\widetilde{p}(r')\,, \\
&\widetilde{p}\equiv m_p\sqrt{\frac{2}{\pi}}\frac{Y_l^0(0)}{\sqrt{\mathcal{A}^2-r^2}} \nonumber \\
&\quad\times\sum_{n \in \mathbb{Z}}  \bigg[\delta_l^\text{even} \cos\left(2 n \arcsin\frac{r}{\mathcal{A}}\right)-i\, \delta_l^\text{odd} \sin\left(2 n \arcsin\frac{r}{\mathcal{A}}\right)\bigg]\,.
\end{align}

One can anticipate that the dominant contribution to the radiation is given by the~$n=1$ mode, which has a frequency~$\omega=2\omega_\text{osc}$. This is the lowest frequency radiated by the perturber and, thus, we expect it to be the one carrying more energy, because the coupling between the perturber and the scalar is stronger for lower frequencies -- as will become evident in the following sections. Indeed, this is in accordance with our numerics. So, we focus on the single~$n=1$ mode. For oscillations deep inside the \acs{NBS} with an amplitude~$\mathcal{A}\ll R$ -- which is where our constant density approximation holds -- we find that the following semi-analytic expression is a good description of our numerical results, 
\begin{align}
\dot{E}^\text{rad}&=\frac{2 \sqrt{2}}{\pi} (m_p\mu_S)^2 \sqrt{ \frac{2\omega_\text{osc}-\gamma}{\mu_S}}\sum_l c_l\, \left(\frac{\mathcal{A}}{R}\right)^{2(l+1)},
\end{align}
with the numerical constants~$c_l$. For the first multipoles we find
\begin{align*}
&c_0\simeq 0.852\,, \qquad c_1 \simeq 67.7 \,, \qquad c_2 \simeq 30.4\,, \nonumber \\
&c_3 \simeq 438\,, \qquad\;\;\; c_4 \simeq 13.6\,, \qquad c_5\simeq 3.85\,.
\end{align*}
The above expression describes our numerics with less than~$1\%$ of error for~$\mathcal{A}/R \lesssim 0.09$. These amplitudes correspond to kicks of~$v_0\lesssim 0.1 M_\text{NBS}\mu_S$, which contains astrophysical relevant velocities; for a \acs{DM} core with mass $\sim10^{10} M_\odot$ and a scalar with mass~$\hbar \mu_S \sim 10^{-22} \text{eV}$, our expression covers $v_0\lesssim 300\, \text{km/s}$, which contains typical recoil velocities imparted by \acs{GW} emission in gravitational collapse. Larger kicks, like the ones delivered in a merger of two supermassive \acsp{BH}, have larger amplitudes and are out of our approximation. However, the framework of Section~\ref{sec:SmallPert} (without the constant density approximation) can still be applied to those cases.

Using the same reasoning that we applied to the \emph{orbiting particles} in Section~\ref{sec:SmallPert} to obtain~\eqref{eq:E_loss_circular}, we can estimate the energy that is lost by the compact object per unit of time to be
\begin{align}
&\hspace{-0.78cm}\dot{E}^\text{lost}=\frac{2}{\pi} \sum_{l, n}\left(2n\omega_\text{osc}-\gamma\right) \Re\left[\sqrt{2 \mu_S (2 n \omega_\text{osc}-\gamma)}\right]|\widetilde{Z}_1^\infty(2n \omega_\text{osc},l,0)|^2\,.
\end{align}
Considering the single (dominant)~$n=1$ mode, the numerical evaluation of the last expression is well described by the semi-analytic formula
\begin{align}
\dot{E}^\text{lost}&=\frac{2 \sqrt{2}}{\pi} (m_p\mu_S)^2 \left( \frac{2\omega_\text{osc}-\gamma}{\mu_S}\right)^{\frac{3}{2}}\sum_l c_l\, \left(\frac{\mathcal{A}}{R}\right)^{2(l+1)}.
\end{align}
Again, this describes our numerics with less than~$1\%$ of error for small amplitude oscillations~$\mathcal{A}/R\leq 0.09$.

One may wonder how long it takes for a kicked \acs{BH} (or star) to settle down at the center of a \acs{DM} core purely due to the \acs{DF} caused by ultralight \acs{DM}. When the condition
\begin{align}\label{AdiabatCond}
\frac{\dot{E}^\text{lost} \left(\frac{2\pi}{\omega_\text{osc}}\right)}{\frac{1}{2}m_p \omega_\text{osc}^2 \mathcal{A}^2} \ll 1
\end{align}
is verified, the system is suited to an adiabatic approximation, and we can compute how the amplitude changes with time by solving
\begin{align}
m_p \omega_\text{osc}^2 \mathcal{A} \,\dot{\mathcal{A}}=-\dot{E}^\text{lost}\,.
\end{align}
Several astrophysical systems fall within this approximation. For example, for a \acs{DM} core with mass~$\sim 10^{10}M \odot$ and a scalar with mass~$\hbar \mu_S \sim 10^{-22} \text{eV}$, we have~$M_\text{NBS} \mu_S \sim 10^{-2}$; so, for an object forming through gravitational collapse and receiving a kick of~$300\,\text{km/s}$, via \acs{GW} emission, the adiabatic approximation is suitable if~$m_p/M_\text{NBS} \ll 0.1$ -- which is verified by all known compact objects. Using only the dominant multipole~$l=0$ (which accounts for more than~$61\%$ of the total energy loss for~$\mathcal{A}/R\leq 0.09$, and more than~$89\%$ for~$\mathcal{A}/R\leq0.04 $) we obtain
\begin{align}
\mathcal{A}=\mathcal{A}_0\, e^{-t/\tau_\text{S}}\,,
\end{align}
with the timescale
\begin{align}
\hspace{-0.4cm}\tau_{S}\simeq \frac{56}{m_p M_ \text{NBS} \mu_S^3} \sim 10^{10} \text{yr} \left(\frac{10^{-22}\, \text{eV}}{\hbar \mu_S}\right)^2\left(\frac{10^5 M_ \odot}{m_p}\right)\left(\frac{0.01}{M_ \text{NBS}\mu_S}\right)\,. 
\end{align}
So, an object kicked at the center of a \acs{NBS}, interacting solely with the scalar, settles down in a timescale smaller than an Hubble time if it has a mass $m_p\gtrsim 10^5 M_\odot$; in other words, if it is a supermassive \acs{BH}. 

The above timescale is in general much larger than the period of oscillation,
\begin{align}
\tau_S  \sim \frac{M_\text{NBS}}{m_p}\, \tau_\text{osc}\,.
\end{align}
This suggest that treating the source as eternal is indeed a good approximation to study this process.
It is interesting to compare this result with the timescale of damping due to \acs{DF} caused by stars in the galactic core. In Ref.~\cite{Gualandris:2007nm} the authors estimate that timescale to be
\begin{align}
\tau^*\sim 0.1\, \frac{M_\text{c}}{m_p} \,\tau_\text{osc} \,,
\end{align}
where~$M_\text{c}$ is the galactic core mass.
For the same mass,~$M_\text{c}=M_\text{NBS}$, we see that~$\tau^* \sim 0.1\,\tau_S$, which is smaller but still comparable to~$\tau_S$. Both ours and Ref.~\cite{Gualandris:2007nm} calculations are only order of magnitude estimates, but our result suggests that \acs{DM} may exert a \acs{DF} comparable to the one due to stars in processes happening in galactic cores.

%%%%%%%%%%%%%%%%%%%%%%%%%%%%%%%%%%%%%%%%%%%%%%%%%%%%%%%%%%%%%%%%%%%%%%%%%%%%%%%%
\section{Low-frequency binaries\label{Orbiting_particle_BS}}
%%%%%%%%%%%%%%%%%%%%%%%%%%%%%%%%%%%%%%%%%%%%%%%%%%%%%%%%%%%%%%%%%%%%%%%%%%%%%%%%

We now focus on orbiting objects within a \acs{NBS}. These will describe binaries, either at an early or late stage in their life, stirring the field and producing disturbances in the local \acs{DM} profile. The matter moments in Eq.~\eqref{p_orbiting} can describe, for instance, stars orbiting around the \emph{SgrA$^*$} \acs{BH} at the center of the Milky Way. This supermassive \acs{BH} has a mass~$\sim 4\times 10^6 M_{\odot}$ with known orbiting companions. The closest known star, \emph{S2}, has a pericenter distance of~$\sim 2800M_\text{BH}$ and a mass~$m_p\sim 20 M_{\odot}$, with large uncertainty~\cite{Abuter:2018drb,Abuter:2020dou}. Its orbit is, however, highly eccentric. Given the mass and sizes of the \acsp{NBS} (modeling the cores of \acs{DM} halos) discussed here, all these systems can be handled via perturbation techniques.
In addition, binaries close to supermassive \acsp{BH}, and therefore to galactic centers, may have been observed recently via electromagnetic counterparts to \acsp{GW}~\cite{Graham:2020gwr}.
%%%%%%%%%%%%%%%%%%%%%%%%%%%%%%%%%%%%%%%%%%%%%%%%%%%%%%%
\subsection{Scalar emission and energy loss}
\begin{figure}
	\centering
	\includegraphics[width=\textwidth]{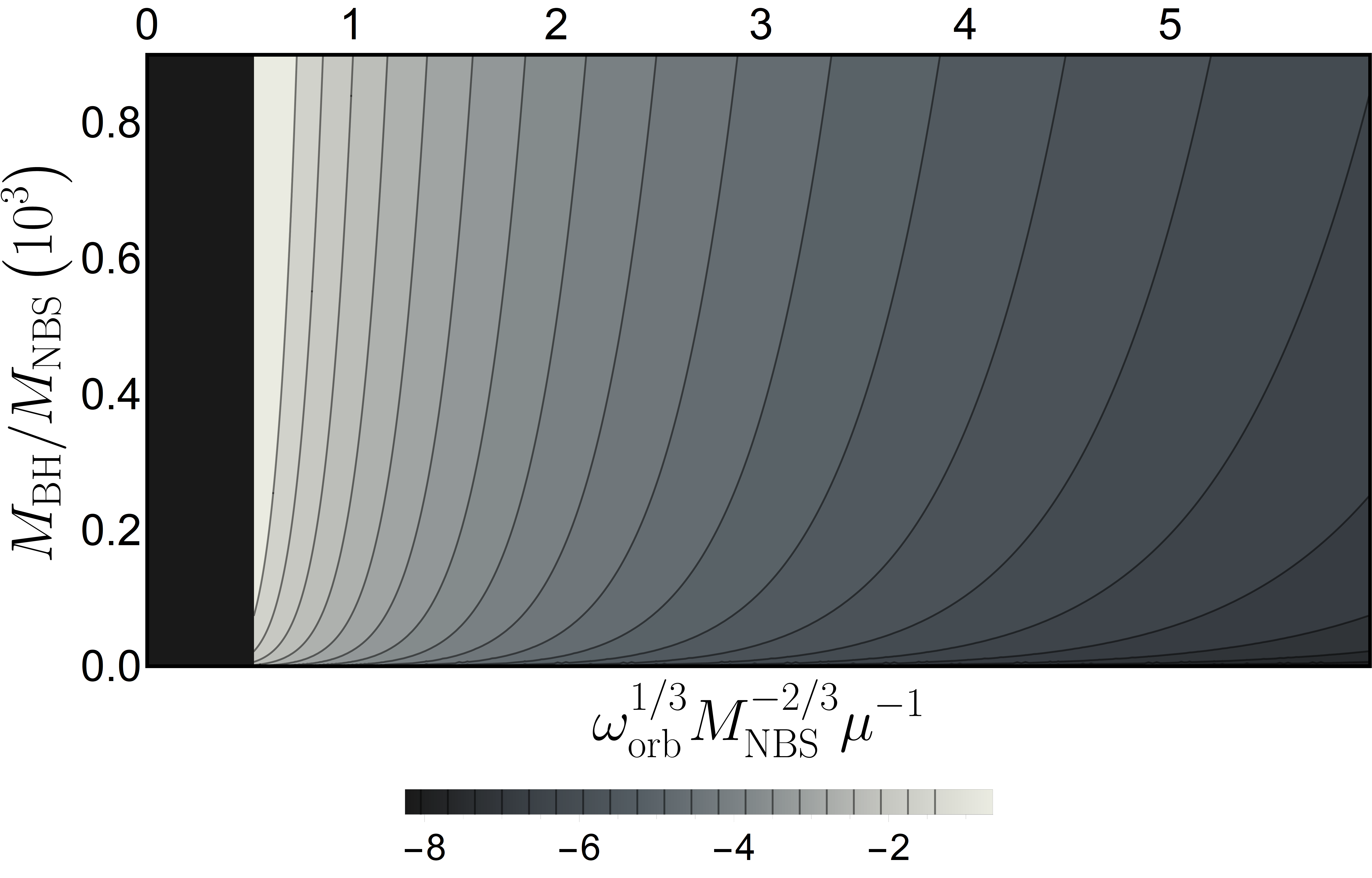} 
	\caption{Rate of scalar field energy radiated by an \acs{EMRI} inside a \acs{NBS}, $\log_{10}\left[\dot{E}_\text{EMRI}^\text{rad}\left(m_p^2 M_\text{NBS}\mu_S^3\right)^{-1}\right]$.
	The \acs{EMRI} is described by a supermassive \acs{BH} of mass $M_\text{BH}$ sitting at the center of the \acs{NBS}, and a star or stellar-mass~\acs{BH} in a circular orbit around it. 
	Note that the maximum energy emitted is associated with the smallest frequency (largest distance). 
	For a \acs{DM} core with~$M_\text{NBS}\sim 10^{10} M_{\odot}$ and mass ratio $m_p/M_\text{BH}\sim 10^{-4}$, the orbital distances corresponding to nonzero fluxes are in the range~$r_\text{orb}\lesssim 10^6 \,M_\text{BH}$. For larger radii, the fluctuation has too low an energy and is confined to the structure. This explains the zero-flux (black) region on the left of the panel, corresponding to the suppression of perturbations with frequency~$\omega \leq\gamma$.}  	
	\label{fig:circularEMRI}
\end{figure}

%%%%%%%%%%%%%%%%%%%%%%%%%%%%%%%%%%%%%%%%%%%%%%%%%%%%
Let us consider first an \acs{EMRI}, \ie, a perturber of mass~$m_p$ orbiting a supermassive \acs{BH} of mass $M_\text{BH}\gg m_p$ placed at the center of a \acs{NBS}. Solving the perturbation equations~\eqref{BS_Perturbation_Matrix_Sourced} with the source~\eqref{p_orbiting} (applying the transformation rule~$m_p(1+(-1)^m)\to m_p$) we find that, up to $3\%$ accuracy, the rate of energy radiated and lost by the perturber (Eqs.~\eqref{eq:circular_flux_Erad} and~\eqref{eq:E_loss_circular}) are described by~\footnote{Note that, for each multipole~$l$, there is emission for orbital frequencies larger than~$\gamma/l$. However, since the emission in multipoles higher than dipole is suppressed by roughly a factor~$10^3$, we consider only~$l=1$ in \eqref{Erad_circular_BS}.}
\begin{align}
\dot{E}^\text{rad}_\text{EMRI}&= 10^{-2}\, m_p^2 M_\text{BH}^{2/3}M_\text{NBS}^{4} \mu^{17/2}\omega_\text{orb}^{-11/6}\Theta(\omega_\text{orb}-\gamma)\nonumber\\
&\quad\times\Big[2.66- 0.49 \, M_\text{NBS}^{4/3}\mu_S^2\omega_\text{orb}^{-2/3} +0.054 \, M_\text{NBS}^{8/3}\mu_S^4\omega_\text{orb}^{-4/3} \Big],\label{Erad_circular_BS}\\
\dot{E}^\text{lost}_\text{EMRI}&= 10^{-2}\, m_p^2 M_\text{BH}^{2/3}M_\text{NBS}^{4} \mu_S^{15/2}\omega_\text{orb}^{-5/6}\Theta(\omega_\text{orb}-\gamma)\nonumber\\
&\quad\times \Big[2.70- 0.96 \, M_\text{NBS}^{4/3}\mu_S^2\omega_\text{orb}^{-2/3} +0.043 \, M_\text{NBS}^{8/3}\mu_S^4\omega_\text{orb}^{-4/3} \Big].\label{Elost_circular_BS}
\end{align}
The results described by~\eqref{Erad_circular_BS} and~\eqref{Elost_circular_BS} were obtained for a non-relativistic scalar field perturbation, and therefore are valid for orbital periods~$T_\text{orb}=2\pi/\omega_\text{orb}\gg 2\pi/\mu_S\sim 10^{-22}\text{eV}/(\hbar\mu_S)\, \text{yr}$. 
We show in Fig.~\ref{fig:circularEMRI} the flux of radiated energy ($E^\text{rad}$) as a function of the orbital period and of the~\acs{BH}-\acs{NBS} mass ratio. Once the orbital frequency is fixed, our results are consistent with exponential convergence in~$l$ for the flux. 

The above calculation is easy to adapt to other systems. Consider an equal mass binary system ($M=2m_p$). Looking at the matter moments in Eq.~\eqref{p_orbiting}, it is clear that the first multipole moment contributing to the scalar emission is the quadrupole $l=|m|=2$. Solving numerically the perturbation equations, we find that the following expressions provide a very good description of the rate of energy emitted in scalar waves and lost by the orbiting particle (up to $3\%$ of accuracy)
\begin{align}
\dot{E}^\text{rad}&= 10^{-2}\, M^{4/3} m_p^{2}M_\text{NBS}^{4} \mu_S^{19/2}\omega_\text{orb}^{-13/6}\Theta\left(2\omega_\text{orb}-\gamma\right) \nonumber\\
&\quad \times\left[1.45- 0.16 \, M_\text{NBS}^{4/3}\mu_S^2\omega_\text{orb}^{-2/3} +0.015 \, M_\text{NBS}^{8/3}\mu_S^4\omega_\text{orb}^{-4/3} \right],\label{Erad_circular_BS_equalmass}\\
\dot{E}^\text{lost}&=10^{-2}\, M^{4/3} m_p^{2}M_\text{NBS}^{4} \mu_S^{17/2}\omega_\text{orb}^{-7/6}\Theta\left(2\omega_\text{orb}-\gamma\right)  \nonumber\\
&\quad\times \left[2.97- 0.58 \, M_\text{NBS}^{4/3}\mu_S^2\omega_\text{orb}^{-2/3} +0.0051 \, M_\textrm{NBS}^{8/3}\mu_S^4\omega_\text{orb}^{-4/3} \right].\label{Elost_circular_BS_equalmass}
\end{align}
These expressions are valid, \eg, for solar mass \acsp{BH}, or \acs{BH} masses of the order~$\sim 10^4 M_{\odot}$.

In the limit~$\omega_\text{orb} \gg \gamma$, but still for non-relativistic $\omega_\text{orb} \ll \mu_S$ excitations, the relevant equations~\eqref{Sourced_SP_System11} and \eqref{Sourced_SP_System22} can be solved analytically in closed form. 
Equation~\eqref{Sourced_SP_System22} has the simple solution
\begin{align}
&\delta U=\frac{2}{\sqrt{2 \pi}}\sum_{l,m} \frac{u(r)}{r} Y_{lm}(\theta, 0) e^{-i m \left(\omega_\text{orb} t -\varphi\right)} \,,
\end{align}
with
\begin{align}
u&=-\left(2 \pi\right)^{3/2} m_p \left[1+(-1)^m\right]\frac{Y_{lm}\left(\frac{\pi}{2},0\right)}{2l+1} \nonumber \\
&\qquad\times  \left[\left(\frac{r}{r_\text{orb}}\right)^{-l} \Theta(r-r_\text{orb})+\left(\frac{r}{r_\text{orb}}\right)^{l+1} \Theta(r_\text{orb}-r)\right]\,.
\end{align}
Then, using the decomposition
\begin{align}
\delta \Psi = \frac{2}{\sqrt{2 \pi}}\sum_{l,m} \frac{Z(r)}{r} Y_{lm}(\theta, 0)e^{-i m\left(\omega_\text{orb} t-\varphi\right)}\,,
\end{align}
equation~\eqref{Sourced_SP_System1} becomes
\begin{align}
\partial_r^2 Z+\left(2 \mu_S m\omega_\text{orb}-\frac{l(l+1)}{r^2}\right)Z=2 \mu_S^2 \Psi_0  u\,.
\end{align}
Using the method of variation of parameters, one can solve the last equation imposing the Sommerfeld radiation condition at large distances and regularity at the origin. The obtained solution is, at large distances, 
\begin{align} \label{vop_hf}
Z(r \to \infty)= i \pi\mu_S^2Z_\infty(r\to \infty) \int_{0}^{\infty} dr' Z_0\Psi_0 u\,,
\end{align}
where $Z_0$ and $Z_\infty$ are homogeneous solutions satisfying, respectively, regularity at the origin and the Sommerfeld radiation condition at large distances, and are given by
\begin{align}
Z_0&=\sqrt{r}\, J_{l+1/2}\left(\sqrt{2 \mu_S m \omega_\text{orb}}r\right)\,, \\
Z_\infty&= \sqrt{r} H^{(1)}_{l+1/2}\,\left(\sqrt{2\mu_S m \omega_\text{orb}}r\right)\,,
\end{align}
with~$J_\nu(x)$,~$H^{(1)}_{\nu}(x)$ Bessel and Hankel functions~\cite{Abramowitz:1970as}.
Using the asymptotic form 
\begin{align}
Z_\infty (r\to \infty)\simeq (-i)^{l+1}\sqrt{\frac{2}{\pi}} \frac{e^{i \sqrt{2\mu_S m \omega_\text{orb}}\,r }}{\left(2\mu_S m \omega_\text{orb}\right)^{1/4}}\,,
\end{align} 
and assuming that~$r_\text{orb}\ll R$, and~$\omega_\text{orb}/\mu_S\gg \left(r_\text{orb} \mu_S\right)^{-2}$, the integration in~\eqref{vop_hf} converges a few wavelengths from the binary and gives
\begin{align}
Z(r \to \infty) &\simeq -(-i)^l \left(2 \pi\right)^2 \mu_S^{2} m_p \Psi_0(0)r_\text{orb}^l\nonumber \\
&\quad\times \left[1+(-1)^m\right]  \frac{2^{-\frac{l}{2}-\frac{3}{2}}\,e^{i \sqrt{2\mu_S m \omega_\text{orb}}\,r }}{\left(\mu_S m \omega_\text{orb}\right)^{1-\frac{l}{2}}} \frac{Y_{lm}\left(\frac{\pi}{2},0\right)}{\Gamma\left(l+\frac{3}{2}\right)}\,.
\end{align}
So, the dominant~$|m|=l$ modes result in the scalar perturbation
\begin{align}
\delta \Psi (r\to \infty) &\simeq -8 \pi^{\frac{3}{2}}\mu_S^2 m_p \Psi_0(0) \sum_{m=1}^{+\infty}(-i)^m\left[1+(-1)^m\right] \nonumber \\
&\quad\times  \frac{Y_{lm}\left(\frac{\pi}{2},0\right)}{\Gamma\left(m+\frac{3}{2}\right)} \frac{(\mu_S m)^{\frac{m}{2}-1}(M \omega_\text{orb})^{\frac{m}{3}}}{2^{2+\frac{m}{2}}\omega_\text{orb}^{\left(1+\frac{m}{2}\right)}}e^{i \sqrt{2\mu_S m \omega_\text{orb}}\,r }\,,
\end{align}
where we have used Kepler's law $r_\text{orb}^3=M/\omega_\text{orb}^2$.
Then, the flux of radiated energy is given by
\begin{align}
\dot{E}^\text{rad}&= r^2 \lim_{r \to \infty}\int d\theta d\varphi \sin \theta \,\delta T_S^{\;\;rt}\nonumber \\
&= 0.28\, \pi^{3} \left(\mu_S m_p\right)^2 \left(\mu_S M_\text{NBS}\right)^4 \sum_{m=1}^{+\infty}\left[1+(-1)^m\right]^2 \nonumber\\ &\times\left(1+\frac{m \omega_ \text{orb}}{\mu_S}\right) \left(\frac{Y_{mm}\left(\frac{\pi}{2},0\right)}{\Gamma\left(m+\frac{3}{2}\right)} \frac{m^{\left(\frac{m}{2}-\frac{3}{4}\right)}(M\omega_\text{orb})^{\frac{m}{3}}}{2^{\left(\frac{7}{4}+\frac{m}{2}\right)}(\omega_\text{orb}/\mu_S)^{\left(\frac{3}{4}+\frac{m}{2}\right)}}\right)^2\,.
\end{align}
The last expression can be further simplified using~$\left(1+m\omega_ \text{orb}/\mu_S\right)\simeq 1$, since we are considering non-relativistic excitations of the scalar field.
The same reasoning that we used to derive~\eqref{eq:E_loss_circular} can be applied here to find that the binary loses energy at a rate
\begin{align}
\dot{E}^\text{lost}&\simeq 0.28\, \pi^{3} \left(\mu_S m_p\right)^2 \left(\mu_S M_\text{NBS}\right)^4 \sum_{m=1}^{+\infty}\left[1+(-1)^m\right]^2 \nonumber\\ &\quad\times \left(\frac{Y_{mm}\left(\frac{\pi}{2},0\right)}{\Gamma\left(m+\frac{3}{2}\right)} \frac{m^{\left(\frac{m}{2}-\frac{1}{4}\right)}(M\omega_\text{orb})^{\frac{m}{3}}}{2^{\left(\frac{7}{4}+\frac{m}{2}\right)}(\omega_\text{orb}/\mu_S)^{\left(\frac{1}{4}+\frac{m}{2}\right)}}\right)^2\,.
\end{align}

Remarkably, these analytic results are in excellent agreement with our numerics for both \acsp{EMRI} (Eq.~\eqref{Erad_circular_BS}) and equal mass binaries (Eqs.~\eqref{Erad_circular_BS_equalmass}); the leading order terms agree with the numerical results within $4\%$. Such agreement is a cross-check both on our numerical routine and our simple analytical description.

%%%%%%%%%%%%%%%%%%%%%%%%%%%%%%%%%%%%%%%%%%%%%%%%%%%%%%%%%%%
\subsection{Comparison with gravitational wave emission}
%%%%%%%%%%%%%%%%%%%%%%%%%%%%%%%%%%%%%%%%%%%%%%%%%%%%%%%%%%%
In vacuum, the orbit of a binary system shrinks in time, due to the emission of \acsp{GW}. At leading order, loss via \acsp{GW} is described by the quadrupole formula~\cite{Peters:1963ux,Peters:1964zz,Poisson:1993vp},
\begin{align}
\dot{E}^\text{GW}=\frac{32}{5}\eta^2\left(M\omega_\text{orb}\right)^{10/3}\,,
\label{eq:quadrupole}
\end{align}
where $\eta=m_1m_2/(m_1+m_2)^2$ is the symmetric mass ratio of a binary of component masses~$m_1$,~$m_2$ and total mass~$M=m_1+m_2$. To estimate the flux of energy radiated in the scalar channel, we consider the orbit to be circular, with the radius equal to the semi-major axis ($\sim 970$ au) of the \emph{S2} star. The scalar field in a \acs{NBS} provides an extra channel for energy loss. For \acsp{EMRI} ($m_p=\eta M$ and $M_\text{BH}=M$), combining together Eqs.~\eqref{Elost_circular_BS} and~\eqref{eq:quadrupole}. we get~\footnote{Since the total scalar field mass contained in a sphere of radius~$r_\text{orb}\ll R$ is negligible with respect to the mass of the central \acs{BH}~$M_\text{NBS}(r_\text{orb})/M\sim 10^{-10}$, we can consider that the entire \acs{GW} flux emitted is due to the quadrupole moment of the binary alone, neglecting the gravitational field of the \acs{DM} halo.}
\begin{align}
\hspace{-0.6cm}\frac{\dot{E}^\text{lost}}{\dot{E}^\text{GW}}\simeq   10^{-3}\, \left[\frac{M_\text{NBS}}{10^{10}M_{\odot}}\right]^4\left[\frac{10^{6}M_{\odot}}{M}\right]^{2/3}\left[\frac{T}{16 \text{ yr}}\right]^{31/6}  \left[\frac{\hbar \mu_S}{10^{-22}\text{ eV}}\right]^{17/2},
\end{align}
where we normalized the ratio to typical values for an \acs{EMRI} composed by \emph{Sagittarius $\text{A}^*$} and \emph{S2} star surrounded by a \acs{DM} core.

The energy balance equation imposes that the loss in the orbital energy of the binary is due to the energy carried away by scalar field and gravitational waves~\cite{1989ApJ...345..434T,Stairs:2003eg} 
\begin{align}
\frac{d E^\text{orb}}{dt}=-\left(\dot{E}^\text{lost}+\dot{E}^\text{GW}\right)\,.\label{eq:energy_balance}
\end{align}

Thus, energy loss leads to a secular change in orbital period
\begin{align}
\dot{T}\simeq-\frac{192\pi\left(2\pi\right)^{5/3}\eta M^{5/3}}{5T^{5/3}}-\frac{5 \eta M M_\text{NBS}^{4} T^{5/2}}{ 10^{3}\mu_S^{-15/2}}\,.
\end{align}

It is amusing to estimate such secular change for astrophysical parameters similar to those of \emph{S2} star orbiting around \emph{SgrA$^*$}, 
\begin{align}
\dot{T}&\simeq \, -\frac{2.42}{10^{15}}  \left[\frac{M}{10^{6}M_{\odot}}\right]^{2/3}\left[\frac{T}{16 \text{ yr}}\right]^{-5/3}\left[\frac{m_p}{20 M_{\odot}}\right]\nonumber\\
&-\frac{4}{10^{17}}\left[\frac{M_\text{NBS}\mu_S}{0.01}\right]^4\left[\frac{\hbar\mu_S}{10^{-22}\text{eV}}\right]^{7/2}\left[\frac{T}{16 \text{ yr}}\right]^{5	/2}\left[\frac{m_p}{20 M_{\odot}}\right]\,,
\end{align}
which seems hopelessly small.

The period change for equal-mass binary systems follows through, and it is
\begin{align}
\dot{T}=-\frac{192\pi\left(2\pi\right)^{5/3}M^{5/3}}{20T^{5/3}}-\frac{3.1 M_\text{NBS}^{4} m_p M^{2/3}T^{17/6}}{10^{3}\mu_S^{-17/2}}\,.
\end{align}
%

%%%%%%%%%%%%%%%%%%%%%%%%%%%%%%%%%%%%%%%%%%%%%%%%%%%%%%%%%%%%%%%
\subsection{Backreaction and scalar depletion}
%%%%%%%%%%%%%%%%%%%%%%%%%%%%%%%%%%%%%%%%%%%%%%%%%%%%%%%%%%%%%%%
One cause for concern is that our calculation assumes a fixed scalar field background~$\Psi_0$, but as the binary evolves, scalar radiation is depleting the \acs{NBS} of scalar field around the binary. Assume, conservatively, that the flux above is only removing scalar field within a sphere of radius~$\sim 10 \,\ell$
centered at the binary, with the radiation wavelength~$\ell=2\pi/\omega_\text{orb}$. Then the timescale for total depletion of the scalar in the sphere is
\begin{align}
 \tau \sim \frac{\rho_M R^3}{\dot{E}^\text{rad}}&\sim 10^{24}\,\text{yr}\, \left[\frac{10^{-2}}{\mu_S M_\text{NBS}}\right]^{2/3}\left[\frac{10^4}{\chi}\right]^{2/3} \left[\frac{20\,M_{\odot}}{m_p}\right]^{2}\nonumber\\
&\times\left[\frac{10^{-22}\text{ eV}}{\hbar\mu_S}\right]^{11/6}\left[\frac{T}{16 \text{ yr}}\right]^{7/6}\,,
\end{align}
that is much larger than a Hubble timescale for an~\acs{EMRI}. A similar value can be found for equal mass binary systems. Thus, our results seem to indicate that the background configuration remains unaffected by the emission of scalars by low-frequency binaries.

%%%%%%%%%%%%%%%%%%%%%%%%%%%%%%%%%%%%%%%%%%%%%%%%%%%%%%%%%%%%%%%%%%%%%%%
\section{High-frequency binaries\label{BS_binaries}}
%%%%%%%%%%%%%%%%%%%%%%%%%%%%%%%%%%%%%%%%%%%%%%%%%%%%%%%%%%%%%%%%%%%%%%%

%%%%%%%%%%%%%%%%%%%%%%%%%%%%%%%%%%%%%%%%%%%%%%%%%%%%
\subsection{Scalar emission and energy loss}
%%%%%%%%%%%%%%%%%%%%%%%%%%%%%%%%%%%%%%%%%%%%%%%%%%%%

We now wish to focus on high-frequency binaries, such as those suitable for LIGO or LISA detections. In such system, the assumption of non-relativistic scalar perturbations is not appropriate. Instead, one can show that the description of these systems, with orbital frequencies~$\omega_\text{orb}\gg \mu_S$, may be accounted for by a slight modification of the previous equations (more details are given in Appendix~\ref{app:PN}),
\begin{align}
&\nabla^2\delta U= 4\pi P\,,\label{eq:high_binary1}\\
&\nabla^2 \delta \Phi-\partial^2_t \delta \Phi  =2 \mu_S^2 \Phi_ 0\, \delta U\,.\label{eq:high_binary2}
\end{align}

\begin{figure}
	\centering
	\includegraphics[width= \textwidth]{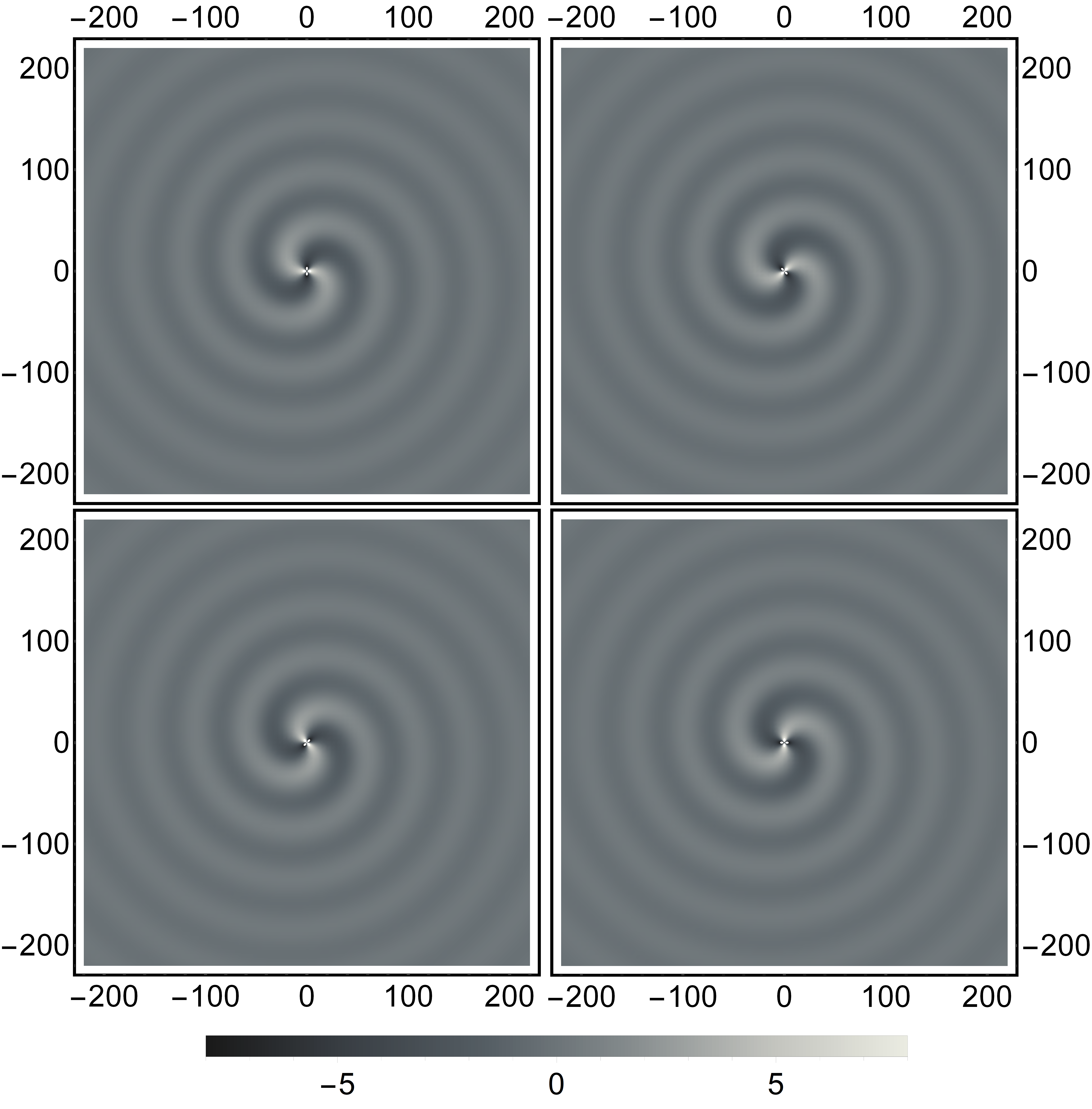}
	\caption{Scalar field perturbation due to an high frequency, equal-mass binary describing a circular orbit of radius~$r_\text{orb}$, and evolving inside an \acs{NBS}. The normalized horizontal and vertical axis are $x/r_\text{orb}$ and~$y/r_\text{orb}$, respectively. Each frame shows an equatorial slice of the scalar field perturbation~$10^{17} \Re \left(\delta\Phi\right)$, induced by a binary orbiting in the equatorial plane. In the upper-left panel, particles are at $(x_1,y_1)=(r_\text{orb},0)$, $(x_2,y_2)=(-r_\text{orb},0)$. Moving clockwise in the panels, the system evolves for an eighth of a period between each frame (the binary is orbiting anti-clockwise). The binary components have the same mass ($m_p\sim 10^6 M_\odot$) and they are orbiting inside a \acs{NBS} of mass $M_\text{NBS}\mu_S\sim 0.01$ with a period of~$\sim 1$ day.
	}
	\label{fig:highfreq_timedomain}
\end{figure}

We consider two equal-mass point particles, each of mass $m_p$, on a circular motion of orbital frequency $\omega_\text{orb}$ and radius $r_\text{orb}$ (described by the source~\eqref{P_orbiting}). We can solve the Poisson equation first, finding
\begin{align}
\delta U&=\sum_{lm}\frac{u_{lm}}{r}Y_{lm} (\theta,0)e^{im\left(\varphi-\varphi_p\right)}\,,\\
u_{lm}&=-\frac{4\pi m_p\left(1+(-1)^m\right)\,Y_{l m}(\pi/2,0)}{2l+1} r_\text{orb}^{-l-1}\nonumber\\
&\qquad\times\left[r_\text{orb}^{2l+1}r^{-l}\Theta(r-r_\text{orb})+r^{l+1}\Theta(r_\text{orb}-r)\right]\,.
\end{align}
Here,~$\varphi_p=\omega_\text{orb}t$ is the azimuthal location of one particle; the other is at $\varphi_p+\pi$.
To solve Eq.~\eqref{eq:high_binary2}, we decompose the scalar field as
\begin{align}
\delta \Phi=\frac{1}{\sqrt{2\pi}}\sum_{l,m}\int d\omega\, \frac{\delta\psi (\omega, r)}{r}e^{-i(\omega+\mu_S-\gamma) t} Y_{lm}(\theta, \varphi)\,,
\label{eq:fourier_deltaphi}
\end{align}
finding the following radial ODE for $\delta\psi$,
\begin{align}
\delta\psi''+\left((\mu_S-\gamma+\omega)^2-\frac{l(l+1)}{r^2}\right)\delta\psi=\sqrt{8\pi}\mu_S^2\Psi_{0}\,\tilde{u}_{lm} \,,
\end{align}
where $\tilde{u}_{lm}=u_{lm}\delta\left(\omega-m \omega_\text{orb}\right)$. Here, primes stand for radial derivatives. We can now solve the last equation using the method of variation of constants, requiring outgoing waves at large distances and regularity at
the origin. The solution is
\begin{align}
\hspace{-0.4cm}\delta\psi=\delta\psi_\infty \int_0^r \frac{2\sqrt{2\pi}\mu_S^2\Psi_{0}\,\delta\psi_H\,\tilde{u}_{lm}}{i\omega}+\delta\psi_H \int_r^\infty \frac{2\sqrt{2\pi}\mu_S^2\Psi_{0}\,\delta\psi_\infty\,\tilde{u}_{lm}}{i\omega}\,,
\label{eq:totaldeltapsi_highfreq}
\end{align}
where~$\omega=m\omega_\text{orb}\gg (\mu_S-\gamma)$ and~$\delta\psi_{H,\infty}$ are the homogeneous solutions,
\begin{align}
\delta\psi_{H}&=\sqrt{\frac{\pi\omega r}{2}}J_{l+1/2}(\omega r)\,,\\
\delta\psi_{\infty}&=\sqrt{\frac{\pi\omega r}{2}}\left(J_{l+1/2}(\omega r)+iY_{l+1/2}(\omega r)\right)\,,
\end{align}
with~$J_\nu(x)$ and~$Y_\nu(x)$ the usual Bessel functions~\cite{Abramowitz:1970as}.
The time domain response of the \acs{NBS} to the perturbations induced by a binary \acs{BH} system can be found by evaluating numerically~\eqref{eq:totaldeltapsi_highfreq} and~\eqref{eq:fourier_deltaphi}. Four snapshots of one period, for two equal mass \acsp{BH} are shown in Fig.~\ref{fig:highfreq_timedomain}.

A binary deep inside the \acs{NBS} ($r_\text{orb}\ll R$) and with high orbital frequency~($\omega_\text{orb}\gg 1/r_\text{orb}$) generates a field at large distances that is independent on the size of the \acs{NBS} -- the first integration in~\eqref{eq:totaldeltapsi_highfreq} converges a few wavelengths~($\sim1/\omega_\text{orb}$) away from the binary. We find the following simple result for the dominant $|m|=l$ modes,
\begin{align}
\delta\psi(r\to \infty)&=i\sqrt{2\pi} m_p\left(1+(-1)^m\right)\Psi_0 \pi^{3/2}\,2^{2-m}m^{m-2}\nonumber\\
&\qquad\times \frac{Y_{mm}(\pi/2,0)}{\Gamma[m+3/2]}\frac{\mu_S^2}{\omega_\text{orb}^2}(M\omega_\text{orb})^{m/3}\,e^{i\omega r}\,.
\end{align}
Here, the total mass of the equal-mass binary is~$M=2m_p$. If we substitute~$m_p\left(1+(-1)^m\right)\to m_p$, these results describe an \acs{EMRI}, where a single particle of mass~$m_p$ is revolving around a massive \acs{BH} of mass~$M$ (note the crucial difference that~$|m|=l=1$ modes are radiated for \acsp{EMRI}, whereas only \emph{even} modes are emitted for equal-mass binaries).
The radiated energy flux is given by
\begin{align}
\dot{E}^\text{rad}&= r^2 \lim_{r \to \infty}\int d\theta d\varphi \sin \theta \,\delta T_S^{\;\;r t}\nonumber \\
&=128 \pi^{3}(\mu_S^2 m_p \Psi_0(0))^2\left(1+(-1)^m\right)^2  \nonumber\\ 
&\qquad\times\sum_{m=1}^{+\infty}\left(\frac{Y_{mm}(\pi/2,0)}{\Gamma(m+3/2)} \frac{m^{m-1}(M\omega_\text{orb})^{m/3}}{2^{m+1}\,\omega_\text{orb}}\right)^2\,.\label{eq:energy_loss_high_binaries}
\end{align}
Since in this section we are considering ultra-relativistic excitations of the scalar ($\omega_\text{orb} \gg \mu_S$) it is easy to see that (at leading order) the rate of change of the \acs{NBS} energy~$\dot{E}_\text{NBS}$ is much smaller than~$\dot{E}^\text{rad}$.~\footnote{Note that, at leading order, $\dot{E}_ \text{NBS}=\hbar\mu_S\, \dot{Q}_ \text{NBS}$}
So, conservation of energy (as expressed in Eq.~\eqref{LossRadE}) implies that~$\dot{E}^\text{lost}\simeq \dot{E}^\text{rad}$.

%%%%%%%%%%%%%%%%%%%%%%%%%%%%%%%%%%%%%%%%%%%%%%%%%%%%%%%%%
\subsection{The phase dependence in vacuum and beyond}
%%%%%%%%%%%%%%%%%%%%%%%%%%%%%%%%%%%%%%%%%%%%%%%%%%%%%%%%%

In vacuum-general relativity, the dynamics of a binary is governed by the energy balance equation~\eqref{eq:energy_balance}, together with the quadrupole formula~\eqref{eq:quadrupole}.
This implies that the orbital energy of the system~$E_\text{orb}=-M^2\eta/(2r_\text{orb})$ must decrease at a rate fixed by such loss. This defines immediately the time-dependence of the \acs{GW} frequency to be~$f_ \text{GW}^{-8/3}=(8\pi)^{8/3}{\cal M}^{5/3}(t_0-t)/5$, where ${\cal M}$ is the \emph{chirp mass} and $f_\text{GW}=\omega_\text{orb}/\pi$. Once the frequency evolution is known, the \acs{GW} phase simply reads
\begin{equation}
\varphi(t)=2\int^t\omega_\text{orb}(t')dt' \,.\label{GWphase}
\end{equation}

To take into account dissipative losses via the scalar channel, we add to the quadrupole formula the energy flux~\eqref{eq:energy_loss_high_binaries}.
In Fourier domain one can write the gauge-invariant metric fluctuations as
\begin{align}
h_+(t)&=A_+(t_\text{ret})\cos\varphi(t_\text{ret}) \,,\\
h_\times(t)&=A_\times(t_\text{ret}) \sin\varphi(t_\text{ret})\,,
\end{align}
where~$t_\text{ret}$ is the retarded time. The Fourier-transformed quantities are
\begin{equation}
\tilde{h}_+= {\cal A}_+e^{i\Upsilon_+}\,,\qquad  \tilde{h}_\times={\cal A}_\times e^{i\Upsilon_\times}\,.
\end{equation}
Dissipative effects are included within the stationary phase approximation, where the secular time evolution is governed by 
the \acs{GW} emission~\cite{Flanagan:1997sx}. In Fourier space, we decompose the phase of the \acs{GW} signal $\tilde{h}={\cal A}(f_\text{GW})e^{i\Upsilon(f_\text{GW})}$ as
\begin{equation}
\Upsilon(f_\text{GW}) =\Upsilon_\text{GR}^{(0)}[1+\text{(PN\ corrections)}+\delta_{\Upsilon}]\,,
\end{equation}
where $\Upsilon_\text{GR}^{(0)}=3/128 ({\cal M}\pi f_\text{GW})^{-5/3}$ represents the leading term of the phase's post-Newtonian expansion. We find the following dominant correction due to~\acs{DF} of the background scalar field,
\begin{align}
\delta_{\Upsilon}=\frac{16\mu_S^4\Psi_0^2(0)}{51\pi^3f_\text{GW}^4}\sim 10^{-24}\left[\frac{\hbar\mu_S}{10^{-22}\,\text{eV}}\right]^4\left[\frac{10^{-4}\text{Hz}}{f_\text{GW}}\right]^4\left[\frac{M_\text{NBS}\mu_S}{0.01}\right]^4\,.
\end{align}
for equal-mass binaries. Such a correction corresponds to a $-6$ PN order correction~\cite{Yunes:2016jcc}. The smallness of the coefficient
makes it hopeless to detect the effect of a scalar with mass~$\sim10^{-22}\text{ eV}$ using a space-based detector like LISA~\cite{Audley:2017drz}. Note, however, that our result is very sensitive to the scalar's and~\acs{DM} core masses; \eg, for~$\hbar \mu_S\sim 10^{-19} \text{ eV}$ and a \acs{DM} core of mass~$\sim10^{12}M_\odot$ we find
\begin{align}
\delta_{\Upsilon}\sim 10^{-8}\left[\frac{\hbar\mu_S}{10^{-19}\,\text{eV}}\right]^4\left[\frac{10^{-4}\text{Hz}}{f_\text{GW}}\right]^4\left[\frac{M_\text{NBS}\mu_S}{0.1}\right]^4\,.
\end{align}
Pulsar timing arrays operate at lower frequencies~\cite{Barack:2018yly}, for which the previous Newtonian non-relativistic analysis is necessary; moreover, our results (\eg, Fig.~\ref{fig:circularEMRI}) indicate that the binary couples more strongly with the scalar for orbital frequencies~$\omega_\text{orb}\sim 2\gamma$, which motivates a more thorough analysis (to be done in the future) of those systems.

%%%%%%%%%%%%%%%%%%%%%%%%%%%%%%%%%%%%%%%%%%%%%%%%%%%%%%%%%
\subsection{Backreaction and scalar depletion}
%%%%%%%%%%%%%%%%%%%%%%%%%%%%%%%%%%%%%%%%%%%%%%%%%%%%%%%%%
During the evolution, the binary radiates scalar field out of the \acs{NBS}. Assuming, again, that the above flux is only removing scalar field within a sphere of radius $\sim 10 \,\ell$ centered at the binary (with the radiation wavelength $\ell=2\pi/\omega_\text{orb}$) the timescale for total depletion of scalar field around the binary is
\begin{align}
\tau\sim 2\times 10^{11}\,\text{yr}\,\left(\frac{0.1}{m_p\omega_\text{orb}}\right)^{7/3} \left(\frac{10^{-2}}{\mu_S M_\text{NBS}}\right)^{2}\left(\frac{\chi}{10^4}\right)^2\frac{m_p}{10^6M_{\odot}}\,,
\end{align}
which is larger than a Hubble timescale, even for binaries close to coalescence. Thus, our results seem to describe well the emission of scalars during the entire lifetime of a compact binary.

\section{Discussion}

This chapter shows how self-gravitating \acsp{NBS} (modeling ultralight \acs{DM} cores) respond to time-varying, localized matter fluctuations.
These are structures that behave classically: they are composed of $N ~\sim 10^{100}\left(10^{-22}\text{ eV}/\hbar\mu_S\right)^2$ particles; a binary of two supermassive \acsp{BH} in the late stages of coalescence emits more than~$10^{60}$ of those particles. Our results show unique features of bosonic ultralight structures. For example, they are not easily depleted by binaries. Even a supermassive \acs{BH} binary close to coalescence would need a Hubble time or more to completely deplete the scalar in a sphere of ten-wavelength radius around the binary. In other words, the perturbative framework is consistent and robust. We have shown how a self-gravitating \acs{NBS} background leads to regular, finite \acs{DF} acting on passing bodies. Our analysis includes both the self-gravity of the scalar field perturbations and the finite-size effects from the~\acs{NBS} background.

Clearly, our results can and should be extended to eccentric motion, or to self-gravitating vectorial configurations, or even other nonlinearly interacting scalars~\cite{Coleman:1985ki}.
Our results should also be a useful benchmark for numerical relativity simulations involving boson stars in the extreme mass ratio regime, when and if the field is able to accommodate such challenging setups. We have considered Newtonian boson stars. Extension of our results to relativistic boson stars is nontrivial, but would provide a full knowledge of the spectrum of boson stars and of their response to external agents. Although we studied \acsp{NBS} only, our methods can be extended to clouds arising from superradiant instabilities of spinning \acsp{BH}~\cite{Brito:2015oca}. We do not expect qualitatively new aspects, at least, when the spatial extent of those clouds is large. In this chapter we neglected the energy (angular momentum) deposited \emph{in}~\acsp{NBS}. This is a very interesting possibility, in particular for \acsp{EMRI} with~$\omega_\text{orb}<\gamma$, which can, in principle, deposit some of their energy in the normal modes of the scalar configuration, and leave some signatures in the orbit of these~\acsp{EMRI}. Another interesting question in this context is what would be the normal mode power spectrum of a~\acs{NBS} excited by hundreds (or millions) of \acsp{BHB} or stars at the galactic center. To solve these problems one needs, however, to obtain the second order perturbation $\delta^2\Phi$, which may turn out not to be an easy task. I intend to study these issues further in the near future.

%*****************************************
%*****************************************
%*****************************************
%*****************************************
%*****************************************

\cleardoublepage
\part{Environmental Effects due to classical fields}\label{pt:environment}
%************************************************
\chapter{Black Holes moving in a scalar field medium}\label{ch:movingBH}
%************************************************
\graffito{''E pur si muove!''}

The response of a \ac{BH} to an incoming wave has been studied for decades, in the frame where the \acs{BH} is at rest~\cite{Matzner:1968,Starobinski:1973,Starobinski2:1973,Teukolsky:1974yv,Unruh:1976fm,Sanchez:1976,Sanchez:1977si,MTB,Glampedakis:2001cx,Macedo:2013afa,Crispino:2009xt,Leite:2016hws,Leite:2017zyb,Leite:2018mon,Benone:2018rtj}.
Such interaction is crucial to understand how \acsp{BH} react to their environment, what types of signatures are imprinted
by strong-field regions and their possible observational effects.
It was shown that non-spinning \acsp{BH} absorb low-frequency plane waves. For a static \acs{BH} of mass~$M$, the low-frequency absorption cross-section of (massless) scalars is equal to the horizon area, $\sigma_{\textrm{abs}}=4\pi (2GM/c^2)^2$~\cite{Das:1996we}. High frequency plane waves, on the other hand, are absorbed with a cross-section~$\sigma_{\textrm{abs}}=\pi (3\sqrt{3}\,GM/c^2)^2$~\cite{MTB,Macedo:2013afa}. Although spinning \acsp{BH} also absorb plane waves, they can amplify certain low-frequency angular modes through superradiance~\cite{zeldovich1,zeldovich2,Brito:2015oca} (which also acts on charged \acsp{BH}~\cite{Benone:2019all}). Superradiance extracts energy from such \acs{BH} and provides important signatures of possible fundamental ultralight fields in nature~\cite{Brito:2015oca,Arvanitaki:2016qwi,Brito:2017wnc,Ikeda:2019fvj}.

A significant fraction of \acsp{BH} are found in binaries, such as those seen by the LIGO/Virgo observatories~\cite{LIGOScientific:2018mvr}.
In addition, most \acsp{BH} are moving at high speeds relative to our own frame. Thus, an understanding of the interaction between waves and moving \acsp{BH} is a necessary ingredient to explore the enormous potential of such sources~\cite{Bernard:2019nkv,Wong:2019kru,Ikeda:2019fvj,Wong:2020qom}. 

It was recently pointed out that \acs{BH} binaries could amplify incoming radiation through a gravitational slingshot mechanism for light~\cite{Bernard:2019nkv}. The argument requires only \emph{one} \acs{BH} moving with velocity~$v$, and a photon reflected at an angle of $180 ^{\circ}$ by the strong-field region (such orbits do exist~\cite{MTB}). Then, a trivial change of frames and consequent blue-shift yields
\begin{equation}
E_f^\textrm{peak}=\frac{1+v/c}{1-v/c}\,E_i\,,\label{max_amp}
\end{equation}
for the energy gain by the photon during the process. This is also the blue-shift of photons reflecting off a mirror moving with velocity $v$. In addition, effective field theory methods were recently used to suggest that \acs{BH} binaries could amplify radiation through superradiance~\cite{Wong:2019kru,Wong:2020qom}. Again, the argument seems to imply that a single moving \acs{BH} is able to amplify incoming radiation.

In this chapter we study the scattering of a scalar plane wave off a moving \acs{BH}. Clearly, such study involves \emph{only} a Lorentz transformation of the well-known results for \acsp{BH} at rest. Here we generalize to \acs{BH} physics the classical problem of scattering off a moving mirror or a sphere addressed by~\citeauthor{Sommerfeld:1964} and others~\cite{Sommerfeld:1964,Restrick:1968}. We obtain also for the first time, from first principles, the \acs{DF} acting on \acsp{BH} moving at possibly relativistic speeds in light scalar field environments. We find several simple analytical expressions valid for different regimes of \acs{BH} velocity. We focused on stationary regimes and extended the Newtonian treatment in Ref.~\cite{Hui:2016ltb}. Our results complement the recent numerical work in Ref.~\cite{Traykova:2021dua}.

\section{Plane wave scattering off a static black hole}

Let us start by considering the classical problem of a (monochromatic) plane wave scattering off a Schwarzschild~\acs{BH} in its proper frame. For concreteness, we will assume that the plane wave is made of a complex scalar field described by the theory~\eqref{theory_action}, interacts only with gravity~$(J_S=0)$ and has a mass potential~\eqref{mass_pot_S}. Therefore, the scalar field satisfies the Klein-Gordon equation~\eqref{KG_EOM},
\begin{align} \label{KG}
	\Box \Phi= \mu_S^2\Phi\,.
\end{align}
This problem was studied previously in, \eg,  Refs.~\cite{Matzner:1968,Starobinski:1973,,Unruh:1976fm,Sanchez:1977si,MTB,Glampedakis:2001cx,Macedo:2013afa,Leite:2016hws}.

For most situations of interest, the scalar is but a small perturbation and can be studied in a \emph{fixed} spacetime geometry -- the so-called \emph{test field} approximation. So, let us consider a fixed background metric describing a static (Schwarzschild)~\acs{BH} with line element
\begin{align}
	ds^2=-f dt^2+f^{-1}dr^2+ r^2 d\Omega^2\,, \qquad f(r)=1-\frac{2M}{r}\,,
\end{align}
where~$d \Omega^2=d \theta^2+ \sin^2 \theta\,d\varphi^2$ is the usual metric on a 2-sphere. 

Let us consider now the multipolar decomposition of a monochromatic scalar field 
\begin{align} \label{scalar_decomp}
	\Phi_{\omega'}=\frac{e^{-i \omega' t}}{r} \sum_{l,m} Y_{lm}(\theta, \varphi)u_{lm}(r)\,.
\end{align}
With the above decomposition the Klein-Gordon equation reduces to a radial equation for the functions~$u_{lm}$,
\begin{align}\label{KG_rad_alt}
&f\frac{d}{d r}\left[f\frac{d}{d r}u_{lm}\right] +\left[\omega'^2-f\left( \frac{l(l+1)}{r^2}+\frac{2 M}{r^3}+\mu_S^2\right)\right]u_{lm}=0\,.
\end{align}
Performing the change of variable~$dr/dr_*=f$, we can write this equation in the form of a (time-independent) Schrödinger-like equation
\begin{align} \label{KG_rad}
	\frac{d^2}{dr_*^2}u_{lm}+\left[\omega'^2-f\left( \frac{l(l+1)}{r^2}+\frac{2 M}{r^3}+\mu_S^2\right)\right]u_{lm}=0\,.
\end{align} 
The regular solutions to the above equations have the asymptotic form~\cite{Unruh:1976fm}
\begin{align} \label{scalar_infin}
	&u_{lm}(r \to +\infty)\sim I_{lm} e^{- i \left[k'r-\eta' \log (2 k'r) \right]}+ R_{lm}e^{ i \left[k'r-\eta' \log (2 k'r)\right] }
\end{align}
at spatial infinity, and~\footnote{The Regge-Wheeler tortoise coordinate~$r_*$ was defined up to a constant; in this chapter we will use~$r_*=r+2M \log\left(\frac{r}{2 M}-1\right)$.}
\begin{align}
	&u_{lm}(r_* \to -\infty)\sim T_{lm} e^{-i \omega' r_*}\sim T_{lm} e^{-2i M\omega' \log \left(\frac{r}{2M}-1\right)}\,
\end{align}
at the~\acs{BH} event horizon~$r=r_\textrm{h}\equiv2M$, with 
\begin{align}
&k'=\sqrt{\omega'^2-\mu_s^2}\,,\\
&\eta'=-M\left(\frac{\omega'^2+k'^2}{k'}\right)\,.
\end{align}
In this chapter we will consider only frequencies~$\omega'>\mu_S$, which can arrive to spatial infinity and, so, allow us to define a well-posed scattering problem.
Note that the ratios~$R_{lm}/I_{lm}$ and~$T_{lm}/I_{lm}$ are fixed by Eq.~\eqref{KG_rad} (or alternatively by~\eqref{KG_rad_alt}).
It is easy to show (\eg, through the conservation of the Wronskian) that the amplitudes satisfy the relation~\footnote{Hereafter we omit the indices~$l$ and~$m$ in~$R/I$,~$T/I$ and~$u(r)$ whenever possible to simplify the notation.}
\begin{align} \label{wronsk}
\left|T/I\right|^2=\frac{k'}{\omega'}\left(1-\left|R/I\right|^2\right)\,.
\end{align}

A monochromatic plane wave of frequency~$\omega'$ and wave vector~$\boldsymbol{k'}=-k'\boldsymbol{e}_z$ deformed by a long-range potential (energy)~$\eta'/r$ can be written in the form~\cite{Matzner:1968,landau1981quantum}
\begin{align}
	&\hspace{-0.5cm}e^{-i \left[\omega't+\left(k'r-\eta' \log (2 k'r) \right) \cos \theta\right]} \simeq \nonumber \\
	&\hspace{-0.5cm}i \frac{e^{-i \left(\omega't+k'r-\eta' \log (2 k'r) \right)}}{2 k' r}\sum_l \sqrt{4\pi(2l+1)} Y_{l0}(\theta, \varphi)
	+\, \textrm{(outgoing wave)} \,,
\end{align}
where~$z=r \cos \theta$.
So, imposing~$I_{lm}=i \delta_m^0 \sqrt{\hbar\rho/\mu_S} \sqrt{4 \pi (2l+1)}/(2 k')$ the asymptotic solution of the Klein-Gordon equation~\eqref{KG} is
\begin{align}
	&\sqrt{\frac{\omega'}{\hbar\rho'}}\,\Phi_{\omega'}(r \to +\infty) \sim\nonumber \\ 
	&e^{-i \left[\omega't+\left(k'r-\eta' \log (2 k'r) \right) \cos \theta\right]}+\frac{e^{-i \left(\omega't-k'r+\eta' \log (2 k'r) \right)}}{r} \sum_l \mathcal{R}_l Y_{l0}(\theta, \varphi) \,,
\end{align}
where~$\mathcal{R}_l$ are complex-valued functions of~$R_{lm}/I_{lm}$. This asymptotic behavior indicates that our choice of~$I$ indeed describes a monochromatic plane wave scattering off a Schwarzschild~\acs{BH}. The scattering is completely determined by the knowledge of the coefficients~$R/I$ (or, equivalently, of~$\mathcal{R}_l$). The constant~$\sqrt{\hbar \rho/\mu_S}$ was included in~$I$ so that the energy density current of the incident plane wave has the form~$\lim_{z \to +\infty}(-T_{tz})=\rho (\hbar\omega') ( k'/\mu_S)$. So,~$\rho$ is the rest number density of scalar particles contained in the plane wave (remember that~$\mu_S=m_S/\hbar$ is the inverse of the scalar's reduced Compton wavelength and~$\rho k'/\mu_S$ is the number density current of scalar particles).

\paragraph{Low frequency limit~($\omega' M \ll 1$)}

This limit was studied in detail in the past by~\citeauthor{Unruh:1976fm} in Ref.~\cite{Unruh:1976fm}, where he found the following approximate expression
\begin{align}
	\frac{T_l}{I_l}\simeq  \frac{2(l!)^2}{(2l)!} (2 M k')^{l+1}c_l(\eta') e^{4 i M \omega' \log\left(\frac{2M\omega'}{l+1}\right)}\,,
\end{align}
where
\begin{align}
	c_l(\eta')=2^le^{-\frac{\pi\eta'}{2}} \frac{|\Gamma(l+1+i \eta')|}{(2l+1)!}
\end{align}
Additionally, using the results in that reference it is possible to obtain the approximated expression
\begin{align} \label{ratio_low_freq}
	\hspace{-0.4cm}\frac{R_l}{I_l} \simeq(-1)^l\left[-1+\frac{2(l!)^4}{(2l)!^2}\frac{\omega'(2M k')^{2l+2}}{k'}[c_l(\eta')]^2\right]e^{2 i \arg\left[\Gamma(l+1+i \eta')\right]}\,.
\end{align}
At leading order in~$\omega'M$ the amplitudes satisfy~\eqref{wronsk}.

The low frequency condition implies that the scalar's de Broglie wavelength is much larger than the~\acs{BH}'s size,~$k' M \ll1$. Note also that, due to the condition for wave propagation~$\omega'>\mu_S$, the results of this section assume implicitly that~$\mu_S M \ll 1$ (which is satisfied, for instance, by light scalars interacting with massive \acsp{BH}).

\paragraph{High frequency limit~($\omega' M \gg 1$)}
In the high frequency limit we will focus only on the ultra-relativistic regime~$\omega' \gg \mu_S$ (which, in particular, is the only possibility for light scalars~$\mu_S M\ll1$). This limit was studied for instance in Refs.~\cite{Sanchez:1976,Andersson:1995vi}. For very large azimuthal numbers~$l \gg \omega' M$ using a WKB approximation one founds that~\cite{landau1981quantum}
\begin{align}
	\frac{R_l}{I_l} \simeq i \exp \left\{-2i\omega'\left[a_*+\int_a^\infty dr\, f^{-1}\left(1-\sqrt{1-\frac{V_\textrm{eff}}{\omega'^2}}\right)\right]\right\}\,,
\end{align}
where
\begin{align}
	V_\textrm{eff} = \frac{f}{r^2}\left(l+\tfrac{1}{2}\right)^2\,,
\end{align}
and~$a$ is the (largest) classical turning point satisfying~$V_\textrm{eff}(a)=\omega'^2$, with the Regge-Wheeler version~$a_*=a+2M \log(a/2M-1)$.
For large azimuthal numbers~$l\sim \omega' M$ it is also possible to use a WKB approximation to compute the absolute value~\cite{landau1981quantum,Sanchez:1976}
\begin{align}
	|R_l/I_l|^2\simeq \frac{1}{1+ \exp\left\{\frac{27 \pi (M \omega' )^2}{l}\bigg[1-\Big(\frac{l}{3 \sqrt{3} M\omega' }\Big)^2\bigg]\right\}}\,,
\end{align}
which for high-frequency~$\omega' M \gg 1$ is a very steep function of~$l$ that vanishes for~$l<3 \sqrt{3} M\omega'$ and is unity for $l>3 \sqrt{3} M\omega'$. In fact it is easy to verify that in the high-frequency limit~$\omega' M \gg 1$ the following is a very good approximation 
\begin{align} \label{ratio_high_freq}
	\frac{R_l}{I_l}\simeq\begin{cases}
	&0 \,,\qquad  l< 3 \sqrt{3} M \omega' \\
	&i\, e^ {-2i\omega'\left[a_*+\int_a^\infty dr\, f^{-1}\left(1-\sqrt{1-\frac{V_\textrm{eff}}{\omega'^2}}\right)\right]}\,, \qquad l> 3 \sqrt{3} M \omega'
	\end{cases}\,.
\end{align}

\subsection{Energy absorption}

The energy of the scalar field contained in a spacelike hypersurface~$\mathcal{S}_1\equiv\{t=t_1\}$ extending from the horizon to infinity is (see~\eqref{energy_angularmom})
\begin{align}
	E(t_1)=\int_{\mathcal{S}_1}dV_3\, T^{\alpha t} N_\alpha\,,
\end{align}
where~$N_\beta=-\delta_\beta^t\sqrt{-g_ {tt}}$ is the unit normal covector and~$dV_3$ is the volume form induced in the hypersurface. Because we are interested in a stationary regime with~$\Phi_{\omega'}\propto e^{-i\omega' t}$ (which results in a static energy-momentum tensor~$T^{\alpha \beta}$) and since the background metric is static, one has
\begin{equation}
	\frac{d}{d t_1} E(t_1)=\int_{\mathcal{S}_1}\mathcal{L}_{\partial_ t} \left(dV_3\, T^{\alpha t} N_\beta\right)=0\,,
\end{equation}
where~$\mathcal{L}_{\partial_t}(\cdot)$ is the Lie derivative with respect to~$(\partial_t)^\alpha$. Then, by applying the divergence theorem it follows that the energy crossing the event horizon per unit time~$t$ (which is the proper time of a stationary observer at infinity) is~$\mathcal{F}_E$, where
\begin{align}
	\mathcal{F}_E=\int_{r \to \infty}  d^2 \Omega \,r^2 T_{rt}
\end{align}
with the 2-sphere element of area~$d^2\Omega=\sin\theta\, d\theta d\varphi$.

Plugging the decomposition~\eqref{scalar_decomp} with the asymptotic solution~\eqref{scalar_infin} in the scalar field's energy-momentum tensor~\eqref{scalarEMT} and using the orthonormality relations of spherical harmonics it is straightforward to show that
\begin{equation}
	\mathcal{F}_E=\omega' k'\sum_l \left(|I_l|^2-|R_l|^2\right)\,.
\end{equation}
For the case of an incident plane wave the last expression becomes
\begin{align}
	\mathcal{F}_E=\frac{\rho \hbar \omega'}{\mu_S k'} \sum_l \pi(2l+1) \left(1-|R_l/I_l|^2\right)\,.
\end{align}

As a consistency check: note that in a flat spacetime (\ie, $M=0$) the plane wave propagates freely and it is easy to show that~$R_l/I_l=(-1)^{l+1}$, which implies~$\mathcal{F}_E=0$ as it must (since there is no~\acs{BH} at all).

To obtain the \acs{BH}'s absorption cross section one just needs to take the ratio between the energy absorbed by the~\acs{BH} per unit of time~$\mathcal{F}_E$ and the energy density current of the (incident) plane wave~$\lim_{z\to +\infty}(-T_{tz})= \rho(\hbar\omega') (k'/\mu_S)$,
\begin{align}
	\sigma_{\textrm{abs}}=\frac{\mathcal{F}_E}{\rho \hbar\omega' k'/\mu_S}=\frac{1}{k'^2} \sum_l \pi(2l+1) \left(1-|R_l/I_l|^2\right)\,.
\end{align}

\paragraph{Low frequency limit~($\omega' M \ll 1$)} In this regime, at leading order in~$\omega' M$, the energy absorbed by the~\acs{BH} is
\begin{align}
	\mathcal{F}_E \simeq 16\pi  (M \omega')^2\left(\frac{\hbar\rho}{\mu_S}\right)\frac{e^{-\pi \eta'}\pi \eta'}{\sinh(\pi \eta')}  \,,
\end{align}
and the \acs{BH}'s absorption cross section is
\begin{align}
	\sigma_{\textrm{abs}}\simeq 16\pi M^2 \left(\frac{\omega'}{k'}\right) \frac{e^{-\pi \eta'}\pi \eta'}{\sinh(\pi \eta')}  \,,
\end{align}
where it was used~$|\Gamma(1+i \eta')|^2=\pi \eta'/\sinh(\pi \eta')$.
At leading order in~$\omega' M$ only the~$l=0$ mode contributes to both~$\mathcal{F}_E$ and~$\sigma_{\textrm{abs}}$. 

In the Newtonian limit~$\omega' \simeq\mu_S$ (which implies also~$k' \ll \mu_S$) the above expressions become
\begin{align}
	&\mathcal{F}_E \simeq 16\pi  (M \mu_S)^2\left(\frac{\hbar\rho}{\mu_S}\right)\left(\frac{\pi M\mu_S^2}{k'\sinh\Big(\frac{\pi M \mu_S^2}{k'}\Big)}\right)e^{\frac{\pi M \mu_S^2}{k'}}  \,, \\
	&\sigma_{\textrm{abs}}\simeq 16\pi M^2 \left(\frac{\mu_S}{k'}\right) \left(\frac{\pi M\mu_S^2}{k'\sinh\Big(\frac{\pi M \mu_S^2}{k'}\Big)}\right)e^{\frac{\pi M \mu_S^2}{k'}} \,.
\end{align}
In the sub-regime~$k'/\mu_S\ll M \mu_S$ these simplify to
\begin{align}
		&\mathcal{F}_E\simeq 32 \pi^2 (M\mu_S )^3\left(\frac{\hbar\rho}{\mu_S}\right) \left(\frac{\mu_S}{k'}\right)\,, \\
	    &\sigma_{\textrm{abs}}\simeq 32 \pi^2 M^2(M\mu_S ) \left(\frac{\mu_S}{k'}\right)^2\,,
\end{align}
while in the sub-regime~$k'/\mu_S \gg M \mu_S$ one finds
\begin{align}
	&\mathcal{F}_E \simeq 16\pi  (M \mu_S)^2\left(\frac{\hbar\rho}{\mu_S}\right) \,, \\
	&\sigma_{\textrm{abs}}\simeq 16\pi M^2 \left(\frac{\mu_S}{k'}\right)  \,.
\end{align}

In the ultra-relativistic limit~$\omega' \gg \mu_S$ (which implies~$k'\simeq \omega'$) the expressions reduce to
\begin{align}
	&\mathcal{F}_E\simeq 16 \pi (M \omega')^2 \left(\frac{\hbar\rho}{\mu_S}\right)\,, \\
	&\sigma_{\textrm{abs}} \simeq 16 \pi M^2\,.
\end{align}
Note that in the latter limit we recover the well-known expressions that are valid for massless scalar fields with low frequency~$\omega' M \ll 1$. All the expressions of this section were obtained decades ago and are in agreement with~Ref.~\cite{Unruh:1976fm}.

\paragraph{High frequency limit~($\omega' M \gg 1$)}
In this regime most of the contribution to the summation in~$l$ comes from large azimuthal numbers~$l\gg1$ and, so, the sum is well approximated by an integral. Then, at leading order, one recovers the well-know results~\cite{Sanchez:1976}
\begin{align}
	&\mathcal{F}_E \simeq 27 \pi (M \omega')^2\left(\frac{\hbar\rho}{\mu_S}\right)\,, \\
	&\sigma_{\textrm{abs}} \simeq 27 \pi M^2\,.
\end{align}
The latter absorption cross section coincides with the one that is obtained using null geodesics~\cite{Misner:1974qy} (\ie, with the geometric optics approximation). Note that~$3 \sqrt{3} M$ is the critical impact parameter below which a null particle falls into the~\acs{BH}.

\subsection{Momentum transfer}

Due to the transfer of momentum from the scalar field to the~\acs{BH}, the latter will feel a "force". Consider the spatial components of the ADM momentum~$P^i$ computed using a 2-sphere with a sufficiently large radius, such that it is in the asymptotically flat region.~\footnote{The ADM charges are defined at infinity, but since we are dealing with monochromatic waves these charges would diverge.} These components can be decomposed into the sum of curvature and scalar field contributions,~$P^i=P_\textrm{curv}^i+ P_S^i$, where~$P_S^i$ is 
\begin{align}
	P_S^i(t_1)=\int_{S_1} dV_3 T^{\alpha i} N_\alpha\,.
\end{align}
The rate of change of~$P^i$ is
\begin{align}
	\frac{d P^i}{d t}=-\int_{r\to \infty} d^2\Omega\, r^2 T^{r i} \,,
\end{align}
and, because we considering a stationary regime, we have
\begin{align}
	\frac{d}{d t_1} P_S^i(t_1)=\int_{\mathcal{S}_1}\mathcal{L}_{\partial_ t} \left(dV_3\, T^{\alpha i} N_\beta\right)=0\,.
\end{align}
Thus, the force acting on the~\acs{BH} is
\begin{align}
	F^i\equiv\frac{d P^i_\textrm{curv}}{d t_1}=\frac{d P^i}{d t_1}=-\int_{r\to \infty} d^2\Omega\, r^2 T^{r i}\,.
\end{align}
Strictly, in the test field approximation one has~$d P^i_\textrm{curv}/dt=0$ (at zero order in the scalar field) and~$d P^i/d t\neq 0$ (at second order in the scalar field). This is not inconsistent with the last equation, which holds at each order in the scalar field amplitude. So, if we compute the backreaction of the scalar field on the metric, we will obtain a second order correction to~$d P^i_\textrm{curv}/dt$ which must be equal to~$d P^i/d t$. For a more thorough discussion, which in particular covers the case where the steady state is attained dynamically see Ref.~\cite{Clough:2021qlv}.

For asymptotic Cartesian coordinates we have
\begin{align}
	&\lim_{r \to \infty}r^2T^{r x} \simeq r^2\sin \theta \cos \varphi T_{rr}\,, \\
	&\lim_{r \to \infty}r^2T^{r y} \simeq r^2\sin \theta \sin \varphi T_{rr}\,, \\
	&\lim_{r \to \infty}r^2T^{r z} \simeq r^2\cos \theta T_{rr}\,. 
\end{align}
Plane waves propagating along~$-\boldsymbol{e}_z$ and scattering off a static black hole are symmetric under rotations around the~$z$ axis. This implies that~$T^{r r}$ is independent of the azimuthal angle~$\varphi$. Thus, as it might have been anticipated, one has~$F^x=F^y=0$. The only non-vanishing component of the force acting on the~\acs{BH} is
\begin{align}
	F^z=-\int_ {r \to \infty} d^2\Omega\, r^2\cos \theta T_{rr}\,.
\end{align}
Now, plugging the decomposition~\eqref{scalar_decomp} with the asymptotic solution~\eqref{scalar_infin} in the scalar field's energy-momentum tensor~\eqref{scalarEMT} and using that
\begin{align}
	\cos \theta=  \sqrt{\frac{4\pi}{3}}Y_{1,0}(\theta)\,, 
\end{align}
and
\begin{align}	
	&\int d^2\Omega \,Y_{1,0}(\theta)Y_{l0}(\theta)Y_{l'0}(\theta)= \nonumber \\
	&=\sqrt{\frac{3}{4 \pi (2l+1)}}\left[\delta_{l'}^{l+1}\left(\frac{l+1}{\sqrt{2l+3}}\right)+\delta_ {l'}^{l-1} \left(\frac{l}{\sqrt{2l-1}}\right)\right]\,,
\end{align}
it is straightforward to show
\begin{align}\label{force_rest}
	F^z=-2 \pi\left(\frac{\hbar\rho}{\mu_S}\right)\sum_l (l+1) \Re \left[1+\frac{R_l^*}{I_l^*}\frac{R_{l+1}}{I_{l+1}}\right] \,.
\end{align}

As a consistency check: note that in a flat spacetime the amplitude ratio is~$R_l/I_l=(-1)^{l+1}$ which implies~$F^z=0$ -- as it should, since in that case the plane wave propagates freely (\ie, its momentum is conserved).

\paragraph{Low frequency limit~($\omega' M \ll 1$)} Using~\eqref{ratio_low_freq} it is straightforward to show that at leading order in~$\omega'M$ the force acting on the \acs{BH} is~\footnote{Although it is not obvious that in the limit~$|\eta'|\gg1$ accretion gives a negligible contribution to the force, it can be seen that this is indeed the case by rewriting~\cite{Unruh:1976fm}
\begin{align}
	c_l^2=\frac{2^{l}(2\pi \eta')}{(2l+1)!^2(1-e^{2 \pi \eta'})} \Pi_{s=1}^l(s^2+ \eta'^2) \sim \mathcal{O}(\eta'^{2l+1})\,, \qquad |\eta'|\gg 1\,.
\end{align}}
\begin{align} \label{force_low}
	F^z=-4\pi \left(\frac{\hbar\rho}{\mu_S}\right)\sum_l (l+1) \sin^2 \left(\frac{\alpha_l}{2}\right) \,,
\end{align}
with
\begin{align} \label{deflection_ang}
	\alpha_l&=2\arg \left(\frac{\Gamma(l+2+i\eta')}{\Gamma(l+1+i\eta')}\right)\nonumber \\
	&=2\arg(l+1+i\eta')=2\arctan\left(\frac{\eta'}{l+1}\right)\,.
\end{align}
The above expression can be rewritten as
\begin{align} \label{force_low_freq}
	F^z=-4 \pi \left(\frac{\hbar\rho}{\mu_S}\right)\sum_l \frac{\eta'^2(l+1)}{\eta'^2+(l+1)^2}\,.
\end{align}

Note that this result is the same as the one it would be obtained for a scalar plane wave scattering off a (weak) Newtonian potential originated by a point particle, in which case it is easy to show that
\begin{align}
	R_l/I_l=\arg \Gamma(l+1+i\eta')\,.
\end{align}
So, remarkably, in the limit~$\omega' M \ll 1$ the force acting on the~\acs{BH} due to the scattering process is indistinguishable of the one that would act on a Newtonian point-like source of gravitational field. In other words, at leading order in~$\omega' M \ll 1$ the strong gravitational field and absorption (\ie, accretion) effects can be neglected, and the~\acs{BH} can be modeled by a Newtonian particle. 

It is worth noting that in the eikonal limit~($l\gg1$) one has~$l\simeq k' b$ and the angle~$\alpha_l$ in Eq.~\eqref{deflection_ang} is the deflection angle of a particle scattering off a Newtonian gravitational field.~\footnote{The eikonal limit~$l\gg1$ can be seen as a manifestation of \emph{Bohr's correspondence principle}, giving us a relation between the \emph{wave} (quantum) number~$l$ and the \emph{particle} (classical) parameter~$b$. For an interesting discussion about the correspondence between wave and particle scattering see Ref.~\cite{FORD1959259}.} In particular, in the Newtonian $\omega'\simeq \mu_S$ limit one has
\begin{align}
	\alpha_l\simeq-2\arctan\left(\frac{M \mu^2_S}{b\,k'^2 }\right)\,,
\end{align}
which is exactly the well-known Newtonian deflection angle, and in the ultra-relativistic limit one finds
\begin{align}
	\alpha_l \simeq -2\arctan \left(\frac{2 M}{b}\right) \simeq - \frac{4 M}{b}\,,
\end{align}
which is equal to the deflection angle of light obtained by~\citeauthor{Einstein1936LENSLIKEAO} using his general theory of relativity~\cite{Einstein1936LENSLIKEAO}. If we compute then the force that would act on a source of gravitational field due to a beam of (classical) particles with impact parameter between~$b$ and $b+\delta b$ being deflected by an angle~$\alpha_l$ we find
\begin{align}
	\frac{\delta F^z}{k'\delta b}=-4 \pi \left(\frac{\hbar \rho}{\mu_S}\right)(k' b) \sin^2 \alpha_ l\,,
\end{align}
which matches~\eqref{force_low} in the eikonal limit. 

It is easy to see that the force~\eqref{force_low_freq} diverges logarithmically in~$l$, which is to be expected due to the long-range ($1/r$) nature of the gravitational potential. In the limit~$\eta'^2\gg1$ the summation in~$l$ is very well approximated by an integral and the force acting on the~\acs{BH} is
\begin{align}
	F^z=-2 \pi \left(\frac{\hbar \rho}{\mu_S}\right) \eta'^2 \log\left(\eta'^2+k'^2 b_\textrm{max}^2\right)\,,
\end{align}
where we have introduced a cutoff~$l_\textrm{max} \gg1$ and used the eikonal limit relation~$l_\textrm{max}\simeq k' b_ \textrm{max}$. By truncating the summation in~$l$ we are in fact considering the scattering of a circular beam of radius~$b_\text{max}$, instead of the original \emph{infinite} plane wave. That the limit~$\eta'^2\gg1$ is well approximated by a continuous~$l$ is a signal that this is in fact a \emph{classical} (or, more correctly, \emph{particle}) limit. Indeed, in the Newtonian regime~$\omega' \simeq \mu_S$ the force is just
\begin{align}
	F^z\simeq - 4 \pi \frac{M^2 (\rho \hbar \mu_S)}{(k'/\mu_S)^2} \log\left(k'\sqrt{\left(\frac{\mu_S^2 M}{k'^2}\right)^2+b_\textrm{max}^2}\,\right)\,,
\end{align}
which matches the famous expression obtained by \citeauthor{Chandrasekhar:1943v1}~\cite{Chandrasekhar:1943v1} for (classical) collisionless media, with the argument of the \emph{Coulomb logarithm} being the ratio between the length-scale $\sqrt{\left(\mu_S^2 M/k'^2\right)^2+b_\textrm{max}^2}$ and the de Broglie wavelength of the scalar particles; note that~$\mu_S^2 M/k'^2$ is the characteristic radius below which classical particles scattering off a weak gravitational field are significantly deflected. 

The ultra-relativistic limit is not consistent with the limit~$\eta'^2 \gg 1$ in the low frequency regime~$\omega' M \ll1$; on the contrary, it must satisfy~$\eta'^2 \ll 1$.
The limit~$\eta'^2\ll1$ corresponds to the regime in which the \emph{quantum} (or, more correctly, \emph{wave}) effects are the more pronounced. It is easy to check that in this limit the force~\eqref{force_low_freq} is very well approximated by
\begin{align}
	F^z =-4 \pi \left(\frac{\hbar \rho}{\mu_S}\right) \eta'^2\left[ \log\left(k' b_\textrm{max}\right)+ \gamma_ \textrm{EM}\right]\,,
\end{align}
where~$\gamma_\textrm{EM}= 0.5772\, ...$ is the Euler-Mascheroni constant. In the Newtonian regime the last expression becomes
\begin{align}
 	F^z\simeq - 4 \pi \frac{M^2 (\rho \hbar \mu_S)}{(k'/\mu_S)^2} \left[\log\left(k'b_\textrm{max}\right)+ \gamma_ \textrm{EM}\right]\,,
\end{align}
and in the ultra-relativistic regime it becomes
\begin{align}
	F^z \simeq -16 \pi \left(\frac{\hbar \rho}{\mu_S}\right) (M \omega')^2 \left[ \log\left(\omega' b_\textrm{max}\right)+ \gamma_ \textrm{EM}\right]\,.
\end{align}

\paragraph{High frequency limit~($\omega' M \gg 1$)} In this limit the accretion of scalar field gives an important contribution to the force acting on the~\acs{BH}. By looking at~\eqref{ratio_high_freq} we see immediately that this contribution is contained in the terms of~\eqref{force_rest} with~$l<3 \sqrt{3} M \omega'$ (for which~$R/I\simeq 0$). In the high frequency limit only large azimuthal numbers~$l\gg1$ contribute significantly and the summation can always be approximated by an integral. So, the accretion of scalar is responsible for the part $-27\pi(M \omega')^2 \left(\hbar \rho/\mu_S\right)$ in the force (which obviously matches~$\mathcal{F}_E$ in absolute value, since we are using units with~$c=1$).
The remaining contribution from larger~$l$'s can be obtained by using the eikonal relation~$l \simeq \omega' b$ and rewrite the summation as
\begin{align}
	\sum_{l>3 \sqrt{3} M \omega'} (l+1) \Re \left[1+\frac{R_l^*}{I_l^*}\frac{R_{l+1}}{I_{l+1}}\right] \simeq 2\omega'^2  \int_{3 \sqrt{3}M}^\infty db\, b \sin^2\left(\frac{\alpha(b)}{2}\right)\,,
\end{align}
where
\begin{align}
	\alpha&=\pi -2 \frac{d}{d b} \left[a_* +\int_a^\infty dr\, f^{-1}\left(1-\sqrt{1-\frac{f}{r^2}b^2}\right)\right] \nonumber \\
	&=\pi-2\int_a^\infty \frac{dr}{\sqrt{\frac{r^2}{b^2}-f}}\,,
\end{align}
with~$a$ given implicitly by~$b=a \sqrt{a/(a-2M)}$ (being the larger real number satisfying this relation). Remarkably, as it happened in the low frequency limit, the angle~$\alpha$ matches exactly the deflection angle of a null particle in a Schwarzschild spacetime~\cite{Darwin,Misner:1974qy,BisnovatyiKogan:2008ts}. 
It is easy to check that for large impact parameters~$b\gg M$ we recover again Einstein's deflection angle
\begin{align}
	\alpha \simeq -\frac{4 M}{b}\,.
\end{align} 
For a beam of scalar particles with maximum (cutoff) impact parameter~$b_{\text{max}}> 20M$ (remember that the integral in~$b$ diverges logarithmically) we find the following approximated expression
\begin{align}
	\int_{3\sqrt{3}M}^{b_\text{max}} db\, b \sin^2\left(\frac{\alpha}{2}\right) \simeq M^2\left[14.5074+4 \log\left(\frac{b_{\textrm{max}}}{20M}\right)\right]\,,
\end{align}
where we have performed a numerical integration between~$3 \sqrt{3}M$ and~$20 M$ to find the numerical factor. Finally, putting all together (including accretion) we find that the force is
\begin{align}
	F^z \simeq -16\pi\left(\frac{\hbar \rho}{\mu_S}\right) (M \omega')^2\left[ \log\left(\frac{b_{\textrm{max}}}{20M}\right)+5.31435\right]\,.
\end{align} 

\section{Moving black hole}

Now that we know the rate at which energy and linear momentum is transfered to a Schwarzschild \acs{BH} from the point of view of an observer at infinity which is stationary with respect to it, one may wonder what changes if the \acs{BH} is moving with respect to the observer. While at rest (and neglecting quantum effects) the~\acs{BH} is a perfect absorver, but when moving it may also transfer translational energy to the scalar field with the interesting possibility of loosing energy globally -- and having a \emph{negative} absorption cross-section. It is also crucial to understand the relativistic corrections to the force acting on the~\acs{BH} when it is moving at large speeds with respect to a distant observer.

Although the framework of this chapter allows for a completely general treatment we will focus here in the case in which the~\acs{BH} is moving with velocity~$\boldsymbol{v}=v \boldsymbol{e}_z$ with respect to the distant observer and: 1) the scalar is at rest with respect to the observer, or 2) the scalar is massless and has momentum~$-\hbar \omega \boldsymbol{e}_ z$ with respect to the observer. The quantities measured by the observer with respect to which the~\acs{BH} is at rest (hereafter "\acs{BH} frame") are primed and the ones measured by the observer with respect to which the~\acs{BH} is moving (hereafter "lab frame") are unprimed. Noting that the curvature part of the ADM four-momentum~$P_\textrm{curv}^\alpha$ is a Lorentz four-vector~\cite{Arnowitt:1962hi} (with its time-component being the energy of the~\acs{BH}, including its mass), it is trivial to find the rate of change of the \acs{BH}'s energy and momentum in the lab frame. One just needs to perform a Lorentz transformation of~$d P_\textrm{curv}^\alpha$ and relate the proper times of the two observers~$d\tilde{t}=dt/\sqrt{1-v^2}$, finding in the lab frame (tilded quantities)~\footnote{Very recently, the same type of reasoning was applied in Ref.~\cite{Traykova:2021dua}. However, there the authors neglected the contribution of~$\mathcal{F}_E$ to~$\widetilde{F}^z$ and so used~$\widetilde{F}^z=F^z$. As we will show next this is not always a good approximation, in particular it fails in the case of~\acsp{BH} moving at relativistic speeds.}
\begin{align}
	&\widetilde{\mathcal{F}}_E=\mathcal{F}_E+v F^z\,,\\
	&\widetilde{F}^z=F^z+ v \mathcal{F}_E\,.
\end{align}

\subsection{Scalar field at rest}

For a scalar field at rest we have~$\omega=\mu_S$ and~$\boldsymbol{k}=0$ in the lab frame and~$\omega'=\mu_S/\sqrt{1-v^2}$ and~$\boldsymbol{k'}=-\mu_S v/\sqrt{1-v^2}\, \boldsymbol{e}_z$ in the~\acs{BH} frame.

\paragraph{Low frequency limit ($\omega' M \ll 1 \Leftrightarrow 1-v^2 \gg (M \mu_S)^2$)}
Again note that this limit is possible only if~$(M \mu_S)^2 \ll1$.
First let us consider the non-relativistic regime~$v\ll1$. If~$v \ll M \mu_S$, we find
\begin{align}
	\frac{\widetilde{\mathcal{F}}_E}{v}\simeq\widetilde{F}^z \simeq F^z\simeq -4 \pi \frac{M^2 \rho m_S}{v^2}\log\left(\mu_S \sqrt{\bigg(\frac{M}{v}\bigg)^2+v^2b_\textrm{max}^2}\,\right)\,,	
\end{align}
where~$m_S=\hbar \mu_S$. On the other hand, for~$M \mu_S\ll v \ll 1$ we obtain
\begin{align}
	\frac{\widetilde{\mathcal{F}}_E}{v}\simeq\widetilde{F}^z \simeq F^z\simeq - 4 \pi \frac{M^2 \rho m_S}{v^2} \left[\log\left(\mu_S v \,b_\textrm{max}\right)+ \gamma_ \textrm{EM}\right]\,.
\end{align}
The above expressions have a similar form (though with some differences) as the ones found by~\citeauthor{Hui:2016ltb}~\cite{Hui:2016ltb} for a Newtonian object (and recently confirmed numerically for a~\acs{BH} by~\citeauthor{Traykova:2021dua}~\cite{Traykova:2021dua}). Some differences are indeed expected since in those works a cutoff in~$r$ was employed, whereas here due to the nature of our framework (which contains a sum in the azimuthal number~$l$) a cutoff in~$b$ was used.~\footnote{A cutoff in~$b$ was also used in Chandrasekhar's works~\cite{Chandrasekhar:1943v1,Chandrasekhar:1943v2,Chandrasekhar:1943v3}. That a difference should be expected between the two approaches was also acknowledge in~\cite{Hui:2016ltb}.}
Now, for relativistic velocities~$v\sim 1$ we find
\begin{align} \label{DF_relativ_lfreq}
	\hspace{-0.85cm}\widetilde{F}_E \simeq \widetilde{F}^z \simeq F^z+ \mathcal{F}_E \simeq -16 \pi \frac{M^2 \rho m_S}{(1-v^2)}\left[ \log\left(\frac{\mu_S b_\textrm{max}}{\sqrt{1-v^2}}\right)+ \gamma_ \textrm{EM}-1\right]\,.
\end{align} 

\paragraph{High frequency limit ($\omega' M \gg 1 \Leftrightarrow 1-v^2 \ll (M \mu_S)^2$)}
In the relativistic regime~$v\sim 1$ we find
\begin{align} \label{DF_relativ_hfreq}
	\hspace{-0.85cm}\widetilde{\mathcal{F}}_E \simeq \widetilde{F}^z \simeq F^z+ \mathcal{F}_E \simeq -16 \pi \frac{M^2 \rho m_S}{(1-v^2)}\left[ \log\left(\frac{b_{\textrm{max}}}{20M}\right)+5.31435-\frac{27}{16}\right]\,.
\end{align}
Remember that this expression is valid for~$b_\text{max}\geq20M$. For a maximum impact parameter~$b_\text{max}<20M$ one needs to perform a (trivial) numerical integration.

Both relativistic expressions~\eqref{DF_relativ_lfreq} and~\eqref{DF_relativ_hfreq} are similar in form with the ones presented recently in Ref.~\cite{Traykova:2021dua}, which were motivated by heuristic arguments and shown to be consistent with numerical evolutions. Again, some differences are to be expected due to the different ways used to truncate the characteristic (Coulombian) logarithmic divergence. 

\subsection{Massless scalar field}

A massless scalar propagating with momentum~$-\hbar \omega \boldsymbol{e}_z$ in the lab frame has~$\omega'=\sqrt{\frac{1+v}{1-v}}\omega$ and~$\boldsymbol{k}'=-\sqrt{\frac{1+v}{1-v}}\omega \,\boldsymbol{e}_z$ in the~\acs{BH} frame. In this case there is obviously no Newtonian limit and the scalar particles are always ultra-relativistic (\ie, move at the speed of light) in any frame. So, the factor~$\rho \omega/\mu_S$ is not well defined (remember that~$\rho$ is the number density in the rest frame) and must be substituted by the number density in the lab frame~$\tilde{\rho}$ (as can be readily seen by continuity, taking the ultra-relativistic limit).

\paragraph{Low frequency limit ($\omega' M \ll 1 \Leftrightarrow \omega M \ll \sqrt{\frac{1-v}{1+v}}$)} In this limit we find that the energy accreted by the~\acs{BH} is 
\begin{align}
	&\hspace{-0.7cm}\widetilde{\mathcal{F}}_E \simeq \mathcal{F}^E+v F^z \nonumber \\
	&\hspace{-0.7cm}\simeq -16 \pi \,v(1+v)\frac{M^2 \tilde{\rho} (\hbar \omega)}{1-v}\left\{\log \bigg(\sqrt{\frac{1+v}{1-v}}\,\omega b_\textrm{max}\bigg)+\gamma_\textrm{EM}-\frac{1}{v}\right\}\,,
\end{align}
and the force acting on it is
\begin{align}
	&\hspace{-0.7cm}\widetilde{F}^z \simeq F^z+ v \mathcal{F}_E \nonumber \\
	&\hspace{-0.7cm}\simeq -16 \pi (1+v)\frac{M^2 \tilde{\rho} (\hbar \omega)}{1-v}\left\{\log \bigg(\sqrt{\frac{1+v}{1-v}}\,\omega b_\textrm{max}\bigg)+\gamma_\textrm{EM}-v\right\}\,.
\end{align}

\paragraph{High frequency limit ($\omega' M \gg 1 \Leftrightarrow \omega M \gg \sqrt{\frac{1-v}{1+v}}$)} Now, in the high-frequency limit we obtain
\begin{align} \label{energy_geomopt}
&\hspace{-0.7cm}\widetilde{\mathcal{F}}_E \simeq \mathcal{F}^E+v F^z \nonumber \\
&\hspace{-0.7cm}\simeq -16 \pi \,v(1+v)\frac{M^2 \tilde{\rho} (\hbar \omega)}{1-v}\left\{\log \bigg(\frac{b_\textrm{max}}{20M}\bigg)+5.31435-\frac{27}{16v}\right\}\,,
\end{align}
and
\begin{align}
	&\hspace{-0.7cm}\widetilde{F}^z \simeq  F^z+ v\mathcal{F}^E \nonumber \\
	&\hspace{-0.7cm}\simeq -16 \pi \,(1+v)\frac{M^2 \tilde{\rho} (\hbar \omega)}{1-v}\left\{\log \bigg(\frac{b_\textrm{max}}{20M}\bigg)+5.31435-\frac{27}{16}v\right\}\,.
\end{align}
Again, these expressions are valid for~$b_\text{max}>20M$. 

\subsubsection{Geometrical optics approximation}

The high-frequency (and large impact parameter) limit corresponds to the geometrical optics (\ie, particle) approximation -- as discussed in previous sections. So, we can use this approximation as an external check to our results. Here, we focus on the case of a massless scalar field, but this approach could equally well be applied to the case of a massive scalar field. The geometrical optics limit is specially interesting in the sense that it provides spin-independent results (\ie, the results hold for scalar, fermion, photon, or graviton \emph{particles}).

In this approximation we study null geodesics in the spacetime of a moving \acs{BH}. In isotropic coordinates the Schwarzschild metric is given by
\begin{align}
ds^2=-\frac{(1-A)^2}{(1+A)^2}dt^2+(1+A)^4\left(dx^2+dy^2+dz^2\right)\,,
\end{align}
where $A\equiv M/(2\rho)$ and $\rho^2\equiv x^2+y^2+z^2$. Here, the standard Schwarzschild radial coordinate is related with $\rho$ via $r=\rho(1+A)^2$. Perform a boost along the $z$ direction, by letting 
\begin{align}
\hat{t}=\gamma (t+v z)\,,\quad \hat{z}=\gamma(z+vt)\,,\quad \hat{y}=y\,,\quad \hat{x}=x\,.\label{boost}
\end{align}
This yields the metric describing a \acs{BH} moving with velocity $v$ and Lorentz factor $\gamma\equiv 1/\sqrt{1-v^2}$.
It is now a simple question to study the scattering of a beam of null particles: follow initially counter-moving null geodesics with impact parameter $b$ (\ie, null geodesics with $\hat{y}(\hat{t}=0)=b$ and $\dot{\hat{x}}=\dot{\hat{y}}=0$ at large distances) and monitor their energy $E=-X^\alpha\,p_\alpha$, where $p_\alpha$ is the four-momentum associated with the geodesic and $X^\alpha=(\partial_{\hat{t}})^\alpha$ (at large distances) the four-velocity of the observer.

\begin{figure}
	\centering
	\includegraphics[width=0.9 \textwidth]{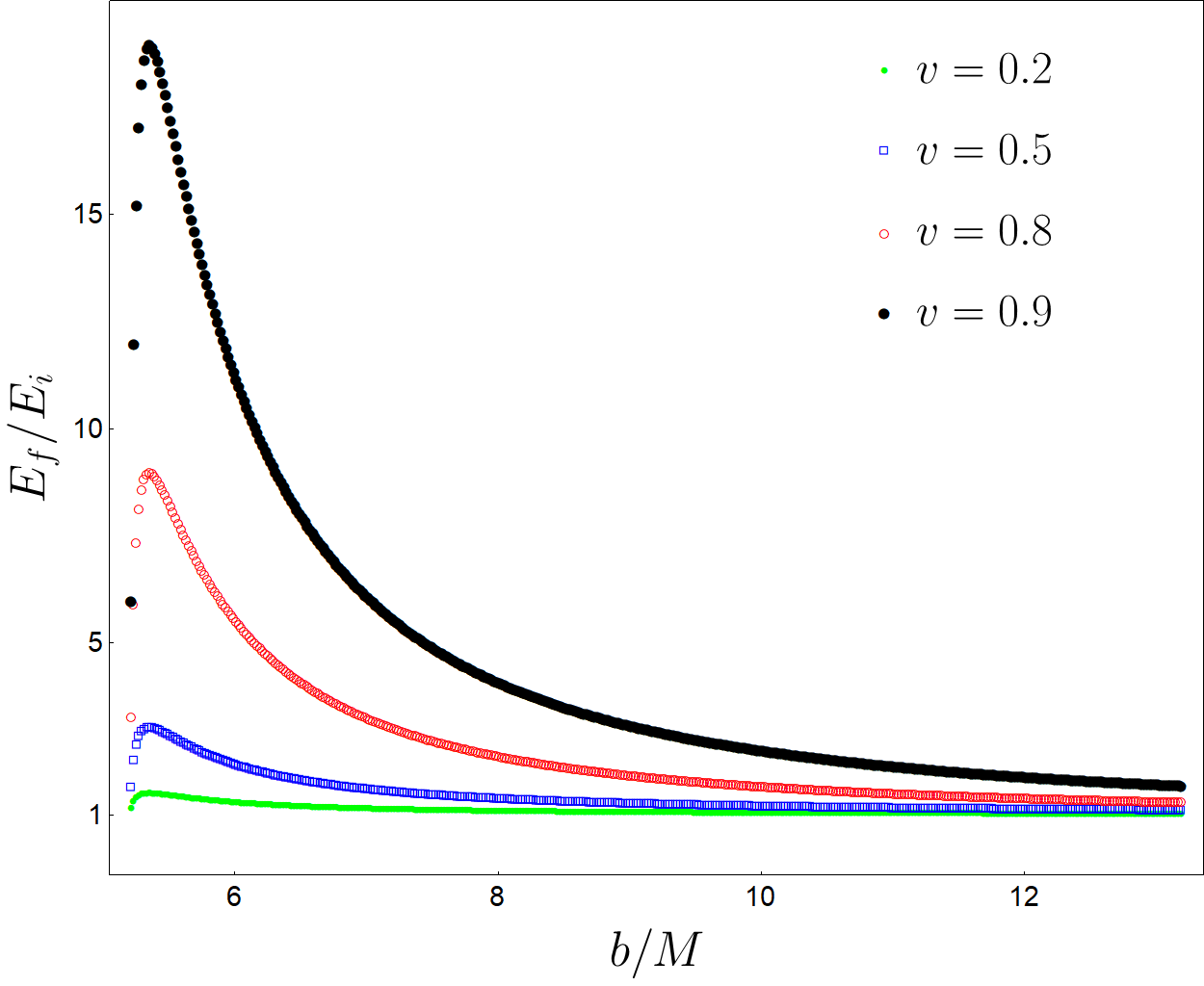}
	\caption{Energy gain of a (high frequency) photon scattered off a moving \acs{BH}. The photon has initial energy~$E_i$, impact parameter~$b$ and scatters off a \acs{BH} moving with velocity~$v$ in the opposite direction; the final energy is~$E_f$. The peak of each curves agrees, to numerical precision, with Eq.~\eqref{max_amp}. For impact parameters~$b<3\sqrt{3}M$ the photon is absorbed by the \acs{BH}.}
	\label{fig:amplification_null}
\end{figure}

Our results are shown in Fig.~\ref{fig:amplification_null} for different velocities $v$. There is a minimum impact parameter~$b=3\sqrt{3}M$, below which the photon simply falls onto the \acs{BH}. As we increase the impact parameter starting from this value, the energy gain peaks very rapidly at a value precisely (to within numerical precision) described by Eq.~\eqref{max_amp} -- these are photons which are reflected back by the geometry. There are in fact a multitude of impact parameters for which photons are reflected back; for, respectively, 
\begin{align}
b/M&=b_1/M=5.356\pm 0.003\,,\\
b/M&=b_2/M=5.199\pm 0.002\,,
\end{align}
the photon circles the \acs{BH} exactly half an orbit (with a distance of minimum approach of~$r/M=3.521\pm 0.001$), and one-and-a-half orbit (with a distance of minimum approach of~$r/M=3.001\pm 0.001$). For impact parameters closer to the critical value, a larger number of orbits around the \acs{BH} are possible. At large impact parameters, our numerical results are perfectly described by the weak-field result
\begin{align}
	E_f^\text{weak}=E_i-\frac{1}{\tilde{\rho}(1+v)}\frac{d \widetilde{\mathcal{F}}_E}{2 \pi b\,db} = \left(1+\frac{8 M^2 v}{b^2 (1-v)}\right) E_i\,,
\end{align}
where we have used Eq.~\eqref{energy_geomopt} with~$E_i=\hbar \omega$, which was obtained for a scalar \emph{field}.

In a scattering experiment, where a plane wave hits a moving \acs{BH} head-on, one can define an absorption cross-section
\begin{align}
\sigma^\text{abs}=\frac{E_\text{in}-E_\text{out}}{E_\text{in}/A_\text{in}}=\frac{\widetilde{\mathcal{F}}_E}{\tilde{\rho}(1+v)(\hbar \omega)}\,,
\end{align}
where $E_\text{in}$ is the total energy in the plane wave, $E_\text{out}$ is the total energy in the outgoing wave after interaction with the \acs{BH}, and $A_\text{in}$ is the surface area that the incident plane occupies. As we discussed in previous sections, due to the long-range ($1/r$) character of gravity, the above absorption cross-section diverges logarithmically~\cite{landau1981quantum}. So, we define instead a finite quantity~$\sigma^\text{abs}_{20}$, computed by sending a constant flux wave centered at the \acs{BH}, but with finite transverse size of radius~$b_\text{max}=20M$ (at large distances). This quantity is shown in Table~\ref{table_basis} for null \emph{particles} and compared with our result for a scalar \emph{field} (Eq.~\eqref{energy_geomopt}),
\begin{align} \label{compare_acs}
	\frac{\sigma^{\textrm{abs}}_{20}}{\pi M^2}\simeq  \frac{16}{1-v}\left(\frac{27}{16 }-5.31435 v\right)\,.
\end{align}

The agreement between the two approaches (expressing the validity of the geometrical optics approximation) is remarkable for most~\acs{BH} velocities. For each cutoff~$b_\text{max}$ there is a critical velocity above which the moving \acs{BH} overall deposits energy in its environment (contrary to the classical idea of being a perfect absorber); for~$b_\text{max}=20M$, this velocity is~$v\simeq 0.32$.

\begin{table}[h]
	\centering
	\begin{tabular}{cc||cc}
		\hline
		\hline
		$v$ & $\sigma^\text{abs}_{20}/(\pi M^2)$ &$v$ & $\sigma^\text{abs}_{20}/(\pi M^2)$    \\ 
		\hline
		\hline
		0.00 & 27.0  (27)                           &0.30  & 2.1 (2.1) \\
		0.01 & 26.4  (26.4)                         &0.50  &-31.1 (-31.0)\\
		0.02 & 25.8    (25.8)                          &0.80  &-205.6 (-205.1)\\
		0.10  & 20.6    (20.6)                         &0.90  &-496.8 (-495.3)\\
		\hline
		\hline
	\end{tabular} 
	\caption{Absorption cross-section for a \acs{BH} moving with velocity~$v$ onto a constant flux wave obtained using the geometric optics approximation, compared with~\eqref{compare_acs} between parenthesis. The incoming wave has a finite spatial extent in the direction transversal to the motion, forming a cylinder of radius $b_\text{max}=20M$. Notice that the absorption cross-section becomes negative at large velocities, indicating that \acs{BH} transfers energy to its environment. }
	\label{table_basis}
\end{table}

\subsubsection{Appearance of a moving black hole}

\begin{figure}
	\centering
	\begin{tabular}{c}
		\includegraphics[width=0.9 \textwidth]{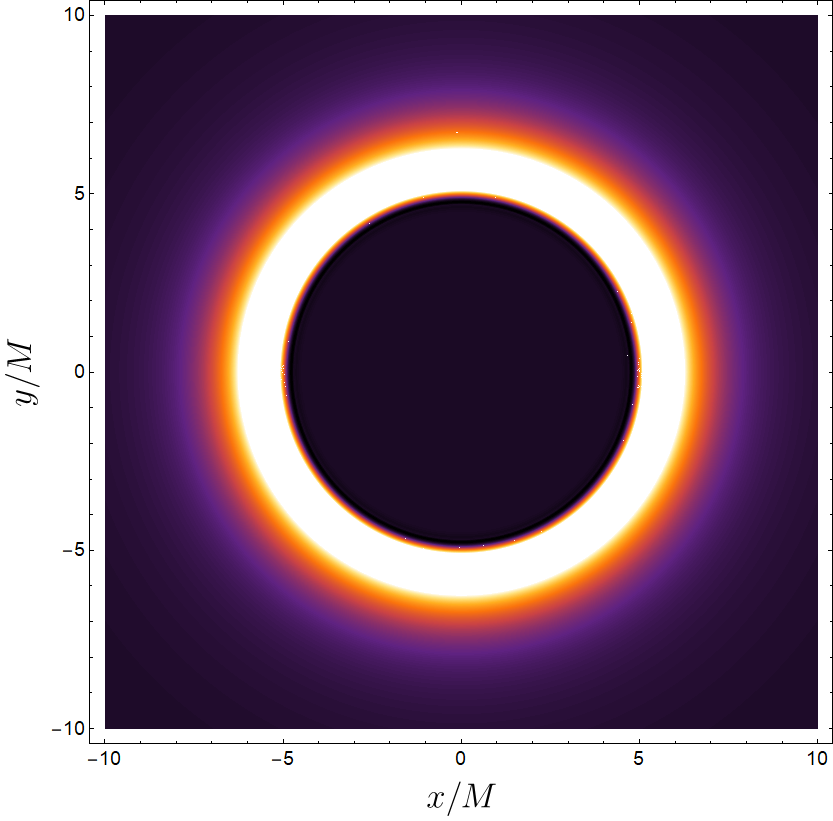} 
	\end{tabular}
	\caption{Appearance of a \acs{BH} moving in a bath of cold (and counter-moving) radiation. The \acs{BH} is moving along the $z$-axis towards us at a speed~$v=0.9$. The colors denote energy flux intensity on a screen placed a short distance away from the \acs{BH}. The peak energy flux is ten times larger than that of the environment. The bright ring has width~$\sim M$ for all boost velocities~$v$. For very large~$v$ even a randomly-moving gas of photons will leave a similar observational imprint, since counter-moving photons will be red-shifted away.}
	\label{fig:image}
\end{figure}
The large amplification for strongly-deflected photons implies that a rapidly moving \acs{BH} looks peculiar. Downstream photons are deflected and blueshifted upstream. Thus a rapidly moving \acs{BH} in a cold gas of radiation will be surrounded by a bright ring of thickness~$\sim M$. A possible image of a moving \acs{BH} is shown in Fig.~\ref{fig:image}. For a stellar-mass \acs{BH} moving at velocities~$v\sim 0.9996$ through the universe, the ambient cosmic microwave background will produce a kilometer-sized ring (locally $\sim 5000$ times hotter and brighter than the cosmic microwave background) in the visible spectrum.

\section{Discussion}

The scattering of waves is a fundamental process in physics. 
In this chapter we showed that the universal nature of gravity, together with the~$1/r$ behavior of Newton's law causes moving \acsp{BH} to amplify plane waves, with a divergent cross-section. This is the only known example of a negative absorption cross section for neutral fields scattering off a \acs{BH}.
We also showed that even a narrow beam of light can extract energy from a rapidly (counter-) moving \acs{BH}.
These results apply to any massless wave (independently of its spin) in the high-frequency regime.  
For \acsp{BH} at rest, the absorption cross-section of low-frequency electromagnetic or gravitational waves vanishes, which may imply that amplification happens sooner at low frequencies, for higher spins. This remains to be understood.
These results may have little practical application, since \acsp{BH} are not expected to be traveling through our universe at relativistic speeds: mergers of \acsp{BH} or neutron stars lead at best to ``kicks'' in the remnant of~$v\lesssim 10^{-2}$~\cite{Campanelli:2007ew,Brugmann:2007zj,Gonzalez:2007hi,Campanelli:2007cga} for astrophysical setups (even the high-energy merger of two \acsp{BH} leads ``only'' to kicks of $v\lesssim 0.05$~\cite{Sperhake:2010uv}). For these velocities, the effects dealt with here are only important when the \acs{BH} moves in very extended media. Nevertheless, our results show how nontrivial strong gravity effects can be. 

The overall result of energy transfer to external radiation echoes that of the inverse Compton scattering for fast-moving electrons in a radiation field~\cite{Dolan:private,Beckmann}. In this latter process, a nearly-isotropic radiation field is seen as extremely anisotropic to the individual ultra-relativistic electrons. Relativistic aberration causes the ambient photons to approach nearly head-on; Thomson scattering of this highly anisotropic radiation reduces the electron's kinetic energy and converts it into inverse-Compton radiation by upscattering radio photons into optical or $X$-ray photons. The process we discuss here, involving \acsp{BH}, is special: \acsp{BH} are natural absorbers, but the universal -- and strong, close to the horizon -- pull of gravity can turn them also into overall amplifiers.

On the other hand, the mechanism for energy extraction can be relevant in the context of fundamental light fields, with confined low-energy excitations~\cite{Bernard:2019nkv}. A \acs{BH} binary in this setup could slow down and transfer some of its energy to the fundamental field, "heating" its environment, and giving rise to potentially observable effects. This mechanism of (kinetic) energy transfer from a moving \acs{BH} to its surroundings is closely related with \acs{DF}. This phenomenon can be of great importance to test, \eg, the nature of \acs{DM} (is it a wave or a particle?~\cite{Hui:2016ltb,Hui:2021tkt}, by studying how \acsp{BHB} evolve. In this chapter, we obtained for the first time, from first principles, the \acs{DF} acting on \acsp{BH} moving at possibly relativistic speeds in light scalar field environments. We found several simple analytical expressions valid for different regimes of \acs{BH} velocity (and our framework can be used to compute numerically the \acs{DF} for any other velocity not covered by the analytical expressions). We focused on stationary regimes and extended the Newtonian treatment in Ref.~\cite{Hui:2016ltb}. Our results complement the recent work in Ref.~\cite{Traykova:2021dua}, where the same problem was solved through numerical time evolutions, which cover also the non-stationary regime and show how the stationary regime is attained. Our analytical expressions should be compared carefully with these numerics (if they allow for a easy change in their cutoff procedure, as discussed previously). Our framework can be easily (and it is presently being) extended to spinning moving \acsp{BH}, where a type of Magnus effect is expected to be at work (causing a bending on the \acs{BH}'s trajectory), which would be an additional potential observable to probe the environment of moving compact objects.

%*****************************************
%*****************************************
%*****************************************
%*****************************************
%*****************************************

%*****************************************
\chapter{Dynamical friction in Slab-like Geometries}\label{ch:DF}
%*****************************************

Dynamical friction is well understood when an object moves in an infinite (collisionless or fluid-like) medium~\citep{Chandrasekhar:1943v1,Chandrasekhar:1943v2,Chandrasekhar:1943v3,Ostriker:1998fa,binney2011galactic}. Most rigorous treatments of this phenomenon in the literature consider infinite three-dimensional media -- some few exceptions are~\citep{Namouni2011,Muto:2011qv,Canto2013}. Clearly, such idealization breaks down in thin accretion or protoplanetary disks, where the geometry of the problem is more "slab-like"~\citep{Novikov:1973kta,Armitage2011}. In this context, \citeauthor{Muto:2011qv}~\cite{Muto:2011qv} obtained estimates for the~\acs{DF} in a steady state using a two-dimensional approximation for the gaseous medium. However, as the authors pointed out, their simplified approach has some limitations and a fully three-dimensional treatment is needed. Also assuming a steady wake, \citeauthor{Canto2013}~\cite{Canto2013} computed the~\acs{DF} acting on a hypersonic perturber moving in the midplane of a gaseous disk with Gaussian vertical density stratification. However, they did not study how (and if) this steady state is dynamically attained.

In this chapter we compute the gravitational wake caused by, and the time-dependent force acting on a massive perturber moving in a three-dimensional medium with a slab-like geometry, subjected to either Dirichlet or Neumann conditions at the boundaries. This setup is a more faithful approximation to the physics of thin disks and I expect some of the main findings to carry over, at least qualitatively, to more generic physical situations where boundaries play a role.
For simplicity, here we consider an inviscid adiabatic medium and neglect the effects of direct collisions between the massive perturber and the gas.

\section{Unbounded gaseous media}

The Euler equations describing the evolution of an inviscid adiabatic gas are~\cite{landau1987fluid}
\begin{align}
	\frac{d\rho}{d t}&=- \rho\,  \boldsymbol{\nabla} \cdot\boldsymbol{v}\,, \\
	\frac{d\boldsymbol{v}}{d t}&=- \frac{\boldsymbol{\nabla} p}{\rho}-\boldsymbol{\nabla} \phi_{\textrm ext} \,, \\
	\frac{d s}{d t}&=0 \label{adiab}\,,
\end{align}
where~$\rho$,~$\boldsymbol{v}$ and~$s$ are, respectively, the mass density, velocity and specific (per unit of mass) entropy of the gas, and~$\phi_{\textrm ext}$ is the potential of some external force. The time-derivative~$d/dt=\partial/\partial t+\boldsymbol{v}\cdot \boldsymbol{\nabla}$ is the usual Lagrangian (or material) derivative of fluid mechanics. 
Assuming that the specific entropy is uniform throughout the medium at some initial instant, then Eq.~\eqref{adiab} implies~$s= \textrm{constant}$ at subsequent times, and the relation~$p \rho^{-\gamma}=\textrm{constant}$ is satisfied, where~$\gamma$ is the heat capacity ratio. we also assume that the gas is calorically perfect, which means that it has a constant~$\gamma$. Its speed of sound is given by the simple expression~$c_s=\sqrt{(\partial p/\partial \rho)_s}=\sqrt{\gamma\, p/\rho}$.

The linearized equations describing the perturbations~$\rho = \rho_0[1+\alpha(t,\boldsymbol{r})]$ and~$ \boldsymbol{v} = c_s \boldsymbol{\beta}(t,\boldsymbol{r})$ in an homogeneous gas of mass density~$\rho_0$ with no velocity are~\cite{Ostriker:1998fa}
\begin{align}
&\frac{1}{c_s}\frac{\partial \alpha}{\partial t} +\boldsymbol{\nabla}\cdot \boldsymbol{\beta}=0 \,, \label{linearizedeq1}\\ 
&\frac{1}{c_s} \frac{\partial \boldsymbol{\beta}}{\partial t}+\boldsymbol{\nabla}\alpha=-\frac{1}{c_s^2}\boldsymbol{\nabla}\phi_{\text{ext}}\,, \label{linearizedeq2}
\end{align}
where $c_s$ is the (constant) speed of sound in the unperturbed medium and the perturbation scheme is valid for $\alpha,|\boldsymbol{\beta}|\ll 1$.
These equations can be combined to obtain the inhomogeneous wave equation 
\begin{equation} 
\nabla^2 \alpha-\frac{1}{c_s^2}\frac{\partial^2 \alpha}{\partial t^2}=-\frac{1}{c_s^2}\nabla^2\phi_{\text{ext}}\,.\label{densdiffeq}
\end{equation}  
If the external influence is due to the gravitational interaction with a massive perturber of mass density $\rho_{\text{ext}}(t,\boldsymbol{r})$, the potential satisfies a Poisson's equation
\begin{equation}
\nabla^2\phi_{\text{ext}}=4 \pi  \rho_{\text{ext}}\,. \label{eq:poisson}
\end{equation}
Equation~\eqref{densdiffeq} can be solved using a Green's function $G(t,\boldsymbol{r};t',\boldsymbol{r}')$, which is a solution of
\begin{equation}
\nabla_{\boldsymbol{r}}^2 G-\frac{1}{c_s^2}\frac{\partial^2 G}{\partial t^2}=- \delta(t-t') \delta^{(3)}(\boldsymbol{r}-\boldsymbol{r}') \,.\label{flat_green}
\end{equation}

The problem of finding the (linearly) perturbed density~$\rho(\boldsymbol{r},t)$ of an \emph{infinite} three-dimensional gaseous medium due to the gravitational pull of a point-like mass~$M$ moving at velocity~$\boldsymbol{V}$ was solved by~\citeauthor{Ostriker:1998fa}~\cite{Ostriker:1998fa}. 
The~\acs{DF} acting on the moving mass was therein computed to be
\begin{align}
\boldsymbol{F}&=\frac{ M^2 \rho_0}{c_s^2} I(\mathcal{M},t)\, \boldsymbol{e_V}\,,\label{3Ddragforce} \\
I&=-\frac{4 \pi}{\mathcal{M}^2}\left[\frac{1}{2} \log\left(\frac{1+\mathcal{M}}{1-\mathcal{M}}\right)- \mathcal{M}\right]\,,\quad \mathcal{M}<1\label{subsonic_ostriker}\\
I&=-\frac{4 \pi}{\mathcal{M}^2} \left[\frac{1}{2} \log\left(1-\frac{1}{\mathcal{M}^2}\right)+ \log\left(\frac{\mathcal{M} c_s t}{r_\text{min}}\right)\right]\,,\quad \mathcal{M}>1
\label{supersonic_ostriker}
\end{align}
where~$\mathcal{M}\equiv V/c_s$ is the Mach number and $r_\text{min}$ is the effective size of the perturber, which is assumed to satisfy $r_\text{min}<(\mathcal{M}-1) c_s t$.~\footnote{The~\acs{DF} acting on a supersonic point-like mass has an \textit{ultraviolet} divergence. Thus, a cutoff $r_\text{min}$ is needed.}
Notice that~\acs{DF} is really a friction force, in the sense that it always opposes the perturber's motion ($\boldsymbol{F} \cdot \boldsymbol{V}<0$).

\section{Gaseous slabs}

In this section we consider an inviscid adiabatic gaseous medium with a slab-like profile: with constant density and extending to arbitrarily large spatial distances in the~$x$ and~$y$ directions, but with compact support in the~$z$ direction, with thickness~$2 L$ (\ie,~$-L\leq z \leq L$).
The linear perturbation in the pressure is $\delta p= (\partial p/\partial \rho) \delta \rho=c_s^2 \rho_0 \alpha(t,\boldsymbol{r})$. Thus, the physically relevant setup $\delta p(t,z=-L)=\delta p(t,z=L)=0$ is associated with Dirichlet conditions in $\alpha(t,\boldsymbol{r})$ at the boundaries of the slab.~\footnote{More about the choice of boundary conditions in Section~\ref{sec:wakerefl}.} For completeness, we study the case of Neumann conditions as well.

\paragraph{Green's function}
Defining $T\equiv t-t'$ and $\boldsymbol{R} \equiv(x-x',y-y')$
the solution of Eq.~\eqref{flat_green} with Dirichlet conditions at~$z=\pm L$ is~\footnote{See Appendix~\ref{app:DF} for a derivation of this expression.}
\begin{align}
\hspace{-0.7cm}G=\frac{c_s}{2 \pi L} \Theta(c_s T - R)\sum_{n=0}^{+\infty}  \frac{\cos\left(m_n D\right)}{D} \sin[m_n (z+L)] \sin[m_n (z'+L)] \,, \label{GreensF}
\end{align}
with $D\equiv \sqrt{c_s^2 T^2-R^2}$ and $m_n\equiv n \pi/(2L)$, and where~$\Theta(x)$ is the Heviside step function. Rewriting the sines and cosines as complex-exponential sums, and using the Dirac comb identity $\sum_{n=-\infty}^{+\infty}e^{i m_n x} = 4 L \delta\left(x \,\, \text{mod}\,\, 4 L \right)$ the last expression reads
\begin{align}
G=\frac{c_s}{4\pi D}\,\Theta(c_s T - R)\sum_{l=-\infty}^{+\infty} i^{2l}  \delta\left(z-i^{2 l} z'-2 l L\pm D\right)\,.	
\end{align}
Using the properties of delta functions this expression can be put into the form
\begin{align}
\hspace{-0.7cm}G=\sum_{l=-\infty}^{+\infty} \frac{i^{2l}}{4\pi  \sqrt{(z-i^{2l} z'-2 l L)^2+R^2}}  \delta\left(\tfrac{\sqrt{(z-i^{2l} z'-2 l L)^2+R^2}}{c_s}-T\right)\,.	
\end{align}

\paragraph{Gravitational wake}
The gravitational interaction between the medium and an external massive perturber is governed by Poisson's equation~\eqref{eq:poisson}. Thus, the solution to Eq.~\eqref{densdiffeq} with Dirichlet boundary conditions is
\begin{align}
\hspace{-0.7cm}\alpha=\sum_{l=-\infty}^{+\infty}\frac{i^{2l}}{c_s^2}\int  \frac{d^3\boldsymbol{r}' dt' \,  \rho_\text{ext}(t',\boldsymbol{r}')}{\sqrt{(z-i^{2l} z'-2 l L)^2+R^2}} \delta\left(\tfrac{\sqrt{(z-i^{2l} z'-2 l L)^2+R^2}}{c_s}-T\right)\,. 
\end{align}
The index $l$ has an interesting physical meaning: it is the number of \emph{reflections} that perturbations have undergone at the boundaries. Therefore, the density perturbation~$\alpha$ is expanded in terms of the number of \emph{echoes} of the Green's function on the slab boundaries. This result is analogous to that of a signal propagating along a four-dimensional brane in a five-dimensional Kaluza-Klein spacetime, except that boundary conditions are of the Neumann type for that problem~\cite{Barvinsky:2003jf}. Notice also that the~$l=0$ term in the expansion corresponds to direct propagation and describes the fluctuations not sensitive to the boundaries. Not surprisingly, this term describes exactly the solution for a three-dimensional infinite medium~\cite{Ostriker:1998fa}. 

Now consider a particle of mass $M$ moving with velocity~$\boldsymbol{V}=V \boldsymbol{e_x}$, with~$V>0$, on a straight-line through the medium. we will assume that the perturbation is turned on at $t=0$.~\footnote{The~$t=0$ can be thought of as the instant when the perturber enters, or forms inside the gaseous medium. This consideration allows a study of how (and if) a stationary regime is attained.} Then, the source is prescribed through the mass density~$\rho_{\text{ext}}=M \delta(x'-V t')\delta(y')\delta(z') \Theta(t')$. Under these conditions it can be shown that the perturbation in the medium density is
\begin{align}
\alpha=&\frac{M}{c_s^2}\sum_{l=-\infty}^{+\infty} i^{2 \eta l}\int_{-\infty}^{+\infty} dw \, \,\Theta(w+x)\tfrac{\delta\left(w+s+\mathcal{M} \sqrt{(z-2 l L)^2+w^2+y^2} \right)}{\sqrt{(z-2 l L)^2+w^2+y^2}}  \,,\label{slabdensity}
\end{align}
with~$w \equiv x'-x$,~$s \equiv x-V t$, and~$\eta \equiv \{1,0\}$ for Dirichlet and Neumann conditions, respectively.

\paragraph{Dynamical friction}
An infinitesimal element of medium~$\rho\, dx\, dy\, dz$ between~$\boldsymbol{r}$ and~$\boldsymbol{r}+d\boldsymbol{r}$ acts gravitationally on a particle of mass~$M$ at position~$\boldsymbol{r}'=\boldsymbol{V} t$ through Newton's law
\begin{equation}
d\boldsymbol{F}(t,\boldsymbol{r})= \frac{(\rho\, dx dy dz) M}{\left[(x-V t)^2+y^2+z^2\right]^{3/2}} (\boldsymbol{r}-\boldsymbol{V} t) \,.
\end{equation} 
Then, the~\acs{DF} acting on the massive perturber is given by
\begin{equation}
	\boldsymbol{F}(t)= \rho_0 M \int d^3\boldsymbol{r}\, \frac{\alpha(t,\boldsymbol{r})}{\left[(x-V t)^2+y^2+z^2\right]^{3/2}} (\boldsymbol{r}-\boldsymbol{V} t)\,. \label{force_drag}
\end{equation}

\subsection{Subsonic regime}
First let me consider the case of a perturber with Mach number $\mathcal{M}<1$. The argument of the delta function in Eq.~\eqref{slabdensity} then vanishes for
\begin{equation}
w=w_l\equiv -\frac{s+\mathcal{M}\sqrt{s^2+(1-\mathcal{M}^2)d_l^2}}{1-\mathcal{M}^2}\,,
\end{equation}  
where we defined~$d_l\equiv \sqrt{y^2+(z-2 l L)^2}$.
Each $l$-contribution to $\alpha(t, \boldsymbol{r})$ vanishes for $w_l+x<0$ or, equivalently,
\begin{equation}
x^2+d_l^2>c_s^2 t^2\,.
\end{equation}
This is a manifestation of the \emph{causality principle}. The perturber is turned on at~$(t,x,y,z)=0$ and moves with velocity~$V<c_s$. The perturbation, on the other hand, propagates with speed~$c_s$. These two facts imply that, at instant~$t$, the maximum domain of influence of the massive particle is the region in the slab defined by~$x^2+y^2+z^2\leq c_s^2t^2$. This is the domain of influence of the~$l=0$ term. At fixed~$t$, each~$l$-term has a different domain of influence. Larger~$l$'s probe smaller regions since the fluctuation is "busy" traveling between the boundaries and is unable to probe larger~$x,y$ directions.
Notice that not all~$l$-terms contribute to~$\alpha$ at an instant~$t$. A given $l$ mode only contributes from~$t_l=(2|l|-1)L/c_s$ onwards. 
Physically, this is due to these terms being echoes and, therefore, requiring a finite time to reach the slab boundaries.
The only exception is the~$l=0$ term, which contributes from~$t=0$ onwards.

\begin{figure} \centering
	\includegraphics[width=1\linewidth]{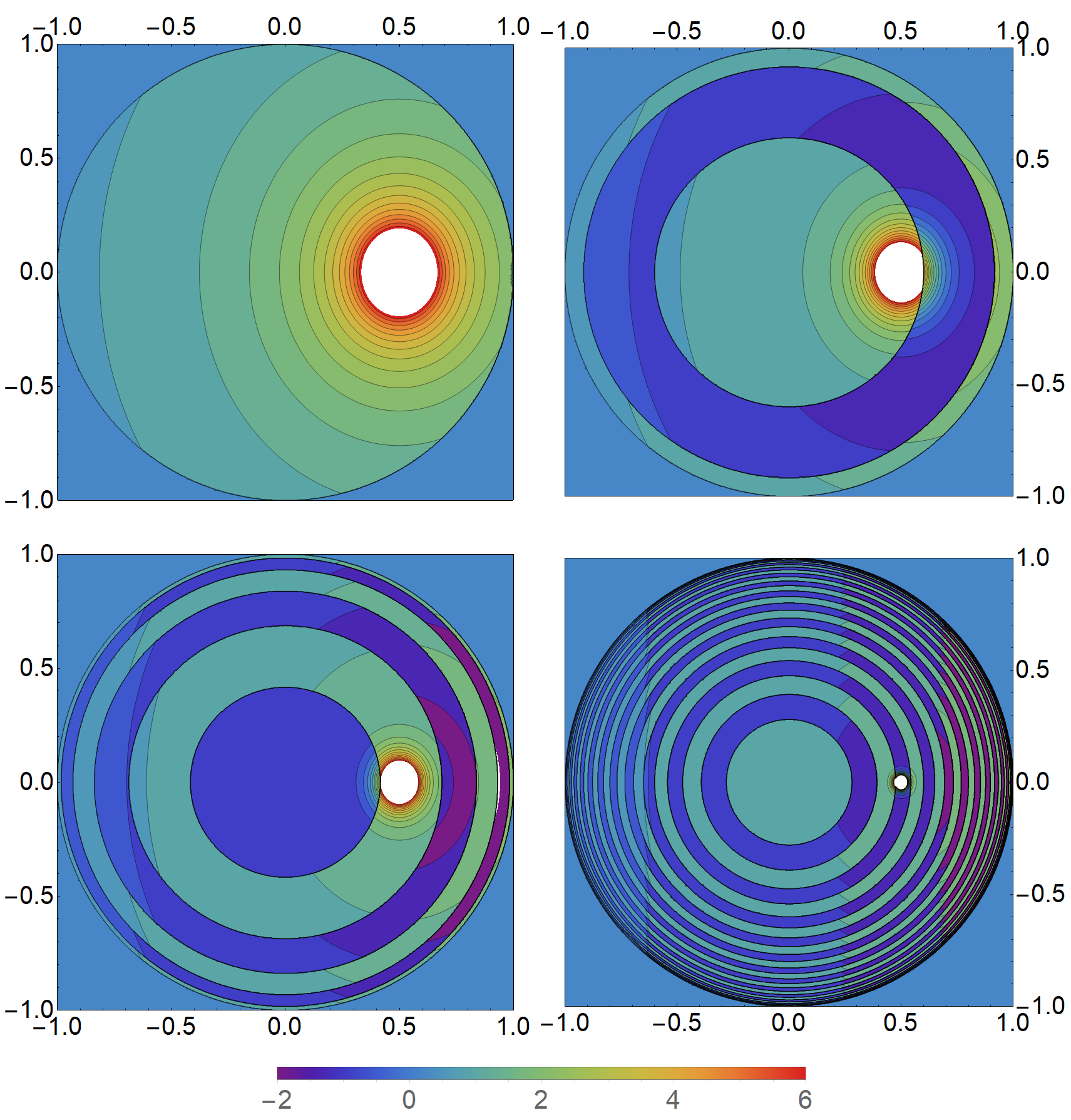}
	\caption{Density wake~$\alpha c_s^3 t/M$ in a gaseous slab along~$z=0$, due to the gravitational interaction with a subsonic particle with Mach number~$\mathcal{M}=0.5$, for~$c_s t/L=(0.5,5,11,50)$ (left to right, top to bottom). The horizontal axis represents the coordinate~$x/(c_s t)$ and the vertical axis~$y/(c_s t)$. The contours represent curves of constant density. The observed ripples centered at the origin -- which turn on at~$c_s t/L\geq1$, but are only seen in~$z=0$ at~$ct/L\geq2$ -- are \emph{echoes} of the original density fluctuation. Each ripple is associated with a different~$l$-term.}
	\label{fig:slabsubsonicz0D}
\end{figure}

In summary, a massive particle moving at subsonic speeds through a gaseous slab causes a density fluctuation
\begin{equation}
\alpha(t,\boldsymbol{r})=\frac{M}{c_s^2}\sum_{l=-\infty}^{+\infty}  i^{2 \eta l} \frac{\Theta \left[c_s^2 t^2-x^2-d_l^2\right]}{\sqrt{s^2+(1-\mathcal{M}^2)d_l^2}} \, \label{wake_sub}
\end{equation}
in the medium, where we used the property~$|A|\delta \left[A(w-w_l) \right]=\delta(w-w_l)$.
A contour plot of the density profile is shown in Fig.~\ref{fig:slabsubsonicz0D} at different instants. The perturber is moving at a subsonic speed with Mach number~$\mathcal{M}=0.5$. The results for~$ct/L=0.5$ coincide \emph{exactly} with the ones obtained for non-compact geometries by~\citeauthor{Ostriker:1998fa}~\cite{Ostriker:1998fa}, since the perturbation did not have time yet to reach the boundaries. 

Using Eq.~\eqref{force_drag} one can see that, by the symmetry of the wake, the net~\acs{DF} acting on the particle points in the $\boldsymbol{e_x}$ direction (\ie,~$\boldsymbol{F}=F\, \boldsymbol{e_x}$). 
For times~$c_s t/L<1$ the only contributing term is the $l=0$ and the~\acs{DF} reduces to
\begin{align} 
\hspace{-1cm}F=  \frac{M^2 \rho_0}{c_s^2} \int  \frac{d^3\boldsymbol{\bar{r}}\,\,\Theta\left(1-\bar{r}^2\right)}{\sqrt{(\bar{x}-\mathcal{M})^2+(1-\mathcal{M}^2)(\bar{y}^2+\bar{z}^2)}} \frac{\bar{x}-\mathcal{M}}{\sqrt[3]{(\bar{x}-\mathcal{M})^2+\bar{y}^2+\bar{z}^2 }}	\label{slabDragforceearly}
\end{align}
for both Dirichlet and Neumann conditions, where we defined barred coordinates~$\bar{x}\equiv x/(c_s t)$. This expression is clearly time-independent and the integration results in~\eqref{subsonic_ostriker}. Thus, for early times~$c_s t/L<1$ the perturbation did not probe the boundary and, once again, one recovers the result of~\citeauthor{Ostriker:1998fa}~\cite{Ostriker:1998fa}.

To find the force at late times~$c_s t/L \gg 1$ one can start by breaking the expansion in even and odd $l$-terms, defining~$l_{\textrm{e}}\equiv 2 l $ and~$l_{\textrm{o}}\equiv 2l+1$. After doing that, Eq.~\eqref{force_drag} with the gravitational wake~\eqref{wake_sub} becomes
\begin{align}
F\simeq& \frac{2 L}{c_s t} \frac{ M^2 \rho_0}{c_s^2} \sum_{|l_{\textrm{e}}|\leq \text{int}\left[c_s t/(4 L)\right]} \int d\bar{x} d\bar{y} \frac{\bar{x}-\mathcal{M}}{\left[(\bar{x}-\mathcal{M})^2+\bar{y}^2\right]^{3/2}} \nonumber\,\, \\
&\times \frac{\Theta \left[1-\bar{x}^2-\bar{y}^2-\left(\frac{4 L}{c_s t}\right)^2 l_{\textrm{e}}^2\right]}{\sqrt{(\bar{x}-\mathcal{M})^2+(1-\mathcal{M}^2)\left[\bar{y}^2+ \left(\frac{4 L}{c_s t}\right)^2 l_{\textrm{e}}^2\right]}} -(l_{\textrm{e}} \to l_{\textrm{o}})\,,\label{slabDragforcelateD}
\end{align}
where $\text{int}(k)$ is the integer part of $k$. Notice that in the last expression we already performed the integration in~$\bar{z}$ (which is trivial, since the integrand is independent of~$\bar{z}$ at late times~$c_s t/L \gg 1$). The last expression shows something remarkable: in slab geometries with Dirichlet boundary conditions the~\acs{DF} is \emph{suppressed} at late times! In particular, at late times $F$ decays linearly with~$\sim L/(c_s t)$ (modulo some residual time-dependence arising from the difference between even and odd~$l$ modes).

\begin{figure}
	\includegraphics[width=1\linewidth]{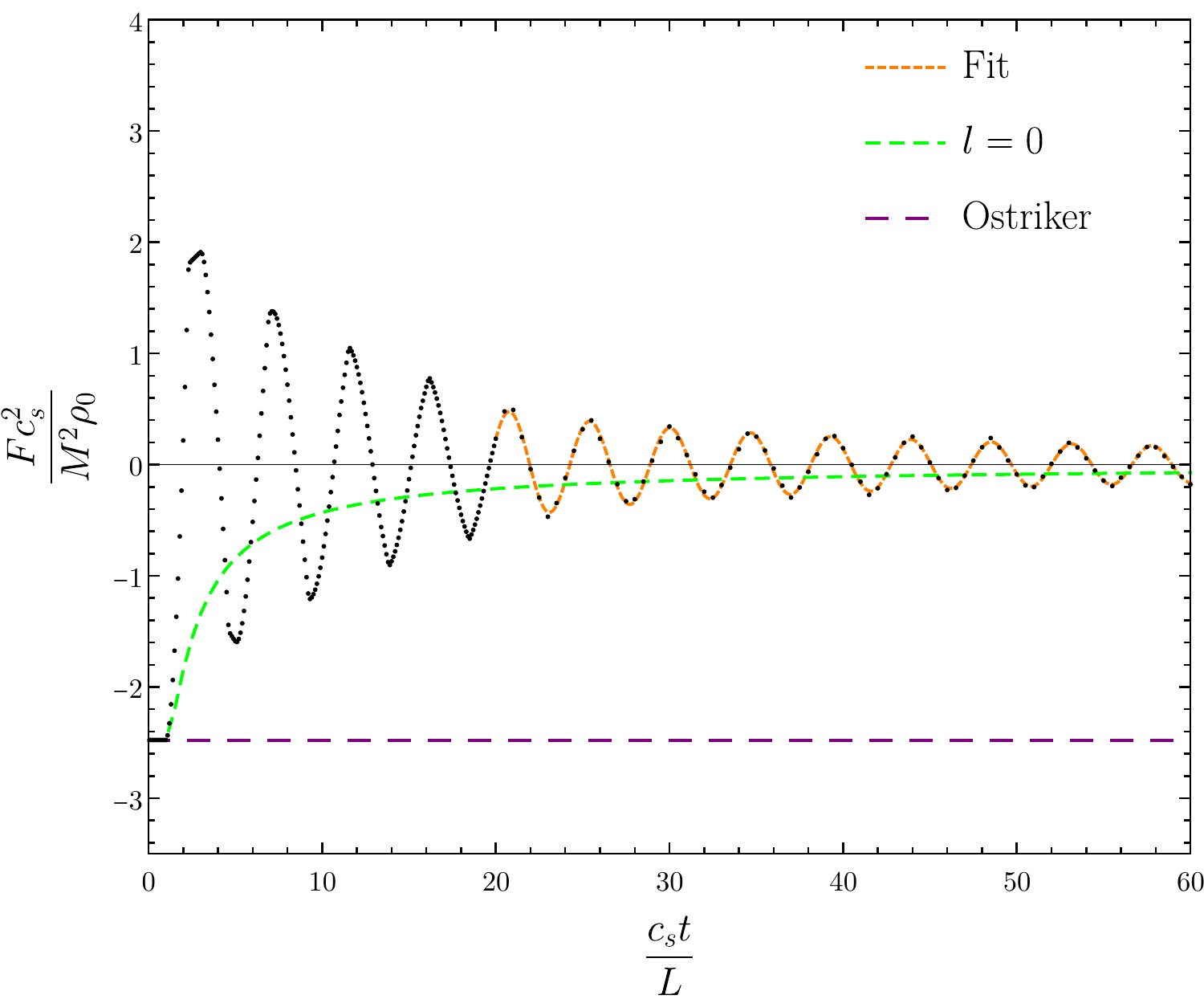} 
	\caption{Dynamical friction~$F c_s^2/\left(M^2 \rho_0\right)$ acting on a particle moving at a subsonic Mach number~$\mathcal{M}=0.5$ as function of time~$c_s t/L$ (black dots). The results are in agreement with the predicted early- and late-time behavior of the force, as described by Eqs.~\eqref{slabDragforceearly} and~\eqref{slabDragforcelateD}, respectively. Notice that the early-time force ($c_s t/L<1$) is independent of the boundary conditions, and, thus, is the same as for non-compact geometries; therefore it is described by well-known results~\cite{Ostriker:1998fa} (purple dashed curve). At late times the force oscillates with a period $\sim \frac{4 L}{c_s}\,e^{ \frac{\mathcal{M}^{2.23}}{2 (1-\mathcal{M})^{0.31}}}$ and decays as $\sim L/(c_s t)$; the orange dashed curve shows the fit expression~\eqref{fitsubD}. In green it is shown the ($l=0$) contribution from the non-reflected wake.
		\label{fig:slabsubsonicforcetD}}
\end{figure}

The numerical results of the integration of Eq.~\eqref{force_drag} with the density perturbation~\eqref{wake_sub} are shown in Fig.~\ref{fig:slabsubsonicforcetD} for a Mach number~$\mathcal{M}=0.5$. The~\acs{DF} is initially the same as that in extended geometries (\ie, Eq.~\eqref{subsonic_ostriker}).
However, after the perturbations reach the boundary such force changes.
It is amusing to see that for some time intervals the force acting on the perturber is \emph{positive} (\ie,~$\boldsymbol{F} \cdot \boldsymbol{V}>0$). This can be traced back to the existence of regions of negative density fluctuation~$\alpha$, which effectively act in a repulsive way on the particle, due to the deficit of matter in such region. "Positive drag" (sometimes called slingshot effect) does not arise with Neumann conditions, nor for an infinite three-dimensional medium, but nothing forbids it from appearing -- and in fact it does in slab geometries.
At late times the~\acs{DF} exhibits damped oscillations well described by (see Fig.~\ref{fig:slabsubsonicforcetD})
\begin{equation} \label{fitsubD}
F \simeq \frac{M^2 \rho_0}{c_s^2} \frac{\mathcal{A}}{(c_s t/L)}  \cos\left(\frac{2 \pi}{\mathcal{T}} \frac{c_s t}{L}+ \varphi \right)\,,
\end{equation}
where~$\mathcal{A}$,~$\mathcal{T}$ and~$\varphi$ are functions of $\mathcal{M}$. The period of oscillation is well approximated by
\begin{equation}
\mathcal{T} \simeq \frac{4 L}{c_s} \exp\left( \dfrac{\mathcal{M}^{2.23}}{2 (1-\mathcal{M})^{0.31}}\right)\,.
\end{equation}

\begin{figure}
	\includegraphics[width=1\linewidth]{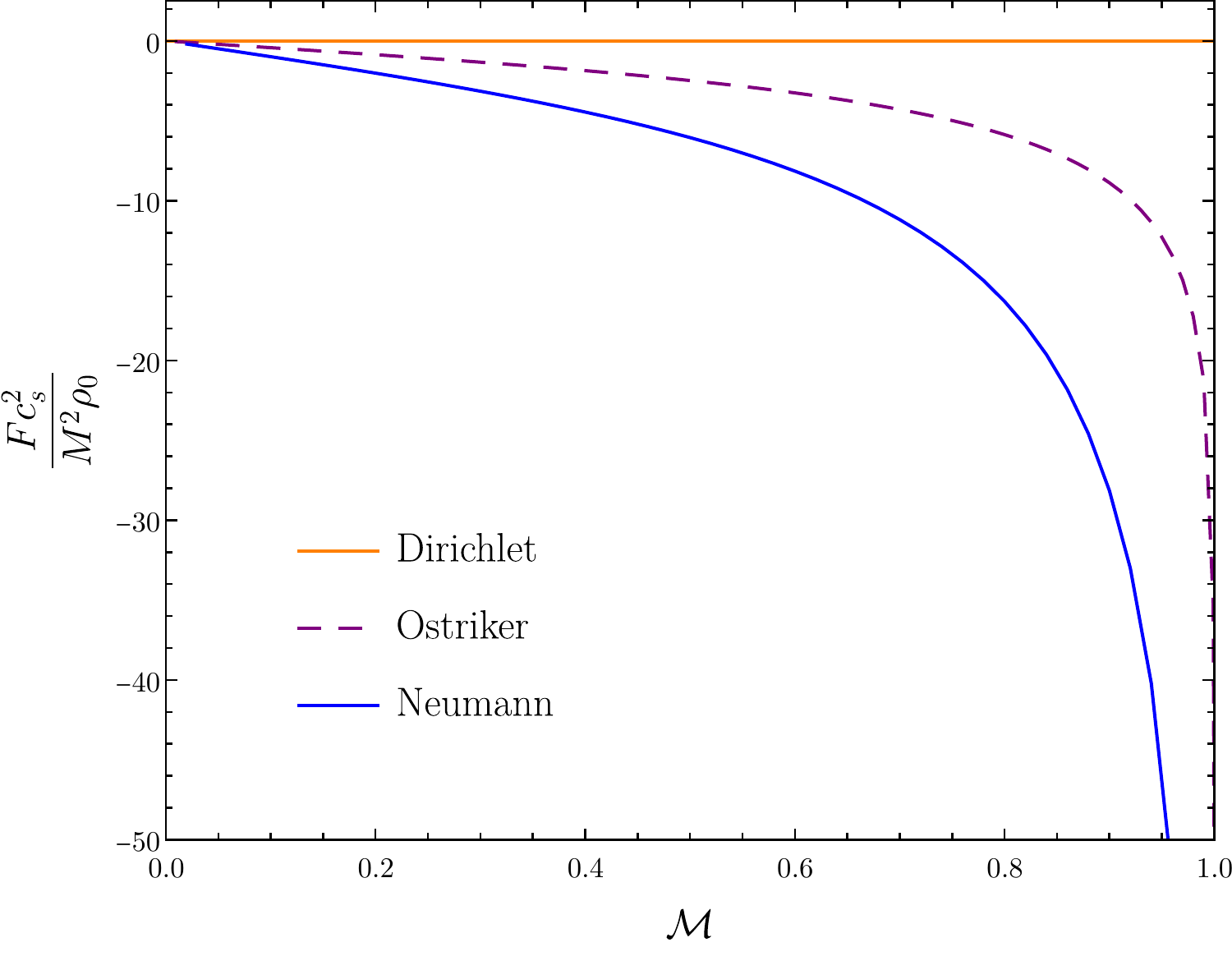} 
	\caption{Comparison between the early- [\citeauthor{Ostriker:1998fa}:~\eqref{slabDragforceearly}] and late-time [Dirichlet:~\eqref{slabDragforcelateD}; Neumann:~\eqref{dragforceapplate}] dynamical friction $F c_s^2/\left(M^2 \rho_0\right)$ as function of the
	Mach number~$\mathcal{M}$. In the subsonic regime, the \acs{DF} in a three-dimensional slab with Dirichlet (Neumann) conditions is always smaller (larger) in magnitude than the one in an infinite three-dimensional medium.}
	\label{fig:slabsubsonicforcem}
\end{figure}

If Neumann conditions are used instead, the numerical results show that at late times the massive particle feels a constant~\acs{DF} well approximated by 
\begin{equation} 
F \simeq - 7.864 \frac{M^2 \rho_0}{c_s^2} \frac{\mathcal{M}}{(1-\mathcal{M})^{3/5}}\,.\label{dragforceapplate}
\end{equation}
The dependence of the early- and late-time~\acs{DF} on the particle's Mach number is shown in Fig.~\ref{fig:slabsubsonicforcem}. The~\acs{DF} in a three-dimensional slab with Dirichlet (Neumann) conditions is always smaller (larger) in magnitude than the one in an infinite three-dimensional medium. 

\subsection{Supersonic regime}
In the case of a massive perturber moving with Mach number~$\mathcal{M}>1$ the argument of the delta function
in Eq.~\eqref{slabdensity} has roots only if
\begin{equation}
s\le-\sqrt{\left(\mathcal{M}^2-1\right)d_l^2}\,,
\end{equation}
and these roots are 
\begin{equation}
w_{l,\mp}\equiv\frac{1}{\mathcal{M}^2-1}\left[s \mp \mathcal{M} \sqrt{s^2-(\mathcal{M}^2-1)d_l^2}\right]\,.
\end{equation}
With some algebra one can show that Eq.~\eqref{slabdensity} becomes
\begin{align} 
	&\hspace{-1cm}\alpha(t,\boldsymbol{r})=\frac{M}{c_s^2}\sum_{l=-\infty}^{+\infty}  \frac{i^{2\eta l}}{\sqrt{s^2-(\mathcal{M}^2-1)d_l^2}} \bigg\{\Theta\left[c_s^2 t^2- x^2-d_l^2\right] \nonumber\\
	&\hspace{-1cm}+2 \Theta\big[c_s^2 t^2\big(1-\tfrac{1}{\mathcal{M}^2}\big)-d_l^2 \big] \Theta\big[x-\sqrt{c_s^2 t^2-d_l^2}\big]\Theta\big[-s-d_l\sqrt{(\mathcal{M}^2-1)}\big]\bigg\} 
\end{align}
(where we are considering the Heaviside function to vanish when evaluated over non-real numbers).

\begin{figure}
	\includegraphics[width=1\linewidth]{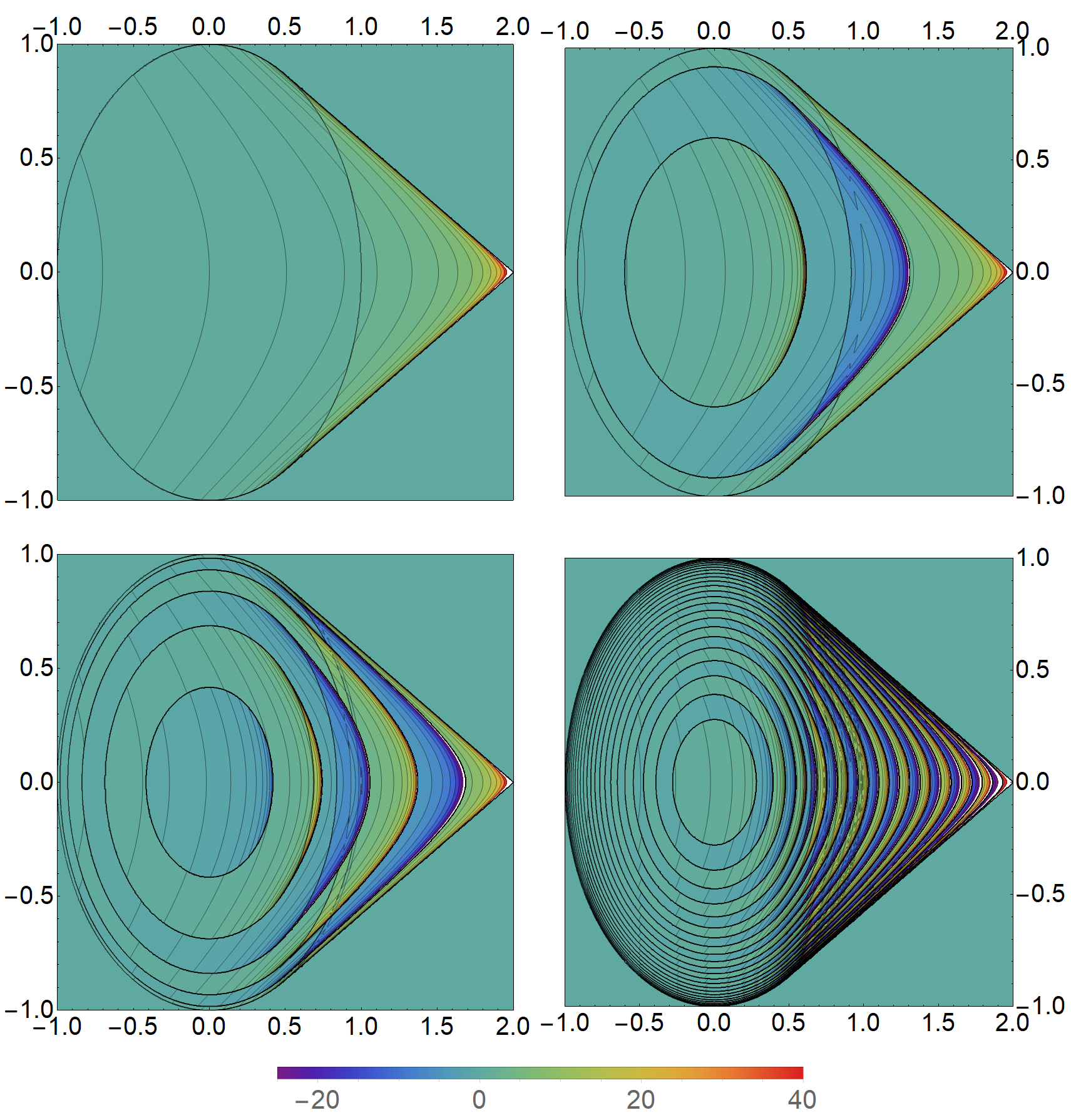}
	\caption{Density perturbation~$\alpha c_s^3 t/M$ in a gaseous slab along $z=0$, due to the gravitational interaction with a supersonic perturber with Mach number~$\mathcal{M}=2$, for~$c_s t/L=0.5,5,11,50$ (left to right, up to down). The horizontal axis represents the coordinate~$x/(c_s t)$ and the vertical axis~$y/(c_s t)$. The contours represent curves of constant density. The observed ripples are echoes of the original density fluctuation. Each ripple is associated with an~$l$-term. There is an infinite-density shock wave with conic shape ($l=0$) and shock wave echoes (coming from other~$l$-terms) located inside the conic surface.}
	\label{fig:slabsupsonicz0D}
\end{figure}

The perturbation in the gas density along the~$z=0$ plane caused by a supersonic particle with Mach number~$\mathcal{M}=2$ is shown in Fig.~\ref{fig:slabsupsonicz0D} at different times. As expected, for early times~$c_s t/L< 1$ all the results are identical to those in infinite media~\cite{Ostriker:1998fa}.~\footnote{Interestingly, the late-time results for the perturbed density profile in a three-dimensional slab with Neumann conditions mimic those obtained in a truly two-dimensional setting (\ie, with the gravitational force falling with~$\sim 1/r$, instead of the usual~$\sim 1/r^2$) in both subsonic and supersonic regimes.} 

In the subsonic regime the density perturbation diverged only \emph{at} the particle location, and surfaces of constant density in the neighborhood of that point were concentric oblate spheroids centered at it, with short-axis along the~$x$ direction (just like in infinite media~\cite{Ostriker:1998fa}). Thus, the front-back symmetry of the density perturbation about the particle suppressed the contribution of this region to the~\acs{DF}, assuring its finiteness~\cite{Ostriker:1998fa,Rephaeli1980ApJ}. That is not the case in the supersonic regime. In fact, it is easy to show that the~\acs{DF} acting on a supersonic point particle is infinite.
Thus, a regularization procedure needs to be introduced. we follow the standard, physically motivated, procedure of describing actual sources via an effective size $r_{\text{min}}$~\cite{Ostriker:1998fa}. This introduces a cutoff in the force integral, describing the effective size of the particle and assuring that~\acs{DF} remains finite.

\begin{figure}
	\includegraphics[width=1\linewidth]{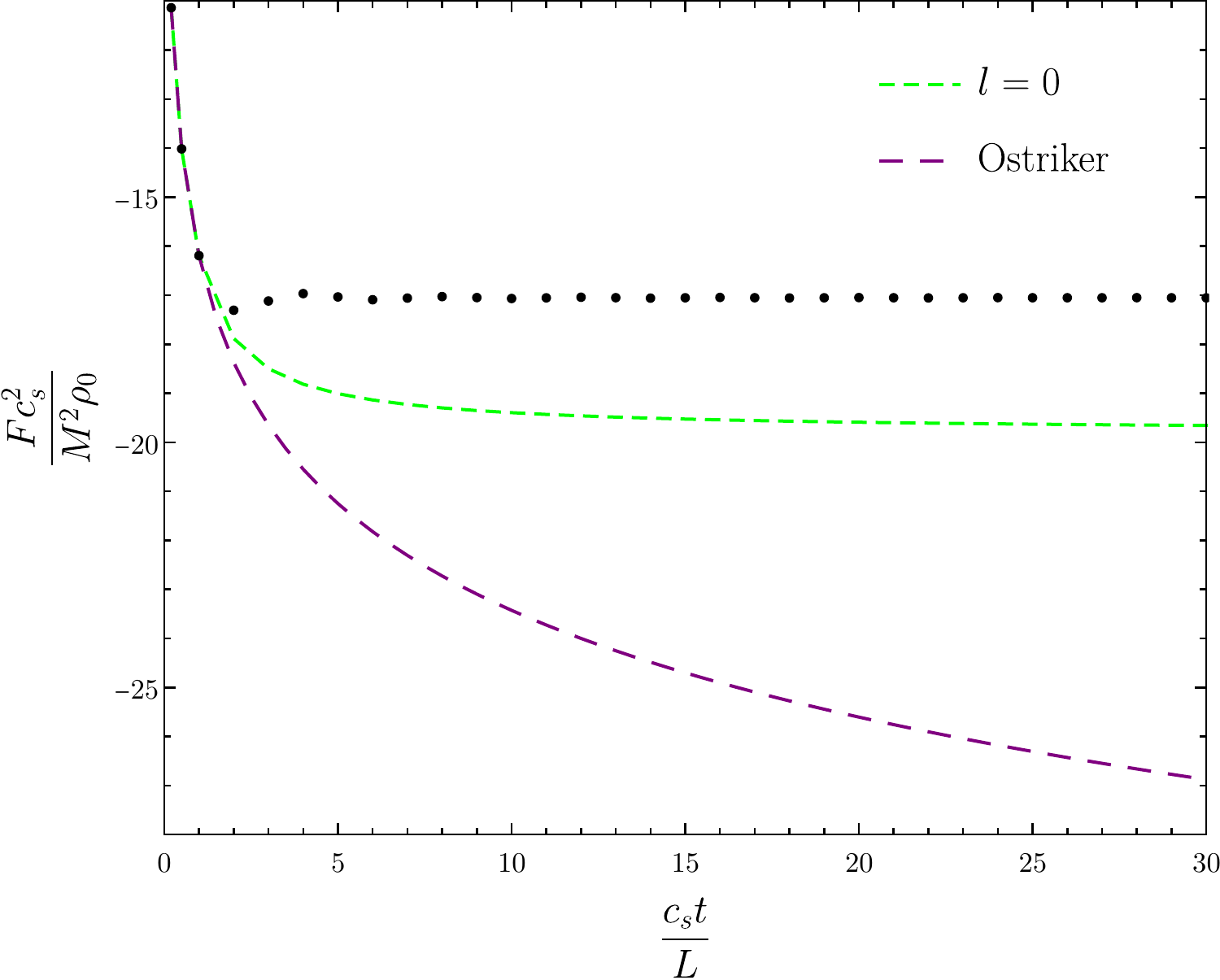} 
	\caption{Time-dependence of the dynamical friction acting on a supersonic particle with Mach number $\mathcal{M}=2$ and size $r_\text{min}/L=10^{-2}$ (black dots). At early times $c_s t/L<1$ the dots are in agreement with Eq.~\eqref{supersonic_ostriker}, which is valid for non-compact mediums (purple dashed curve). At late times $c_s t/L \gg 1$, the numerical results are well approximated by Eq.~\eqref{slabDragforcesupappD}. In green, it is shown the ($l=0$) contribution from the non-reflected wake.}
	\label{fig:slabsupsonicforcetm2D}
\end{figure}

\begin{figure}
	\includegraphics[width=1\linewidth]{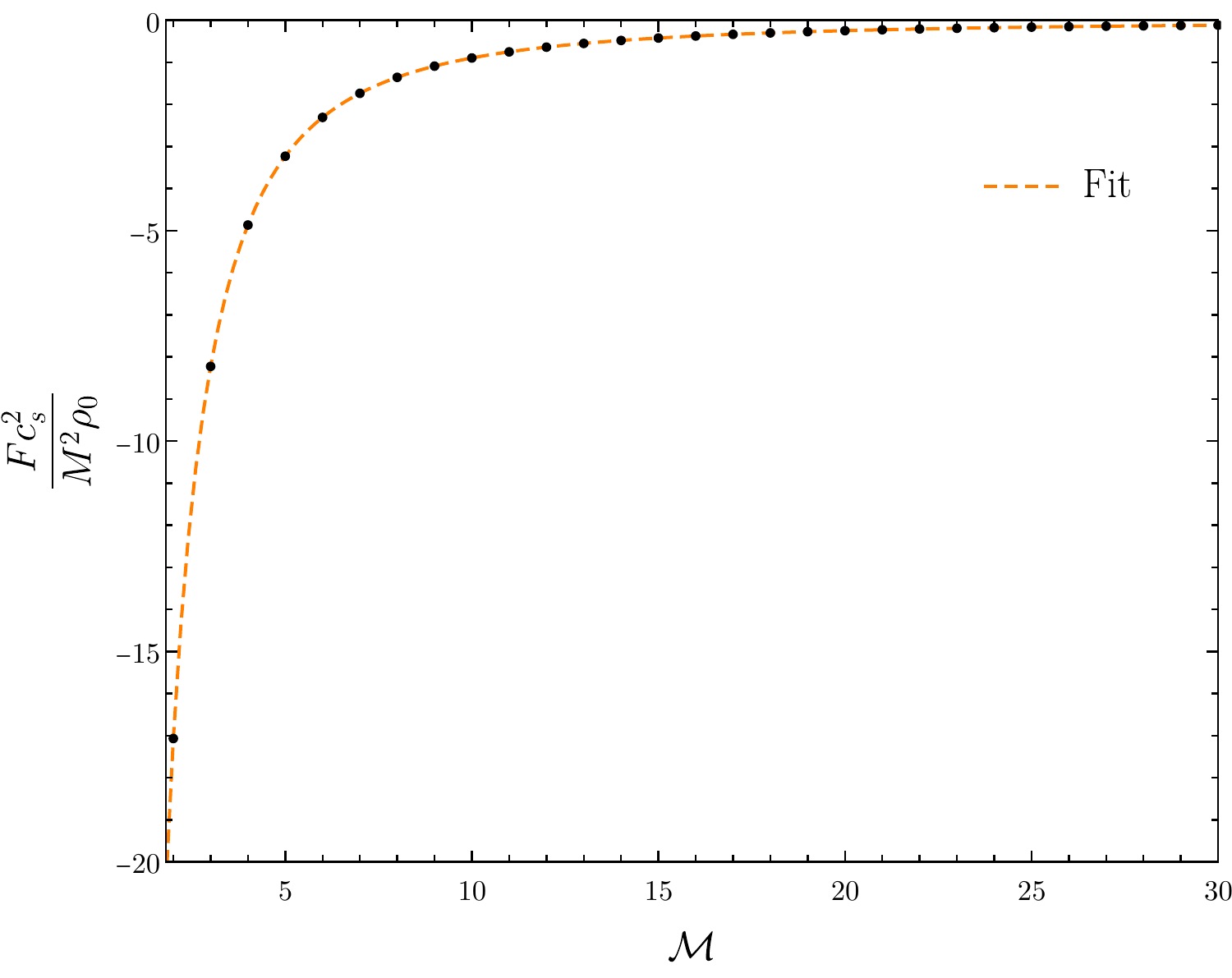} 
	\caption{Late-time dynamical friction acting on a supersonic particle with finite size~$r_\text{min}/L=10^{-2}$ (black dots). The results are well approximated by the fit expression~\eqref{slabDragforcesupappD} for~$\mathcal{M}>2$ (orange dashed curve).}
	\label{fig:slabsupsonicforcelateD}
\end{figure}

Figure~\ref{fig:slabsupsonicforcetm2D} shows the time-dependence of the drag force for a fixed Mach number~$\mathcal{M}=2$ and effective size~$r_\text{min}=10^{-2}L$. At early times~$c_s t/L<1$ we find a friction identical to that computed in infinite three-dimensional gaseous media~\cite{Ostriker:1998fa}.
Surprisingly, at late times~$c_s t/L\gg 1$ the~\acs{DF} in a slab with Dirichlet conditions is time-independent. The numerical results indicate (see Fig.~\ref{fig:slabsupsonicforcelateD}) that this late-time~\acs{DF} is well approximated by
\begin{equation} \label{slabDragforcesupappD}
F\simeq -\frac{M^2 \rho_0}{c_s^2} \left[\mathcal{D}+ \frac{4\pi}{\mathcal{M}^2} \log\left(\frac{L}{r_\text{min}}\right)\right] \,,
\end{equation} 
with $\mathcal{D}\equiv (21.17 \mathcal{M}^{0.83} -22.05)/\mathcal{M}^{2.58}$,
for Mach number $\mathcal{M}>2$. The magnitude of the friction is larger when the size of the particle is smaller, but it is a very mild, logarithmic, dependence. For fixed Mach number~$\mathcal{M}$, the second term in the expression above is dominant for a sufficiently small perturber~$L/r_\text{min}\gg 1$.

For a slab with Neumann conditions the numerical results show that, at late times~$c_s t/L \gg 1$, the~\acs{DF} is well approximated by
\begin{equation}
F\simeq -\frac{M^2 \rho_0}{c_s^2} \left[\mathcal{J}+ \frac{4 \pi}{\mathcal{M}^2} \log\left(\mathcal{M} \frac{c_s t}{r_\text{min}}\right)\right]\,,
\end{equation}
where $\mathcal{J}\sim 1$ is a function of $\mathcal{M}$.
Notice that this is the same late-time ($c_s t/r_{\text{min}}\gg 1$) behavior of the \acs{DF} in an infinite three-dimensional medium in the supersonic regime~\eqref{supersonic_ostriker} which was obtained by~\citeauthor{Ostriker:1998fa}~\cite{Ostriker:1998fa}. Thus, interestingly, in a slab with Neumann boundary conditions both the early- and late-time \acs{DF} have the same behavior as in non-compact geometries, in the supersonic regime.  

As in the subsonic regime, the~\acs{DF} acting on a particle moving at supersonic speed in a three-dimensional slab medium with Dirichlet (Neumann) conditions is always smaller (larger) in magnitude than the one in an infinite three-dimensional medium.

\section{Wake reflections in stratified media} \label{sec:wakerefl}
The results of the previous section show that the late-time~\acs{DF} is strongly dependent on the boundary conditions of the slab.
Since the reflections (echoes) of the wake play an crucial role in this calculation, it is important to understand if these reflections are also present in a more realistic, vertically stratified, open medium. Otherwise, the conclusions obtained with this simple setup cannot be extrapolated to more realistic astrophysical setups. In this section, we show that, indeed, wake reflections are also present in open media, provided that their density falls off sufficiently fast in the vertical direction (compared with the length scale over which the density is nearly constant). {To show that we consider a medium constituted by an homogeneous slab part and a stratified edge. Note, however, that this is still meant to be a toy model; in real astrophysical disks, one may not be able to distinguish between a bulk and an edge part.

Let me start by considering a vertically stratified isothermal gaseous medium with unperturbed density~$\rho_0(z)$. Here, we focus on the $z$-direction (one-dimensional) dynamics. So, the linearized equation describing the relative perturbed density is
\begin{equation} \label{EOMstrat}
\frac{\partial^2}{\partial z^2} \alpha-\frac{1}{c_s^2}\frac{\partial^2}{\partial t^2}\alpha + \left(\frac{\partial}{\partial z} \log \rho_0\right) \frac{\partial}{\partial z}\alpha=0\,.
\end{equation}
%%%
Defining $\bar{\alpha}\equiv \alpha(z,t) e^{-k(z)}$ with
\begin{equation*}
k\equiv-\frac{1}{2}\log \left(\frac{\rho_0(z)}{\rho_0(0)}\right)\,,
\end{equation*}
Eq.~\eqref{EOMstrat} gives
\begin{equation} \label{EOMstrat2}
\frac{\partial^2}{\partial z^2} \bar{\alpha}-\frac{1}{c_s^2}\frac{\partial^2}{\partial t^2}\bar{\alpha}+ \left[k''-\left(k'\right)^2\right]\bar{\alpha}=0\,,
\end{equation}
where $k'$ denotes the derivative of $k$ with respect to $z$. Now, one can write $\bar{\alpha}$ as the Fourier-integral 
\begin{equation}
\bar{\alpha}=\int d\omega\, \bar{\alpha}_\omega(z) e^{-i \omega t} \,,
\end{equation} 
which after substitution in Eq.~\eqref{EOMstrat2} gives
\begin{equation}
\frac{\partial^2}{\partial z^2} \bar{\alpha}_\omega+q_\omega(z) \bar{\alpha}_\omega=0\,,
\end{equation}
with
\begin{equation}
q_\omega \equiv \left(\frac{\omega}{c_s}\right)^2+k''-\left(k'\right)^2\,.
\end{equation}
By looking at the sign of~$q_\omega$, one can identify the regions where~$\omega$-mode fluctuations of the medium density propagate and the ones where they evanesce: propagation happens in regions with positive~$q_\omega$ and evanescence in regions with negative~$q_\omega$~\cite{Kumar}.

As an example consider the unperturbed density profile~\footnote{Throughout this section, the slab edge spans $z>0$; this contrasts with the treatment in the previous section where the boundary was at $z=L$.}
\begin{equation}\label{expsetup}
\rho_0=\rho_0(0)\left[1+(e^{-z/h}-1)\Theta(z)\right]\,,
\end{equation}
with~$h$ the effective width of the medium's edge and $\Theta(z)$ the Heaviside step function. This density profile gives
\begin{equation}
q_\omega=\left(\frac{\omega}{c_s}\right)^2-\Theta(z) \left(\frac{1}{2 h}\right)^2\,.
\end{equation}
We see that any $\omega$-mode can propagate in $z<0$, whereas only the $|\omega|>c_s/(2 h)$ modes can propagate in $z>0$. In other words: an $\omega$-mode coming from $z<0$ and propagating in the positive $z$-direction gets totally reflected at $z=0$, iff $|\omega|\leq c_s/(2 h)$; otherwise, the $\omega$-mode is partially reflected and transmitted. The frequency $\omega=c_s/(2 h)$ is often called \emph{cutoff frequency}~\cite{lamb1945hydrodynamics}.

The gravitational wake produced by a perturber moving at constant velocity can be modeled by a real-valued wave packet with spatial width $\delta z \sim 2 L$, where $2 L$ is the effective thickness of the medium. Thus, its Fourier-transform in~$z$ is centered at~$\omega/c_s=0$ and has width~$\delta \omega /c_s \sim 1/\delta z\sim 1/(2L)$. So, the frequency content of a gravitational wake produced in slab-like media is~$\delta \omega \sim c_s/(2L)$.~\footnote{In other words: the only time scale in this system is~$2 L/c_s$. So, it is natural to expect the frequency content of the wake to be~$\delta \omega \sim c_s/(2 L)$.} Thus, if $h\ll L$ the wake is totally reflected at $z=0$. In that case, in what concerns wake reflections this stratified medium is well modeled by an homogeneous medium ($z<h$) with Dirichlet conditions at a~$z=h$ (cutoff) boundary.~\footnote{Although there is no propagation in $z>0$, the stratified edge introduces a phase shift in the wave packet. So, in order to take this effect into account, we need to choose $z=h$ (and not~$z=0$) as the cutoff boundary.}

Figure~\ref{fig:expedge} shows the results for the time-evolutions of (i) a wave packet propagating in the stratified setup~\eqref{expsetup}, with (open) radiation boundary conditions; and (ii) a wave packet propagating in an homogeneous medium with Dirichlet conditions at~$z=h$. The results are in accordance with the predictions in the previous paragraph.

\begin{figure}
	\begin{tabular}{c}
		\includegraphics[width=0.95\linewidth]{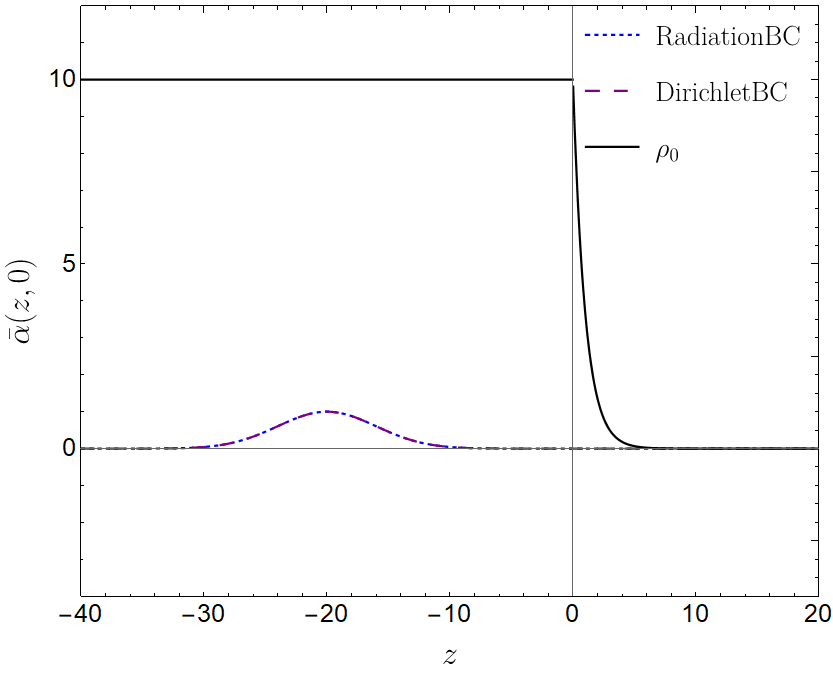} \\
		\includegraphics[width=0.95\linewidth]{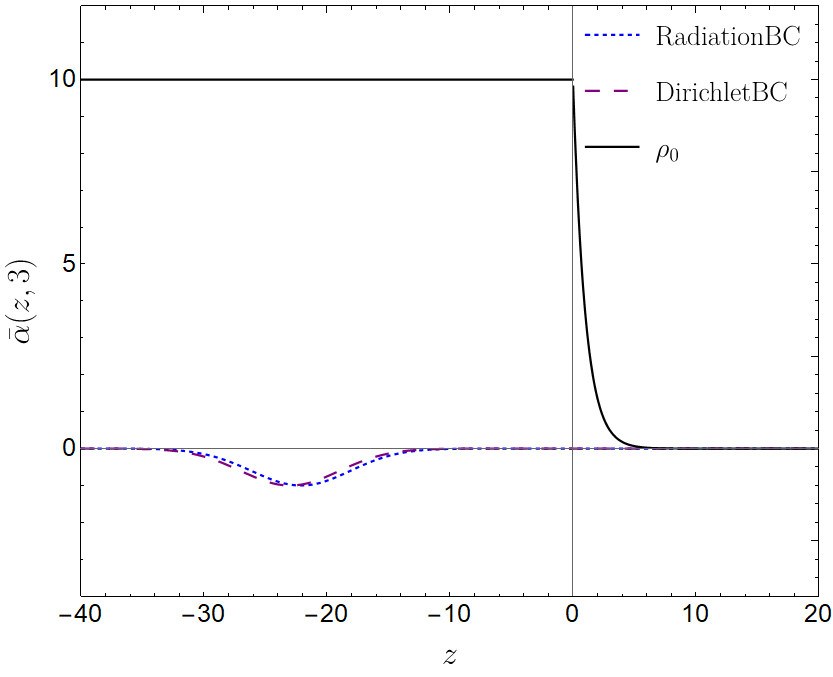} 
	\end{tabular}
	\caption{Time-evolution of a wave packet with initial conditions $\bar{\alpha}(z,0)=e^{-\frac{1}{2}\left(\frac{z+20}{4}\right)^2}$ and $\partial_t \bar{\alpha}(z,0)=-c_s\, \partial_z \bar{\alpha}(z,0)$ using Eq.~\eqref{EOMstrat2}. \textbf{Blue:} wave packet propagating in the stratified setup~\eqref{expsetup} (also represented in the figure), with radiation boundary conditions; \textbf{Purple:} wave packet propagating in an homogeneous medium with Dirichlet conditions at $z=h$. \textbf{Above:} initial (incident) wave packets propagating from left to right; \textbf{Below:} reflected wave packets propagating from right to left. 
	The parameters used were~$c_s=15$,~$\rho_0(0)=10$ and~$h=1$.}
	\label{fig:expedge}
\end{figure}

Finally, let me consider an additional example. For a disk edge in isothermal equilibrium the unperturbed density is~\cite{shakura}
\begin{equation}\label{gausssetup}
\rho_0=\rho_0(0)\left[1+\left(e^{-\frac{1}{2}\left(\frac{z}{h}\right)^2}-1\right)\Theta(z)\right]\,.
\end{equation}
This profile gives
\begin{equation}
q_\omega=\left(\frac{\omega}{c_s}\right)^2-\Theta(z) \left(\frac{z^2}{4 h^4}- \frac{1}{2h^2}\right)\,.
\end{equation}
We see that each $\omega$-mode can only propagate in the region
\begin{equation*}
z<z_\omega \equiv h \sqrt{2+\left(\frac{2 h \omega}{c_s}\right)^2}\,,
\end{equation*}
being evanescent in~$z>z_\omega$. Again, the gravitational wave frequency content is~$\delta \omega \sim c_s/(2L)$. So, if the edges are sufficiently thin ($h\ll L$), then~$z_\omega \sim \sqrt{2} h$ for all frequencies composing the wave packet. In other words: the whole packet is totally reflected at $z=\sqrt{2}\,h$. Thus, once again, concerning the wake reflections this stratified medium is well modeled by an homogeneous medium ($z<h$) with Dirichlet conditions at a~$z=h$ (cutoff) boundary.~\footnote{As in the last example, we need to choose $z=h$ (and not~$z=\sqrt{2} h$) as the cutoff boundary to account for the correct phase shift introduced by the reflection.}

\begin{figure}
	\begin{tabular}{c}
		\includegraphics[width=0.95\linewidth]{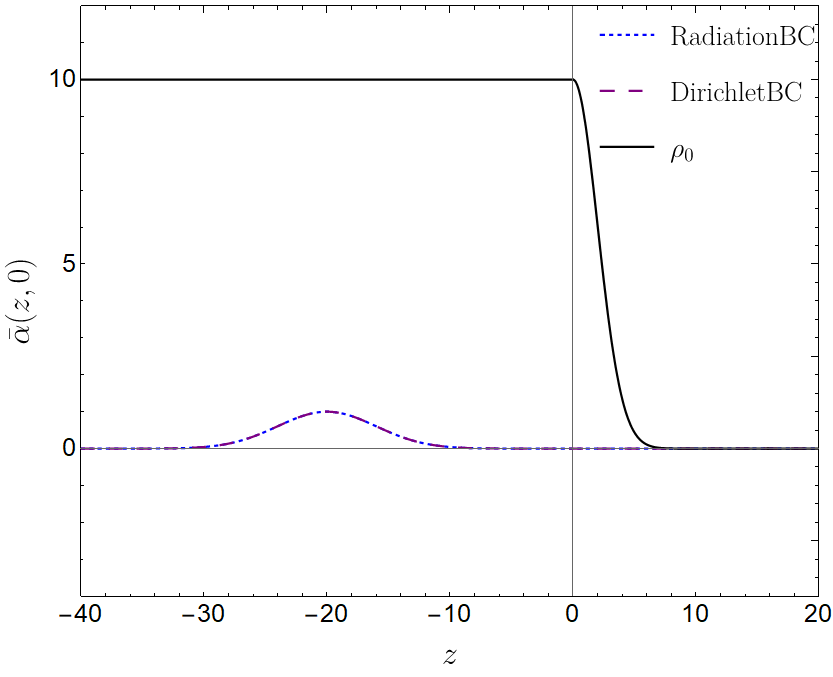} \\
		\includegraphics[width=0.95\linewidth]{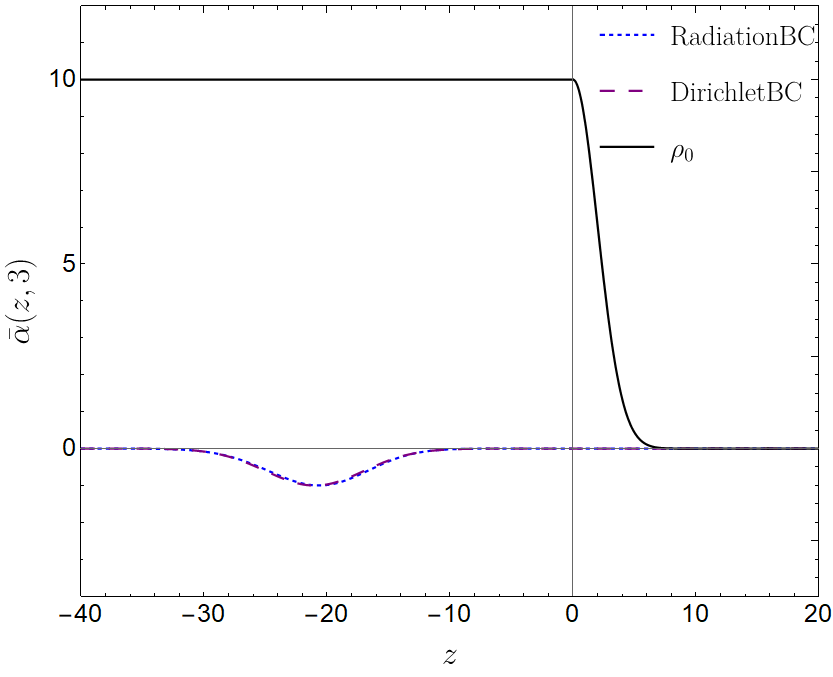} 
	\end{tabular}
	\caption{Time-evolution of a wave packet with initial conditions~$\bar{\alpha}(z,0)=e^{-\frac{1}{2}\left(\frac{z+20}{4}\right)^2}$ and~$\partial_t \bar{\alpha}(z,0)=-c_s\, \partial_z \bar{\alpha}(z,0)$ using Eq.~\eqref{EOMstrat2}. \textbf{Blue:} wave packet propagating in the stratified setup~\eqref{gausssetup} (also represented in the figure), with radiation boundary conditions; \textbf{Purple:} wave packet propagating in an homogeneous medium with Dirichlet conditions at $z=h$. \textbf{Above:} initial (incident) wave packets propagating from left to right; \textbf{Below:} reflected wave packets propagating from right to left. 
	The parameters used were~$c_s=15$,~$\rho_0(0)=10$ and~$h=2$.}
	\label{fig:gaussedge}
\end{figure}

The numerical results also show that the boundary conditions that a physically realistic slab-like medium satisfies are indeed of the Dirichlet (reflection with inversion) type. Had we used Neumann (reflection without inversion) conditions for the time-evolutions, the reflected wave packets would be inverted with respect to the ones in (realistic) stratified media. 

\section{Discussion}\label{sec:conclusions}
In this chapter we computed the gravitational wake due to and the~\acs{DF} acting on a massive particle moving in a straight line through a three-dimensional slab, taking into account reflections of the wake on the boundaries.

I want to highlight that~\citeauthor{namouni}~\cite{namouni} also studied the time-dependence of~\acs{DF} in compact homogeneous media. In particular, the effect of wake reflections on the boundaries was investigated. However, only one wake reflection was considered, and, though not explicitly stated, Neumann boundary conditions were used. As explained before and showed in the previous section, more realistic boundary conditions are of Dirichlet type. Thus, an important conclusion was missed in this previous work: that, generically, wake reflections tend to suppress~\acs{DF}.

It is worth pointing out that an estimate for the steady \acs{DF} acting on a particle moving in a straight line through a very thin disk was made previously by~\citeauthor{Muto:2011qv}~\cite{Muto:2011qv} using a two-dimensional approximation to describe the disk. This approximation is very good in describing the contribution to friction coming from perturbations far from the perturber, which have already felt the slab boundaries. In very thin disks the dominant contribution to the subsonic~\acs{DF} comes from far regions and the approximation is expected to hold in that regime~\cite{Muto:2011qv}. For an inviscid medium (as in this chapter),~\citeauthor{Muto:2011qv} estimated~\acs{DF} to be suppressed with $1/t$. This is in very good agreement with my results (see Eq.~\eqref{fitsubD}). However, the approximation in Ref.~\cite{Muto:2011qv} fails to describe the contribution to~\acs{DF} from the near region, which is the dominant one in the supersonic regime. Nevertheless, though not succeeding in obtaining the correct dependence on~$L/r_\text{min}$, the authors estimated the supersonic late-time drag to be steady and proportional to~$1/\mathcal{M}^2$, which is in agreement with my results for~$L/r_\text{min}\gg1$ (see Eq.~\eqref{slabDragforcesupappD}). In that case (sufficiently small perturber) we recover the well-known estimates for the steady supersonic~\acs{DF} in a three-dimensional medium with effective size $L$, both in collisional media~\cite{Dokuchaev1964,Ruderman1971,Rephaeli1980ApJ,canto} and collisionless media~\cite{Tremaine1987}.~\footnote{In the case of collisionless media there is no notion of sound speed. The analogous regime to the supersonic motion is when the perturber has a velocity much larger than the particle dispersion velocity of the medium~\cite{Ostriker:1998fa}.} Again, this is due to the fact that in the supersonic regime the dominant contribution to~\acs{DF} comes from the near region. So, one does not expect the wake reflections to play an important role in the friction; the more so for a very small particle.

I expect the results of this chapter to be important to the study of the physics of accretion and protoplanetary disks. There is a substantial body of theoretical and numerical studies on the disk-planet gravitational interaction \cite{Tremaine79,Tremaine80,Ward86,Tanaka_2002,Muto:2011qv,Stone2018}. However, in most of them two oversimplifications are used: (i) the disks are assumed to be very thin and a two-dimensional approximation is used to treat the medium; (ii) the gravitational wake is assumed to be completely dissipated at the boundaries, without any reflection. A full three-dimensional treatment of the gravitational interaction between a planet and a disk not assuming (i), but maintaining assumption (ii) finds the following~\cite{Tanaka_2002}: the migration time of an Earth-sized planet at $5$AU is of the order of $10^6$yr, which is $2$ or $3$ times longer than previously obtained results using the two-dimensional approximation~\cite{Hayashi}. Their result is very relevant: since the formation time of a giant planet at $5$AU is of the order of $10^6$yr~\cite{Tanaka}, the planetary migration must happen in a longer, or at least comparable, timescale to explain the existence of giant planets. In that same work the authors also suggested that the reflection of the gravitational wake on the disk edges (which they neglected) could weaken even more the disk-planet interaction, increasing the planet migration time. The results of this chapter clearly support their intuition in the subsonic regime, where~\acs{DF} is strongly suppressed (see Fig.~\ref{fig:slabsubsonicforcetD}). In the supersonic regime the $l=0$ term, which is not sensitive to the boundaries, accounts for most of the late-time~\acs{DF}. Thus, even though the drag is also slightly suppressed in the supersonic regime, I do not expect the effect of wake reflections to be as striking as in the subsonic regime. 

One can argue that all the results derived here assume linear motion and cannot, formally, be applied in setups involving circular motion. Despite this being true,~\citeauthor{Kim:2007zb}~\cite{Kim:2007zb} obtained the remarkable result that the~\acs{DF} formulae derived for linear motion in extended media by~\citeauthor{Ostriker:1998fa}~\cite{Ostriker:1998fa} give reasonably good estimates for the drag acting on circular-orbit perturbers. I expect the same to happen here. In fact, the approach of Refs.~\cite{Kim:2007zb,Kim:2008ab} to extend the~\acs{DF} formulae derived in Ref.~\cite{Ostriker:1998fa} from linear motion to circular-orbit and binary motion, respectively, can, in principle, be applied in a straightforward way to extend results of this chapter to those same motions.

The unbounded-medium approximation derived by~\citeauthor{Kim:2007zb} \cite{Kim:2007zb} was used recently to estimate the impact of~\acs{DF} in thin accretion disks on gravitational-wave observables~\cite{Barausse:2014tra}. It was concluded that~\acs{DF} may indeed be important and lead to degradation of gravitational-wave templates for detection. In a physically realistic setup the accretion disk height is~$L\sim  r c_s/v_K$, where $r$ is the distance from the disk center and $v_K\equiv (M/r)^{1/2}$ is the local Keplerian velocity at which the perturber is moving in its circular-orbit motion. In Ref.~\cite{Barausse:2014tra} the authors assumed that the relative velocity of the perturber with respect to the disk is~$\mathcal{M}\simeq v_K/c_s \sim r/L$, and so, for a thin accretion disk $r/L\gg 1$, the motion is supersonic. Then, as discussed above, the dominant contribution to~\acs{DF} comes from the region near the perturber. So, even though the toy model considered in this chapter neglects variations of $c_s$ and $L$ (which are present in realistic disks), I still expect it to describe appropriately the present setup. Moreover, the sound travel time to the disk edges is of the same order of the orbital-motion period (\ie, $c_s/L \sim v_K/r$). Thus, the boundary effects of the disk may be relevant for~\acs{DF} in thin accretion disks (by suppressing it) and can, possibly, change the conclusion of Ref.~\cite{Barausse:2014tra}.  

%*****************************************
%*****************************************
%*****************************************
%*****************************************
%*****************************************

%************************************************
\chapter{Eccentricity evolution of compact binaries}\label{ch:eccentricity}
%************************************************

\def\be{\begin{equation}}
\def\ee{\end{equation}}
\def\beq{\begin{align}}
\def\eeq{\end{align}}

Merging black hole binaries (\acsp{BHB}) are now ``visible'', thanks to gravitational wave (\acs{GW}) astronomy~\cite{Abbott:2016blz,Barack:2018yly}.
A good modeling of the dynamics of such compact binaries is important to increase our ability to actually see them,
to infer the properties of the merging objects and to impose constraints on the underlying gravitational theory, or other fundamental interactions~\cite{Barack:2018yly}.

It has long been known that orbits which are initially eccentric will quickly circularize on relatively short timescales~\cite{Peters:1964zz,Krolak:1987ofj,Key:2010tc}.
This is true \textit{in vacuum}, and thought to describe well stellar mass \acsp{BHB}, which form substantially prior to merger and evolve mostly only via \acs{GW} emission.
However, a re-appreciation of eccentricity evolution is required for different reasons. To begin with, the formation of supermassive \acsp{BHB} is poorly understood. Some of the mechanisms that contribute to such binaries forming and merging actually may also impart a substantial eccentricity,
specially in their initial stages~\cite{Barack:2018yly}. In addition, observations are progressively indicating that large eccentricities may not be rare. One known supermassive BHB (\textit{OJ287}) was reported to have eccentricity $e\sim 0.65$, while evolving around the disk of the massive component~\cite{Laine:2020dnr}. Such observations were made in the electromagnetic spectrum, but there are indications that some of the \acs{GW} events, such as \textit{GW190521}~\cite{Abbott:2020tfl,Abbott:2020mjq} could also originate from eccentric orbits~\cite{Gayathri:2020coq,CalderonBustillo:2020odh}. It is interesting to note that this same event may have an associated electromagnetic counterpart, product of a nontrivial surrounding environment~\cite{Graham:2020gwr}.
A nontrivial environment leads to large center-of-mass drift velocities~\cite{Cardoso:2020lxx} and may lead to large eccentricities during evolution. Even in vacuum, spin-spin couplings at the second post-Newtonian order may induce a nontrivial eccentricity evolution~\cite{Gergely:1998sr,Klein:2010ti,Klein:2018ybm,Phukon:2019gfh}.

The understanding of eccentricity evolution is also important to constrain the presence of new fields. Under the assumption of circular motion, it has been shown that \acs{GW} observations can impose severe limits on the dipolar moment and charge of the inspiralling objects~\cite{Barausse:2016eii,Cardoso:2016olt}. When the binary components are charged under new fields, emission
in such channels dominates over \acs{GW} emission at sufficiently low frequencies; hence the assumption that circular remains circular (\ie, that radiative processes conspire to circularize the orbit) must be
proved. The purpose of this chapter is precisely to address the issues above.

%%%%%%%%%%%%%%%%%%%%%%%%%%%%%%%%%%%%%%%%%%%%%%%%%%%%%%%%%%%%%%
\section{Evolution driven by fundamental fields\label{sec:ff}}
%%%%%%%%%%%%%%%%%%%%%%%%%%%%%%%%%%%%%%%%%%%%%%%%%%%%%%%%%%%%%%
The problem of eccentricity and orbital radius evolution is tightly connected to the ratio of energy to angular momentum loss during the binary evolution.
Take a compact binary of two
objects of mass $m_1, m_2$, and define the total mass and mass ratio
\begin{align}
M\equiv m_1+m_2\,,\qquad q=\frac{m_2}{m_1}\,.
\end{align}
For binaries dominated by the gravitational interaction, the (Newtonian) orbital frequency $\omega_0$ satisfies Kepler's law
\begin{align}
\omega_0=\sqrt{\frac{GM}{a^3}}\,,
\end{align}
where $a$ is the orbital semi-major axis.
In this case, the conserved energy and angular momentum on Keplerian motion are
\begin{align}
E&=-\frac{Gm_1m_2}{2a}\,,\label{Kepler_Energy}\\
L^2&=\frac{Gm_1^2m_2^2a(1-e^2)}{M}\,, \label{Kepler_AngMom}
\end{align}
where $e$ is the eccentricity. 

Suppose now that the only decay channel available for the binary evolution is a 
(possibly massive) field of frequency $\omega$ and azimuthal dependence $e^{im\phi}$. This could be a \acs{GW}, but could include also a scalar or even a vector field. In this circumstance, then the emitted angular momentum and energy satisfy~\cite{Brito:2015oca}
\begin{align}
\frac{\dot{L}^{\textrm{rad}}}{\dot{E}^{\textrm rad}}=\frac{m}{\omega}=\frac{1}{\omega_0}\,.\label{special_massless} 
\end{align}
How do the eccentricity and semi-major axis of the binary
evolve? Energy and angular momentum balance yield
\begin{align}
\dot{E}=-\dot{E}^{\textrm{rad}}\leq 0 \,, \quad \dot{L}=-\dot{L}^{\textrm{rad}}\,,
\end{align}
so we find
\begin{align}
\dot{a}&=-\frac{2a^2\dot{E}^{\textrm{rad}}}{Gm_1m_2} \leq 0\,,\\
\dot{e}&=\sqrt{\frac{M}{Ga}}\frac{\sqrt{1-e^2}}{e}\frac{\dot{E}^{\textrm {rad}}}{m_1m_2}\left(\frac{\dot{L}^{\textrm{rad}}}{\dot{E}^{\textrm {rad}}}-\frac{\sqrt{1-e^2}}{\omega_0}\right)\,.
\end{align}
We see immediately that, if~$\dot{L}^{\textrm{rad}}/\dot{E}^{\textrm{rad}}$ have eccentricity-dependence starting at order higher than $e^2$, then circular orbits are unstable (\textit{i.e.} $\dot{e}\geq0$ for~$e\sim 0$) on account
of condition \eqref{special_massless}. In case of~$\dot{L}^{\textrm {rad}}/\dot{E}^{\textrm{rad}}$ having eccentricity-dependence starting at order~$e^2$, circular orbits will also be unstable if the coefficient multiplying~$e^2$ is larger than~$-\tfrac{1}{2 \omega_0}$.

We therefore start our analysis by asking how does the emission of fundamental massless fields affect eccentricity evolution.~\footnote{This analysis could be extended to massive fields in a straightforward way.}

%%%%%%%%%%%%%%%%%%%%%%%%%%%%%%%%%%%%%%%%%%%%%%%%
\subsection{GW radiation}
%%%%%%%%%%%%%%%%%%%%%%%%%%%%%%%%%%%%%%%%%%%%%%%%

Let us first assume that our system is in vacuum, isolated from all other sources in the universe.
In this case, the evolution is driven solely by \acs{GW} emission.
Eccentricity in vacuum-general relativity can be calculated in a two-step procedure.
Take a binary of point-like objects of mass $m_1, m_2$. To lowest post-Newtonian order,
their motion is elliptical, of semi-major axis $a$ and eccentricity $e$.
Their binding energy $E$ and angular momentum $L$ are simply described by Eqs.~\eqref{Kepler_Energy}-\eqref{Kepler_AngMom}.
Now, when relativistic effects are included, the system radiates energy and angular momentum, via \acsp{GW}, at a rate
\begin{align}
\langle\dot{E}\rangle&=-\frac{32}{5}\frac{G^4m_1^2m_2^2M}{a^5(1-e^2)^{7/2}}\left(1+\frac{73}{24}e^2+\frac{37}{96}e^4\right)\,,\\
\langle\dot{L}\rangle&=-\frac{32}{5}\frac{G^{7/2}m_1^2m_2^2M^{1/2}}{a^{7/2}(1-e^2)^{2}}\left(1+\frac{7}{8}e^2\right)\,.
\end{align}
%
%We see that the energy and angular momentum fluxes have eccentricity dependence which when Taylor expanded begins at order $e^2$, and from the previous discussion we expect circular orbits to be stable.
Assuming a slow, adiabatic evolution, one can now follow \citeauthor{Peters:1964zz}~\cite{Peters:1964zz} and compute the major axis and eccentricity evolution. For small eccentricity, one finds
\begin{align}
\langle\dot{a}\rangle&=-\frac{64 G^3}{5}\frac{m_1m_2M}{a^3}<0\,,\\
\langle\dot{e}\rangle&=-\frac{304G^3}{15}\frac{m_1m_2M}{a^4}\,e \leq 0\,.
\end{align}
In other words, the major axis decreases with time due to energy loss in \acsp{GW}. So does the eccentricity, thus orbits tend to become circular on long timescales.
Note, however, that eccentricity evolution is very sensitive, in particular, it hardly evolves for quasi-circular orbits.
One is thus forced to consider what happens when other physics sets in.

%*****************************************
%*****************************************
%*****************************************
%*****************************************
%*****************************************

%%%%%%%%%%%%%%%%%%%%%%%%%%%%%%%%%%%%%%%%%%%%%%%%%%%%%%%%%%%%%%%%%%%%%
\subsection{Scalar and vector radiation}
%%%%%%%%%%%%%%%%%%%%%%%%%%%%%%%%%%%%%%%%%%%%%%%%%%%%%%%%%%%%%%%%%%%%%
Consider, then, binary components carrying some additional charge. The simplest examples include scalar charge, as is the case in scalar-tensor theories, or 
electromagnetic charge (the theory below also describes some dark matter models with mili-charged components~\cite{Cardoso:2016olt}). We model this via the theory of massless fields~\eqref{theory_action} with currents
\begin{align}
	J_S&= \sum_{n=1}^2 q_n^0  \int d \tau_n \frac{\delta^{(4)}\left(x^\alpha-x_n^{\;\; \alpha}(\tau_n)\right)}{\sqrt{-g}}\,, \\
	J_V^{\;\;\alpha}&=\sum_{n=1}^2 q_n^1 \int d \tau_n \,\frac{d x_n^{\;\;\alpha}}{d \tau_n} \frac{\delta^{(4)}\left(x^\delta-x_n^{\;\; \delta}(\tau_n)\right)}{\sqrt{-g}}\,.
\end{align}
Each of the binary components carries a charge $q^{s}_n$ of the corresponding spin-$s$ field ($s=0,1$ for scalar and vectors, respectively).

The details of the calculation are shown in Appendix~\ref{app:SVrad}. As might be anticipated, in the weak field regime the motion is Keplerian with energy and angular momentum 
\begin{align}
E=-\frac{\tilde{G} m_1m_2}{2a}\,,\qquad L^2=\frac{\tilde{G}m_1^2m_2^2a(1-e^2)}{M}\,,\label{eq:energy_angularmomentum_main}
\end{align}
where the effective Newton's constant is now
\begin{align}
\tilde{G}\equiv G-\frac{1}{4 \pi} \frac{q^s_1 q^s_2}{m_1 m_ 2}\,,
\end{align}
where we assume (without loss of generality) that only one further interaction ($s=0$ \textit{or} $s=1$) is turned on.

In the Newtonian approximation, radiation propagates in flat space and the Green's function for the problem is well known.
Averaging over an orbit, we find the surprisingly compact expressions for the rate of energy and angular momentum emission
\begin{align}
&\langle\dot{E}^{\textrm{rad}}\rangle=\frac{(s+1)}{24\pi}\frac{\tilde{G}^2}{a^4}(q^s_1 m_2-q^s_2 m_1)^2 \left(\frac{2+e^2}{(1-e^2)^{\frac{5}{2}}}\right) \label{EnergySavp} \,, \\
&\langle\dot{L}^{\textrm{rad}}\rangle=\frac{(s+1)}{12 \pi} \frac{\tilde{G}^\frac{3}{2}}{\sqrt{M}a^{\frac{5}{2}}(1-e^2)}(q^s_1 m_2-q^s_2 m_1)^2  \label{AngMomSavp}\,,
\end{align}
resulting in the spin-independent dipolar ratio
\begin{align}
&\frac{\langle\dot{L}^{\textrm{rad}} \rangle}{\langle\dot{E}^{\textrm{rad}} \rangle} = \frac{\sqrt{1-e^2}}{\omega_0} \left(\frac{1-e^2}{1+\frac{e^2}{2}}\right) \,.\label{LEratioSp}
\end{align}
The flux of scalar energy in the circular orbit limit agrees with that of Refs.~\cite{Cardoso:2011xi,Yunes:2011aa,Cardoso:2019nis}.
Our results for the electromagnetic flux of energy and angular momentum agree with those in Refs.~\cite{Christiansen:2020pnv,Liu:2020cds}.
In the adiabatic approximation the major semi-axis and the eccentricity follow
\begin{align}
\langle\dot{a}\rangle&=-\frac{2a^2\langle\dot{E}^{\textrm{rad}}\rangle}{\tilde{G} m_1m_2}<0\,,\label{shrinkp}\\
\langle\dot{e}\rangle&=\sqrt{\frac{M}{ \tilde{G}a}}\frac{\sqrt{1-e^2}}{e}\frac{\langle\dot{E}^{\textrm{rad}}\rangle}{m_1m_2}\left(\frac{\langle\dot{L}^{\textrm{rad}}\rangle}{\langle\dot{E}^{\textrm{rad}}\rangle}-\frac{\sqrt{1-e^2}}{\omega_0}\right) \nonumber \\
&=-\sqrt{\frac{M}{\tilde{G}a}}\left(\frac{1-e^2}{e\, \omega_0}\right)\frac{\langle\dot{E}^{\textrm{rad}}\rangle}{m_1m_2}\left(\frac{3 e^2}{2+e^2}\right)\leq 0\,.\label{circularp}
\end{align}
Thus, the emission of massless radiation by a binary causes the major semi-axis and the eccentricity to decrease in time: the orbit shrinks and circularizes. 
Although we will not explore the subject further, it is important to realize that electromagnetic fields couple strongly to plasmas. Thus, when applied to the Maxwell sector, the previous results should be taken with care~\cite{Cardoso:2020nst}.

%%%%%%%%%%%%%%%%%%%%%%%%%%%%%%%%%%%%%%%%%%%%%%%%%%%%%%%%%%%%%%%%%%%%%%%%%%%%%%%%%%%%%%%%%%%%%%%%%%%%
\section{Eccentricity evolution in constant-density environments}
%%%%%%%%%%%%%%%%%%%%%%%%%%%%%%%%%%%%%%%%%%%%%%%%%%%%%%%%%%%%%%%%%%%%%%%%%%%%%%%%%%%%%%%%%%%%%%%%%%%%

The presence of surrounding dust or plasma affects the above picture in different ways. Binaries, such as the event \textit{GW190521}~\cite{Abbott:2020tfl,Abbott:2020mjq}, may in fact evolve within accretion disks, where the density of the surrounding environment may play an important role.
The presence of matter surrounding a \acs{BHB} will cause accretion to occur~\cite{Bondi:1944jm,Macedo:2013qea,Edgar:2004mk}. A second mechanism at play is \acs{DF}, whereby the moving \acsp{BH} get dragged down by the surrounding matter~\cite{Chandrasekhar:1943v1,Ostriker:1998fa,Vicente:2019ilr,Annulli:2020lyc,Macedo:2013qea}. 

Consider first accretion. We assume that the surrounding medium has constant density. This implies in particular that there is a supply mechanism
that keeps the density constant even as the binary sweeps through and accretes some of the particles.
We neglect here the gravitational potential generated by the accretion disk or surrounding matter; this approximation is expected to be extremely good for \acsp{BHB} close to merger. We focus on Bondi-Hoyle accretion~\cite{Edgar:2004mk}. The mass flux at the horizon is
\begin{align}
\dot{m}_i=4\pi G^2\rho \frac{m_i^2}{(v_i^2+c_s^2)^{3/2}} \label{eq:accretionm}\,,
\end{align}
when the binary components are \acsp{BH}. These are Newtonian formulas, expected to be valid up to factors of order 1 when the binary is non-compact. Here, $v_i$ is the relative velocity between \acs{BH} ``$i$'' and the environment, and $c_s$ is the sound speed in the medium. We will always consider regimes for which $v_i\gg c_s$. Numerical studies indicate that the above description is solid, even in the presence of wake instabilities~\cite{Edgar:2004mk}.

Binaries in a medium are also subject to the gravitational force due to the wakes generated by the moving bodies, as we mentioned. This \acs{DF} depends on the characteristics of the fluid and on the moving bodies. In summary, \acs{DF} can usually be represented by an external force of the type
\begin{align}
\boldsymbol{F}_{{\textrm{d}},i}=-G^2m_i^2\rho I_{\textrm{d}}(v_i)\dot{\boldsymbol{r}}_i\,,
\end{align}
where the form of the function $I_{\textrm{d}}$ depends on the specifics of the \acs{DF} model at hand.
We consider the~\acs{DF} in a fluid (collisional) medium in the supersonic regime ($v_i\gg c_s$), for which~\cite{Dokuchaev1964,Ruderman1971,Rephaeli1980ApJ,Ostriker:1998fa}~\footnote{This expression assumes linear motion in an extended medium. The fact that the binary components do not follow a linear motion and are inside a (possibly thin) disk introduces some modifications to the \acs{DF}, which we neglect here for simplicity. For a more careful analysis of the \acs{DF} in these type of systems see~\textit{e.g.} Refs.~\cite{Kim:2008ab,Antoni:2019pgq,Vicente:2019ilr}.}
\begin{align}
I_{\textrm{d}}(v_i)= \frac{4\pi\lambda}{v_i^3}\,,
\end{align}
where $\lambda$ is the Coulomb logarithm. It is easy to see that, for large velocities, the Chandrasekhar formula for collisionless media~\cite{Chandrasekhar:1943v1} reduces also to the last expression. 
We adopt $\lambda\sim20$, unless stated otherwise, but note that changing $\lambda$ is equivalent to re-normalizing the density in the \acs{DF} expression. As we show below, even a factor 10 variation in this parameter has only a mild effect on the overall evolution of the system.

Taking then a binary evolving under the influence of accretion and \acs{DF}, the \acs{EOM} can be written as
\begin{align}
m_i \ddot{\boldsymbol{r}}_i+\dot{m}_i \dot{\boldsymbol{r}}_i=\pm\frac{G m_1m_2}{r^3}\boldsymbol{r}+\boldsymbol{F}_{{\textrm{d}},i}\,,\label{eq:r1r2}
\end{align}
where $\boldsymbol{r}=\boldsymbol{r}_2-\boldsymbol{r}_1$ is the orbital separation vector of the binary. Introducing the center of mass of the binary
\begin{equation}
\boldsymbol{R}=\frac{m_1\boldsymbol{r}_1+m_2\boldsymbol{r}_2}{m_1+m_2}\,,
\end{equation}
we can write a system of equations describing the vectors $\boldsymbol{r}$ and $\boldsymbol{R}$, namely
\begin{align}
\ddot{\boldsymbol{r}}&=f_{1}\dot{\boldsymbol{r}}+f_{2}\dot{\boldsymbol{R}}+f_{3}{\boldsymbol{r}}\,,\label{eq:eqr}\\
\ddot{\boldsymbol{R}}&=f_{4}\dot{\boldsymbol{r}}+f_{5}\dot{\boldsymbol{R}}+f_{6}{\boldsymbol{r}}\,,\label{eq:eqR}
\end{align}
where the functions $f_i$ are given by
\begin{align}
	f_1&=-\frac{G^2 M q \rho  (I_{\text{a1}}+I_{\text{a2}}+I_{\text{d1}}+I_{\text{d2}})}{(q+1)^2}\,,\\
	f_2&=\frac{G^2 M \rho  [I_{\text{a1}}+I_{\text{d1}}-q (I_{\text{a2}}+I_{\text{d2}})]}{q+1}\,,\\
	f_3&=G M \left\{\frac{G^3 M q \rho ^2 (I_{\text{a1}}-q I_{\text{a2}}) [I_{\text{a1}}+I_{\text{d1}}-q
		(I_{\text{a2}}+I_{\text{d2}})]}{(q+1)^4}-\frac{1}{r^3}\right\}\,,\\
	f_4&=\frac{G^2 M q \rho  [q (I_{\text{a2}}-I_{\text{d2}})-I_{\text{a1}}+I_{\text{d1}}]}{(q+1)^3}\,,\\
	f_5&=-\frac{G^2 M \rho  \left[q^2 (I_{\text{a2}}+I_{\text{d2}})+I_{\text{a1}}+I_{\text{d1}}\right]}{(q+1)^2}\,,\\
	f_6&=-\frac{G^4 M^2 q \rho ^2 (I_{\text{a1}}-q I_{\text{a2}}) \left[q^2 (I_{\text{a2}}+I_{\text{d2}})+2 q
		(I_{\text{a1}}+I_{\text{a2}})+I_{\text{a1}}+I_{\text{d1}}\right]}{(q+1)^5}\,.
\end{align}
Here, we defined
\begin{equation}
I_{\text{a}i}=\frac{4\pi}{(v_i^2+c_s^2)^{3/2}},~I_{\text{d}i}=I_\text{d}(v_i)\,.
\end{equation}
Note that due to accretion, both the mass-ratio and the total mass evolve in time. We can compute their evolution via Eq.~\eqref{eq:accretionm}, obtaining
\begin{align}
\dot{q}&=\frac{G^2 M q \rho  (q I_{\text{a2}}-I_{\text{a1}})}{q+1},\label{eq:q}\\
\dot{M}&=\frac{G^2 M^2 \rho  \left(q^2 I_{\text{a2}}+I_{\text{a1}}\right)}{(q+1)^2}.\label{eq:M}
\end{align}

To investigate the evolution of the system, Eqs.~\eqref{eq:eqr}, \eqref{eq:eqR}, \eqref{eq:q}, and \eqref{eq:M} must be solved together. Note that the equations for the center of mass vector predict a boost, as can be seen in~\cite{Cardoso:2020lxx}. To analyze the eccentricity evolution, however, we have to focus into $\boldsymbol{r}$ instead. Before going into the full regime, it is instructive to focus on some particular cases.

%%%%%%%%%%%%%%%%%%%%%%%%%%%%%%%%
\subsection{Equal-mass binaries}
%%%%%%%%%%%%%%%%%%%%%%%%%%%%%%%
For equal mass ratio binaries, $q=1$ during the whole evolution, due to symmetry [c.f. Eq.~\eqref{eq:q}].~\footnote{We note that we are considering a homogeneous medium. Density lumps in the medium can introduce asymmetries that can affect the outcome of the motion.} In this case, the center of mass remains at rest (or at constant velocity) and the equations simplify considerably. Considering $\boldsymbol{R}=0$, we have
\begin{equation}
\ddot{\boldsymbol{r}}=-\frac{G^2M\rho}{2}(I_\text{a}+I_\text{d})\dot{\boldsymbol{r}} - \frac{G M}{r^3}\boldsymbol{r}\,,\label{eq:r_eq}
\end{equation}
where we dropped the particle label index because drag and accretion forces are the same for both particles. Additionally, the total mass of the binary also evolves because of accretion. The total mass evolution is given by
\begin{equation}
\dot{M}=\frac{G^2M^2\rho I_\text{a}}{2}\,.\label{eq:M_eq}
\end{equation}

To track the eccentricity of the system it is useful to describe the evolution of the total mechanical energy and the angular moment per reduced mass. The evolution of the mechanical energy can be found by analyzing the power extracted by the external force. We have that the evolution of the energy per reduced mass~$(\varepsilon)$ is determined by
\begin{equation}
\dot{\varepsilon}=-\frac{G^2M\rho(I_\text{a}+I_\text{v})}{2}\dot{\boldsymbol{r}}\cdot\dot{\boldsymbol{r}} =-\frac{G^2 M\rho k}{2v}\,,\label{eq:dote}
\end{equation}
where $v=|\dot{\boldsymbol{r}}|$, and we considered $(I_\text{a}+I_\text{v})\simeq k/v^3$, which is valid even for collisional \acs{DF} in the limit $v/c_s\gg1$.
\footnote{
	For the model adopted here, considering only \acs{DF}, we have $k=32\pi \lambda$ (note that $v_i=v/2$ for symmetric binaries).
} The evolution of the angular momentum per reduced mass ($h=|\mathbf{r}\times\dot{\mathbf{r}}|$) follows from Eq.~\eqref{eq:r_eq},
\begin{equation}
\dot{h}=-\frac{G^2 M\rho k}{2v^3}h\,.\label{eq:doth}
\end{equation}
Finally, the eccentricity can be found by tracking
\begin{equation}
e=\sqrt{1+2\frac{\varepsilon h^2}{G^2M^2}}\,.
\label{eq:ecc}
\end{equation}

%%%%%%%%%%%%%%%%%%%%%%%%%%%%%%%%%%%%%%%%%%%%%%%%%%%%%%%%%%%%%%%%%%%%%%%%%%%%%%%%%%%%%%%
\subsubsection{Averaging the energy and angular momentum evolution for elliptic orbits}
%%%%%%%%%%%%%%%%%%%%%%%%%%%%%%%%%%%%%%%%%%%%%%%%%%%%%%%%%%%%%%%%%%%%%%%%%%%%%%%%%%%%%%%
In a similar fashion to that of Section~\ref{sec:ff} where we dealt with fundamental fields, we can consider Eqs.~\eqref{eq:dote} and \eqref{eq:doth} as ``fluxes" in which the right-hand side is computed for a fixed orbit. For simplicity, let us consider only \acs{DF}, \ie,~$M$ is constant during the evolution. For an elliptical orbit, using the average procedure defined in Appendix~\ref{app:SVrad}, we find the energy and angular momentum loss for one complete period
\begin{align}
\left<\dot{\varepsilon}\right>&=-\frac{a \left(1-e^2\right)^2 G k \rho  \sqrt{\frac{G M}{a}}}{4 \pi }\int_0^{2\pi}d\varphi\, g_\varepsilon\,,\label{eq:vare}\\
\left<\dot{h}\right>&=-\frac{a^2(1-e^2)^{7/2}G k \rho}{4\pi}\int_0^{2\pi}d\varphi\, g_h\,,\label{eq:varh}\\
g_\varepsilon&=(1+e\cos\varphi)^{-2}(1+e^2+2e\cos\varphi)^{-1/2}\,,\\
g_h&=(1+e\cos\varphi)^{-2}(1+e^2+2e\cos\varphi)^{-3/2}\,.
\end{align}
Finally, we can use the following relations
\begin{equation}
a=-\frac{GM}{2\varepsilon}\,,\quad e^2=1-2\frac{\varepsilon \,h^2}{G^2M^2}\,,
\end{equation}
to rewrite Eqs.~\eqref{eq:vare}-\eqref{eq:varh} in terms of $a$ and $e$. For low-eccentricity orbits we find
\begin{align}
\left<\dot{a}\right>&=-k \rho  \sqrt{\frac{G\,a^5 }{M}}\left(1+\frac{3 e^2}{4}+{\cal O}(e^4)\right)\,,\\
\left<\dot{e}\right>&=\frac{3}{2} k \rho  \sqrt{\frac{G\,a^3}{M}}e\left(1+\frac{3 e^2}{8}+{\cal O}(e^4)\right)\,.
\end{align}
From the above relations we see that eccentricity \textit{increases} in time under the effect of the dissipative environmental forces. This has been observed in some works considering motion under the influence of drag~\cite{Gair:2010iv,Macedo:2013qea,Cardoso:2020lxx}.

Using the formalism of adiabatic invariants (see \eg,~\cite{landau1982mechanics}) one may be led to expect eccentricity to be constant under the adiabatic approximation (which would contradict some of the results discussed here). While eccentricity is a constant at leading order, the semi-major axis does evolve on this time scale, and some conclusions can be drawn for \acs{GW} binary systems~\cite{DeLuca:2020qqa}. Although eccentricity is indeed an adiabatic invariant at leading order, it does not need to be (and it is not, in general) a constant of motion at next-to-leading order~\cite{1985Salmassi,1993Djukic}. However, under the regime of validity of the adiabatic approximation, it is true that the eccentricity must change over a timescale much larger than, for instance, the semi-major axis (which is not a constant of motion at leading order). We have verified that eccentricity indeed increases by considering, for instance, a system subject to only accretion-driven forces (which is subdominant over \acs{DF}), with the evolution of $e(a)$ converging for $\rho\to 0$, indicating that indeed eccentricity does change in the adiabatic regime.

%%%%%%%%%%%%%%%%%%%%%%%%%%%%%%%%%%%%%%%%%%%%%%%%%%%%%%%%%%%%%%%%%%%%%%
\subsubsection{Dissipative forces, GWs and the eccentricity evolution}
%%%%%%%%%%%%%%%%%%%%%%%%%%%%%%%%%%%%%%%%%%%%%%%%%%%%%%%%%%%%%%%%%%%%%%
%
\begin{figure*}
	\begin{tabular}{c}
	\includegraphics[width=\linewidth]{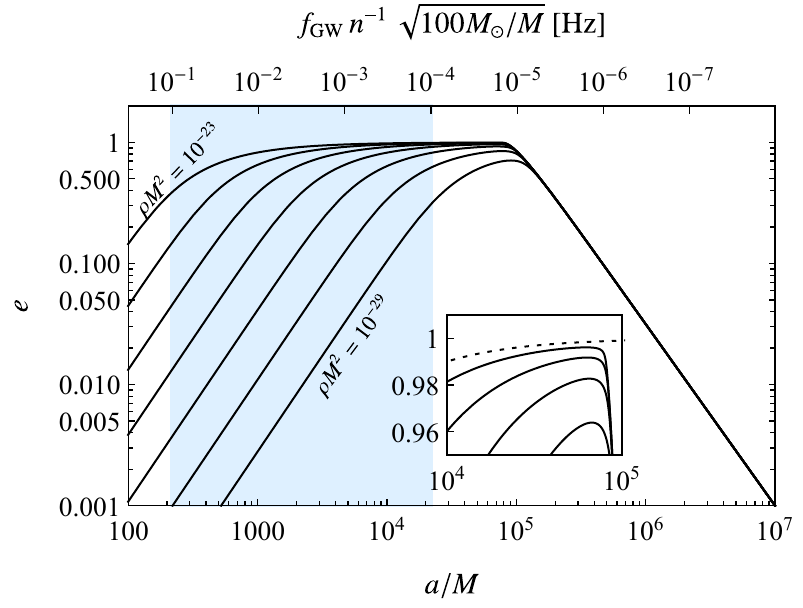} \\\\
	\includegraphics[width=\linewidth]{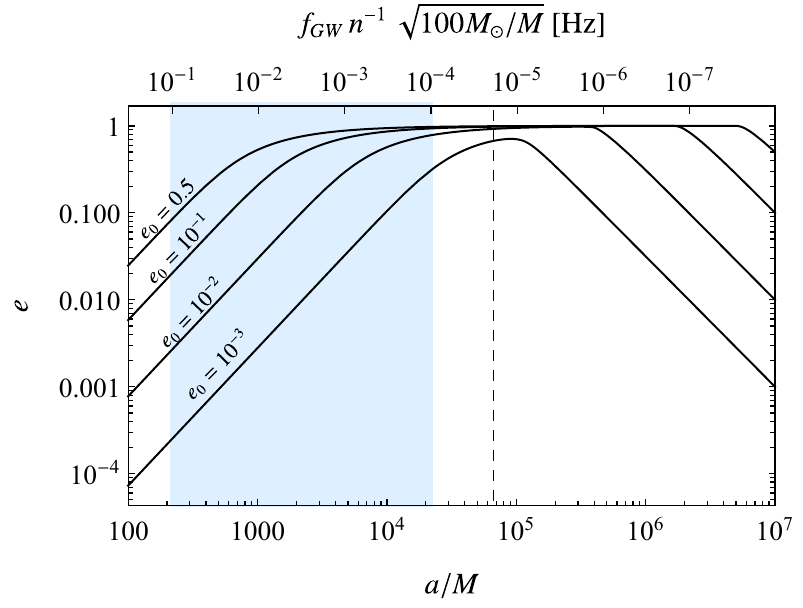}
	\end{tabular}
	\caption{Eccentricity evolution of a binary system, with an initial semi-axis~$a/M=10^{7}$. Bottom axis shows the semi-major axis as function of eccentricity, top axis shows the \acs{GW} frequency. We run the binary up to a distance of~$a=100M$. Blue bands indicate LISA's frequency range~\cite{Audley:2017drz}. \textit{Up panel:} We consider a system with an initial eccentricity of~$e=10^{-3}$ and different values of the environment density. Dashed line in inset shows threshold values for which periastron is~$100M$. \textit{Down panel:} We fix the density to be~$\rho M^2=10^{-29}$, changing the initial eccentricity of the system. The vertical line indicates the critical distance given by Eq.~\eqref{eq:criticala}. }
	\label{fig:ecc_evol}
\end{figure*}
As seen above, dissipative forces such as \acs{DF} increase the orbital eccentricity of the binary. On the other hand, radiative mechanisms, such as \acs{GW} emission, act to decrease the orbital eccentricity. We now quantify the combined effect, to understand how binaries behave in astrophysical environments, focusing in the \acs{GW} channel only. 
We can use the equations for~$\left<\dot{a}\right>$ and~$\left<\dot{e}\right>$ to compute~$da/de$. When only \acs{GW} emission contributes~\cite{Peters:1964zz,MaggioreBook},
\begin{align}
\frac{da}{de}=\frac{12a}{19e}\left(1+\frac{3323}{912}e^2+{\cal O}(e^4)\right) \qquad (\text{GW-only})\,.
\label{eq:gravonly}
\end{align}
On the other hand, \acs{DF} alone produces
\begin{equation}
\frac{da}{de}=-\frac{2a}{3e}\left(1+\frac{3}{8}e^2+{\cal O}(e^3)\right) \qquad (\text{DF-only})\,.
\label{eq:dfecc}
\end{equation}
Curiously, the \acs{DF} result (expressed in this way) does not depend explicitly on the medium density. At linear order we can combine the effects of \acs{GW} emission and \acs{DF} by simply adding the energy and angular momentum loss, and find, up to terms of order ${\cal O}(e^0)$,
\begin{equation}
\frac{da}{de}=\frac{6 a \left(5 c^5 k \rho  \sqrt{G M a^{11}}+32 G^3 M^4\right)}{e \left(304 G^3 M^4-45 c^5 k \rho  \sqrt{G M a^{11}}\right)} \qquad(\text{GW+DF})\,.	\label{eq:gwdf}
\end{equation} 
Interestingly, when the two effects are combined the density of the medium manifests itself. This is because the density balances the contribution from the energy and angular momentum loss. For~$\rho=0$ we recover the standard vacuum \acs{GW} case. Clearly, there is a critical value for the distance as function of the medium density at which~$da/de$ changes sign. We have
\begin{align}
\frac{a_{\textrm{c}}}{\left( \frac{ 100 G M_\odot}{c^2}\right)}&=3 \times 10^{4}\,k^{-2/11}\left(\frac{M}{100 M_\odot}\right)^{7/11}\left(\frac{\rho_{10}}{\rho}\right)^{2/11}\,,
\label{eq:criticala}
\end{align}
where $\rho_{10}=10^{-10}\,{\textrm{g}/\textrm{cm}^{3}}$.
For~$a\lesssim a_c$, \acs{GW} emission is dominant over \acs{DF} and the eccentricity decreases. For most reasonable scenarios we have~$k^{-2/11} \in [0.1,0.5]$.~\footnote{Considering $\lambda\in [0.5,2000]$.}

The critical distance given by Eq.~\eqref{eq:criticala} dictates the balance between environmental forces and \acs{GW} emission, indicative of whether quasi-circular orbits are indeed expected close to coalescence. However, other factors may be important. One of them is the adiabatic assumption (which we explore in Section~\ref{sec:adiabatic}, where we show evidence that it does not impact our findings substantially), the other concerns the eccentricity evolution, which depends on the initial conditions and which may lead to extremely small periastron distances.

Figure~\ref{fig:ecc_evol} shows the result of the integration of Eq.~\eqref{eq:gwdf}, including corrections for the \acs{DF} part up to order~${\cal O}(e^{12})$. We focus on initial semi-major axis of~$a(e_0)=10^7M$, for different values of the medium density and the initial eccentricity of the system, but the results hold for other initial distances, observing how the density scales with the separation of the system. Note that
\begin{align}
\frac{G^3}{c^6}\rho M^2=1.6\times 10^{-24}\frac{\rho}{\rho_{10}}\left(\frac{M}{100M_{\odot}}\right)^2\,,\label{eq:density_normalization}
\end{align}
where we used typical values of event \textit{GW190521}~\cite{Abbott:2020tfl,Abbott:2020mjq,Graham:2020gwr} as reference values.

It is clear from the figure that the eccentricity increases when the environmental effects dominate, for separations larger than those in Eq.~\eqref{eq:criticala}. In this region~$e\propto (a/M)^{-3/2}$, regardless of the medium density and of the initial eccentricity, as predicted by Eq.~\eqref{eq:dfecc}. It is also important to note that, while for small separations \acs{GW} drives the process with~$e\propto (a/M)^{19/12}$, the eccentricity inherited from the environment-dominated phase may be substantial. Thus, the system could still be observed with a considerable eccentricity in a wide range of binary evolution stages. 
Note that~$\rho M^2\sim 10^{-22}$ or larger are possible close to the inner edge of thin accretion disks, thus eccentricities larger than~$e\sim 0.1$ are expected during a substantial portion of the time-in band for a detector such as LISA.

It is instructive to understand the initial and final stages of the binary evolution analytically. As indicated previously, the \acs{GW} and medium dominated regions can be estimated by looking into their respective solutions for low eccentricities [\ie, Eqs.~\eqref{eq:gravonly} and~\eqref{eq:dfecc}]. The link between the two regimes can be estimated by analyzing Eq.~\eqref{eq:gwdf}, imposing the initial eccentricities $e_0=e(a_0)$. Let us assume that the motion starts far from the critical distance~\eqref{eq:criticala}. We obtain the following simple expressions for the two regimes
\begin{equation}
e=\left\{
\begin{array}{ll}
e_0\left(\frac{a}{a_0}\right)^{-3/2}, & a\gg a_c\,,\\
0.35\,e_0\,\tilde{a}_0^{3/2} \tilde{a}^{19/12}(k \,\tilde{\rho})^{37/66}, \quad& a\ll a_c\,,
\end{array}
\right.
\end{equation}
with~$\tilde{a}=a/(G M/c^2)$, and~$\tilde{\rho}=G^3M^2\rho/c^{6}$. 	The above solutions are valid mostly for low densities and low initial eccentricities. These expressions can be used to understand all of the peculiarities of Fig.~\ref{fig:ecc_evol}.

For very large eccentricities it is conceivable that the distance of closest approach would be so small that the components would effectively collide. For the systems we explored this possibility is not realized. The minimum distance~$r_{\textrm{min}}$ obeys
\begin{equation}
r_{\textrm{min}}>100\frac{G M}{c^2},
\end{equation}
which can be translated to maximum eccentricity of~$e=1-100(G M/c^2)/a$, represented by the dashed line in the inset of the left panel of Fig.~\ref{fig:ecc_evol}. This indicates that we can expect the objects to pass relatively close to each other without colliding during the evolution, for the density range investigated in the figure. Interestingly, this collision avoidance is only possible due to the \acs{GW} effect of decreasing the binary eccentricity; if only the medium effects were in play, the objects would collide much sooner and during a highly eccentric motion.

Newtonian circular binaries emit \acsp{GW} at a frequency $f_{\textrm{GW}}=\omega_0/\pi$. Eccentricity makes the spectrum more complex.
Elliptical orbits will in general generate a spectrum
\be
f_{\textrm{GW}}=n\frac{\omega_0}{2\pi},~{\textrm{with}}~n\geq1.
\ee
Therefore, in general, all harmonics of the orbital frequency contribute to the \acs{GW} frequency. The dominant frequency $n=\bar{n}$ depends on the eccentricity of the system. The higher the eccentricity, the higher the value of~$\bar{n}$. In other words, high-frequency bursts are emitted at periastron~\cite{Hopper:2017qus}, which means in practice that the source can enter the LISA band much sooner than what seems to be implied by the figure. In Fig.~\ref{fig:ecc_evol} we also show the frequency of the system normalized by the value of $n$. We highlight that the frequencies fall into the LISA band while having a considerable eccentricity.

%%%%%%%%%%%%%%%%%%%%%%%%%%%%%%%%%%%%%%%%%%%%%%%%%%%%%%%%%%%%%%%%
\subsection{Asymmetric binaries and accretion}
%%%%%%%%%%%%%%%%%%%%%%%%%%%%%%%%%%%%%%%%%%%%%%%%%%%%%%%%%%%%%%%%
To implement the simple adiabatic approximation described in the previous sections, we have focused on symmetric binaries and neglected accretion. This approximation enabled us to understand the evolution under the effects of both \acs{DF} and \acs{GW} backreaction. However, asymmetry leads to novel, important effects. It was realized recently that unequal-mass binaries may acquire a large center-of-mass velocity as the evolution proceeds~\cite{Cardoso:2020lxx}. We can also verify here that accretion might not play a central role in the earlier stages of eccentricity gain.

\begin{figure}
	\includegraphics[width=\linewidth]{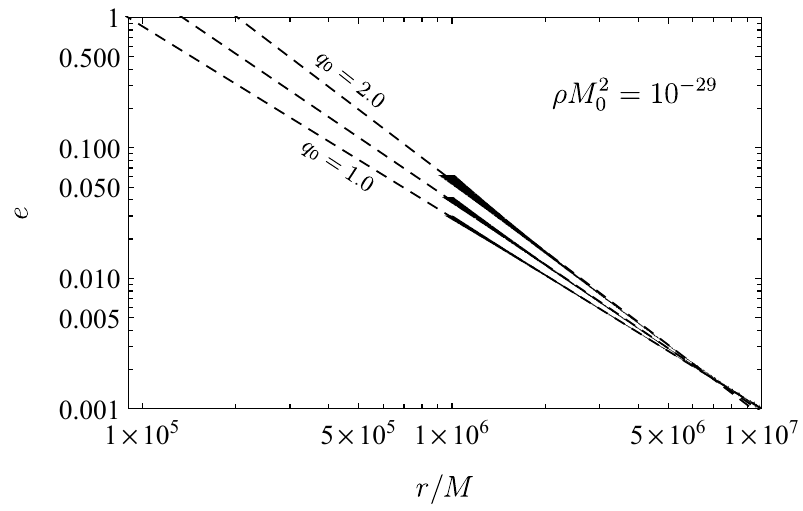}
	\caption{Eccentricity evolution for different initial mass-ratios ($q=1.0,~1.5$ and~$2.0$) when accretion is included. The dashed line is an analytical fit that enable us to predict at which distance the system will reach highly eccentric motion.}
	\label{fig:eccenti}
\end{figure}
In order to understand asymmetric binaries and the influence of accretion, we integrate the full system of equations given by Eqs.~\eqref{eq:eqr}-\eqref{eq:eqR} and~\eqref{eq:q}-\eqref{eq:M}, neglecting possible \acs{GW} backreaction into the system. This approximation should be valid far from the critical distance \eqref{eq:criticala}, where the environmental effects dominate over \acs{GW}. We also focus in a regime in which the adiabatic approximation is valid for symmetric binaries in the absence of accretion.

In Fig.~\ref{fig:eccenti} we plot the eccentricity as function of the orbital distance for a medium with density~$\rho M^2=10^{-29}$, with initial separation major semi-axis~$a_0=10^7M$ and eccentricity~$e=0.001$. We verify that the results remain essentially the same for~$\rho M^2\in [10^{-28},10^{-30}]$, indicating that we are in the regime in which the adiabatic approximation is valid (more about that in the following section). We also consider initial mass-ratios~$q=1,~1.5$, and~$2$. For higher mass-ratios eccentricity grows faster as the distance decreases, which is evident by analyzing the slope of the curves in Fig.~\ref{fig:eccenti}. We also display this eccentricity growth by using a fit (dashed lines in Fig.~\ref{fig:eccenti}) to extrapolate the evolution data up to higher eccentricities. This implies that asymmetric binaries will reach highly eccentric motion faster than symmetric ones. 

Accretion has little impact in the evolution of eccentricity, when compared to~\acs{DF}, at least for the density range considered in this paper. However, we should highlight that this is model-dependent; to perform the computations, we fix the \acs{DF} model with~$\lambda=20$. In general, in the high-velocity limit, the ratio between the \acs{DF} force and accretion force is~$\lambda$ and, as such,~$\lambda=20$ indicates a medium in which \acs{DF} generally dominates over accretion. Additionally, because~$\lambda$ appears combined with the medium density in the \acs{DF} force, it also influences the density scales in which the orbits evolve adiabatically.

%%%%%%%%%%%%%%%%%%%%%%%%%%%%%%%%%%%%%%%%%%%%%%%%%%%%%%%%%%%%%%%%%%%%%
\section{When the adiabatic assumption fails}\label{sec:adiabatic}
%%%%%%%%%%%%%%%%%%%%%%%%%%%%%%%%%%%%%%%%%%%%%%%%%%%%%%%%%%%%%%%%%%%%%
We have made extensive use of the adiabatic approximation in the previous sections to analyze the evolution of the eccentricity of the system subjected to \acs{GW} radiation-reaction and environmental forces. However, depending on the environmental density and the initial separation of the binary, this approximation may not be valid. In this subsection, we address how much the adiabatic approximation may underestimate the eccentricity increase in the system.
In order to investigate the validity of the adiabatic approximation for equal mass binaries we integrate Eq.~\eqref{eq:r_eq} (neglecting accretion), considering specific initial conditions. With the numerical solution, we construct the eccentricity as function of the orbital distance, by tracking the expression~\eqref{eq:ecc}. Since this system only takes into account the environmental effects, we compare this solution to the one obtained from the adiabatic approach by integrating Eq.~\eqref{eq:dfecc} under similar conditions (with higher order terms of eccentricity included). Using these results, we compute the relative deviation of the eccentricity, \ie,
\begin{equation}
\frac{\delta e}{e_\text{a}}=\frac{|e_\text{n}-e_\text{a}|}{e_\text{a}}\,,
\end{equation}
where $e_\text{n}$ is the result from Eq.~\eqref{eq:r_eq} and $e_\text{a}$ the one from the adiabatic approximation (considering terms up to ${\cal O}(e^{12})$). The deviation depends on the medium density and the initial conditions, but we expect it to approach zero as the medium density decreases.

In Fig.~\ref{fig:deviation} we plot the eccentricity deviation, considering an initial separation of~$a=10^7M$ and initial eccentricity~$e_0=0.001$. For the \acs{DF}, we consider $\lambda=20$. We can see that for densities of~$\rho M^2=10^{-27}$ the adiabatic approximation fails to quantitatively describe the eccentricity evolution of the system, underestimating the eccentricity increase from the \acs{DF}. For densities as small as~$\rho M^2=10^{-29}$ the adiabatic approach works mostly in the initial stages of the binary evolution. At late times, meaning short distances, we can see that the eccentricity deviation increases, indicating a possible breaking of the adiabatic approximation. 

Going beyond the adiabatic approximation shows that the eccentricity increases even further; this effect is enhanced for asymmetric binaries and accretion, as we discussed in the previous sections.

\begin{figure}
	\includegraphics[width=\linewidth]{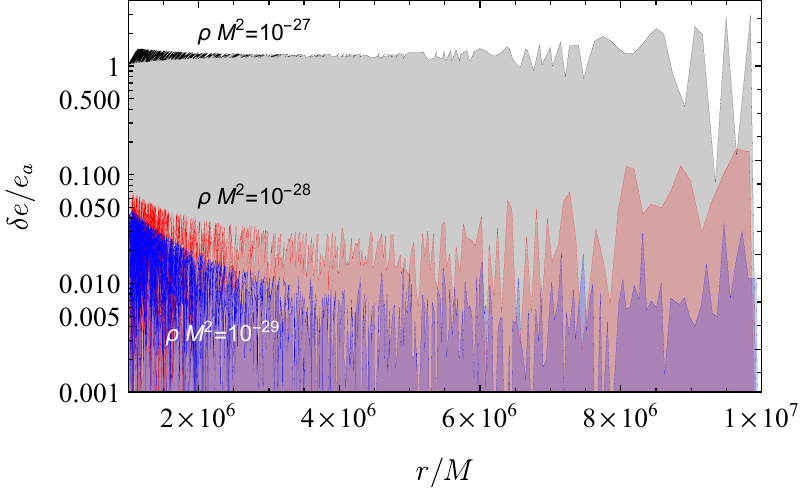}
	\caption{Comparison between the numerical integration of Eq.~\eqref{eq:r_eq} and the result from the adiabatic approach. We plot the deviation normalized by the adiabatic result.}
	\label{fig:deviation}
\end{figure}

%%%%%%%%%%%%%%%%%%%%%%%%%%%%%%%%%%%%%%%%%%%%%%%%%%%%%%%%%%%%%%%%%%%%%%%%%%%%%
\section{Discussion}
%%%%%%%%%%%%%%%%%%%%%%%%%%%%%%%%%%%%%%%%%%%%%%%%%%%%%%%%%%%%%%%%%%%%%%%%%%%%%

In this chapter we studied the evolution of eccentricity of compact binaries, evolving via emission of massless fields and of environmental
accretion and gravitational drag. We proved that the emission of massless scalars, vectors of tensors circularizes the orbits.
In particular, the critical distance at which the orbits start to circularize is larger when additional scalar or vector charges are considered. The integration of 
Eqs.~\eqref{shrinkp}-\eqref{circularp} shows that
\begin{equation}
\frac{a}{M}\propto \frac{e^{4/3}}{1-e^2}\,,
\end{equation}
for scalar or vector-driven binaries. Compare this against the gravitational-driven result, $a/M\propto \kappa e^{12/19}/(1-e^2)$, at small eccentricities~\cite{Peters:1964zz}.
The eccentricity for these channels thus decays less quickly than in vacuum. Nevertheless, even when additional massless fields are considered,
circular orbits remain stable.

By contrast, we show that sources of interest for \acs{GW} detectors evolving in thin accretion disks or other relatively large-density environment may inherit
a substantial eccentricity by the time they reach the mHz band. As we showed, high eccentricity is also a key feature of large mass ratio binaries, which is one possible explanation of
the \textit{GW190521} event~\cite{Nitz:2020mga}.
Together with previous results on the center-of-mass velocity of asymmetric binaries~\cite{Cardoso:2020lxx},
these results show that modeling binaries in accretion disks or nontrivial environments is challenging but crucial. In particular, these effects may have an important impact in attempts to constrain environmental properties~\cite{Barausse:2014tra,Cardoso:2019rou,Annulli:2020lyc,Toubiana:2020drf} or on testing fundamental properties of compact binaries~\cite{Cardoso:2019upw,Cardoso:2020nst}.

Our results complement previous findings~\cite{Roedig:2011rn,Zrake:2020zkw}. In particular, eccentricity excitation via asymmetric torques from circumbinary discs was found to keep supermassive black holes on eccentric orbits for a relevant fraction of their evolutionary phase~\cite{Roedig:2011rn}. Along the same line, it was recently shown that circumbinary disk torques may lead an equal-mass binary to evolve towards an equilibrium orbital eccentricity of~$e\simeq 0.45$~\cite{Zrake:2020zkw}. Interestingly, in that same analysis it was found that, when the circumbinary gas is in a thin disk, \acs{DF} causes a \textit{damping} in the eccentricity if the orbital eccentricity is~$e>0.45$. This effect is not captured by our model, as we do not consider the full modeling of the fluid perturbations and its gravitational effects.

%************************************************
\chapter{Test fields cannot destroy extremal black holes}\label{ch:wcc}
%************************************************

In the wake of the proofs of the singularity theorems in general relativity~\cite{Penrose:1964wq,Hawking:1967ju,Hawking:1970zqf}, Penrose formulated the weak cosmic censorship conjecture~\cite{Penrose:1969pc,Wald:1997wa}, according to which, generically, the singularities resulting from gravitational collapse are hidden from the observers at infinity by a black hole event horizon. Penrose's expectation was that, independently of what might happen inside black holes, the evolution of the outside universe would proceed undisturbed.

To test this conjecture, Wald~\cite{WALD1974548} devised a thought experiment to destroy extremal Kerr-Newman black holes, already on the verge of becoming naked singularities, by dropping charged and/or spinning test particles into the event horizon. Both him and subsequent authors~\cite{Tod:1976ud,Needham:1980fb} found that if the parameters of the infalling particle (energy, angular momentum, charge and/or spin) were suited to overspin/overcharge the black hole then the particle would not go in, in agreement with the cosmic censorship conjecture. Similar conclusions were reached by analyzing scalar and electromagnetic test fields propagating in extremal Kerr-Newman black hole backgrounds~\cite{Semiz:2005gs,Toth:2012vvy,Duztas:2013wua,Duztas:2013gza}. In this case, the fluxes of energy, angular momentum and charge across the event horizon were found to be always insufficient to overspin/overcharge the black hole. Some of these results have been extended to higher dimensions~\cite{Bouhmadi-Lopez:2010yjy} and also to the case when there is a negative cosmological constant~\cite{Gwak:2015fsa,Rocha:2014jma}.

More recently, it was noticed that Wald's thought experiment may produce naked singularities when applied to nearly extremal black holes~\cite{Hubeny:1998ga,Matsas:2007bj,Jacobson:2009kt,Saa:2011wq}. However, in this case the perturbation cannot be assumed to be infinitesimal, and so backreaction effects have to be taken into account; when this is done, the validity of the cosmic censorship conjecture appears to be restored~\cite{Hod:2008zza,Barausse:2010ka,Zimmerman:2012zu,Shaymatov:2014dla,Colleoni:2015ena}. It can also be argued that the third law of black hole thermodynamics~\cite{Bardeen:1973gs}, for which there is some evidence~\cite{Israel:1986gqz,Dadhich:1997rq,Chirco:2010rq}, forbids subextremal black holes from ever becoming extremal, and so, presumably, from being destroyed. Nonetheless, this cannot be taken as a definitive argument, since, for instance, extremal Reissner-Nordstr\"om black holes can be formed by collapsing charged thin shells~\cite{Boulware73}.

In this chapter, we consider arbitrary (possibly charged) test fields propagating in extremal Kerr-Newman, Kerr-Newman-anti de Sitter (AdS), or Kerr-Newman-de Sitter (dS) black hole backgrounds. Apart from ignoring their gravitational and electromagnetic backreaction, we make no further hypotheses on these fields: they could be any combination of scalar, vector or tensor fields, charged fluids, sigma models, elastic media, or other types of matter. This also includes test particles, since they can be seen as singular limits of continuous media~\cite{Geroch:1975uq,Lasota:2013kia}. We give a general proof that if the test fields satisfy the null energy condition at the event horizon then they cannot overspin/overcharge the  black hole. This is done by first establishing, in Section~\ref{section2}, a test field version of the second law of black hole thermodynamics for extremal Kerr-Newman or Kerr-Newman-AdS black holes (which does not assume cosmic censorship). We use this result in Section~\ref{section3}, together with the Smarr formula and the first law, to conclude the proof. This last step requires the black hole to be extremal, and cannot be extended to near-extremal black holes. In the same section, we discuss generalizations of this result to other extremal black holes, including higher dimensions and alternative theories of gravity. Finally, in Section~\ref{section4} determine the timelike Killing vector field that gives the correct definition of energy for test fields propagating in a Kerr-Newman-de Sitter spacetime, and use this to extend the previous result to extremal Kerr-Newman-de Sitter black holes.

\section{Second law for test fields}\label{section2}
In this section we prove that a version of the second law of \acs{BH} thermodynamics holds in the case of (possibly charged) test fields propagating on a background Kerr-Newman or Kerr-Newman-AdS \acs{BH} (either subextremal or extremal). This calculation is similar to the one in~\cite{Gao:2001ut}, but we do not assume cosmic censorship, \ie, we do not assume that the \acs{BH} is not destroyed by interacting with the test field.

We start by recalling the Kerr-Newman-(A)dS metric, given in Boyer-Lindquist coordinates by
\begin{align}
ds^2 = & - \frac{\Delta_r}{\rho^2}\left( dt - \frac{a \sin^2 \theta}{\Xi} d\varphi \right)^2 + \frac{\rho^2}{\Delta_r} dr^2 \nonumber \\
& + \frac{\rho^2}{\Delta_\theta} d \theta^2 + \frac{\Delta_\theta \sin^2\theta}{\rho^2}\left( a \, dt - \frac{r^2 + a^2}{\Xi} d\varphi \right)^2\,, \label{KNAdS}
\end{align}
where 
\begin{align}
& \rho^2 = r^2 + a^2 \cos^2 \theta\,, \\
& \Xi = 1 \pm \frac{a^2}{l^2}\,, \\
& \Delta_r = (r^2 + a^2)\left(1\mp\frac{r^2}{l^2}\right) - 2mr + q^2\,, \\
& \Delta_\theta = 1 -pm \frac{a^2}{l^2}\cos^2\theta
\end{align}
(see for instance Ref.~\cite{Caldarelli:1999xj}). In what follows, the upper sign will always refer to a positive cosmological constant, and the lower sign to a negative cosmological constant, given in terms of the parameter $l$ by~\footnote{Note that the Kerr-Newman metric can be obtained by taking the limit $l^2\to\infty$.}
\begin{equation}
\Lambda = \pm \frac{3}{l^2}.
\end{equation} 
Here $m$, $a$ and $q$ denote the mass, rotation and electric charge parameters, respectively. These parameters are related to the physical mass $M$, angular momentum $L$ and electric charge $Q$ by
\begin{equation}\label{physical}
M = \frac{m}{\Xi^2}\,, \qquad L = \frac{ma}{\Xi^2}\,, \qquad Q = \frac{q}{\Xi}\,.
\end{equation}
To avoid repetition, we will present all calculations below for the Kerr-Newman-AdS metric only; the corresponding formulae for the Kerr-Newman metric can be easily retrieved by making $l \to +\infty$.

The Kerr-Newman-AdS metric, together with the electromagnetic $4$-potential
\begin{equation}
A = - \frac{qr}{\rho^2}\left(dt - \frac{a\sin^2\theta}{\Xi} d\varphi\right)\,,
\end{equation}
is a solution of the Einstein-Maxwell equations with cosmological constant $\Lambda$. It admits a two-dimensional group of isometries, generated by the Killing vector fields $X^\alpha = (\partial_t)^\alpha$ and $Y^\alpha=(\partial_\varphi)^\alpha$. 

We consider arbitrary (possibly charged) test fields propagating in this background. Apart from ignoring their gravitational and electromagnetic backreaction, we make no further hypotheses on the fields: they could be any combination of scalar, vector or tensor fields, charged fluids, sigma models, elastic media, or other types of matter. Since the fields may be charged, their energy-momentum tensor $T^{\alpha \beta}$ satisfies the generalized Lorentz law~\footnote{See the Appendix~\ref{app:Lorentz} for a complete explanation of the origin and meaning of this equation.}
\begin{equation}\label{motion}
\nabla_\alpha T^{\alpha \beta} = F^{\beta\alpha} J_\alpha\,,
\end{equation}
where $F=dA$ is the Faraday tensor of the background electromagnetic field and $J_\alpha$ is the charge current density $4$-vector associated to the test fields. Using the symmetry of $T^{\alpha \beta}$ and the Killing equation, 
\begin{equation}\label{Killing}
\nabla_\alpha X_\beta + \nabla_\beta X_\alpha = 0\,,
\end{equation}
we have
\begin{equation}
\nabla_\alpha (T^{\alpha\beta} X_\beta) = F^{\beta\alpha} J_\alpha X_\beta\,. \label{div1}
\end{equation}
On the other hand, using the charge conservation equation,
\begin{equation}\label{charge}
\nabla_\alpha J^\alpha = 0\,,
\end{equation}
we obtain
\begin{align}
\nabla_\alpha (J^\alpha A^\beta X_\beta) & =  J^\alpha (\nabla_\alpha A^\beta) X_\beta + J^\alpha A^\beta \nabla_\alpha X_\beta \nonumber \\
& = J^\alpha (F_\alpha^{\,\,\,\,\beta} + \nabla^\beta A_\alpha) X_\beta - J^\alpha A^\beta \nabla_\beta X_\alpha \nonumber \\
& = F^{\alpha\beta} J_\alpha X_\beta + J_\alpha (X^\beta \nabla_\beta A^\alpha - A^\beta \nabla_\beta X^\alpha)\,. \label{div2}
\end{align}
Since $A^\alpha$ is invariant under time translations, we have
\begin{equation}\label{Lie}
\mathcal{L}_X A^\alpha = 0 \Leftrightarrow [X,A]^\alpha= 0 \Leftrightarrow X^\beta \nabla_\beta A^\alpha - A^\beta \nabla_\beta X^\alpha = 0\,,
\end{equation}
and so from \eqref{div1} and \eqref{div2} we obtain
\begin{equation}
\nabla_\alpha (T^{\alpha\beta} X_\beta + J^\alpha A^\beta X_\beta) = 0\,. \label{div3}
\end{equation}
This conservation law suggests that the total field energy on a given spacelike hypersurface~$\mathcal{S}$ extending from the \acs{BH} event horizon~$\mathscr{H^+}$ to infinity (Fig.~\ref{Penrose}) should be
\begin{equation}
\widetilde{E}' = \int_\mathcal{S} dV_{3}(T^{\alpha \beta} + J^\alpha A^\beta) X_\beta N_\alpha \,,
\end{equation}
where $N_\alpha$ is the future-pointing unit normal to $\mathcal{S}$. However, in the Kerr-Newman-AdS case the non-rotating observers at infinity are rotating with respect to the Killing vector field $X^\alpha$ with angular velocity
\begin{equation}
\Omega_\infty = - \frac{a}{l^2}\,,
\end{equation}
and so, as shown in Ref.~\cite{Olea:2005gb}, the physical energy should be computed with respect to the non-rotating Killing vector field
\begin{equation}
K^\alpha = X^\alpha + \Omega_\infty Y^\alpha = X^\alpha - \frac{a}{l^2} Y^\alpha\,,
\end{equation}
that is, the physical energy is actually
\begin{equation} \label{energy}
\widetilde{E} = \int_S dV_{3}(T^{\alpha \beta} + J^\alpha A^\beta) K_\beta N_\alpha \,.
\end{equation}
This correction was implemented for test particles in Ref.~\cite{Gwak:2015fsa}. The need for the corresponding correction in the calculation of the physical \acs{BH} mass has been stressed in \cite{Gibbons:2004ai,McInnes:2015vga}. Note that in the Kerr-Newman case~$\Omega_\infty = 0$ and no correction is needed.

\begin{figure}
	\includegraphics[width=\textwidth]{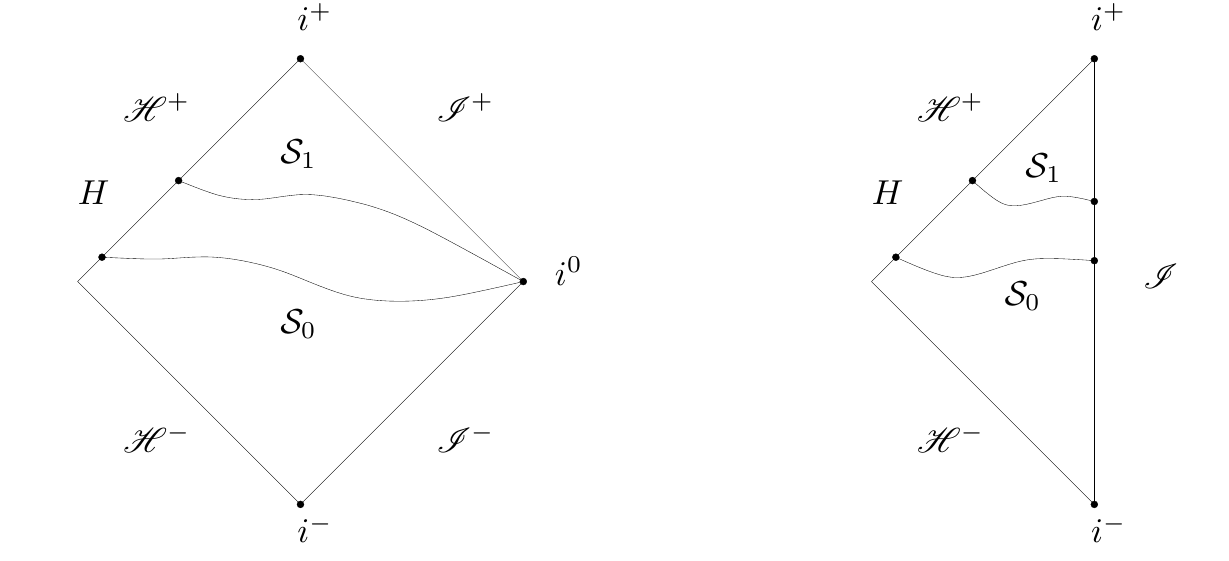}
	\caption{Penrose diagrams for the region of outer communication of the Kerr-Newman (left) and Kerr-Newman-AdS (right) spacetimes.} \label{Penrose}
\end{figure}

Analogously, but now without ambiguity, the total field angular momentum on a spacelike hypersurface $\mathcal{S}$ extending from the event horizon to infinity is
\begin{equation}
\widetilde{L} = - \int_\mathcal{S} dV_{3} (T^{\alpha \beta} + J^\alpha A^\beta) Y_\beta N_\alpha \,,
\end{equation}
where the minus sign accounts for the timelike unit normal. 

Consider now two such spacelike hypersurfaces,~$\mathcal{S}_0$ and~$\mathcal{S}_1$, with~$\mathcal{S}_1$ to the future of~$\mathcal{S}_0$ (Fig.~\ref{Penrose}). We assume reflecting boundary conditions in the Kerr-Newman-AdS case, so that all fluxes vanish at infinity. The energy absorbed by the~\acs{BH} across the subset~$H$ of~$\mathscr{H^+}$ between~$\mathcal{S}_0$ and~$\mathcal{S}_1$ is then
\begin{equation}
\Delta M = \int_{\mathcal{S}_0} dV_{3}(T^{\alpha \beta} + J^\alpha A^\beta) K_\beta N_\alpha  - \int_{\mathcal{S}_1}dV_{3} (T^{\alpha \beta} + J^\alpha A^\beta) K_\beta N_\alpha \,,
\end{equation}
whereas the angular momentum absorbed by the \acs{BH} across~$H$ is
\begin{equation}
\Delta L = - \int_{\mathcal{S}_0}dV_{3} (T^{\alpha \beta} + J^\alpha A^\beta) Y_\beta N_\alpha  + \int_{S_1}dV_{3} (T^{\alpha \beta} + J^\beta A^\alpha) Y_\beta N_\alpha \,.
\end{equation}

Recall that the angular velocity of the black hole horizon is
\begin{equation}
\Omega_H = \frac{a \Xi}{r_+^2 +a^2}\,,
\end{equation}
where~$r_+$ is the largest root of~$\Delta_r=0$. This means that the (future-pointing) Killing generator of~$\mathscr{H^+}$ is
\begin{equation}
Z^\alpha = X^\alpha + \Omega_H Y^\alpha = K^\alpha + \Omega Y^\alpha\,,
\end{equation}
where
\begin{equation}
\Omega = \Omega_H - \Omega_\infty
\end{equation}
is precisely the thermodynamic angular velocity, \ie, the angular velocity that occurs in the first law for Kerr-Newman-AdS \acs{BH}~\cite{Gibbons:2004ai}. Therefore, we have
\begin{equation}
\Delta M  - \Omega \Delta L = \int_\mathcal{{S}_0} dV_{3}(T^{\alpha \beta} + J^\alpha A^\beta) Z_\beta N_\alpha  - \int_{\mathcal{S}_1}dV_{3} (T^{\alpha \beta} + J^\alpha A^\beta) Z_\beta N_\alpha \,.
\end{equation}
Because $Z^\alpha$ is also a Killing vector field,
\begin{equation}
\nabla_\alpha (T^{\alpha \beta} Z_\beta + J^\alpha A^\beta Z_\beta) = 0\,,
\end{equation}
and so the divergence theorem, applied to the region bounded by $\mathcal{S}_0$, $\mathcal{S}_1$ and $\mathcal{H}$, yields
\begin{align} \label{Delta}
\Delta M  - \Omega \Delta L = \int_{H}dV_{3} (T^{\alpha \beta} + J^\alpha A^\beta) Z_\beta Z_\alpha 
\end{align}
(we use $-Z_\alpha$ as the null normal~\footnote{Recall that the divergence theorem on a Lorentzian manifold requires that the unit normal is outward-pointing when spacelike and inward-pointing when timelike. When the normal is null it is non-unique, and the volume element depends on the choice of normal; it should be past-pointing in the future null subset of the boundary, and future-pointing in the past null subset of the boundary.} on~$H$).  Since on $\mathscr{H^+}$ one has
\begin{equation}
A^\alpha Z_\beta = - \frac{e r_+}{r_+^2 +a^2} = - \Phi,
\end{equation}
where $\Phi$ is the horizon's electric potential, then we have
\begin{align}
\int_{H} dV_{3}J^\alpha A^\beta Z_\beta Z_\alpha  = - \Phi \int_{H} dV_{3} J^\alpha Z_\alpha \,.
\end{align}
Using again the divergence theorem, this time together with the charge conservation equation \eqref{charge}, we obtain
\begin{align}
\int_{H} dV_{3}J^\alpha A^\beta Z_\beta Z_\alpha  = -  \Phi \int_{\mathcal{S}_0} dV_{3} J^\alpha N_\alpha   +  \Phi \int_{\mathcal{S}_1}dV_{3} J^\alpha N_\alpha  \,.
\end{align}
Now, the total charge on a spacelike hypersurface $\mathcal{S}$ extending from the event horizon to infinity is
\begin{equation}
\widetilde{Q}=- \int_S dV_{3} J^\alpha N_\alpha \,,
\end{equation}
where the minus sign accounts for the timelike unit normal. Therefore, denoting by $\Delta Q$ the electric charge absorbed by the \acs{BH} across $H$, we have
\begin{equation}
\int_{H} dV_{3}J^\alpha A^\beta Z_\beta Z_\alpha  =  \Phi \Delta Q,
\end{equation}
and so equation \eqref{Delta} can then be written as
\begin{equation}
\Delta M  - \Omega \Delta L - \Phi \Delta Q  = \int_{H}dV_{3} (T^{\alpha \beta} Z_\alpha Z_\beta) \,.
\end{equation}

Since $Z$ is null on $H$, we have the following test field version of the second law of \acs{BH} thermodynamics:

\paragraph{Second law for test fields} \emph{If the energy-momentum tensor~$T^{\alpha \beta}$ corresponding to any collection of test fields propagating on a Kerr-Newman or Kerr-Newman-AdS \acs{BH} background satisfies the null energy condition at the event horizon and appropriate boundary conditions at infinity, then the energy~$\Delta M$, angular momentum~$\Delta L$ and electric charge~$\Delta Q$ absorbed by the \acs{BH} satisfy}
\begin{equation} \label{second}
\Delta M  \geq \Omega \Delta L + \Phi \Delta Q\,.
\end{equation}

It should be stressed that \eqref{second} is valid for extremal \acsp{BH} and it does not assume cosmic censorship, \ie, it does not assume that the Kerr-Newman-AdS metric with physical mass~$M + \Delta M$, angular momentum~$L + \Delta L$ and electric charge is~$Q + \Delta Q$ represents a \acs{BH} rather than a naked singularity. Note that this scenario where the test fields interact with the geometry and change the values of the black hole charges is not in contradiction with the test field approximation, since the change is supposed to be infinitesimal.

\section{Proof of the main result}\label{section3}

The physical mass of a Kerr-Newman or Kerr-Newman-AdS \acs{BH}, given in~\eqref{physical}, is completely determined by the \acs{BH}'s event horizon area~$A$, angular momentum~$L$ and electric charge $Q$~through a Smarr formula
\begin{equation} \label{one}
M=M(A,L,Q)\,.
\end{equation}
From the first law of \acs{BH} thermodynamics we know that this function satisfies
\begin{equation} \label{two}
dM = \frac{\kappa}{8 \pi} dA + \Omega dL + \Phi dQ\,,
\end{equation}
where $\kappa$ is the surface gravity of the event horizon \cite{Bardeen:1973gs,Caldarelli:1999xj,Gibbons:2004ai}. The condition for the \acs{BH} to be extremal is
\begin{equation}
\kappa = 0 \Leftrightarrow \frac{\partial M}{\partial A}(A,L,Q)=0\,,
\end{equation}
which can be solved to yield the area of an extremal \acs{BH} as a function of its angular momentum and charge,
\begin{equation}
A=A_{\text{ext}}(L,Q)\,.
\end{equation}
The mass of an extremal \acs{BH} with angular momentum~$L$ and electric charge~$Q$ is then
\begin{equation}
M_{\text{ext}}(L,Q) = M(A_{\text{ext}}(L,Q),L,Q)\,.
\end{equation}
A Kerr-Newman-AdS metric characterized by~$M$,~$L$ and~$Q$ will represent a black hole if~$M \geq M_{\text{ext}}(L,Q)$, and a naked singularity if~$M < M_{\text{ext}}(L,Q)$.
We have
\begin{align}
dM_{\text{ext}} & = \left(\frac{\partial M}{\partial A} \frac{\partial A_{\text{ext}}}{\partial L} + \frac{\partial M}{\partial L}\right) dL + \left(\frac{\partial M}{\partial A} \frac{\partial A_{\text{ext}}}{\partial Q} + \frac{\partial M}{\partial Q}\right) dQ \nonumber \\
& = \left(\frac{\kappa}{8 \pi} \frac{\partial A_{\text{ext}}}{\partial L} + \Omega \right) dL + \left(\frac{\kappa}{8 \pi} \frac{\partial A_{\text{ext}}}{\partial Q} + \Phi\right) dQ \nonumber \\
& = \Omega dL + \Phi dQ\,, \label{dMext}
\end{align}
where all quantities are evaluated at the extremal \acs{BH}. 

Consider now an extremal \acs{BH} with angular momentum~$L$, electric charge~$Q$ and mass~$M = M_{\text{ext}}(L,Q)$. After interacting with the test fields, its angular momentum is~$L + \Delta L$, its electric charge is~$Q + \Delta Q$ and its mass is, using \eqref{second} and \eqref{dMext},
\begin{align}
M + \Delta M & \geq M + \Omega \Delta L + \Phi \Delta Q \nonumber \\
& = M_{\text{ext}}(L,Q) + \Delta M_{\text{ext}} \nonumber \\
& = M_{\text{ext}}(L + \Delta L,Q + \Delta Q)\,.
\end{align}
In other words, the final mass is above the mass of an extremal \acs{BH} with the same angular momentum and electric charge, meaning that the final metric does \emph{not} represent a naked singularity, \ie, the \acs{BH} has not been destroyed.

Thus, we have just proved the following result:
\paragraph{Test fields cannot destroy extremal Kerr-Newman BHs} \emph{Test fields satisfying the null energy condition at the event horizon and appropriate boundary conditions at infinity cannot destroy extremal Kerr-Newman or Kerr-Newman-AdS \acsp{BH}. More precisely, if an extremal \acs{BH} is characterized by the physical quantities $(M,L,Q)$, and absorbs energy, angular momentum and electric charge $(\Delta M,\Delta L,\Delta Q)$ by interacting with the test fields, then the metric corresponding to the physical quantities $(M+\Delta M, L + \Delta L, Q+\Delta Q)$ represents either a subextremal or an extremal \acs{BH}.}

Our proof depends only on certain generic features of the Kerr-Newman or Kerr-Newman-AdS metric and can therefore be adapted to other \acsp{BH}. In fact, the above result can be generalized as follows.

\paragraph{Test fields cannot destroy extremal BHs} \emph{Consider a family of charged and spinning \acsp{BH} in some metric theory of gravity, with suitable asymptotic regions, and test fields propagating in these backgrounds, such that:}
\begin{enumerate}
	\item
	\emph{There exists an asymptotically timelike Killing vector field $K^\alpha$, determining the \acs{BH}'s physical mass, and angular Killing vector fields $(Y_i)^\alpha$, yielding the \acs{BH}'s angular momenta, such that the event horizon's Killing generator is}
	\begin{equation}
	Z = K + \sum_i \Omega_i Y_i\,,
	\end{equation}
	\emph{where $\Omega_i$ are the thermodynamic angular velocities (that is, the angular velocities that occur in the first law).}
	\item
	\emph{There exists a Smarr formula relating the \acs{BH}'s physical mass~$M$, its entropy~$S$, its angular momenta~$L_i$ and its electric charge~$Q$,}
	\begin{equation} \label{one}
	M=M(S,L_i,Q)\,,
	\end{equation}
	\emph{yielding the first law of \acs{BH} thermodynamics,}
	\begin{equation}
	dM = T dS + \sum_i \Omega_i dL_i + \Phi dQ\,,
	\end{equation}
	\emph{where~$T$ is the \acs{BH}'s temperature and~$\Phi$ is the event horizon's electric potential.}
	\item
	\emph{Extremal black holes (that is, black holes with $T=0$) are characterized by $M=M_{\text{ext}}(L_i,Q)$, and subextremal black holes by $M>M_{\text{ext}}(L_i,Q)$.}
	\item
	\emph{The test fields satisfy the null energy condition at the event horizon and appropriate boundary conditions at infinity.}
\end{enumerate}
\emph{Then the test fields cannot destroy extremal black holes. More precisely, if an extremal \acs{BH} is characterized by the physical quantities~$(M,L_i,Q)$, and absorbs energy, angular momenta and electric charge $(\Delta M,\Delta L_i,\Delta Q)$ by interacting with the test fields, then the metric corresponding to the physical quantities $(M+\Delta M, L_i + \Delta L_i, Q+\Delta Q)$ represents either a subextremal or an extremal \acs{BH}.}

It is easy to check that this result applies to \acsp{BH} in higher dimensions~\cite{Emparan:2008eg}, including the case of a negative cosmological constant~\cite{Gibbons:2004ai}. It can also be used for other \acsp{BH}, like accelerated \acsp{BH} with conical singularities~\cite{Appels:2016uha} or \acsp{BH} in alternative theories of gravity~\cite{Faraoni:2010yi}. There is, however, no {\em a priori} reason why it should apply to arbitrary parametrized deformations of the Kerr metric~\cite{Cardoso:2015xtj}. It does not apply directly to the case of a \emph{positive} cosmological constant, because the first hypothesis is not strictly satisfied. However, in the following section we determine the timelike Killing vector field that gives the correct definition of energy for test fields propagating in a Kerr-Newman-de Sitter spacetime, and use this to extend the above result to extremal Kerr-Newman-de Sitter \acsp{BH}. 

\section{Energy in positive cosmological constant spacetimes} \label{section4}

In all the gedanken experiments to destroy a~\acs{BH} (described in the beginning of this chapter) one must be very careful with what is meant by the energy of the test matter, and how it relates to the increase in the \acs{BH} mass. In fact, from a logical point of view, these are independent concepts: the energy of the test matter is computed with respect to a given timelike Killing vector field, whereas the \acs{BH} mass is a parameter in a \acs{BH} solution of the Einstein-Maxwell field equations. In the asymptotic flat case, the two can be related via the ADM mass: indeed, the ADM mass of a spacetime containing an isolated \acs{BH} is precisely the \acs{BH} mass, whereas the energy of test matter located in the asymptotically flat region (measured with respect to the unique timelike Killing vector field) simply adds to the ADM mass; since this energy is conserved as the test matter moves into the black hole spacetime, the \acs{BH} mass should increase by precisely that amount when the test matter is absorbed. In the non-asymptotically flat cases, however, there is no ADM mass, and there may exist many or no timelike Killing vector fields in the asymptotic region. In the asymptotically anti-de Sitter (AdS) case there are notions of total mass available \cite{Wang01,Chrusciel:2001qr,Chrusciel:2003qr}, and these were used in \cite{Natario:2016bay}, together with the results in \cite{Olea:2005gb}, to determine which of the infinitely many stationary Killing fields should be used to compute the energy of the test matter.~\footnote{This Killing vector field turns out to be the one corresponding to the zero angular momentum observers at infinity; it is singled out by the property that its charge, as defined in \cite{Olea:2005gb}, is precisely the \acs{BH}'s physical mass. Note that in this case the connection between the test matter energy and the variation in the \acs{BH} mass is not as clearcut as in the asymptotically flat case; moreover, it does not extend to the asymptotically de Sitter case, where the construction in \cite{Olea:2005gb} does not apply.} Notice that this choice is critical, and in fact incorrect choices have lead to erroneous claims of violations of weak cosmic censorship in the literature, as pointed out in \cite{McInnes:2015vga}; such claims have been disproved by \cite{Gwak:2015fsa}. In the asymptotically de Sitter (dS) case, on the other hand, there exists neither a generally accepted notion of total mass~\footnote{For example, certain scalar curvature rigidity results in Riemannian geometry directly inspired by the positive mass theorem, which hold both for asymptotically flat~\cite{Miao02} and asymptotically hyperbolic (AdS) manifolds~\cite{Min1989,Andersson1998}, are false when transposed to the positive curvature (dS) setting \cite{Brendle2011}.} (see however \cite{Kastor:2002fu,Luo:2007se}) nor a Killing vector field which is timelike in the asymptotic region, and so it is not clear how one should compute the energy of the test matter falling into the \acs{BH}. The main purpose of this section is to address this issue, and, as a consequence, to extend the result of the last section to asymptotically dS \acsp{BH}. As an added bonus, we will confirm that the choice of timelike Killing vector field used in the last section for the asymptotically AdS case is indeed correct.

\subsection{Kerr-(A)dS}\label{subsection1}

Here we construct a metric that interpolates between two Kerr-(A)dS regions of different (physical) masses~$M_1$ and~$M_2$ by letting the mass parameter become a function of the radial coordinate~$r$. We then determine, from the Einstein equations, the energy-momentum tensor of the (unphysical) field generating this metric, and use it to compute the corresponding energy with respect to a given timelike Killing vector field. This energy is seen to be precisely the difference $M_2-M_1$ between the two physical masses for a particular choice of Killing vector field.

We consider the stationary spacetime constructed as follows (Figure~\ref{interpolating}): for $r \leq r_1$ it coincides with a Kerr-(A)dS solution with mass parameter $m_1$; for $r \geq r_2 > r_1$ it corresponds to a Kerr-(A)dS solution with mass parameter $m_2 > m_1$; and for $r_1 < r < r_2$ it is the solution of the Einstein equations obtained by taking $m=m(r)$ (and $q=0$) in \eqref{KNAdS}, corresponding to some (unphysical) field which generates the energy-momentum tensor $T^{\alpha \beta}$ dictated by the Einstein equations. We assume that $r_1$ is larger than the radius of the event horizon corresponding to the mass parameter $m_1$, and that $r_2$ is smaller than the radius of the cosmological horizon corresponding to the mass parameter $m_2$ in the Kerr-dS case. In other words, we take the metric \eqref{KNAdS} with $m=m(r)$ satisfying $m(r)\equiv m_1$ for $r\leq r_1$, $m(r)\equiv m_2$ for $r\geq r_2$, and $\Delta_r(r)>0$ for $r_1 \leq r \leq r_2$; to avoid thin shells, we assume that $m(r)$ is at least $C^1$, implying in particular that $m'(r_1)=m'(r_2)=0$. For this spacetime it is fairly obvious what the energy of the field should be: since the physical masses, $M_1=\dfrac{m_ 1}{\Xi^2}$ and $M_2=\dfrac{m_2}{\Xi^2}$, correspond to the total energy contained in the regions $r<r_1$ and $r<r_2$, respectively, the energy of the field should be $E=\Delta M\equiv M_2-M_1$. We would like to calculate this energy as an integral on a given spacelike hypersurface $\mathcal{S}$ extending from $r=r_1$ to $r=r_2$. In fact, it turns out that this is possible in Kerr-AdS, where it is known that (at least for test fields)
\begin{equation}\label{energyKerr}
\widetilde{E}=\int_\mathcal{S} dV_3 T^{\mu \nu}K_\mu N_\nu \,,
\end{equation}
with $N_\alpha$ the future-pointing unit normal to~$\mathcal{S}$, and~$K^\alpha$ the Killing vector field
\begin{equation}\label{killing}
K^\alpha=X^\alpha-\frac{a}{l^2}Y^\alpha\,.
\end{equation}
It is interesting to note that (as mentioned in the previous section) $K^\alpha$ has zero rotation with respect to the zero-angular momentum observers at infinity. There are some works in the literature (\eg,~\cite{Gwak:2018akg,Gwak:2018tmy}) where an expression analogous to Eq.~\eqref{energyKerr} is used to calculate the energy of test fields propagating on Kerr-dS, but, this time, using the Killing vector field 
\begin{equation}\label{killing2}
K^\alpha=X^\alpha+\dfrac{a}{l^2}Y^\alpha\,. 
\end{equation}
However, to the best of my knowledge, in the literature there is neither a rigorous proof nor a clear physical motivation for the use of this definition of energy. In what follows we will show that, in our particular setup, the definition of Eq.~\eqref{energyKerr} gives $\Delta M$ in both asymptotically AdS and dS spacetimes, if one uses the corresponding Killing vector field $K^\alpha$, defined by either \eqref{killing} or \eqref{killing2}, respectively.

\begin{figure}
	\includegraphics[width=\textwidth]{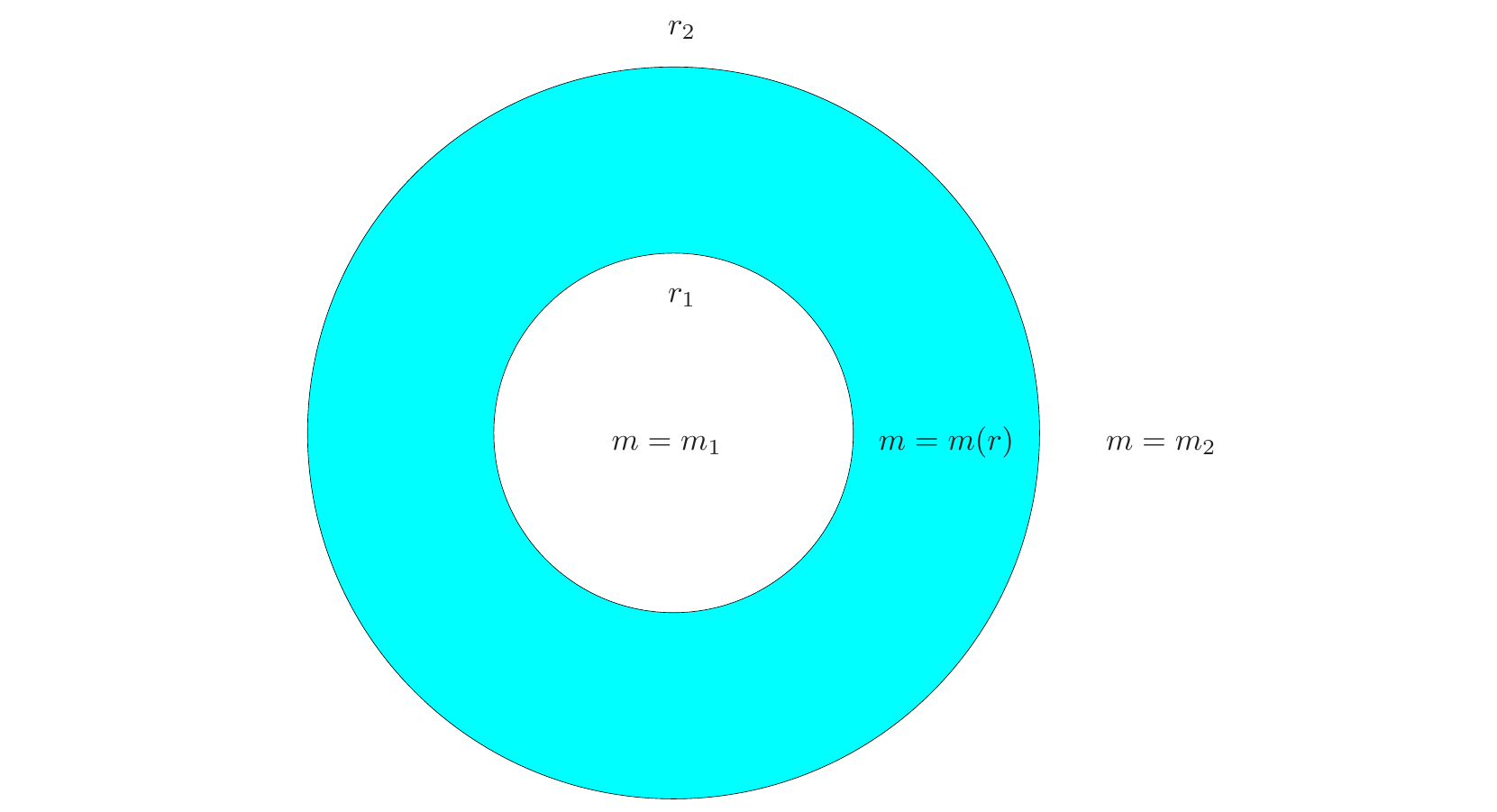}
	\caption{Schematic diagram for the spacetime interpolating between Kerr-Newman-(A)dS metrics with different masses.} \label{interpolating}
\end{figure}

Since the metric $g_{\alpha \beta}$ is known (by construction), the energy-momentum tensor $T^{\alpha \beta}$ of the field is obtained from the Einstein equations as
\begin{equation} \label{stresstensor}
T^{\alpha \beta}=\frac{1}{8 \pi}(G^{\alpha \beta} + \Lambda g^{\alpha \beta})\,,
\end{equation}
where $G^{\alpha \beta}$ is the Einstein tensor. Computing $G^{\alpha \beta}$ explicitly, and substituting the last expression in Eq.~\eqref{energyKerr}, we obtain
\begin{equation}\label{energyr}
\widetilde{E}=\int_{r_1}^{r_2} \left[\mathcal{A}(r) m'(r)+ \mathcal{B}(r) m''(r)\right]dr\,,
\end{equation}
where we have chosen a hypersurface $\mathcal{S}$ of constant~$t$ extending from~$r_1$ to~$r_2$, and performed the integrations in~$\theta$ and~$\varphi$. The radial functions~$\mathcal{A}$ and~$\mathcal{B}$ are given by
\begin{align}
&\hspace{-0.6cm}\mathcal{A}= \mp \frac{l^2}{ a (a^2\pm l^2)^2}\left[2 a (r^2\mp l^2) -   \arctan\left(\frac{a}{r}\right)r\left( a^2 \mp l^2+2r^2\right)\right]\,,\\
&\hspace{-0.6cm}\mathcal{B}= \mp \frac{l^2}{2 a (a^2\pm l^2)^2}\left[a r \left(r^2 \mp l^2\right)-\arctan\left(\frac{a}{r}\right)\left(a^2+r^2\right)\left(r^2 \mp l^2\right) \right]\,.
\end{align}
Integrating Eq.~\eqref{energyr} by parts, we obtain
\begin{align}
\widetilde{E}&=\int_{r_1}^{r_2} \left[\mathcal{B}''-\mathcal{A}'\right] m\, dr +\biggl[\left(\mathcal{A}-\mathcal{B}'\right)m+\mathcal{B} m'\biggr]_{r_1}^{r_2}\,.
\end{align}
Using~$\mathcal{B}''(r)=\mathcal{A}'(r)$,~$m'(r_2)=m'(r_1)=0$, and~$\mathcal{A}(r)-\mathcal{B}'(r)=\dfrac{1}{\Xi^2}$, the last expression becomes
\begin{equation} \label{energyKerrconsistent}
\widetilde{E}=\frac{m_2-m_1}{\Xi^2}=M_2-M_1\equiv \Delta M \, ,
\end{equation}
as we wanted to show. 

We can also calculate the field angular momentum~$L$ as an integral on a given spacelike hypersurface~$\mathcal{S}$ extending from~$r=r_1$ to~$r=r_2$. This can be done in Kerr-AdS (at least for test fields), where it is known that
\begin{equation}\label{angularmKerr}
\widetilde{L}=-\int_\mathcal{S} dV_3 T^{\mu \nu}Y_\mu N_\nu 
\end{equation}
(note the minus sign in the integral, since we are using the future-pointing unit timelike normal but now the Killing vector field is spacelike). In our particular setup, we know what the angular momentum of the field should be; since the physical angular momenta, $L_1=a M_1$ and $L_2=a M_2$, correspond to the total angular momentum contained in the regions $r<r_1$ and $r<r_2$, respectively, the angular momentum of the field should be $\widetilde{L}=\Delta L\equiv L_2-L_1$. We will now show that, in our setup, the definition of Eq.~\eqref{angularmKerr} does indeed give~$\Delta L$ in both asymptotically AdS and dS spacetimes.
Computing $G^{\alpha \beta}$ explicitly, and substituting Eq.~\eqref{stresstensor} in the definition of Eq.~\eqref{angularmKerr}, we obtain
\begin{equation}\label{angulamr}
\widetilde{L}=\int_{r_1}^{r_2}\left[\mathcal{C}(r) m'(r)+ \mathcal{D}(r) m''(r)\right]dr\,,
\end{equation}
where again we have chosen a hypersurface~$\mathcal{S}$ of constant~$t$ extending from~$r_1$ to~$r_2$, and performed the integrations in~$\theta$ and~$\varphi$. The radial functions~$\mathcal{C}$ and~$\mathcal{D}$ are given by 
\begin{align}
\mathcal{C}&= 2 l^4 \frac{a^2+r^2}{ a^2 (a^2\pm l^2)^2}\left[a-r  \arctan\left(\frac{a}{r}\right)\right]\,,\\
\mathcal{D}&= l^4 \frac{a^2+r^2}{2 a^2 (a^2\pm l^2)^2}\left[a r -\arctan\left(\frac{a}{r}\right)\left(a^2+r^2\right) \right]\,.
\end{align}
Integrating Eq.~\eqref{angulamr} by parts, we obtain
\begin{align}
\widetilde{L}&=\int_{r_1}^{r_2}\left[\mathcal{D}''-\mathcal{C}'\right]m\,dr+\biggl[\left(\mathcal{C}-\mathcal{D}'\right)m+\mathcal{D} m'\biggr]_{r_1}^{r_2}\,.
\end{align}
Using $\mathcal{D}''(r)=\mathcal{C}'(r)$, $m'(r_2)=m'(r_1)=0$, and $\mathcal{C}(r)-\mathcal{D}'(r)=\dfrac{a}{\Xi^2}$, the last expression becomes
\begin{equation} \label{angularmKerrconsistent}
\widetilde{L}=a\frac{m_2-m_1}{\Xi^2}=a (M_2-M_1)\equiv \Delta L \, ,
\end{equation}
as we wanted to show.  As a consequence, the energy of the unphysical field computed by using any timelike Killing vector field of the form 
\begin{equation}
K^\alpha + \omega Y^\alpha = X^\alpha + \left(\omega \pm \frac{a}{l^2}\right) Y^\alpha
\end{equation}
is
\begin{equation}
\widetilde{E} + \omega \widetilde{L} = (1+\omega a) \Delta M\,,
\end{equation}
strongly suggesting that $K^\alpha$ (that is, $\omega=0$) is in fact the correct choice. We will have more to say about the uniqueness of~$K$ in Subsection~\ref{subsection4}.

\subsection{Kerr-Newman-(A)dS}\label{subsection2}
Here we construct a metric that interpolates between two Kerr-Newman-(A)dS regions of different (physical) masses~$M_1$ and~$M_2$ and (physical) charges~$Q_1$ and~$Q_2$ by letting both the mass and the charge parameters become functions of the radial coordinate~$r$. We then determine, from the Einstein equations, the energy-momentum tensor of the (unphysical) field generating this metric, and use it to compute the corresponding energy with respect to a given timelike Killing vector field. This energy, appropriately corrected by the electromagnetic field energy, is seen to be precisely the difference~$M_2-M_1$ between the two physical masses for the particular choice of Killing vector field given by Eqs.~\eqref{killing} and~\eqref{killing2}, thus generalizing the results in~Subsection~\ref{subsection1}.

Let us then take the charge parameter~$q(r)$ to be changing in the region~$r_1<r<r_2$, with~$q(r)\equiv q_1$ for~$r \leq r_1$ and~$q(r)\equiv q_2$ for~$r \geq r_2$. Moreover, assume that~$q'(r_1)= q'(r_2)=0$, and again that~$\Delta_r(r)>0$~for~$r_1\leq r\leq r_2$. In this case we have an electromagnetic field with energy-momentum tensor $T^{\alpha \beta}_\text{EM}$, and it is not obvious what the mass contained on a spacelike hypersurface $\mathcal{S}$ extending from~$r_1$ to~$r_2$ should be. In the asymptotically flat case, it is well known that the physical mass accounts also for the electromagnetic energy in the whole spacetime. By analogy, the energy contained on a spacelike hypersurface~$\mathcal{S}$ extending from~$r_1$ to~$r_2$ should then be
\begin{equation} \label{EKNAsS}
\widetilde{E}=\left(M_2-\int_{r>r_2} T_\text{EM,2}^{\mu \nu}K_\nu N_\mu dV_3\right)-\left(M_1-\int_{r>r_1} T_\text{EM,1}^{\mu \nu}K_\nu N_\mu dV_3\right)\,,
\end{equation}
where the first term is the mass contained in~$r<r_2$, and the second term is the mass in~$r<r_1$. Here,~$T^{\alpha \beta}_\text{EM,1}$ and $T^{\alpha \beta}_\text{EM,2}$ are the energy-momentum tensors of the electromagnetic field in a Kerr-Newman-(A)dS spacetime with mass parameters~$m_1$ and~$m_2$, and charge parameters~$q_1$ and~$q_2$, respectively. Note that in~\eqref{EKNAsS} we have already made use of the Killing vector field~$K^\alpha$ to calculate the electromagnetic energy. On the other hand, the energy contained on $\mathcal{S}$ should be directly
\begin{equation} \label{energyKerrN}
\widetilde{E}=\int_\mathcal{S} \left(T^{\mu \nu}+T_\text{EM}^{\mu \nu}\right)K_\mu N_\nu dV_3\,,
\end{equation}
where $T^{\alpha \beta}_\text{EM}$ is the energy-momentum tensor of the electromagnetic field in the Kerr-Newman-(A)dS spacetime with varying mass parameter~$m(r)$ and varying charge parameter~$q(r)$. Thus, if our definition of energy is to be consistent, we must have
\begin{align}\label{deltam}
\Delta M&=\int_\mathcal{S} \left(T^{\mu \nu}+T_\text{EM}^{\mu \nu}\right)K_\mu N_\nu dV_3 \nonumber\\
&+\int_{r>r_2} T_\text{EM,2}^{\mu \nu}K_\nu N_\mu dV_3-\int_{r>r_1} T_\text{EM,1}^{\mu \nu}K_\nu N_\mu dV_3 \,.
\end{align}
Again, since the metric $g_{\alpha \beta}$ is known, the Einstein equations imply that
\begin{equation} \label{stresstensorem}
T^{\alpha \beta}+T^{\alpha \beta}_\text{EM}=\frac{1}{8 \pi}\left(G^{\alpha \beta}+\Lambda g^{\alpha \beta}\right)\,.
\end{equation}
Computing $G^{\alpha \beta}$ explicitly, and using Eq.~\eqref{stresstensorem}, allows us to write the first integral in Eq.~\eqref{deltam} as
\begin{align} \label{FirstTermEn}
&\hspace{-0.7cm}\int_\mathcal{S} \left(T^{\mu \nu}+T_\text{EM}^{\mu \nu}\right)K_\mu N_\nu dV_3\nonumber\\ &\hspace{-0.7cm}=\int_{r_1}^{r_2} dr \left[\mathcal{A}(r) m'+ \mathcal{B}(r) m''+\mathcal{E}(r) q^2-r \mathcal{E}(r) (q^2)'-\frac{\mathcal{B}(r)}{2 r}(q^2)''\right]\,,
\end{align}
where again we have chosen an hypersurface~$\mathcal{S}$ of constant~$t$ extending from~$r_1$ to~$r_2$, and performed the integrations in~$\theta$ and~$\varphi$. The radial functions~$\mathcal{A}$ and~$\mathcal{B}$ are defined as in the last subsection, and
\begin{equation}
\mathcal{E}= \mp \frac{l^2}{2 a r^3 (a^2 \pm l^2)^2}\left[a r \left(r^2 \mp l^2\right)-\arctan\left(\frac{a}{r}\right)\left(r^4\pm a^2 l^2\right) \right]\, .
\end{equation}
Integrating by parts, and using the results of the last section, we have
\begin{align} 
&\hspace{-1cm}\int_\mathcal{S} \left(T^{\mu \nu}+T_\text{EM}^{\mu \nu}\right)K_\mu N_\nu dV_3 =\Delta M \nonumber\\
&\hspace{-1cm}\quad+\int_{r_1}^{r_2} dr \bigg[\mathcal{E}+\left(r \mathcal{E}\right)'-\bigg(\frac{\mathcal{B}}{2 r}\bigg)'' \bigg]q^2+  \bigg[\bigg(\bigg[\frac{\mathcal{B}}{2 r}\bigg]'-r \mathcal{E}\bigg)q^2-\frac{\mathcal{B}}{2 r}(q^2)'\bigg]_{r_1}^{r_2} .
\end{align}
Using $q'(r_1)= q'(r_2)=0$, and
\begin{equation}\label{primc}
\mathcal{E}=\bigg[\bigg(\frac{\mathcal{B}}{2 r}\bigg)'-r \mathcal{E}\bigg]'\,,
\end{equation}
we obtain
\begin{equation}
\int_\mathcal{S} \left(T^{\mu \nu}+T_\text{EM}^{\mu \nu}\right)K_\mu N_\nu dV_3=\Delta M+\bigg[\bigg(\left[\frac{\mathcal{B}}{2 r}\right]'-r \mathcal{E}\bigg)q^2\bigg]_{r_1}^{r_2} \,.
\end{equation}
Furthermore, the last two terms of Eq.~\eqref{deltam} are
\begin{align} \label{last}
&\int_{r>r_2} T_\text{EM,2}^{\mu \nu}K_\nu N_\mu dV_3-\int_{r>r_1} T_\text{EM,1}^{\mu \nu}K_\nu N_\mu dV_3\nonumber\\
&\quad=(q_2)^2\int_{r_2}^{\infty} dr \,\mathcal{E}(r)-(q_1)^2\int_{r_1}^{\infty} dr \,\mathcal{E}(r)\,,
\end{align}
where we used Eq.~\eqref{FirstTermEn}, with $m\equiv m_ 2$ ($m\equiv m_1$), $q\equiv q_2$ ($q\equiv q_1$) in the first (second) term, but integrating on a spacelike hypersurface of constant $t$ with $r>r_2$ ($r>r_1$).  In the Kerr-Newman-dS case, a hypersurface of constant $t$ is not spacelike beyond the cosmological horizon; nevertheless, since we are integrating a divergenceless quantity, any unbounded spacelike hypersurface can be deformed into the union of a spacelike hypersurface of constant~$t$ within the cosmological horizon and a timelike hypersurface of constant~$t$ beyond the cosmological horizon (see Figure~\ref{Penrose1}).

\begin{figure}
	\includegraphics[width=\textwidth]{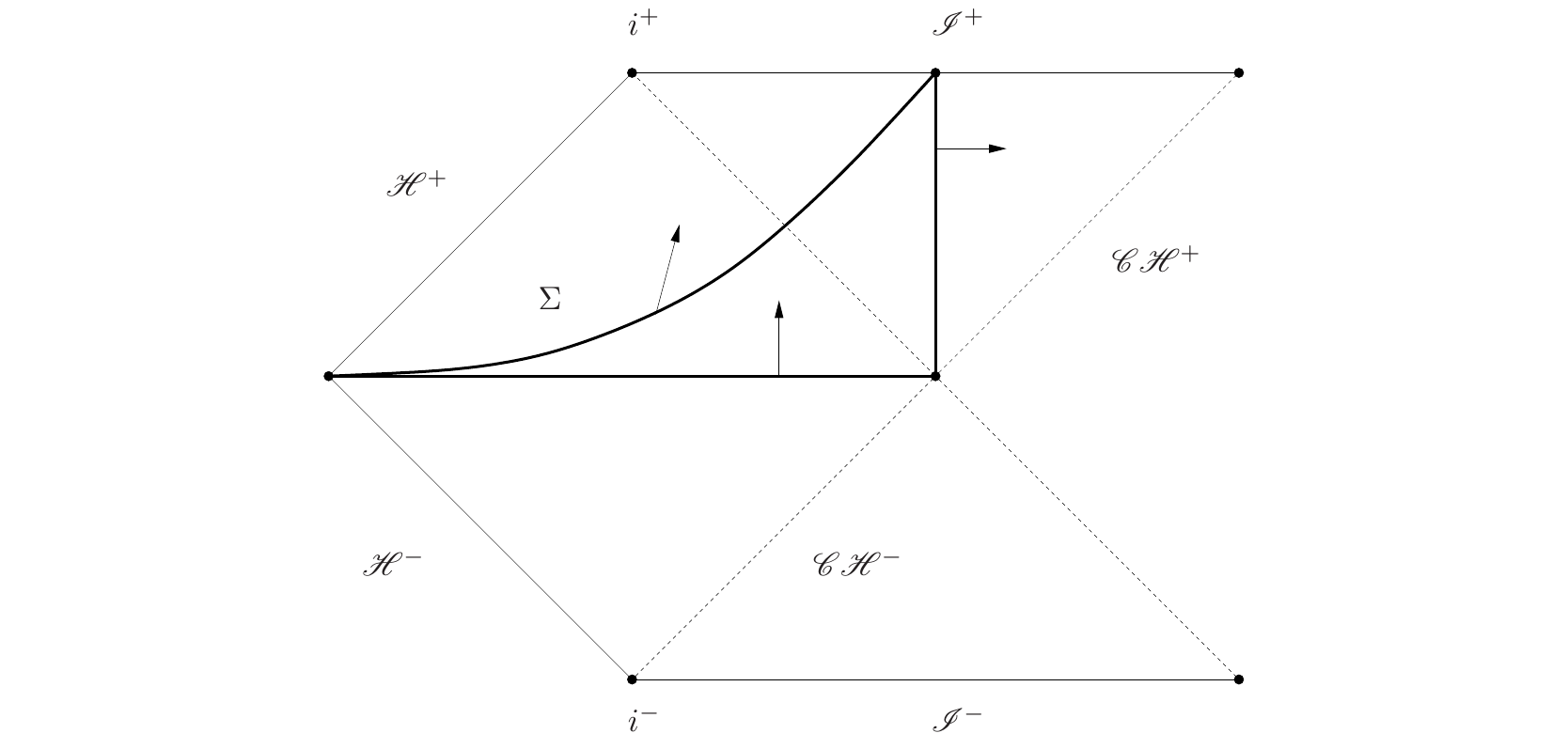}
	\caption{Penrose diagram illustrating the deformation of an unbounded spacelike hypersurface $\Sigma$ into the union of two hypersurfaces of constant $t$, with the corresponding unit normals depicted.} \label{Penrose1}
\end{figure}

Using~\eqref{primc}, Eq.~\eqref{last} becomes
\begin{align}
&\int_{r>r_2} T_\text{EM,2}^{\mu \nu}K_\nu N_\mu dV_3-\int_{r>r_1} T_\text{EM,1}^{\mu \nu}K_\nu N_\mu dV_3\nonumber\\
&\qquad =(q_2)^2\bigg[\bigg(\bigg[\frac{\mathcal{B}}{2 r}\bigg]'-r \mathcal{E}\bigg)\bigg]_{r_2}^{\infty}-(q_1)^2\bigg[\bigg(\bigg[\frac{\mathcal{B}}{2 r}\bigg]'-r \mathcal{E}\bigg)\bigg]_{r_1}^{\infty} \nonumber \\
&\qquad =-\bigg[\bigg(\bigg[\frac{\mathcal{B}}{2 r}\bigg]'-r \mathcal{E}\bigg)q^2\bigg]_{r_1}^{r_2}\,,
\end{align}
where in the last equality we used 
\begin{equation*}
\lim_{r\to \infty}\bigg(\bigg[\frac{\mathcal{B}(r)}{2 r}\bigg]'-r \mathcal{E}(r)\bigg)=0\,.
\end{equation*}
Putting everything together, we finally obtain
\begin{align} \label{energyKerrNconsistent}
\int_\mathcal{S} \left(T^{\mu \nu}+T_\text{EM}^{\mu \nu}\right)K_\mu N_\nu dV_3&+\int_{r>r_2} T_\text{EM,2}^{\mu \nu}K_\nu N_\mu dV_3 \nonumber \\
&-\int_{r>r_1} T_\text{EM,1}^{\mu \nu}K_\nu N_\mu dV_3=\Delta M\,,
\end{align}
showing that our definition of energy is indeed consistent.

In the same way, the angular momentum contained on $\mathcal{S}$ should be
\begin{equation}
\widetilde{L}=\left(L_2+\int_{r>r_2} T_\text{EM,2}^{\mu \nu}Y_\nu N_\mu dV_3\right)-\left(L_1+\int_{r>r_1} T_\text{EM,1}^{\mu \nu}Y_\nu N_\mu dV_3\right)\,,
\end{equation}
where the first term is the angular momentum contained in $r<r_2$, and the second term is the angular momentum contained in $r<r_1$ (note the minus sign in the integral, since we are using the future-pointing unit timelike normal but now the Killing vector field is spacelike). On the other hand, the angular momentum contained on $\mathcal{S}$ should be directly
\begin{equation} \label{angularmKerrN}
\widetilde{L}=-\int_\mathcal{S} \left(T^{\mu \nu}+T_\text{EM}^{\mu \nu}\right)Y_\mu N_\nu dV_3\,.
\end{equation}
Thus, if this definition of angular momentum is to be consistent, the relation
\begin{align}\label{deltaj}
\Delta L&=-\int_\mathcal{S} \left(T^{\mu \nu}+T_\text{EM}^{\mu \nu}\right)Y_\mu N_\nu dV_3\nonumber \\
&-\int_{r>r_2} T_\text{EM,2}^{\mu \nu}Y_\nu N_\mu dV_3+\int_{r>r_1} T_\text{EM,1}^{\mu \nu}Y_\nu N_\mu dV_3
\end{align}
must hold. Computing $G^{\alpha \beta}$ explicitly, and using Eq.~\eqref{stresstensorem}, allows us to write the first integral in Eq.~\eqref{deltaj} as
\begin{align} \label{FirstTermAn}
&\hspace{-0.7cm}-\int_\mathcal{S} \left(T^{\mu \nu}+T_\text{EM}^{\mu \nu}\right)Y_\mu N_\nu dV_3\nonumber\\ &\hspace{-0.7cm}=\int_{r_1}^{r_2} dr \left[\mathcal{C}(r) m'+ \mathcal{D}(r) m''+\mathcal{F}(r) q^2-r \mathcal{F}(r) (q^2)'-\frac{\mathcal{D}(r)}{2 r}(q^2)''\right]\,,
\end{align}
where again we have chosen a hypersurface $\mathcal{S}$ of constant~$t$ extending from~$r_1$ to~$r_2$, and performed the integrations in~$\theta$ and~$\varphi$. The radial functions~$\mathcal{C}$ and~$\mathcal{D}$ are defined as in the last subsection, and
\begin{equation}
\mathcal{F}= l^4 \frac{a^2+r^2}{2 a^2 r^3 (a^2 \pm l^2)^2}\left[a r +\arctan\left(\frac{a}{r}\right)\left(a^2- r^2\right) \right]\,.
\end{equation}
Integrating by parts, and using the results in the last subsection, we have
\begin{align} 
&\hspace{-1cm}-\int_\mathcal{S} \left(T^{\mu \nu}+T_\text{EM}^{\mu \nu}\right)Y_\mu N_\nu dV_3=\Delta L\nonumber\\
&\hspace{-0.9cm}\;\;+\int_{r_1}^{r_2} dr \bigg[\mathcal{F}+\left(r \mathcal{F}\right)'-\bigg(\frac{\mathcal{D}}{2 r}\bigg)'' \bigg]q^2+  \bigg[\bigg(\bigg[\frac{\mathcal{D}}{2 r}\bigg]'-r \mathcal{F}\bigg)q^2-\frac{\mathcal{D}}{2 r}(q^2)'\bigg]_{r_1}^{r_2} .
\end{align}
Using $q'(r_1)= q'(r_2)=0$, and
\begin{equation}\label{primf}
\mathcal{F}=\bigg[\bigg(\frac{\mathcal{D}}{2 r}\bigg)'-r \mathcal{F}\bigg]'\,,
\end{equation}
we have
\begin{equation}
\hspace{-0.3cm}-\int_\mathcal{S} \left(T^{\mu \nu}+T_\text{EM}^{\mu \nu}\right)Y_\mu N_\nu dV_3=\Delta L+\bigg[\bigg(\bigg[\frac{\mathcal{D}}{2 r}\bigg]'-r \mathcal{F}\bigg)q^2\bigg]_{r_1}^{r_2} \,.
\end{equation}
Moreover, the last two integrals of Eq.~\eqref{deltaj} are
\begin{align}
&-\int_{r>r_2} T_\text{EM,2}^{\mu \nu}Y_\nu N_\mu dV_3+\int_{r>r_1} T_\text{EM,1}^{\mu \nu}Y_\nu N_\mu dV_3 \nonumber \\
&\qquad=(q_2)^2\int_{r_2}^{\infty} dr \,\mathcal{F}(r)-(q_1)^2\int_{r_1}^{\infty} dr \,\mathcal{F}(r)\,,
\end{align}
where we have used Eq.~\eqref{FirstTermAn}, with~$m\equiv m_ 2$ ($m\equiv m_1$),~$q\equiv q_2$ ($q\equiv q_1$) in the first (second) term, but integrating on a spacelike hypersurface of constant~$t$ with~$r>r_2$ ($r>r_1$). 
Using Eq.~\eqref{primf}, the last expression becomes
\begin{align}
&-\int_{r>r_2} T_\text{EM,2}^{\mu \nu}Y_\nu N_\mu dV_3+\int_{r>r_1} T_\text{EM,1}^{\mu \nu}Y_\nu N_\mu dV_3\nonumber\\
&\qquad=(q_2)^2\bigg[\bigg(\bigg[\frac{\mathcal{D}}{2 r}\bigg]'-r \mathcal{F}\bigg)\bigg]_{r_2}^{\infty}-(q_1)^2\bigg[\bigg(\bigg[\frac{\mathcal{D}}{2 r}\bigg]'-r \mathcal{F}\bigg)\bigg]_{r_1}^{\infty}\nonumber \\
&\qquad=-\bigg[\bigg(\bigg[\frac{\mathcal{D}}{2 r}\bigg]'-r \mathcal{F}\bigg)q^2\bigg]_{r_1}^{r_2}\,,
\end{align}
where, in the last equality, we used 
\begin{equation*}
\lim_{r\to \infty}\bigg(\bigg[\frac{\mathcal{D}(r)}{2 r}\bigg]'-r \mathcal{F}(r)\bigg)=0\,.
\end{equation*}
Putting everything together, we finally obtain
\begin{align} \label{angularmKerrNconsistent}
-\int_\mathcal{S} \left(T^{\mu \nu}+T_\text{EM}^{\mu \nu}\right)Y_\mu N_\nu dV_3&-\int_{r>r_2} T_\text{EM,2}^{\mu \nu}Y_\nu N_\mu dV_3\nonumber \\
&+\int_{r>r_1} T_\text{EM,1}^{\mu \nu}Y_\nu N_\mu dV_3=\Delta L\,,
\end{align}
showing that our definition of angular momentum is indeed consistent. As a consequence, a timelike Killing vector field of the form 
\begin{equation}
K^\alpha + \omega Y^\alpha = X^\alpha + \left(\omega \pm \frac{a}{l^2}\right) Y^\alpha
\end{equation}
will again only satisfy Eq.~\eqref{energyKerrNconsistent} if $\omega a =0$, strongly suggesting that $K^\alpha$ (that is,~$\omega=0$) is in fact the correct choice. The uniqueness of~$K^\alpha$ will be further discussed in Subsection~\ref{subsection4}.

\subsection{Linearized calculation}\label{subsection3}
In the previous subsections we showed that there exists a timelike Killing vector field~$K^\alpha$, given by Eqs.~\eqref{killing} and~\eqref{killing2}, such that the definitions in Eqs.~\eqref{energyKerr} and~\eqref{energyKerrN} give the correct total energy~$\widetilde{E}$ contained in the (unphysical) field that is generated by allowing the mass and charge parameters to become functions of the radial coordinate. This energy is related to the variation~$\Delta M = M_2 - M_1$ of the physical mass by Eqs.~\eqref{energyKerrconsistent} and~\eqref{energyKerrNconsistent}. However, the Killing vector field~$K^\alpha$ is defined on a unphysical stationary spacetime that coincides with Kerr-Newman-(A)dS spacetimes of mass and charge parameters~$m_1$ and~$q_1$ for~$r \leq r_1$, and mass and charge parameters~$m_2$ and~$q_2$ for~$r \geq r_2$, whereas our aim is to identify the timelike Killing vector field that gives the correct definition of energy of {\em test fields} on a {\em fixed} Kerr-Newman-(A)dS background. 

To achieve this goal, we consider a solution of the linearized Einstein-Maxwell equations, possibly coupled to matter, on a Kerr-Newman-(A)dS background of mass and charge parameters~$m_1$ and~$q_1$, vanishing for~$r \leq r_1$ and coinciding with the linearized Kerr-Newman-(A)dS solution of mass and charge parameters~$m_2 = m_1 + \Delta m$ and~$q_2 = q_1 + \Delta q$ for~$r \geq r_2$ (and the same spin parameter~$a$); if the energy computed from the linearized energy-momentum tensor with respect to the Killing vector field~$K^\alpha$ (which is now defined on the fixed Kerr-Newman-(A)dS background of mass and charge parameters~$m_1$ and~$q_1$) is~$\Delta M=\Delta m / \Xi^2$ then $K^\alpha$ does indeed give the correct definition of energy. Note that one such linearized solution, albeit for unphysical matter, can be obtained by linearizing the spacetime constructed in the previous sections; as we have shown, the Killing vector field~$K^\alpha$ does give the correct energy in this case.  A simple application of the divergence theorem then shows that~$K^\alpha$ will give the same energy for any other linearized solution, including solutions corresponding to physical matter fields. Indeed, if~$\delta g_{\alpha \beta}(t,r,\theta,\varphi)$ is an arbitrary linearized metric, $\delta g_{\alpha \beta}^0(r,\theta)$ is the linearization of the metric constructed in the previous subsections, and~$\rho(t)$ is a smooth function satisfying~$\rho(t) \equiv 1$ for~$t \leq 0$ and~$\rho(t) \equiv 0$ for~$t \geq 1$, consider the linearized metric~$\rho(t-t_0) \delta g_{\alpha \beta} + (1-\rho(t-t_0)) \delta g_{\alpha \beta}^0$. The linearized energy-momentum tensor corresponding to this metric has zero divergence in the Kerr-Newman-(A)dS background, coincides with the energy-momentum tensor of the arbitrary linearized metric for~$t=t_0$, and with the energy-momentum tensor of~$\delta g_{\alpha \beta}^0$ for $t=t_0+1$. Moreover, it vanishes for~$r \leq r_1$ and it is time-independent for~$r \geq r_2$ (so in particular does not depend on the choice of~$\delta g_{\alpha \beta}$ in those regions). Applying the divergence theorem to the vector field~$J^\alpha=\big(T^{\alpha \beta}+T_\text{EM}^{\alpha \beta}\big)K_\beta$ in the hollow cylinder defined by $r_1 \leq r \leq r_2$ and $t_0 \leq t \leq t_0 + 1$ (see Fig.~\ref{divergence}), we obtain
\begin{align}
& \int_{r_1 < r < r_2} \left(T^{\mu \nu}+T_\text{EM}^{\mu \nu}\right)K_\mu N_\nu dV_3 - \int_{r_1 < r < r_2} \left(T_0^{\mu \nu}+T_{\text{EM},0}^{\mu \nu}\right)K_\mu N_\nu dV_3 \nonumber \\
& - \int_{r = r_1} \left(T^{\mu \nu}+T_\text{EM}^{\mu \nu}\right)K_\mu N_\nu dV_3 + \int_{r = r_2} \left(T^{\mu \nu}+T_\text{EM}^{\mu \nu}\right)K_\mu N_\nu dV_3 = 0\,,
\end{align}
where the unit normal $N_\alpha$ is future-pointing when timelike and outward-pointing when spacelike, and the energy-momentum tensor~$T_0^{\alpha \beta}+T_{\text{EM},0}^{\alpha \beta}$ refers to~$\delta g_{\alpha \beta}^0$. Since the last two integrals do not depend on the choice of~$\delta g_{\alpha \beta}$, and their sum clearly vanishes when one chooses~$\delta g_{\alpha \beta} = \delta g_{\alpha \beta}^0$ (because the first two integrals cancel in that case), it always vanishes; therefore we obtain
\begin{equation}
\int_{r_1 < r < r_2} \left(T^{\mu \nu}+T_\text{EM}^{\mu \nu}\right)K_\mu N_\nu dV_3 = \int_{r_1 < r < r_2} \left(T_0^{\mu \nu}+T_{\text{EM},0}^{\mu \nu}\right)K_\mu N_\nu dV_3\,,
\end{equation}
showing tha~ $K^\alpha$ does indeed yield the correct energy for any linearized solution.

\begin{figure}
	\includegraphics[width=\textwidth]{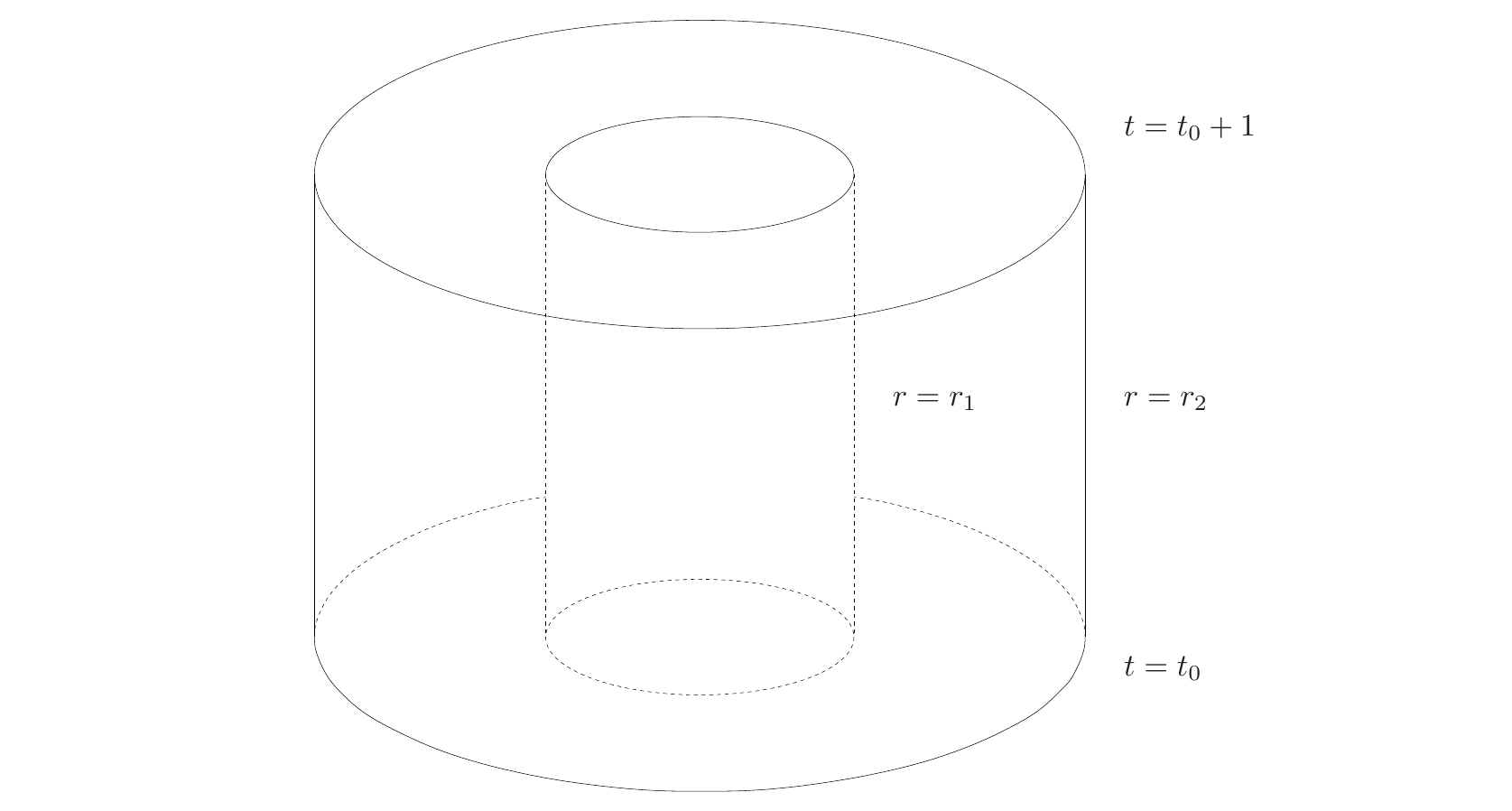}
	\caption{Domain for the application of the divergence theorem.} \label{divergence}
\end{figure}

\subsection{Uniqueness of~$K^\alpha$}\label{subsection4}
We have now identified a timelike Killing vector field~$K^\alpha$ in the Kerr-Newman-(A)dS spacetime, given by Eqs.~\eqref{killing} and~\eqref{killing2}, such that the definitions in~Eqs.~\eqref{energyKerr} and~\eqref{energyKerrN} give the correct total energy~$\widetilde{E}$ contained in linearized (test) fields. This energy is related to the variation~$\Delta M = M_2 - M_1$ of the physical mass by Eqs.~\eqref{energyKerrconsistent} and~\eqref{energyKerrNconsistent}. Similarly, the definitions in Eqs.~\eqref{angularmKerr} and~\eqref{angularmKerrN} give the correct total angular momentum $\widetilde{L}$ in the test fields, which is related to the variation~$\Delta L = a \Delta M$ of the angular momentum by Eqs.~\eqref{angularmKerrconsistent} and~\eqref{angularmKerrNconsistent}. However, because the variations of energy and angular momentum are related through the spin parameter~$a$, which we did not vary, the possibility that~$K^\alpha$ is not unique remains.

To understand this, we note that any other future-pointing timelike Killing vector field can be written in the form
\begin{equation}
\tilde{K}^\alpha = \gamma \left(K^\alpha+\epsilon Y^\alpha\right) \,,
\end{equation}
with~$\gamma > 0$ and~$\epsilon \in \mathbb{R}$ appropriately chosen. Combining Eqs.~\eqref{energyKerrNconsistent} and~\eqref{angularmKerrNconsistent}, we see that~$\tilde{K}^\alpha$ will also give the correct total energy~$\widetilde{E}$ contained in the unphysical field if and only if
\begin{equation}
\gamma \Delta M - \gamma \epsilon \Delta L = \Delta M \Leftrightarrow \gamma (1 - \epsilon a) = 1 \, ,
\end{equation}
that is, if and only if
\begin{equation}
\tilde{K}^\alpha = \frac{1}{1-a \epsilon(a)}\left[K^\alpha+\epsilon(a) Y^\alpha\right] \,,
\end{equation}
where we made it explicit that~$\epsilon$ is an unknown function of~$a$. To show that~$\epsilon(a)$ must be identically zero, and therefore that~$K^\alpha$ is unique, we allow the spin parameter~$a$ to become a function of~$r$ in the region~$r_1\leq r\leq r_2$, while keeping the mass and charge parameters fixed. To perform the linearization, we assume that~$a(r)= a_0+\delta a(r)$ varies infinitesimally between~$a(r_1)= a_0$ and~$a(r_2)= a_0+\Delta a$ (\ie,~$|\delta a(r)| \ll a_0$). Since in this case the calculations are much more involved than in the previous subsections, we assume that all quantities are analytic functions of~$a$ and expand them as power series of~$a_0$ (and to linear order in~$\delta a(r)$). In particular, we have
\begin{equation}
\epsilon(a)=\sum_{n=0}^{+\infty} \epsilon_n a^n \,.
\end{equation} 
In what follows we will show that~$\epsilon_0=\epsilon_1=0$. Due to the complexity of the calculations, we have not computed the higher order coefficients~$\epsilon_n$ with~$n\geq 2$, but we expect them to also vanish.

To further simplify calculations we consider only the Kerr-(A)dS case~$q_1=q_2=0$. Using the definition of Eq.~\eqref{energyKerr} with the Killing vector field~$\tilde{K}(a_0)$, and applying the same procedure of the previous subsections, we obtain the radial integral
\begin{equation} \label{energyreps}
\widetilde{E} = \int_\mathcal{S} T^{\mu \nu}(\tilde{K}(a_0))_\mu N_\nu dV_3=\int_{r_1}^{r_2}dr[\mathcal{G}(r)\delta a'(r)+\mathcal{H}(r)\delta a''(r)]\,,
\end{equation}
with the radial functions
\begin{align}
\mathcal{G}&= -\frac{1}{15 l^4 r^3}\left[\left(10 \epsilon_0 l^4 m r^3 \mp 10 \epsilon_0 l^2 r^6\right)\nonumber\right.\\ &\left.\quad+a_0 \Big(10 \epsilon_0^2 l^4 m r^3\mp 10 \epsilon_0^2 l^2 r^6+10 \epsilon_1 l^4 m r^3\nonumber\right.\\
&\left.\quad\mp 10 \epsilon_1 l^2 r^6+20 l^4 m^2+10 l^4 m r\pm 75 l^2 m r^3\mp 40 l^2 r^4+40 r^6\Big)\nonumber\right.\\ &\left.\quad+a_0^2 \Big(20 \epsilon_0 \epsilon_1 l^4 m r^3\mp 20 \epsilon_0 \epsilon_1 l^2 r^6+4 \epsilon_0 l^4 m^2\nonumber \right.\\
&\left.\quad+2 \epsilon_0 l^4 m r\mp 5 \epsilon_0 l^2 m r^3\mp 50 \epsilon_0 l^2 r^4+52 \epsilon_0 r^6\Big)+ \mathcal{O}(a_0^3)  \right]\,,\\
\mathcal{H}&= \frac{1}{30 l^4 r^2}\left[\left(10 \epsilon_0 l^4 m r^3\pm 5 \epsilon_0 l^2 r^6\right)\nonumber \right.\\&\left.\quad+a_0 \Big(10 \epsilon_0^2 l^4 m r^3\pm 5 \epsilon_0^2 l^2 r^6+10 \epsilon_1 l^4 m r^3\pm 5 \epsilon_1 l^2 r^6\nonumber \right. \\
&\left. \quad+20 l^4 m^2+20 l^4 m r-20 l^4 r^2\mp 30 l^2 m r^3\pm 40 l^2 r^4-20 r^6\Big)\nonumber \right.\\&\left.\quad+a_0^2 \Big(20 \epsilon_0 \epsilon_1 l^4 m r^3\pm 10 \epsilon_0 \epsilon_1 l^2 r^6+4 \epsilon_0 l^4 m^2+4 \epsilon_0 l^4 m r\nonumber \right. \\
&\left.\quad-20 \epsilon_0 l^4 r^2\mp 50 \epsilon_0 l^2 m r^3\pm 50 \epsilon_0 l^2 r^4-26 \epsilon_0 r^6\right)+\mathcal{O}(a_0^3)\Big]\,.
\end{align}
Integrating Eq.~\eqref{energyreps} by parts, we obtain 
\begin{align}
\widetilde{E}&=\int_{r_1}^{r_2}dr  \left(\mathcal{H}''-\mathcal{G}'\right)\delta a+\left[\left(\mathcal{G}-\mathcal{H}'\right)\delta a+\mathcal{H} \delta a'\right]_{r_1}^{r_2}\,.
\end{align}
Using $\mathcal{H}''(r)=\mathcal{G}'(r)$, $\delta a'(r_2)=\delta a'(r_1)=0$, and
\begin{equation} \nonumber
\mathcal{G}-\mathcal{H}'=-\epsilon_0 m-a_0 m \left(\epsilon_0^2+ \epsilon_1 \pm\frac{4}{l^2}\right)- 2 a_0^2 \epsilon_0 m\left(\epsilon_1\mp \frac{1}{l^2}\right)+\mathcal{O}(a_0^3)\,, 
\end{equation}
we get
\begin{equation}
\widetilde{E}=-\left[\epsilon_0 m+a_0 m \left(\epsilon_0^2+ \epsilon_1 \pm \frac{4}{l^2}\right)+ 2 a_0^2 \epsilon_0 m\left(\epsilon_1\mp \frac{1}{l^2}\right)\right]\Delta a+\mathcal{O}(a_0^3)\,.
\end{equation}

On the other hand, it is easily seen from \eqref{physical} that
\begin{equation}
\Delta M = \mp \frac{4 a_0 m}{l^2}\Delta a+ \mathcal{O}(a_0^3) \,.
\end{equation}
Finally, imposing $\widetilde{E}=\Delta M$ as an equality of power series in $a_0$ we obtain $\epsilon_0=\epsilon_1=0$. 

\subsection{Extension of the main result}\label{subsection5}
In the previous subsections we have shown that the timelike Killing vector field~$K^\alpha$ given by Eq.~\eqref{killing2} is the correct choice to compute the energy of a test field in a Kerr-Newman-de Sitter background, at least in what concerns its interaction with the \acs{BH}. On the other hand, it is well known that the null generator of the event horizon is~$Z^\alpha=K^\alpha+\Omega Y^\alpha$, where $\Omega$ is the thermodynamic angular velocity, that is, the angular velocity that occurs in the first law (see for instance \cite{Dolan:2013ft,Kubiznak:2015bya}). Therefore, we can apply the main result of the previous section to conclude that test fields cannot destroy extremal Kerr-Newman-dS \acsp{BH}.~\footnote{The statement of that result requires~$K^\alpha$ to be asymptotically timelike; however, it is clear from the proof that all that is in fact needed is that~$K^\alpha$ determines the correct notion of energy.}

\section{Discussion}\label{section5}

In this chapter we proved that extremal Kerr-Newman or Kerr-Newman-AdS \acsp{BH} cannot be destroyed by interacting with (possibly charged) test fields satisfying the null energy condition at the event horizon and appropriate boundary conditions at infinity. This includes as particular cases all previous results of this kind obtained for scalar and electromagnetic test fields~\cite{Semiz:2005gs,Toth:2012vvy,Duztas:2013wua,Duztas:2013gza}. The corresponding results for test particles~\cite{WALD1974548,Tod:1976ud,Needham:1980fb} can also be considered particular cases, since particles can be seen as singular limits of continuous media~\cite{Geroch:1975uq,Lasota:2013kia}. It is interesting to note that if the null energy condition is not satisfied then the weak cosmic censorship conjecture may indeed be violated, as shown in Refs.~\cite{Duztas:2014sga,Toth:2015cda} for Dirac fields.

We have also shown that the timelike Killing vector field~$K^\alpha$ given by Eq.~\eqref{killing2} gives the correct definition of energy for test fields propagating in the Kerr-Newman-dS spacetime. Additionally, we have confirmed that the timelike Killing vector field $K^\alpha$ given by Eq.~\eqref{killing} gives the correct definition of energy for test fields propagating in the Kerr-Newman-AdS spacetime. Finally, we used this definition of energy to extend our main result to extremal Kerr-Newman-dS \acsp{BH}. 
The technique employed in the last subsection, namely allowing parameters in the metric to become functions in order to interpolate between \acs{BH} spacetimes with different physical masses, can be useful in other situations where the choice of the timelike Killing vector field with which to compute the energy of test fields is not clear. It is also possible that these ideas may play a role in determining an appropriate definition of mass for asymptotically de Sitter spacetimes.

%*****************************************
%*****************************************
%*****************************************
%*****************************************
%*****************************************

% ********************************************************************
% Backmatter
%*******************************************************
\appendix
\cleardoublepage
\part{Appendix}
%*****************************************
\chapter{PN expansion of Einstein-Klein-Gordon}\label{app:PN}
%*****************************************

%%%%%%%%%%%%%%%%%%%%%%%%%%%%%%%%%%%%%%%%%%%%%%%%%%%%%%%%%%%%%%%%%%%%%%%%%%%%%%%%%%%%%%%%%%%%%%%%%%%%%%%%
In this appendix, we show that the Einstein-Klein-Gordon system reduces to the Schr\"{o}dinger-Poisson system in the Newtonian limit. Then, we obtain the equations describing a perturbation to the Newtonian fields up to first post-Newtonian corrections. Finally, we consider perturbations caused by a point particle. In this section we follow the treatment in Chapter 8.2 of Ref.~\cite{poisson_will_2014}.  

The Einstein-Klein-Gordon system is the set of field equations for~$\Phi$ and~$g_{\mu \nu}$ which is obtained through the variation of action~\eqref{theory_action} with respect to $\Phi^*$ and $g_{\mu \nu}$, and reads
\begin{align} 
&\frac{1}{\sqrt{-g}}\partial_{\mu}\left(\sqrt{-g}g^{\mu\nu}\partial_{\nu}\Phi\right)= \mu_S^2 c^2\Phi\,,  \\
&R_{\mu\nu}= \frac{8\pi}{c^4} \widetilde{T}_{\mu\nu}^S\,,\label{EOM_BosonSapp}
\end{align}
where the Einstein equations are written in an alternative form using the trace-reversed stress-energy tensor of the scalar field
\begin{align}
\widetilde{T}^S_{\mu \nu }\equiv T^S_{\mu \nu}-\frac{1}{2}T^S g_{\mu \nu}=\partial_{(\mu}\Phi^* \partial_{\nu)}\Phi+\frac{1}{2}g_{\mu \nu}\mu_S^2c^2|\Phi|^2  \,.\nonumber
\end{align}
In the last equations we used $\mathcal{U}_S\sim \mu_S^2c^2|\Phi|^2/2$, since we want to consider a (Newtonian) diluted scalar field $|\Phi|\ll1$.
Moreover, in the Newtonian limit, we consider the spacetime metric ansatz
\begin{align}
g_{00}&=-1-\frac{2}{c^2} U +\mathcal{O}(c^{-4})\,, \\
g_{0j}&=\mathcal{O}(c^{-3})\,,\quad	g_{jk}=\mathcal{O}(c^{-2})\,,
\end{align}
with~$x^0=ct$ and the Cartesian coordinates $x^j=\{x,y,z\}$. This gives the Ricci tensor components
\begin{align}
R_{00}&=\frac{1}{c^2} \nabla^2U+ \mathcal{O}(c^{-4})\,,  \\
R_{0j}&=\mathcal{O}(c^{-3})\,,\quad R_{jk}= \mathcal{O}(c^{-2})\,.
\end{align}
The non-relativistic limit of the scalar field $\Phi$ is incorporated in our perturbation scheme by considering that~\footnote{It corresponds to the assertion that, in the non-relativistic limit, the energy-momentum relation is $E\sim \hbar\mu_S c^2+\frac{1}{2 \hbar\mu_S}p^2+\hbar\mu_S U$, with $p^2 \ll (\hbar \mu_S c)^2$ and $|U|\ll 1$. Here, the order parameter~$\epsilon$ is~$\epsilon\sim \mathcal{O}(p/(\hbar \mu_S c))\sim \mathcal{O}(\sqrt{U}/c)$.}
\begin{align}
c^{-1}|\partial_j \phi|/|\phi| \sim \mathcal{O}(\epsilon)\,, \quad c^{-2}|\partial_t \phi|/|\phi| \sim \mathcal{O}(\epsilon^2)\,,
\end{align}
where we introduced an auxiliary scalar field $\phi$ such that
\begin{align}
\Phi= \frac{1}{\sqrt{\mu_S }} e^{-i\mu_S c^2 t}\phi\,.
\end{align}
Then, the components of the trace-reversed stress-energy tensor of the scalar field are
\begin{align}
\widetilde{T}_{00}^S&=\frac{1}{2} \mu_S c^2 |\phi|^2+\mathcal{O}(c^0)\,, \\
\widetilde{T}_{0j}^S&=\mathcal{O}(c)\,, \quad\widetilde{T}_{jk}^S= \mathcal{O}(c^2)\,.
\end{align}
Therefore, at Newtonian order, the Einstein equations reduce to the Poisson equation
\begin{equation}
\nabla^2U=4\pi \mu_S |\phi|^2\,,
\end{equation}
which implies that~$|\phi|\sim \mathcal{O}(\epsilon^2)$.
On the other hand, it is easy to show that, at leading order in $\epsilon$ (and in the Newtonian limit), the Klein-Gordon equation reduces to the Schr\"{o}dinger equation
\begin{equation}
i \partial_t \phi=-\frac{1}{2 \mu_S} \nabla^2\phi+\mu_S U \phi\,.
\end{equation}
So, we have showed that, in the Newtonian limit, the Einstein-Klein-Gordon system for $\Phi$ and $g_{\mu \nu}$ reduces to the Schr\"{o}dinger-Poisson system for $\phi$ and $U$.

Let us now extend our perturbation scheme to first post-Newtonian order. We start by considering the spacetime metric ansatz
\begin{align}
g_{00}&=-1-\frac{2}{c^2}(U+\delta U) -\frac{2}{c^4}\left(\psi+ U^2\right)+\mathcal{O}(c^{-6})\,, \\
g_{0j}&=-\frac{4}{c^3} U_j+\mathcal{O}(c^{-5})\,, \\
g_{jk}&=\left(1-\frac{2}{c^2}(U+\delta U)\right) \delta_{jk}+\mathcal{O}(c^{-4})\,,
\end{align}
with the post-Newtonian terms~$U_j$,~$\psi$ and the perturbation $\delta U$. This results in the Ricci tensor components
\begin{align}
R_{00}=& \frac{1}{c^2}\nabla^2(U+ \delta U)+ \frac{1}{c^4}(3 \partial_t^2 U+4 U \nabla^2 U+ \nabla^2 \psi)+ \mathcal{O}(c^{-6})\,,\\
R_{0j}=&\frac{2}{c^3} \nabla^2U_j+ \mathcal{O}(c^{-5})\,,  \\
R_{jk}=&\frac{1}{c^2}\nabla^2(U+ \delta U)\delta_{jk}+\mathcal{O}(c^{-4})\,,
\end{align}
where we imposed the harmonic coordinate condition, which results in
\begin{equation}
\partial_tU+\partial_j U^j=0\,.
\end{equation}
Now, we introduce a perturbation $\delta \Phi$ to the Newtonian scalar field, such that
\begin{equation}
\delta \Phi=\frac{1}{\sqrt{\mu_S}}e^{-i \mu_S c^2 t}\delta \phi\,,
\end{equation}
treated in our perturbation scheme with
\begin{align}
\hspace{-1cm}|\delta \phi|\sim \mathcal{O}(\xi \epsilon^{-2})\,, \quad	c^{-1}|\partial_j\delta \phi|/|\delta \phi| \sim \mathcal{O}(\epsilon)\,, \quad c^{-2}|\partial_t \delta \phi|/|\delta\phi| \sim \mathcal{O}(\epsilon^2)\,,
\end{align}
and~$\delta U/c^2 \sim \mathcal{O}(\xi \epsilon^{-2})$.
Then, the components of the trace-reversed stress-energy tensor of the scalar field are
\begin{align}
&\hspace{-0.7cm}\widetilde{T}_{00}^S=\frac{1}{2} \mu_S c^2 |\phi|^2+\Im\left(\phi\,\partial_t \phi^* \right)-\mu_S U|\phi|^2+\mu_S c^2\Re\left(\phi^* \delta \phi\right)+\mathcal{O}(c^{-2})\,, \\
&\hspace{-0.7cm}\widetilde{T}_{0j}^S=c\Im\left(\phi\,\partial_j \phi^*\right)+\mathcal{O}(c^{-1})\,,  \\
&\hspace{-0.7cm}\widetilde{T}_{jk}^S=\frac{1}{2} \mu_S c^2 |\phi|^2+\mu_S c^2\Re\left(\phi^* \delta \phi\right)+\mathcal{O}(c^0)\,.
\end{align}
Thus, it is possible to show that, at first post-Newtonian order, the Einstein equations reduce to
\begin{align}
&\nabla^2 \psi=8\pi \Big[\Im\left(\phi\, \partial_t \phi^*\right)-3\mu_S U |\phi|^2\Big]\,, \\
&\nabla^2U_j=4 \pi \,\Im\left(\phi\,\partial_j \phi^*\right)\,, \\
&\nabla^2 \delta U=8 \pi \mu_S\, \Re \left(\phi^* \delta \phi\right)\,,\label{PN_Einstein}
\end{align}	
where we used the equations that are satisfied at Newtonian order and we assumed~$\partial^2_t U=0$, since this happens to be always the case in this work.
On the other hand, at next-to-leading order in $\epsilon$ (but still in the Newtonian limit), the Klein-Gordon equation reduces to
\begin{align}
\hspace{-0.9 cm}i \partial_t \delta \phi=-\frac{1}{2 \mu_S} \nabla^2 \delta \phi+\mu_S U \delta \phi+\mu_S \phi \,\delta U +\frac{1}{2 \mu_S} \partial_t^2 \phi+ i U \partial_t \phi -\frac{U}{\mu_S}  \nabla^2 \phi\,.\label{KG_all}
\end{align}	
Finally, note that, in the case $\mathcal{O}(\epsilon^6)<\mathcal{O}(\xi) <\mathcal{O}(\epsilon^4)$, the last equation becomes simply 
\begin{equation}
i \partial_t \delta \phi=-\frac{1}{2 \mu_S} \nabla^2 \delta \phi+\mu_S U \delta \phi+\mu_S \phi \,\delta U\,.
\end{equation}

In the case of a perturbation caused by a point particle, one just needs to include the trace-reversed stress energy tensor of the point particle~\eqref{Stress_energy_particle} in the Einstein equation~\eqref{EOM_BosonSapp}. This is given by
\begin{align}
\widetilde{T}^p_{\mu \nu}&\equiv T^p_{\mu \nu}-\frac{1}{2}T^p g_{\mu \nu}\nonumber \\
&=\frac{m_p c}{2 u^0} \left(2 u_\mu u_\nu+g_{\mu \nu}c^2\right) \frac{\delta (r-r_p)}{r^2} \frac{\delta(\theta-\theta_p)}{\sin \theta}\delta(\varphi-\varphi_p)\,,
\end{align}
with the particle's 4-velocity $u^\mu\equiv d x_p^\mu/d \tau$.
We consider that $m_p \sim \mathcal{O}(\xi)$ and that the particle is non-relativistic, in particular, we consider $u^i\sim \sqrt{U}\sim \mathcal{O}(\epsilon)$ in our perturbation scheme.
Then, the components of the trace-reversed stress-energy tensor of the particle are
\begin{align}
\widetilde{T}_{00}^p&=\frac{m_p c^2}{2} \frac{\delta (r-r_p)}{r^2} \frac{\delta(\theta-\theta_p)}{\sin \theta}\delta(\varphi-\varphi_p)+\mathcal{O}(c^0)\,,\\
\widetilde{T}_{tj}^p&=\mathcal{O}(c)\,, \quad\widetilde{T}_{jk}^p=\mathcal{O}(c^2)\,.
\end{align}
Thus, we conclude that we just need to add an extra term to the last equation in~\eqref{PN_Einstein}, which becomes
\begin{equation}
\nabla^2 \delta U=4\pi \left[2\mu_S \, \Re\left(\phi^* \delta \phi\right)+P\right]\,,
\end{equation}
with
\begin{equation}
P(t,r,\theta,\varphi)\equiv m_p \frac{\delta (r-r_p(t))}{r^2} \frac{\delta(\theta-\theta_p(t))}{\sin \theta}\delta(\varphi-\varphi_p(t))\,.
\end{equation}

Let us now consider the case of a non-relativistic point particle sourcing ultra-relativistic scalar perturbations to the Newtonian background. 
At Newtonian order, the Einstein equations describing the perturbation reduce to the Poisson equation~\footnote{The assumption of a non-relativistic perturber sourcing an ultra-relativistic scalar perturbation is consistent as long as the scalar is sufficiently light.}
\begin{equation}
\nabla^2\delta U= 4\pi P\,. \label{eq_UR_1b}
\end{equation} 
Finally, at leading order, the Klein-Gordon reduces to
\begin{align}
\nabla^2 \delta \Phi -\partial^2_t \delta \Phi=2 \mu_S^2 \Phi\, \delta U\,.\label{eq_UR_2b}
\end{align}
%

%*****************************************
%*****************************************
%*****************************************
%*****************************************
%*****************************************

%*****************************************
\chapter{Constancy of fundamental matrix determinant}\label{app:detF}
%*****************************************

%%%%%%%%%%%%%%%%%%%%%%%%%%%%%%%%%%%%%%%%%%%%%%%%%%%%%%%%%%%%%%%%%%%%%%%%%%%%%%%%%%%%%%%%%%%%%%%%%%%%%%%%
Consider a first-order matrix ordinary differential equation
\begin{equation}\label{matrixsystem}
\frac{d \boldsymbol{X}(r)}{dr} -V(r)\boldsymbol{X}(r)=0\,,
\end{equation}
with $\boldsymbol{X}$ a $N$-dimensional column vector and $V$ a $N\times N$ matrix.
A fundamental matrix of this system is a matrix of the form $F(r)\equiv \big(\boldsymbol{X_{(1)}},...,\boldsymbol{X_{(N)}}\big)$, where $\{\boldsymbol{X_{(1)}},...,\boldsymbol{X_{(N)}}\}$ is a set of $N$ independent solutions of Eq.~\eqref{matrixsystem}. The determinant of this $N\times N$ matrix can be written as
\begin{equation}
\det F(r)=\epsilon^{i_1\,...\,i_N} X_{(i_1)}^1\,...\,X_{(i_N)}^N\,,
\end{equation}
where $\epsilon$ is the Levi-Civita symbol, and $X_{(k)}^j$ is the $j$-th component of the vector $\boldsymbol{X_{(k)}}$. Using Eq.~\eqref{matrixsystem} it is easy to see that
\begin{equation}
\frac{d }{dr}\det F=\sum_{k=1}^N \epsilon^{i_1\,...\,i_N}V^k_{\;\;\; j} \,X_{(i_1)}^1\,...\,X_{(i_k)}^j\,...\,X_{(i_N)}^N\,.
\end{equation}
Using the relation 
\begin{equation}
\epsilon^{i_1\,...\,i_N} \,X_{(i_1)}^1\,...\,X_{(i_k)}^j\,...\,X_{(i_N)}^N=\delta_k^j \det F\,,
\end{equation}
one gets
\begin{equation}
\frac{d }{dr}\det F= \text{Tr}(V) \det F\,.
\end{equation}
If the trace $\text{Tr}(V)\equiv V^k_{\;\;\; k}$ is identically zero (which is always the case in Part~\ref{pt:ultralight} of this thesis), the determinant of the fundamental matrix is constant.

%*****************************************
%*****************************************
%*****************************************
%*****************************************
%*****************************************

%*****************************************
\chapter{Scalar field accretion by a static black hole}\label{app:acc}
%*****************************************

%%%%%%%%%%%%%%%%%%%%%%%%%%%%%%%%%%%%%%%%%%%%%%%%%%%%%%%%%%%%%%%%%%%%%%%%%%%%%%
\section{Incoming flux of energy at the center of a NBS} \label{app:incoming_flux}
%%%%%%%%%%%%%%%%%%%%%%%%%%%%%%%%%%%%%%%%%%%%%%%%%%%%%%%%%%%%%%%%%%%%%%%%%%%%%%
Here, we compute the incoming flux of energy over a tiny spherical surface at the center of a fundamental \acs{NBS}.
Consider a stationary \acs{NBS} of the form
\begin{equation}
\Phi= \Psi(r)e^{-i\left(\mu_S- \gamma\right) t}\,,
\end{equation}
where $\Psi$ is a solution of~\eqref{EOM_BS_radial1} and~\eqref{EOM_BS_radial2}.
This stationary field can be written as a sum of incoming and outgoing parts, $\Phi=\Phi_\text{in}+\Phi_\text{out}$, where
\begin{align}
&\Phi_\text{in}\equiv e^{-i\left(\mu_S- \gamma\right) t} \int_{-\infty}^{0}ds\,\overline{\Psi}(s) e^{i s r}\,, \nonumber \\
&\Phi_\text{out}\equiv e^{-i\left(\mu_S- \gamma\right) t} \int_{0}^{+\infty}ds\,\overline{\Psi}(s) e^{i s r}\,,	
\end{align}
with
\begin{equation}
\overline{\Psi}(s)=\frac{1}{2 \pi} \int_{-\infty}^{+\infty}dr\, \Psi(r)e^{-i s r}\,,
\end{equation}
and where we are using an even extension of $\Psi$ to negative values of $r$. Note that $\overline{\Psi}$ is a real-valued function, since $\Psi$ is real-valued. 
Now, the incoming flux of energy over a tiny spherical surface of radius $r_+\ll R$ is given by
\begin{equation}\label{Ein}
\dot{E}_\text{in}\simeq 4 \pi r_+^2 T_{tr}^\text{in}(r=0)\,.
\end{equation}
At leading order, one has
\begin{align}
T_{tr}^\text{in}(r=0) &\simeq \mu_S\, \Im\left(\Phi_\text{in}\partial_r \Phi_\text{in}^*\right) \nonumber \\
&=-\frac{\mu_S}{2} \int_{-\infty}^{0} ds' \int_{-\infty}^{0} ds \left(s'+s\right)\overline{\Psi}(s') \overline{\Psi}(s)\,.
\end{align}
Numerical evaluation of the last expression for a fundamental \acs{NBS} gives
\begin{equation}
T_{tr}^\text{in}(r=0) \sim 2.69\times 10^{-4}\,\mu_S^7 M_\text{NBS}^5\,.
\end{equation}
Finally, the incoming flux of energy is
\begin{equation}
\dot{E}_\text{in}\sim 3.38\times 10^{-3}\,r_+^2 \mu_S^7 M_\text{NBS}^5\,.
\end{equation}

%%%%%%%%%%%%%%%%%%%%%%%%%%%%%%%%%%%%%%%%%%%%%%%%%%%%%%%%%%%%%%%%%%%%%%%%%%%%%%
\section{Introducing a dissipative boundary} 
%%%%%%%%%%%%%%%%%%%%%%%%%%%%%%%%%%%%%%%%%%%%%%%%%%%%%%%%%%%%%%%%%%%%%%%%%%%%%%
In this section we look at two toy models, aimed at understanding the evolution
of an \acs{NBS} with a small \acs{BH} at its center. The main effect that the \acs{BH} produces is, naturally, dissipation at the horizon. This dissipative boundary condition can also be mimicked with some toy models.
%%%%%%%%%%%%%%%%%%%%%%%%%%%%%%%%%%%%%%%%%%%%%%%%%%%%%%%%%%%%%%%%%%%%%%%%%%%%%%
\subsection{A string absorptive at one end} \label{app:string_toy}
%%%%%%%%%%%%%%%%%%%%%%%%%%%%%%%%%%%%%%%%%%%%%%%%%%%%%%%%%%%%%%%%%%%%%%%%%%%%%%
Here, we wish to study a one-dimensional model of absorption of a scalar structure
when the boundary conditions suddenly change. Consider then a string, initially fixed at $x=0,\,L$, described by the wave equation
\begin{align}
\partial^2_x\Phi-\partial^2_t\Phi=0\,.
\end{align}
A normal mode satisfying~$\Phi(x=0)=\Phi(x=L)=0$ is
\begin{align}
&\Phi=e^{-i\omega_n t}\sin\omega_n x\,,\\
&\omega_n=\frac{(n+1)\pi}{L}\,, \quad n=0,1,2,...\,.
\end{align}
We take a configuration with~$\omega_n=\omega_0$ and use this as initial data for a problem where the boundary condition at the origin becomes absorptive. In particular, Laplace-transforming the wave equation gives
\begin{align}
&\frac{d^2\Psi}{dx^2}+\omega^2\Psi=-\dot{\Phi}(0,x)+i\omega\Phi(0,x)\,,\label{eq_inh}\\
&\Psi(\omega,x)=\int dt e^{i\omega t}\Phi(t,x)\,.
\end{align}
As boundary conditions, we require that 
\begin{align}
\Psi(\omega,L)=0\,,\quad \Psi(\omega,x\sim 0)=\sin\omega x-\epsilon e^{-i\omega x}\,. 
\end{align}
These conditions maintain the mirror-like boundary at one extreme $x=L$, while providing an absorption
of energy at $x=0$. The flux of absorbed energy scales like $\epsilon^2\ll1$.
The solution of Eq.~\eqref{eq_inh} subjected to the above boundary conditions is
\begin{align}
\Psi&=i\frac{\cos^2(\omega x)\sin(\pi x/L)+\sin^2(\omega x)\sin (\pi x/L)}{\omega-\pi/L}\nonumber\\
&\qquad+\epsilon\frac{\pi\sin[\omega(L-x)]}{\omega(\pi-L\omega)(i\epsilon\cos(\omega L)+(\epsilon-i)\sin(\omega L))}\,.\label{sol_nonh_string}
\end{align}
The original time-domain field is given by the inverse
\begin{align}
\Phi(t,x)=\frac{1}{2\pi}\int d\omega e^{-i\omega t}\Psi(\omega, x)\,.
\end{align}
The integral can be done with the help of the residue theorem. We separate the response~$\Phi=\Phi_1+\Phi_2$. The first term in Eq.~\eqref{sol_nonh_string} has a simple, \emph{real} pole
at $\omega=\omega_0=\pi/L$, and it evaluates to
\begin{align}
\Phi_1(t,x)=\sin(\pi x/L)e^{-i\pi t/L}\,,
\end{align}
\ie, it corresponds to the initial data.
The second term has poles at complex values of the frequency, which are also the \acsp{QNM} of the dissipative system,
\begin{align}
\omega\approx \frac{n\pi+\epsilon -i\epsilon^2}{L}\,,
\end{align}
These poles lie close to the normal modes of the system, including those not present in the initial data.
They dictate an exponential decay $\sim e^{-\epsilon^2 t}$, and a consequent lifetime $\tau \sim \epsilon^{-2}$.
Note that this simple exercise shows that all modes are excited when new boundary conditions are turned on.
For \acsp{NBS}, all the modes cluster around $\omega \sim \mu_S$, thus we expect to always be in the low-frequency regime
used to estimate the lifetime.

%%%%%%%%%%%%%%%%%%%%%%%%%%%%%%%%%%%%%%%%%%%%%%%%%%%%%%%%%%%%%%%%%%%%%%%%%%%%%%
\subsection{A black hole in a scalar-filled sphere} \label{app:bh_bomb}
%%%%%%%%%%%%%%%%%%%%%%%%%%%%%%%%%%%%%%%%%%%%%%%%%%%%%%%%%%%%%%%%%%%%%%%%%%%%%%
%
\begin{figure}
	\begin{tabular}{cc}
		\centering
		\includegraphics[width=0.5 \textwidth]{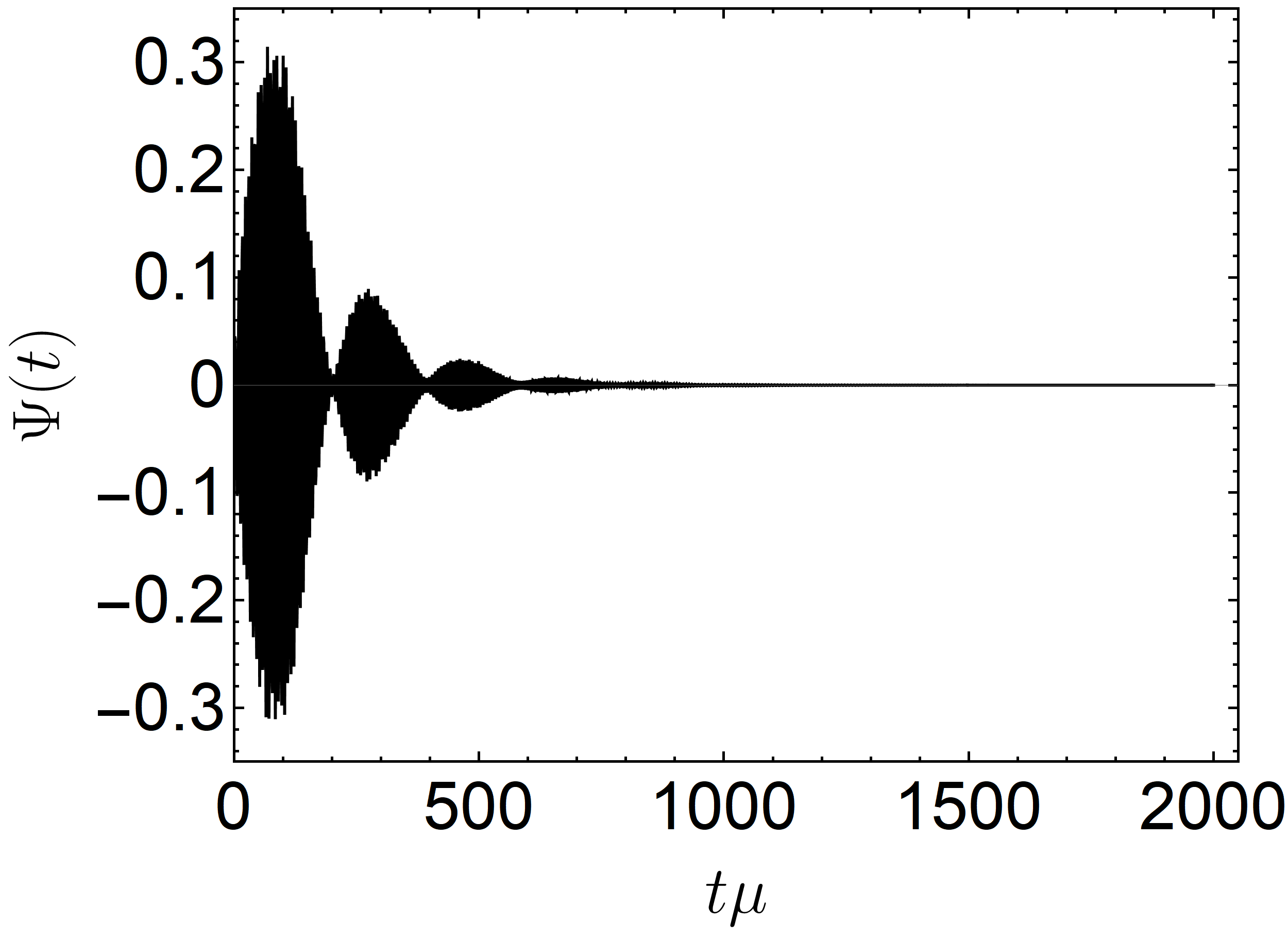} & \includegraphics[width=0.5\textwidth]{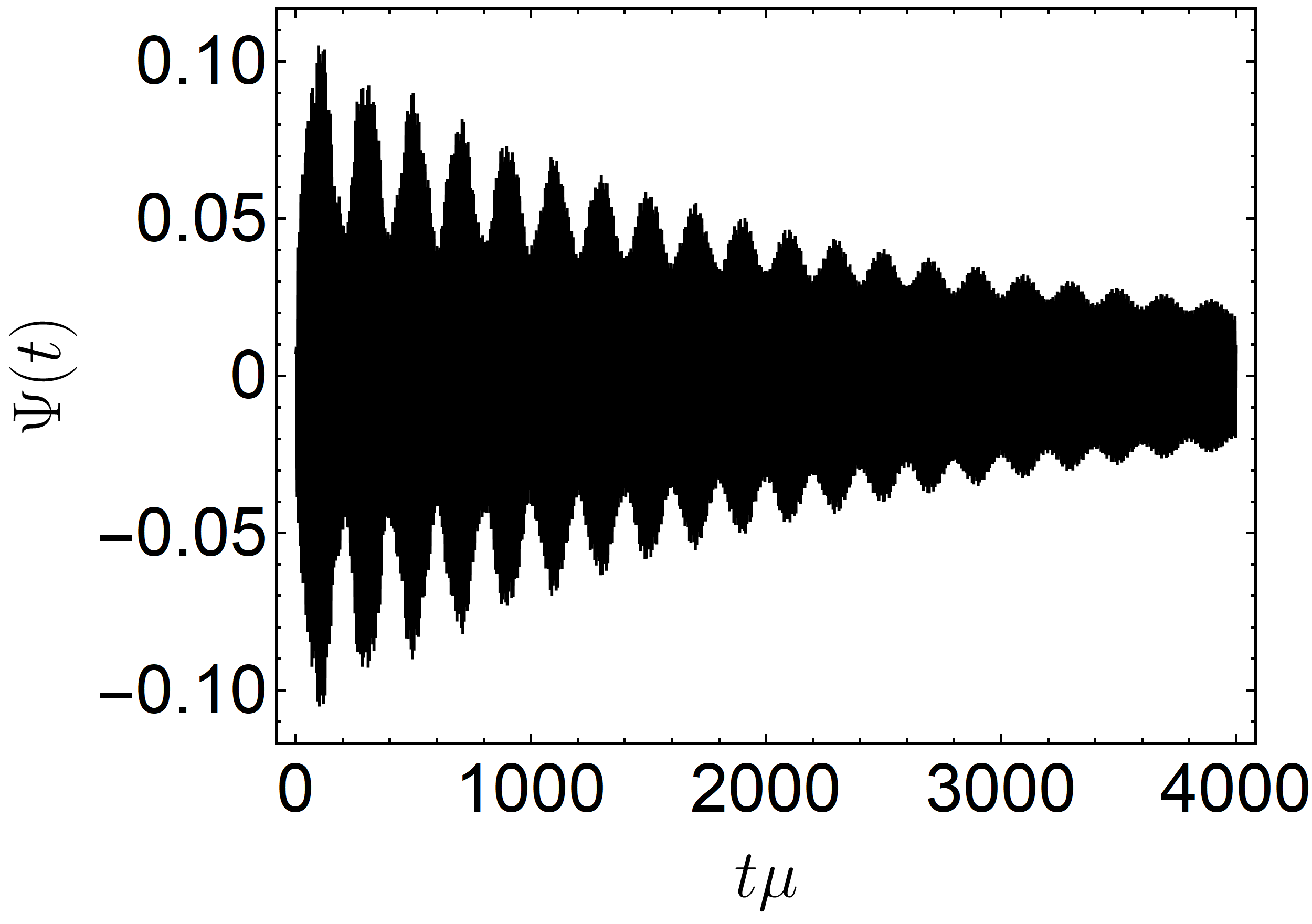} \\
		\includegraphics[width=0.5 \textwidth]{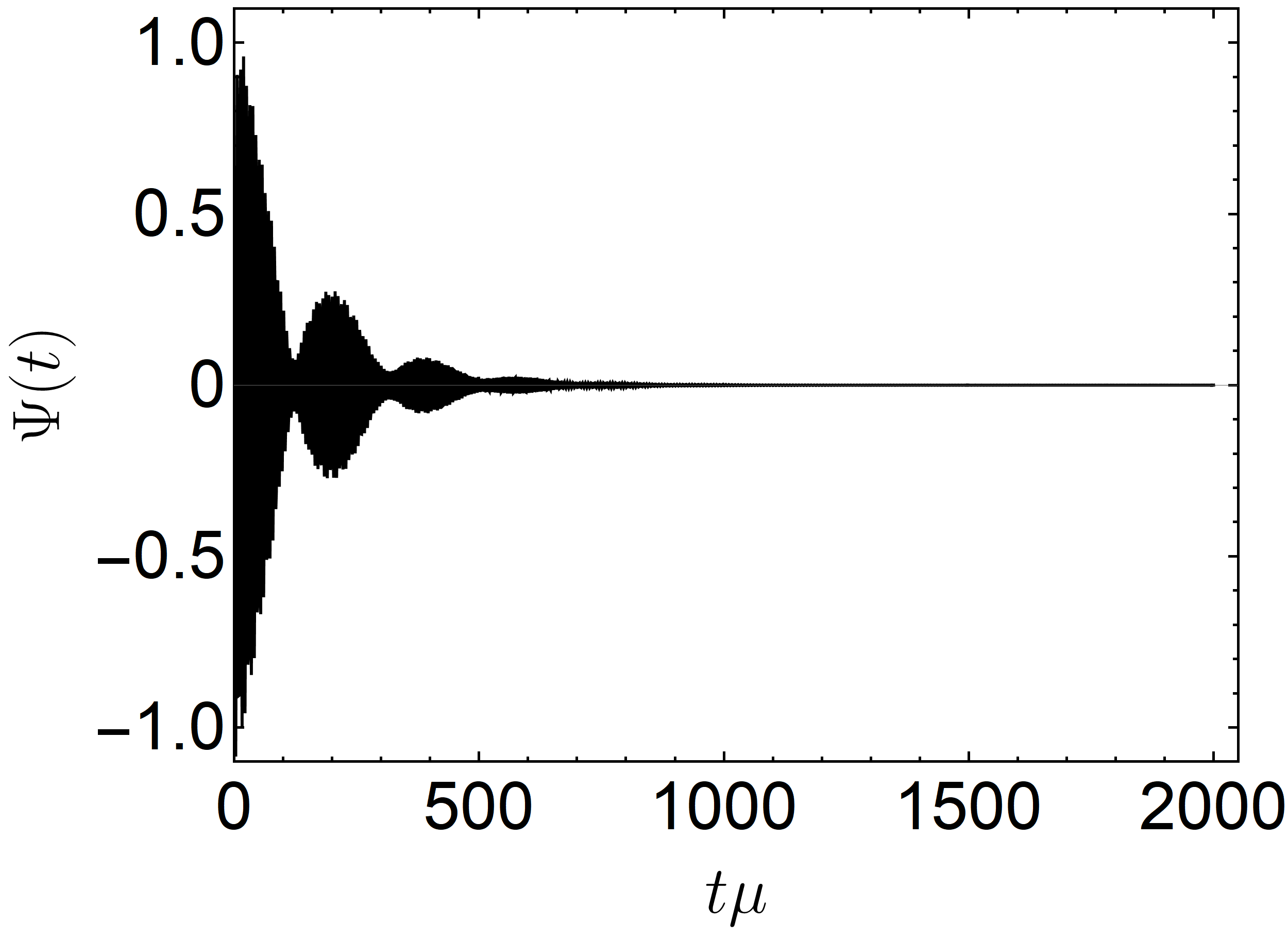} &
		\includegraphics[width=0.5 \textwidth]{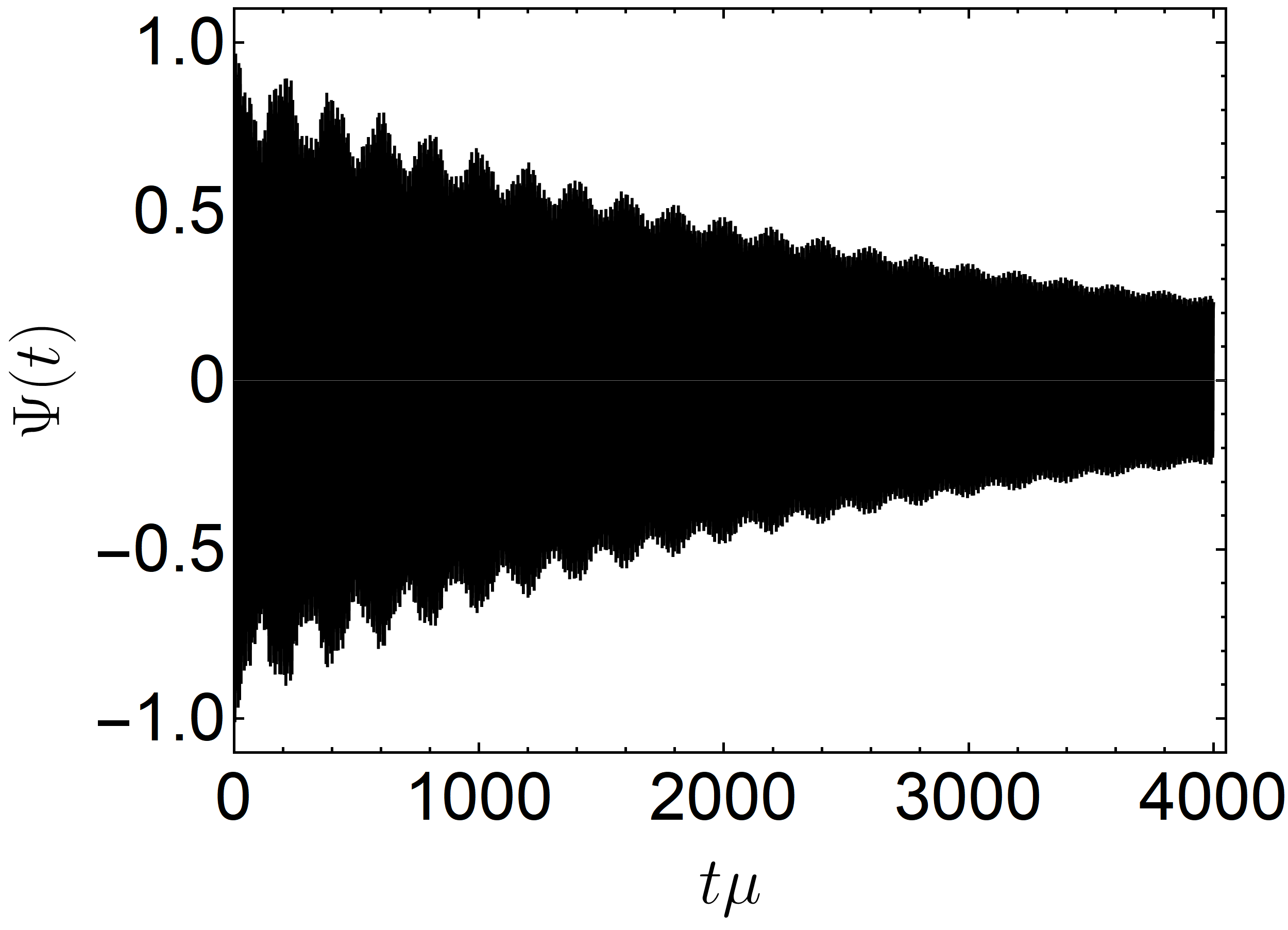} \\
		\includegraphics[width=0.5 \textwidth]{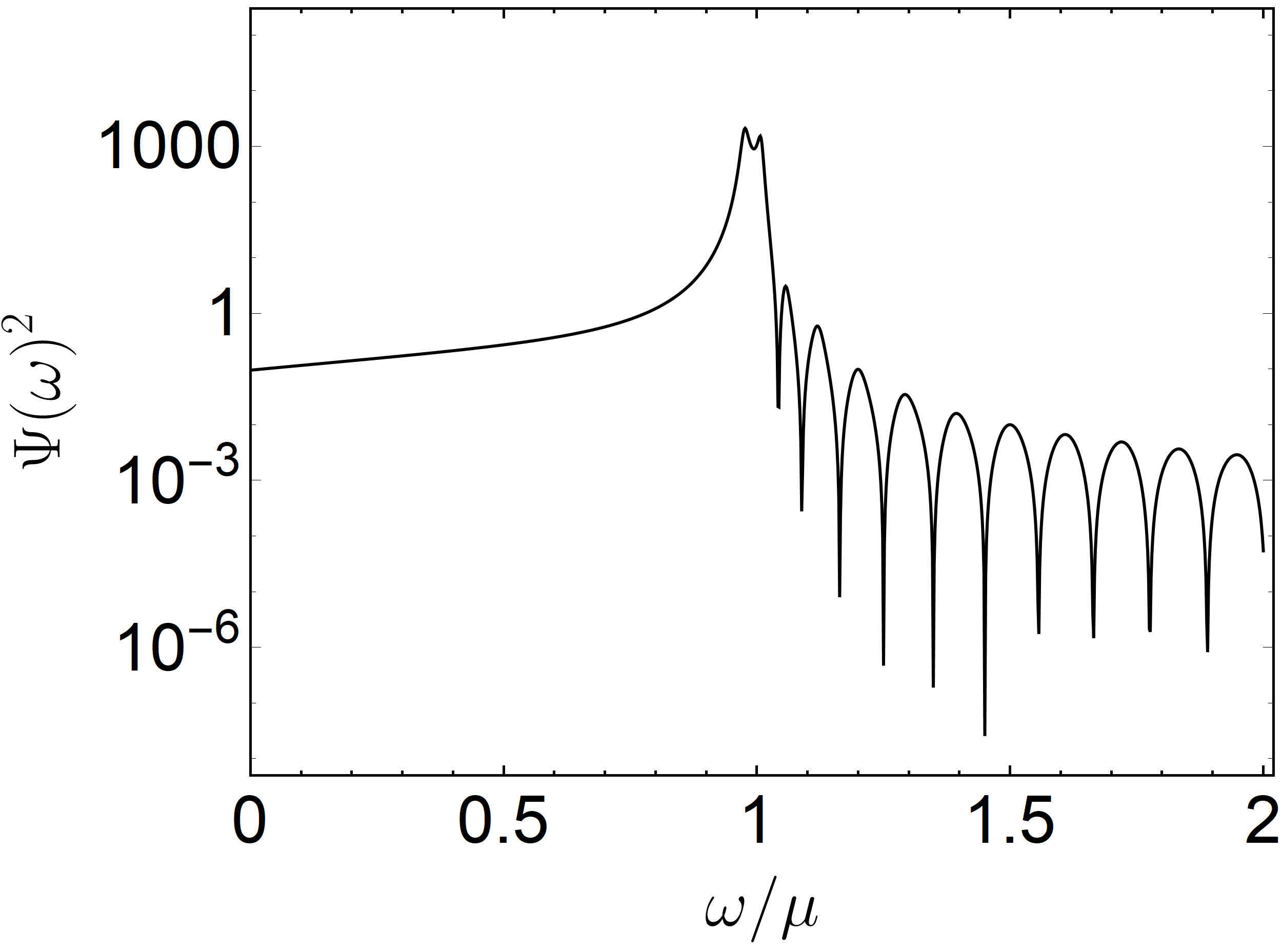} &
		\includegraphics[width=0.5 \textwidth]{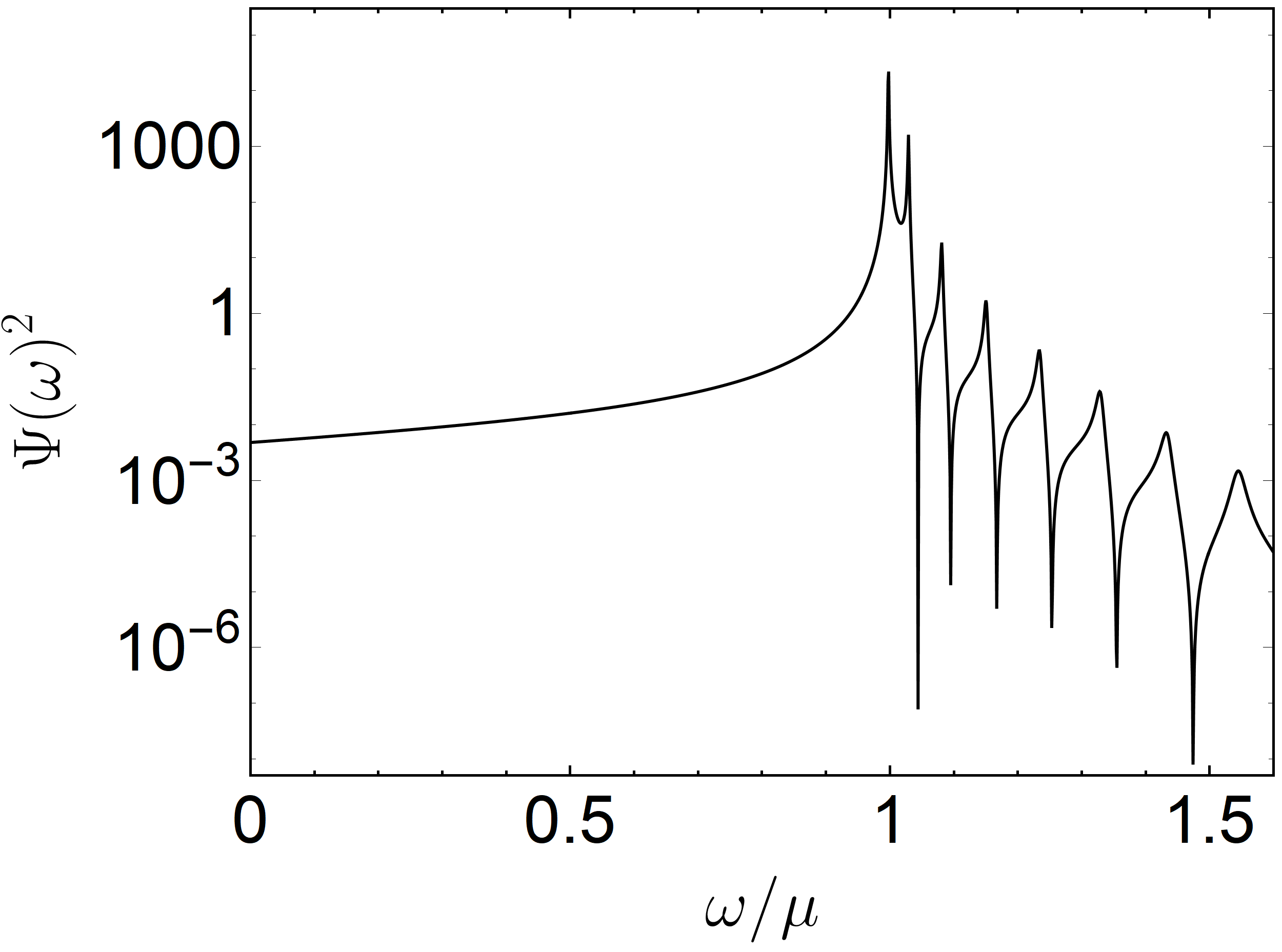}
	\end{tabular}
	\caption{
		The evolution of a massive scalar field inside a perfectly reflecting spherical surface of radius $R\mu_S=20$.
		In the center of such a sphere, there sits a \acs{BH} of mass $M_\text{BH}\mu_S=0.2$ (left panels) and $M_\text{BH}\mu_S=0.1$ (right panels). 
		\textbf{First row:} Scalar field measured on the horizon. 
		\textbf{Second row:} Scalar field measured at $r\mu_S =10$.
		\textbf{Third row:} Flux measured at the horizon.
		\label{fig:BHBomb_evolution}
	}
\end{figure}
A toy model more similar to the problem we wish to study is that of a \acs{BH} of mass~$M_\text{BH}$, at the center of a sphere of radius~$R$, which was filled with a massive scalar field.
The profile for the scalar is, initially, that of a normal mode (the Klein-Gordon field $\Phi=\Psi/r$),
\begin{align}
\Psi=\sin \omega_0r\,,
\end{align}
with $\omega_0=\sqrt{\mu_S^2+\pi^2/R^2}$. The problem simplifies enormously when the scalar is non self-gravitating and is but a small disturbance in the background of the~\acs{BH} spacetime. This is what we assume from now onwards. In such a case all one has to do is evolve the Klein-Gordon equation in a Schwarzschild geometry, subjected to Dirichlet conditions at the surface of the sphere. The results are summarized in Fig.~\ref{fig:BHBomb_evolution}. While they do not mimic entirely the process of accretion of a self-gravitating \acs{NBS} by a central \acs{BH}, these results illustrate some of the possible physics in the more realistic setup.

The figures show the scalar extracted at the horizon (first row), at a midpoint inside the sphere (second raw) and the flux per frequency bin (third raw).
The scalar, measured either at the horizon or somewhere within the sphere, decays exponentially. 
The first noteworthy aspect is the sensitive dependence of the decay rate on the size of the \acs{BH}. Our results are consistent with a decay timescale~$\tau \sim (M_\text{BH}\mu_S)^{-\beta}$, with $\beta \sim 4-5$, in agreement with our analysis in Section~\ref{sec_sitting_bh} and also with a quasinormal mode ringdown of such fields~\cite{Brito:2015oca}.
Note that such suppressed decay for small~$M\mu_S$ couplings happens due to the filtering properties of small \acsp{BH}, keeping out most of the low-frequency field.
This also explains why the ratio between the field measured at~$r=10$ and at the horizon increases when the \acs{BH} size decreases.
Note also that, in accordance with the previous one-dimensional toy model, overtones are also excited. This is clearly seen in the Fourier analysis
(third raw panels in Fig.~\ref{fig:BHBomb_evolution}), showing local peaks at all the subsequent overtones, which were absent in the initial data. These correspond to frequencies $\omega=\sqrt{\mu_S^2+\pi^2n^2/R^2}\,,n=0,\,1,...$. This is one important difference between this system and \acsp{NBS}, for which overtones are all bounded in frequency.

%*****************************************
%*****************************************
%*****************************************
%*****************************************
%*****************************************

%*****************************************
\chapter{Scalar Q-balls}\label{app:Qball}
%*****************************************

Here, we generalize the calculations done in Chapter~\ref{ch:nbs} to Q-balls, where gravity is absent and self-interactions are necessary.

%%%%%%%%%%%%%%%%%%%%%%%%%%%%%%%%%%%%%%
\section{Background configurations}
%%%%%%%%%%%%%%%%%%%%%%%%%%%%%%%%%%%%%%
%
\begin{figure}	
	\centering
	\includegraphics[width=0.9\textwidth]{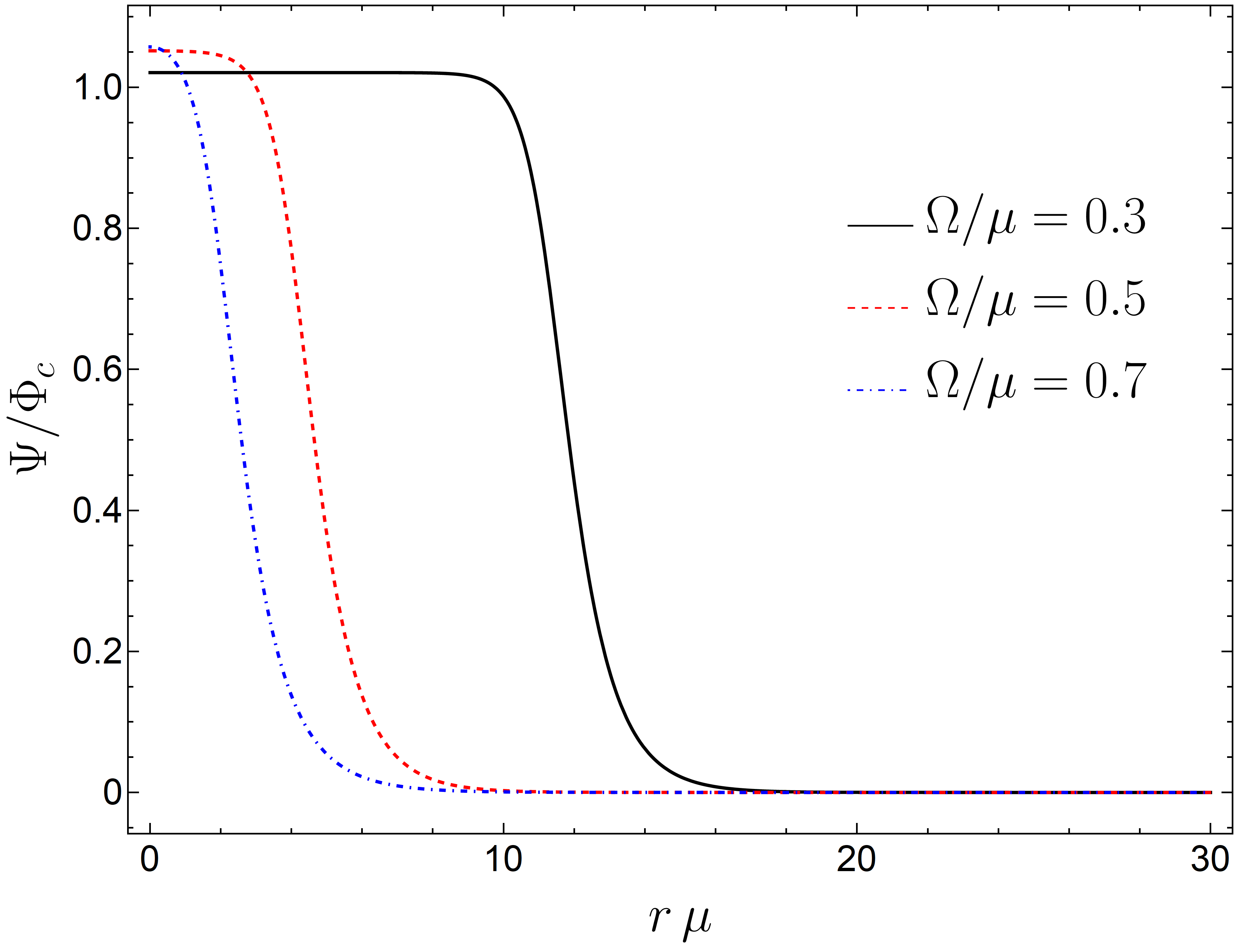} 
	\caption{Three radial profiles $\Psi(r)/\Phi_c$ obtained through numerical integration of Eq.~\eqref{EOM_Qball_radial1} with appropriate boundary conditions ($\Psi(\infty)\rightarrow 0$ and $\partial_r\Psi(0)=0$). Each curve corresponds to a different Q-ball. }\label{fig:QBalls}
\end{figure}
The field equation for $\Phi$ is obtained through the variation of action~\eqref{theory_action} with respect to $\Phi^*$ and reads
\begin{align}
&\nabla^\mu\partial_\mu  \Phi-2 \frac{d \mathcal{U}_\text{Q}}{d|\Phi|^2} \Phi=0\,,\label{EOM_Qball}
\end{align}
where we used $g_{\mu \nu}=\eta_{\mu \nu}$ and the self-interaction potential $\mathcal{U}_\text{Q}$ defined in Eq.~\eqref{Potential_Qball}.
We now look for localized solutions of this model with the form~\eqref{BKG_ansatz} -- the so-called Q-balls. This ansatz yields the radial equation
\begin{align} 
\partial^2_r \Psi+\frac{2}{r}\partial_r \Psi+\left[\Omega^2-2 \frac{d \mathcal{U}_\text{Q}}{d|\Phi|^2}\right]\Psi=0\,.\label{EOM_Qball_radial}
\end{align}
For the class of nonlinear potentials~\eqref{Potential_Qball}, the last equation becomes  
\begin{align} 
\partial^2_r \Psi+\frac{2}{r}\partial_r \Psi+\left[\Omega^2-\mu^2 \left(1-\frac{\Psi^2}{\Phi_c^2}\right)\left(1-3\frac{\Psi^2}{\Phi_c^2}\right)\right]\Psi=0\,.\label{EOM_Qball_radial1}
\end{align}
According to the results of Ref.~\cite{Coleman:1985ki}, there exist stable Q-ball solutions for any $0<\Omega<\mu_S$, independently of the free parameter $\Phi_c$. 
Additionally, it is known that, in the limit $\Omega/\mu_S \ll1$, the radial function $\Psi$ mimics an Heaviside step function (the so-called \textit{thin-wall} Q-balls)~\cite{Coleman:1985ki,Ioannidou,Tsumagari2008}. On the other hand, in the regime $\Omega/\mu_S \sim 1$, the function $\Psi$ starts to fall earlier and drops very slowly (\textit{thick-wall} Q-ball)~\cite{Ioannidou,Tsumagari2008}. In particular, using the results of Ref.~\cite{Tsumagari2008} one can show that, in the thin-wall limit,
\begin{equation}\label{Psi_thin}
\Psi(r) \simeq\Phi_c\left[1+\left(\frac{\Omega}{2 \mu_S}\right)^2\right]\Theta\left(\frac{\mu_S}{\Omega^2}-r\right)\,.
\end{equation}
Notice that the Q-ball radius is approximately given by $R_Q\simeq\mu_S/\Omega^2$.

A few examples of radial profiles $\Psi(r)$ constructed numerically from Eq.~\eqref{EOM_Qball_radial1} are shown in Fig.~\ref{fig:QBalls}. 
From these results it is already evident that, when $\Omega/\mu_S \to 0$, the scalar does acquire a Heaviside-type profile. In such a limit the scalar drops to zero on the outside, on a lengthcale
$\sim 1/\mu_S$. These results also indicate that the radius of the Q-ball grows when $\Omega/\mu_S \to 0$. This is made more explicit in
Fig.~\ref{fig:Radius_Omega}, showing the numerical results for the dependence of the Q-ball radius $R_Q$ on the frequency $\Omega$.~\footnote{We define the Q-ball radius $R_Q$ to be such that $\dfrac{\Psi(R_Q)}{\Psi(0)}=1/2$.} The dashed line, corresponding to the thin-wall limit \eqref{Psi_thin}, agrees remarkably well with the numerics. 
\begin{figure}	
	\centering
	\includegraphics[width=0.9 \textwidth]{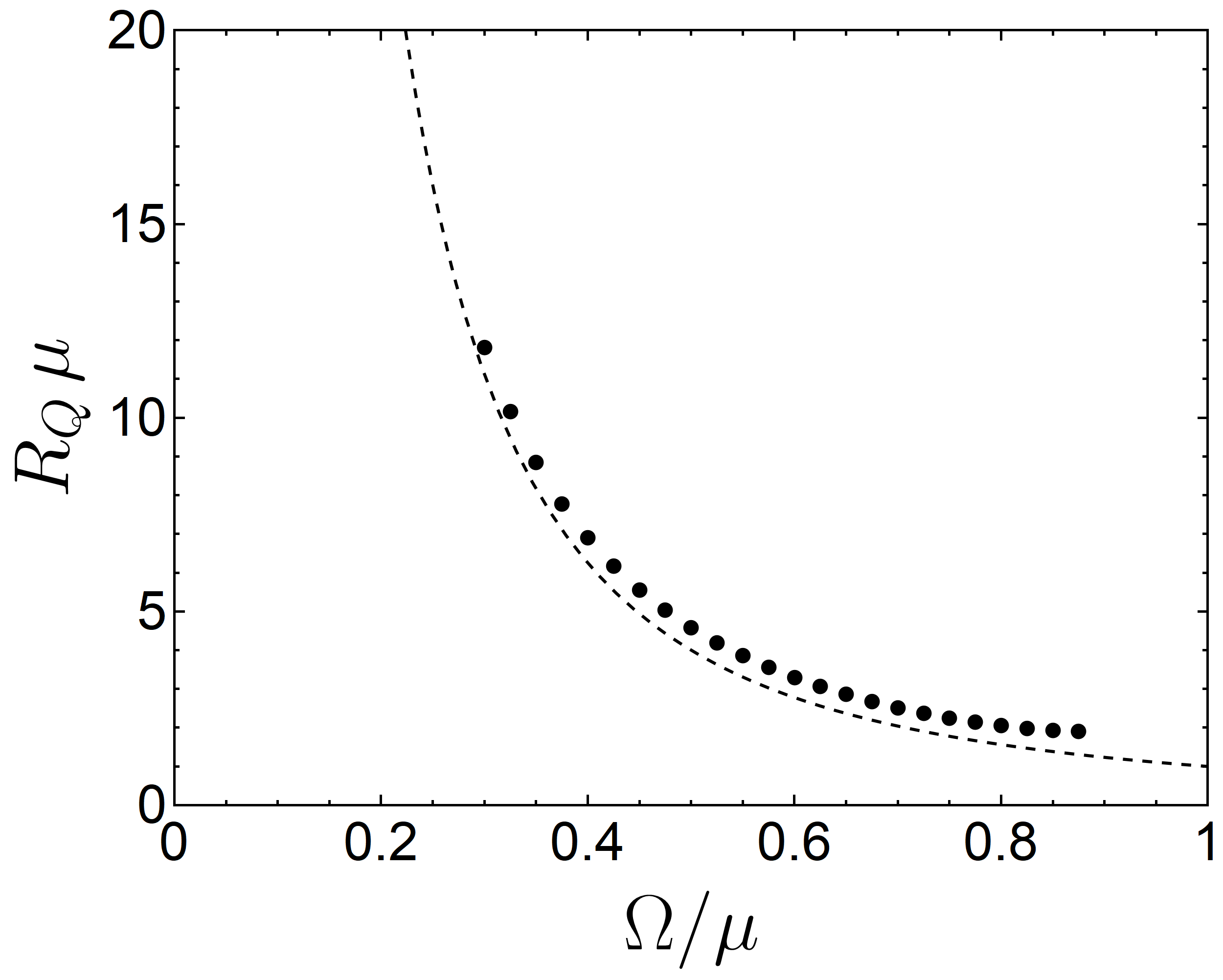} 
	\caption{Numerical results for the dependence of the Q-ball radius $R_Q \mu_S$ on the internal frequency $\Omega/\mu_S$, obtained through direct integration of Eq.~\eqref{EOM_Qball_radial1}. The dashed line is the thin-wall limit prediction, Eq.~\eqref{Psi_thin}. A fit on the numerical results gives $R_Q\sim 1.08 \mu_S \Omega^{-2}$, within $2\%$ of error, showing a good accordance with the predicted behavior (Eq.~\eqref{Psi_thin}).
	} \label{fig:Radius_Omega}
\end{figure}

The Q-ball charge $Q$ and mass $M_Q$ are obtained through~\eqref{NoetherCharge} and \eqref{energy_spacelike}, respectively, and read  
\begin{align}
& Q=\frac{4\pi}{\hbar} \Omega \int dr\, r^2 \Psi^2(r)\,,\\
& M_Q=\frac{1}{2} Q \Omega+4\pi \int dr\,r^2\left(\frac{(\partial_r \Psi)^2}{2}+\mathcal{U}(\Psi^2)\right)\,.
\end{align}
For thin-wall Q-balls these become
\begin{align}
&Q=\frac{4 \pi}{3} \frac{\Omega^4}{\mu_S^6} \Psi_c^2\,,\\
&M_Q=\frac{2 \pi}{3} \frac{\Omega^5}{\mu_S^6} \Psi_c^2\,.
\end{align}
We are using a flat background spacetime, which requires that $M_Q/R_Q \ll 1$. In the thin-wall limit, this corresponds to
\begin{align}
\Omega/\mu_S \ll \Psi_c^{-2/7}\,.
\end{align}
%%%%%%%%%%%%%%%%%%%%%%%%%%%%%%%%%%%%%%%%%%%%%%%%%%%%
\section{Small perturbations}
%%%%%%%%%%%%%%%%%%%%%%%%%%%%%%%%%%%%%%%%%%%%%%%%%%%%

We now wish to understand the effect of a small perturbation on such Q-ball configurations. They can be either (free) sourceless small deformations of the background, or sourced by an external particle. Such perturber could be another (much smaller) Q-ball, or simply some scalar charge piercing the Q-ball or orbiting around it. In the following, the external probe is modelled as point-like, which means that our results are valid only for objects whose spatial extent are~$\ll R_\text{Q}$.
We consider an interaction between the perturber and the Q-ball described by the action~\eqref{theory_action} with $J_S=T_p$, where~$T_p\equiv \eta_{\mu \nu} T_p^{\mu \nu}$ is the trace of the particle's energy-momentum tensor defined in~\eqref{Stress_energy_particle}. This coupling allows for \acsp{EOM} that are both simple enough to be handled via our perturbation scheme, described in Chapter~\ref{ch:framew}, and that show interesting dynamical features, as we shall see later. In the present analysis, we neglect the backreaction on the particle motion, therefore, the particle's world-line $x_p^\mu(\tau)$ is considered to be known. 

An external particle sources a scalar field fluctuation of the form~\eqref{Perturbation} in the Q-ball background, which satisfies the linearized equation
\begin{align}
&\nabla^2 \delta\Psi-\partial_t^2\delta\Psi +\left[\Omega^2-\mu_S^2\left(1- 8\frac{\Psi^2}{\Phi_c^2}+9 \frac{\Psi^4}{\Phi_c^4}\right)\right]\delta\Psi\nonumber\\
&\qquad+2 i \Omega \partial_t \delta\Psi+2\mu_S^2 \frac{\Psi^2}{\Phi_c^2} \left(2-3 \frac{\Psi^2}{\Phi_c^2}\right)\delta\Psi^*=T_p e^{i \Omega t}\,.
\label{eq:Qball_pert_eq}
\end{align}
The sourceless case is recovered simply by setting $T_p=0$.
Decomposing the particle stress-energy trace as
\begin{align}
T_p e^{i \Omega t}&=\sum_{l,m} \int \frac{d \omega}{\sqrt{2 \pi} r}\Big[T_1^{\omega l m} Y_l^m e^{-i \omega t}+\left(T_2^{\omega l m }\right)^*\left(Y_l^m\right)^* e^{i \omega t}\Big]\,,\label{MatterDecomposition}
\end{align}
where $T_1^{\omega l m}$ and $T_2^{\omega l m}$ are radial complex-functions defined by
\begin{align}
&T_1^{\omega l m}\equiv \frac{r}{2 \sqrt{2 \pi}}\int dt d\theta d\varphi \sin \theta\, T_p e^{i (\omega+ \Omega)t} \left(Y_l^m\right)^*\,, \hspace{0.5cm} \label{eq:source_decomp_T1}\\
&T_2^{\omega l m}\equiv \frac{r}{2 \sqrt{2 \pi}}\int dt d\theta d\varphi \sin \theta\, T_p e^{i (\omega- \Omega)t} \left(Y_l^m\right)^*\,.\hspace{0.5cm}\label{eq:source_decomp_T2}
\end{align}
Plugging the decompositions~\eqref{Decomposition} and~\eqref{MatterDecomposition} in Eq.~\eqref{eq:Qball_pert_eq}, one obtains the matrix equation~\footnote{The symmetry of this system implies that the radial functions satisfy $Z_2(\omega,l;r)=Z_1(-\omega,l;r)^*$. The functions $Z_1$ and $Z_2$ are clearly independent of the azimuthal number $m$.}
\begin{equation} 
\partial_r \boldsymbol{Z} -V_Q(r) \boldsymbol{Z}=\boldsymbol{T}\,,\label{Qball_Perturbation_Matrix_Sourced}
\end{equation}
with the vector $\boldsymbol{Z}\equiv (Z_1, Z_2, \partial_r Z_1, \partial_r Z_2)^T$, the matrix $V_Q$ given by
\begin{equation*}
V_Q\equiv
\begin{pmatrix} 
0 & 0 & 1 & 0 \\
0 & 0 & 0 & 1 \\
V_s-(\omega+\Omega)^2 & V_c & 0 & 0  \\
V_c & V_s-(\omega-\Omega)^2 & 0 & 0  
\end{pmatrix}\,,
\end{equation*}
where we defined the radial potentials
\begin{align}
V_s(r)&\equiv \frac{l(l+1)}{r^2}+\mu_S^2\left(1- 8\frac{\Psi_0^2}{\Phi_c^2}+9 \frac{\Psi_0^4}{\Phi_c^4}\right)\,, \\
V_c(r)&\equiv - 2\mu_S^2 \frac{\Psi_0^2}{\Phi_c^2} \left(2-3 \frac{\Psi_0^2}{\Phi_c^2}\right)\,.
\end{align}
and the source term~\footnote{To simplify the notation, we omit the labels $\omega$, $l$ and $m$ in the functions $T_1^{\omega l m}$ and $T_2^{\omega l m}$.}
\begin{align}
\boldsymbol{T}(r)\equiv \big(0,0,T_1,T_2\big)^T\,.\label{stress_energy_decomposition_qball}
\end{align}
To solve for the small perturbations, either in the sourced or sourceless case, we need to establish suitable boundary conditions. We require regular solutions at the origin,
\begin{equation}
\boldsymbol{Z}(r \to 0)\sim \left(a r^{l+1},b r^{l+1},a (l+1)r^l,b (l+1)r^l\right)^T\,,\nonumber
\end{equation} 
with (complex) constants $a$ and $b$, and the Sommerfeld radiation condition at infinity
\begin{align}
\label{Qball_FO_Inf}
\boldsymbol{Z}(r \to \infty)\sim \left(Z_1^\infty e^{i k_1 r},
Z_2^\infty e^{i k_2 r},
i k_1 Z_1^\infty e^{i k_1r},
i k_2 Z_2^\infty e^{i k_2r}\right)\,,
\end{align} 
with 
\begin{align}
k_1&\equiv \epsilon_1 \sqrt{\left(\omega+\Omega\right)^2-\mu_S^2}\,, \\ 
k_2&\equiv \epsilon_2 \left(\sqrt{\left(\omega-\Omega\right)^2-\mu_S^2}\right)^*\,,
\end{align}
where we are using the principal complex square root.

Consider then the set of independent solutions $\{\boldsymbol{Z_{(1)}},\boldsymbol{Z_{(2)}},\boldsymbol{Z_{(3)}},\boldsymbol{Z_{(4)}}\}$ uniquely determined by
\begin{align}
\boldsymbol{Z_{(1)}}(r \to 0)&\sim \Big(r^{l+1},0,(l+1)r^l,0\Big)^T\,,\nonumber \\
\boldsymbol{Z_{(2)}}(r \to 0)&\sim \Big(0,r^{l+1},0,(l+1)r^l\Big)^T\,,\nonumber \\
\boldsymbol{Z_{(3)}}(r \to \infty)&\sim \Big(e^{i k_1 r},0,i k_1 e^{i k_1 r},0\Big)^T\,,\nonumber \\
\boldsymbol{Z_{(4)}}(r \to \infty)&\sim \Big(0,e^{i k_2 r},0,i k_2 e^{i k_2 r}\Big)^T\,.
\end{align}
The $4 \times 4$ matrix $F(r)\equiv\big(\boldsymbol{Z_{(1)}},\boldsymbol{Z_{(2)}},\boldsymbol{Z_{(3)}},\boldsymbol{Z_{(4)}}\big)$ is the fundamental matrix of the system~\eqref{Qball_Perturbation_Matrix_Sourced}. As shown in Appendix~\ref{app:detF}, for a system of the form~\eqref{Qball_Perturbation_Matrix_Sourced}, the determinant $\text{det}(F)$ is independent of $r$.
%%%%%%%%%%%%%%%%%%%%%%%%%%%%%%%%%%%%%%%%%%%%%%%%%%%%
\subsection{Sourceless perturbations}
%%%%%%%%%%%%%%%%%%%%%%%%%%%%%%%%%%%%%%%%%%%%%%%%%%%%%%%
Free oscillations of Q-ball configurations are regular scalar fluctuations satisfying the Sommerfeld radiation condition at infinity. They correspond to scalar perturbations of the form 
\begin{equation}
\delta \Psi=\frac{1}{\sqrt{2 \pi} r} \left[Z_1 Y_l^m e^{-i\omega t}+Z_2^*\left(Y_l^m\right)^* e^{i\omega^* t}\right]\,,\label{field_decomposition_QNM_qball}
\end{equation}
where $Z_1$ and $Z_2$ are solutions of system~\eqref{Qball_Perturbation_Matrix_Sourced} with $\boldsymbol{T}=0$. For complex-valued $\omega$, the free oscillations are \acsp{QNM}. For a real $\omega$, these are termed normal modes. Notice that for the discrete set $\{\omega_\text{QNM}\}$ of \acs{QNM} frequencies, the solutions $\{\boldsymbol{Z_{(1)}},\boldsymbol{Z_{(2)}},\boldsymbol{Z_{(3)}},\boldsymbol{Z_{(4)}}\}$ are not linearly independent. In fact, it is easy to see that the condition $\text{det}(F)=0$ holds if and only if $\omega$ is a \acs{QNM} frequency (\ie,} $\omega \in \{\omega_\text{QNM}\}$).

%%%%%%%%%%%%%%%%%%%%%%%%%%%%%%%%%%%%%%%%%%%%%%%%%%%%
\subsection{External perturbers}
%%%%%%%%%%%%%%%%%%%%%%%%%%%%%%%%%%%%%%%%%%%%%%%%%%%%%%%
Let us now turn to the perturbations induced by an external particle, which interacts with the background scalar field. How is such a body exciting the Q-ball, how much radiation does the interaction give rise to, what backreaction does the Q-ball
exert on the perturber? These are all questions that can be raised in this context, and that we wish to answer here.
In order to do that, one needs to find the solutions of system~\eqref{Qball_Perturbation_Matrix_Sourced} that are regular at the origin and satisfy the Sommerfeld condition at infinity. These can be obtained through the method of variation of parameters, 
\begin{align}
Z_1(r)&=\sum_{k=3}^4 \Bigg[\sum_{n=1}^2 F_{1,n}(r) \int_\infty^r dr' F^{-1}_{n,k} \boldsymbol{T}_k+\sum_{n=3}^4 F_{1,n}(r) \int_0^r dr' F^{-1}_{n,k} \boldsymbol{T}_k \Bigg]\,,\\
Z_2(r)&=\sum_{k=3}^4 \Bigg[\sum_{n=1}^2 F_{2,n}(r) \int_\infty^r dr' F^{-1}_{n,k} \boldsymbol{T}_k+\sum_{n=3}^4 F_{2,n}(r) \int_0^r dr' F^{-1}_{n,k} \boldsymbol{T}_k \Bigg]\,.
\end{align}
The total scalar field energy, linear and angular momenta radiated during a given process can be found using solely the amplitudes $Z_1^\infty$ and $Z_2^\infty$. These are given by
\begin{align}
Z_1^\infty&=\sum_{k=3}^4 \int_{0}^{\infty} dr' F^{-1}_{3,k}(r')\boldsymbol{T}_k(r') \,,\label{Z1inf} \\
Z_2^\infty&=\sum_{k=3}^4 \int_{0}^{\infty} dr' F^{-1}_{4,k}(r')\boldsymbol{T}_k(r') \,. 
\end{align} 
Let us now apply our framework to two physically interesting setups: a particle plunging into a Q-ball configuration, and a particle in a circular orbit within the Q-ball.
%%%%%%%%%%%%%%%%%%%%%%%%%%%%%%%%%%%%%%%%%%%%%%%%%%%%%%%%%%%
\paragraph{Plunging particle.}
%%%%%%%%%%%%%%%%%%%%%%%%%%%%%%%%%%%%%%%%%%%%%%%%%%%%%%%%%%%
Consider a particle moving at a constant velocity~$\boldsymbol{v}=-v\boldsymbol{e}_z$ (with $v>0$), plunging into a Q-ball, and crossing its center at $t=0$. 
In this case, the trace of the particle's energy-momentum tensor reads
\begin{align}
T_p=&-\left[\delta\left(r+v t\right)\delta\left(\theta\right) \Theta(-t)+ \delta\left(r-v t\right)\delta\left(\theta-\pi\right) \Theta(t)\right]\nonumber \\
&\qquad\times m_p \,\delta(\varphi)\frac{\sqrt{1-v^2}}{r^2 \sin \theta}\,.
\label{T_p_plunging}
\end{align}
Therefore, the source decompositions~\eqref{eq:source_decomp_T1} and~\eqref{eq:source_decomp_T2} read 
\begin{align}
T_1=&-\left[\cos\left(\frac{(\omega+\Omega)r}{v}\right) \delta_l^\text{even}-i \sin\left(\frac{(\omega+\Omega)r}{v}\right) \delta_l^\text{odd}\right]\nonumber \\
&\qquad\times m_p \, Y_l^0(0,0) \delta_m^0\frac{\sqrt{1-v^2}}{\sqrt{2\pi}r v}\,,	\\
T_2=&-\left[\cos\left(\frac{(\omega-\Omega)r}{v}\right) \delta_l^\text{even}-i \sin\left(\frac{(\omega-\Omega)r}{v}\right) \delta_l^\text{odd}\right]\nonumber \\
&\qquad\times m_p \, Y_l^0(0,0) \delta_m^0\frac{\sqrt{1-v^2}}{\sqrt{2\pi}r v}\,.
\end{align}
These satisfy the property 
\begin{align}
T_2(\omega,l,0;r)= T_1(-\omega,l,0;r)^*\,.
\end{align}
Thus, due to the form of the system~\eqref{Qball_Perturbation_Matrix_Sourced}, one has
\begin{align}
&Z_2(\omega,l,0;r)=Z_1(-\omega,l,0;r)^*\,,\\
&Z_2^\infty(\omega,l,0)=Z_1^\infty(-\omega,l,0)^*\,.\label{property_Z1Z2_plunge_qball}
\end{align}
Finally, the spectral fluxes~\eqref{Energy_flux}, \eqref{Momentum_flux} and \eqref{AngularMomentum_flux} become, respectively,
\begin{align}
\frac{d E^\text{rad}}{d \omega}&=4 \left|\omega+\Omega\right| \Re\left[\sqrt{(\omega+\Omega)^2-\mu_S^2}\right] \sum_l \left|Z_1^\infty(\omega,l,0)\right|^2\,, \label{Energy_flux_qball}\\
\frac{d P_z^\text{rad}}{d \omega}&=\sum_{l}\frac{8(l+1) \Theta\left[\left(\omega+\Omega\right)^2-\mu_S^2\right]\left|(\omega+\Omega)^2-\mu_S^2\right|}{\sqrt{(2l+1)(2l+3)}} \nonumber \\
&\qquad\times\Re\left[Z_1^\infty(\omega,l,0) Z_1^\infty(\omega,l+1,0)^*\right]\,,\label{Momentum_flux_qball}\\
\frac{d L_z^\text{rad}}{d \omega}&=0\,.
\end{align}
%
%%%%%%%%%%%%%%%%%%%%%%%%%%%%%%%%%%%%%%%%%%%%%%%%%%%%%%%%%%%
\paragraph{Orbiting particle}
%%%%%%%%%%%%%%%%%%%%%%%%%%%%%%%%%%%%%%%%%%%%%%%%%%%%%%%%%%%
Now we consider a system composed by a particle describing a circular orbit of radius $r_\text{orb}$ and angular frequency $\omega_\text{orb}$ inside a Q-ball and in its equatorial plane. The trace of the particle's stress-energy tensor is
\begin{align}
T_p&=-\frac{m_p}{r_\text{orb}^2}\sqrt{1-\left(\omega_\text{orb} r_\text{orb}\right)^2}\, \delta(r-r_\text{orb})\delta\left(\theta-\frac{\pi}{2}\right)\delta(\varphi-\omega_\text{orb} t)\,, \label{T_p_orbiting}
\end{align}
which implies
\begin{align}
T_{1,2}&=-m_p \sqrt{\frac{\pi}{2}}\, Y_l^m\left(\frac{\pi}{2},0\right) \frac{\sqrt{1-\left(\omega_\text{orb} r_\text{orb}\right)^2}}{r_\text{orb}}  \nonumber \\
&\qquad\times\delta\left(r-r_\text{orb}\right)\delta\left(\omega \pm\Omega-m \omega_\text{orb}\right)\,. \label{T12_orbiting}
\end{align}
Notice that $T_2(\omega,l,m)=(-1)^m T_1(-\omega,l,-m)$, hence due to the form of system~\eqref{Qball_Perturbation_Matrix_Sourced}, we have
\begin{align}
&Z_2(\omega,l,m;r)=(-1)^m Z_1(-\omega,l,-m;r)^*\,,\\
&Z_2^\infty(\omega,l,m)=(-1)^m Z_1^\infty(-\omega,l,-m)^*\,.
\end{align}
Then, the emission rate expressions~\eqref{Energy_flux_rate} and~\eqref{AngularMomentum_flux_rate} imply, omitting the arguments $(\omega,l,m)$, 
\begin{align}
&\dot{E}^\text{rad}= \frac{2}{\pi}\int d\omega \left|\omega+\Omega\right| \Re\left[\sqrt{(\omega+\Omega)^2-\mu_S^2}\right]\sum_{l,m} \left|Z_1^\infty\right|^2  \,, \nonumber\\
&\dot{L}_z^\text{rad}=\frac{2}{\pi}\int d\omega \,\epsilon_1(\omega)\Re\left[\sqrt{(\omega+\Omega)^2-\mu_S^2}\right]\sum_{l,m} m \left|Z_1^\infty\right|^2. \nonumber
\end{align}
where we remind that $\epsilon_1\equiv \text{sign}(\omega+\Omega+ \mu_S)$. Re-writing expression~\eqref{T12_orbiting} in the form
\begin{align}
T_{1,2}=\widetilde{T}(\omega_\text{orb},r_\text{orb}) \,\delta\left(r-r_\text{orb}\right) \delta\left(\omega\pm\Omega-m \omega_\text{orb}\right)\,,
\end{align}
the previous expressions for the rate of emission read
\begin{align}
\dot{E}^\text{rad}&=\frac{2}{\pi}  \sum_{l,m}\widetilde{T}^2\Big[a_1\left|F_{3,3}^{-1}\left(m\omega_\text{orb}-\Omega;\,r_\text{orb}\right)\right|^2\nonumber\\
&\qquad+a_2\left|F_{3,4}^{-1}\left(m\omega_\text{orb}+\Omega;\,r_\text{orb}\right)\right|^2\Big]\,,\label{Energy_flux_orbiting}
\end{align}
\begin{align}
\dot{L}_z^\text{rad}&=\frac{2}{\pi}  \sum_{l,m}m \widetilde{T}^2\Big[\epsilon_1 a_1 \left|F_{3,3}^{-1}\left(m\omega_\text{orb}-\Omega;\,r_\text{orb}\right)\right|^2\nonumber\\
&\qquad+\epsilon_1 a_2\left|F_{3,4}^{-1}\left(m\omega_\text{orb}+\Omega;\,r_\text{orb}\right)\right|^2\Big]\,.\label{AngularMomentum_flux_orbiting}
\end{align}
where 
\begin{align}
&a_1=|m \omega_\text{orb}|\Re\left[\sqrt{\left(m \omega_\text{orb}\right)^2- \mu_S^2}\right]\,,\nonumber \\
& a_2=\left|m \omega_\text{orb}+2\Omega\right|\Re\left[\sqrt{\left(m \omega_\text{orb}+2\Omega\right)^2- \mu_S^2}\right].
\end{align}
%%%%%%%%%%%%%%%%%%%%%%%%%%%%%%%%%%%%%%%%%%%%%%%%%%%%%%%
\section{Free oscillations}
%%%%%%%%%%%%%%%%%%%%%%%%%%%%%%%%%%%%%%%%%%%%%%%%%%%%
%
%
The numerical search for \acs{QNM} frequencies for Q-balls is summarized in Table~\ref{table:QNM_Qball}, for the particular configuration with $\Omega=0.3\mu_S$. When $\omega_\text{QNM}$ are pure real numbers, they are normal modes of the object. For a mode to be normal, it must not be dispersed to infinity, hence the condition $\omega<\mu_S-\Omega$ is necessary, which also implies that such modes are screened from far-away observers by the Q-ball background itself. This means that perturbations associated with the real-valued frequencies in Table~\ref{table:QNM_Qball} do not reach spatial infinity. Such modes are the analogs of the \acs{NBS} modes found in Chapter~\ref{ch:nbs}, which were \emph{all} normal (Table~\ref{table:QNM_BS_invariant}). Q-balls have, in addition to such modes, also quasi-normal modes, which decay in time since they have a sufficiently large energy to disperse to infinity.
\begin{table}[h] 
	\centering
	\begin{tabular}{ccccc}
		\hline
		\hline
		$l$ &  \multicolumn{4}{c}{$\omega_\text{QNM}/\mu_S$} \\ 
		\hline
		\hline
		%		0 & $0.186$    & $0.439$ & $0.689$ & $0.931 - 1.2 \times 10^{-4} i$& $1.153 - 1.6\times 10^{-2} i$ &$1.381 - 2.0 \times 10^{-2} i$\\
		0 &     $0.439$ & $0.689$ & $0.931 - 1.2 \times 10^{-4} i$& $1.153 - 1.6\times 10^{-2} i$ \\
		1 & $0.300$  &  $0.555$  &   $0.806 - 9.8 \times 10^{-4} i$ &  $1.04 - 3.3\times 10^{-3}i$\\
		%		0 & $0.185$    & $0.93 - 1\times 10^{-4} i$      &$2.27 - 3\times 10^{-3} i$  \\
		%		1 & $0.300$    & $1.04 - 4\times 10^{-3}i$       &$2.30 - 3\times 10^{-3} i$\\
		%		2 & $0.403$    & $0.92 - 3\times 10^{-3}i$       &$2.29 - 1\times 10^{-2} i$\\
		\hline
		\hline
	\end{tabular} 
	\caption{Some \acs{QNM} frequencies of a Q-ball configuration with $\Omega/\mu_S=0.3$, for $l=\{0,1,2\}$. Note that the first column corresponds to normal modes, with $\omega<\mu_S$, hence screened from distant observers; they are confined to a spatial extent~$\sim R_\text{Q}$, the radius of the Q-ball (these modes are the analogs of the \acs{NBS} modes in Table \ref{table:QNM_BS_invariant}). There is an infinity of \acs{QNM} frequencies, parametrized by an integer overtone index~$n$. At large $n$, $\Re\left(\omega_\text{QNM}\right)\sim 0.22n\sim \pi n/R_\text{Q}$, as might be anticipated by a WKB analysis. Our results for the imaginary part of $\omega_\text{QNM}$ carry a large uncertainty, and should be taken only as an order of magnitude estimate.
	}
	\label{table:QNM_Qball}
\end{table}

%%%%%%%%%%%%%%%%%%%%%%%%%%%%%%%%%%%%%%%%%%%%%%%%%%%%%%%%%%%%%%%%%%%%%%%%%
\subsection{Particles plunging into Q-balls\label{sec:Plunging_particle}}
%%%%%%%%%%%%%%%%%%%%%%%%%%%%%%%%%%%%%%%%%%%%%%%%%%%%%%%%%%%%%%%%%%%%%%%%%
%
\begin{figure}
	\centering
	\includegraphics[width=0.9 \textwidth]{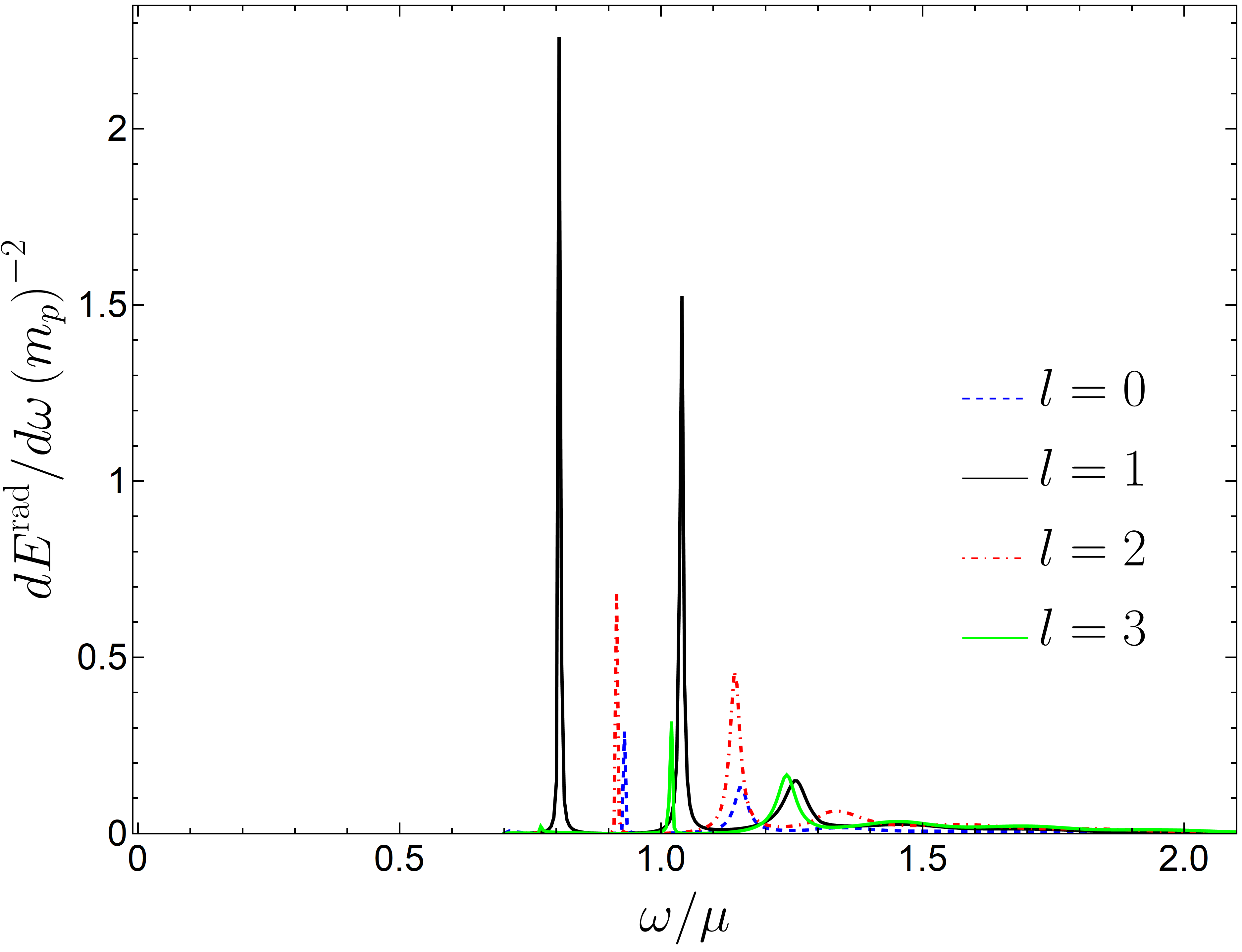} 
	\caption{Energy spectra of scalar radiation emitted when a particle of rest-mass $m_p$ plunges through a Q-ball with $\Omega=0.3\mu_S$, with a large velocity $v=0.8c$.
	The spectrum was decomposed into multipoles (Eq.~\eqref{Energy_flux_qball}). 
	The sharp peaks correspond to the excitation of \acs{QNM} frequencies $\omega_\text{QNM}$ (Table~\ref{table:QNM_Qball}).
	}
	\label{fig:Plunging_Spectra_Qball}
\end{figure}
\begin{figure}
	\centering
	\includegraphics[width=0.93 \textwidth]{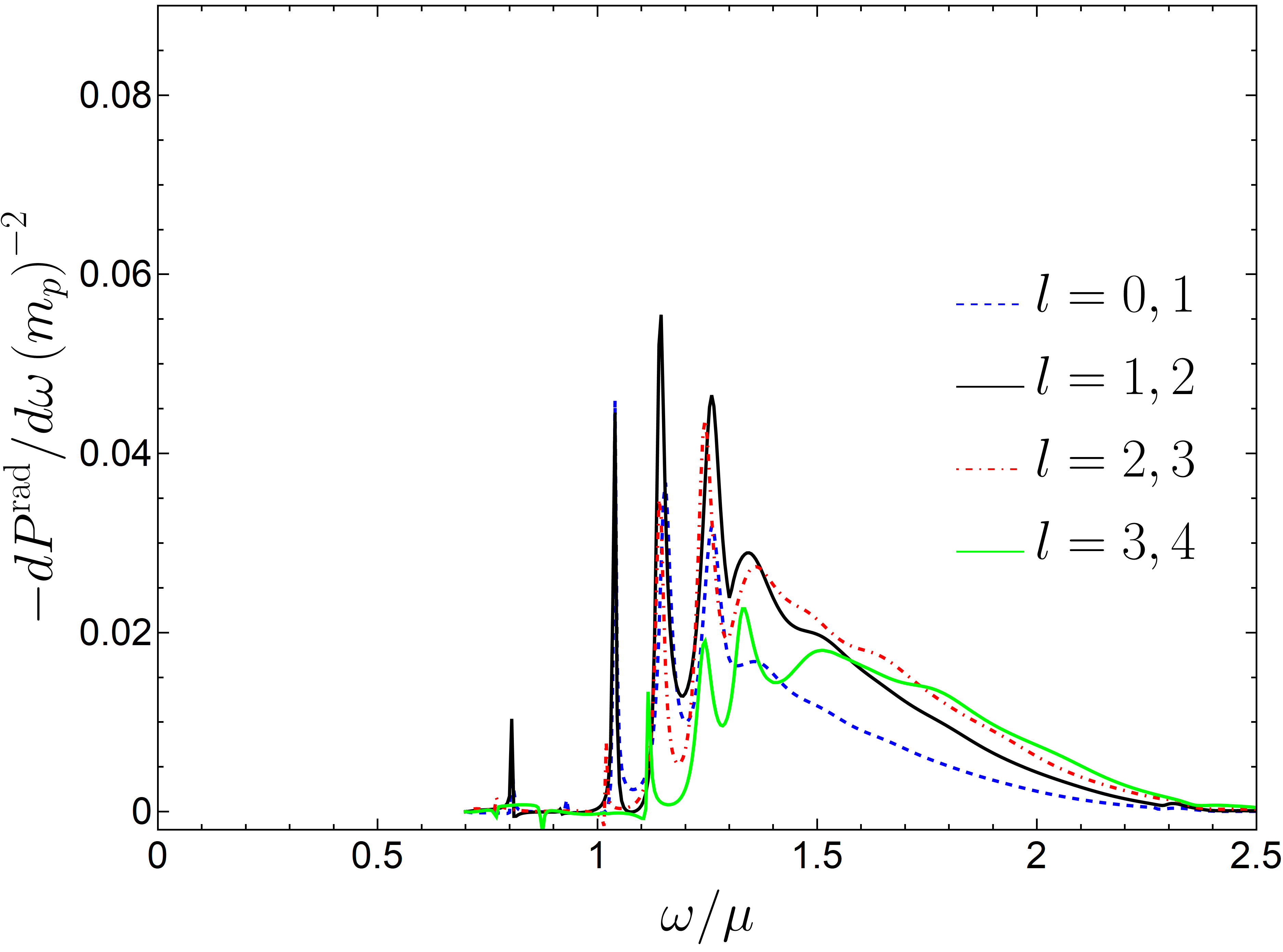}
	\caption{Linear momentum spectra of scalar radiation emitted when a particle plunges through a Q-ball with $\Omega=0.3\mu_S$, with a velocity $v=0.8$. 
		Different lines correspond to the different multipolar cross-terms in Eq.~\eqref{Momentum_flux_qball}.}
	\label{fig:Plunging_momentum_Qball}
\end{figure}

For concreteness, here we restrict the discussion to a large-velocity plunge $v=0.8c$.
The multipolar energy spectrum $d E^\text{rad}_l/d \omega$ of radiation released during such process is shown in Fig.~\ref{fig:Plunging_Spectra_Qball} for the first lowest multipoles, obtained through numerical evaluation of Eq.~\eqref{Energy_flux_qball}. Just like a hammer hitting a bell excites its characteristic vibration modes, the effect of a plunging particle is to excite the \acsp{QNM} of a Q-ball.
Figure~\ref{fig:Plunging_Spectra_Qball} illustrates this feature very clearly, the peaks in the energy spectrum are all coincident with the \acsp{QNM}, some of them identified in Table~\ref{table:QNM_Qball}. This feature was absent in the dynamics of \acsp{NBS} simply because the modes of \acsp{NBS} (Table~\ref{table:QNM_BS_invariant}) are all normal and confined within the \acs{NBS}; they do not arrive at spatial infinity. We see that most of the radiation is dipolar, looking at Fig.~\ref{fig:Plunging_Spectra_Qball}, but a substantial amount is also emitted in higher multipoles.
For example,~the $l=4$ mode still carries roughly~$10\%$ of the total radiated energy. Our results are compatible with an exponential suppression at large~$l$, of the form
$E^\text{rad}_l\sim 0.085 e^{-0.39 l}$. We can use this to sum over all multipoles and find the total energy radiated, 
\begin{align}
E^\text{rad}\sim 0.188 \, m_p^2 \,\mu_S\,.
\end{align}

The emitted radiation carries linear momentum, which is due to an interference term between multipoles (Eq.~\eqref{Momentum_flux_qball}). 
Figure~\ref{fig:Plunging_momentum_Qball} shows the contribution of the multipoles $l\leq 4$ to the spectral flux of linear momentum $d P_z^\text{rad}/d \omega$, obtained through numerical evaluation of~\eqref{Momentum_flux_qball}. Again, most of the contribution comes from the excitation of the Q-ball's \acsp{QNM}. Note the interesting feature that, although not shown in the figures, in some frequency ranges and for some interference terms, the radiated momentum is actually positive along~$\boldsymbol{e}_z$, \ie, opposite to the direction of the motion. 
We observed numerically that the total flux of linear momentum~$P_z^\text{rad}$ converge exponentially in $l$, for sufficiently large $l$. The total radiated momentum is always negative, and, thus, represents a slowing-down of the moving point particle. Using a similar fitting procedure to sum over all multipoles, we find for this particular configuration,
\begin{align}
P^\text{rad}\sim-0.088 \, m_p^2\, \mu_S\,.
\end{align}
%

%%%%%%%%%%%%%%%%%%%%%%%%%%%%%%%%%%%%%%%%%%%%%%%%%%%%%%%%%%%%%%%%%
\subsection{Orbiting particles\label{sec:Orbiting_particle}}
%%%%%%%%%%%%%%%%%%%%%%%%%%%%%%%%%%%%%%%%%%%%%%%%%%%%%%%%%%%%%%%%%
%
\begin{figure}
	\begin{tabular}{c}
	\includegraphics[width=0.9 \textwidth]{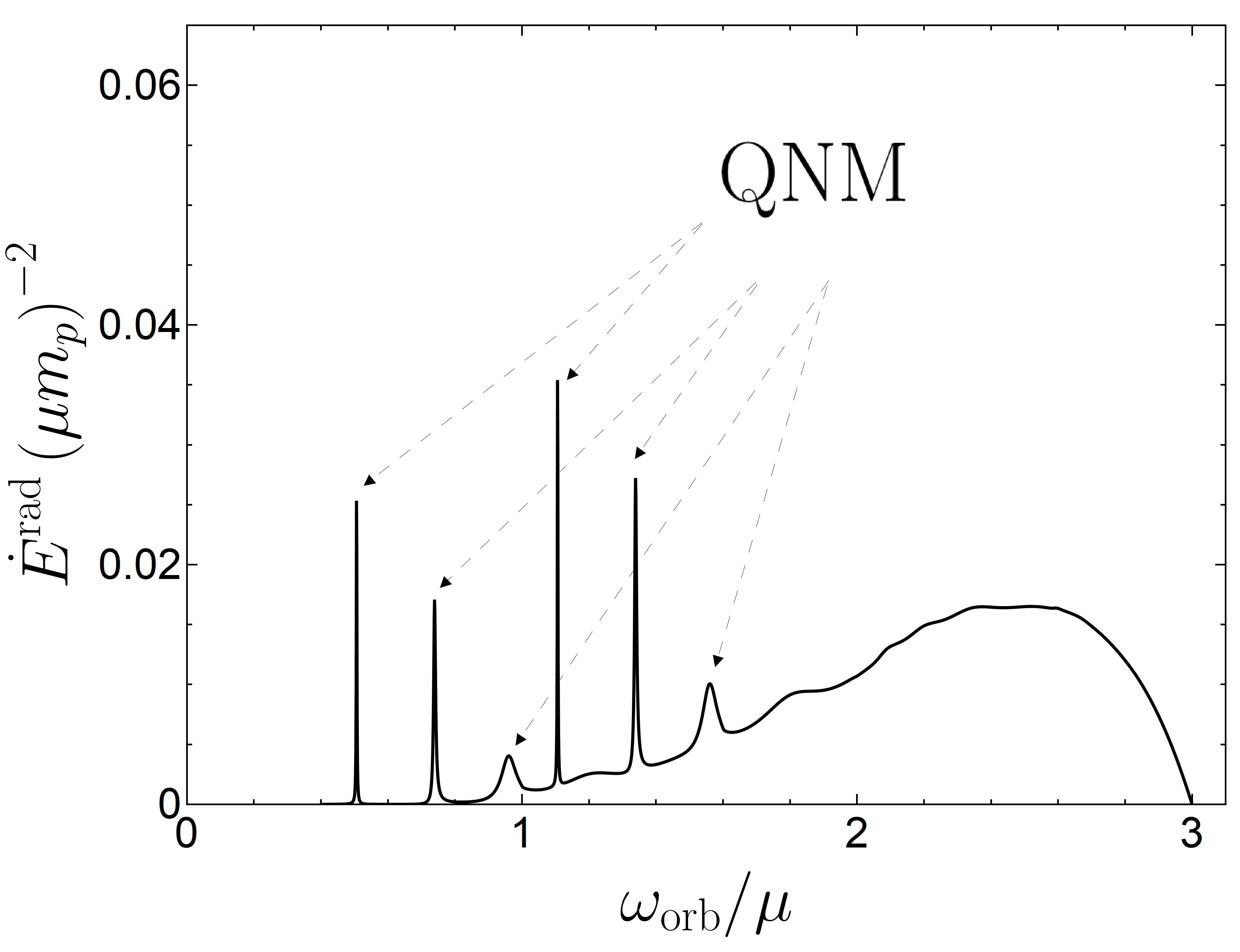}\\ \includegraphics[width=0.9 \textwidth]{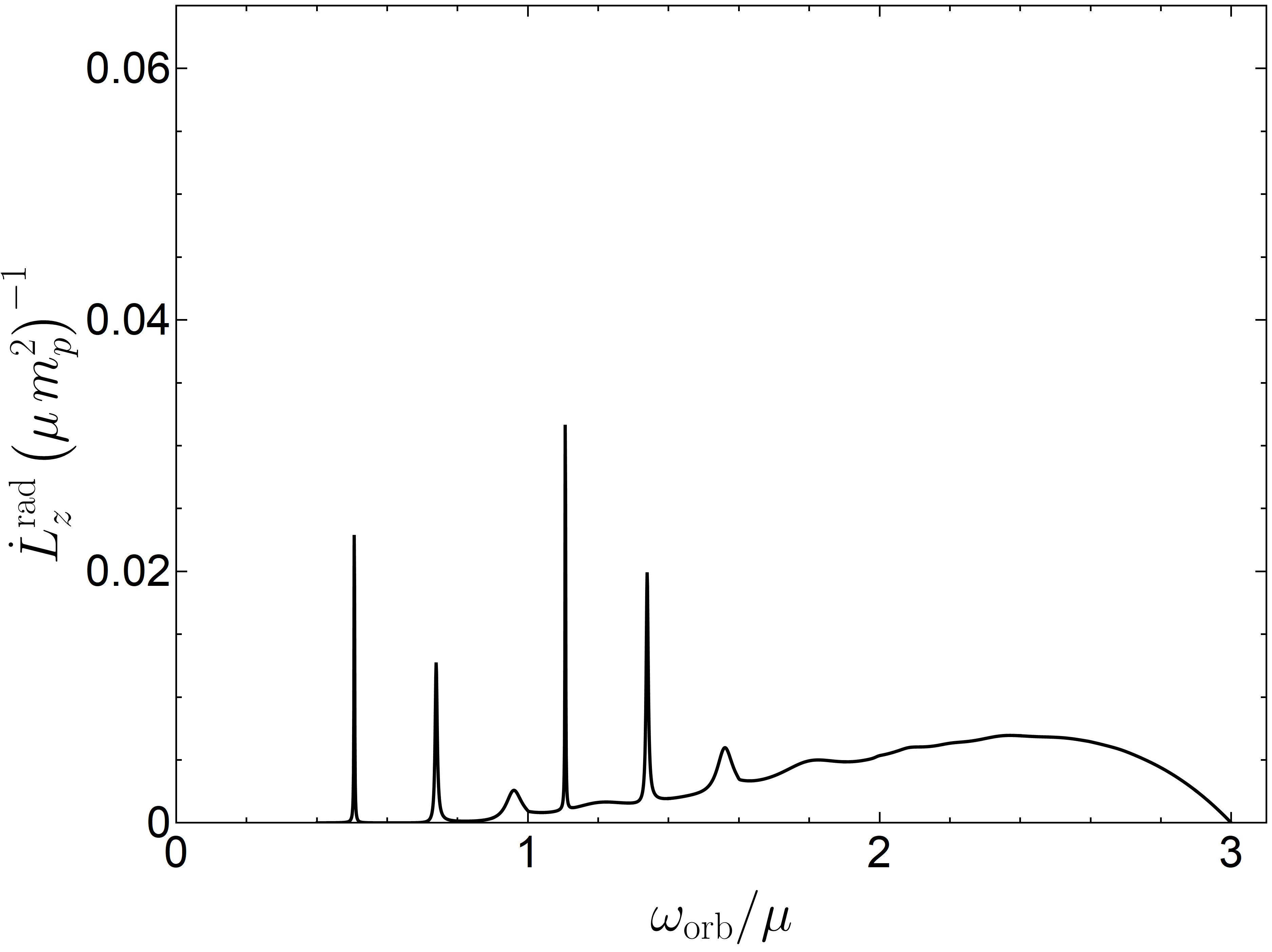}
	\end{tabular}
	\caption{Average dipolar ($|m|=l=1$) rate of energy (up), and angular momentum (down) radiated by a particle describing a circular orbit around a Q-ball with $\Omega=0.3\mu_S$, at radius $r_\text{orb}\mu_S=1/3$ and with orbital frequency $\omega_\text{orb}$. The peaks are associated with the excitation of \acs{QNM} frequencies $\omega_\text{QNM}$ for $\omega_\text{orb}=\Re\left(\omega_\text{QNM}\right)\pm \Omega$ -- each \acs{QNM} frequency is excited by two different $\omega_\text{orb}$ spaced by $2\Omega$. The excitation of the \acs{QNM} frequencies with $\Re(\omega_\text{QNM})=\{0.806, \,1.04$ (in Table~\ref{table:QNM_Qball})$,\,1.298\}\mu_S$ is clearly seen from these plots. However, it seems that not all the \acs{QNM} frequencies can be efficiently excited; \eg, $\Re(\omega_\text{QNM})/\mu_S=2.30$ (in Tab.~\ref{table:QNM_Qball}).}
	\label{fig:OrbitingFluxesQball}
\end{figure}

\begin{figure}
	\centering
	\includegraphics[width=0.9 \textwidth]{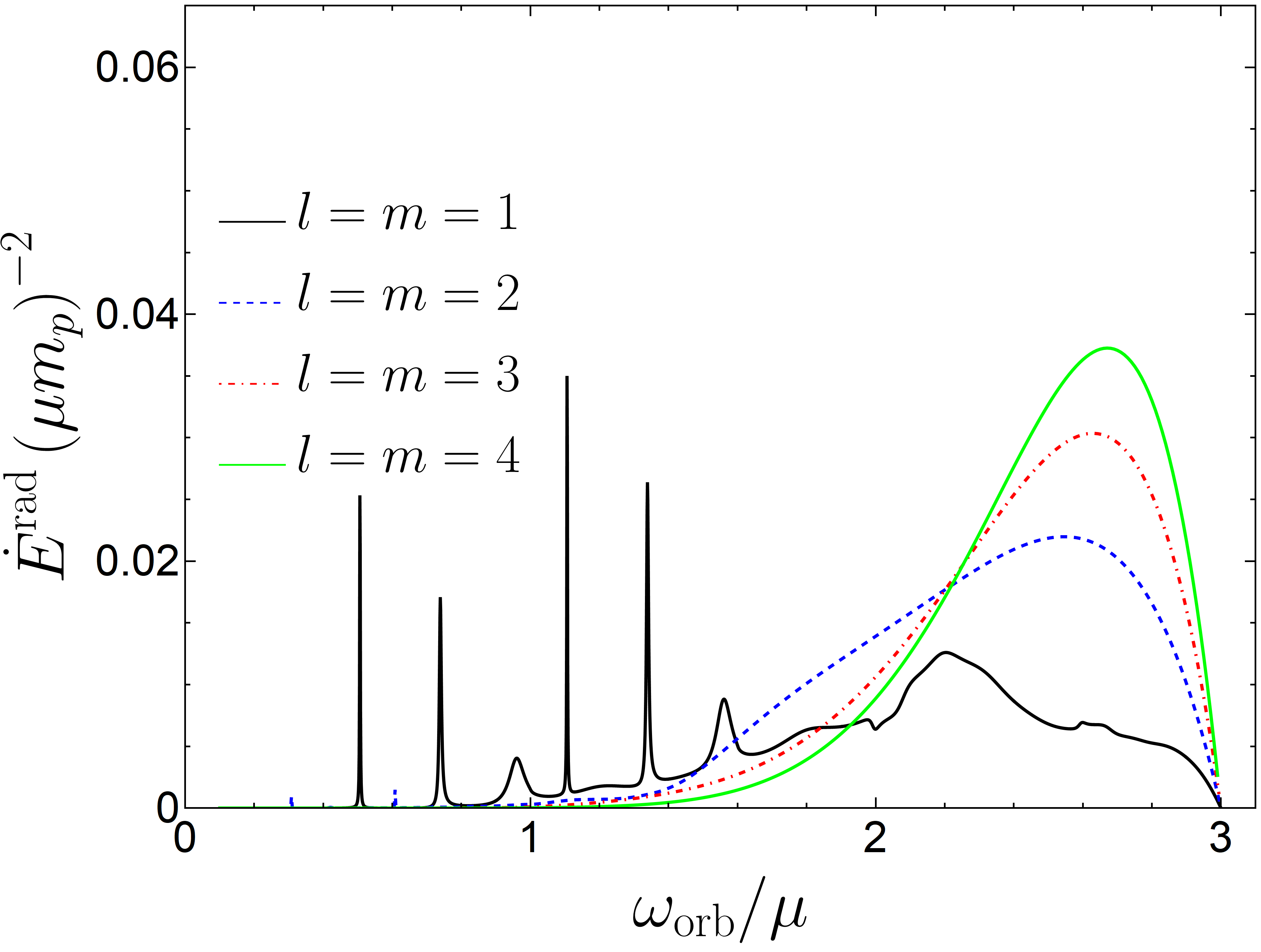} \caption{Average rate of energy radiated by a particle describing a circular orbit around a Q-ball with $\Omega=0.3\mu_S$, at a radius $r_\text{orb}\mu_S=1/3$ and with orbital frequency $\omega_\text{orb}$ for different values of $l=m$. At low frequencies the radiation is mostly dipolar. At large orbital frequencies the radiation is synchrotron-like and peaked at large $l=m$. In the high-frequency regime, there is a critical multipole $m$ beyond which the energy radiated decreases exponentially
		(see main text for further details). 
		There are QNM peaks for all multipoles, but they are visible only for the dipolar and quadrupolar.} 	
	\label{fig:OrbitingFluxesQballMorels}
\end{figure}
The average dipolar flux of energy and angular momentum emitted by a particle in circular orbit inside a Q-ball ($\Omega=0.3\mu_S$), at an orbital distance $r_\text{orb}\mu_S=1/3$, 
are shown in Fig.~\ref{fig:OrbitingFluxesQball}. The point-like source is assumed to be orbiting due to some external force, and its orbital frequency is varied, scanning possible resonant behavior with the Q-ball. As expected, and verified numerically, the quantity $\dot{E}^\text{rad}$ is an even function of $\omega_\text{orb}$, whereas $\dot{L}_z^\text{rad}$ is an odd one. A few features are apparent in the results above (obtained evaluating Eqs.~\eqref{Energy_flux_orbiting}-\eqref{AngularMomentum_flux_orbiting}).
The fluxes have clear peaks, which correspond to the resonant excitation of the \acsp{QNM} of the Q-ball. It is worth to note that for each \acs{QNM} frequency listed in Table~\ref{table:QNM_Qball} there are two peaks associated with different orbital frequencies separated by a distance~$2\Omega$; the resonances now occur at $\omega_\text{orb}=\Omega\pm\omega_\text{QNM}$. This comes directly from the decomposition~\eqref{MatterDecomposition}.

In flat space, a scalar charge on a circular orbit also emits radiation~\cite{Cardoso:2007uy,Cardoso:2011xi}. For small orbital frequencies and massless fields, the flux is dipolar and of order~$\dot{E}\sim q^2r_\text{orb}^2\omega_\text{orb}^4/(12\pi)$~\cite{Cardoso:2007uy,Cardoso:2011xi} (with a scalar charge $q=m_p$, in our coupling).
This explains the rise of the dipolar flux when the orbital frequency increases. However, at large frequencies, the radiation becomes of synchrotron type, and it is emitted preferentially in higher multipoles~\cite{Misner:1972jf,Breuer}. This is apparent in Fig.~\ref{fig:OrbitingFluxesQballMorels} where we show the contribution of higher multipoles to the flux. 
Note that all other multipoles also have resonant peaks, but these are less pronounced than the dipolar. At large Lorentz factors~$\gamma$, there is a critical~$m$ mode after which the fluxes becomes exponentially suppressed. The critical multipole is of order $m_\text{crit}\propto \gamma^2$~\cite{Misner:1972jf,Breuer}. Thus an evaluation of a large number of multipoles is necessary to have an accurate estimate of fluxes at large velocities. Our results are consistent with such a prediction. We find that as $\omega_\text{orb}$ increases, the flux peaks at higher and higher $m$, but there is always a threshold $m$ beyond which the radiation output is exponentially suppressed. Finally, since this process is not axially symmetric, one cannot use expression~\eqref{Momentum_flux} to compute the flux of linear momentum along $z$. Nevertheless, it is straightforward to show that the average rate of linear momentum radiated $\dot{P}_z^\text{rad}$ vanishes.

\begin{figure}
	\centering
	\includegraphics[width=0.9 \textwidth]{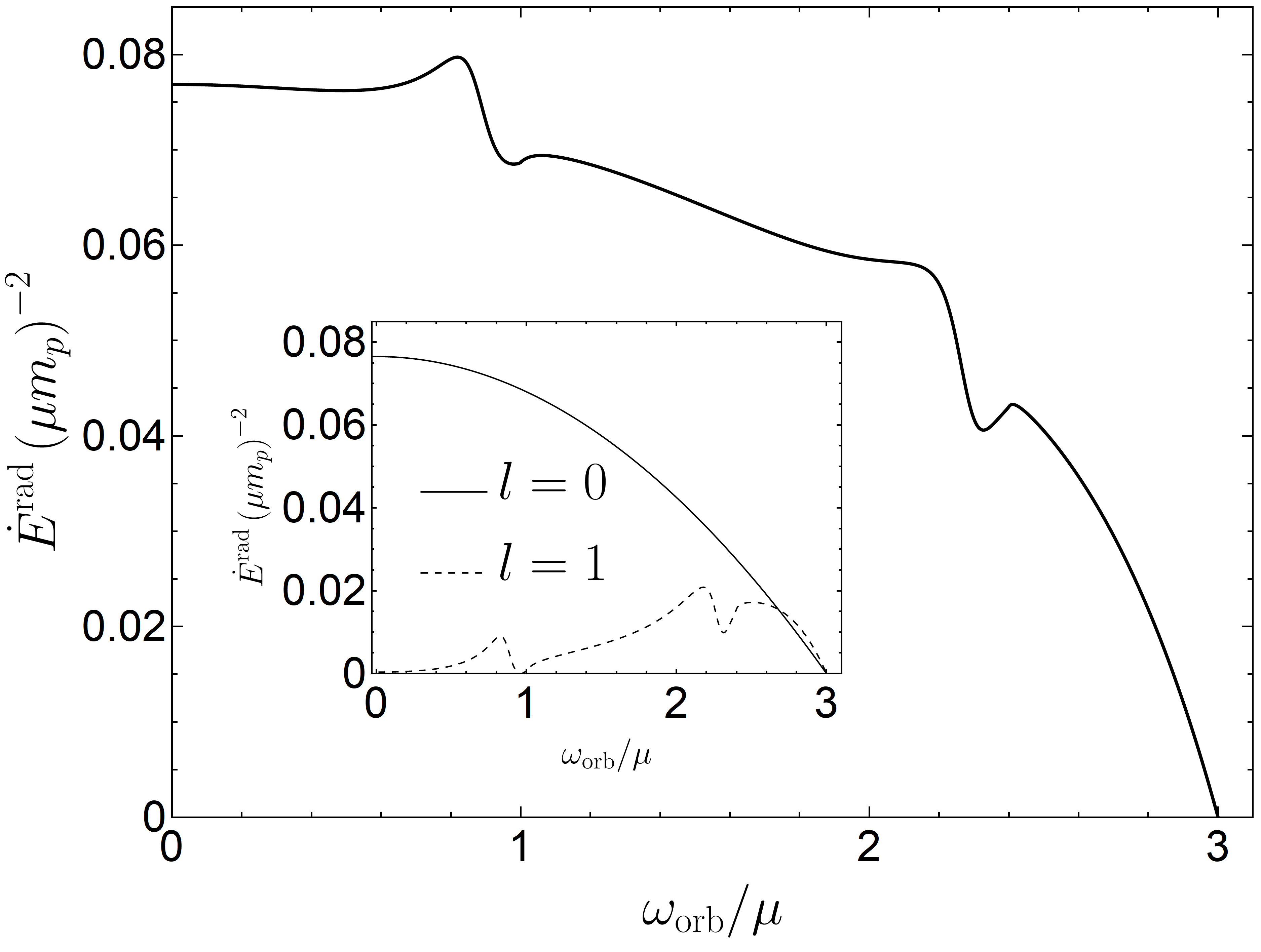} \caption{Average rate of energy radiated by a particle describing a circular orbit around a Q-ball with $\Omega=0.7\mu_S$, at radius $r_\text{orb}\mu_S=1/3$ and with orbital frequency $\omega_\text{orb}$. For such a scalar configuration there is radiation emitted also in the monopole mode, and it dominates the emission, as seen in the inset.} 	\label{fig:OrbitingFluxesQballOmega0p7}
\end{figure}
\begin{figure}
	\centering
	\includegraphics[width=0.9 \textwidth]{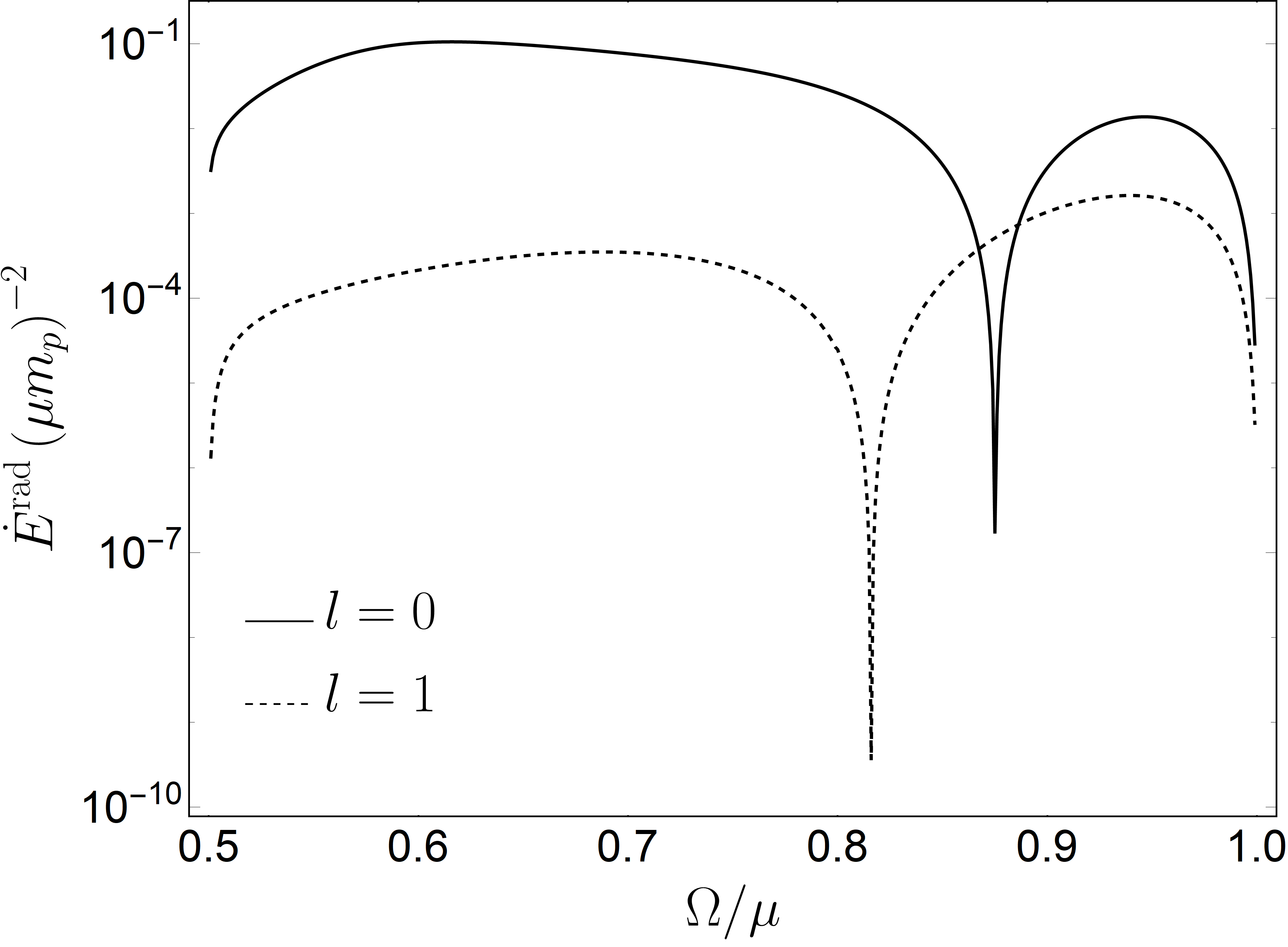} \caption{Average rate of energy radiated in the case of a particle standing at a fixed radius $r_\text{orb}\mu_S=1/3$ as function of $\Omega/\mu_S$. It is shown the dominant contributions from the modes $l=0$ and $l=1$. The average rate of angular momentum radiated in this case vanishes.} 	\label{fig:NotOrbitingFluxesQball}
\end{figure}
One interesting aspect, not seen in the study of \acsp{NBS}, concerns \emph{monopolar} emission and emission from particles \emph{at rest}. Both features are usually absent.
It follows from Eq.~\eqref{Energy_flux_orbiting}, that for Q-ball configurations with $\Omega\leq \mu_S/2$ there is no emission of $l=0$, and the first mode contributing to the radiation is $l=1$. For these objects there is no radiation emitted if the particle is at rest, with $\omega_\text{orb}=0$. However, for Q-balls
with $\Omega/\mu_S>1/2$ there is indeed emission from $l=0$ modes, contributing more than (or, at least as much as) the~$l=1$ modes to the radiation (Fig.~\ref{fig:OrbitingFluxesQballOmega0p7}). Interestingly, for these Q-balls there is also radiation emitted even when the particle is at rest (Fig.~\ref{fig:NotOrbitingFluxesQball}). This type of behavior is due to the \emph{scalar} coupling between two dynamical entities: the external perturber (through $T_p$) and the Q-ball configuration (through $\Phi$). The different coupling (purely gravitational, with no scalar charge) considered in the treatment of \acsp{NBS} led to the absence of these features.

%*****************************************
%*****************************************
%*****************************************
%*****************************************
%*****************************************

%*****************************************
\chapter{Green's Function of Gaseous Slabs }\label{app:DF}
%*****************************************

Here we derive expression~\eqref{GreensF}. The Green's function~$G(t,\boldsymbol{r};t',\boldsymbol{r}')$ of a three-dimensional gaseous slab is a solution of
\begin{equation}
\nabla_{\boldsymbol{r}}^2 G-\frac{1}{c^2}\frac{\partial^2 G}{\partial t^2}=- \delta(t-t') \delta^{(3)}(\boldsymbol{r}-\boldsymbol{r}') \,,
\end{equation}
with Dirichlet boundary conditions
\begin{equation} \label{DirichletBC}
	G(t,x,y,z=\pm L;\,t', \boldsymbol{r}')=G(t, \boldsymbol{r};\,t', x',y',z'=\pm L)=0\,.
\end{equation} 

Since the slab is homogeneous in the~$x$ and~$y$ directions, the Green's function is translation-invariant in these directions and can only depend on~$\boldsymbol{R}=(x-x',y-y')$. In the same way, since the medium was assumed to be static, the Green's function is time translation-invariant and only depends on~$T=t-t'$.
Now, note that the most general function of~$T$,~$\boldsymbol{R}$,~$z$ and~$z'$ satisfying~\eqref{DirichletBC} can be expanded as
\begin{align} \label{GreenExpan}
	G=\frac{1}{(2\pi)^3L} \sum_{n,n'>0}\int d\omega\, &d^2 \boldsymbol{K}\,\widetilde{G}_{n n'}(\omega, \boldsymbol{K})e^{-i (\omega T- \boldsymbol{K}\cdot \boldsymbol{R})}\nonumber \\&\times\sin[m_{n'}(z'+L)]\sin[m_{n}(z+L)]\,,
\end{align}
with~$m_n=n \pi/(2 L)$. Using the relations
\begin{align}
	&\delta(T)\delta^2(\boldsymbol{R})=\frac{1}{(2 \pi)^3} \int d\omega\, d^2 \boldsymbol{K}\,e^{-i (\omega T- \boldsymbol{K}\cdot \boldsymbol{R})}\,, \\
	&\delta(z-z')=\frac{1}{L} \sum_{n>0}\sin[m_{n}(z'+L)]\sin[m_{n}(z+L)]\,,
\end{align}
where the second expression is the \emph{completeness relation} of the orthonormal basis~$\{\frac{1}{\sqrt{L}}\sin[m_{n}(z+L)]\}_{n>0}$ of the function space~$\mathbb{L}^2(-L,L)$ with Dirichlet conditions at the boundaries.~\footnote{The completeness relation comes directly from the fact that any function~$f(z)$ in~$\mathbb{L}^2(-L,L)$ with Dirichlet conditions at the boundaries can be expanded as
\begin{align*}
	f(z)=\frac{1}{\sqrt{L}}\sum_{n>0}\left\{\frac{1}{\sqrt{L}}\int_{-L}^{L} dz'f(z')\sin[m_{n}(z'+L)] \right\}\sin[m_{n}(z+L)]\,.
\end{align*}
Interchanging the order of the sum and the integration one arrives at the relation.
}
Then, Eq.~\eqref{GreensF} implies
\begin{align}\label{GreenFour}
	\widetilde{G}=\frac{\delta_{n n'}}{k^2+m_n^2-\left(\frac{\omega}{c}\right)^2}\,.
\end{align}
Now, note that
\begin{align}
	&\hspace{-0.8cm}\int d\omega d^2 \boldsymbol{K}\frac{e^{-i (\omega T- \boldsymbol{K}\cdot \boldsymbol{R})}}{K^2+m_n^2-\left(\frac{\omega}{c}\right)^2}=-2\pi c^2 \int_0^{\infty} dK\, K J_0(KR) \int_{-\infty}^{\infty}d\omega\frac{e^{-i \omega T}}{\omega^2-\omega_K^2}\nonumber\\
	&=(2 \pi)^2 c^2\, \Theta(T) \int_0^\infty dK\,K J_0(KR) \frac{\sin(\omega_K T)}{\omega_K} \nonumber \\
	&=(2 \pi)^2 c\, \Theta(cT-R)\frac{\cos\Big(m_ n \sqrt{(c T)^2-R^2}\Big)}{\sqrt{(c T)^2-R^2}}\,,
\end{align}
where~$\omega_K=c \sqrt{K^2+m_n^2}$ and~$J_0$ is a Bessel function of the first kind. The first equality results from the integration in the angle between~$\boldsymbol{K}$ and~$\boldsymbol{R}$ (note that both vectors are two-dimensional), the second equality results from an integration in~$\omega$ over the complex-plane using Cauchy's integral formula, and for the third equality we used formula~$6.737.5.$ of Ref.~\cite{gradshteyn2007}.

Finally, plugging~\eqref{GreenFour} back in the Fourier expansion~\eqref{GreenExpan} and using the result of the last integration one finds
\begin{align}
	\hspace{-0.6cm} G=\frac{c}{2 \pi L} \Theta(c T- R) \sum_ {n>0} \frac{\cos(m_ n D)}{D}\sin[m_{n}(z'+L)]\sin[m_{n}(z+L)]\,,
\end{align}
where~$D=\sqrt{(c T)^2-R^2}$. That is exactly expression~\eqref{GreensF}. All the steps could be repeated in a similar fashion to derive
\begin{align}
\hspace{-0.6cm}G=\frac{c}{2 \pi L} \Theta(c T- R) \sum_ {n\geq 0} \frac{\cos(m_ n D)}{D}\cos[m_{n}(z'+L)]\cos[m_{n}(z+L)]\,
\end{align}
for Neumann boundary conditions.

%*****************************************
%*****************************************
%*****************************************
%*****************************************
%*****************************************

%*****************************************
\chapter{Scalar and vector charged binaries}\label{app:SVrad}
%*****************************************

In addition to \acs{GW} emission, many theories predict that binaries could also emit through other channels, such as scalar and vector radiation. These additional emission can take place, for instance, if the \acsp{BH} composing the binaries have scalar charges, as it is the case for self-interacting scalar fields, or even electromagnetic charges, as predicted by the Kerr-Newman class of \acsp{BH}. In what follows, we explore the consequences of additional radiative sectors for the evolution of binaries.

%%%%%%%%%%%%%%%%%%%%%%%%%%
\subsection{Scalar charged}
%%%%%%%%%%%%%%%%%%%%%%%%%%
%%%%%%%%%%%%%%%%%%%%%%%%%%
\subsubsection{The theory}
%%%%%%%%%%%%%%%%%%%%%%%%%%
Consider the theory~\eqref{theory_action} with currents
\begin{align}
J_S&= \sum_{n=1}^2 q_n^0  \int d \tau_n \frac{\delta^{(4)}\left(x^\alpha-x_n^{\;\; \alpha}(\tau_n)\right)}{\sqrt{-g}}\,, \\
J_V^{\;\;\alpha}&=0\,,
\end{align}
describing a massless real scalar field sourced by two particles moving on a curved spacetime with metric~$g_{\alpha \beta}$. 
Here $x_n^{\;\;\alpha}(\tau_n)$ is the world-line of the particle~$n=\{1,2\}$ parametrized by its proper time~$\tau_n$. Particle~$n$ has mass and scalar charge, respectively,~$m_n$ and $q^0_n$.
This theory has been extensively studied in the past (see, e.g., Refs.~\cite{Burko:2002ge,Quinn:2000wa}).

The Einstein~\acs{EOM}~\eqref{Eins_EOM} is
\begin{align} \label{Einstein_EOM}
G^{\alpha \beta}=8 \pi G &\Bigg[T_S^{\;\;\alpha \beta}+ \sum_{n=1}^{2}(m_n+ q^0_n \Phi)\nonumber\\ 
&\times\int d\tau_n \,\frac{dx_n^{\;\;\alpha}}{d \tau_n} \frac{dx_n^{\;\;\beta}}{d \tau_n}\frac{\delta^{(4)}\left(x^\delta-x_n^{\;\; \delta}(\tau_n)\right)}{{\sqrt{-g}}}\Bigg]\,,
\end{align}
with the scalar energy-momentum tensor given in~\eqref{scalarEMT}, the scalar~\acs{EOM}~\eqref{KG_EOM} is
\begin{align} 
\frac{1}{\sqrt{-g}}\partial_\alpha\left(\sqrt{-g}\,g^{\alpha \beta} \partial_\beta \Phi\right)=\sum_{n=1}^2 q_n^0  \int d \tau_n \frac{\delta^{(4)}\left(x^\alpha-x_n^{\;\; \alpha}(\tau_n)\right)}{\sqrt{-g}}\,,\label{Scalar_EOM}
\end{align}
and~\acs{EOM} of the particles is
\begin{align} 
\left(m_n+q^0_n \Phi\right) u_n^{\;\;\alpha} \nabla_\alpha u_n^{\;\;\beta}=- q^0_n\left(g^{\alpha \beta}+u_n^{\;\;\alpha} u_n^{\;\;\beta}\right)\Phi_{,\alpha}\,,\label{Particle_EOM}
\end{align}
where~$u_n^\alpha \equiv dx_n^{\;\;\alpha}/d\tau_n$ is the 4-velocity of particle~$n$.

%%%%%%%%%%%%%%%%%%%%%%%%%%%%%%%%%%%%%%%%%%%%%%%%%%
\subsubsection{Newtonian binary with no radiation}
%%%%%%%%%%%%%%%%%%%%%%%%%%%%%%%%%%%%%%%%%%%%%%%%%%
Consider a slowly-moving, Newtonian binary, such that energy and angular momentum fluxes can be neglected at leading order. In this limit~Eq.~\eqref{Einstein_EOM} becomes a simple Poisson equation~\cite{poisson_will_2014}
\begin{align}
\nabla^2 U=4\pi G \sum_{n=1}^2 m_j \delta^{(3)}(\boldsymbol{x}-\boldsymbol{r}_n(t))\,, \label{Poisson_eq}
\end{align}
where~$x_n^{\;\;\alpha}\equiv(t,\boldsymbol{r}_n(t))$. The gravitational potential~$U(t,\boldsymbol{x})$ is weak, \textit{i.e.}~$|U|\ll1$, and enters in the Newtonian metric
\begin{align} 
ds^2=-(1+2 U)dt^2+dr^2+r^2\left(d \theta^2+ \sin^2\theta d\varphi^2\right)\,.\label{Newtonian_metric}
\end{align}
There is a (slowly time-varying) scalar field sourced by the point charges described by Eq.~\eqref{Scalar_EOM}, which in this limit becomes also a Poisson equation
\begin{align}
\nabla^2 \Phi_0= \sum_{n=1}^2 q^0_n \delta^{(3)}(\boldsymbol{x}-\boldsymbol{r}_n(t))\,, \label{Poisson_eq1}
\end{align}
The equation of motion of the particles~\eqref{Particle_EOM} simplifies to 
\begin{align}
u_n^{\;\;\alpha} \nabla_\alpha u_n^{\;\;\beta}=-\frac{q^0_j}{m_j+q^0_j \Phi_0}g^{\alpha \beta} \Phi_{,\alpha}\,.
\end{align}
We see that the particles are accelerated by the scalar. With the Newtonian metric~\eqref{Newtonian_metric} and assuming $q_1,q_2 \ll |\boldsymbol{r}_2-\boldsymbol{r}_1|$, this equation can be written in a familiar form~\footnote{One can see this directly by plugging the Newtonian metric~\eqref{Newtonian_metric} inside the particles action~\eqref{matter_particles_Lagr} obtaining 
\begin{align}
\mathcal{S}_{\textrm{part}}&=\sum_ j m_j\int dt \sqrt{(1+2 U)-|d\boldsymbol{r}_j/dt|^2} \nonumber \\
&\simeq \sum_ j m_j\int dt\left( 1+U-\tfrac{1}{2}|d\boldsymbol{r}_j/dt|^2\right)\,.
\end{align}
This is just the action describing a non-relativistic system of particles in a gravitational potential~$U$.}

\begin{align}\label{Newton_2law}
\frac{d^2}{d t^2} \boldsymbol{r}_n=-\boldsymbol{\nabla} U(t,\boldsymbol{r}_n)- \frac{q^0_n}{m_n} \boldsymbol{\nabla} \Phi_0(t,\boldsymbol{r}_n)\,,
\end{align}
where $\boldsymbol{\nabla}$ is the usual $3$-dimensional gradient operator. 
Using equation~\eqref{Poisson_eq} we obtain~\footnote{Actually, in this step we cannot really consider point sources, otherwise we would find problems with a diverging ``self-force''. Fortunately, this is not a real problem, and we can proceed by assuming that the particles have a small, but finite size.}
\begin{align}
&U(t,\boldsymbol{r}_1)=\frac{G m_2}{|\boldsymbol{r}_2(t)-\boldsymbol{r}_1|}\,, \quad &U(t,\boldsymbol{r}_2)=\frac{G m_1}{|\boldsymbol{r}_2-\boldsymbol{r}_1(t)|}\,, \\
&\Phi_0(t,\boldsymbol{r}_1)=\frac{q_2^0}{4\pi|\boldsymbol{r}_2(t)-\boldsymbol{r}_1|}\,, \quad &\Phi_0(t,\boldsymbol{r}_2)=\frac{q_1^0}{4 \pi|\boldsymbol{r}_2-\boldsymbol{r}_1(t)|}\,.\label{Coloumb_pot}
\end{align}
%
%%%%%%%%%%%%%%%%%%%%%%%%%%%%%%%%%%%%%%%%%%%%%%%%%%
\subsubsection{Elliptic motion and orbit-averaging}
%%%%%%%%%%%%%%%%%%%%%%%%%%%%%%%%%%%%%%%%%%%%%%%%%%
As one expects, Eqs.~\eqref{Newton_2law} with~\eqref{Coloumb_pot} describes a Keplerian orbital motion with energy and angular momentum given by Eq.~\eqref{eq:energy_angularmomentum_main}.
%
%\begin{align}
%E&=-\frac{\tilde{G} m_1m_2}{2a}\,,\\
%%
%L^2&=\frac{\tilde{G}m_1^2m_2^2a(1-e^2)}{M}\,,
%\end{align}
%%
%where
%%
%\begin{align}
%\tilde{G}\equiv 1-4 \pi \frac{q_1 q_2}{m_1 m_ 2}\,.
%\end{align}
%
These differ from~\eqref{Kepler_Energy} and~\eqref{Kepler_AngMom} due to the scalar interaction. Using spherical coordinates with origin at the system's center of mass the trajectories can be written as~$\boldsymbol{r}_1=\left(r_1(\varphi_p),\varphi_p,\pi/2\right)$ and~$\boldsymbol{r}_2=\left(r_2(\varphi_p),\varphi_p+\pi,\pi/2\right)$ with
\begin{align}
r_1=\frac{m_2}{M}r_p \,, \qquad r_2=\frac{m_1}{M} r_p\,, 
\end{align}
\begin{align}
r_p(\varphi_p)=\frac{a(1-e^2)}{1+e \cos \varphi_p} \,.
\end{align}
Their angular velocity is
\begin{align}
\dot{\varphi}_p=\sqrt{\frac{\tilde{G} M}{a^3}}(1-e^2)^{-3/2}(1+e\cos\varphi)^2\,.
\end{align}
%

%%
%\be
%v=\sqrt{\frac{\tilde{G} M(1+e^2+2e\cos\varphi)}{a(1-e^2)}}\,.
%\ee
%%

Finally, we define the average of a quantity $X$ over one period $T$ as
\begin{equation}
\left<X\right>=\frac{\omega_0}{2\pi}\int_0^{2\pi}\frac{d\varphi}{\dot{\varphi}}X(\varphi)\,.
\end{equation}
where $\omega_0$ is the (Keplerian) orbital frequency.
%%%%%%%%%%%%%%%%%%%%%%%%%%%%%%%%%%%%%%%%%%%%%%%%%%%%%%%%
\subsubsection{Radiation emitted by a Newtonian binary}
%%%%%%%%%%%%%%%%%%%%%%%%%%%%%%%%%%%%%%%%%%%%%%%%%%%%%%%%
A Newtonian binary sources a scalar field described by Eq.~\eqref{Scalar_EOM}, which can be put in the form
\begin{align} \label{Scalar_EOM1}
\Box \Phi= \rho(t,\boldsymbol{x}) \equiv\frac{1}{\sqrt{-g}} \sum_{n=1}^2 q^0_n\delta^{(3)}\left(\boldsymbol{x}-\boldsymbol{r}_n(t)\right)\,.
\end{align}
Thus, the binary will lose energy and angular momentum through this channel and the motion will not be truly Keplerian; the radiation reaction force entering~\eqref{Particle_EOM} (which we are neglecting in the computation of the radiation, since we are using an adiabatic approximation) will be responsible for a deviation to the Keplerian orbit. Let us compute the radiation emitted by this binary of scalar charges at the (leading) dipole approximation.

In the Newtonian approximation the scalar radiation propagates in flat space. So, the solution of the (sourced) scalar wave equation is
\begin{align}
\Phi(t, \boldsymbol{x})=\int d^3 \boldsymbol{x}'\sqrt{-g'}\, \frac{\rho(t-|\boldsymbol{x}-\boldsymbol{x}'|,\boldsymbol{x}')}{4 \pi|\boldsymbol{x}-\boldsymbol{x}'|}\,.
\end{align}
In the dipole approximation it is straightforward to show that
\begin{align}
\Phi(t,r\to \infty, \theta, \varphi)\simeq \frac{1}{4\pi r}\boldsymbol{e}_ r\cdot \dot{\boldsymbol{p}}(t-r)\,, 
\end{align}
with the dipole moment
\begin{align}
\boldsymbol{p}(t)\equiv \int d^3 \boldsymbol{x}' \sqrt{-g'} \rho(t, \boldsymbol{x}')\, \boldsymbol{x}'=\left(\frac{q^0_1 m_2-q^0_2 m_1}{M}\right) \boldsymbol{r}_p(t)\,.\nonumber
\end{align}
This approximation is valid for scalar waves with frequency~$\omega\sim \omega_0 \ll 1/a$, where~$\omega_0$ is the orbital frequency (which is compatible with the Newtonian approximation).
%
%Now we have all we need to compute the energy and angular momentum radiated with the scalar field. 
The radiated energy flux is
\begin{align}
\dot{E}^{\textrm{rad}}=-\lim_{r \to \infty} r^2 \int d \Omega\, T^S_ {rt}\,,
\end{align}
and the angular momentum through
\begin{align}
\dot{L}^{\textrm{rad}}=\lim_{r \to \infty} r^2 \int d \Omega\, T^S_ {r\varphi}\,.
\end{align}
Plugging the dipole approximation in the (real) scalar's energy-momentum tensor~\eqref{scalarEMT} we can write the last two expressions in the form
\begin{align}
\dot{E}^{\textrm{rad}}&=\left(\frac{q^0_1 m_2-q^0_2 m_1}{4 \pi M}\right)^2 \int d\Omega \, \left[\boldsymbol{e}_ r\cdot \ddot{\boldsymbol{r}}_p \right]^2 \nonumber \\
&=\frac{1}{12 \pi}\frac{\tilde{G}^2}{r_p^4}(q^0_1 m_2-q^0_2 m_1)^2\,, \label{EnergyS}
\end{align}
where we used~$\ddot{\boldsymbol{r}}_p=-\tilde{G} M \boldsymbol{r}_p/r_p^3$ and integrated over the sphere, and
\begin{align}
&\dot{L}^{\textrm{rad}}=-\left(\frac{q^0_1 m_2-q^0_2 m_1}{4 \pi M}\right)^2 \int d\Omega \, \left(\boldsymbol{e}_ r\cdot \ddot{\boldsymbol{r}}_p \right)\partial_\varphi \left(\boldsymbol{e}_ r\cdot \dot{\boldsymbol{r}}_p \right) \nonumber \\
&=\frac{1}{12 \pi} \tilde{G}^{\frac{3}{2}}\frac{\sqrt{a (1-e^2)}}{\sqrt{M}r_p^3}(q^0_1 m_2-q^0_2 m_1)^2\,.\label{AngMomS}
\end{align}
Averaging over an orbit we find
\begin{align}
&\langle\dot{E}^{\textrm{rad}}\rangle=\frac{1}{24 \pi}\frac{\tilde{G}^2}{a^4}(q^0_1 m_2-q^0_2 m_1)^2 \left(\frac{2+e^2}{(1-e^2)^{\frac{5}{2}}}\right) \label{EnergySav} \,, \\
&\langle\dot{L}^{\textrm{rad}}\rangle=\frac{1}{12 \pi} \frac{\tilde{G}^\frac{3}{2}}{\sqrt{M}a^{\frac{5}{2}}(1-e^2)}(q^0_1 m_2-q^0_2 m_1)^2  \label{AngMomSav}\,,
\end{align}
resulting in the ratio
\begin{align}\label{LEratioS}
&\frac{\langle\dot{L}^{\textrm{rad}} \rangle}{\langle\dot{E}^{\textrm{rad}} \rangle} = \frac{\sqrt{1-e^2}}{\omega_0} \left(\frac{1-e^2}{1+\frac{e^2}{2}}\right) \,.
\end{align}
In the adiabatic approximation the major semi-axis and the eccentricity follow
\begin{align}
\langle\dot{a}\rangle&=-\frac{2a^2\langle\dot{E}^{\textrm{rad}}\rangle}{\tilde{G} m_1m_2}<0\,,\label{shrink}\\
\langle\dot{e}\rangle&=\sqrt{\frac{M}{ \tilde{G}a}}\frac{\sqrt{1-e^2}}{e}\frac{\langle\dot{E}^{\textrm{rad}}\rangle}{m_1m_2}\left(\frac{\langle\dot{L}^{\textrm{rad}}\rangle}{\langle\dot{E}^{\textrm{rad}}\rangle}-\frac{\sqrt{1-e^2}}{\omega_0}\right) \nonumber \\
&=-\sqrt{\frac{M}{ \tilde{G}a}}\left(\frac{1-e^2}{e\, \omega_0}\right)\frac{\langle\dot{E}^{\textrm{rad}}\rangle}{m_1m_2}\left(\frac{3 e^2}{2+e^2}\right)\leq 0\,.\label{circular}
\end{align}
Thus, the emission of scalar radiation by a binary causes the major semi-axis and the eccentricity to decrease in time: the orbit shrinks and circularizes. 
In the circular orbit limit our results are in agreement with those of Refs.~\cite{Cardoso:2011xi,Yunes:2011aa,Cardoso:2019nis}.
%%%%%%%%%%%%%%%%%%%%%%%%%%%%%%%%%%%%%%%%%%%%%%%%%%%%%%%%%%%%%%%%%%%%%%%%%%%%%%%
\subsection{Vector charged}
%%%%%%%%%%%%%%%%%%%%%%%%%%%%%%%%%%%%%%%%%%%%%%%%%%%%%%%%%%%%%%%%%%%%%%%%%%%%%%%
%%%%%%%%%%%%%%%%%%%%%
\subsubsection{Theory}
%%%%%%%%%%%%%%%%%%%%%
Here we consider the theory~\eqref{theory_action} with currents
\begin{align}
	J_S&=0\,,\\
	J_V^{\;\;\alpha}&=\sum_{n=1}^2 q_n^1 \int d \tau_n \,\frac{d x_n^{\;\;\alpha}}{d \tau_n} \frac{\delta^{(4)}\left(x^\delta-x_n^{\;\; \delta}(\tau_n)\right)}{\sqrt{-g}}\,.
\end{align}
Particle~$n$ has mass~$m_n$ and electric charge~$q^1_n$.

The sourced Maxwell~\acs{EOM}~\eqref{Maxw_EOM}
\begin{align}
&\nabla_\mu F^{\alpha \beta}= J_V^{\;\;\beta}\,.
\end{align} 
In the Newtonian approximation and neglecting radiation (valid for slowly moving charges) we can repeat the exact same steps that we applied to the scalar charges to find that the electric charges also describe a Keplerian orbit; the only difference being that in the definition of~$\tilde{G}$ we have now electric charges instead of scalar charges. 
The energy-momentum tensor of the electromagnetic field is given in Eq.~\eqref{vectorEMT}.
%
%%%%%%%%%%%%%%%%%%%%%%%%%%%%%%%%%%%%%%%%%%%%%%%%%%%%%%%
\subsubsection{Radiation emitted by a Newtonian binary}
%%%%%%%%%%%%%%%%%%%%%%%%%%%%%%%%%%%%%%%%%%%%%%%%%%%%%%%
Again, the binary will radiate energy and angular momentum -- in this case through electromagnetic waves -- and the motion will not be truly Keplerian; we are considering the regime in which the orbits change adiabatically.

Using the Lorenz gauge~$\nabla_\mu A^\mu=0$ the sourced Maxwell equations become
\begin{align}
\Box A^\alpha= J_V^{\;\;\alpha}\,,
\end{align}
which we can decompose into
\begin{align}
&\Box \Phi= \rho(t,\boldsymbol{x})\equiv\frac{1}{\sqrt{-g}}\sum_{n=1}^2q^1_n \delta^{(3)}\left(\boldsymbol{x}-\boldsymbol{r}_ n\right)\,, \\
&\Box \boldsymbol{A}=  \boldsymbol{j}(t, \boldsymbol{x})\equiv\frac{1}{\sqrt{-g}}\sum_{n=1}^2q^1_n \boldsymbol{v}_n \delta^{(3)}\left(\boldsymbol{x}-\boldsymbol{r}_n\right)\,,
\end{align}
where we used that the sources are non-relativistic. In the Newtonian approximation we consider that the electromagnetic waves propagate in flat space. So, the solution to the (sourced) Maxwell equations is
\begin{align}
&\Phi(t, \boldsymbol{x})=\int d^3 \boldsymbol{x}'\sqrt{-g'}\, \frac{\rho(t-|\boldsymbol{x}-\boldsymbol{x}'|,\boldsymbol{x}')}{4\pi|\boldsymbol{x}-\boldsymbol{x}'|}\,, \\
&\boldsymbol{A}(t, \boldsymbol{x})=\int d^3 \boldsymbol{x}'\sqrt{-g'}\, \frac{\boldsymbol{j}(t-|\boldsymbol{x}-\boldsymbol{x}'|,\boldsymbol{x}')}{4\pi|\boldsymbol{x}-\boldsymbol{x}'|}\,.
\end{align}
In the dipole approximation one can show that
\begin{align}
&\Phi(t,r\to \infty, \theta, \varphi)\simeq \frac{1}{4\pi r}\boldsymbol{e}_ r\cdot \dot{\boldsymbol{p}}(t-r)\,, \\
&\boldsymbol{A}(t,r\to \infty, \theta, \varphi)\simeq \frac{1}{4\pi r} \dot{\boldsymbol{p}}(t-r)\,,
\end{align}
with the dipole moment
\begin{align}
\boldsymbol{p}(t)\equiv \int d^3 \boldsymbol{x}' \sqrt{-g'} \rho(t, \boldsymbol{x}')\, \boldsymbol{x}' =\left(\frac{q^1_1 m_2-q^1_2 m_1}{M}\right) \boldsymbol{r}_p(t)\,.\nonumber
\end{align}
Now, the magnetic field is
\begin{align}
\boldsymbol{B}(t, r\to\infty,\theta,\varphi)\equiv \boldsymbol{\nabla} \times \boldsymbol{A}\simeq -\frac{1}{4 \pi r} \boldsymbol{e}_r \times \ddot{\boldsymbol{p}}(t-r)
\end{align}
and using Ampère-Maxwell's law we have
\begin{align}
\dot{\boldsymbol{E}}(t, r\to \infty,\theta,\varphi)=\boldsymbol{\nabla}\times \boldsymbol{B}=\dot{\boldsymbol{B}}\times\boldsymbol{e}_ r\,,
\end{align}
which, integrating in time, gives the electric field
\begin{align}
\boldsymbol{E}(t, r\to \infty,\theta,\varphi)=\boldsymbol{B}\times\boldsymbol{e}_ r\,.
\end{align}
These result in the Poynting vector
\begin{align}
\boldsymbol{S}(t,r\to \infty,\theta,\varphi)\equiv \boldsymbol{E}\times \boldsymbol{B}=|\boldsymbol{B}|^2\boldsymbol{e}_r\,, 
\end{align}
where we used Lagrange's rule for the triple cross product and that~$(\boldsymbol{B}\cdot \boldsymbol{e}_r)=0$.
Now using the scalar quadruple product identity we have
\begin{align}
&|\boldsymbol{B}|^2=\frac{1}{16 \pi^2r^2}\left(|\ddot{\boldsymbol{p}}|^2-\left(\ddot{\boldsymbol{p}}\cdot \boldsymbol{e}_ r\right)^2\right) \,.
\end{align}
So the radiated energy flux is
\begin{align}
\dot{E}^{\textrm{rad}}&=-\lim_{r \to \infty} r^2 \int d \Omega\, T^{V}_ {rt}=\lim_{r \to \infty} r^2 \int d\Omega\, \boldsymbol{S}\cdot \boldsymbol{e}_r \nonumber \\
&=\left(\frac{q^1_1 m_2-q^1_2 m_1}{4 \pi M}\right)^2 \int d\Omega \, \left[|\ddot{\boldsymbol{r}}_p|^2-\left(\ddot{\boldsymbol{r}}_p\cdot \boldsymbol{e}_ r \right)^2\right] \nonumber \\
&=\frac{1}{6 \pi}\frac{\tilde{G}^2}{r_p^4}(q^1_1 m_2-q^1_2 m_1)^2\,,
\end{align}
where we used~$\ddot{\boldsymbol{r}}_p=-\tilde{G} M \boldsymbol{e}_ r/r^2$ and integrated over the sphere.
The radiated angular momentum flux
\begin{align}
\dot{L}^{\textrm{rad}}&=\lim_{r \to \infty} r^2 \int d \Omega\, T^{V}_ {r\varphi}\nonumber \\
&=2 \left(\frac{q^1_1 m_2-q^1_2m_1}{4 \pi M}\right)^2 \nonumber\\ 
&\times\int d\Omega \, \left[\left(\boldsymbol{e}_ r\cdot \ddot{\boldsymbol{r}}_p \right)\partial_\varphi \left(\boldsymbol{e}_ r\cdot \dot{\boldsymbol{r}}_p \right)-\left(\boldsymbol{e}_ r\cdot \ddot{\boldsymbol{r}}_p \right)\left(\boldsymbol{e}_ \varphi\cdot \ddot{\boldsymbol{r}}_p \right)\right]\nonumber \\
&=\frac{1}{6 \pi} \tilde{G}^\frac{3}{2}\frac{\sqrt{a (1-e^2)}}{\sqrt{M}r_p^3}(q^1_1 m_2-q^1_2 m_1)^2\,.
\end{align}
Thus, averaging over one orbital period, we conclude that electric charges radiate twice the energy and twice the angular momentum per unit of time in comparison with scalar charges (compare with Eqs.~\eqref{EnergyS} and~\eqref{AngMomS}). So, the ratio between the angular momentum and energy carried by the radiated electromagnetic field~$\langle\dot{L}^{\textrm{rad}} \rangle/\langle\dot{E}^{\textrm{rad}} \rangle$ is the same as for the scalar field and is given by~\eqref{LEratioS}. So, the emission of electromagnetic waves by a binary causes both the major semi-axis and eccentricity to decrease in time: the orbit shrinks and circularizes (see~\eqref{shrink} and~\eqref{circular}). 
Our results for the electromagnetic radiation emitted by a binary are in agreement with the ones of Refs.~\cite{Christiansen:2020pnv,Liu:2020cds}.

%*****************************************
\chapter{Generalized Lorentz law}\label{app:Lorentz}
%*****************************************

To obtain equation~\eqref{motion}, we observe that the charged test fields (with charge current density~$J^\alpha$) generate an extra electromagnetic field with Faraday tensor $f$ satisfying the Maxwell equations~$df=0$ and
\begin{equation} \label{Maxwell}
\nabla_\mu f^{\mu\nu} = - J^\nu\,. 
\end{equation}
The total electromagnetic energy-momentum tensor is then
\begin{equation}
T^{EM}_{\mu\nu} = (F_{\mu\alpha}+f_{\mu\alpha})(F_{\nu}^{\,\,\,\,\alpha}+f_{\nu}^{\,\,\,\,\alpha}) - \frac14 g_{\mu\nu} (F_{\alpha\beta}+f_{\alpha\beta}) (F^{\alpha\beta}+f^{\alpha\beta})\,.
\end{equation}
Besides the stationary part, due solely to $F$, one has to consider, in the test field approximation, the cross terms
\begin{equation}\label{cross}
t_{\mu\nu} = f_{\mu\alpha} F_{\nu}^{\,\,\,\,\alpha} + F_{\mu\alpha} f_{\nu}^{\,\,\,\,\alpha} - \frac12 g_{\mu\nu} F_{\alpha\beta} f^{\alpha\beta}\,.
\end{equation}
We have
\begin{align}
\hspace{-1 cm}\nabla^\mu t_{\mu\nu} & = - J_\alpha F_{\nu}^{\,\,\,\,\alpha} + f_{\mu\alpha} \nabla^\mu F_{\nu}^{\,\,\,\,\alpha} + F_{\mu\alpha} \nabla^\mu f_{\nu}^{\,\,\,\,\alpha} - \frac12 (\nabla_\nu F_{\alpha\beta}) f^{\alpha\beta} - \frac12 F_{\alpha\beta} \nabla_\nu  f^{\alpha\beta} \nonumber \\
& = - F_{\nu\alpha} J^{\alpha}, \label{tmunu}
\end{align}
where we used \eqref{Maxwell}, the Maxwell equation $\nabla^\mu F_{\mu\alpha} = 0$, the fact that
\begin{equation}
f_{\mu\alpha} \nabla^\mu F^{\nu\alpha} - \frac12 (\nabla^\nu F_{\alpha\beta}) f^{\alpha\beta} = \frac12 f_{\alpha\beta} \left( \nabla^\alpha F^{\nu\beta} + \nabla^\beta F^{\alpha\nu} - \nabla^\nu F^{\alpha\beta}\right) = 0
\end{equation}
(because of the Maxwell equation $dF=0$), and the fact that
\begin{equation}
F_{\mu\alpha} \nabla^\mu f^{\nu\alpha} - \frac12 F_{\alpha\beta} \nabla^\nu f^{\alpha\beta} = \frac12 F_{\alpha\beta} \left( \nabla^\alpha f^{\nu\beta} + \nabla^\beta f^{\alpha\nu} - \nabla^\nu f^{\alpha\beta}\right) = 0
\end{equation}
(because of the Maxwell equation $df=0$). Therefore, in the test field approximation, we have
\begin{equation}
\nabla^\mu \left(T_{\mu\nu} + T^{EM}_{\mu\nu} \right) = 0 \Leftrightarrow \nabla^\mu \left(T_{\mu\nu} + t_{\mu\nu} \right) = 0  \Leftrightarrow \nabla^\mu T_{\mu\nu} = F_{\nu\alpha} J^{\alpha},
\end{equation}
which is equation~\eqref{motion}.

One may wonder why not use the conserved current
\begin{equation}
\nabla_\mu (T^{\mu\nu} K_\nu + t^{\mu\nu} K_\nu) = 0
\end{equation}
to define the energy of the test field as
\begin{equation} \label{energy1}
E'' = \int_\mathcal{S} dV_{3}(T^{\mu\nu} + t^{\mu\nu}) K_\nu N_\mu \,.
\end{equation}
The reason is that this expression accounts for the energy of the interaction between the charged field and the background electromagnetic field through the electromagnetic cross terms \eqref{cross}, whereas \eqref{energy} localizes it on the charges. As is well known, the physical mass of a charged \acs{BH} includes the energy of its background electromagnetic field; when charge enters the \acs{BH}, the interaction energy should be transferred from the energy of the electromagnetic field to the \acs{BH}'s mass. This accounting is accomplished by \eqref{energy}, but not by \eqref{energy1}.~\footnote{As a toy model of this situation, consider a distribution of test charges with density~$\rho$ on a background electrostatic field $\boldsymbol{E}=-\boldsymbol{\nabla} \phi$ generated by a closed surface kept at a constant potential $\Phi$. Using Gauss's law, it is easily seen that the total electrostatic energy outside the surface is $\int_{\text{out}} \rho \phi = \int_{\text{out}} \boldsymbol{E} \cdot \boldsymbol{e} - q_{\text{in}} \Phi$, where $\boldsymbol{e}$ is the electric field generated by the test charges and $q_{\text{in}}$ is the total test charge inside the surface.}

One might also worry that the presence of the extra energy-momentum tensor $t^{\mu \nu}$ with nonzero divergence \eqref{tmunu} could invalidate our previous conclusions. That is not the case, however, because $t^{\mu \nu}$ does not contribute to the flux across the horizon. In fact, using \eqref{Lie} and the Killing equation \eqref{Killing}, we have
\begin{align}
 \int_H t_{\mu\nu} Z^\mu Z^\nu &= \int_H  2 f_{\mu}^{\,\,\,\,\alpha} F_{\nu\alpha} Z^\mu Z^\nu = \int_H  2 f_{\mu}^{\,\,\,\,\alpha} (\nabla_\nu A_\alpha - \nabla_\alpha A_\nu) Z^\mu Z^\nu \nonumber \\
& = \int_H  2 f_{\mu}^{\,\,\,\,\alpha} (A^\nu \nabla_\nu Z_\alpha - Z^\nu \nabla_\alpha A_\nu) Z^\mu \nonumber \\
&= \int_H 2 f_{\mu}^{\,\,\,\,\alpha} (- A^\nu \nabla_\alpha Z_\nu - Z^\nu \nabla_\alpha A_\nu) Z^\mu \nonumber \\
& = - \int_H  2 Z^\mu f_{\mu}^{\,\,\,\,\alpha} \nabla_\alpha (A^\nu Z_\nu) = 0\,,
\end{align}
since the vector field $Z^\mu f_{\mu}^{\,\,\,\,\alpha}$ is tangent to the event horizon, \ie,
\begin{equation}
Z^\mu f_{\mu}^{\,\,\,\,\alpha} Z_\alpha = 0,
\end{equation}
and $A^\nu Z_\nu = \Phi$ is constant along the event horizon.

%
%*****************************************
%*****************************************
%*****************************************
%*****************************************
%*****************************************

%********************************************************************
% Other Stuff in the Back
%*******************************************************
\cleardoublepage%********************************************************************
% Bibliography
%*******************************************************
% work-around to have small caps also here in the headline
% https://tex.stackexchange.com/questions/188126/wrong-header-in-bibliography-classicthesis
% Thanks to Enrico Gregorio
\defbibheading{bibintoc}[\bibname]{%
  \phantomsection
  \manualmark
  \markboth{\spacedlowsmallcaps{#1}}{\spacedlowsmallcaps{#1}}%
  \addtocontents{toc}{\protect\vspace{\beforebibskip}}%
  \addcontentsline{toc}{chapter}{\tocEntry{#1}}%
  \chapter*{#1}%
}
\printbibliography[heading=bibintoc]

@article{Natario:2016bay,
    author = "Natario, J. and Queimada, L. and Vicente, R.",
    title = "{Test fields cannot destroy extremal black holes}",
    eprint = "1601.06809",
    archivePrefix = "arXiv",
    primaryClass = "gr-qc",
    doi = "10.1088/0264-9381/33/17/175002",
    journal = "Class. Quant. Grav.",
    volume = "33",
    number = "17",
    pages = "175002",
    year = "2016"
}

@article{Vicente:2019ilr,
    author = "Vicente, R. and Cardoso, V. and Zilh\~ao, M.",
    title = "{Dynamical friction in slab geometries and accretion disks}",
    eprint = "1905.06353",
    archivePrefix = "arXiv",
    primaryClass = "astro-ph.GA",
    doi = "10.1093/mnras/stz2526",
    journal = "Mon. Not. R. Astron. Soc.",
    volume = "489",
    number = "4",
    pages = "5424--5435",
    year = "2019"
}

@article{Cardoso:2019dte,
    author = "Cardoso, V. and Vicente, R.",
    title = "{Moving black holes: energy extraction, absorption cross-section and the ring of fire}",
    eprint = "1906.10140",
    archivePrefix = "arXiv",
    primaryClass = "gr-qc",
    doi = "10.1103/PhysRevD.100.084001",
    journal = "Phys. Rev.",
    volume = "D100",
    number = "8",
    pages = "084001",
    year = "2019"
}

@article{Natario:2019iex,
    author = "Natario, J. and Vicente, R.",
    title = "{Test fields cannot destroy extremal de Sitter black holes}",
    eprint = "1908.09854",
    archivePrefix = "arXiv",
    primaryClass = "gr-qc",
    doi = "10.1007/s10714-020-2658-3",
    journal = "Gen. Relativ. Gravit.",
    volume = "52",
    number = "1",
    pages = "5",
    year = "2020"
}

@article{Annulli:2020ilw,
    author = "Annulli, L. and Cardoso, V. and Vicente, R.",
    title = "{Stirred and shaken: Dynamical behavior of boson stars and dark matter cores}",
    eprint = "2007.03700",
    archivePrefix = "arXiv",
    primaryClass = "astro-ph.HE",
    doi = "10.1016/j.physletb.2020.135944",
    journal = "Phys. Lett.",
    volume = "B811",
    pages = "135944",
    year = "2020"
}

@article{Annulli:2020lyc,
    author = "Annulli, L. and Cardoso, V. and Vicente, R.",
    title = "{Response of ultralight dark matter to supermassive black holes and binaries}",
    eprint = "2009.00012",
    archivePrefix = "arXiv",
    primaryClass = "gr-qc",
    doi = "10.1103/PhysRevD.102.063022",
    journal = "Phys. Rev.",
    volume = "D102",
    number = "6",
    pages = "063022",
    year = "2020"
}

@article{Cardoso:2020iji,
    author = "Cardoso, V. and Macedo, C. F. B. and Vicente, R.",
    title = "{Eccentricity evolution of compact binaries and applications to gravitational-wave physics}",
    eprint = "2010.15151",
    archivePrefix = "arXiv",
    primaryClass = "gr-qc",
    doi = "10.1103/PhysRevD.103.023015",
    journal = "Phys. Rev.",
    volume = "D103",
    number = "2",
    pages = "023015",
    year = "2021"
}

@article{Chandrasekhar:1943v1,
       author = {{Chandrasekhar}, S.},
        title = "{Dynamical Friction. I. General Considerations: the Coefficient of Dynamical Friction.}",
      journal = {Astrophys. J.},
         year = 1943,
        month = mar,
       volume = {97},
        pages = {255},
          doi = {10.1086/144517},
       adsurl = {https://ui.adsabs.harvard.edu/abs/1943ApJ....97..255C}
}

@article{Chandrasekhar:1943v2,
   author = {{Chandrasekhar}, S.},
    title = "{Dynamical Friction. II. The Rate of Escape of Stars from Clusters and the Evidence for the Operation of Dynamical Friction.}",
  journal = {Astrophys. J.},
     year = 1943,
    month = mar,
   volume = 97,
    pages = {263},
      doi = {10.1086/144518},
   adsurl = {http://adsabs.harvard.edu/abs/1943ApJ....97..263C}
}

@article{Chandrasekhar:1943v3,
   author = {{Chandrasekhar}, S.},
    title = "{Dynamical Friction. III. a More Exact Theory of the Rate of Escape of Stars from Clusters.}",
  journal = {Astrophys. J.},
     year = 1943,
    month = jul,
   volume = 98,
    pages = {54},
      doi = {10.1086/144544},
   adsurl = {http://adsabs.harvard.edu/abs/1943ApJ....98...54C}
}

@article{Ostriker:1998fa,
      author         = "Ostriker, E. C.",
      title          = "{Dynamical friction in a gaseous medium}",
      eprint         = "astro-ph/9810324",
      archivePrefix  = "arXiv",
      primaryClass   = "astro-ph",
             journal = {Astrophys. J.},
                year = 1999,
               month = mar,
              volume = 513,
               pages = {252-258},
                 doi = {10.1086/306858},
              adsurl = {http://adsabs.harvard.edu/abs/1999ApJ...513..252O}
}

@book{binney2011galactic,
  title={Galactic Dynamics: (Second Edition)},
  author={Binney, J. and Tremaine, S.},
  isbn={9781400828722},
  series={Princeton Series in Astrophysics},
  url={https://books.google.ch/books?id=6mF4CKxlbLsC},
  year={2011},
  publisher={Princeton University Press}
}

@ARTICLE{Namouni2011,
       author = {{Namouni}, F.},
        title = "{On dynamical friction in a gaseous medium with a boundary}",
      journal = {Astrophy. Space Sci.},
         year = "2011",
        month = "Feb",
       volume = {331},
        pages = {575-595},
          doi = {10.1007/s10509-010-0482-z},
archivePrefix = {arXiv},
       eprint = {0911.4891},
 primaryClass = {astro-ph.GA},
       adsurl = {https://ui.adsabs.harvard.edu/abs/2011Ap&SS.331..575N}
}

@article{Muto:2011qv,
      author         = "Muto, T. and Takeuchi, T. and Ida, S.",
      title          = "{On the Interaction between a Protoplanetary Disk and a
                        Planet in an Eccentric Orbit: Application of Dynamical
                        Friction}",
      journal        = {Astrophys. J.},
      volume         = "737",
      year           = "2011",
      pages          = "37",
      doi            = "10.1088/0004-637X/737/1/37",
      eprint         = "1106.0417",
      archivePrefix  = "arXiv",
      primaryClass   = "astro-ph.EP",
      SLACcitation   = "%%CITATION = ARXIV:1106.0417;%%"
}

@ARTICLE{Canto2013,
       author = {{Cant{\'o}}, J. and {Esquivel}, A. and {S{\'a}nchez-Salcedo}, F.~J. and
         {Raga}, A.~C.},
        title = "{Gravitational Drag on a Point Mass in Hypersonic Motion within a Gaussian Disk}",
      journal = {Astrophys. J.},
         year = "2013",
        month = "Jan",
       volume = {762},
          eid = {21},
        pages = {21},
          doi = {10.1088/0004-637X/762/1/21},
archivePrefix = {arXiv},
       eprint = {1211.3988},
 primaryClass = {astro-ph.GA},
       adsurl = {https://ui.adsabs.harvard.edu/abs/2013ApJ...762...21C}
}

@inproceedings{Novikov:1973kta,
      author         = "Novikov, I. D. and Thorne, K. S.",
      title          = "{Astrophysics and black holes}",
      booktitle      = "{Proceedings, Ecole d'Ete de Physique Theorique: Les
                        Astres Occlus: Les Houches, France, August, 1972}",
      year           = "1973",
      pages          = "343-550",
      SLACcitation   = "%%CITATION = INSPIRE-1361968;%%"
}

@ARTICLE{Armitage2011,
   author = {{Armitage}, P.~J.},
    title = "{Dynamics of Protoplanetary Disks}",
  journal = {Annu. Rev. Astron. Astrophys.},
archivePrefix = "arXiv",
   eprint = {1011.1496},
 primaryClass = "astro-ph.SR",
     year = 2011,
    month = sep,
   volume = 49,
    pages = {195-236},
      doi = {10.1146/annurev-astro-081710-102521},
   adsurl = {http://adsabs.harvard.edu/abs/2011ARA%26A..49..195A}
}

@book{landau1987fluid,
  title={Fluid Mechanics},
  author={Landau, L. D. and Lifshitz, E. M.},
  isbn={9780080570730},
  series={Course of Theoretical Physics},
  url={https://books.google.pt/books?id=eVKbCgAAQBAJ},
  year={1987},
  publisher={Elsevier Science}
}

@article{Barvinsky:2003jf,
      author         = "Barvinsky, A. O. and Solodukhin, S. N.",
      title          = "{Echoing the extra dimension}",
      journal        = "Nucl. Phys.",
      volume         = "B675",
      year           = "2003",
      pages          = "159-178",
      doi            = "10.1016/j.nuclphysb.2003.10.011",
      eprint         = "hep-th/0307011",
      archivePrefix  = "arXiv",
      primaryClass   = "hep-th",
      SLACcitation   = "%%CITATION = HEP-TH/0307011;%%"
}

@ARTICLE{Rephaeli1980ApJ,
       author = {{Rephaeli}, Y. and {Salpeter}, E.~E.},
        title = "{Flow past a massive object and the gravitational drag}",
      journal = {Astrophys. J.},
         year = 1980,
        month = Aug,
       volume = {240},
        pages = {20-24},
          doi = {10.1086/158202},
       adsurl = {https://ui.adsabs.harvard.edu/\#abs/1980ApJ...240...20R}
}

@INPROCEEDINGS{Kumar,
   author = {{Kumar}, P.},
    title = "{Solar Oscillations with Frequencies above the Acoustic Cutoff Frequency}",
booktitle = {GONG 1992. Seismic Investigation of the Sun and Stars},
     year = 1993,
   series = {Astronomical Society of the Pacific Conference Series},
   volume = 42,
   editor = {{Brown}, T.~M.},
    month = jan,
    pages = {15},
   adsurl = {http://adsabs.harvard.edu/abs/1993ASPC...42...15K}
}

@book{lamb1945hydrodynamics,
  title={Hydrodynamics: 6th Ed},
  author={Lamb, H.},
  lccn={46001891},
  url={https://books.google.pt/books?id=WsE6xwEACAAJ},
  year={1945},
  publisher={Dover}
}

@ARTICLE{shakura,
   author = {{Shakura}, N.~I. and {Sunyaev}, R.~A.},
    title = "{Black holes in binary systems. Observational appearance.}",
  journal = {Astron. Astrophys.},
     year = 1973,
   volume = 24,
    pages = {337-355},
   adsurl = {http://adsabs.harvard.edu/abs/1973A%26A....24..337S}
}

@ARTICLE{Namouni,
       author = {{Namouni}, F.},
        title = "{On dynamical friction in a gaseous medium with a boundary}",
      journal = {Astrophy. Space Sci.},
         year = "2011",
        month = "Feb",
       volume = {331},
        pages = {575-595},
          doi = {10.1007/s10509-010-0482-z},
archivePrefix = {arXiv},
       eprint = {0911.4891},
 primaryClass = {astro-ph.GA},
       adsurl = {https://ui.adsabs.harvard.edu/abs/2011Ap&SS.331..575N}
}

@ARTICLE{Dokuchaev1964,
   author = {{Dokuchaev}, V.~P.},
    title = "{Emission of Magnetoacoustic Waves in the Motion of Stars in Cosmic Space.}",
  journal = {Sov. Astron.},
     year = 1964,
    month = aug,
   volume = 8,
    pages = {23},
   adsurl = {http://adsabs.harvard.edu/abs/1964SvA.....8...23D}
}

@ARTICLE{Ruderman1971,
   author = {{Ruderman}, M.~A. and {Spiegel}, E.~A.},
    title = "{Galactic Wakes}",
  journal = {Astrophys. J.},
     year = 1971,
    month = apr,
   volume = 165,
    pages = {1},
      doi = {10.1086/150870},
   adsurl = {http://adsabs.harvard.edu/abs/1971ApJ...165....1R}
}

@ARTICLE{canto,
       author = {{Cant{\'o}}, J. and {Esquivel}, A. and {S{\'a}nchez-Salcedo}, F.~J. and
         {Raga}, A.~C.},
        title = "{Gravitational Drag on a Point Mass in Hypersonic Motion within a Gaussian Disk}",
      journal = {Astrophys. J.},
         year = "2013",
        month = "Jan",
       volume = {762},
          eid = {21},
        pages = {21},
          doi = {10.1088/0004-637X/762/1/21},
archivePrefix = {arXiv},
       eprint = {1211.3988},
 primaryClass = {astro-ph.GA},
       adsurl = {https://ui.adsabs.harvard.edu/abs/2013ApJ...762...21C}
}

@BOOK{Tremaine1987,
   author = {{Binney}, J. and {Tremaine}, S.},
    title = "{Galactic Dynamics}",
booktitle = {Princeton, NJ, Princeton University Press, 1987, 747 p.},
publisher = {Princeton University Press},
     year = 1987,
   adsurl = {http://adsabs.harvard.edu/abs/1987gady.book.....B}
}

@ARTICLE{Tremaine79,
   author = {{Goldreich}, P. and {Tremaine}, S.},
    title = "{The excitation of density waves at the Lindblad and corotation resonances by an external potential}",
  journal = {Astrophys. J.},
     year = 1979,
    month = nov,
   volume = 233,
    pages = {857-871},
      doi = {10.1086/157448},
   adsurl = {http://adsabs.harvard.edu/abs/1979ApJ...233..857G}
}

@ARTICLE{Tremaine80,
   author = {{Goldreich}, P. and {Tremaine}, S.},
    title = "{Disk-satellite interactions}",
  journal = {Astrophys. J.},
     year = 1980,
    month = oct,
   volume = 241,
    pages = {425-441},
      doi = {10.1086/158356},
   adsurl = {http://adsabs.harvard.edu/abs/1980ApJ...241..425G}
}

@ARTICLE{Ward86,
   author = {{Ward}, W.~R.},
    title = "{Density waves in the solar nebula - Differential Lindblad torque}",
  journal = {Icarus},
     year = 1986,
    month = jul,
   volume = 67,
    pages = {164-180},
      doi = {10.1016/0019-1035(86)90182-X},
   adsurl = {http://adsabs.harvard.edu/abs/1986Icar...67..164W}
}

@article{Tanaka_2002,
	doi = {10.1086/324713},
	url = {https://doi.org/10.1086%2F324713},
	year = 2002,
	month = {feb},
	publisher = {{IOP} Publishing},
	volume = {565},
	number = {2},
	pages = {1257--1274},
	author = {H. Tanaka and T. Takeuchi and W. R. Ward},
	title = {Three-dimensional Interaction between a Planet and an Isothermal Gaseous Disk. I. Corotation and Lindblad Torques and Planet Migration},
	journal = {Astrophys. J.}
}

@article{Stone2018,
    author = {Stone, J. M. and Arzamasskiy, L. and Zhu, Z.},
    title = "{Three-dimensional disc–satellite interaction: torques, migration, and observational signatures}",
    journal = {Mon. Not. R. Astron. Soc.},
    volume = {475},
    number = {3},
    pages = {3201-3212},
    year = {2018},
    month = {01},
    issn = {0035-8711},
    doi = {10.1093/mnras/sty001},
    url = {https://dx.doi.org/10.1093/mnras/sty001}
}

@INPROCEEDINGS{Hayashi,
   author = {{Hayashi}, C. and {Nakazawa}, K. and {Nakagawa}, Y.},
    title = "{Formation of the solar system}",
booktitle = {Protostars and Planets II},
     year = 1985,
   editor = {{Black}, D.~C. and {Matthews}, M.~S.},
    pages = {1100-1153},
   adsurl = {http://adsabs.harvard.edu/abs/1985prpl.conf.1100H}
}

@ARTICLE{Tanaka,
   author = {{Tanaka}, H. and {Ida}, S.},
    title = "{Growth of a Migrating Protoplanet}",
  journal = {Icarus},
     year = 1999,
    month = jun,
   volume = 139,
    pages = {350-366},
      doi = {10.1006/icar.1999.6107},
   adsurl = {http://adsabs.harvard.edu/abs/1999Icar..139..350T}
}

@article{Kim:2007zb,
      author         = "Kim, H. and Kim, W.",
      title          = "{Dynamical Friction of a Circular-Orbit Perturber in a
                        Gaseous Medium}",
      journal        = {Astrophys. J.},
      volume         = "665",
      pages          = "432-444",
      doi            = "10.1086/519302",
      year           = "2007",
      eprint         = "0705.0084",
      archivePrefix  = "arXiv",
      primaryClass   = "astro-ph",
      SLACcitation   = "%%CITATION = ARXIV:0705.0084;%%",
}

@article{Kim:2008ab,
      author         = "Kim, H. and Kim, W. and Sanchez-Salcedo, F.
                        J.",
      title          = "{Dynamical Friction of Double Perturbers in a Gaseous
                        Medium}",
      journal        = {Astrophys. J. Lett.},
      volume         = "679",
      year           = "2008",
      pages          = "L33",
      doi            = "10.1086/589149",
      eprint         = "0804.2010",
      archivePrefix  = "arXiv",
      primaryClass   = "astro-ph",
      SLACcitation   = "%%CITATION = ARXIV:0804.2010;%%"
}

@article{Barausse:2014tra,
      author         = "Barausse, E. and Cardoso, V. and Pani, P.",
      title          = "{Can environmental effects spoil precision
                        gravitational-wave astrophysics?}",
      journal        = "Phys. Rev.",
      volume         = "D89",
      number         = "10"
      year           = "2014",
      pages          = "104059",
      doi            = "10.1103/PhysRevD.89.104059",
      eprint         = "1404.7149",
      archivePrefix  = "arXiv",
      primaryClass   = "gr-qc",
      SLACcitation   = "%%CITATION = ARXIV:1404.7149;%%"
}

@incollection{gradshteyn2007,
title = {Definite Integrals of Special Functions},
editor = {A. Jeffrey and D. Zwillinger and I.S. Gradshteyn and I.M. Ryzhik},
booktitle = {Table of Integrals, Series, and Products},
publisher = {Academic Press},
edition = {Seventh Edition},
address = {Boston},
pages = {631-857},
year = {2007},
isbn = {978-0-12-373637-6},
doi = {https://doi.org/10.1016/B978-0-08-047111-2.50015-7},
url = {https://www.sciencedirect.com/science/article/pii/B9780080471112500157}
}

@book{Misner:1974qy,
    author = "Misner, C. W. and Thorne, K. S. and Wheeler, J. A.",
    title = "{Gravitation}",
    isbn = "978-0-7167-0344-0, 978-0-691-17779-3",
    publisher = "W. H. Freeman",
    address = "San Francisco",
    year = "1973"
}

@book{Wald:1984rg,
    author = "Wald, R. M.",
    title = "{General Relativity}",
    doi = "10.7208/chicago/9780226870373.001.0001",
    publisher = "Chicago Univ. Pr.",
    address = "Chicago, USA",
    year = "1984"
}

@book{poisson_will_2014, 
	place={Cambridge}, 
	title={Gravity: Newtonian, Post-Newtonian, Relativistic}, 
	DOI={10.1017/CBO9781139507486}, 
	publisher={Cambridge University Press}, 
	author={Poisson, E. and Will, C. M.}, 
	year={2014}
}

@article{Matzner:1968,
author = {Matzner,R. A. },
title = {Scattering of Massless Scalar Waves by a Schwarzschild ``Singularity''},
journal = {J. Math. Phys.},
volume = {9},
number = {1},
pages = {163-170},
year = {1968},
doi = {10.1063/1.1664470},

URL = { 
        https://doi.org/10.1063/1.1664470   
}
}

@article{Starobinski:1973,
      author         = "Starobinski, A. A.",
      title          = "{Amplification of waves during reflection from a rotating black hole}",
      journal        = "Zh. Eksp. Teor. Fiz.",
      volume         = "64",
      pages          = "48",
      year           = "1973"
}

@article{Starobinski2:1973,
      author         = "Starobinski, A. A. and Churilov, S. M.",
      title          = "{Amplification of electromagnetic and gravitational waves scattered by a rotating black hole}",
      journal        = "Zh. Eksp. Teor. Fiz.",
      volume         = "65",
      pages          = "3",
      year           = "1973"
}

@article{Teukolsky:1974yv,
      author         = "Teukolsky, S. A. and Press, W. H.",
      title          = "{Perturbations of a rotating black hole. III -
                        Interaction of the hole with gravitational and
                        electromagnetic radiation}",
      journal        = "Astrophys. J.",
      volume         = "193",
      year           = "1974",
      pages          = "443-461",
      doi            = "10.1086/153180",
      SLACcitation   = "%%CITATION = ASJOA,193,443;%%"
}

@article{Unruh:1976fm,
      author         = "Unruh, W. G.",
      title          = "{Absorption Cross-Section of Small Black Holes}",
      journal        = "Phys. Rev.",
      volume         = "D14",
      year           = "1976",
      pages          = "3251-3259",
      doi            = "10.1103/PhysRevD.14.3251",
      SLACcitation   = "%%CITATION = PHRVA,D14,3251;%%"
}

@article{Sanchez:1977si,
      author         = "Sanchez, N. G.",
      title          = "{Absorption and Emission Spectra of a Schwarzschild Black
                        Hole}",
      journal        = "Phys. Rev.",
      volume         = "D18",
      year           = "1978",
      pages          = "1030",
      doi            = "10.1103/PhysRevD.18.1030",
      reportNumber   = "Print-77-0242 (MEUDON)",
      SLACcitation   = "%%CITATION = PHRVA,D18,1030;%%"
}

@article{Sanchez:1976,
author = {Sanchez,N. G. },
title = {Scattering of scalar waves from a Schwarzschild black hole},
journal = {J. Math. Phys.},
volume = {17},
number = {5},
pages = {688-692},
year = {1976},
doi = {10.1063/1.522949}
}

@Book{MTB,
     author    =  "Chandrasekhar, S.",
     title     =  "The Mathematical Theory of Black Holes",
     publisher =  "Oxford University Press",
     address   =  "New York",
     year      =  "1983"
}

@article{Glampedakis:2001cx,
      author         = "Glampedakis, K. and Andersson, N.",
      title          = "{Scattering of scalar waves by rotating black holes}",
      journal        = "Class. Quantum Gravity",
      volume         = "18",
      year           = "2001",
      pages          = "1939-1966",
      doi            = "10.1088/0264-9381/18/10/309",
      eprint         = "gr-qc/0102100",
      archivePrefix  = "arXiv",
      primaryClass   = "gr-qc",
      SLACcitation   = "%%CITATION = GR-QC/0102100;%%"
}

@article{Macedo:2013afa,
      author         = "Macedo, C. F. B. and Leite, L. C. S. and Oliveira,
                        E. S. and Dolan, S. R. and Crispino, L. C. B.",
      title          = "{Absorption of planar massless scalar waves by Kerr black
                        holes}",
      journal        = "Phys. Rev.",
      volume         = "D88",
      year           = "2013",
      number         = "6",
      pages          = "064033",
      doi            = "10.1103/PhysRevD.88.064033",
      eprint         = "1308.0018",
      archivePrefix  = "arXiv",
      primaryClass   = "gr-qc",
      SLACcitation   = "%%CITATION = ARXIV:1308.0018;%%"
}

@article{Crispino:2009xt,
      author         = "Crispino, L. C. B. and Dolan, S. R. and Oliveira,
                        E. S.",
      title          = "{Electromagnetic wave scattering by Schwarzschild black
                        holes}",
      journal        = "Phys. Rev. Lett.",
      volume         = "102",
      year           = "2009",
      pages          = "231103",
      doi            = "10.1103/PhysRevLett.102.231103",
      eprint         = "0905.3339",
      archivePrefix  = "arXiv",
      primaryClass   = "gr-qc",
      SLACcitation   = "%%CITATION = ARXIV:0905.3339;%%"
}

@article{Leite:2016hws,
      author         = "Leite, L. C. S. and Crispino, L. C. B. and
                        Oliveira, E. S. and Macedo, C. F. B. and Dolan,
                        S. R.",
      title          = "{Absorption of massless scalar field by rotating black
                        holes}",
      booktitle      = "{Proceedings, 3rd Amazonian Symposium on Physics: Belem,
                        Brazil, September 28-October 2, 2015}",
      journal        = "Int. J. Mod. Phys.",
      volume         = "D25",
      year           = "2016",
      number         = "09",
      pages          = "1641024",
      doi            = "10.1142/S0218271816410248",
      SLACcitation   = "%%CITATION = IMPAE,D25,1641024;%%"
}

@article{Leite:2018mon,
      author         = "Leite, L. C. S. and Dolan, S. and Crispino, L., C.
                        B.",
      title          = "{Absorption of electromagnetic plane waves by rotating
                        black holes}",
      journal        = "Phys. Rev.",
      volume         = "D98",
      year           = "2018",
      number         = "2",
      pages          = "024046",
      doi            = "10.1103/PhysRevD.98.024046",
      eprint         = "1805.07840",
      archivePrefix  = "arXiv",
      primaryClass   = "gr-qc",
      SLACcitation   = "%%CITATION = ARXIV:1805.07840;%%"
}

@article{Leite:2017zyb,
      author         = "Leite, L. C. S. and Dolan, S. R. and Crispino, L.
                        C. B.",
      title          = "{Absorption of electromagnetic and gravitational waves by
                        Kerr black holes}",
      journal        = "Phys. Lett.",
      volume         = "B774",
      year           = "2017",
      pages          = "130-134",
      doi            = "10.1016/j.physletb.2017.09.048",
      eprint         = "1707.01144",
      archivePrefix  = "arXiv",
      primaryClass   = "gr-qc",
      SLACcitation   = "%%CITATION = ARXIV:1707.01144;%%"
}

@article{Benone:2018rtj,
      author         = "Benone, C. L. and Leite, L. C. S. and Crispino,
                        L., C. B. and Dolan, S. R.",
      title          = "{On-axis scalar absorption cross section of Kerr–Newman
                        black holes: Geodesic analysis, sinc and low-frequency
                        approximations}",
      booktitle      = "{Proceedings, 4th Amazonian Symposium on Physics:
                        Celebrating 100 years of the de Sitter solution and 60
                        years of Atsushi Higuchi: Belem, Brazil, September 18-22,
                        2017}",
      journal        = "Int. J. Mod. Phys.",
      volume         = "D27",
      year           = "2018",
      number         = "11",
      pages          = "1843012",
      doi            = "10.1142/S0218271818430125",
      eprint         = "1809.08275",
      archivePrefix  = "arXiv",
      primaryClass   = "gr-qc",
      SLACcitation   = "%%CITATION = ARXIV:1809.08275;%%"
}

@article{Das:1996we,
      author         = "Das, S. R. and Gibbons, G. W. and Mathur, S. D.",
      title          = "{Universality of low-energy absorption cross-sections for
                        black holes}",
      journal        = "Phys. Rev. Lett.",
      volume         = "78",
      year           = "1997",
      pages          = "417-419",
      doi            = "10.1103/PhysRevLett.78.417",
      eprint         = "hep-th/9609052",
      archivePrefix  = "arXiv",
      primaryClass   = "hep-th",
      reportNumber   = "TIFR-TH-96-49, MIT-CTP-2565",
      SLACcitation   = "%%CITATION = HEP-TH/9609052;%%"
}

@article{zeldovich1,
      author    = "Zel'dovich,Ya. B.",
      journal   = "Pis'ma Zh. Eksp. Teor. Fiz.",
      volume    = "14", 
      pages     = "270",
      year      = "1971"
}

@article{zeldovich2,
      author    = "Zel'dovich,Ya. B.",
      journal   = "Zh. Eksp. Teor. Fiz",
      volume    = "62", 
      pages     = "2076",
      year      = "1972"
}

@article{Brito:2015oca,
      author         = "Brito, R. and Cardoso, V. and Pani, P.",
      title          = "{Superradiance}",
      journal        = "Lect. Notes Phys.",
      volume         = "906",
      year           = "2015",
      pages          = "pp.1-237",
      doi            = "10.1007/978-3-319-19000-6",
      eprint         = "1501.06570",
      archivePrefix  = "arXiv",
      primaryClass   = "gr-qc",
      SLACcitation   = "%%CITATION = ARXIV:1501.06570;%%"
}

@article{Benone:2019all,
      author         = "Benone, C. L. and Crispino, L. C. B.",
      title          = "{Massive and charged scalar field in Kerr-Newman
                        spacetime: Absorption and superradiance}",
      journal        = "Phys. Rev.",
      volume         = "D99",
      year           = "2019",
      number         = "4",
      pages          = "044009",
      doi            = "10.1103/PhysRevD.99.044009",
      eprint         = "1901.05592",
      archivePrefix  = "arXiv",
      primaryClass   = "gr-qc",
      SLACcitation   = "%%CITATION = ARXIV:1901.05592;%%"
}

@article{Arvanitaki:2016qwi,
      author         = "Arvanitaki, A. and Baryakhtar, M. and Dimopoulos,
                        S. and Dubovsky, S. and Lasenby, R.",
      title          = "{Black Hole Mergers and the QCD Axion at Advanced LIGO}",
      journal        = "Phys. Rev.",
      volume         = "D95",
      year           = "2017",
      number         = "4",
      pages          = "043001",
      doi            = "10.1103/PhysRevD.95.043001",
      eprint         = "1604.03958",
      archivePrefix  = "arXiv",
      primaryClass   = "hep-ph",
      SLACcitation   = "%%CITATION = ARXIV:1604.03958;%%"
}

@article{Brito:2017wnc,
      author         = "Brito, R. and Ghosh, S. and Barausse, E.
                        and Berti, E. and Cardoso, V. and Dvorkin, I.
                        and Klein, A. and Pani, P.",
      title          = "{Stochastic and resolvable gravitational waves from
                        ultralight bosons}",
      journal        = "Phys. Rev. Lett.",
      volume         = "119",
      year           = "2017",
      number         = "13",
      pages          = "131101",
      doi            = "10.1103/PhysRevLett.119.131101",
      eprint         = "1706.05097",
      archivePrefix  = "arXiv",
      primaryClass   = "gr-qc",
      SLACcitation   = "%%CITATION = ARXIV:1706.05097;%%"
}

@article{Ikeda:2019fvj,
      author         = "Ikeda, T. and Brito, R. and Cardoso, V.",
      title          = "{Blasts of Light from Axions}",
      journal        = "Phys. Rev. Lett.",
      volume         = "122",
      year           = "2019",
      number         = "8",
      pages          = "081101",
      doi            = "10.1103/PhysRevLett.122.081101",
      eprint         = "1811.04950",
      archivePrefix  = "arXiv",
      primaryClass   = "gr-qc",
      SLACcitation   = "%%CITATION = ARXIV:1811.04950;%%"
}

@article{LIGOScientific:2018mvr,
      author         = "Abbott, B. P. and others",
      title          = "{GWTC-1: A Gravitational-Wave Transient Catalog of
                        Compact Binary Mergers Observed by LIGO and Virgo during
                        the First and Second Observing Runs}",
      collaboration  = "LIGO Scientific, Virgo",
      year           = "2018",
      eprint         = "1811.12907",
      archivePrefix  = "arXiv",
      primaryClass   = "astro-ph.HE",
      reportNumber   = "LIGO-P1800307",
      SLACcitation   = "%%CITATION = ARXIV:1811.12907;%%"
}

@article{Bernard:2019nkv,
    author = "Bernard, L. and Cardoso, V. and Ikeda, T. and Zilh\~ao, M.",
    title = "{Physics of black hole binaries: Geodesics, relaxation modes, and energy extraction}",
    eprint = "1905.05204",
    archivePrefix = "arXiv",
    primaryClass = "gr-qc",
    doi = "10.1103/PhysRevD.100.044002",
    journal = "Phys. Rev.",
    volume = "D100",
    number = "4",
    pages = "044002",
    year = "2019"
}

@article{Ikeda:2020xvt,
    author = "Ikeda, T. and Bernard, L. and Cardoso, V. and Zilh\~ao, M.",
    title = "{Black hole binaries and light fields: Gravitational molecules}",
    eprint = "2010.00008",
    archivePrefix = "arXiv",
    primaryClass = "gr-qc",
    doi = "10.1103/PhysRevD.103.024020",
    journal = "Phys. Rev.",
    volume = "D103",
    number = "2",
    pages = "024020",
    year = "2021"
}

@ARTICLE{Wong:2019kru,
       author = {{Wong}, L. K.},
        title = "{Superradiant scattering by a black hole binary}",
      journal = {Phys. Rev.},
         year = 2019,
        month = aug,
       volume = {D100},
       number = {4},
          eid = {044051},
        pages = {044051},
          doi = {10.1103/PhysRevD.100.044051},
archivePrefix = {arXiv},
       eprint = {1905.08543},
 primaryClass = {hep-th},
       adsurl = {https://ui.adsabs.harvard.edu/abs/2019PhRvD.100d4051W}
}

@article{Wong:2020qom,
    author = "Wong, L. K.",
    title = "{Evolution of diffuse scalar clouds around binary black holes}",
    eprint = "2004.03570",
    archivePrefix = "arXiv",
    primaryClass = "hep-th",
    doi = "10.1103/PhysRevD.101.124049",
    journal = "Phys. Rev.",
    volume = "D101",
    number = "12",
    pages = "124049",
    year = "2020"
}

@Book{Sommerfeld:1964,
     author    =  "Sommerfeld, A.",
     title     =  "Optics; Lectures on Theoretical Physics, IV ",
     publisher =  "Academic Press",
     address   =  "New York",
     year      =  "1964"
}

@article{Restrick:1968,
      author         = "Restrick, R. C.",
      title          = "{Electromagnetic Scattering by a Moving Conducting Sphere}",
      journal        = "Radio Science",
      volume         = "3",
      year           = "1968",
      number         = "12",
      pages          = "1144",
}

@book{landau1981quantum,
  title={Quantum Mechanics: Non-Relativistic Theory},
  author={Landau, L. D. and Lifshitz, E. M.},
  isbn={9780080503486},
  series={Course of Theoretical Physics},
  url={https://books.google.pt/books?id=SvdoN3k8EysC},
  year={1981},
  publisher={Elsevier Science}
}

@article{Clough:2021qlv,
    author = "Clough, K.",
    title = "{Continuity equations for general matter: applications in numerical relativity}",
    eprint = "2104.13420",
    archivePrefix = "arXiv",
    primaryClass = "gr-qc",
    month = "4",
    year = "2021"
}

@article{Einstein1936LENSLIKEAO,
  title={Lens-like action of a star by the deviation of light in the gravitational field},
  author={A. Einstein},
  journal={Science},
  year={1936},
  volume={84 2188},
  pages={506-7
        }
}

@article{Andersson:1995vi,
    author = "Andersson, N.",
    title = "{Scattering of massless scalar waves by a Schwarzschild black hole: A Phase integral study}",
    doi = "10.1103/PhysRevD.52.1808",
    journal = "Phys. Rev.",
    volume = "D52",
    pages = "1808--1820",
    year = "1995"
}

@article{FORD1959259,
title = {Semiclassical description of scattering},
journal = {Ann. Phys.},
volume = {7},
number = {3},
pages = {259-286},
year = {1959},
issn = {0003-4916},
doi = {https://doi.org/10.1016/0003-4916(59)90026-0},
url = {https://www.sciencedirect.com/science/article/pii/0003491659900260},
author = {K. W. Ford and J. A. Wheeler}
}

@article{BisnovatyiKogan:2008ts,
    author = "Bisnovatyi-Kogan, G. S. and Tsupko, O. Y.",
    title = "{Strong Gravitational Lensing by Schwarzschild Black Holes}",
    eprint = "0803.2468",
    archivePrefix = "arXiv",
    primaryClass = "astro-ph",
    doi = "10.1007/s10511-008-0011-8",
    journal = "Astrophysics",
    volume = "51",
    pages = "99--111",
    year = "2008"
}

@article{Darwin,
author = {Darwin, C. G. },
title = {The gravity field of a particle},
journal = {Proc. R. Soc. London, Ser. A},
volume = {249},
number = {1257},
pages = {180-194},
year = {1959},
doi = {10.1098/rspa.1959.0015},

URL = {https://royalsocietypublishing.org/doi/abs/10.1098/rspa.1959.0015},
eprint = {https://royalsocietypublishing.org/doi/pdf/10.1098/rspa.1959.0015}
}

@article{Arnowitt:1962hi,
    author = "Arnowitt, R. L. and Deser, S. and Misner, C. W.",
    title = "{The Dynamics of general relativity}",
    eprint = "gr-qc/0405109",
    archivePrefix = "arXiv",
    doi = "10.1007/s10714-008-0661-1",
    journal = "Gen. Rel. Grav.",
    volume = "40",
    pages = "1997--2027",
    year = "2008"
}

@article{Traykova:2021dua,
    author = "Traykova, D. and Clough, K. and Helfer, T. and Ferreira, P. G. and Berti, E. and Hui, L.",
    title = "{Dynamical friction from scalar dark matter in the relativistic regime}",
    eprint = "2106.08280",
    archivePrefix = "arXiv",
    primaryClass = "gr-qc",
    month = "6",
    year = "2021"
}

@article{Hui:2016ltb,
    author = "Hui, L. and Ostriker, J. P. and Tremaine, S. and Witten, E.",
    title = "{Ultralight scalars as cosmological dark matter}",
    eprint = "1610.08297",
    archivePrefix = "arXiv",
    primaryClass = "astro-ph.CO",
    doi = "10.1103/PhysRevD.95.043541",
    journal = "Phys. Rev.",
    volume = "D95",
    number = "4",
    pages = "043541",
    year = "2017"
}

@article{Abbott:2016blz,
    author = "Abbott, B.P. and others",
    collaboration = "LIGO Scientific, Virgo",
    title = "{Observation of Gravitational Waves from a Binary Black Hole Merger}",
    eprint = "1602.03837",
    archivePrefix = "arXiv",
    primaryClass = "gr-qc",
    reportNumber = "LIGO-P150914",
    doi = "10.1103/PhysRevLett.116.061102",
    journal = "Phys. Rev. Lett.",
    volume = "116",
    number = "6",
    pages = "061102",
    year = "2016"
}

@article{Barack:2018yly,
    author = "Barack, L. and others",
    title = "{Black holes, gravitational waves and fundamental physics: a roadmap}",
    eprint = "1806.05195",
    archivePrefix = "arXiv",
    primaryClass = "gr-qc",
    doi = "10.1088/1361-6382/ab0587",
    journal = "Class. Quant. Grav.",
    volume = "36",
    number = "14",
    pages = "143001",
    year = "2019"
}

@article{Peters:1964zz,
    author = "Peters, P.C.",
    title = "{Gravitational Radiation and the Motion of Two Point Masses}",
    doi = "10.1103/PhysRev.136.B1224",
    journal = "Phys. Rev.",
    volume = "136",
    pages = "B1224--B1232",
    year = "1964"
}

@article{Key:2010tc,
    author = "Shapiro K., Joey and Cornish, N. J.",
    title = "{Characterizing Spinning Black Hole Binaries in Eccentric Orbits with LISA}",
    eprint = "1006.3759",
    archivePrefix = "arXiv",
    primaryClass = "gr-qc",
    doi = "10.1103/PhysRevD.83.083001",
    journal = "Phys. Rev.",
    volume = "D83",
    pages = "083001",
    year = "2011"
}

@article{Krolak:1987ofj,
    author = "Krolak, A. and Schutz, B. F.",
    title = "{Coalescing binaries --- Probe of the universe}",
    doi = "10.1007/BF00759095",
    journal = "Gen. Rel. Grav.",
    volume = "19",
    pages = "1163--1171",
    year = "1987"
}

@article{Laine:2020dnr,
    author = "Laine, S. and others",
    title = "{Spitzer Observations of the Predicted Eddington Flare from Blazar OJ 287}",
    eprint = "2004.13392",
    archivePrefix = "arXiv",
    primaryClass = "astro-ph.HE",
    doi = "10.3847/2041-8213/ab79a4",
    journal = "Astrophys. J. Lett.",
    volume = "894",
    number = "1",
    pages = "L1",
    year = "2020"
}

@article{Abbott:2020tfl,
    author = "Abbott, R. and others",
    collaboration = "LIGO Scientific, Virgo",
    title = "{GW190521: A Binary Black Hole Merger with a Total Mass of 150\,\,M\ensuremath{\odot}}",
    eprint = "2009.01075",
    archivePrefix = "arXiv",
    primaryClass = "gr-qc",
    doi = "10.1103/PhysRevLett.125.101102",
    journal = "Phys. Rev. Lett.",
    volume = "125",
    number = "10",
    pages = "101102",
    year = "2020"
}

@article{Abbott:2020mjq,
    author = "Abbott, R. and others",
    collaboration = "LIGO Scientific, Virgo",
    title = "{Properties and Astrophysical Implications of the 150 M$_\odot$ Binary Black Hole Merger GW190521}",
    eprint = "2009.01190",
    archivePrefix = "arXiv",
    primaryClass = "astro-ph.HE",
    reportNumber = "LIGO-P2000021",
    doi = "10.3847/2041-8213/aba493",
    journal = "Astrophys. J.",
    volume = "900",
    number = "1",
    pages = "L13",
    year = "2020"
}

@article{Gayathri:2020coq,
    author = "Gayathri, V. and Healy, J. and Lange, J. and O'Brien, B. and Szczepanczyk, M. and Bartos, I. and Campanelli, M. and Klimenko, S. and Lousto, C. and O'Shaughnessy, R.",
    title = "{GW190521 as a Highly Eccentric Black Hole Merger}",
    eprint = "2009.05461",
    archivePrefix = "arXiv",
    primaryClass = "astro-ph.HE",
    month = "9",
    year = "2020"
}

@article{CalderonBustillo:2020odh,
    author = "Calder\'on Bustillo, J. and Sanchis-Gual, N. and Torres-Forn\'e, A. and Font, J. A.",
    title = "{Confusing head-on and precessing intermediate-mass binary black hole mergers}",
    eprint = "2009.01066",
    archivePrefix = "arXiv",
    primaryClass = "gr-qc",
    reportNumber = "LIGO-P1900363",
    month = "9",
    year = "2020"
}

@article{Graham:2020gwr,
    author = "Graham, M.J. and others",
    title = "{Candidate Electromagnetic Counterpart to the Binary Black Hole Merger Gravitational Wave Event S190521g}",
    eprint = "2006.14122",
    archivePrefix = "arXiv",
    primaryClass = "astro-ph.HE",
    doi = "10.1103/PhysRevLett.124.251102",
    journal = "Phys. Rev. Lett.",
    volume = "124",
    number = "25",
    pages = "251102",
    year = "2020"
}

@article{Cardoso:2020lxx,
    author = "Cardoso, V. and Macedo, C. F.B.",
    title = "{Drifting through the medium: kicks and self-propulsion of binaries within accretion disks and other environments}",
    eprint = "2008.01091",
    archivePrefix = "arXiv",
    primaryClass = "astro-ph.HE",
    doi = "10.1093/mnras/staa2396",
    month = "8",
    year = "2020"
}

@article{Gergely:1998sr,
	author = "Gergely, L. A. and Perjes, Z. I. and Vasuth, M.",
	title = "{Spin effects in gravitational radiation back reaction. 3. Compact binaries with two spinning components}",
	eprint = "gr-qc/9808063",
	archivePrefix = "arXiv",
	doi = "10.1103/PhysRevD.58.124001",
	journal = "Phys. Rev.",
	volume = "D58",
	pages = "124001",
	year = "1998"
}

@article{Klein:2010ti,
    author = "Klein, A. and Jetzer, P.",
    title = "{Spin effects in the phasing of gravitational waves from binaries on eccentric orbits}",
    eprint = "1005.2046",
    archivePrefix = "arXiv",
    primaryClass = "gr-qc",
    doi = "10.1103/PhysRevD.81.124001",
    journal = "Phys. Rev.",
    volume = "D81",
    pages = "124001",
    year = "2010"
}

@article{Klein:2018ybm,
    author = "Klein, A. and Boetzel, Y. and Gopakumar, A. and Jetzer, P. and de Vittori, L.",
    title = "{Fourier domain gravitational waveforms for precessing eccentric binaries}",
    eprint = "1801.08542",
    archivePrefix = "arXiv",
    primaryClass = "gr-qc",
    doi = "10.1103/PhysRevD.98.104043",
    journal = "Phys. Rev.",
    volume = "D98",
    number = "10",
    pages = "104043",
    year = "2018"
}

@article{Phukon:2019gfh,
    author = "Phukon, K. S. and Gupta, A. and Bose, S. and Jain, P.",
    title = "{Effect of orbital eccentricity on the dynamics of precessing compact binaries}",
    eprint = "1904.03985",
    archivePrefix = "arXiv",
    primaryClass = "gr-qc",
    reportNumber = "LIGO-P1900084",
    doi = "10.1103/PhysRevD.100.124008",
    journal = "Phys. Rev.",
    volume = "D100",
    number = "12",
    pages = "124008",
    year = "2019"
}

@article{Barausse:2016eii,
    author = "Barausse, E. and Yunes, N. and Chamberlain, K.",
    title = "{Theory-Agnostic Constraints on Black-Hole Dipole Radiation with Multiband Gravitational-Wave Astrophysics}",
    eprint = "1603.04075",
    archivePrefix = "arXiv",
    primaryClass = "gr-qc",
    doi = "10.1103/PhysRevLett.116.241104",
    journal = "Phys. Rev. Lett.",
    volume = "116",
    number = "24",
    pages = "241104",
    year = "2016"
}

@article{Cardoso:2016olt,
    author = "Cardoso, V. and Macedo, C. F. B. and Pani, P. and Ferrari, V.",
    title = "{Black holes and gravitational waves in models of minicharged dark matter}",
    eprint = "1604.07845",
    archivePrefix = "arXiv",
    primaryClass = "hep-ph",
    doi = "10.1088/1475-7516/2016/05/054",
    journal = "JCAP",
    volume = "05",
    pages = "054",
    year = "2016",
    note = "[Erratum: JCAP 04, E01 (2020)]"
}

@article{Burko:2002ge,
    author = "Burko, L. M. and Harte, A. I. and Poisson, E.",
    title = "{Mass loss by a scalar charge in an expanding universe}",
    eprint = "gr-qc/0201020",
    archivePrefix = "arXiv",
    doi = "10.1103/PhysRevD.65.124006",
    journal = "Phys. Rev.",
    volume = "D65",
    pages = "124006",
    year = "2002"
}

@article{Quinn:2000wa,
    author = "Quinn, T. C.",
    title = "{Axiomatic approach to radiation reaction of scalar point particles in curved space-time}",
    eprint = "gr-qc/0005030",
    archivePrefix = "arXiv",
    doi = "10.1103/PhysRevD.62.064029",
    journal = "Phys. Rev.",
    volume = "D62",
    pages = "064029",
    year = "2000"
}

@article{Cardoso:2011xi,
    author = "Cardoso, V. and Chakrabarti, S. and Pani, P. and Berti, E. and Gualtieri, L.",
    archivePrefix = "arXiv",
    doi = "10.1103/PhysRevLett.107.241101",
    eprint = "1109.6021",
    journal = "Phys.\ Rev.\ Lett.",
    pages = "241101",
    primaryClass = "gr-qc",
    title = "{Floating and sinking: The Imprint of massive scalars around rotating black holes}",
    volume = "107",
    year = "2011"
}

@article{Yunes:2011aa,
    author = "Yunes, N. and Pani, P. and Cardoso, V.",
    title = "{Gravitational Waves from Quasicircular Extreme Mass-Ratio Inspirals as Probes of Scalar-Tensor Theories}",
    eprint = "1112.3351",
    archivePrefix = "arXiv",
    primaryClass = "gr-qc",
    doi = "10.1103/PhysRevD.85.102003",
    journal = "Phys. Rev.",
    volume = "D85",
    pages = "102003",
    year = "2012"
}

@article{Cardoso:2019nis,
      author         = "Cardoso, V. and del Rio, A. and Kimura, M.",
      title          = "{Distinguishing black holes from horizonless objects
                        through the excitation of resonances during inspiral}",
      year           = "2019",
      eprint         = "1907.01561",
      archivePrefix  = "arXiv",
      primaryClass   = "gr-qc",
      SLACcitation   = "%%CITATION = ARXIV:1907.01561;%%"
}

@article{Christiansen:2020pnv,
    author = "Christiansen, \O{}y. and Jim\'enez, J. B. and Mota, D. F.",
    title = "{Charged Black Hole Mergers: Orbit Circularisation and Chirp Mass Bias}",
    eprint = "2003.11452",
    archivePrefix = "arXiv",
    primaryClass = "gr-qc",
    month = "3",
    year = "2020"
}

@article{Liu:2020cds,
    author = "Liu, L. and Guo, Z. and Cai, R. and Kim, S. P.",
    title = "{Merger rate distribution of primordial black hole binaries with electric charges}",
    eprint = "2001.02984",
    archivePrefix = "arXiv",
    primaryClass = "astro-ph.CO",
    doi = "10.1103/PhysRevD.102.043508",
    journal = "Phys. Rev.",
    volume = "D102",
    number = "4",
    pages = "043508",
    year = "2020"
}

@article{Cardoso:2020nst,
    author = "Cardoso, V. and Guo, W. and Macedo, C. F.B. and Pani, P.",
    title = "{The tune of the universe: the role of plasma in tests of strong-field gravity}",
    eprint = "2009.07287",
    archivePrefix = "arXiv",
    primaryClass = "gr-qc",
    month = "9",
    year = "2020"
}

@article{Bondi:1944jm,
    author = "Bondi, H. and Hoyle, F.",
    title = "{On the mechanism of accretion by stars}",
    journal = "Mon. Not. Roy. Astron. Soc.",
    volume = "104",
    pages = "273",
    year = "1944"
}

@article{Macedo:2013qea,
      author         = "Macedo, C. F. B. and Pani, P. and Cardoso, V. and
                        Crispino, L. C. B.",
      title          = "{Into the lair: gravitational-wave signatures of dark
                        matter}",
      journal        = "Astrophys. J.",
      volume         = "774",
      year           = "2013",
      pages          = "48",
      doi            = "10.1088/0004-637X/774/1/48",
      eprint         = "1302.2646",
      archivePrefix  = "arXiv",
      primaryClass   = "gr-qc",
      SLACcitation   = "%%CITATION = ARXIV:1302.2646;%%"
}

@article{Edgar:2004mk,
    author = "Edgar, R. G.",
    title = "{A Review of Bondi-Hoyle-Lyttleton accretion}",
    eprint = "astro-ph/0406166",
    archivePrefix = "arXiv",
    doi = "10.1016/j.newar.2004.06.001",
    journal = "New Astron. Rev.",
    volume = "48",
    pages = "843--859",
    year = "2004"
}

@article{Antoni:2019pgq,
    author = "Antoni, A. and MacLeod, M. and Ramirez-Ruiz, E.",
    title = "{The Evolution of Binaries in a Gaseous Medium: Three-Dimensional Simulations of Binary Bondi-Hoyle-Lyttleton Accretion}",
    eprint = "1901.07572",
    archivePrefix = "arXiv",
    primaryClass = "astro-ph.HE",
    doi = "10.3847/1538-4357/ab3466",
    journal = "Astrophys. J.",
    volume = "884",
    pages = "22",
    year = "2019"
}

@article{Gair:2010iv,
	author = "Gair, J. R. and Flanagan, E. E. and Drasco, S. and Hinderer, T. and Babak, S.",
	title = "{Forced motion near black holes}",
	eprint = "1012.5111",
	archivePrefix = "arXiv",
	primaryClass = "gr-qc",
	doi = "10.1103/PhysRevD.83.044037",
	journal = "Phys. Rev.",
	volume = "D83",
	pages = "044037",
	year = "2011"
}

@Book{landau1982mechanics,
  title={Mechanics},
  author={Landau, L. D. and Lifshitz, E. M.},
  number={v. 1},
  isbn={9780080503479},
  url={https://books.google.pt/books?id=bE-9tUH2J2wC},
  year={1982},
  publisher={Elsevier Science}
}

@article{DeLuca:2020qqa,
    author = "De Luca, V. and Franciolini, G. and Pani, P. and Riotto, A.",
    title = "{Primordial Black Holes Confront LIGO/Virgo data: Current situation}",
    eprint = "2005.05641",
    archivePrefix = "arXiv",
    primaryClass = "astro-ph.CO",
    doi = "10.1088/1475-7516/2020/06/044",
    journal = "JCAP",
    volume = "06",
    pages = "044",
    year = "2020"
}

@ARTICLE{1985Salmassi,
       author = {{Salmassi}, M.},
        title = "{Second Order Adiabatic Invariants Associated with the Two-Body Problem with Slowly Varying Mass}",
      journal = {Celest. Mech. Dyn. Astron.},
         year = 1985,
        month = dec,
       volume = {37},
       number = {4},
        pages = {359-369},
          doi = {10.1007/BF01261625},
       adsurl = {https://ui.adsabs.harvard.edu/abs/1985CeMec..37..359S}
}

@ARTICLE{1993Djukic,
       author = {{Djukic}, D. S.},
        title = "{Adiabatic Invariants for the Nonconservative Kepler's Problem}",
      journal = {Celest. Mech. Dyn. Astron.},
         year = 1993,
        month = aug,
       volume = {56},
       number = {4},
        pages = {523-540},
          doi = {10.1007/BF00696184},
       adsurl = {https://ui.adsabs.harvard.edu/abs/1993CeMDA..56..523D}
}

@article{Audley:2017drz,
    author = "Amaro-Seoane, P. and others",
    collaboration = "LISA",
    title = "{Laser Interferometer Space Antenna}",
    eprint = "1702.00786",
    archivePrefix = "arXiv",
    primaryClass = "astro-ph.IM",
    month = "2",
    year = "2017"
}

@Book{MaggioreBook,
     author    =  "Maggiore, M.",
     title     =  "Gravitational Waves: Volume 1: Theory and Experiments",
     publisher =  "Oxford University Press",
     address   =  "Oxford",
     year      =  "2008"
}

@article{Hopper:2017qus,
    author = "Hopper, S. and Cardoso, V.",
    title = "{Scattering of point particles by black holes: gravitational radiation}",
    eprint = "1706.02791",
    archivePrefix = "arXiv",
    primaryClass = "gr-qc",
    doi = "10.1103/PhysRevD.97.044031",
    journal = "Phys. Rev.",
    volume = "D97",
    number = "4",
    pages = "044031",
    year = "2018"
}

@article{Nitz:2020mga,
    author = "Nitz, A. H. and Capano, C. D.",
    title = "{GW190521 may be an intermediate mass ratio inspiral}",
    eprint = "2010.12558",
    archivePrefix = "arXiv",
    primaryClass = "astro-ph.HE",
    month = "10",
    year = "2020"
}

@article{Cardoso:2019rou,
    author = "Cardoso, V. and Maselli, A.",
    archivePrefix = "arXiv",
    eprint = "1909.05870",
    month = "9",
    primaryClass = "astro-ph.HE",
    title = "{Constraints on the astrophysical environment of binaries with gravitational-wave observations}",
    year = "2019"
}

@article{Toubiana:2020drf,
    author = "Toubiana, A. and others",
    title = "{Detectable environmental effects in GW190521-like black-hole binaries with LISA}",
    eprint = "2010.06056",
    archivePrefix = "arXiv",
    primaryClass = "astro-ph.HE",
    month = "10",
    year = "2020"
}

@article{Cardoso:2019upw,
    author = "Cardoso, V. and Duque, F.",
    title = "{Environmental effects in gravitational-wave physics: Tidal deformability of black holes immersed in matter}",
    eprint = "1912.07616",
    archivePrefix = "arXiv",
    primaryClass = "gr-qc",
    doi = "10.1103/PhysRevD.101.064028",
    journal = "Phys. Rev.",
    volume = "D101",
    number = "6",
    pages = "064028",
    year = "2020"
}

@article{Roedig:2011rn,
    author = "Roedig, C. and Sesana, A.",
    editor = "Hannam, M. and Sutton, P. and Hild, S. and van den Broeck, C.",
    title = "{Origin and Implications of high eccentricities in massive black hole binaries at sub-pc scales}",
    eprint = "1111.3742",
    archivePrefix = "arXiv",
    primaryClass = "astro-ph.CO",
    doi = "10.1088/1742-6596/363/1/012035",
    journal = "J. Phys. Conf. Ser.",
    volume = "363",
    pages = "012035",
    year = "2012"
}

@article{Zrake:2020zkw,
    author = "Zrake, J. and Tiede, C. and MacFadyen, A. and Haiman, Z.",
    title = "{Equilibrium eccentricity of accreting binaries}",
    eprint = "2010.09707",
    archivePrefix = "arXiv",
    primaryClass = "astro-ph.HE",
    month = "10",
    year = "2020"
}

@article{Penrose:1964wq,
    author = "Penrose, R.",
    title = "{Gravitational collapse and space-time singularities}",
    doi = "10.1103/PhysRevLett.14.57",
    journal = "Phys. Rev. Lett.",
    volume = "14",
    pages = "57--59",
    year = "1965"
}

@article{Hawking:1967ju,
    author = "Hawking, S. W.",
    title = "{The occurrence of singularities in cosmology. III. Causality and singularities}",
    doi = "10.1098/rspa.1967.0164",
    journal = "Proc. Roy. Soc. Lond. A",
    volume = "300",
    pages = "187--201",
    year = "1967"
}

@article{Hawking:1970zqf,
    author = "Hawking, S. W. and Penrose, R.",
    title = "{The Singularities of gravitational collapse and cosmology}",
    doi = "10.1098/rspa.1970.0021",
    journal = "Proc. Roy. Soc. Lond. A",
    volume = "314",
    pages = "529--548",
    year = "1970"
}

@article{Penrose:1969pc,
    author = "Penrose, R.",
    title = "{Gravitational collapse: The role of general relativity}",
    doi = "10.1023/A:1016578408204",
    journal = "Riv. Nuovo Cim.",
    volume = "1",
    pages = "252--276",
    year = "1969"
}

@inproceedings{Wald:1997wa,
    author = "Wald, R. M.",
    title = "{Gravitational collapse and cosmic censorship}",
    eprint = "gr-qc/9710068",
    archivePrefix = "arXiv",
    reportNumber = "EFI-97-43",
    doi = "10.1007/978-94-017-0934-7_5",
    month = "10",
    year = "1997"
}

@article{WALD1974548,
title = {Gedanken experiments to destroy a black hole},
journal = {Ann. Phys.},
volume = {82},
number = {2},
pages = {548-556},
year = {1974},
issn = {0003-4916},
doi = {https://doi.org/10.1016/0003-4916(74)90125-0},
url = {https://www.sciencedirect.com/science/article/pii/0003491674901250},
author = {R. M. Wald}
}

@article{Tod:1976ud,
    author = "Tod, K. P. and de Felice, F. and Calvani, M.",
    title = "{Spinning test particles in the field of a black hole}",
    doi = "10.1007/BF02728614",
    journal = "Nuovo Cim. B",
    volume = "34",
    pages = "365",
    year = "1976"
}

@article{Needham:1980fb,
    author = "Needham, T.",
    title = "{Cosmic Censorship and test particles}",
    doi = "10.1103/PhysRevD.22.791",
    journal = "Phys. Rev.",
    volume = "D22",
    pages = "791--796",
    year = "1980"
}

@article{Semiz:2005gs,
    author = "Semiz, I.",
    title = "{Dyonic Kerr-Newman black holes, complex scalar field and cosmic censorship}",
    eprint = "gr-qc/0508011",
    archivePrefix = "arXiv",
    doi = "10.1007/s10714-010-1108-z",
    journal = "Gen. Rel. Grav.",
    volume = "43",
    pages = "833--846",
    year = "2011"
}

@article{Toth:2012vvy,
    author = "Toth, G. Z.",
    title = "{Test of the weak cosmic censorship conjecture with a charged scalar field and dyonic Kerr-Newman black holes}",
    eprint = "1112.2382",
    archivePrefix = "arXiv",
    primaryClass = "gr-qc",
    doi = "10.1007/s10714-012-1374-z",
    journal = "Gen. Rel. Grav.",
    volume = "44",
    pages = "2019--2035",
    year = "2012"
}

@article{Duztas:2013wua,
    author = {D\"uzta\c{s}, K. and Semiz, \.I.},
    title = "{Cosmic Censorship, Black Holes and Integer-spin Test Fields}",
    eprint = "1307.1481",
    archivePrefix = "arXiv",
    primaryClass = "gr-qc",
    doi = "10.1103/PhysRevD.88.064043",
    journal = "Phys. Rev.",
    volume = "D88",
    number = "6",
    pages = "064043",
    year = "2013"
}

@article{Duztas:2013gza,
    author = {D\"uzta\c{s}, Koray},
    title = "{Electromagnetic field and cosmic censorship}",
    eprint = "1312.7361",
    archivePrefix = "arXiv",
    primaryClass = "gr-qc",
    doi = "10.1007/s10714-014-1709-z",
    journal = "Gen. Rel. Grav.",
    volume = "46",
    pages = "1709",
    year = "2014"
}

@article{Bouhmadi-Lopez:2010yjy,
    author = "Bouhmadi-Lopez, M. and Cardoso, V. and Nerozzi, A. and Rocha, J. V.",
    title = "{Black holes die hard: can one spin-up a black hole past extremality?}",
    eprint = "1003.4295",
    archivePrefix = "arXiv",
    primaryClass = "gr-qc",
    doi = "10.1103/PhysRevD.81.084051",
    journal = "Phys. Rev.",
    volume = "D81",
    pages = "084051",
    year = "2010"
}

@article{Gwak:2015fsa,
    author = "Gwak, Bogeun and Lee, Bum-Hoon",
    title = "{Cosmic Censorship of Rotating Anti-de Sitter Black Hole}",
    eprint = "1509.06691",
    archivePrefix = "arXiv",
    primaryClass = "gr-qc",
    doi = "10.1088/1475-7516/2016/02/015",
    journal = "JCAP",
    volume = "02",
    pages = "015",
    year = "2016"
}

@article{Rocha:2014jma,
    author = "Rocha, J. V. and Santarelli, R.",
    title = "{Flowing along the edge: spinning up black holes in AdS spacetimes with test particles}",
    eprint = "1402.4840",
    archivePrefix = "arXiv",
    primaryClass = "gr-qc",
    doi = "10.1103/PhysRevD.89.064065",
    journal = "Phys. Rev.",
    volume = "D89",
    number = "6",
    pages = "064065",
    year = "2014"
}

@article{Hubeny:1998ga,
    author = "Hubeny, V. E.",
    title = "{Overcharging a black hole and cosmic censorship}",
    eprint = "gr-qc/9808043",
    archivePrefix = "arXiv",
    doi = "10.1103/PhysRevD.59.064013",
    journal = "Phys. Rev.",
    volume = "D59",
    pages = "064013",
    year = "1999"
}

@article{Matsas:2007bj,
    author = "Matsas, G. E. A. and da Silva, A. R. R.",
    title = "{Overspinning a nearly extreme charged black hole via a quantum tunneling process}",
    eprint = "0706.3198",
    archivePrefix = "arXiv",
    primaryClass = "gr-qc",
    doi = "10.1103/PhysRevLett.99.181301",
    journal = "Phys. Rev. Lett.",
    volume = "99",
    pages = "181301",
    year = "2007"
}

@article{Jacobson:2009kt,
    author = "Jacobson, T. and Sotiriou, T. P.",
    title = "{Over-spinning a black hole with a test body}",
    eprint = "0907.4146",
    archivePrefix = "arXiv",
    primaryClass = "gr-qc",
    doi = "10.1103/PhysRevLett.103.141101",
    journal = "Phys. Rev. Lett.",
    volume = "103",
    pages = "141101",
    year = "2009",
    note = "[Erratum: Phys.Rev.Lett. 103, 209903 (2009)]"
}

@article{Saa:2011wq,
    author = "Saa, A. and Santarelli, R.",
    title = "{Destroying a near-extremal Kerr-Newman black hole}",
    eprint = "1105.3950",
    archivePrefix = "arXiv",
    primaryClass = "gr-qc",
    doi = "10.1103/PhysRevD.84.027501",
    journal = "Phys. Rev.",
    volume = "D84",
    pages = "027501",
    year = "2011"
}

@article{Hod:2008zza,
    author = "Hod, Shahar",
    title = "{Weak Cosmic Censorship: As Strong as Ever}",
    eprint = "0805.3873",
    archivePrefix = "arXiv",
    primaryClass = "gr-qc",
    doi = "10.1103/PhysRevLett.100.121101",
    journal = "Phys. Rev. Lett.",
    volume = "100",
    pages = "121101",
    year = "2008"
}

@article{Barausse:2010ka,
    author = "Barausse, E. and Cardoso, V. and Khanna, G.",
    title = "{Test bodies and naked singularities: Is the self-force the cosmic censor?}",
    eprint = "1008.5159",
    archivePrefix = "arXiv",
    primaryClass = "gr-qc",
    doi = "10.1103/PhysRevLett.105.261102",
    journal = "Phys. Rev. Lett.",
    volume = "105",
    pages = "261102",
    year = "2010"
}

@article{Zimmerman:2012zu,
    author = "Zimmerman, P. and Vega, I. and Poisson, E. and Haas, R.",
    title = "{Self-force as a cosmic censor}",
    eprint = "1211.3889",
    archivePrefix = "arXiv",
    primaryClass = "gr-qc",
    doi = "10.1103/PhysRevD.87.041501",
    journal = "Phys. Rev.",
    volume = "D87",
    number = "4",
    pages = "041501",
    year = "2013"
}

@article{Shaymatov:2014dla,
    author = "Shaymatov, S. and Patil, M. and Ahmedov, B. and Joshi, P. S.",
    title = "{Destroying a near-extremal Kerr black hole with a charged particle: Can a test magnetic field serve as a cosmic censor?}",
    eprint = "1409.3018",
    archivePrefix = "arXiv",
    primaryClass = "gr-qc",
    doi = "10.1103/PhysRevD.91.064025",
    journal = "Phys. Rev.",
    volume = "D91",
    number = "6",
    pages = "064025",
    year = "2015"
}

@article{Colleoni:2015ena,
    author = "Colleoni, M. and Barack, L. and Shah, A. G. and van de Meent, M.",
    title = "{Self-force as a cosmic censor in the Kerr overspinning problem}",
    eprint = "1508.04031",
    archivePrefix = "arXiv",
    primaryClass = "gr-qc",
    doi = "10.1103/PhysRevD.92.084044",
    journal = "Phys. Rev.",
    volume = "D92",
    number = "8",
    pages = "084044",
    year = "2015"
}

@article{Bardeen:1973gs,
    author = "Bardeen, J. M. and Carter, B. and Hawking, S. W.",
    title = "{The Four laws of black hole mechanics}",
    doi = "10.1007/BF01645742",
    journal = "Commun. Math. Phys.",
    volume = "31",
    pages = "161--170",
    year = "1973"
}

@article{Israel:1986gqz,
    author = "Israel, W.",
    title = "{Third Law of Black-Hole Dynamics: A Formulation and Proof}",
    doi = "10.1103/PhysRevLett.57.397",
    journal = "Phys. Rev. Lett.",
    volume = "57",
    number = "4",
    pages = "397",
    year = "1986"
}

@article{Dadhich:1997rq,
    author = "Dadhich, N. and Narayan, K.",
    title = "{On the third law of black hole dynamics}",
    eprint = "gr-qc/9704070",
    archivePrefix = "arXiv",
    reportNumber = "IUCAA-31-97",
    doi = "10.1016/S0375-9601(97)00337-X",
    journal = "Phys. Lett. A",
    volume = "231",
    pages = "335--338",
    year = "1997"
}

@article{Chirco:2010rq,
    author = "Chirco, G. and Liberati, S. and Sotiriou, T. P.",
    title = "{Gedanken experiments on nearly extremal black holes and the Third Law}",
    eprint = "1006.3655",
    archivePrefix = "arXiv",
    primaryClass = "gr-qc",
    doi = "10.1103/PhysRevD.82.104015",
    journal = "Phys. Rev.",
    volume = "D82",
    pages = "104015",
    year = "2010"
}

@article{Boulware73,
  title = {Naked Singularities, Thin Shells, and the Reissner-Nordstr\"om Metric},
  author = {Boulware, D. G.},
  journal = {Phys. Rev.},
  volume = {D8},
  issue = {8},
  pages = {2363--2368},
  numpages = {0},
  year = {1973},
  month = {Oct},
  publisher = {American Physical Society},
  doi = {10.1103/PhysRevD.8.2363},
  url = {https://link.aps.org/doi/10.1103/PhysRevD.8.2363}
}

@article{Geroch:1975uq,
    author = "Geroch, R. P. and Jang, P. S.",
    title = "{Motion of a body in general relativity}",
    doi = "10.1063/1.522416",
    journal = "J. Math. Phys.",
    volume = "16",
    pages = "65--67",
    year = "1975"
}

@article{Lasota:2013kia,
    author = "Lasota, J. -P. and Gourgoulhon, E. and Abramowicz, M. and Tchekhovskoy, A. and Narayan, R.",
    title = "{Extracting black-hole rotational energy: The generalized Penrose process}",
    eprint = "1310.7499",
    archivePrefix = "arXiv",
    primaryClass = "gr-qc",
    doi = "10.1103/PhysRevD.89.024041",
    journal = "Phys. Rev.",
    volume = "D89",
    number = "2",
    pages = "024041",
    year = "2014"
}

@article{Gao:2001ut,
    author = "Gao, S. and Wald, R. M.",
    title = "{The 'Physical process' version of the first law and the generalized second law for charged and rotating black holes}",
    eprint = "gr-qc/0106071",
    archivePrefix = "arXiv",
    doi = "10.1103/PhysRevD.64.084020",
    journal = "Phys. Rev.",
    volume = "D64",
    pages = "084020",
    year = "2001"
}

@article{Caldarelli:1999xj,
    author = "Caldarelli, M. M. and Cognola, G. and Klemm, D.",
    title = "{Thermodynamics of Kerr-Newman-AdS black holes and conformal field theories}",
    eprint = "hep-th/9908022",
    archivePrefix = "arXiv",
    reportNumber = "UTF-434",
    doi = "10.1088/0264-9381/17/2/310",
    journal = "Class. Quant. Grav.",
    volume = "17",
    pages = "399--420",
    year = "2000"
}

@article{Olea:2005gb,
    author = "Olea, R.",
    title = "{Mass, angular momentum and thermodynamics in four-dimensional Kerr-AdS black holes}",
    eprint = "hep-th/0504233",
    archivePrefix = "arXiv",
    doi = "10.1088/1126-6708/2005/06/023",
    journal = "JHEP",
    volume = "06",
    pages = "023",
    year = "2005"
}

@article{Gibbons:2004ai,
    author = "Gibbons, G. W. and Perry, M. J. and Pope, C. N.",
    title = "{The First law of thermodynamics for Kerr-anti-de Sitter black holes}",
    eprint = "hep-th/0408217",
    archivePrefix = "arXiv",
    reportNumber = "DAMTP-2004-87, MIFP-04-17",
    doi = "10.1088/0264-9381/22/9/002",
    journal = "Class. Quant. Grav.",
    volume = "22",
    pages = "1503--1526",
    year = "2005"
}

@article{McInnes:2015vga,
    author = "McInnes, B. and Ong, Y. C.",
    title = "{A Note on Physical Mass and the Thermodynamics of AdS-Kerr Black Holes}",
    eprint = "1506.01248",
    archivePrefix = "arXiv",
    primaryClass = "gr-qc",
    doi = "10.1088/1475-7516/2015/11/004",
    journal = "JCAP",
    volume = "11",
    pages = "004",
    year = "2015"
}

@article{Emparan:2008eg,
    author = "Emparan, R. and Reall, H. S.",
    title = "{Black Holes in Higher Dimensions}",
    eprint = "0801.3471",
    archivePrefix = "arXiv",
    primaryClass = "hep-th",
    doi = "10.12942/lrr-2008-6",
    journal = "Living Rev. Rel.",
    volume = "11",
    pages = "6",
    year = "2008"
}

@article{Faraoni:2010yi,
    author = "Faraoni, V.",
    title = "{Black hole entropy in scalar-tensor and f(R) gravity: An Overview}",
    eprint = "1005.2327",
    archivePrefix = "arXiv",
    primaryClass = "gr-qc",
    doi = "10.3390/e12051246",
    journal = "Entropy",
    volume = "12",
    pages = "1246",
    year = "2010"
}

@article{Appels:2016uha,
    author = "Appels, M. and Gregory, R. and Kubiznak, D.",
    title = "{Thermodynamics of Accelerating Black Holes}",
    eprint = "1604.08812",
    archivePrefix = "arXiv",
    primaryClass = "hep-th",
    reportNumber = "DCPT-16-15",
    doi = "10.1103/PhysRevLett.117.131303",
    journal = "Phys. Rev. Lett.",
    volume = "117",
    number = "13",
    pages = "131303",
    year = "2016"
}

@article{Cardoso:2015xtj,
    author = "Cardoso, V. and Queimada, L.",
    title = "{Cosmic Censorship and parametrized spinning black-hole geometries}",
    eprint = "1511.00690",
    archivePrefix = "arXiv",
    primaryClass = "gr-qc",
    doi = "10.1007/s10714-015-1990-5",
    journal = "Gen. Rel. Grav.",
    volume = "47",
    number = "12",
    pages = "150",
    year = "2015"
}

@article{Wang01,
author = {X. Wang},
title = {{The Mass of Asymptotically Hyperbolic Manifolds}},
volume = {57},
journal = {Journal of Differential Geometry},
number = {2},
publisher = {Lehigh University},
pages = {273 -- 299},
year = {2001},
doi = {10.4310/jdg/1090348112},
URL = {https://doi.org/10.4310/jdg/1090348112}
}

@article{Chrusciel:2001qr,
    author = "Chrusciel, P. T. and Nagy, G.",
    title = "{The Mass of space - like hypersurfaces in asymptotically anti-de Sitter space-times}",
    eprint = "gr-qc/0110014",
    archivePrefix = "arXiv",
    doi = "10.4310/ATMP.2001.v5.n4.a3",
    journal = "Adv. Theor. Math. Phys.",
    volume = "5",
    pages = "697--754",
    year = "2002"
}

@article{Chrusciel:2003qr,
    author = "Chrusciel, P. T. and Herzlich, M.",
    title = "{The Mass of asymptotically hyperboloidal Riemannian manifolds}",
    doi = "10.2140/pjm.2003.212.231",
    journal = "Pacific J. Math.",
    volume = "212",
    pages = "231--264",
    year = "2003"
}

@article{Miao02,
title = "Positive Mass Theorem on manifolds admitting corners along a hypersurface",
author = "P. Miao",
year = "2002",
month = nov,
doi = "10.4310/ATMP.2002.v6.n6.a4",
language = "English (US)",
volume = "6",
pages = "1163--1182",
journal = "Adv. Theor. Math. Phys.",
issn = "1095-0761",
number = "6"
}

@article{Min1989,
author = {Min-Oo, M.},
journal = {Math. Ann.},
number = {4},
pages = {527-540},
title = {Scalar curvature rigidity of asymptotically hyperbolic spin manifods},
url = {http://eudml.org/doc/164616},
volume = {285},
year = {1989},
}

@article{Andersson1998,
  title={Scalar curvature rigidity for asymptotically locally hyperbolic manifolds},
  author={Andersson, L. and Dahl, M.},
  journal={Ann. Global Anal. Geom.},
  volume={16},
  number={1},
  pages={1--27},
  year={1998},
  publisher={Springer}
}

@article{Brendle2011,
  title={Deformations of the hemisphere that increase scalar curvature},
  author={Brendle, S. and Marques, F. C. and Neves, A.},
  journal={Invent. Math.},
  volume={185},
  number={1},
  pages={175--197},
  year={2011},
  publisher={Springer}
}

@article{Kastor:2002fu,
    author = "Kastor, D. and Traschen, J. H.",
    title = "{A Positive energy theorem for asymptotically de Sitter space-times}",
    eprint = "hep-th/0206105",
    archivePrefix = "arXiv",
    doi = "10.1088/0264-9381/19/23/302",
    journal = "Class. Quant. Grav.",
    volume = "19",
    pages = "5901--5920",
    year = "2002"
}

@article{Luo:2007se,
    author = "Luo, M. and Xie, N. and Zhang, X.",
    title = "{Positive mass theorems for asymptotically de Sitter spacetimes}",
    eprint = "0712.4113",
    archivePrefix = "arXiv",
    primaryClass = "math.DG",
    doi = "10.1016/j.nuclphysb.2009.09.017",
    journal = "Nucl. Phys. B",
    volume = "825",
    pages = "98--118",
    year = "2010"
}

@article{Gwak:2018akg,
    author = "Gwak, B.",
    title = "{Weak Cosmic Censorship Conjecture in Kerr-(Anti-)de Sitter Black Hole with Scalar Field}",
    eprint = "1807.10630",
    archivePrefix = "arXiv",
    primaryClass = "gr-qc",
    doi = "10.1007/JHEP09(2018)081",
    journal = "JHEP",
    volume = "09",
    pages = "081",
    year = "2018"
}

@article{Gwak:2018tmy,
    author = "Gwak, B.",
    title = "{Thermodynamics and Cosmic Censorship Conjecture in Kerr\textendash{}Newman\textendash{}de Sitter Black Hole}",
    doi = "10.3390/e20110855",
    journal = "Entropy",
    volume = "20",
    number = "11",
    pages = "855",
    year = "2018"
}

@article{Dolan:2013ft,
    author = "Dolan, B. P. and Kastor, D. and Kubiznak, D. and Mann, R. B. and Traschen, J.",
    title = "{Thermodynamic Volumes and Isoperimetric Inequalities for de Sitter Black Holes}",
    eprint = "1301.5926",
    archivePrefix = "arXiv",
    primaryClass = "hep-th",
    reportNumber = "DIAS-STP-13-1, PI-STRONGGRV-314",
    doi = "10.1103/PhysRevD.87.104017",
    journal = "Phys. Rev.",
    volume = "D87",
    number = "10",
    pages = "104017",
    year = "2013"
}

@article{Kubiznak:2015bya,
    author = "Kubiznak, D. and Simovic, F.",
    title = "{Thermodynamics of horizons: de Sitter black holes and reentrant phase transitions}",
    eprint = "1507.08630",
    archivePrefix = "arXiv",
    primaryClass = "hep-th",
    doi = "10.1088/0264-9381/33/24/245001",
    journal = "Class. Quant. Grav.",
    volume = "33",
    number = "24",
    pages = "245001",
    year = "2016"
}

@article{Duztas:2014sga,
    author = {D\"uzta\c{s}, K.},
    title = "{Stability of event horizons against neutrino flux: The classical picture}",
    eprint = "1408.1735",
    archivePrefix = "arXiv",
    primaryClass = "gr-qc",
    doi = "10.1088/0264-9381/32/7/075003",
    journal = "Class. Quant. Grav.",
    volume = "32",
    number = "7",
    pages = "075003",
    year = "2015"
}

@article{Toth:2015cda,
    author = "T\'oth, G. Z.",
    title = "{Weak cosmic censorship, dyonic Kerr-Newman black holes and Dirac fields}",
    eprint = "1509.02878",
    archivePrefix = "arXiv",
    primaryClass = "gr-qc",
    doi = "10.1088/0264-9381/33/11/115012",
    journal = "Class. Quant. Grav.",
    volume = "33",
    number = "11",
    pages = "115012",
    year = "2016"
}

@article{Derrick:1964ww,
      author         = "Derrick, G. H.",
      title          = "{Comments on nonlinear wave equations as models for
                        elementary particles}",
      journal        = "J. Math. Phys.",
      volume         = "5",
      year           = "1964",
      pages          = "1252-1254",
      doi            = "10.1063/1.1704233",
      SLACcitation   = "%%CITATION = JMAPA,5,1252;%%"
}

@article{Herdeiro:2014goa,
      author         = "Herdeiro, C. A. R. and Radu, E.",
      title          = "{Kerr black holes with scalar hair}",
      journal        = "Phys. Rev. Lett.",
      volume         = "112",
      pages          = "221101",
      doi            = "10.1103/PhysRevLett.112.221101",
      year           = "2014",
      eprint         = "1403.2757",
      archivePrefix  = "arXiv",
      primaryClass   = "gr-qc",
      SLACcitation   = "%%CITATION = ARXIV:1403.2757;%%",
}

@article{Cardoso:2019rvt,
      author         = "Cardoso, V. and Pani, P.",
      title          = "{Testing the nature of dark compact objects: a status
                        report}",
      journal        = "Living Rev. Rel.",
      volume         = "22",
      year           = "2019",
      number         = "1",
      pages          = "4",
      doi            = "10.1007/s41114-019-0020-4",
      eprint         = "1904.05363",
      archivePrefix  = "arXiv",
      primaryClass   = "gr-qc",
      SLACcitation   = "%%CITATION = ARXIV:1904.05363;%%"
}

@article{Giudice:2016zpa,
    author = "Giudice, G. F. and McCullough, M. and Urbano, A.",
    archivePrefix = "arXiv",
    doi = "10.1088/1475-7516/2016/10/001",
    eprint = "1605.01209",
    journal = "JCAP",
    pages = "001",
    primaryClass = "hep-ph",
    reportNumber = "CERN-TH-2016-106",
    title = "{Hunting for Dark Particles with Gravitational Waves}",
    volume = "10",
    year = "2016"
}

@article{Ellis:2017jgp,
      author         = "Ellis, J. and Hektor, A. and Hütsi, G. and
                        Kannike, K. and Marzola, L. and Raidal, M. and
                        Vaskonen, V.",
      title          = "{Search for Dark Matter Effects on Gravitational Signals
                        from Neutron Star Mergers}",
      journal        = "Phys. Lett.",
      volume         = "B781",
      year           = "2018",
      pages          = "607-610",
      doi            = "10.1016/j.physletb.2018.04.048",
      eprint         = "1710.05540",
      archivePrefix  = "arXiv",
      primaryClass   = "astro-ph.CO",
      reportNumber   = "CERN-TH-2017-208, KCL-PH-TH-2017-50",
      SLACcitation   = "%%CITATION = ARXIV:1710.05540;%%"
}

@article{Robles:2012uy,
    author = "Robles, V. H. and Matos, T.",
    archivePrefix = "arXiv",
    doi = "10.1111/j.1365-2966.2012.20603.x",
    eprint = "1201.3032",
    journal = "Mon. Not. Roy. Astron. Soc.",
    pages = "282--289",
    primaryClass = "astro-ph.CO",
    title = "{Flat Central Density Profile and Constant DM Surface Density in Galaxies from Scalar Field Dark Matter}",
    volume = "422",
    year = "2012"
}

@article{Bar:2019bqz,
      author         = "Bar, N. and Blum, K. and Eby, J. and Sato,
                        R.",
      title          = "{Ultralight dark matter in disk galaxies}",
      journal        = "Phys. Rev.",
      volume         = "D99",
      year           = "2019",
      number         = "10",
      pages          = "103020",
      doi            = "10.1103/PhysRevD.99.103020",
      eprint         = "1903.03402",
      archivePrefix  = "arXiv",
      primaryClass   = "astro-ph.CO",
      reportNumber   = "DESY-19-036",
      SLACcitation   = "%%CITATION = ARXIV:1903.03402;%%"
}

@article{Bar:2018acw,
      author         = "Bar, N. and Blas, D. and Blum, K. and
                        Sibiryakov, S.",
      title          = "{Galactic rotation curves versus ultralight dark matter:
                        Implications of the soliton-host halo relation}",
      journal        = "Phys. Rev.",
      volume         = "D98",
      year           = "2018",
      number         = "8",
      pages          = "083027",
      doi            = "10.1103/PhysRevD.98.083027",
      eprint         = "1805.00122",
      archivePrefix  = "arXiv",
      primaryClass   = "astro-ph.CO",
      reportNumber   = "CERN-TH-2018-102, INR-TH-2018-008, KCL-PH-TH/2018-16",
      SLACcitation   = "%%CITATION = ARXIV:1805.00122;%%"
}

@article{Desjacques:2019zhf,
      author         = "Desjacques, V. and Nusser, A.",
      title          = "{Axion core–halo mass and the black hole–halo mass
                        relation: constraints on a few parsec scales}",
      journal        = "Mon. Not. Roy. Astron. Soc.",
      volume         = "488",
      year           = "2019",
      number         = "4",
      pages          = "4497-4503",
      doi            = "10.1093/mnras/stz1978",
      eprint         = "1905.03450",
      archivePrefix  = "arXiv",
      primaryClass   = "astro-ph.CO",
      SLACcitation   = "%%CITATION = ARXIV:1905.03450;%%"
}

@article{Davoudiasl:2019nlo,
    author = "Davoudiasl, H. and Denton, P. B",
    title = "{Ultralight Boson Dark Matter and Event Horizon Telescope Observations of M87*}",
    eprint = "1904.09242",
    archivePrefix = "arXiv",
    primaryClass = "astro-ph.CO",
    doi = "10.1103/PhysRevLett.123.021102",
    journal = "Phys. Rev. Lett.",
    volume = "123",
    number = "2",
    pages = "021102",
    year = "2019"
}

@article{Kaup:1968zz,
      author         = "Kaup, D. J.",
      title          = "{Klein-Gordon Geon}",
      journal        = "Phys.Rev.",
      volume         = "172",
      pages          = "1331-1342",
      doi            = "10.1103/PhysRev.172.1331",
      year           = "1968",
      SLACcitation   = "%%CITATION = PHRVA,172,1331;%%",
}

@article{Ruffini:1969qy,
      author         = "Ruffini, R. and Bonazzola, S.",
      title          = "{Systems of selfgravitating particles in general
                        relativity and the concept of an equation of state}",
      journal        = "Phys.Rev.",
      volume         = "187",
      pages          = "1767-1783",
      doi            = "10.1103/PhysRev.187.1767",
      year           = "1969",
      SLACcitation   = "%%CITATION = PHRVA,187,1767;%%",
}

@article{Liebling:2012fv,
      author         = "Liebling, S. L. and Palenzuela, C.",
      title          = "{Dynamical Boson Stars}",
      journal        = "Living Rev. Rel.",
      volume         = "15",
      year           = "2012",
      pages          = "6",
      doi            = "10.12942/lrr-2012-6, 10.1007/s41114-017-0007-y",
      note           = "[Living Rev. Rel.20,no.1,5(2017)]",
      eprint         = "1202.5809",
      archivePrefix  = "arXiv",
      primaryClass   = "gr-qc",
      SLACcitation   = "%%CITATION = ARXIV:1202.5809;%%"
}

@article{Coleman:1985ki,
    author = "Coleman, S. R.",
    title = "{Q Balls}",
    reportNumber = "HUTP-85/A050",
    doi = "10.1016/0550-3213(86)90520-1",
    journal = "Nucl. Phys. B",
    volume = "262",
    pages = "263",
    year = "1985",
    note = "[Erratum: Nucl.Phys.B 269, 744 (1986)]"
}

@article{Kusenko:1997si,
    author = "Kusenko, A. and Shaposhnikov, M. E.",
    title = "{Supersymmetric Q balls as dark matter}",
    eprint = "hep-ph/9709492",
    archivePrefix = "arXiv",
    reportNumber = "CERN-TH-97-259",
    doi = "10.1016/S0370-2693(97)01375-0",
    journal = "Phys. Lett. B",
    volume = "418",
    pages = "46--54",
    year = "1998"
}

@article{Frieman:1988,
  title = {Primordial Origin of Nontopological Solitons},
  author = {Frieman, J. A. and Gelmini, G. B. and Gleiser, M. and Kolb, E. W.},
  journal = {Phys. Rev. Lett.},
  volume = {60},
  issue = {21},
  pages = {2101--2104},
  numpages = {0},
  year = {1988},
  month = {May},
  publisher = {American Physical Society},
  doi = {10.1103/PhysRevLett.60.2101},
  url = {https://link.aps.org/doi/10.1103/PhysRevLett.60.2101}
}

@article{Khlopov:1985,
       author = {{Khlopov}, M. I. and {Malomed}, B.~A. and {Zeldovich}, I. B.},
        title = "{Gravitational instability of scalar fields and formation of primordial black holes}",
      journal = {Mon. Not. Roy. Astron. Soc.},
         year = 1985,
        month = aug,
       volume = {215},
        pages = {575-589},
          doi = {10.1093/mnras/215.4.575},
       adsurl = {https://ui.adsabs.harvard.edu/abs/1985MNRAS.215..575K}
}

@ARTICLE{1973ApJ...183..657B,
       author = {{Bekenstein}, J. D.},
        title = "{Gravitational-Radiation Recoil and Runaway Black Holes}",
      journal = {Astrophys. J.},
         year = 1973,
        month = jul,
       volume = {183},
        pages = {657-664},
          doi = {10.1086/152255},
       adsurl = {https://ui.adsabs.harvard.edu/abs/1973ApJ...183..657B}
}

@article{Hui:2021tkt,
    author = "Hui, L.",
    title = "{Wave Dark Matter}",
    eprint = "2101.11735",
    archivePrefix = "arXiv",
    primaryClass = "astro-ph.CO",
    month = "1",
    year = "2021"
}

@article{Palenzuela:2017kcg,
      author         = "Palenzuela, C. and Pani, P. and Bezares, M.
                        and Cardoso, V. and Lehner, L. and Liebling, S.",
      title          = "{Gravitational Wave Signatures of Highly Compact Boson
                        Star Binaries}",
      journal        = "Phys. Rev.",
      volume         = "D96",
      year           = "2017",
      number         = "10",
      pages          = "104058",
      doi            = "10.1103/PhysRevD.96.104058",
      eprint         = "1710.09432",
      archivePrefix  = "arXiv",
      primaryClass   = "gr-qc",
      SLACcitation   = "%%CITATION = ARXIV:1710.09432;%%"
}

@article{Eda:2013gg,
      author         = "Eda, K. and Itoh, Y. and Kuroyanagi, S.
                        and Silk, J.",
      title          = "{A new probe of dark matter properties: gravitational
                        waves from an intermediate mass black hole embedded in a
                        dark matter mini-spike}",
      journal        = "Phys. Rev. Lett.",
      volume         = "110",
      pages          = "221101",
      doi            = "10.1103/PhysRevLett.110.221101",
      year           = "2013",
      eprint         = "1301.5971",
      archivePrefix  = "arXiv",
      primaryClass   = "gr-qc",
      SLACcitation   = "%%CITATION = ARXIV:1301.5971;%%",
}

@article{Hannuksela:2018izj,
    author = "Hannuksela, O. A. and Wong, K. W. K. and Brito, R. and Berti, E. and Li, T. G. F.",
    title = "{Probing the existence of ultralight bosons with a single gravitational-wave measurement}",
    eprint = "1804.09659",
    archivePrefix = "arXiv",
    primaryClass = "astro-ph.HE",
    doi = "10.1038/s41550-019-0712-4",
    journal = "Nature Astron.",
    volume = "3",
    number = "5",
    pages = "447--451",
    year = "2019"
}

@article{Baumann:2019ztm,
    author = "Baumann, D. and Chia, H. S. and Porto, R. A. and Stout, J.",
    title = "{Gravitational Collider Physics}",
    eprint = "1912.04932",
    archivePrefix = "arXiv",
    primaryClass = "gr-qc",
    reportNumber = "DESY-19-221, DESY 19-221",
    doi = "10.1103/PhysRevD.101.083019",
    journal = "Phys. Rev.",
    volume = "D101",
    number = "8",
    pages = "083019",
    year = "2020"
}

@article{Kavanagh:2020cfn,
    author = "Kavanagh, B. J. and Nichols, D. A. and Bertone, G. and Gaggero, D.",
    archivePrefix = "arXiv",
    eprint = "2002.12811",
    month = "2",
    primaryClass = "gr-qc",
    title = "{Detecting dark matter around black holes with gravitational waves: Effects of dark-matter dynamics on the gravitational waveform}",
    year = "2020"
}

@article{Babichev:2013usa,
    author = "Babichev, E. and Deffayet, C.",
    archivePrefix = "arXiv",
    doi = "10.1088/0264-9381/30/18/184001",
    eprint = "1304.7240",
    journal = "Class.\ Quant.\ Grav.",
    pages = "184001",
    primaryClass = "gr-qc",
    reportNumber = "LPT-ORSAY-13-45",
    title = "{An introduction to the Vainshtein mechanism}",
    volume = "30",
    year = "2013"
}

@article{Brito:2014ifa,
    author = "Brito, R. and Terrana, A. and Johnson, M. and Cardoso, V.",
    archivePrefix = "arXiv",
    doi = "10.1103/PhysRevD.90.124035",
    eprint = "1409.0886",
    journal = "Phys.\ Rev.",
    pages = "124035",
    primaryClass = "hep-th",
    title = "{Nonlinear dynamical stability of infrared modifications of gravity}",
    volume = "D90",
    year = "2014"
}

@article{Zerilli:1971wd,
    author = "Zerilli, F. J.",
    doi = "10.1103/PhysRevD.2.2141",
    journal = "Phys.\ Rev.",
    pages = "2141--2160",
    title = "{Gravitational field of a particle falling in a schwarzschild geometry analyzed in tensor harmonics}",
    volume = "D2",
    year = "1970"
}

@article{Davis:1971gg,
    author = "Davis, M. and Ruffini, R. and Press, W.H. and Price, R.H.",
    doi = "10.1103/PhysRevLett.27.1466",
    journal = "Phys.\ Rev.\ Lett.",
    pages = "1466--1469",
    title = "{Gravitational radiation from a particle falling radially into a schwarzschild black hole}",
    volume = "27",
    year = "1971"
}

@article{Barack:2018yvs,
    author = "Barack, L. and Pound, A.",
    archivePrefix = "arXiv",
    doi = "10.1088/1361-6633/aae552",
    eprint = "1805.10385",
    journal = "Rept.\ Prog.\ Phys.",
    number = "1",
    pages = "016904",
    primaryClass = "gr-qc",
    title = "{Self-force and radiation reaction in general relativity}",
    volume = "82",
    year = "2019"
}

@Book{Ari,
  title={Seismic Waves and Sources},
  author={Ben-Menahem, A. and Singh, S. J. },
  isbn={ 978-0486404615},
  url={https://www.amazon.com/Seismic-Waves-Sources-Ari-Ben-Menahem/dp/0486404617},
  year={1982},
  publisher={Dover publications}
}

@article{Cardoso:2016oxy,
    author = "Cardoso, V. and Hopper, S. and Macedo, C. F. B. and Palenzuela, C. and Pani, P.",
    title = "{Gravitational-wave signatures of exotic compact objects and of quantum corrections at the horizon scale}",
    eprint = "1608.08637",
    archivePrefix = "arXiv",
    primaryClass = "gr-qc",
    doi = "10.1103/PhysRevD.94.084031",
    journal = "Phys. Rev.",
    volume = "D94",
    number = "8",
    pages = "084031",
    year = "2016"
}

@article{Helfer:2018vtq,
    author = "Helfer, T. and Lim, E. A. and Garcia, M. A.G. and Amin, M. A.",
    title = "{Gravitational Wave Emission from Collisions of Compact Scalar Solitons}",
    eprint = "1802.06733",
    archivePrefix = "arXiv",
    primaryClass = "gr-qc",
    doi = "10.1103/PhysRevD.99.044046",
    journal = "Phys. Rev.",
    volume = "D99",
    number = "4",
    pages = "044046",
    year = "2019"
}

@article{Sanchis-Gual:2019ljs,
    author = "Sanchis-Gual, N. and Di Giovanni, F. and Zilhão, M. and Herdeiro, C. and Cerdá-Durán, P. and Font, J.A. and Radu, E.",
    title = "{Nonlinear Dynamics of Spinning Bosonic Stars: Formation and Stability}",
    eprint = "1907.12565",
    archivePrefix = "arXiv",
    primaryClass = "gr-qc",
    doi = "10.1103/PhysRevLett.123.221101",
    journal = "Phys. Rev. Lett.",
    volume = "123",
    number = "22",
    pages = "221101",
    year = "2019"
}

@article{Bezares:2018qwa,
    author = "Bezares, M. and Palenzuela, C.",
    title = "{Gravitational Waves from Dark Boson Star binary mergers}",
    eprint = "1808.10732",
    archivePrefix = "arXiv",
    primaryClass = "gr-qc",
    doi = "10.1088/1361-6382/aae87c",
    journal = "Class. Quant. Grav.",
    volume = "35",
    number = "23",
    pages = "234002",
    year = "2018"
}

@article{Sanchis-Gual:2018oui,
    author = "Sanchis-Gual, N. and Herdeiro, C. and Font, J. A. and Radu, E. and Di Giovanni, F.",
    title = "{Head-on collisions and orbital mergers of Proca stars}",
    eprint = "1806.07779",
    archivePrefix = "arXiv",
    primaryClass = "gr-qc",
    doi = "10.1103/PhysRevD.99.024017",
    journal = "Phys. Rev.",
    volume = "D99",
    number = "2",
    pages = "024017",
    year = "2019"
}

@article{Widdicombe:2019woy,
    author = "Widdicombe, J. Y. and Helfer, T. and Lim, E. A.",
    title = "{Black hole formation in relativistic Oscillaton collisions}",
    eprint = "1910.01950",
    archivePrefix = "arXiv",
    primaryClass = "astro-ph.CO",
    doi = "10.1088/1475-7516/2020/01/027",
    journal = "JCAP",
    volume = "01",
    pages = "027",
    year = "2020"
}

@ARTICLE{Chavanis1,
       author = {{Chavanis}, P.},
        title = "{Mass-radius relation of Newtonian self-gravitating Bose-Einstein condensates with short-range interactions. I. Analytical results}",
      journal = {Phys. Rev.},
         year = 2011,
        month = aug,
       volume = {D84},
       number = {4},
          eid = {043531},
        pages = {043531},
          doi = {10.1103/PhysRevD.84.043531},
archivePrefix = {arXiv},
       eprint = {1103.2050},
 primaryClass = {astro-ph.CO},
       adsurl = {https://ui.adsabs.harvard.edu/abs/2011PhRvD..84d3531C}
}

@article{Boskovic:2018rub,
    author = "Boskovic, M. and Duque, F. and Ferreira, M. C. and Miguel, F. S. and Cardoso, V.",
    archivePrefix = "arXiv",
    doi = "10.1103/PhysRevD.98.024037",
    eprint = "1806.07331",
    journal = "Phys. Rev.",
    pages = "024037",
    primaryClass = "gr-qc",
    title = "{Motion in time-periodic backgrounds with applications to ultralight dark matter haloes at galactic centers}",
    volume = "D98",
    year = "2018"
}

@article{Membrado,
  title = {Hartree solutions for the self-Yukawian boson sphere},
  author = {Membrado, M. and Pacheco, A. F. and Sa\~nudo, J.},
  journal = {Phys. Rev.},
  volume = {A39},
  issue = {8},
  pages = {4207--4211},
  numpages = {0},
  year = {1989},
  month = {Apr},
  publisher = {American Physical Society},
  doi = {10.1103/PhysRevA.39.4207},
  url = {https://link.aps.org/doi/10.1103/PhysRevA.39.4207}
}

@ARTICLE{Chavanis2,
       author = {{Chavanis}, P. and {Delfini}, L.},
        title = "{Mass-radius relation of Newtonian self-gravitating Bose-Einstein condensates with short-range interactions. II. Numerical results}",
      journal = {Phys. Rev.},
         year = 2011,
        month = aug,
       volume = {D84},
       number = {4},
          eid = {043532},
        pages = {043532},
          doi = {10.1103/PhysRevD.84.043532},
archivePrefix = {arXiv},
       eprint = {1103.2054},
 primaryClass = {astro-ph.CO},
       adsurl = {https://ui.adsabs.harvard.edu/abs/2011PhRvD..84d3532C}
}

@article{Kling:2017mif,
    author = "Kling, F. and Rajaraman, A.",
    archivePrefix = "arXiv",
    doi = "10.1103/PhysRevD.96.044039",
    eprint = "1706.04272",
    journal = "Phys. Rev.",
    number = "4",
    pages = "044039",
    primaryClass = "hep-th",
    reportNumber = "UCI-HEP-TR-2017-04",
    title = "{Towards an Analytic Construction of the Wavefunction of Boson Stars}",
    volume = "D96",
    year = "2017"
}

@article{Guzman:2004wj,
    author = "Guzman, F. Siddhartha. and Urena-Lopez, L. A.",
    title = "{Evolution of the Schrodinger-Newton system for a selfgravitating scalar field}",
    eprint = "gr-qc/0404014",
    archivePrefix = "arXiv",
    doi = "10.1103/PhysRevD.69.124033",
    journal = "Phys. Rev.",
    volume = "D69",
    pages = "124033",
    year = "2004"
}

@article{Guzman:2018bmo,
    author = "Guzman, F.S.",
    title = "{Oscillation modes of ultralight BEC dark matter cores}",
    eprint = "1812.11612",
    archivePrefix = "arXiv",
    primaryClass = "astro-ph.CO",
    doi = "10.1103/PhysRevD.99.083513",
    journal = "Phys. Rev.",
    volume = "D99",
    number = "8",
    pages = "083513",
    year = "2019"
}

@article{Yoshida:1994xi,
    author = "Yoshida, S. and Eriguchi, Y. and Futamase, T.",
    doi = "10.1103/PhysRevD.50.6235",
    journal = "Phys.\ Rev.",
    pages = "6235--6246",
    title = "{Quasinormal modes of boson stars}",
    volume = "D50",
    year = "1994"
}

@article{Kojima:1991np,
    author = "Kojima, Y. and Yoshida, S. and Futamase, T.",
    doi = "10.1143/PTP.86.401",
    journal = "Prog.\ Theor.\ Phys.",
    pages = "401--410",
    title = "{Nonradial pulsation of a boson star. 1: Formulation}",
    volume = "86",
    year = "1991"
}

@article{Macedo:2013jja,
    author = "Macedo, C. F. B. and Pani, P. and Cardoso, V. and Crispino, L. C. B.",
    archivePrefix = "arXiv",
    doi = "10.1103/PhysRevD.88.064046",
    eprint = "1307.4812",
    journal = "Phys.\ Rev.",
    number = "6",
    pages = "064046",
    primaryClass = "gr-qc",
    title = "{Astrophysical signatures of boson stars: quasinormal modes and inspiral resonances}",
    volume = "D88",
    year = "2013"
}

@article{Macedo:2016wgh,
    author = "Macedo, C. F. B. and Cardoso, Vitor. and Crispino, L. C. B. and Pani, P.",
    archivePrefix = "arXiv",
    doi = "10.1103/PhysRevD.93.064053",
    eprint = "1603.02095",
    journal = "Phys.\ Rev.",
    number = "6",
    pages = "064053",
    primaryClass = "gr-qc",
    title = "{Quasinormal modes of relativistic stars and interacting fields}",
    volume = "D93",
    year = "2016"
}

@article{GRITJHU,
journal = "Ringdown data and routines",
year = "2021",
note         = "\noindent\url{http://blackholes.ist.utl.pt/?page=Files} \\
\noindent\url{https://pages.jh.edu/~eberti2/ringdown/}",
}

@article{Kimura:2018eiv,
    author = "Kimura, M. and Tanaka, T.",
    title = "{Robustness of the $S$-deformation method for black hole stability analysis}",
    eprint = "1805.08625",
    archivePrefix = "arXiv",
    primaryClass = "gr-qc",
    reportNumber = "KUNS-2723, YITP-18-52",
    doi = "10.1088/1361-6382/aadc13",
    journal = "Class. Quant. Grav.",
    volume = "35",
    number = "19",
    pages = "195008",
    year = "2018"
}

@article{Kimura:2017uor,
    author = "Kimura, M.",
    title = "{A simple test for stability of black hole by $S$-deformation}",
    eprint = "1706.01447",
    archivePrefix = "arXiv",
    primaryClass = "gr-qc",
    doi = "10.1088/1361-6382/aa903f",
    journal = "Class. Quant. Grav.",
    volume = "34",
    number = "23",
    pages = "235007",
    year = "2017"
}

@article{Guzman:2006yc,
    author = "Guzman, F. S. and Urena-Lopez, L. A",
    title = "{Gravitational cooling of self-gravitating Bose-Condensates}",
    eprint = "astro-ph/0603613",
    archivePrefix = "arXiv",
    doi = "10.1086/504508",
    journal = "Astrophys. J.",
    volume = "645",
    pages = "814--819",
    year = "2006"
}

@article{Seidel1994,
  title = {Formation of solitonic stars through gravitational cooling},
  author = {Seidel, E. and Suen, W.},
  journal = {Phys. Rev. Lett.},
  volume = {72},
  issue = {16},
  pages = {2516--2519},
  numpages = {0},
  year = {1994},
  month = {Apr},
  publisher = {American Physical Society},
  doi = {10.1103/PhysRevLett.72.2516},
  url = {https://link.aps.org/doi/10.1103/PhysRevLett.72.2516}
}

@article{Balakrishna:2006ru,
    author = "Balakrishna, J. and Bondarescu, R. and Daues, G. and S. Guzman, F. and Seidel, E.",
    title = "{Evolution of 3-D boson stars with waveform extraction}",
    eprint = "gr-qc/0602078",
    archivePrefix = "arXiv",
    doi = "10.1088/0264-9381/23/7/024",
    journal = "Class. Quant. Grav.",
    volume = "23",
    pages = "2631--2652",
    year = "2006"
}

@article{Caio:2020comment,
    author = "Macedo, C. F. B.",
    journal = "{private communication}",
    year = "2020"
}

@article{Mendes:2016vdr,
    author = "Mendes, R. F. P. and Yang, H.",
    archivePrefix = "arXiv",
    doi = "10.1088/1361-6382/aa842d",
    eprint = "1606.03035",
    journal = "Class.\ Quant.\ Grav.",
    number = "18",
    pages = "185001",
    primaryClass = "astro-ph.CO",
    title = "{Tidal deformability of boson stars and dark matter clumps}",
    volume = "34",
    year = "2017"
}

@article{Cardoso:2017cfl,
    author = "Cardoso, V. and Franzin, E. and Maselli, A. and Pani, P. and Raposo, G.",
    archivePrefix = "arXiv",
    doi = "10.1103/PhysRevD.95.084014",
    eprint = "1701.01116",
    journal = "Phys.\ Rev.",
    note = "[Addendum: Phys.Rev.D 95, 089901 (2017)]",
    number = "8",
    pages = "084014",
    primaryClass = "gr-qc",
    title = "{Testing strong-field gravity with tidal Love numbers}",
    volume = "D95",
    year = "2017"
}

@article{Sennett:2017etc,
    author = "Sennett, N. and Hinderer, T. and Steinhoff, J. and Buonanno, A. and Ossokine, S.",
    archivePrefix = "arXiv",
    doi = "10.1103/PhysRevD.96.024002",
    eprint = "1704.08651",
    journal = "Phys.\ Rev.",
    number = "2",
    pages = "024002",
    primaryClass = "gr-qc",
    title = "{Distinguishing Boson Stars from Black Holes and Neutron Stars from Tidal Interactions in Inspiraling Binary Systems}",
    volume = "D96",
    year = "2017"
}

@article{Gondolo:1999ef,
    author = "Gondolo, P. and Silk, J.",
    archivePrefix = "arXiv",
    doi = "10.1103/PhysRevLett.83.1719",
    eprint = "astro-ph/9906391",
    journal = "Phys. Rev. Lett.",
    pages = "1719--1722",
    reportNumber = "MPI-PHT-99-10, OUAST-99-9",
    title = "{Dark matter annihilation at the galactic center}",
    volume = "83",
    year = "1999"
}

@article{Sadeghian:2013laa,
    author = "Sadeghian, L. and Ferrer, F. and Will, C. M.",
    archivePrefix = "arXiv",
    doi = "10.1103/PhysRevD.88.063522",
    eprint = "1305.2619",
    journal = "Phys. Rev.",
    number = "6",
    pages = "063522",
    primaryClass = "astro-ph.GA",
    title = "{Dark matter distributions around massive black holes: A general relativistic analysis}",
    volume = "D88",
    year = "2013"
}

@article{Merritt:2002vj,
    author = "Merritt, D. and Milosavljevic, M. and Verde, L. and Jimenez, R.",
    archivePrefix = "arXiv",
    doi = "10.1103/PhysRevLett.88.191301",
    eprint = "astro-ph/0201376",
    journal = "Phys. Rev. Lett.",
    pages = "191301",
    reportNumber = "RUTGERS-AP-337",
    title = "{Dark matter spikes and annihilation radiation from the galactic center}",
    volume = "88",
    year = "2002"
}

@article{Bertone:2005hw,
    author = "Bertone, G. and Merritt, D.",
    archivePrefix = "arXiv",
    doi = "10.1103/PhysRevD.72.103502",
    eprint = "astro-ph/0501555",
    journal = "Phys. Rev. D",
    pages = "103502",
    reportNumber = "FERMILAB-PUB-05-013-A",
    title = "{Time-dependent models for dark matter at the Galactic Center}",
    volume = "72",
    year = "2005"
}

@article{Merritt:2003qk,
    author = "Merritt, D.",
    archivePrefix = "arXiv",
    doi = "10.1103/PhysRevLett.92.201304",
    eprint = "astro-ph/0311594",
    journal = "Phys. Rev. Lett.",
    pages = "201304",
    title = "{Evolution of the dark matter distribution at the galactic center}",
    volume = "92",
    year = "2004"
}

@article{Herdeiro:2015waa,
    author = "Herdeiro, C. A.R. and Radu, E.",
    archivePrefix = "arXiv",
    doi = "10.1142/S0218271815420146",
    eprint = "1504.08209",
    journal = "Int. J. Mod. Phys. D",
    number = "09",
    pages = "1542014",
    primaryClass = "gr-qc",
    title = "{Asymptotically flat black holes with scalar hair: a review}",
    volume = "24",
    year = "2015"
}

@article{Cardoso:2016ryw,
      author         = "Cardoso, V. and Gualtieri, L.",
      title          = "{Testing the black hole ‘no-hair’ hypothesis}",
      journal        = "Class. Quant. Grav.",
      volume         = "33",
      year           = "2016",
      number         = "17",
      pages          = "174001",
      doi            = "10.1088/0264-9381/33/17/174001",
      eprint         = "1607.03133",
      archivePrefix  = "arXiv",
      primaryClass   = "gr-qc",
      SLACcitation   = "%%CITATION = ARXIV:1607.03133;%%"
}

@article{Bamber:2021knr,
    author = "Bamber, J. and Tattersall, O. J. and Clough, K. and Ferreira, P. G.",
    title = "{Quasinormal modes of growing dirty black holes}",
    eprint = "2103.00026",
    archivePrefix = "arXiv",
    primaryClass = "gr-qc",
    doi = "10.1103/PhysRevD.103.124013",
    journal = "Phys. Rev.",
    volume = "D103",
    number = "12",
    pages = "124013",
    year = "2021"
}

@article{Giddings:2008gr,
    author = "Giddings, S. B. and Mangano, M. L.",
    archivePrefix = "arXiv",
    doi = "10.1103/PhysRevD.78.035009",
    eprint = "0806.3381",
    journal = "Phys. Rev.",
    pages = "035009",
    primaryClass = "hep-ph",
    reportNumber = "CERN-PH-TH-2008-025",
    title = "{Astrophysical implications of hypothetical stable TeV-scale black holes}",
    volume = "D78",
    year = "2008"
}

@article{Clough:2019jpm,
    author = "Clough, K. and Ferreira, P. G. and Lagos, M.",
    title = "{Growth of massive scalar hair around a Schwarzschild black hole}",
    eprint = "1904.12783",
    archivePrefix = "arXiv",
    primaryClass = "gr-qc",
    doi = "10.1103/PhysRevD.100.063014",
    journal = "Phys. Rev.",
    volume = "100",
    number = "D6",
    pages = "063014",
    year = "2019"
}

@article{Hui:2019aqm,
    author = "Hui, L. and Kabat, D. and Li, X. and Santoni, L. and Wong, S. S.C.",
    title = "{Black Hole Hair from Scalar Dark Matter}",
    eprint = "1904.12803",
    archivePrefix = "arXiv",
    primaryClass = "gr-qc",
    doi = "10.1088/1475-7516/2019/06/038",
    journal = "JCAP",
    volume = "06",
    pages = "038",
    year = "2019"
}

@article{Detweiler:1980uk,
      author         = "Detweiler, S. L.",
      title          = "{Klein-Gordon equation and rotating black holes}",
      journal        = "Phys. Rev.",
      volume         = "D22",
      year           = "1980",
      pages          = "2323-2326",
      doi            = "10.1103/PhysRevD.22.2323",
      SLACcitation   = "%%CITATION = PHRVA,D22,2323;%%"
}

@article{Lancaster:2019mde,
    author = "Lancaster, L. and Giovanetti, C. and Mocz, P. and Kahn, Y. and Lisanti, M. and Spergel, D. N.",
    title = "{Dynamical Friction in a Fuzzy Dark Matter Universe}",
    eprint = "1909.06381",
    archivePrefix = "arXiv",
    primaryClass = "astro-ph.CO",
    doi = "10.1088/1475-7516/2020/01/001",
    journal = "JCAP",
    volume = "01",
    pages = "001",
    year = "2020"
}

@article{Gualandris:2007nm,
    author = "Gualandris, A. and Merritt, D.",
    title = "{Ejection of Supermassive Black Holes from Galaxy Cores}",
    eprint = "0708.0771",
    archivePrefix = "arXiv",
    primaryClass = "astro-ph",
    doi = "10.1086/586877",
    journal = "Astrophys. J.",
    volume = "678",
    pages = "780",
    year = "2008"
}

@article{Abuter:2018drb,
      author         = "Abuter, R. and others",
      title          = "{Detection of the gravitational redshift in the orbit of
                        the star S2 near the Galactic centre massive black hole}",
      collaboration  = "GRAVITY",
      journal        = "Astron. Astrophys.",
      volume         = "615",
      year           = "2018",
      pages          = "L15",
      doi            = "10.1051/0004-6361/201833718",
      eprint         = "1807.09409",
      archivePrefix  = "arXiv",
      primaryClass   = "astro-ph.GA",
      SLACcitation   = "%%CITATION = ARXIV:1807.09409;%%"
}

@article{Abuter:2020dou,
    author = "Abuter, R. and others",
    collaboration = "GRAVITY",
    title = "{Detection of the Schwarzschild precession in the orbit of the star S2 near the Galactic centre massive black hole}",
    eprint = "2004.07187",
    archivePrefix = "arXiv",
    primaryClass = "astro-ph.GA",
    doi = "10.1051/0004-6361/202037813",
    month = "4",
    year = "2020"
}

@Book{Abramowitz:1970as,
     author    = "Abramowitz, M. and Stegun, I.~A.",
     title     = "Handbook of Mathematical Functions with Formulas, Graphs, and Mathematical Tables",
     address   = "New York",
     publisher = "Dover",
     year      = "1972"
}

@article{Peters:1963ux,
    author = "Peters, P.C. and Mathews, J.",
    doi = "10.1103/PhysRev.131.435",
    journal = "Phys. Rev.",
    pages = "435--439",
    title = "{Gravitational radiation from point masses in a Keplerian orbit}",
    volume = "131",
    year = "1963"
}

@article{Poisson:1993vp,
    author = "Poisson, E.",
    doi = "10.1103/PhysRevD.47.1497",
    journal = "Phys. Rev.",
    pages = "1497--1510",
    title = "{Gravitational radiation from a particle in circular orbit around a black hole. 1: Analytical results for the nonrotating case}",
    volume = "D47",
    year = "1993"
}

@ARTICLE{1989ApJ...345..434T,
       author = {{Taylor}, J.~H. and {Weisberg}, J.~M.},
        title = "{Further Experimental Tests of Relativistic Gravity Using the Binary Pulsar PSR 1913+16}",
      journal = {Astrophys. J.},
         year = 1989,
        month = oct,
       volume = {345},
        pages = {434},
          doi = {10.1086/167917},
       adsurl = {https://ui.adsabs.harvard.edu/abs/1989ApJ...345..434T}
}

@article{Stairs:2003eg,
    author = "Stairs, I. H.",
    title = "{Testing general relativity with pulsar timing}",
    eprint = "astro-ph/0307536",
    archivePrefix = "arXiv",
    doi = "10.12942/lrr-2003-5",
    journal = "Living Rev. Rel.",
    volume = "6",
    pages = "5",
    year = "2003"
}

@article{Flanagan:1997sx,
    author = "Flanagan, E. E. and Hughes, S. A.",
    title = "{Measuring gravitational waves from binary black hole coalescences: 1. Signal-to-noise for inspiral, merger, and ringdown}",
    eprint = "gr-qc/9701039",
    archivePrefix = "arXiv",
    reportNumber = "GRP-456",
    doi = "10.1103/PhysRevD.57.4535",
    journal = "Phys. Rev.",
    volume = "D57",
    pages = "4535--4565",
    year = "1998"
}

@article{Yunes:2016jcc,
    author = "Yunes, N. and Yagi, K. and Pretorius, F.",
    title = "{Theoretical Physics Implications of the Binary Black-Hole Mergers GW150914 and GW151226}",
    eprint = "1603.08955",
    archivePrefix = "arXiv",
    primaryClass = "gr-qc",
    doi = "10.1103/PhysRevD.94.084002",
    journal = "Phys. Rev.",
    volume = "D94",
    number = "8",
    pages = "084002",
    year = "2016"
}

@article{Ioannidou,
author = {Ioannidou, T. A.  and Kouiroukidis, A.  and Vlachos, N. D.},
title = {Universality in a class of Q-ball solutions: An analytic approach},
journal = {J. Math. Phys.},
volume = {46},
number = {4},
pages = {042306},
year = {2005},
doi = {10.1063/1.1851972},
URL = { 
        https://doi.org/10.1063/1.1851972  
},
eprint = { 
        https://doi.org/10.1063/1.1851972}
}

@article{Tsumagari2008,
  title = {Some stationary properties of a $Q$-ball in arbitrary space dimensions},
  author = {Tsumagari, M. I. and Copeland, E. J. and Saffin, P. M.},
  journal = {Phys. Rev.},
  volume = {D78},
  issue = {6},
  pages = {065021},
  numpages = {21},
  year = {2008},
  month = {Sep},
  publisher = {American Physical Society},
  doi = {10.1103/PhysRevD.78.065021},
  url = {https://link.aps.org/doi/10.1103/PhysRevD.78.065021}
}

@article{Cardoso:2007uy,
    author = "Cardoso, V. and Cavaglia, M. and Guo, J.",
    archivePrefix = "arXiv",
    doi = "10.1103/PhysRevD.75.084020",
    eprint = "hep-th/0702138",
    journal = "Phys.\ Rev.",
    pages = "084020",
    title = "{Gravitational Larmor formula in higher dimensions}",
    volume = "D75",
    year = "2007"
}

@article{Misner:1972jf,
      author         = "Misner, C. W. and Breuer, R. A. and Brill, D. R. and
                        Chrzanowski, P. L. and Hughes, H. G. and Pereira, C. M.",
      title          = "{Gravitational synchrotron radiation in the schwarzschild
                        geometry}",
      journal        = "Phys. Rev. Lett.",
      volume         = "28",
      year           = "1972",
      pages          = "998-1001",
      doi            = "10.1103/PhysRevLett.28.998",
      SLACcitation   = "%%CITATION = PRLTA,28,998;%%"
}

@Book{Breuer,
  title={Gravitational Perturbation Theory and Synchrotron Radiation},
  author={Breuer, R. A. },
  isbn={978-3-540-07530-1},
  url={https://www.springer.com/gp/book/9783540075301},
  year={1975},
  publisher={Springer}
}

@article{Campanelli:2007ew,
      author         = "Campanelli, M. and Lousto, C. O. and Zlochower,
                        Y. and Merritt, D.",
      title          = "{Large merger recoils and spin flips from generic
                        black-hole binaries}",
      journal        = "Astrophys. J.",
      volume         = "659",
      year           = "2007",
      pages          = "L5-L8",
      doi            = "10.1086/516712",
      eprint         = "gr-qc/0701164",
      archivePrefix  = "arXiv",
      primaryClass   = "gr-qc",
      SLACcitation   = "%%CITATION = GR-QC/0701164;%%"
}

@article{Brugmann:2007zj,
      author         = "Bruegmann, B. and Gonzalez, J. A. and Hannam, M.
                        and Husa, S. and Sperhake, U.",
      title          = "{Exploring black hole superkicks}",
      journal        = "Phys. Rev.",
      volume         = "D77",
      year           = "2008",
      pages          = "124047",
      doi            = "10.1103/PhysRevD.77.124047",
      eprint         = "0707.0135",
      archivePrefix  = "arXiv",
      primaryClass   = "gr-qc",
      SLACcitation   = "%%CITATION = ARXIV:0707.0135;%%"
}

@article{Gonzalez:2007hi,
      author         = "Gonzalez, J. A. and Hannam, M. D. and Sperhake, U. and
                        Bruegmann, B. and Husa, S.",
      title          = "{Supermassive recoil velocities for binary black-hole
                        mergers with antialigned spins}",
      journal        = "Phys. Rev. Lett.",
      volume         = "98",
      year           = "2007",
      pages          = "231101",
      doi            = "10.1103/PhysRevLett.98.231101",
      eprint         = "gr-qc/0702052",
      archivePrefix  = "arXiv",
      primaryClass   = "GR-QC",
      SLACcitation   = "%%CITATION = GR-QC/0702052;%%"
}

@article{Campanelli:2007cga,
      author         = "Campanelli, M. and Lousto, C. O. and Zlochower,
                        Y. and Merritt, D.",
      title          = "{Maximum gravitational recoil}",
      journal        = "Phys. Rev. Lett.",
      volume         = "98",
      year           = "2007",
      pages          = "231102",
      doi            = "10.1103/PhysRevLett.98.231102",
      eprint         = "gr-qc/0702133",
      archivePrefix  = "arXiv",
      primaryClass   = "GR-QC",
      SLACcitation   = "%%CITATION = GR-QC/0702133;%%"
}

@article{Sperhake:2010uv,
      author         = "Sperhake, U. and Berti, E. and Cardoso, V.
                        and Pretorius, F. and Yunes, N.",
      title          = "{Superkicks in ultrarelativistic encounters of spinning
                        black holes}",
      journal        = "Phys. Rev.",
      volume         = "D83",
      year           = "2011",
      pages          = "024037",
      doi            = "10.1103/PhysRevD.83.024037",
      eprint         = "1011.3281",
      archivePrefix  = "arXiv",
      primaryClass   = "gr-qc",
      SLACcitation   = "%%CITATION = ARXIV:1011.3281;%%"
}

@misc{Dolan:private,
note = {I thank Sam Dolan for pointing this parallel to my collaborators and me.}
}

@misc{Beckmann,
note = {\url{https://eud.gsfc.nasa.gov/Volker.Beckmann/school/download/Longair_Radiation3.pdf}}
}

@article{Baibhav:2019rsa,
    author = "Baibhav, V. and others",
    title = "{Probing the Nature of Black Holes: Deep in the mHz Gravitational-Wave Sky}",
    eprint = "1908.11390",
    archivePrefix = "arXiv",
    primaryClass = "astro-ph.HE",
    month = "8",
    year = "2019"
}

@article{Bertone:2004pz,
    author = "Bertone, G. and Hooper, D. and Silk, J.",
    title = "{Particle dark matter: Evidence, candidates and constraints}",
    eprint = "hep-ph/0404175",
    archivePrefix = "arXiv",
    reportNumber = "FERMILAB-PUB-04-047-A",
    doi = "10.1016/j.physrep.2004.08.031",
    journal = "Phys. Rept.",
    volume = "405",
    pages = "279--390",
    year = "2005"
}

@article{PecceiQuinn,
  title = "$\mathrm{CP}$ Conservation in the Presence of Pseudoparticles",
  author = {Peccei, R. D. and Quinn, H. R.},
  journal = {Phys. Rev. Lett.},
  volume = {38},
  issue = {25},
  pages = {1440--1443},
  numpages = {0},
  year = {1977},
  month = {Jun},
  publisher = {American Physical Society},
  doi = {10.1103/PhysRevLett.38.1440},
  url = {https://link.aps.org/doi/10.1103/PhysRevLett.38.1440}
}

@article{PecceiQuinn1,
  title = "Constraints imposed by $\mathrm{CP}$ conservation in the presence of pseudoparticles",
  author = {Peccei, R. D. and Quinn, H. R.},
  journal = {Phys. Rev.},
  volume = {D16},
  issue = {6},
  pages = {1791--1797},
  numpages = {0},
  year = {1977},
  month = {Sep},
  publisher = {American Physical Society},
  doi = {10.1103/PhysRevD.16.1791},
  url = {https://link.aps.org/doi/10.1103/PhysRevD.16.1791}
}

@ARTICLE{Peebles,
       author = {{Peebles}, P.~J.~E.},
        title = "{Large-scale background temperature and mass fluctuations due to scale-invariant primeval perturbations}",
      journal = {{Astrophys. J. Lett}},
         year = 1982,
        month = dec,
       volume = {263},
        pages = {L1-L5},
          doi = {10.1086/183911},
       adsurl = {https://ui.adsabs.harvard.edu/abs/1982ApJ...263L...1P}
}

@article{Blumenthal:1984bp,
    author = "Blumenthal, G. R. and Faber, S. M. and Primack, J. R. and Rees, M. J.",
    editor = "Srednicki, M.A.",
    title = "{Formation of Galaxies and Large Scale Structure with Cold Dark Matter}",
    reportNumber = "SLAC-PUB-3307",
    doi = "10.1038/311517a0",
    journal = "Nature",
    volume = "311",
    pages = "517--525",
    year = "1984"
}

@article{Weinberg:2013aya,
    author = "Weinberg, D. H. and Bullock, J. S. and Governato, F. and Kuzio de Naray, R. and Peter, A. H. G.",
    title = "{Cold dark matter: controversies on small scales}",
    eprint = "1306.0913",
    archivePrefix = "arXiv",
    primaryClass = "astro-ph.CO",
    doi = "10.1073/pnas.1308716112",
    journal = "Proc. Nat. Acad. Sci.",
    volume = "112",
    pages = "12249--12255",
    year = "2015"
}

@article{Flores:1994gz,
    author = "Flores, R. A. and Primack, J. R.",
    title = "{Observational and theoretical constraints on singular dark matter halos}",
    eprint = "astro-ph/9402004",
    archivePrefix = "arXiv",
    reportNumber = "SCIPP-93-01-REV, SCIPP-93-01",
    doi = "10.1086/187350",
    journal = "Astrophys. J. Lett.",
    volume = "427",
    pages = "L1--4",
    year = "1994"
}

@article{Moore:1999gc,
    author = "Moore, B. and Quinn, T. R. and Governato, F. and Stadel, J. and Lake, G.",
    title = "{Cold collapse and the core catastrophe}",
    eprint = "astro-ph/9903164",
    archivePrefix = "arXiv",
    doi = "10.1046/j.1365-8711.1999.03039.x",
    journal = "Mon. Not. Roy. Astron. Soc.",
    volume = "310",
    pages = "1147--1152",
    year = "1999"
}

@article{Navarro:1996gj,
    author = "Navarro, J. F. and Frenk, C. S. and White, S. D. M.",
    title = "{A Universal density profile from hierarchical clustering}",
    eprint = "astro-ph/9611107",
    archivePrefix = "arXiv",
    doi = "10.1086/304888",
    journal = "Astrophys. J.",
    volume = "490",
    pages = "493--508",
    year = "1997"
}

@article{Del_Popolo_2017,
   title={Small Scale Problems of the $\Lambda$CDM Model: A Short Review},
   volume={5},
   ISSN={2075-4434},
   url={http://dx.doi.org/10.3390/galaxies5010017},
   DOI={10.3390/galaxies5010017},
   number={1},
   journal={Galaxies},
   publisher={MDPI AG},
   author={Del Popolo, A. and Le Delliou, M.},
   year={2017},
   month={Feb},
   pages={17}
}

@article{Press:1989id,
    author = "Press, W. H. and Ryden, B. S. and Spergel, D. N.",
    title = "{Single Mechanism for Generating Large Scale Structure and Providing Dark Missing Matter}",
    reportNumber = "CFA-3031",
    doi = "10.1103/PhysRevLett.64.1084",
    journal = "Phys. Rev. Lett.",
    volume = "64",
    pages = "1084",
    year = "1990"
}

@article{Sin:1992bg,
    author = "Sin, S.",
    title = "{Late time cosmological phase transition and galactic halo as Bose liquid}",
    eprint = "hep-ph/9205208",
    archivePrefix = "arXiv",
    reportNumber = "UFIFT-HEP-92-11",
    doi = "10.1103/PhysRevD.50.3650",
    journal = "Phys. Rev.",
    volume = "D50",
    pages = "3650--3654",
    year = "1994"
}

@article{Hu:2000ke,
    author = "Hu, W. and Barkana, R. and Gruzinov, A.",
    title = "{Cold and fuzzy dark matter}",
    eprint = "astro-ph/0003365",
    archivePrefix = "arXiv",
    doi = "10.1103/PhysRevLett.85.1158",
    journal = "Phys. Rev. Lett.",
    volume = "85",
    pages = "1158--1161",
    year = "2000"
}

@article{Marsh:2015xka,
    author = "Marsh, D. J. E.",
    title = "{Axion Cosmology}",
    eprint = "1510.07633",
    archivePrefix = "arXiv",
    primaryClass = "astro-ph.CO",
    reportNumber = "KCL-PH-TH-2015-50",
    doi = "10.1016/j.physrep.2016.06.005",
    journal = "Phys. Rept.",
    volume = "643",
    pages = "1--79",
    year = "2016"
}

@article{Arvanitaki:2009fg,
    author = "Arvanitaki, A. and Dimopoulos, S. and Dubovsky, S. and Kaloper, N. and March-Russell, J.",
    title = "{String Axiverse}",
    eprint = "0905.4720",
    archivePrefix = "arXiv",
    primaryClass = "hep-th",
    doi = "10.1103/PhysRevD.81.123530",
    journal = "Phys. Rev.",
    volume = "D81",
    pages = "123530",
    year = "2010"
}

@article{Freitas:2021cfi,
    author = "Freitas, F. F. and Herdeiro, C. A. R. and Morais, A. P. and Onofre, A. and Pasechnik, R. and Radu, E. and Sanchis-Gual, N. and Santos, R.",
    title = "{Ultralight bosons for strong gravity applications from simple Standard Model extensions}",
    eprint = "2107.09493",
    archivePrefix = "arXiv",
    primaryClass = "hep-ph",
    month = "7",
    year = "2021"
}

@article{Bovy:2012tw,
    author = "Bovy, J. and Tremaine, S.",
    title = "{On the local dark matter density}",
    eprint = "1205.4033",
    archivePrefix = "arXiv",
    primaryClass = "astro-ph.GA",
    doi = "10.1088/0004-637X/756/1/89",
    journal = "Astrophys. J.",
    volume = "756",
    pages = "89",
    year = "2012"
}

@article{Sivertsson:2017rkp,
    author = "Sivertsson, S. and Silverwood, H. and Read, J. I. and Bertone, G. and Steger, P.",
    title = "{The localdark matter density from SDSS-SEGUE G-dwarfs}",
    eprint = "1708.07836",
    archivePrefix = "arXiv",
    primaryClass = "astro-ph.GA",
    doi = "10.1093/mnras/sty977",
    journal = "Mon. Not. Roy. Astron. Soc.",
    volume = "478",
    number = "2",
    pages = "1677--1693",
    year = "2018"
}

@article{McKee_2015,
   title={Stars, gas, and dark matter in the solar neighborhood},
   volume={814},
   ISSN={1538-4357},
   url={http://dx.doi.org/10.1088/0004-637X/814/1/13},
   DOI={10.1088/0004-637x/814/1/13},
   number={1},
   journal={The Astrophysical Journal},
   publisher={American Astronomical Society},
   author={McKee, C. F. and Parravano, A. and Hollenbach, D. J.},
   year={2015},
   month={Nov},
   pages={13}
}

@article{Copeland:1995fq,
    author = "Copeland, E. J. and Gleiser, M. and Muller, H. R.",
    title = "{Oscillons: Resonant configurations during bubble collapse}",
    eprint = "hep-ph/9503217",
    archivePrefix = "arXiv",
    reportNumber = "SUSX-TH-95-3-3, FERMILAB-PUB-95-021-A, DART-HEP-95-01",
    doi = "10.1103/PhysRevD.52.1920",
    journal = "Phys. Rev.",
    volume = "D52",
    pages = "1920--1933",
    year = "1995"
}

@article{Khmelnitsky:2013lxt,
    author = "Khmelnitsky, A. and Rubakov, V.",
    title = "{Pulsar timing signal from ultralight scalar dark matter}",
    eprint = "1309.5888",
    archivePrefix = "arXiv",
    primaryClass = "astro-ph.CO",
    reportNumber = "LMU-ASC-66-13, INR-TH-2013-26",
    doi = "10.1088/1475-7516/2014/02/019",
    journal = "JCAP",
    volume = "02",
    pages = "019",
    year = "2014"
}

@ARTICLE{zeldovich,
       author = {{Zel'Dovich}, Y. B.},
        title = "{Generation of Waves by a Rotating Body}",
      journal = {JETP Lett.},
         year = 1971,
        month = aug,
       volume = {14},
        pages = {180},
       adsurl = {https://ui.adsabs.harvard.edu/abs/1971JETPL..14..180Z}
}

@article{Misner:1972kx,
    author = "Misner, C. W.",
    title = "{Interpretation of gravitational-wave observations}",
    doi = "10.1103/PhysRevLett.28.994",
    journal = "Phys. Rev. Lett.",
    volume = "28",
    pages = "994--997",
    year = "1972"
}

@article{Baumann:2019eav,
    author = "Baumann, D. and Chia, H. S. and Stout, J. and ter Haar, L.",
    title = "{The Spectra of Gravitational Atoms}",
    eprint = "1908.10370",
    archivePrefix = "arXiv",
    primaryClass = "gr-qc",
    doi = "10.1088/1475-7516/2019/12/006",
    journal = "JCAP",
    volume = "12",
    pages = "006",
    year = "2019"
}

@article{Arvanitaki:2010sy,
    author = "Arvanitaki, A. and Dubovsky, S.",
    title = "{Exploring the String Axiverse with Precision Black Hole Physics}",
    eprint = "1004.3558",
    archivePrefix = "arXiv",
    primaryClass = "hep-th",
    doi = "10.1103/PhysRevD.83.044026",
    journal = "Phys. Rev.",
    volume = "D83",
    pages = "044026",
    year = "2011"
}

@article{Brito:2014wla,
    author = "Brito, R. and Cardoso, V. and Pani, P.",
    title = "{Black holes as particle detectors: evolution of superradiant instabilities}",
    eprint = "1411.0686",
    archivePrefix = "arXiv",
    primaryClass = "gr-qc",
    doi = "10.1088/0264-9381/32/13/134001",
    journal = "Class. Quant. Grav.",
    volume = "32",
    number = "13",
    pages = "134001",
    year = "2015"
}

% Old version, will be removed later
% work-around to have small caps also here in the headline
%\manualmark
%\markboth{\spacedlowsmallcaps{\bibname}}{\spacedlowsmallcaps{\bibname}} % work-around to have small caps also
%\phantomsection
%\refstepcounter{dummy}
%\addtocontents{toc}{\protect\vspace{\beforebibskip}} % to have the bib a bit from the rest in the toc
%\addcontentsline{toc}{chapter}{\tocEntry{\bibname}}
%\label{app:bibliography}
%\printbibliography

%\cleardoublepage\include{FrontBackmatter/Declaration}
\cleardoublepage\pagestyle{empty}

\hfill

\vfill

\pdfbookmark[0]{Colophon}{colophon}
\section*{Colophon}
Most symbolic and numerical calculations in this thesis were carried out in \texttt{Mathematica}. The same software with the package \texttt{MaTex} was used to build the plots.
This document was typeset using the typographical look-and-feel \texttt{classicthesis} developed by Andr\'e Miede and Ivo Pletikosić.
The style was inspired by Robert Bringhurst's seminal book on typography ``\emph{The Elements of Typographic Style}''.

\bigskip

\noindent\finalVersionString

\end{document}